\documentclass[11pt,a4paper]{cernrep}
\pdfoutput=1
\usepackage{graphicx}
\usepackage{overpic}
\usepackage{hyperref}
\usepackage{cite}
\usepackage{hepunits}
\usepackage{xspace}
\usepackage{amsmath,amssymb,bm}
\usepackage{rotating}
\usepackage{relsize}
\usepackage{booktabs}
\usepackage[labelfont=bf]{caption}
\usepackage{subfigure}
\usepackage{sidecap}
\usepackage{fancyvrb}
\usepackage{axodraw}
\usepackage[utf8]{inputenc}
\usepackage{textcomp}

\input{lhdefs}

\begin{document}

\title{\centerline{THE TOOLS AND MONTE CARLO WORKING GROUP}\centerline{Summary Report}}

\author{
J.~Alwall$^{1}$,
A.~Arbey$^{3}$, 
L.~Basso$^{4,5}$, 
S.~Belov$^{6}$,
A.~Bharucha$^{7}$, 
F.~Braam$^{9}$,  
A.~Buckley$^{8}$,
J.~M.~Butterworth$^{10*}$,
M.~Campanelli$^{10}$, 
R.~Chierici$^{12}$,
A.~Djouadi$^{15}$, 
L.~Dudko$^{16}$,
C.~Duhr$^{7}$,    
F.~Febres~Cordero$^{17}$,
P.~Francavilla$^{18}$,
B.~Fuks$^{19}$, 
L.~Garren$^{11}$,
T.~Goto$^{20}$, 
M.~Grazzini$^{21,22}$  
T.~Hahn$^{23}$,    
U.~Haisch$^{24}$, 
K.~Hamilton$^{25}$,
S.~Heinemeyer$^{26}$,
G.~Hesketh$^{2}$,
S.~H{\"o}che$^{27}$,
H.~Hoeth$^{7}$,
J.~Huston$^{28}$,
J.~Kalinowski$^{29}$,
D.~Kekelidze$^{6}$,
S.~Kraml$^{30}$,
H.~Lacker$^{31}$,
P.~Lenzi$^{13}$,
P.~Loch$^{32}$,
L.~L\"onnblad$^{33}$,
F.~Mahmoudi$^{34}$, 
E.~Maina$^{35,36}$,
D.~Majumder$^{37}$, 
F.~Maltoni$^{38*}$, 
M.~Mangano$^{2}$,
K.~Mazumdar$^{37}$, 
A.~Martin$^{11,39}$,
J.~Monk$^{10}$,
F.~Moortgat$^{21*}$,
M.~Muhlleitner$^{40}$,
C.~Oleari$^{41}$,
S.~Ovyn$^{38}$, 
R.~Pittau$^{42}$, 
S.~Pl\"atzer$^{40}$,
G. Piacquadio$^{2,9}$,
L.~Reina$^{43}$,
J.~Reuter$^{9}$,  
P.~Richardson$^{7*}$,
X.~Rouby$^{38}$,
C.~Robinson$^{10}$,
T.~Roy$^{44}$,
M.~D.~Schwartz$^{45}$,
H.~Schulz$^{31}$,
S.~Schumann$^{46*}$,
E.~von~Seggern$^{31}$, 
A.~Sherstnev$^{16,48}$ 
F.~Siegert$^{7,10}$,
T.~Sj\"ostrand~$^{33}$,
P.~Skands$^{2*}$,
P.~Slavich$^{49}$,
M.~Spira$^{50}$,  
C.~Taylor$^{10}$,
M.~Vesterinen$^{54}$, 
S.~de Visscher$^{27}$,
D.~Wackeroth$^{51}$,
S.~Weinzierl$^{52}$, 
J.~Winter$^{53}$, 
T.~R.~Wyatt$^{54}$
}
\institute{$^*$ Session converners.
\\$^{1}$SLAC National Accelerator Laboratory, Theoretical Physics Group, Mail Stop 81, 2575 Sand Hill Road, Menlo Park, CA 94025, USA
\\$^{2}$ CERN, CH-1211 Geneva 23, Switzerland
\\$^{3}$Universit\'e de Lyon, France; Universit\'e Lyon 1, F--69622; CRAL, Observatoire de Lyon, F--69561 Saint-Genis-Laval; CNRS, UMR 5574; ENS de Lyon, France
\\$^{4}$ School of Physics \& Astronomy, University of Southampton, Highfield, Southampton SO17 1BJ, UK
\\$^{5}$ Particle Physics Department, Rutherford Appleton Laboratory, Chilton, Didcot, Oxon OX11 0QX, UK
\\$^{6}$ Joint Institute for Nuclear Research, Dubna, Moscow region, Russia, 141980
\\$^{7}$ IPPP, Physics Department, Durham University, DH1 3LE, UK
\\$^8$ School of Physics and Astronomy, University of Edinburgh, EH9 3JZ, UK
\\$^9$ University of Freiburg, Institute of Physics, Hermann-Herder-Str. 3, 79104 Freiburg, Germany
\\$^{10}$ Department of Physics and Astronomy, University College London, WC1E 6BT, UK
\\$^{11}$ Fermi National Accelerator Laboratory, P.O Box 500, Batavia, IL 60510, USA
\\$^{12}$Institut de Physique Nucleaire de Lyon, IN2P3-CNRS, Universit\'e Claude Bernard Lyon 1, Villeurbanne, France
\\$^{13}$ Universit\`a degli Studi di Firenze \& INFN Firenze, via Sansone 1, 50019 Sesto F.no, Firenze, Italy
\\$^{14}$ Department of Physics and Astronomy, UCLA, Los Angeles, CA 90095-1547, USA
\\$^{15}$Laboratoire de Physique Th\'eorique, Universit\'e Paris XI, F--91405 Orsay Cedex, France
\\$^{16}$Skobeltsyn Institute of Nuclear Physics, Lomonosov Moscow State University, Vorob'evy Gory, Moscow 119992, Russia
\\$^{17}$ Universidad Sim\'on Bol\'{\i}var, Departamento de F\'{\i}sica, Apartado 89000, Caracas 1080A, Venezuela.
\\$^{18}$INFN - Pisa, Universita' di Pisa, G. Galilei Graduate School, Pisa, Italy
\\$^{19}$ Institut Pluridisciplinaire Hubert Curien/D\'epartement Recherche Subatomique, Universit\'e de Strasbourg/CNRS-IN2P3, 23 Rue du Loess, F-67037 Strasbourg, France
\\$^{20}$KEK Theory Center, Institute of Particle and Nuclear Studies, KEK, Tsukuba, 305-0801 Japan 
\\$^{21}$Dept. of Physics, ETH Z\"urich, Z\"urich, Switzerland 
\\$^{22}$ INFN, Sezione di Firenze, 50019 Firenze, Italy.
\\$^{23}$Max-Planck-Institut f\"ur Physik, F\"ohringer Ring 6, D--80805 Munich, Germany
\\$^{24}$Institut f\"ur Physik (WA THEP), Johannes Gutenberg-Universit\"at, D--55099 Mainz, Germany
\\$^{25}$ INFN, Sezione di Milano-Bicocca, 20126 Milan, Italy.
\\$^{26}$Instituto de F\'isica de Cantabria (CSIC-UC), Santander, Spain
\\$^{27}$ Universit{\"a}t Z{\"u}rich, CH-8057 Z{\"u}rich, Switzerland
\\$^{28}$Dept. of Physics and Astronomy, Michigan State University, East Lansing (MI), USA
\\$^{29}$ Instytut Fizyki Teoretycznej UW, Hoza 69, PL-00681 Warsaw, Poland
\\$^{30}$Laboratoire de Physique Subatomique et de Cosmologie (LPSC), UJF Grenoble 1, CNRS/IN2P3, 53 Avenue des Martyrs, 38026 Grenoble, France
\\$^{31}$ Inst. f. Physik, Humboldt-Universit\"at zu Berlin, Berlin, Germany
\\$^{32}$Department of Physics, University of Arizona, Tucson, Arizona, USA
\\$^{33}$Department of Theoretical Physics, Lund University, S\"olvegatan14A, S-223~62, Sweden.
\\$^{34}$Clermont Universit\'e, Universit\'e Blaise Pascal, CNRS/IN2P3, LPC, BP 10448, 63000 Clermont-Ferrand, France
\\$^{35}$Dipartimento di Fisica Teorica, Universita' di Torino, Via Giuria 1, 10125 Torino, Italy
\\$^{36}$ INFN, Sezione di Torino, Via Giuria 1, 10125 Torino, Italy.
\\$^{37}$Tata Institute of Fundamental Research, Homi Bhabha Road, Mumbai 400 005, India.
\\$^{38}$Centre for Particle Physics and Phenomenology (CP3), Universit\'{e} Catholique de Louvain, 
Chemin du Cyclotron 2, B--1348 Louvain-la-Neuve, Belgium
\\$^{39}$ Department of Physics, Sloane Laboratory, Yale University, New Haven, CT 06520 USA
\\$^{40}$ Institut f\"ur Theoretische Physik, KIT, 76128 Karlsruhe, Germany
\\$^{41}$ Universit\`a di Milano-Bicocca, 20126 Milano, Italy.
\\$^{42}$Departamento de Física Teórica y del Cosmos Campus Fuentenueva, Universidad de Granada E-18071 Granada, Spain
\\$^{43}$ Florida State University
\\$^{44}$ Department of Physics and Institute of Theoretical Science, University of Oregon, Eugene, OR 97403 USA
\\$^{45}$Department of Physics, Harvard University, Cambridge, MA, USA
\\$^{46}$ Institut f{\"u}r Theoretische Physik, Universit{\"a}t Heidelberg, Philosophenweg 16, D-69120 Heidelberg, Germany
\\$^{48}$ R.~Peierls Centre for Theoretical Physics, University of Oxford, OX1 3NP, UK
\\$^{49}$LPTHE, 4, Place Jussieu, 75252 Paris, France
\\$^{50}$ Paul Scherrer Institut, CH--5232 Villigen PSI, Switzerland.
\\$^{51}$ University at Buffalo, SUNY
\\$^{52}$ Institut f{\"u}r Physik, Universit{\"a}t Mainz, D - 55099 Mainz, Germany
\\$^{53}$Theoretical Physics Department, Fermi National Accelerator Laboratory, Batavia, IL 60510, USA
\\$^{54}$Particle Physics Group, School of Physics and Astronomy, University of Manchester, UK.
}

\maketitle

\begin{abstract}
This is the summary and introduction to the proceedings contributions for the Les Houches 2009 
``Tools and Monte Carlo'' working group.
\end{abstract}

\setcounter{tocdepth}{1}
\tableofcontents
\setcounter{footnote}{0}

\section{FOREWORD}

The working group on ``Tools'' and Monte Carlos for TeV-scale physics held discussions throughout the 
two-week period of the Les Houches meetings. The topics covered herein span both sessions. Several of the 
topics followed on from those discussed in the ``Standard Model Handles and Candles'' session of the 
previous workshop~\cite{Buttar:2008jx}; there has been substantial progress and several new topics were 
introduced.

The contributions here in fact include substantial physics results derived from the programmes, interfaces and 
techniques discussed, as well as status reports on existing projects and proposals for new standards and 
interfaces.

In Part I we have collected the more technical proposals for common standards and interfaces, mostly required 
because of the rapid progress in higher order calculations. Part II collects results on the tuning of MC simulations,
a critical topic for understanding LHC data. Part III contains several contributions comparing various all-order 
calculations with fixed-order results, with and without matching between the two. Part IV reflects some key issues on 
the communicability of results between experiment and theory. Part V discusses recent progress and ideas in using
jets and jet substructure at the LHC to study QCD and search for new physics, and finally Part VI discusses progress
in some key modeling tools for beyond-the-standard-model physics.

The productivity and pleasure of this workshop is overshadowed by the dreadful loss of our friend and colleague
Thomas Binoth, and we dedicate these proceedings to him.

\clearpage

\part{INTERFACES}


\section[A STANDARD FORMAT FOR LES HOUCHES EVENT FILES, VERSION 2]
{A STANDARD FORMAT FOR LES HOUCHES EVENT FILES, VERSION 2}\protect\footnote{Contributed by L.~L\"onnblad Email
  \protect\href{mailto:leif.lonnblad@thep.lu.se}{\protect\texttt{leif.lonnblad@thep.lu.se}},
  J.~Alwall, S.~Belov, L.~Dudko, L.~Garren, K.~Hamilton, J.~Huston,
  D.~Kekelidze, E.~Maina, F.~Maltoni, M.~Mangano, R.~Pittau,
  S.~Pl\"atzer, A.~Sherstnev, T.~Sj\"ostrand, P.~Skands.}
\label{sec:LHEF2}



\subsection{INTRODUCTION}

The Les Houches Accord (LHA) for user-defined processes
\cite{Boos:2001cv} has been immensely successful. It is routinely used
to pass information from matrix-element-based generators (MEGs) to
general-purpose ones (here, somewhat unfairly referred to as parton
shower generators --- PSGs), in order to generate complete events for
a multitude of processes. The original standard was in terms of two
\textit{Fortran common blocks} where information could be stored, while
the actual usage has tended to be mainly in terms of \textit{files}
with parton-level events. For this purpose a new accord --- the Les
Houches Event File (LHEF) accord\cite{Alwall:2006yp} --- was
introduced in 2006, which standardized the way such event files should
be structured.

The LHEF was constructed using XML tags in order to make it flexible
and easy to extend (although some additional structure is assumed
inside some tags which is not formulated in XML). The format has been
extremely useful, and has basically become the standard way to
interface matrix element generators and parton shower programs.

As the matching and merging of tree-level matrix elements and parton
showers are now becoming the state-of-art, it is reasonable to let
this be reflected in an updated file format to standardize how
relevant information should be given by the matrix element generators
in a usable fashion for the parton shower programs. Furthermore, with
the matching of next-to-leading order (NLO) calculations and parton
showers becoming increasingly widespread, it is worth considering how
the LHEF can be augmented in order to facilitate this.

For the CKKW-type merging algorithms
\cite{Catani:2001cc,Lonnblad:2001iq} it is convenient to allow the
Sudakov-reweighting to be done in the MEG, as this will automatically
regularize soft- and collinear divergencies. Hence it would be
desirable if the LHEF could include information about this. This does
not only mean that a weight needs to be added, but also information
about which cuts has been imposed in the MEG as well as information on
how the generated event was clustered to obtain the relevant scales
and Sudakov form factors.

In the case the events are produced by a NLO MEG, the situation is a
bit more complicated. Here a subtraction scheme is typically used to
handle the cancellation between real and virtual corrections. This
means that, besides loop-level events, each tree-level real event with
one extra parton will need to be supplemented by \textit{counter
  events} corresponding to the assumed projections of the tree-level
event to born-level events with one parton less. To allow for matching
or merging with a PSG, these events need to be considered together in
a group of events, something that was not forseen in the original
file format.

Independent of these ME-PS matching considerations, we also wish to
introduce some further, minor, additions to assist the determination
of errors arising from the the parton density function (PDF)
parameterizations used in the MEG. Normally these error estimates are
given as a set of different PDFs where the parameters have been varied
around the best fit value. Hence, a given event may be associated with
several weights corresponding to the different PDFs used.

Note that the scope of the format suggested here is somewhat different
from the HepML schema \cite{Belov:2010xm} (used by eg.\ the MCDB
project \cite{Belov:2007qg}). The LHEF format is specialized in the
interface between matrix element generators and parton shower
programs, while HepML is intended to give more general
meta-information on how events have been produced by an event
generator. However, there is nothing that prevents the LHEF format to
be included in the HepML structure in the future.

The outline of this article is as follows. In section \ref{sec:recap}
we review the structure of the original LHEF accord and of the Les
Houches common block structure on which it is based. Then in
section~\ref{sec:new-file-format} we present the additional XML tags
which may be used to specify additional global, and per-event
information. Finally we give a brief summary and outlook.

\begin{figure}
  \centering
\begin{Verbatim}[frame=single,fontsize=\relsize{-2}]
<LesHouchesEvents version="1.0">
  <!--
    # optional information in completely free format,
    # except for the reserved end tag (see next line)
  -->
  <header>
    <!-- individually designed XML tags, in fancy XML style -->
  </header>
  <init>
    compulsory initialization information
    # optional initialization information
  </init>
  <event>                            
    compulsory event information       
    # optional event information       
  </event>                           
  (further <event> ... </event> blocks, one for each event)
</LesHouchesEvents>
\end{Verbatim}  
  \caption{The original structure of a Les Houches event file.}
  \label{fig:lhef1}
\end{figure}

\subsection{THE ORIGINAL EVENT FILE FORMAT AND
  COMMON BLOCK STRUCTURE}
\label{sec:recap}

The first version of the Les Houches event file format was a simple
structure specifying how to write the Les Houches common blocks to a
text file. A few XML tags were defined to simplify parsing but not
much more than the information in the common blocks was
formalized. The structure of a file is outlined in
figure~\ref{fig:lhef1}, where the tags are as follows.
\begin{itemize}
\item \verb:LesHouchesEvents:: which contains the whole file and
  which mandates a \verb:version: attribute set to \verb:"1.0":.
\item \verb:header:: which may contain any number of unspecified XML
  tags describing how the events were generated.
\item \verb:init:: This is the tag which specifies the information in
  the \verb:HEPRUP: common block. The start tag must be alone on a
  line and the following line must contain the information which is in
  common for all processes in the file. The lines following this must
  contain the per-process information from the common block, one
  process per line. If there are any other lines before the end tag,
  they must be preceded by a \verb:#:-sign (c.f.\
  figure~\ref{fig:initlhef1}).
\item \verb:event:: The \verb:init: tag may be followed by any number
  of \verb:event: tags, one for each event generated. Also the
  \verb:event: start tag must be alone on a line and the following
  line must contain the general event information from the
  \verb:HEPEUP: common block. The lines following this must contain
  the per-particle information, one line per particle. Also here
  additional lines may be included before the end tag if they are
  preceded by a \verb:#:-sign. (c.f.\ figure~\ref{fig:eventlhef1}).
\item Before the \verb:init: tag one may, optionally, include arbitrary
  text enclosed in XML comment tags, \verb:<!-- ... -->:, but no other
  text is allowed in the enclosing \verb:LesHouchesEvents: tag.
\end{itemize}

For a more detailed description of the LHEF format we refer to
\cite{Alwall:2006yp}.

\begin{figure}
  \centering
\begin{Verbatim}[frame=single,fontsize=\relsize{-2}]
<init>
IDBMUP(1) IDBMUP(2) EBMUP(1) EBMUP(2) PDFGUP(1) PDFSUP(1) PDFSUP(2) IDWTUP NPRUP
XSECUP(1) XERRUP(1) XMAXUP(1) LPRUP(1)
XSECUP(2) XERRUP(2) XMAXUP(2) LPRUP(2)
...
XSECUP(NPRUP) XERRUP(NPRUP) XMAXUP(NPRUP) LPRUP(NPRUP)
# Additional
# information
</init>
\end{Verbatim}  
\caption{The structure of the \texttt{init} tag in the original LHEF
  format. See \cite{Boos:2001cv} for the meaning of the different
  common block variables.}
  \label{fig:initlhef1}
\end{figure}

\begin{figure}
  \centering
\begin{Verbatim}[frame=single,fontsize=\relsize{-3}]
<event>
NUP IDPRUP XWGTUP SCALUP AQEDUP AQCDUP
IDUP(1) ISTUP(1) MOTHUP(1,1) MOTHUP(2,1) ICOLUP(1,1) ICOLUP(2,1) PUP(1,1) PUP(2,1) PUP(3,1) PUP(4,1) PUP(5,1)
IDUP(2) ISTUP(2) MOTHUP(1,2) MOTHUP(2,2) ICOLUP(1,2) ICOLUP(2,2) PUP(1,2) PUP(2,2) PUP(3,2) PUP(4,2) PUP(5,2)
...
# In total 1+NUP lines after the <event> tag
# Additional
# information
</event>
\end{Verbatim}  
\caption[dummy]{The structure of the \texttt{event} tag in the
  original LHEF format. See \cite{Boos:2001cv} for the meaning of the
  different common block variables.}
  \label{fig:eventlhef1}
\end{figure}

\subsection{THE NEW FILE FORMAT}
\label{sec:new-file-format}

We now describe our suggestion for an updated file format which
includes the additional information mentioned in the introduction. All
such information is encoded in XML tags with optional attributes given
in the usual way:
\begin{verbatim}
<tag attribute1="value" attribute2="value">content</tag>
\end{verbatim}
or, for a tag without content,
\begin{verbatim}
<tag attribute1="value" attribute2="value" attribute3="value" />
\end{verbatim}
The new tags can either be given in the \verb:init: block, should they
refer to the whole file, or in the \verb:event: block, if they only
refer to an individual event. In addition \verb:group: tags can be
inserted to group events together.

\subsubsection{GLOBAL INFORMATION }
\label{sec:init}

The following tags may be included inside the \verb:init: tag and
contain additional global information about how the events in the file
were produced. They must be placed after the mandatory lines containing
\verb:HEPRUP: common block information (see
figure~\ref{fig:initlhef1}), but otherwise the order is
unimportant. Only tags which are not marked optional below need to be
supplied.

\paragraph{The \texttt{\rm generator} tag (optional)}
This is just added to give easy access to the name of the program
which has generated the file. The content of the tag is simply the
name and the only allowed attribute is
\begin{itemize}
\item[-] \verb:version:: a string describing the version of the generator
  used.
\end{itemize}

\paragraph{The \texttt{\rm xsecinfo} tag (required)}
The information in the \verb:HEPRUP: common block is in principle
sufficient to determine the cross sections of the processes
involved. Currently, the way in which this information is specified is
a bit complicated and sometimes confusing, since it was assumed to be
used to pass information between the MEG and PSG in both
directions. For the event file, the communication is per definition
one-way, and the information can be made more easily accessible. The
tag itself has no content, and the information is given in the
following attributes.
\begin{itemize}
\item[-] \verb:neve: (R)\footnote{(R) means the attribute is
    mandatory}: the number of events\footnote{Note that if the file
    contains events inside \texttt{group} tags (see section
    \ref{sec:eventgroup} below), \texttt{neve} must refer to the
    number of event groups (plus the events which are outside the
    groups).} in the file.
\item[-] \verb:totxsec: (R): the total cross section (in units of pb)
  of all processes in the file.
\item[-] \verb:maxweight: (D=1)\footnote{For attributes which are not
    mandatory, (D=\ldots) indicates which value is assumed it not
    present}: the maximum weight of any event\footnote{Note that if
    the file contains events inside \texttt{group} tags (see section
    \ref{sec:eventgroup} below), \texttt{maxweight},
    \texttt{minweight} and \texttt{meanweight} must refer to the
    weights of the groups (and the weights of the events which are
    outside the groups).} in the file (in an arbitrary unit).
\item[-] \verb:minweight: (D=\verb:-maxweight:): if the file contains
  negative weights, the \verb:minweight: is the most negative
  of the negative weights in the file. (Must obviously be the same
  unit as \verb:maxweight:.)
\item[-] \verb:meanweight: (D=1): The average weight of the events in
  the file (same unit as maxweight).
\item[-] \verb:negweights: (D=\verb:no:): If \verb:yes:, then the file
  may contain negative weights.
\item[-] \verb:varweights: (D=\verb:no:): If \verb:yes:, then the file
  may contain varying event weights. If \verb:no:, all events are
  weighted with maxweight (or, if \verb:negweights=yes:, with
  minweight).
\item[-] \verb:eventgroups: (D=\verb:no:): If \verb:yes:, the events
  in the file may be grouped together with \verb:group: tags, in
  which case the attributes above count an event group as one event
  rather than several separate ones.
\item[-] \verb:maxingroupweight: (D=\verb:maxweight:): If
  \verb:eventgroups=yes:, this gives the maximum weight among the
  events inside groups.
\item[-] \verb:miningroupweight: (D=\verb:-maxingroupweight:): If
  \verb:eventgroups=yes:, this gives the minimum weight among the
  events inside groups.
\end{itemize}
Note that it is assumed that all processes in the file are weighted
with respect to a common total cross section, such that summing the
weights for the events of a given process and multiplying with
\verb:totxsec/maxweight/neve: will give the cross section for that
process. In this way, the per-process information in the \verb:HEPRUP:
common block can be safely ignored.

\paragraph{The \texttt{\rm cutsinfo} tag (optional)}

This tag is used to supply information about which kinematical cuts
were used to generate the events in the file. Several different cuts
can be given with \verb:cut: tags and it is possible to specify which
particles should be affected by each cut using \verb:ptype: tags.

The \verb:cut: tag contains an actual cut made in the generation of
the events. In general, all events in the file will pass this cut. The
cut is defined in terms of a kinematical variable and the particles
which are affected. The content of the tag is one or two numbers
giving the allowed range of value of the kinematical variable
according to the attribute \verb:limit: (see below).

The variable is defined according to the following attributes of the
\verb:cut: tag:
\begin{itemize}
\item[-] \verb:p1: (D=0): Lists the particle types for which this cut
  applies. This can be either a number, corresponding to a given
  particle PDG\cite{Amsler:2008zzb} code, or a string corresponding to
  a group of particles previously defined with a \verb:ptype: tag (see
  below). The default is zero which means \textit{any particle type}.
\item[-] \verb:p2:, \ldots, \verb:p9:: Allows the specification of
  additional sets of particle types, by analogy to \verb:p1:, in order
  to facilitate the application of different classes of cuts to
  different classes of particles.
\item[-] \verb:type: (R): This defines the variable which is cut. The
  following values are predefined, but also other variables may be
  specified. (Where relevant, the laboratory frame is assumed, and all
  energy units are in GeV.)
  \begin{itemize}
  \item \verb:m:: the invariant mass of a particle of type
    \verb:p1:. If additional particle types are specified the cut
    applies to the invariant mass of the corresponding number of
    particles, i.e.\ if \verb:p1:, \verb:p2: and \verb:p3: are
    specified the cut is on the invariant mass of any set of
    three matching particles.
  \item \verb:pt:: the transverse momentum of a particle matching
    \verb:p1:.
  \item \verb:eta:: the pseudo-rapidity of a particle matching \verb:p1:.
  \item \verb:y:: the true rapidity of a particle matching \verb:p1:.
  \item \verb:deltaR:: the pseudo-rapidity--azimuthal-angle
    difference ($\sqrt{\Delta\eta^2+\Delta\phi^2}$) between two
    particles matching \verb:p1: and \verb:p2: respectively.
  \item \verb:E:: the energy of a particle matching \verb:p1:.
  \item \verb:ETmiss:: the norm of the vectorial sum of the pt of
    final state particles matching \verb:p1: and \textit{not} matching
    \verb:p2: (Note that an empty \verb:p2: defaults to the empty set
    here and for \verb:HT: below).
  \item \verb:HT:: the scalar sum of the transverse momentum of
    final state particles matching \verb:p1: and \textit{not} matching
    \verb:p2:.
  \end{itemize}
\item[-] \verb:limit: (D=\verb:min:): If set to \verb:min:
  (\verb:max:) only one number should be marked by the tag and give
  the minimum (maximum) for the kinematical variable, while if it is
  set to \verb:minmax:, there should be two numbers corresponding to
  the minimum and maximum (in that order).
\end{itemize}

The groups of particles to be considered in the \verb:p1: and
\verb:p2: attributes of the \verb:cut: tag are specified by
\verb:ptype: tags, which simply contains the PDG codes of the
particle types belonging to the group. The only allowed attribute in
the \verb:ptype: tag is
\begin{itemize}
\item[-] \verb:name: (R): the name of this group of particle types.
\end{itemize}

Here is a short example on how to specify a cut where a charged
electron or muon is required to have a transverse momentum of at least
20~GeV and a minimum of 25~GeV missing transverse energy is
required:
\begin{verbatim}
<cutsinfo>
  <ptype name="l+-">11 -11 13 -13</ptype>
  <ptype name="nu">12 -12 14 -14 16 -16</ptype>
  <cut type="pt" p1="l+-">20</cut>
  <cut type="ETmiss" p1="0" p2="nu">25</cut>
</cutsinfo>
\end{verbatim}

\paragraph{The \texttt{\rm procinfo} tag (optional)}

The \verb:procinfo: tag is used to supply additional per-process
information in addition to what is given in the \verb:HEPRUP: common
block part of the \verb:init: tag. The content of the tag is simply an
arbitrary string describing the process. The attributes are the
following:
\begin{itemize}
\item[-] \verb:iproc: (D=0): The process number for which the
  information is given. This must correspond to the \verb:LPRUP: code
  in the \verb:HEPRUP: common block for the corresponding
  process. Also zero can be given, in which case it refers to all
  processes in the file (except those with a separate \verb:procinfo:
  tag).
\item[-] \verb:loops: (D=0): The number of loops used in calculating
  this process.
\item[-] \verb:qcdorder:: The power of $\alpha_S$ used in calculating
  this process.
\item[-] \verb:eworder:: The power of the electro-weak coupling used
  in calculating this process.
\item[-] \verb:rscheme: (D=\verb:MSbar:): The renormalization scheme used in
  calculating this process.
\item[-] \verb:fscheme: (D=\verb:MSbar:): The factorization scheme used in
  calculating this process.
\item[-] \verb:scheme: (D=\verb:tree:): Information about the scheme
  used to calculate the matrix elements to NLO. If absent, a pure
  tree-level calculation is assumed. Possible values could be
  \verb:CSdipole: (NLO calculation with Catani--Seymour
  subtraction\cite{Catani:1996vz}),
  \verb:FKS:\cite{Frixione:1995ms,Frixione:1997np},
  \verb:MC@NLO:\cite{Frixione:2002ik,Frixione:2006gn},
  \verb:POWHEG:\cite{Nason:2004rx,Frixione:2007vw} and
  \verb:NLOexclusive: (NLO calculation according to the exclusive
  cross section (see eg.\ \cite{Nagy:2003tz}) within the given cuts).
\end{itemize}

\paragraph{The \texttt{\rm mergetype} tag (optional)}

For some merging schemes (eg. for CKKW) it is possible to reweight the
the events with Sudakov form factors already in the MEG. If this has
been done the content of the \verb:mergetype: tag for the corresponding
process should give a name corresponding to the scheme used. The
attributes are:
\begin{itemize}
\item[-] \verb:iproc:: The process number for which the information is
  given. A zero means all processes except those with a separate
  \verb:mergeinfo: tag.
\item[-] \verb:mergingscale: (R): The value of the merging scale in GeV. 
\item[-] \verb:maxmult: (D=\verb:no:): If \verb:yes:, the
  corresponding process is reweighted as if it is the maximum
  multiplicity process, i.e.\ the Sudakov form factor associated with
  evolution between the smallest clustering scale and the merging
  scale is not included.
\end{itemize}

\subsubsection{PER-EVENT INFORMATION}
\label{sec:event}

Information about a given event may be given with XML tags after the
mandatory lines containing \verb:HEPEUP: common block information (see
figure~\ref{fig:eventlhef1}).

\paragraph{The \texttt{\rm weight} tag (optional)}

An event can be associated with a number of different weights given in
\verb:weight: tags. The content of these tags is simply a sequence of
weights corresponding to the cross section for the event using
different PDFs, $\alpha_S$ values, etc.\ which can be used to estimate
the systematic errors due to, e.g., PDF uncertainties. Each
\verb:weight: tag should be given a name for identification. Only one
\verb:weight: tag per event can be without a name and should then only
contain one weight, which is the one for which the statistics in the
\verb:xsecinfo: tag is given. The attributes of the \verb:weight: tag
are as follows
\begin{itemize}
\item[-] \verb:name:: An arbitrary string describing this set of
  weights. If no \verb:name: is given this is the main weight for the
  event.
\item[-] \verb:born: (D=1): If this is not a normal tree-level event
  but reweighted in some way (eg.\ by Sudakov reweighting or using
  loop contributions), this should be set to the relative weight of
  the tree-level cross section.
\item[-] \verb:sudakov: (D=1): If this event has been reweighted by a
  Sudakov form factor, the size of this factor should be given here.
\end{itemize}
The last two attributes will probably only be given for the main
weight. If an event has only been reweighted by a Sudakov form factor
then these attributes are related by \verb:born*sudakov=1:. The
total Born cross section is obtained by summing the weights multiplied
by \verb:born: for each event of the given process, and multiplying
with \verb:totxsec/maxweight/neve: from the \verb:xsecinfo: tag.

\paragraph{The \texttt{\rm clustering} tag (optional)}

If an event has eg. been reweighted with Sudakov form factors, it is
possible to specify how the current event has been clustered to find
the scales involved. The contents of this this tag should be a series
of \verb:clus: tags. The clustering should be defined from the final state
backwards in terms inverse time-like splittings, in the end defining a
"bare" ladder diagram. This is then followed by a sequence of
space-like splittings.

\begin{figure}
  \centering
  \begin{minipage}[]{0.5\linewidth}
    \begin{center}
      \includegraphics[width=0.4\linewidth]{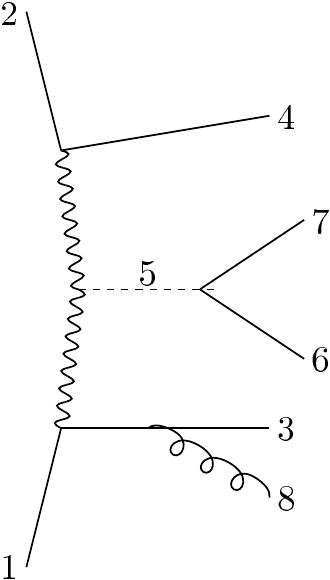}
    \end{center}
  \end{minipage}%
  \begin{minipage}[]{0.5\linewidth}
\begin{verbatim}
<clustering>
<clus>6 7 5</clus>
<clus>3 8</clus>
<clus>1 3</clus>
<clus>1 5</clus>
<clus>1 4</clus>
</clustering>
\end{verbatim}
  \end{minipage}

  \caption{Example of how a Feynman diagram can be encoded using a
    clustering tag. The numbering in the diagram corresponds to the
    particle entries in the \texttt{HEPEUP common block}}
  \label{fig:clus}
\end{figure}

The \verb:clustering: tag contains a number of \verb:clus: tags
corresponding to each of the splittings. Each \verb:clus: tag contains
two or three integers. The first two numbers indicate which particles
entries in the \verb:HEPEUP: common block are clustered. If a third
number is given it should correspond to an actual particle entry which
corresponds to the combined object (if eg. a decayed resonance is
explicitly present in the \verb:HEPEUP: common block). If no third
number is given, the clustered object is in the following referred to
by the first number. The attributes of the \verb:clus: tag are:
\begin{itemize}
\item[-] \verb:scale:: The scale (in GeV) associated with the clustering.
\item[-] \verb:alphas:: If the event has been reweighted with an
  $\alpha_S$ at the scale of this clustering, the value of this
  $\alpha_S$ should be supplied here.
\end{itemize}
We also wish to draw attention to the fact that the \verb:clustering:
tag can equally be used to encode the Feynman diagram (or the most
likely of the ones) used to produce the event. See
figure~\ref{fig:clus} for an example.

\paragraph{The \texttt{\rm pdfinfo} tag (optional)}

The \verb:pdfinfo: tag contains the values of the PDFs used when
generating this event, given by two numbers, $xf_1(x_1,Q^2)$ and
$xf_2(x_2,Q^2)$, for the two incoming partons. The attributes are:
\begin{itemize}
\item[-] \verb:p1:: The PDG code of the first incoming parton.
\item[-] \verb:p2:: The PDG code of the second incoming parton.
\item[-] \verb:x1:: The momentum fraction of the first incoming parton.
\item[-] \verb:x2:: The momentum fraction of the second incoming parton.
\item[-] \verb:scale:: (D=\verb:SCALUP:) The scale in GeV used in the
  PDFs (the default is taken from the \verb:HEPEUP: common block).
\end{itemize}

\subsubsection{GROUPING OF EVENTS}
\label{sec:eventgroup}

If we have a NLO calculation using eg.\ Catani--Seymour subtraction of a
process with $N$ particles in the Born level, each $N+1$ tree-level
event will come with a number of counter events with $N$ particles. For
this reason there is a need to group events together. Such a group of
events should be included in a \verb:group: tag.

\paragraph{The \texttt{\rm group} tag (optional)}

The content of this tag is a number of \verb:event: tags which for
should be considered together. If this is a $N+1$ tree-level event
with a number of $N$-particle counter events, the first \verb:event:
must always be the $N+1$-particle event even if this event fails the
cuts. The rest of the events are then the counter events. The
\verb:group: tag must also contain at least one main \verb:weight: tag
(without name attribute) which is the one for which the statistics in
the \verb:xsecinfo: tag is given. The individual weights of the events
in the group should sum up to the weight of the whole group. The only
allowed attribute is
\begin{itemize}
\item[-] \verb:n: (R): The number of events in the group.
\end{itemize}

Note that if event groups are present, the \verb:neve: attribute in
the \verb:xsecinfo: tag should count an event group as a single event.
Also, it is the \verb:weight: of the event group which relates to the
\verb:maxweight: and \verb:meanweight: attributes in the
\verb:xsecinfo: tag. To be compatible with the previous standard,
where the \verb:<event>: and \verb:</event>: tags are required to be
alone on a single line, also the \verb:<group>: and \verb:</group>:
tags are required to be alone on a single line.

\subsection{OUTLOOK}
\label{sec:summary}

Event files which follows this new standard should have their
\verb:version: attribute in the \texttt{LesHouches\-Events} (see
figure~\ref{fig:lhef1}) tag set to \verb:2.0:. The web page
\href{http://home.thep.lu.se/~leif/LHEF/}%
{\texttt{http://home.thep.lu.se/$\sim$leif/LHEF/}} contains a number
of \verb:C++: classes implementing the reading and writing of files
according to the new standard.

We have tried to make the new standard backward compatible with the
previous one, and although all existing parsers may not be able to
read new files we propose to keep the old preferred \texttt{.lhe} file name
extension.

As with the previous standard, the current proposal must not be viewed
as the end of the road.  There may be further information exchange
that ought to be standardized.  It is allowed to use/promote a
``private standard'' of tags in the \texttt{header} block or of
additional event information, and experience with such could point the
way towards an extended standard at a later date.

Note that a formal description of the proposed new standard can be
found at the Les Houches 09
\href{http://www.lpthe.jussieu.fr/LesHouches09Wiki/index.php/LHEF_for_Matching}%
{wiki pages}.%
\footnote{\href{http://www.lpthe.jussieu.fr/LesHouches09Wiki/images/6/60/Grammar.pdf}%
{\texttt{http://www.lpthe.jussieu.fr/LesHouches09Wiki/images/6/60/Grammar.pdf}}}

\clearpage

\section[A DRAFT RUNTIME INTERFACE TO
COMBINE PARTON SHOWERS AND NEXT-TO-LEADING ORDER QCD PROGRAMS]
{A DRAFT RUNTIME INTERFACE TO
COMBINE PARTON SHOWERS AND NEXT-TO-LEADING ORDER QCD PROGRAMS~\protect
\footnote{Contributed by: S.~Pl\"atzer}}


\subsection{INTRODUCTION}

Parton shower simulation programs like \pythiasix and
\herwigsix \cite{Corcella:2002jc,Sjostrand:2006za} have been the
workhorses for high energy physics experiments for a long time. With
the advent of the Large Hadron Collider (LHC) at CERN, the FORTRAN
generators have been completely rewritten and extended as \pythiaeight and
\herwigpp \cite{Gieseke:2003hm,Sjostrand:2007gs,Bahr:2008pv}
and the program \sherpa has been established
\cite{Gleisberg:2003xi,Gleisberg:2008ta}. 

Many new developments have been made in order to refine these simulations and
to extend their applicability. Particularly, progress is now being made towards
automation of schemes combining parton shower Monte Carlos and NLO QCD corrections
consistently along the lines of now established schemes 
\cite{Frixione:2002ik,Nason:2004rx}. Especially the POWHEG scheme
exhibits close connections
to methods correcting the hardest emission in parton shower simulations
to the relevant exact real emission matrix element, \cite{Seymour:1994df}.
Dedicated codes performing just the matching to NLO QCD are as well under development,
\cite{Frixione:2007nu}.

In order to use existing and future tools for fixed-order calculations,
especially in view of efforts towards automation present for these problems
as well, it is desirable to specify an interface to the generic building
blocks of fixed-order, particularly NLO QCD calculations carried out within 
the subtraction formalism.

The approach of communicating simulation results between different programs taken so
far by exchanging event files of a definite format \cite{Alwall:2006yp} is not appropriate for the
task being addressed here. Rather a {\it runtime interface} between two distinct
codes is desirable. Here, one code links to another either statically or dynamically
and makes use of the implemented functionality through an interface which is now
a set of definite function calls.

The interface should however not limit the particular ways in which fixed-order
calculations are carried out, neither how the matching is actually performed by
the parton shower Monte Carlo or a dedicated matching code. It should further
not limit which programming language is actually being used and is therefore
formulated in a language independent way. Any binding to a particular language
is an implementational detail.

\subsection{PRELIMINARIES}

\subsubsection{OBJECTS HANDLED BY THE INTERFACE}

The interface aims at having each individual building block of a
leading- or next-to-leading order QCD calculation at hand. In particular
\begin{itemize}
\item phase space generation
\item tree-level matrix elements
\item subtraction terms
\item finite parts of one loop/Born interferences
\item finite remainder terms originating from collinear factorization
\end{itemize}
are addressed as individual objects. These objects will be defined in detail
in the following sections.

\subsubsection{DIFFERENTIAL CROSS SECTIONS}

The interface is formulated to handle all pieces inherent to a leading-
or next-to-leading order differential cross section to be available in
the following form, where $\alpha$ denotes any contribution as listed
in the previous subsection:
\begin{multline}
\label{eq:xsec}
{\rm d}\sigma^\alpha_{ij} = f_i(x_i,\mu_F)f_j(x_j,\mu_F) \ \times \\
F^\alpha_{\{i,j,k_1,...,k_n\}}(\{p_i,p_j,p_1,...,p_n\},\mu_R)
\left|\frac{\partial\phi (\{p_i,p_j,p_1,...,p_n\},x_i,x_j)}{\partial \vec{r}}\right|{\rm d}^d r\ .
\end{multline}
Here $\{i,j,k_1,...,k_n\}$ identifies a particular subprocess of which the
term indexed by $\alpha$ contributes to, $\{p_i,p_j,p_1,...,p_n\}$ is
the physical phase space point, and $\vec{r}$ is a set of $d$ random numbers
in the $d$-dimensional unit hypercube.

In the following the Jacobian determinant will be referred to as a phase space
generator. The corresponding functionality will be addressed as an individual part of
differential cross sections within the interface. Note that this does not
reference the precise way in which the Monte Carlo integration or the generation
of unweighted events is performed by any client code\footnote{Throughout this work
the NLO code itself is considered to be `client' code of the interface.}.

The parton density functions (PDFs) $f$ as well as the value of renormalization scale $\mu_R$,
factorization scale $\mu_F$ and the strong coupling will be provided by
the client code, such that the part of the interface representing the contribution
$\alpha$ is completely specified as a function evaluating $F^\alpha$. Additionally,
few bookkeeping specifications will be introduced.

For tree-level matrix elements (including the real-emission part of a NLO
calculation, $F$ just represents the corresponding tree-level matrix element
squared times the appropriate flux and symmetry factors.

\subsubsection{NLO DIFFERENTIAL CROSS SECTIONS AND SUBTRACTION}

The interface focuses on NLO QCD corrections being carried out within
the subtraction formalism to obtain finite contributions from both
virtual and real-emission corrections to any infrared safe observable.

Examples of schemes widely in use are \cite{Catani:1996vz,Catani:2002hc,Frixione:1995ms}, 
the interface to be specified in detail in the next section should however not 
limit any implementation to a particular scheme.

Few universal properties are however assumed to be common to any subtraction
scheme at next-to-leading order. In particular it will be assumed that the auxiliary
cross section introduced to subtract the divergent behaviour of the real emission
contribution to NLO corrections for a $2\to n$ process is of the form
\begin{equation}
\sum_{\alpha=1}^k {\rm d}\sigma_{sub}^{\alpha} (p_{n+1},i_{n+1}) {\cal O}(p_n^\alpha(p_{n+1}),i_n^\alpha) \ .
\end{equation}
Here, ${\cal O}$ denotes an infrared safe observable (sometimes also referred to
as a jet defining function) and $p_n^\alpha(p_{n+1})$ is a unique and invertible momentum
mapping from a real emission to an {\it underlying Born configuration}, which is identified
by a flavour mapping $i_{n+1}\to i_n^\alpha$.

The number of individual subtraction terms $k$ is not assumed to be fixed. Besides
the unique association of a subtraction term to an underlying Born configuration 
each subtraction term is expected to subtract of a particular set of one or more
collinear divergences, which at NLO are identified by just labelling the pair of partons
becoming collinear. This information is crucial for a parton shower simulation
to assign an additional parton to a unique emitter in a meaningful way thereby 
fixing the initial conditions for subsequent showering.

\subsubsection{`COLLINEAR REMAINDERS'}

After analytical integration of the subtraction terms, the encountered
divergences cancel these present in the virtual correction and counter terms
needed to renormalize the parton distribution functions. The finite remainder
left from cancelling the latter divergences will then additionally depend on the
factorization scale, the momentum fractions $x_{1,2}$ and convolutions in one or
two variables $z\in [0,1]$ are in general present. Further, these 
convolutions are assumed to be casted in a form which is suitable to be done by Monte Carlo
methods as well. The interface therefore assumes a modified version of the contribution $F$
associated to these finite remainders,
\begin{equation}
F^{coll}_{\{i,j,k_1,...,k_n\}} = F^{coll}_{\{i,j,k_1,...,k_n\}}(\{p_i,p_j,p_1,...,p_n\},\mu_R,\mu_F,x_1,x_2,z_1,z_2) \ ,
\end{equation}
where two additional random numbers $z_{1,2}$ on the unit interval are provided
such that the convolution is performed on averaging over all events.

\subsubsection{COLOUR DECOMPOSITION}

In order to determine parton shower initial conditions but as well for the
purpose of implementing independent subtraction schemes, the knowledge of
partial amplitudes for the Born and real emission contribution is crucial.

The basis being used in this colour decomposition is however not unique.
To the extent that parton shower initial conditions are usually determined
by colour flows in the large-$N$ limit, it would be desirable to have
this decomposition in the fundamental representation of $SU(N=3)$ available.
Here, the basis tensors are just strings of Kronecker $\delta$'s,
\begin{equation}
\delta^{i_1}_{j_1}\cdots \delta^{i_n}_{j_n} \ ,
\end{equation}
where an upstairs index transforms according to the anti-fundamental, a downstairs
index according to the fundamental representation, indicating outgoing colour (incoming
anticolour) and outgoing anticolour (incoming colour), respectively.
Note that this representation is not limited to, but well suited, for the large-$N$
limit. The interface assumes this representation being available, but may be extended
to other representations leaving it up to the client code to change basis from
one to another decomposition.

\subsubsection{NOTATION}

The interface is formulated as a set of functions, taking any number of arguments 
and returning a result. The result may be a composite object of several types, which, 
depending on the language binding, may individually be passed as references to the function call.

A function of name \program{{\bf F}} taking arguments of type \program{A1}, \program{A2},..., 
returning a result of type \program{T} is denoted by \program{T {\bf F} (A1,A2,...)}, 
where each argument may be followed by a name identifying the meaning of the argument.
The required types are defined in the following section, except for obvious simple types.
Semantic definitions of each interface part are accompanied by pre- and postconditions
where needed. Examples of language bindings are given to explain the relation between
the abstract notation and implementational details.

\subsubsection{TYPES USED}

The purpose of this section is to introduce the more complex types used to specify the
interface components. Few basic types with obvious semantics are used without further
documentation.

\program{vector} -- A list of objects of the same type. Elements in the list
can be accessed randomly by an integer index. The type of objects stored is 
indicated in angle brackets, e.g. \program{vector$<$double$>$} is a vector storing doubles. 
The size of the vector is to be asserted in the context of a particular 
piece of the interface. Examples are \program{INTEGER V(4)} or \program{std::vector$<$int$>$ v(4);}
for FORTRAN and C++, respectively.

\program{pair} -- A pair stores to values of potentially different type. 
This is an auxiliary concept which does not have to be supported by a 
particular language. A pair storing values of type \program{T1} and \program{T2}, 
respectively, is denoted \program{pair$<$T1,T2$>$}. 

\program{union} -- A union generalizes \program{pair} to more than two entries.

\program{processid} -- Defined to be \program{vector$<$int$>$}. Identifies a 
subprocess giving PDG ids for incoming partons in the first two entries, 
and PDG ids for outgoing partons in the following entries. The size of 
the vector is guaranteed to be greater or equal to three. The subprocess
$gg\to d \bar{d} g$, for example, would be identified by the entries
$\{21,21,1,-1,21\}$.

\program{momentum} -- Defined to be \program{vector$<$double$>$} of length five. 
The first four entries contain the four-momentum components $p_x,p_y,p_z,E$ in 
units of GeV. The fifth component is optional containing the invariant mass 
squared in units of GeV squared. The metric is agreed to be mostly-minus,
$p\cdot q= E_p E_q - \vec{p}\cdot \vec{q}$.

\program{pspoint} -- Defined to be \program{union$<$double,pair$<$double,double$>$,vector$<$momentum$>$ $>$}. 
Represents a phase space point as the phase space weight (Jacobian from unit-hypercube 
random numbers to phase space measure), momentum fractions of incoming partons and a set of momenta. 
The first two momentum 
entries specify the momenta of incoming partons, subsequent these of outgoing partons. 
The size of the momentum vector is greater or equal to three. The weight is given
in units of the proper power of GeV to obtain a dimensionless quantity.

\program{colourflow} -- Defined to be \program{vector$<$pair$<$int,int$>$ $>$}. 
Represents a colour flow assigned to a particular subprocess in the 
following convention: in association to a processid, the first and second members 
of an entry of a colourflow at position $i$ contain an integer id for the colour 
and anticolour line, the corresponding parton at position $i$ in the process id 
is connected to. By convention, a triplet always has its second colour index set 
to zero, an antitriplet its first index. A singlet uses the pair $\{0,0\}$. Note 
that this notation is not limited to large-$N$ flows, but may represent any 
basis tensor in the fundamental representation by the following identification: 
Each non-zero entry in the first entry of a pair in a colour flow vector identifies 
an index transforming according to the anti-fundamental, each entry at the second 
position an index transforming according to the fundamental representation of $SU(N)$
and entries of the same id are attached to a Kronecker-$\delta$. 
Thus, the $1/N$ suppressed singlet contribution to a gluon (cf. the Fierz identity
for the fundamental representation generators) carries the pair $\{k,k\}$, 
where $k$ is not zero. Example: Consider the process $e^+e^-\to d \bar{d} g$,
which would be identified by the \program{processid} $\{-11,11,1,-1,21\}$.
Here two colour flows are possible, the leading part $\delta^{i_d}_{j_g}\delta^{i_g}_{j_{\bar{d}}}$,
and the $1/N$ suppressed $\delta^{i_d}_{j_{\bar{d}}}\delta^{i_g}_{j_g}$ contribution.
The first one would be identified by the \program{colourflow} $\{\{0,0\},\{0,0\},\{k_1,0\},\{0,k_2\},\{k_2,k_1\}\}$,
the latter by $\{\{0,0\},\{0,0\},\{k_1,0\},\{0,k_1\},\{k_2,k_2\}\}$, where $k_1\ne k_2$ are non-zero, positive
integers.

\subsection{SPECIFICATION OF THE INTERFACE}

\subsubsection*{}
{\bf Initialization and Bookkeeping}

\program{bool {\bf initialize} ()} -- The fixed-order code performs 
initialization and reads any relevant parameters from its preferred input mechanism. 
It returns true on success and false on failure. 

\program{pair$<$int,string$>$ {\bf alphasinfo} ()} -- Return information on the running 
of the strong coupling to be used as a combination of the number of loops 
contributing to the QCD $\beta$-function, and a string identifying the renormalization
scheme used. Values for the latter have to be agreed on.

\program{bool {\bf haveleadingorder} (processid)} -- Return true, if the process 
identified by the given processid can be calculated at leading order.

\program{vector$<$colourflow$>$ {\bf colourflows} (processid)} -- Return the possible 
colourflows which could be selected for the given processid.

\program{vector$<$colourflow$>$ {\bf largencolourflows} (processid)} -- Return the possible 
colourflows in the large-$N$ limit which could be selected for the given processid.

\program{bool {\bf haveoneloop} (processid)} -- Return true, if one loop QCD corrections 
to the given process can be calculated.

\program{string {\bf havesubtraction} (processid)} -- Return a non-empty string 
identifying a subtraction scheme, if real emission subtraction terms are available for the
given process. This does not require that the real emission process itself can be calculated. 
Return an empty string, if subtraction terms are not present. Other return values have to be
agreed on.

\program{vector$<$processid$>$ {\bf realemissions} (processid)} -- Assuming the given processid 
identifies a Born process, return the real emission processes to be considered 
for a NLO QCD correction.

\program{vector$<$int$>$ {\bf subtractions} (processid)} -- Assuming the given processid 
identifies a real emission process, return a list of ids for the subtraction terms to
be considered. The ids need to be unique to the fixed-order code and are not used except
for identifying a subtraction term. For dynamically allocated vectors, the function may
return an empty vector to indicate that the process considered is non-singular, 
for fixed-size vectors it should indicate this by filling the vector with zeroes.

\program{processid {\bf underlyingborn} (processid, int)} -- Given a real emission 
process id and subtraction term id through the first and second arguments, respectively, 
return the underlying Born process the subtraction term maps to.

\program{vector$<$pair$<$int,int$>$ $>$ {\bf collinearlimits} (processid, int)} -- Given a real 
emission process id and subtraction term id through the first and second arguments, 
respectively, return the positions of the partons in the processid given, for which the 
identified subtraction term subtracts collinear singularities. The ordering of the returned 
value entries is irrelevant and by convention fixed such that the first entry is less 
than the second. 

\subsubsection*{}
{\bf Kinematics and Phase Space}

\program{int {\bf ndim} (processid)} -- Return the number of random numbers needed to 
generate a phase space point for the given process.

\program{pspoint {\bf phasespace} (pair$<$momentum,momentum$>$, vector$<$double$>$, processid)} -- 
Generate a phase space point given incoming particle's momenta, a list of 
random numbers $\in ]0,1[$ and a process id. 

\subsubsection*{}
{\bf Dynamics}

Each function call defined here represents the associated contribution $F^\alpha$ to
differential cross section as defined in eq. \ref{eq:xsec}. Results are assumed
to be scaled by the appropriate power of GeV such as to return a dimensionless quantity.

\program{double {\bf me2} (processid, pspoint, double)} -- Return the helicity and 
colour summed matrix element squared for the given process, phase space point and value of
the renormalization scale in GeV.

\program{double {\bf partialme2} (processid, pspoint, double, colourflow)} -- 
Return the helicity summed partial amplitude squared, identified by the given colour flow, 
evaluated for the given process, phase space point and value of the renormalization scale in GeV. 

\program{double {\bf bornvirt} (processid, pspoint, double)} -- 
Return the helicity and colour summed Born-virtual interference plus the integrated subtraction 
terms ({\it i.e}. the finite part remaining after carrying out subtraction), 
consistent with the subtraction scheme chosen for the real emissions. Arguments in order are the 
process to be considered, the phase space point and the renormalization scale
in GeV.

\program{double {\bf collinear} (processid, pspoint, double, double, pair$<$double,double$>$)} -- 
Return the finite collinear remainder contribution consistent with the subtraction scheme chosen
for the process identified by the given \program{processid} and phase space point.
Further arguments in order are renormalization and factorization scales in units of GeV,
and two further random variables on the unit interval to perform Monte Carlo integration over 
convolutions present.

\program{double {\bf subtraction} (processid, int, pspoint, double)} -- 
Return the subtraction term identified by real emission process and subtraction term id, 
given a phase space point and renormalization scale in units of GeV.

\subsection*{CONCLUSIONS}

A general runtime interface has been outlined to access the individual building blocks
of a next-to-leading order (NLO) QCD calculation carried out within the subtraction
formalism.

Such an interface is an indispensable tool for `client' programs dedicated to the
matching of parton showers and higher order corrections. A subset of the interface
may as well be used to implement so-called matrix element corrections within
parton shower Monte Carlos, correcting the hardest shower emission to the exact
real emission matrix element squared.

\subsection*{ACKNOWLEDGEMENTS}

I would like to thank Stephen Mrenna, Peter Skands, Leif L\"onnblad and  Nicolas Greiner 
for many fruitful discussions and encouraging comments.

\clearpage

\section[STATUS OF THE FLAVOUR LES HOUCHES ACCORD]{STATUS OF THE FLAVOUR LES HOUCHES ACCORD\protect \footnote{Contributed by: F.~Mahmoudi, S.~Heinemeyer, A.~Arbey, A.~Bharucha, T.~Goto, T.~Hahn, U. Haisch, S.~Kraml, M.~Muhlleitner, J.~Reuter, P.~Skands, P.~Slavich}}



{

\subsection{INTRODUCTION}

In addition to the increasing number of refined approaches in the literature 
for calculating flavour-related observables, advanced programs dedicated to 
the calculation of such quantities, e.g.\ Wilson coefficients, branching
ratios, mixing amplitudes, 
renormalisation group equation (RGE) running including flavour effects 
have recently been
developed~\cite{Mahmoudi:2007vz,Mahmoudi:2008tp,Mahmoudi:2009zz,Degrassi:2007kj,sufla}. 
Flavour-related observables are also implemented by many other
non-dedicated public codes to provide additional checks for the models under
investigation 
\cite{Belanger:2008sj,Arbey:2009gu,Heinemeyer:1998yj,Heinemeyer:1998np,Lee:2003nta,Ellwanger:2005dv,Paige:2003mg}.
The results are often subsequently used by other codes, e.g.\ as
constraints on the parameter space of the model under consideration
\cite{Lafaye:2004cn,Bechtle:2004pc,deAustri:2006pe,Master3}.

At present, a small number of specialised interfaces exist between
the various codes. Such tailor-made interfaces are not easily generalised
and are time-consuming to construct and test for each specific
implementation. A universal interface would clearly be an advantage
here. 
A similar problem arose some time ago in the context of Supersymmetry
(SUSY). The solution took the form of the SUSY Les Houches Accord
(SLHA)~\cite{Skands:2003cj,Allanach:2008qq}, which is nowadays
frequently used to exchange 
information between SUSY related codes, such as soft SUSY-breaking
parameters, particle masses and mixings, branching ratios etc.  
The SLHA is a robust solution, allowing information to be exchanged between 
different codes via ASCII files.
The detailed structure of these input and output files is described in
\citeres{Skands:2003cj,Allanach:2008qq}.  

The goal of this work is to exploit the existing organisational structure
of the SLHA and use it to define an accord for the exchange of flavour
related quantities, which we refer to as the ``Flavour Les Houches
Accord'' (FLHA). In brief, the purpose of this Accord is thus to present
a set of generic definitions for an input/output file structure which
provides a universal framework for interfacing flavour-related
programs. Furthermore, the standardised format will provide the users
with a clear and well-structured result that could eventually be used
for other purposes. 

The structure is set up in such a way that the SLHA and the FLHA can be
used together or independently.  
Obviously, some of the SLHA entries, such as measured parameters in the
Standard Model (SM) 
and the Cabibbo-Kobayashi-Maskawa (CKM) matrix elements are
also needed for flavour observable 
calculations. Therefore, a FLHA file can indeed contain a SLHA block if
necessary. 
Also, in order to avoid any confusion,
the SLHA blocks are not modified or redefined in the FLHA.
If a block needs to be extended to meet the requirements of flavour
physics, a new ``\texttt{F}'' block is defined instead.  

Note that different codes may \emph{technically} achieve the FLHA
input/output in different ways. The details of how to
`switch on' the FLHA input/output for a particular program should be
described in the manual of that program and are not covered here. 
For the SLHA, libraries have been developed to permit an easy
implementation of the input/output routines~\cite{Hahn:2006nq}. In
principle these programs could be extended to include the FLHA as well. 

It should be noted that, while the SLHA was developed especially for the
case of SUSY, the FLHA is, at least in principle, model
independent. While it is possible to indicate the model used in a
specific block, the general structure for the information exchange can
be applied to any model. 

This report summarizes the current status of the FLHA. Several issues are not
defined in an unambigous way yet. This will be indicated in the text below.


\subsection{CONVENTIONS AND DEFINITIONS}
\label{sec:conventions}

The structure of the Flavour Les Houches Accord input and output files is
based on the existing SUSY Les Houches Accord structure and flavour quantities
are defined in blocks. The general conventions for the blocks are very similar
to the SLHA blocks \cite{Skands:2003cj} and they are not reproduced here. 

Since a FLHA file can also contain SLHA blocks, to clearly identify the
blocks of the FLHA, the first letter of the name of a block is an
``\texttt{F}''. There are two exceptions to this rule: blocks borrowed from
the SLHA, which keep their original name, and blocks containing imaginary
parts, which start with ``\texttt{IMF}''. 

The following general structure for the FLHA file is proposed:
\begin{itemize}

\item \texttt{BLOCK FCINFO}:
Information about the flavour code used to prepare the FLHA file.

\item \texttt{BLOCK FMODSEL}:
Information about the underlying model used for the calculations.

\item \texttt{BLOCK SMINPUTS}: 
Measured values of SM parameters used for the calculations.

\item \texttt{BLOCK VCKMIN}: 
Input parameters of the CKM matrix in the Wolfenstein parameterisation.

\item \texttt{BLOCK UPMNSIN}: 
Input parameters of the PMNS neutrino mixing matrix in the PDG parameterisation.

\item \texttt{BLOCK VCKM}: 
Real part of the CKM matrix elements.

\item \texttt{BLOCK IMVCKM}: 
Imaginary part of the CKM matrix elements.

\item \texttt{BLOCK UPMNS}: 
Real part of the PMNS matrix elements.

\item \texttt{BLOCK IMUPMNS}: 
Imaginary part of the PMNS matrix elements.

\item \texttt{BLOCK FMASS}:
Masses of quarks, mesons, hadrons, etc.

\item \texttt{BLOCK FLIFE}:
Lifetime (in seconds) of flavour-related mesons, hadrons, etc.

\item \texttt{BLOCK FCONST}:
Decay constants.

\item \texttt{BLOCK FCONSTRATIO}:
Ratios of decay constants.

\item \texttt{BLOCK FBAG}:
Bag parameters.

\item \texttt{BLOCK FWCOEF}: 
Real part of the Wilson coefficients.

\item \texttt{BLOCK IMFWCOEF}: 
Imaginary part of the Wilson coefficients.

\item \texttt{BLOCK FOBS}:
Prediction of flavour observables.

\item \texttt{BLOCK FOBSERR}:
Theory error on the prediction of flavour observables.

\item \texttt{BLOCK FOBSSM}:
SM prediction for flavour observables.

\item \texttt{BLOCK FFORM}:
Form factors.\\

\end{itemize}
More details on several blocks are given in the following.
The blocks \ttt{SMINPUTS}, \ttt{VCKMIN}, \ttt{UPMNSIN}, \ttt{VCKM}, 
\ttt{IMVCKM}, \ttt{UPMNS}, \ttt{IMUPMNS} are defined exactly as in the SLHA(2)
and not further discussed here.


\subsubsection*{\texttt{BLOCK FCINFO}}

Flavour code information, including the name and the version of the program:\\
\numentry{1}{Name of the flavour calculator}
\numentry{2}{Version number of the flavour calculator}
  Optional warning or error messages can also be specified:\\
\numentry{3}{If this entry is present, warning(s) were produced by the
  flavour calculator. The resulting file may still be used. The entry
  should contain a description of the problem (string).}   
\numentry{4}{If this entry is present, error(s) were produced by the
  flavour calculator. The resulting file should not be used. The entry
  should contain a description of the problem (string).}  
  This block is purely informative, and is similar to 
  \texttt{BLOCK SPINFO} in the SLHA. 


\subsubsection*{\texttt{BLOCK MODSEL}}

This block provides switches and options for the model selection.
The SLHA2 \texttt{BLOCK MODSEL} is extended to allow more
flexibility. 

\numentry{1}{Choice of SUSY breaking model or indication of other
  model. By default, a
minimal type of model will always be assumed. Possible
values are:\\
\snumentry{-1}{SM}
\snumentry{0}{General MSSM simulation} 
\snumentry{1}{(m)SUGRA model} 
\snumentry{2}{(m)GMSB model}
\snumentry{3}{(m)AMSB model}
\snumentry{4}{...}
\snumentry{31}{THDM}
\snumentry{99}{other model. This choice requires a string given in the
  entry \ttt{99}}
}\\
\numentry{3}{(Default=0) Choice of particle content, only used for SUSY
  models. The defined switches are:\\ 
\snumentry{0}{MSSM}
\snumentry{1}{NMSSM}
\snumentry{2}{...}
}\\
\numentry{4}{(Default=\ttt{0}) R-parity violation. Switches defined are:\\
\snumentry{0}{R-parity conserved. This corresponds to the SLHA1.}
\snumentry{1}{R-parity violated. 
} 
}\\
\numentry{5}{(Default=\ttt{0}) CP violation. Switches defined are:\\
\snumentry{0}{CP is conserved. No information on the CKM phase
is used.}
\snumentry{1}{CP is violated, but only by the standard CKM
phase. All other phases are assumed zero.}
\snumentry{2}{CP is violated. Completely general CP phases
allowed.}
}\\
\numentry{6}{(Default=\ttt{0}) Flavour violation. Switches defined are:\\
\snumentry{0}{No flavour violation. 
}
\snumentry{1}{Quark flavour is violated.}
\snumentry{2}{Lepton flavour is violated.}
\snumentry{3}{Lepton and quark flavour is violated.}
}
\numentry{31}{defines the type of THDM, is used only if entry~\ttt{1} is 
  given as~\ttt{31}, otherwise it is ignored.\\
\snumentry{1}{type I}
\snumentry{2}{type II}
\snumentry{3}{type III}
\snumentry{4}{type IV}
}
\numentry{99}{a string that defines other models is used only if
  entry~\ttt{1} is given as~\ttt{99}, otherwise it is ignored.}


\subsubsection*{\texttt{BLOCK FMASS}}

The block \texttt{BLOCK FMASS} contains the mass spectrum for the
involved particles. It is an addition to the 
SLHA \texttt{BLOCK MASS} which contained only pole masses and to the
SLHA \ttt{BLOCK SMINPUTS} which contains quark masses.
If a mass is given in two blocks the block \ttt{FMASS} overrules the
other blocks.
In \ttt{FMASS} we
specify additional information concerning the renormalisation scheme as
well as the scale at which the masses are given and thus allow for
larger flexibility. The standard for each
line in the block should correspond to the following FORTRAN format 
\begin{center} 
\texttt{(1x,I9,3x,1P,E16.8,0P,3x,I2,3x,1P,E16.8,0P,3x,'\#',1x,A)},
\end{center} 
where the first nine-digit integer should be the PDG code of a particle,
followed by a double precision number for its mass. The next integer
corresponds to the renormalisation scheme, and finally the last double
precision number points to the energy scale (0 if not relevant). 
An additional comment can be given after \texttt{\#}.

The schemes are defined as follows:\\
\numentry{0}{pole}
\numentry{1}{$\overline{\mathrm{MS}}$}
\numentry{2}{$\overline{\mathrm{DR}}$}
\numentry{3}{1S}
\numentry{4}{kin}
\numentry{5}{\ldots}


\subsubsection*{\texttt{BLOCK FLIFE}}

The block \texttt{BLOCK FLIFE} contains the lifetimes of mesons and
hadrons in seconds. The standard for each line 
in the block should correspond to the FORTRAN format 
\begin{center} 
\texttt{(1x,I9,3x,1P,E16.8,0P,3x,'\#',1x,A)},
\end{center} 
where the first nine-digit integer should be the PDG code of a particle
and the double precision number its lifetime.  


\subsubsection*{\texttt{BLOCK FCONST}}

The block \texttt{BLOCK FCONST} contains the decay constants in GeV. The
standard for each line in the block should 
correspond to the FORTRAN format 
\begin{center} 
\texttt{(1x,I9,3x,I2,3x,1P,E16.8,0P,3x,'\#',1x,A)},
\end{center} 
where the first nine-digit integer should be the PDG code of a particle,
the second integer the number of the decay constant, and the double
precision number its decay constant. 


\subsubsection*{\texttt{BLOCK FCONSTRATIO}}

The block \texttt{BLOCK FCONSTRATIO} contains the ratios of decay
constants, which often have less uncertainty than the decay constants
themselves. The ratios are specified by the two PDG codes in the form
f(code1)/f(code2). The standard for each line in the block should
correspond to the FORTRAN format 
\begin{center}
\texttt{(1x,I9,3x,I9,3x,I2,3x,I2,3x,1P,E16.8,0P,3x,'\#',1x,A)},
\end{center}
where the two nine-digit integers should be the two PDG codes of
particles,
the third and fourth integers the numbers of the decay constants, which
correspond to
the second index of the entry in \texttt{BLOCK FCONST},
and the double precision number the ratio of the decay
constants.  


\subsubsection*{\texttt{BLOCK FBAG}}

The block \texttt{BLOCK FBAG} contains the bag parameters. The standard
for each line in the block should correspond to the FORTRAN format
\begin{center}
\texttt{(1x,I9,3x,I2,3x,1P,E16.8,0P,3x,'\#',1x,A)},
\end{center} 
where the
first nine-digit integer should be the PDG code of a particle, the
second integer the number of the bag parameter, and the double
precision number its bag parameter.  

So far no normalisation etc.\ has been defined, which at this stage has
to be taken care of by the user. An unambiguous definition will be given
elsewhere.


\subsubsection*{\texttt{BLOCK FWCOEF Q= \ldots}}

The block \texttt{BLOCK FWCOEF Q= \ldots} contains the real part of the
Wilson coefficients at the scale \texttt{Q}.

The different orders $C^{(k)}_i$ have to be given separately according
to the following convention for the perturbative expansion: 
\begin{eqnarray}
  C_{i}(\mu) &=&
  C^{(0)}_{i}(\mu)
  + \dfrac{\alpha_s(\mu)}{4\pi} C^{(1)}_{i,s}(\mu)
  + \left( \dfrac{\alpha_s(\mu)}{4\pi} \right)^2 C^{(2)}_{i,s}(\mu)
\nonumber\\&&
  + \dfrac{\alpha(\mu)}{4\pi} C^{(1)}_{i,e}(\mu)
  + \dfrac{\alpha(\mu)}{4\pi}
    \dfrac{\alpha_s(\mu)}{4\pi} C^{(2)}_{i,es}(\mu)
  + \cdots.
\label{eq:WCexpansion}
\end{eqnarray}
The couplings should therefore not be included in the Wilson coefficients.     

The entries in \texttt{BLOCK FWCOEF} should consist of two integers
defining the fermion structure of the operator and the operator
structure itself. These two numbers are not thought to give a full
representation including normalisation etc.\ of the operator, but
merely correspond to a unique identifier
for any possible Wilson coefficient. 
Consequently, the user has to take care that a consistent
normalisation including prefactors etc.\ is indeed fulfilled.
As an example, for the operator $O_1$,
\begin{align}
O_1 &= (\bar{s} \gamma_{\mu} T^a P_L c)
       (\bar{c} \gamma^{\mu} T^a P_L b)
\label{O1}
\end{align}
the definition of the two numbers is given as follows. 
The appearing fermions are encoded by a two-digit number originating
from their PDG code, where no difference is made between particles and
antiparticles, as given in Table~\ref{tab:pdgcodes}. 
Correspondingly, the first integer number defining $O_1$, containing the
fermions $\bar s c \bar c b$, is given by~03040405.
The various operators are defined in Table~\ref{tab:opcodes}. 
Correspondingly, the second integer number defining $O_1$, containing
the operators $\gamma_\mu T^a P_L \, \gamma^\mu T^a P_L$ is given by~6161.

A few more rules are needed for an unambigous definition.
\begin{itemize}
\item
If an operators appears without fermions (as it is possible, e.g., for
$F_{\mu\nu}$) it should appear right-most, so that the encoded fermions
correspond to the left-most operators.
\item
In the case of a possible ambiguity, for instance
$O_1 = (\bar s \gamma_\mu T^a P_L c) (\bar c \gamma^\mu T^a P_L b)$
corresponding to 03040405~6161 and
$O_1 =  (\bar c \gamma_\mu T^a P_L b) (\bar s \gamma^\mu T^a P_L c)$
corresponding to 04050304~6161
the ``smaller'' number, i.e.\ in this case 03040405~6161 should be
used.
\end{itemize}

\begin{table}[!ht]
\begin{center}
\begin{tabular}{|c|c|c||c|c|c|}
\hline
name & PDG code & two-digit number & 
name & PDG code & two-digit number \\
\hline
    $d$ & 1 & 01 & $e$     & 11 & 11 \\
    $u$ & 2 & 02 & $\nu_e$ & 12 & 12 \\
    $s$ & 3 & 03 & $\mu$   & 13 & 13 \\  
    $c$ & 4 & 04 & $\nu_\mu$ & 14 & 14 \\
    $b$ & 5 & 05 & $\tau$    & 15 & 15 \\
    $t$ & 6 & 06 & $\nu_\tau$ & 16 & 16 \\
    $\sum_q q$ &  & 07 & $\sum_l l$ &   & 17\\ 
    $\sum_q q Q_q$ & & 08 & $\sum_l l Q_l$ & & 18\\
\hline
\end{tabular}
\caption{PDG codes and two-digit number identifications of quarks and
  leptons. The summations are over active fermions.\label{tab:pdgcodes}}
\end{center}
\end{table}

\begin{table}[!htb]
\begin{center}
\begin{tabular}{|c|c||c|c||c|c|}
\hline
operator & number & 
operator & number & 
operator & number \\
\hline
    $1$                   & 30 & $T^a$    & 50 & $\delta_{ij}$     & 70 \\
    $P_L$                 & 31 & $P_L T^a$ & 51 & $P_l \delta_{ij}$ & 71 \\
    $P_R$                 & 32 & $P_R T^a$ & 52 & $P_R \delta_{ij}$ & 72 \\
    $\gamma^\mu$          & 33 & $\gamma^\mu T^a$ & 53 &
                                 $\gamma^\mu \delta_{ij}$ & 73 \\
    $\gamma_5$            & 34 & $\gamma_5 T^a$ & 54 &
                                 $\gamma_5 \delta_{ij}$ & 74 \\
    $\sigma^{\mu\nu}$     & 35 & $\sigma^{\mu\nu} T^a$ & 55 &
                                $\sigma^{\mu\nu} \delta_{ij}$ & 75 \\
    $\gamma^\mu \gamma^\nu \gamma^\rho$ & 36 &  
    $\gamma^\mu \gamma^\nu \gamma^\rho T^a$ & 56 &
    $\gamma^\mu \gamma^\nu \gamma^\rho \delta_{ij}$ & 76 \\
    $\gamma^\mu \gamma_5$ & 37 &
    $\gamma^\mu \gamma_5 T^a$ & 57 &
     $\gamma^\mu \gamma_5 \delta_{ij}$ & 77 \\
    $\gamma^\mu P_L$      & 41 & $\gamma^\mu T^a P_L$ & 61 &
                                 $\gamma^\mu \delta_{ij} P_L$ & 81 \\  
    $\gamma^\mu P_R$      & 42 & $\gamma^\mu T^a P_R$ & 62 &
                                 $\gamma^\mu \delta_{ij} P_R$ & 82 \\
    $\sigma^{\mu\nu} P_L$ & 43 & $\sigma^{\mu\nu} T^a P_L$ & 63 &
                                $\sigma^{\mu\nu} \delta_{ij} P_L$ & 83 \\
    $\sigma^{\mu\nu} P_R$ & 44 & $\sigma^{\mu\nu} T^a P_R$ & 64 &
                                $\sigma^{\mu\nu} \delta_{ij} P_R$ & 84 \\
    $\gamma^\mu \gamma^\nu \gamma^\rho P_L$ & 45 &
    $\gamma^\mu \gamma^\nu \gamma^\rho T^a P_L$ & 65 &
    $\gamma^\mu \gamma^\nu \gamma^\rho \delta_{ij} P_L$ & 85 \\
    $\gamma^\mu \gamma^\nu \gamma^\rho P_R$ & 46 &
    $\gamma^\mu \gamma^\nu \gamma^\rho T^a P_R$ & 66 &
    $\gamma^\mu \gamma^\nu \gamma^\rho \delta_{ij} P_R$ & 86 \\
    $F_{\mu\nu}$          & 22 & $G_{\mu\nu}^a$ & 21 & & \\ 
\hline
\end{tabular}
\caption{Two-digit number definitions for the operators.
$T^a$ ($a = 1 \ldots 8$) denote the $SU(3)_C$ generators,
$P_{L,R} = \frac{1}{2} (1 \mp \gamma_5)$, and
$(T^a)_{ij} (T^a)_{kl} = \frac{1}{2} (\delta_{il} \delta_{kj} 
                      - 1/N_c \, \delta_{ij} \delta_{kl})$, where
                      $i,j,k,l$ are colour indices. 
\label{tab:opcodes}}
\end{center}
\end{table}
The third index corresponds to each term in Eq.~(\ref{eq:WCexpansion}):\\[2mm]
\numentry{00}{$C^{(0)}_{i}(\mu)$}
\numentry{01}{$C^{(1)}_{i,s}(\mu)$}
\numentry{02}{$C^{(2)}_{i,s}(\mu)$}
\numentry{10}{$C^{(1)}_{i,e}(\mu)$}
\numentry{11}{$C^{(2)}_{i,es}(\mu)$}
\numentry{99}{total}

The information about the order is given by a two-digit number $xy$, where
$x$ indicates $\mathcal{O}(\alpha^x)$ and $y$ indicates
$\mathcal{O}(\alpha_s^y)$, and 0 indicates $C_i^{(0)}$. 

The Wilson coefficients can be provided either via separate new
physics and SM contributions, or as a total contribution
of both new physics and SM, depending on the code generating them. To
avoid any confusion, the fourth entry must specify whether the given
Wilson coefficients correspond to the SM contributions, new physics
contributions or to the sum of them, using the following definitions:\\ 
\numentry{0}{SM}
\numentry{1}{NPM}
\numentry{2}{SM+NPM}
The new Physics model is the model specified in the \texttt{BLOCK FMODSEL}.

\noindent The standard for each line in the block should thus correspond to the
FORTRAN format
\begin{center}
\texttt{(1x,I8,1x,I4,3x,I2,3x,I1,3x,1P,E16.8,0P,3x,'\#',1x,A)},
\end{center}
where the eight-digit integer specifies the fermion content, the
four-digit integer the operator structure, the two-digit integer the
order at which the Wilson coefficients are calculated followed by the
one-digit integer specifying the model, and finally the double precision
number gives the real part of the Wilson coefficient. 

Note that there can be several such blocks for different scales \texttt{Q}.


\subsubsection*{\texttt{BLOCK IMFWCOEF Q= \ldots}}

The block \texttt{BLOCK IMFWCOEF} contains the imaginary part of the
Wilson coefficients at the scale \texttt{Q}. 
The structure is exactly the same as for the \texttt{BLOCK FWCOEF}.


\subsubsection*{\texttt{BLOCK FOBS}}

The block \texttt{BLOCK FOBS} contains the flavour observables. The
structure of this block is based on the decay 
table in SLHA format. The decay is defined by the PDG number of the
parent, the type of the observable, the value of the observable, the
number of daughters and PDG IDs of the daughters.\\ 
The types of the observables are defined as follows:\\
\numentry{1}{Branching ratio}
\numentry{2}{Ratio of the branching ratio to the SM value}
\numentry{3}{Asymmetry -- CP}
\numentry{4}{Asymmetry -- isospin}
\numentry{5}{Asymmetry -- forward-backward}
\numentry{6}{Asymmetry -- lepton-flavour}
\numentry{7}{Mixing}
\numentry{8}{\ldots}
The standard for each line in the block should correspond to the FORTRAN
format
\begin{center}
\texttt{(1x,I9,3x,I2,3x,1P,E16.8,0P,3x,I1,3x,I9,3x,I9,3x,\ldots,3x,'\#',1x,A)},
\end{center}
where the first nine-digit integer should be the PDG code of the parent
decaying particle, the second integer the type of the observable, the
double precision number the value of the observable, the next integer
the number of daughters, and the following nine-digit integers the PDG
codes of the daughters. It is strongly advised to give the descriptive
name of the observable as comment. 


\subsubsection*{\texttt{BLOCK FOBSERR}}

The block \texttt{BLOCK FOBSERR} contains the theoretical error for
flavour observables, with the structure similar to 
\texttt{BLOCK FOBS}, where the double precision number for the value of
the observable is replaced by two double precision numbers for the minus
and plus uncertainties. 

In a similar way, for every block, a corresponding error block with the
name \texttt{BLOCK FnameERR} can be defined. 


\subsubsection*{\texttt{BLOCK FOBSSM}}

The block \texttt{BLOCK FOBSSM} contains the SM values of the flavour
observables in the same format as in 
\texttt{BLOCK FOBS}. The given SM values may be very helpful as a
comparison reference. 


\subsubsection*{\texttt{BLOCK FFORM}}

The block \texttt{BLOCK FFORM} contains the form factors for a specific decay. 
This decay should be defined as in 
\texttt{BLOCK FOBS}, but replacing the type of the observable by the
number of the form factor. It is essential here to describe the variable
in the comment area. The dependence on $q^2$ can be specified as a comment. 
A more unambiguous definition will be given elsewhere.


\subsection{CONCLUSION}

The interplay of collider and flavour physics is entering a new era with
the start-up of the LHC. In the future more and more programs will be
interfaced in order to exploit maximal information from both
collider and flavour data. Towards this end, an accord will play a
crucial role. The accord presented specifies a unique set of conventions
in ASCII file format for most commonly investigated flavour-related
observables and provides a universal framework for interfacing different
programs. 

The number of flavour related codes is growing constantly, while the
connection between results from flavour physics and high $p_T$ physics
becomes more relevant to disentangle the underlying physics model. 
Using the
lessons learnt from the SLHA, we hope the FLHA will prove useful for studies
related to flavour physics.
It is planned to update/correct the FLHA after more experience with its
application will have been gathered.


\subsubsection*{ACKNOWLEDGEMENTS}

The work of S.H.\ was partially supported by CICYT (grant FPA 2007--66387).
Work supported in part by the European Community's Marie-Curie Research
Training Network under contract MRTN-CT-2006-035505
`Tools and Precision Calculations for Physics Discoveries at Colliders'.
The work of T.G.\ is supported in part by the Grant-in-Aid for Science
Research, Japan Society for the Promotion of Science, No.\ 20244037.





\clearpage

\part[TUNING]{TUNING}

\section[STATUS OF RIVET AND PROFESSOR MC VALIDATION \& TUNING TOOLS]
{STATUS OF RIVET AND PROFESSOR MC VALIDATION \& TUNING TOOLS~\protect
  \footnote{Contributed by: A.~Buckley, J. M.~Butterworth, H.~Hoeth, H.~Lacker, J.~Monk,
   H.~Schulz, J.\,E.~von~Seggern, F.~Siegert}}
\label{sec:rivet}

The \rivet~\cite{Buckley:2010rivet} package for MC generator validation and the
\professor~\cite{Buckley:2009bj} system for generator tuning have become
established tools for systematically verifying event simulations and optimising
their parameters, where required and physically sensible. In this short report,
we summarise the status and development of these tools.

\subsection{\rivet}

\rivet is an MC \emph{validation} tool: it encodes MC equivalents of an
increasingly comprehensive set of HEP collider analyses which are useful for
testing the physics of MC generators. \rivet does not itself produce tunings,
but provides a standard set of analyses by which to verify the accuracy of a
given generator with a given tuning.

Several fundamental design principles have been derived from the experience on
\rivet's predecessor system, \hztool~\cite{Bromley:1995np,Waugh:2006ip}, and from iteration of the \rivet
design:
\begin{itemize}
\item No generator steering: \rivet relies entirely on being provided, by
  unspecified means, with events represented by the \hepmc~\cite{Dobbs:2001ck} event record.
\item No generator-specific analyses: all \rivet analyses are specifically not
  allowed to use the generator-specific portions of the supplied event
  records. Apart from a few limited (and deprecated) exceptions, all analyses
  are based solely on physical observables, i.e. those constructed from stable
  particles (those with status 1) and physical decayed particles (those with
  status 2).
\item \rivet can be used either as a C++ library to be interfaced with generator
  author or experiment analysis frameworks, or as a command line tool (which
  itself makes use of the library interface). This is an example of the general
  philosophy to keep things simple and flexible, since we do not \textit{a
    priori} know every task to which our tools will be employed.
\end{itemize}

Internally, \rivet analyses are based on a comprehensive set of calculational
tools called \emph{projections}, which perform standard computations such as jet
algorithms (using \fastjet~\cite{Cacciari:2006sm}), event shape tensors, and a variety of other
standard tasks. Use of projections makes analysis code much simpler,
encapsulates any complexities arising from the ban on use of event record
internal entities (the summation of photon momenta around charged leptons during
$Z$-finding is a good example, see Section~\ref{sec:zqed}), and is more efficient than just using library
functions, due to a complex (but hidden) system of automatic result caching.

Users can write their own analyses using the \rivet components and use them via
the \rivet API or command-line tool without re-compiling \rivet, due to use of
an analysis ``plugin'' system. Separation between generator and \rivet on the
command-line is most simply achieved by using the \hepmc plain text
\texttt{IO\_GenEvent} format via a UNIX pipe (a.k.a. FIFO): this avoids disk
access and writing of large files, and the CPU penalty in converting event
objects to and from a text stream is in many cases outweighed by the
general-purpose convenience. For generator-specific use of \rivet, the
programmatic interface allows \hepmc objects to be passed directly in code,
without this computational detour. A sister tool, \agile~\cite{Buckley:2007hi}, is provided
for convenience control of several \fortran-based generators, with command-line
and parameter file based run-time steering of generator parameters. This is a
convenient tool when exploring generator parameter space as part of a tuning.

Reference data for the standard analyses is included in the \rivet package as a
set of XML files in the AIDA format. After several years of re-development as
part of the CEDAR~\cite{Buckley:2007hi} project, the HepData~\cite{Buckley:2006np} database of HEP
experimental results can be used to directly export data files usable by \rivet
from its Web interface at \url{http://hepdata.cedar.ac.uk/}. Analysis histograms
are directly booked using the reference data as a binning template, ensuring
that data and MC histograms are always maximally consistent.

\rivet is in use within the MC generator development community, particularly in
general-purpose shower MC programs, and the LHC experimental community, for MC
validation and MC analysis studies which do not require detector simulation.

\subsubsection{Recent developments}

\rivet~1.1.3 was released during the first week of this Les~Houches workshop, in
June~2009: this release includes many new analyses and fixes to existing
analyses. Since the workshop, a huge number of extra improvements and
developments have taken place in the run-up to the~1.2.0 release. Aside from
many technical improvements, and the addition of a large number of QCD analyses
(primarily for minimum bias and multi-jet physics) the major conceptual
developments have been an emphasis on automated testing and validation of \rivet
code, and the removal of hard-coded cross-section normalisations whenever
possible. This latter step required more development than may be expected, due
to the separation of generator and analysis: the \hepmc record had to be
enhanced to store cross-section information in a way which can be passed to
\rivet. This has now been done, and recent versions of major generators such as
\herwigpp~\cite{Gieseke:2003hm,Bahr:2008pv}, 
\sherpa~\cite{Gleisberg:2003xi,Gleisberg:2008ta}, and 
\pythiaeight~\cite{Sjostrand:2007gs},  
support this \hepmc feature ``out of the
box''. \agile's generator interfaces have also been updated to write
cross-section information into their \hepmc output. Determining scaling
$K$-factors where required is now performed via post-processing scripts, which
automatically support common approaches to constraining this remaining degree of
freedom.

A technical development, but one worth mentioning, has been the emphasis on
making \rivet analyses ``self-documenting'': each analysis has a structured set
of metadata specifying name, authors, run conditions, a description, etc., which
is used to provide interactive help, HTML documentation, and a reference section
in the \rivet manual.

At the time of writing, the final stage of systematic validation of \rivet for
the 1.2.0 release is underway. The validation scripts used for this checking
will henceforth be included in automatic build tests, to ensure that future
developments do not unexpectedly change existing analysis functionality. The
final major stage of development is the upgrade of \rivet's histogramming and
data analysis code, which is currently rather basic. The upgrade will enable
statistically accurate combination of runs, allowing for greater parallelisation
of \rivet analyses which require large event statistics.

\subsection{\professor}

The \professor system builds on the output of MC validation analyses such as
those in \rivet, by optimising generator parameters to achieve the best possible
fit to reference data. The main description of \professor's details is found in
reference~\cite{Buckley:2009bj}, and we will not significantly replicate it
here, except in the most high-level sense.

Fundamentally, generator tuning is an example of the more general problem of
optimising a very expensive function with many parameters: the volume of the
space grows exponentially with the number of parameters and the CPU requirements
of even a single evaluation of the function means that any attempt to scan the
parameter space will fail for more than a few parameters. Here, the expensive
function is running a generator with a particular parameter set to recreate a
wide range of analysis observables, using a package such as \rivet. The approach
adopted by \professor is to parameterise the expensive function based on a
non-exhaustive scan of the space: it is therefore an approximate method, but its
accuracy is systematically verifiable and it is currently the best approach that
we have.

The parameterisation is generated by independently fitting a function to each of
the observable bin values, approximating how they vary in response to changes in
the parameter vector. One approach to fitting the functions would be to make
each function a linear combination of algebraic terms with $n$ coefficients
$\alpha_i$, then to sample $n$ points in the parameter space. A matrix inversion
would then fix the values of $\alpha_i$. However, use of a pseudoinverse for
rectangular matrices allows a more robust coefficient definition with many more
samples than are required, with an automatic least-squares fit to each of the
sampled ``anchor points'': this is the method used by \professor. By aggregating
the parameterisations of all the observable bins under a weighted goodness of
fit measure -- usually a heuristic \chisq -- a numerical optimisation can be
used to create an ``optimal'' tune. In practice, many different semi-independent
combinations of MC runs are used to provide a systematic handle on the degree of
variation expected in tunes as a result of the inputs, avoiding the problem that
a single ``maximum-information'' tune may not be typical of the parameter space.

The first application of \professor, due to its popularity and fairly
well-understood steering parameters, was the \pythiasix MC generator\cite{Sjostrand:2006za}. This was
tuned in reference~\cite{Buckley:2009bj}, using both of the available parton
shower and multi-parton interaction (MPI) models, to data from LEP, SLD, and
Tevatron Runs~I and~II. It was found that the parameterisation method worked
well in all cases, and a range of systematic methods and tools were developed to
check the accuracy of the approximations, such as line-scans through the
parameter space. It was found that a sensible maximum number of parameters to be
included in a single tune was $\sim\!10$, and hence, there being $\sim\!20$
\pythia6 parameters relevant to the studied observables, we separated the tune
into an initial stage of final state shower and fragmentation tuning using $e^+
e^-$ observables, and then a second stage based on tuning initial state parton
shower and MPI parameters to best describe hadron collider data. In these tunes,
a quadratic parameterisation was used throughout, this being the simplest
suitable function to account for parameter correlations.

\subsubsection{Recent developments}

The main development since the initial publication and use of \professor has
been the application to more MC tunings. A first extra application was to use
the same \chisq weightings as for the \pythiasix tune to obtain additional tunings
of \pythiasix with different PDFs. The results of this study, shown at the PDF4LHC
meeting in April~2009, indicated that modified leading order PDFs, developed for
use with LO MC generators and characterised by a larger than normal low-$x$
gluon component, drive statistical generator tunings in a physically expected
direction: the main effect was to increase the screening of MPI effects
proportional to the size of the gluon PDF at low $x$ values.

Moving away from \pythiasix, substantial tuning effort has been expended with the
\jimmy~\cite{Butterworth:1996zw,jimmy} MPI simulation (used with 
the \fortran \herwigsix code \cite{Corcella:2002jc}), and the \pythiaeight
and \sherpa generators. In the case of \pythiaeight, the default fragmentation
settings are now those fixed by \professor, and use of \professor has identified
a problem with description of underlying event (UE) observables in QCD events:
this problem is being addressed. An extremely useful tool in this study was the
\texttt{prof-I} GUI, which makes use of the \professor parameterisations not for
minimisation, but for interactive mimicking of generator responses to parameter
changes: this tool makes exploration of speculative tuning ideas easily
testable, and helped to verify that no combination of parameters would achieve
the desired effect in \pythiaeight's description of UE observables.

Another major effort has been the use of \professor to tune and develop the
\sherpa generator's simulation of hadronisation and soft initial-state QCD
physics. This collaboration has helped to rapidly iterate model improvements and
debugging, due to the fast turnaround of tune information.

\professor is additionally being used within the ATLAS, CMS, and LHCb LHC
experimental collaborations for tuning studies of the main generators used for
their MC simulation, and in plans for re-tuning to first LHC data at new centre
of mass energies.

Development of the \professor framework and application to different
implications of tuning continues: the next contribution details how ensembles of
tunes created by \professor may be used for estimations of tune uncertainties in
MC predictions. Other suggested extensions to optimisation of observable
definitions or parameterisation of other expensive functions, e.g. observables
in SUSY parameter space, remain open to exploration.


\subsection*{Acknowledgements}
The \rivet and \professor collaborations acknowledge support from the EU MCnet
Marie Curie Research Training Network (funded under Framework Programme 6
contract MRTN-CT-2006-035606) for financial support and for many useful
discussions and collaborations with its members.  A.~Buckley additionally
acknowledges support from the Scottish Universities Physics Alliance (SUPA);
H.~Schulz acknowledges the support of the German Research Foundation (DFG).


\clearpage

\section[QUANTITATIVE ERROR ESTIMATION IN MC TUNES]
{QUANTITATIVE ERROR ESTIMATION IN MC TUNES~\protect
  \footnote{Contributed by: A.~Buckley, H.~Hoeth, H.~Lacker, H.~Schulz, J.\,E.~von~Seggern}}

\newcommand{\p}{\ensuremath{\vec{p}}\xspace}
\newcommand{\pzero}{\ensuremath{\vec{p}_0}\xspace}
\newcommand{\pprime}{\ensuremath{\vec{p}^{\,\prime}}\xspace}
\newcommand{\N}[1]{\ensuremath{N^{\mspace{0.5mu}(\mspace{-0.5mu}#1)}}\xspace}
\newcommand{\Nn}[2]{\ensuremath{N_{\,#1}^{\mspace{0.5mu}(\mspace{-0.5mu}#2)}}\xspace}
\newcommand{\Nmin}[1]{\ensuremath{N_\text{min}^{\mspace{0.5mu}(\mspace{-0.5mu}#1)}}\xspace}
\newcommand{\Nmax}[1]{\ensuremath{N_\text{max}^{\mspace{0.5mu}(\mspace{-0.5mu}#1)}}\xspace}
\newcommand{\Ndf}{\ensuremath{N_\text{df}}\xspace}
\newcommand{\Ptilde}{\ensuremath{\mathbf{\tilde{P}}}\xspace}
\newcommand{\ptilde}{\ensuremath{\tilde{p}}\xspace}
\newcommand{\vptilde}{\ensuremath{\vec{\tilde{p}}}\xspace}
\newcommand{\Ipseudo}{\ensuremath{\tilde{\mathcal{I}}}\xspace}
\newcommand{\vcoeff}{\ensuremath{\vec{c}}\xspace}
\newcommand{\vcoeffb}{\ensuremath{\vec{c}^{(b)}}\xspace}
\newcommand{\coeffb}[1]{\ensuremath{c^{(b)}_{#1}}\xspace}
\newcommand{\vval}{\ensuremath{\vec{v}}\xspace}
\newcommand{\vvalb}{\ensuremath{\vec{v}^{(b)}}\xspace}

Recent developments in Monte Carlo generator tuning have led to more robust and
general-purpose ``optimal'' tunes to existing data, and there is a clear hope
that when existing data is well-described, extrapolations to future collider
energies will also be reliable. However, especially in the area of soft QCD,
generator predictions are never expected to be exactly accurate: there is no
single ``best'' tune for a given model, but rather one or more \emph{regions} of
parameter space which contains reasonable tunes. The size of these regions
reflects the degree of constraint which existing data is able to place on the
model -- discrete choices of model are also crucial to obtain a true sense of
the total uncertainty on any generator prediction. In this contribution we
present studies of how the mechanisms used for systematic generator tuning can
be used to quantitively estimate the contributions to the uncertainty of MC
predictions from several sources.

A key example of generator uncertainty is the extrapolation of minimum bias and
underlying event QCD physics to LHC design energies, i.e. $\sqrt{s} \gg
\unit{2}{\TeV}$. The physics that drives the rise in activity is a combination
of the non-perturbative total $pp$ cross-section; the non-perturbative physics
of beam remnants, diffraction, and multiple partonic scattering; and
perturbative low-\pT QCD scattering. Accordingly, the most-used models are
highly phenomenological and have many tweakable parameters: one interesting
consequence of first LHC data in the multi-\TeV{} energy regime will be the
testing of whether these models extrapolate in agreement with nature. The expectation
is that some models will fail the test!


Previous studies of prediction uncertainty have been necessarily qualitative and
subjective, since MC tunes have themselves been somewhat approximate
affairs. The most common approach to assigning a systematic uncertainty has been
to compare predictions from two different models, such as \pythia and \phojet in
the case of Minimum Bias/Underlying Event extrapolation.  Discrete choices of model remain a major
source of uncertainty, even when the range of historic \pythia tunes with a
Tevatron-excluded energy extrapolation is excluded\footnote{This includes the
  default \pythia tune, the ATLAS MC08 tune, and (by construction) the ~Tune~*T
  series from Rick~Field. Essentially, any setup with $\text{PARP}(90) < 0.2$ is
  in contradiction with Tevatron data.} -- we encourage all extrapolated studies
to make use of as many distinct (non-excluded) models as possible. In this
contribution, however, we describe how to quantitively assess major sources of
uncertainty arising from the tuning process itself, i.e. the reasonable scatter
to be expected around a tune for a discrete model. Our baseline for this
approach is the \professor tuning system, which we now summarise.


\subsection{MC tuning with \professor}

The ``\professor'' approach to MC tuning constitutes both a numerical method and
a suite of tools which implement it. Fundamentally, \professor attempts to
parameterise expensive functions -- the bin values in a set of MC observables --
by least-squares fitting of the parameterisation coefficients. The least-squares
minimisation is made more approachable by use of the pseudoinverse method,
implemented via a matrix singular value decomposition. Armed with a fast
analytic model of how every bin of a large set of observables will respond to
variations of the generator parameters, numerical optimisation of the
generator's fit to reference data may be efficiently computed. A detailed
description may be found in reference~\cite{Buckley:2009bj}.

The benefit of this approach is clear for more than two parameters: \professor
requires as input the values of observables for a moderately large number of MC
runs distributed suitably in the generator parameter space, each point in the
space perhaps requiring \order{48} CPU hours to complete. A serial optimisation
approach such as Markov chain sampling would hence require thousands of CPU days to
have a chance of converging, if the generator itself is not
batch-parallelised. \professor, given sufficiently large batch computing
resources, can trivially parallelise the generation of the input MC points for
any generator and thereafter complete the parameterisation and fit optimisation
in negligible time, allowing for scaling to higher numbers of tune parameters
than could be attempted by methods which require iteration of the time-limiting
step.

\subsection{Qualitative error estimation in \professor}

An important feature of the \professor method is that it has always allowed for
\emph{qualitative} assessment of the tune robustness. A per-bin parameterisation
in $p$ parameters will require a minimum number of MC runs, \Nmin{p}, for the
least-squares pseudoinversion to be performed. For robustness it is advisable to
oversample this minimum requirement by a factor \order{3} (or more, especially
for large $p$) such that a tune will in fact use $N \gg \Nmin{p}$ input
runs. Additionally, since the parameterisation and optimisation steps are fast,
we take the opportunity to make many such overconstrained tunes by in fact
sampling an even greater number of runs, $N_\text{sampled} \gg N$. We can then
randomly sample a large number of mostly-independent $N$-run tunes to obtain an
ensemble of reasonable tunes -- again, this step can be trivially
batch-parallelised. The spread of this tune ensemble as projected on each
parameter has been used in several \professor MC tunes as a heuristic for
determining whether a parameter is well or poorly constrained, for detecting
parameterisation problems, and for ensuring that the ``maximum information''
tune is typical of the ensemble. Several other checking methods, such as
eigenvector line scans, are also used to ensure that the details of the tune,
and particularly the generator parameterisation, are reliable.

\begin{figure}[t]
  \centering
  
  \includegraphics[width=0.325\textwidth]{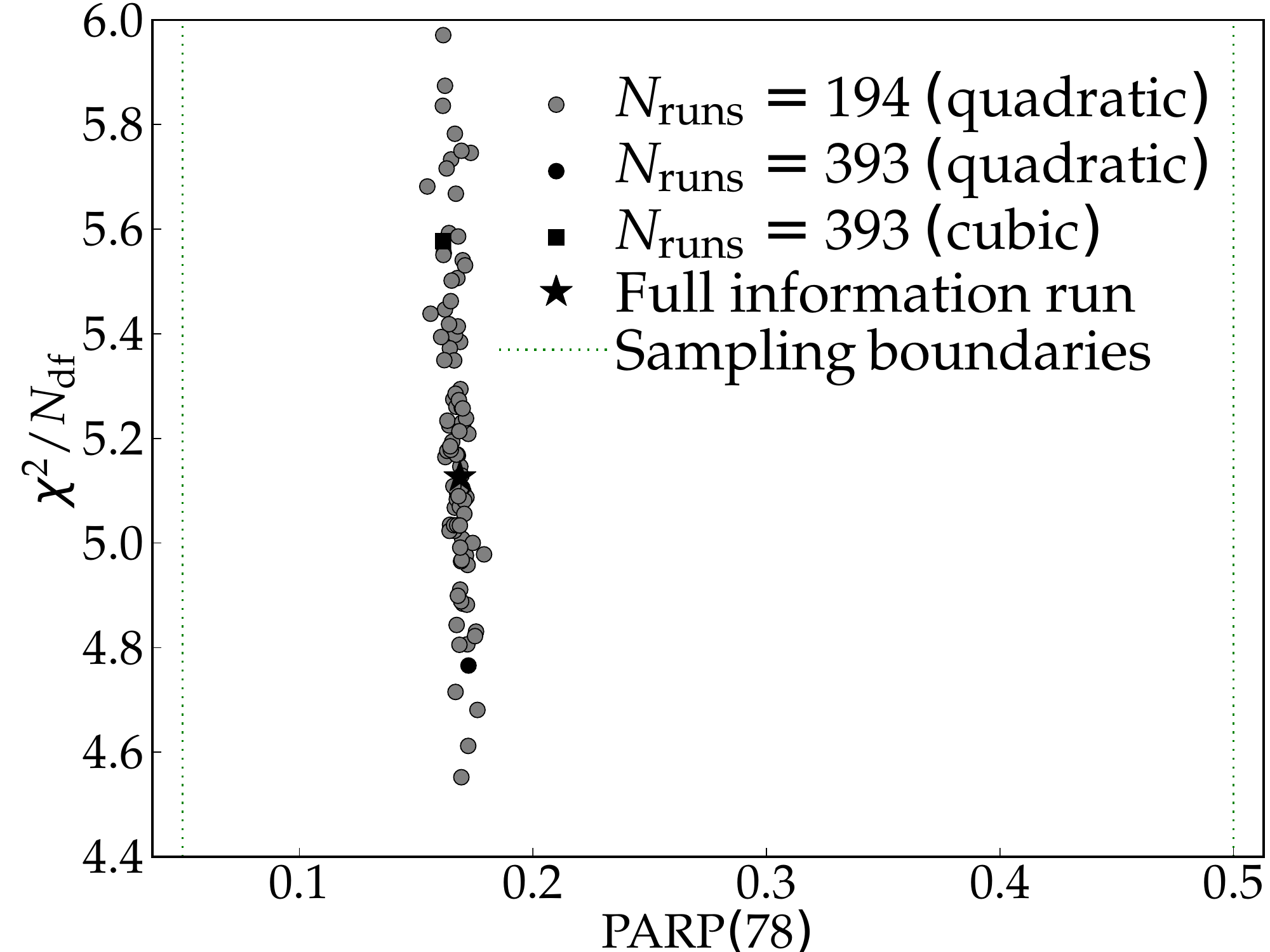}
  \includegraphics[width=0.325\textwidth]{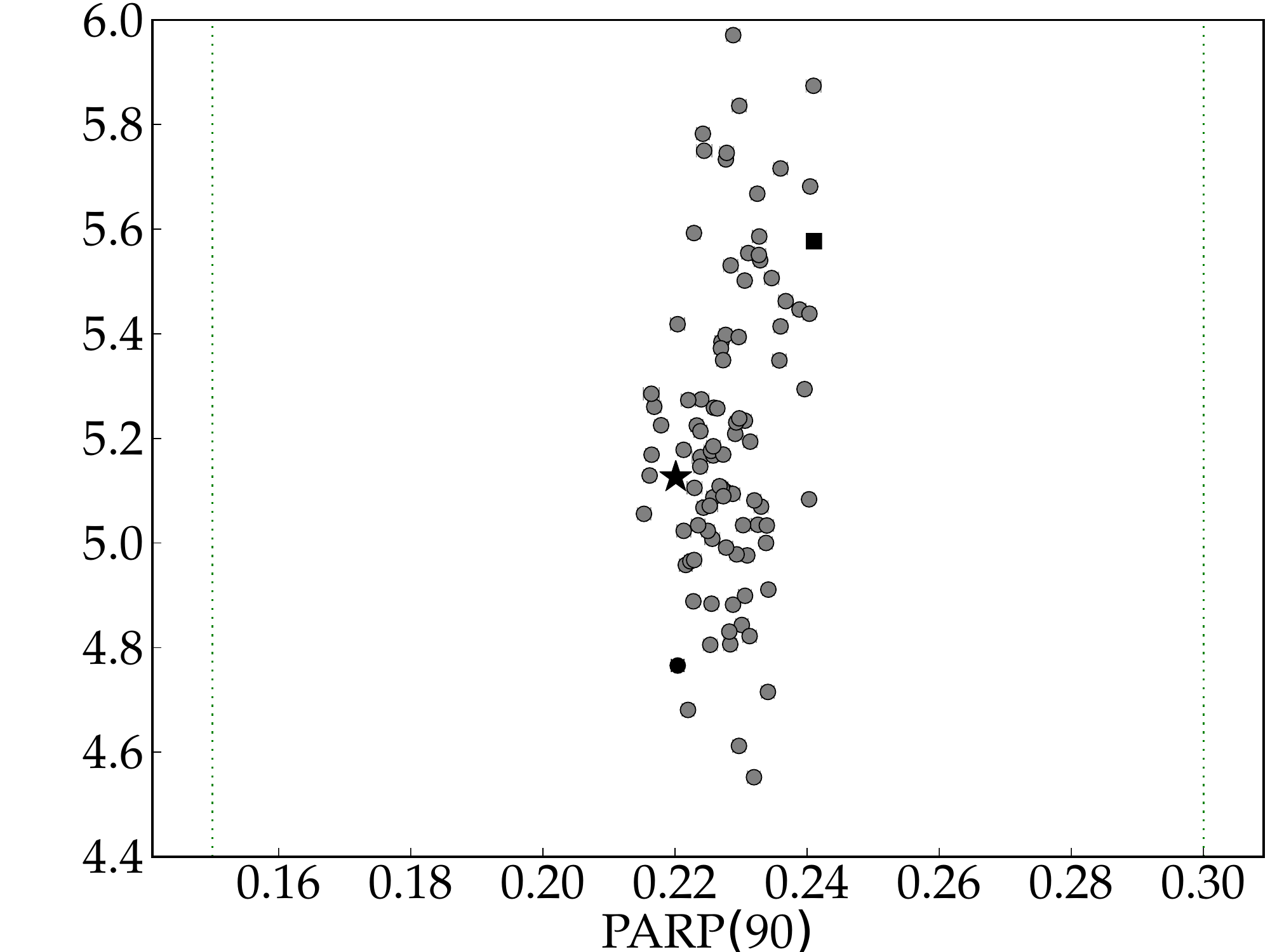}
  \includegraphics[width=0.325\textwidth]{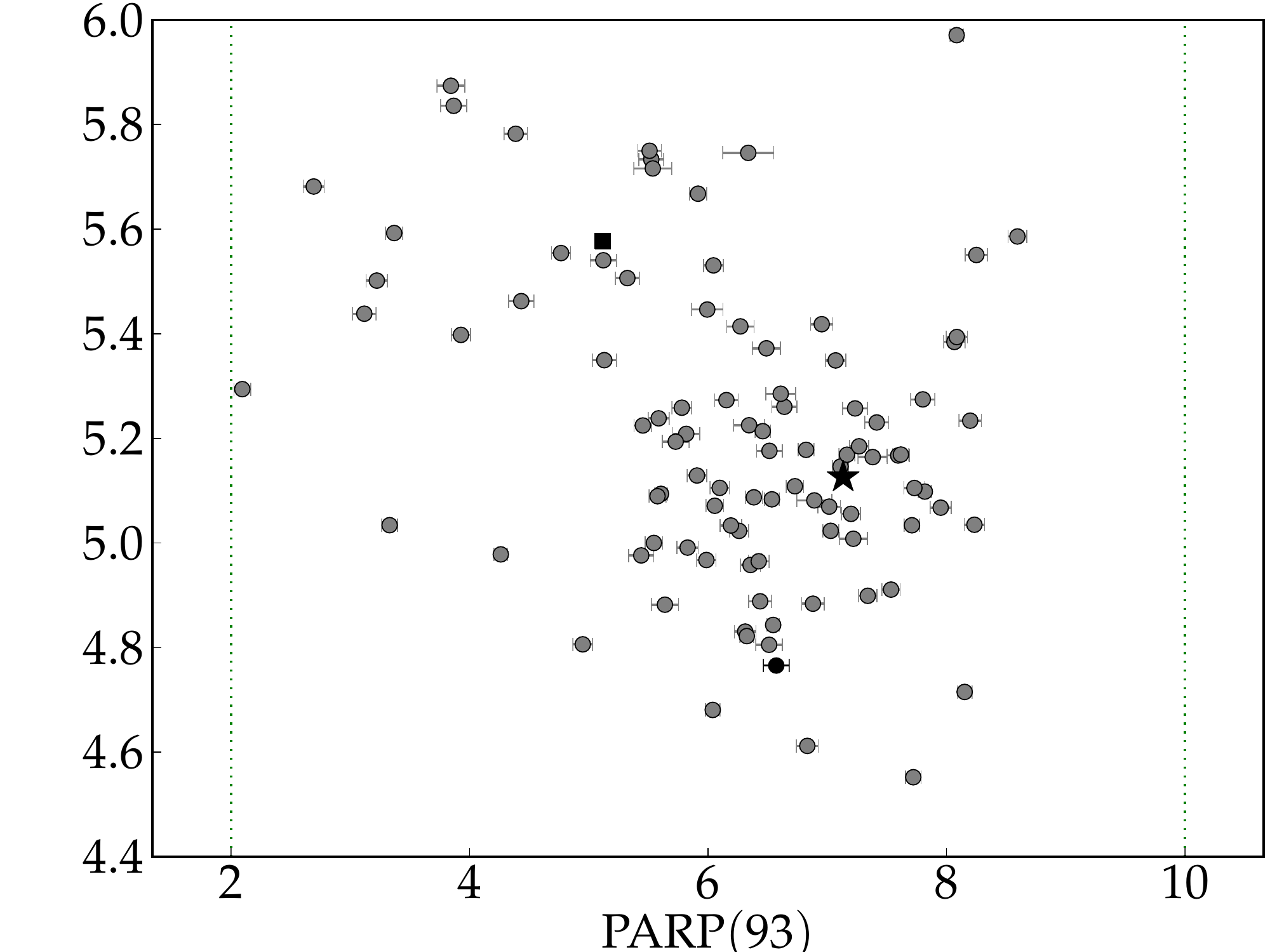}

  \caption{A scatter plot of \chisq vs. parameter value for a set of \professor
    run combinations on three \pythia\!6 parameters. Implicitly, this projection
    of tune parameter vectors on to parameter axes gives a qualitative measure
    of whether or not a parameter is well-constrained: these parameters become
    increasingly ill-defined from left to right.  The different markers
    represent different degrees of oversampling, with the star representing the
    maximum information run --- the points are for the same run combinations in
    all three scatter plots.}
  \label{fig:paramscatter}
\end{figure}
\textbf{}

To make a prediction of an observable for which there is no reference data (this
may be an energy extrapolation, or simply an unmeasured feature at existing
energies), the simplest approach is of course to run the generator with the
obtained tune(s) and compute the observable. Using the many tunes resulting from
the different run combinations will give a spread in the observable prediction,
reflecting one part of tune uncertainty. In practice, since we can build a fast
parameterisation of the generator behaviour (in fact many of them) on the unfitted
observable as part of the main \professor tuning process, this offers a much
faster turnaround than processing another large (perhaps \emph{very} large) set
of generator runs -- with the proviso that the parameterisations are of course
non-exact.

\subsection{Sources of tune uncertainty}

Before making this method more quantitive, we now consider the various sources
of uncertainty in the procedure outlined so far. This will help us to understand
which sources of uncertainty are computationally controllable and which will
have to, for now, remain more nebulous. These main sources of tuning error are
as follows:
\begin{enumerate}
\item Error on experimental reference data.
\item Statistical error on the MC at the anchor points from which the
  parameterisation is constructed.
\item Systematic limitations of the parameterising function to describe bin
  responses to parameter variations -- pathological MC parameters with
  discontinuous or critical behaviour are particularly hard to generically
  parameterise, since a Jacobian transformation to a suitable meta-parameter is
  not always easily available.
\item Choice of run combinations to make the parameterisation.
\item Goodness of fit definition, including both the type of GoF measure and the
  choice and relative weighting of different data.
\item Reasonable minimiser scatter within the \chisq valley containing the
  optimal tune point for a given parameterisation. Note that this cannot be
  completely disentangled from the role that error sources 1, 2 and 3 play in
  defining the \chisq valley.%
\item Limitations of the parameterisation(s) used to compute the parameterised
  MC value in extrapolated/unfitted observables. Of course, this error doesn't
  exist if the less convenient strategy of re-running the MC generator is used.
\item For completeness, we again highlight the systematic error associated with
  the discrete physics model being tuned: the total error is far from complete
  without considering more than one viable model. Within a given model there are
  also systematic uncertainties, some of which may be quantified,
  e.g. cross-section integration uncertainty and PDFs: the second of these is
  particularly quantifiable due to the existence of error or replica PDF sets,
  themselves expressing reasonable variations in PDF fitting.
\end{enumerate}

Note that, for example, these sources of uncertainty such as the variation
between members of the ensemble of run combinations are not unique errors
introduced by the \professor approach: failing to test different run
combinations does not \emph{eliminate} the error associated with the choice of
anchor runs used! A similar, but unquantifiable, error exists for any form of
manual tuning.

\subsection{Construction of tune uncertainty confidence belts}

Our approach to quantifying the uncertainties from (combinations of) the sources
listed in the previous section is to construct central confidence belts for
observable bins from the various ensembles of tune results, parameterisations,
etc. we have described. Explicitly, given a large number of reasonable and
equivalent predictions for an observable bin value, we construct a band of given
$P$-value as being the region containing fraction $P$ of predictions, with equal
fractions above and below.

There are many ways to construct such ensembles -- for the purposes of this
study we identify three:

\paragraph{Combination error:} The ensemble from which we construct the
confidence belt is simply the ensemble of predictions from different run
combinations. This will hence represent the variation due to error sources 1, 2,
and 4 -- the other sources of uncertainty exist, but are not quantified by this
approach\footnote{MC error is currently not \emph{explicitly} propagated into
  \professor's fit measure, due to stability problems: it enters implicitly via
  the statistical scatter of MC samples. This is being remedied.}.%
This approach requires that correlations between different run combinations are
small, which is ensured by the $N_\text{sampled} \gg N$ requirement.
If parameterisation (as opposed to explicit MC runs) is used for the translation
of this tune ensemble into bin value predictions, then source 7 also
applies. This can be included into the band construction by using many
parameterisations, again constructed from run combinations. Different
parameterisations should be used for minimisation and prediction to avoid
reinforcing parameterisation systematics.

\paragraph{Statistical error:} The obvious failure of the combination error
approach is that error source 6 -- the measure of reasonable tune variation
within the \chisq valley -- is left unquantified. Hence it will be no surprise
that our ``statistical'' error band is constructed explicitly from an ensemble
of samples from this valley. This is obtained in the simplest case by only using
the maximum information \professor tune -- that which is constructed from all
the available MC runs. The covariance matrix of the parameters in the vicinity
of the tune point is obtained -- in principle directly from the
parameterisation, in practice from the minimiser -- and used to define a rotated
hyper-Gaussian probability distribution in the parameter space. Sampling
parameter points from this distribution gives us another ensemble of tune points
and, as for the combination error, they can be mapped into observable
predictions either by direct MC runs or by using one or many constructed
parameterisations. In practice, we take advantage of the combined potential of
the hyper-Gaussian sampling and MC parameterisation to build a confidence belt
from \order{10,000} samples. The quantified sources of error are hence 1, 2, 6,
and 7.

These approaches to error band building will be used in the next section to
construct sample error bands for underlying event predictions. Although not
currently implemented in \professor, we also highlight the most complete form of
quantifiable error band within the \professor approach:

\paragraph{Combined error:} This extension is an obvious fusion of the above two
ensemble/band constructions: as in the combination error, we construct an
ensemble of points from run combinations, then for each run-combination point we
construct a statistical ensemble. The combined ensemble of tune ensembles, and a
variety of parameterisations to transform them into predictions will lead to
error bands quantifying error sources 1, 2, 4, 6, and 7. 

The remaining error sources are 3, 5 and 8: limitations of the parameterisation,
the observable weights/goodness of fit definition, and -- most importantly --
the uncertainty due to different physics models. These missing systematics
remain qualitative in this scheme, and reliable MC predictions should take care
to include estimates of their influence, albeit in a more ad hoc fashion than
for the more statistically-induced errors.


\subsection{Results}

We now briefly present results using the first two confidence belt definitions
presented above -- ongoing work is addressing the ``combined'' belt definition
and inclusion of PDF uncertainties.

\begin{SCfigure}[1.8][t]
  \scalebox{.25}{
    \input{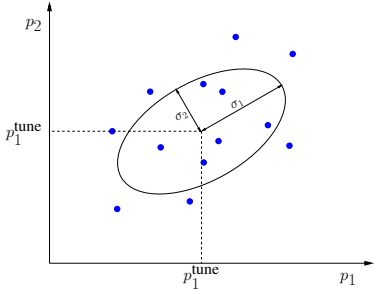}
  }
  \hspace{0.5cm}
  \caption{Two dimensional illustration of the parameter point sampling used for
    the statistical uncertainty estimate. We exploit the covariance matrix returned
    by the minimiser for a Gaussian sampling from the corresponding $p$-dimensional
    hyper-ellipsoid. The $\sigma_i$ ($i=1\ldots p$) are the eigenvalues obtained from
    an eigen-decomposition of the covariance matrix.}
  \label{fig:stat-sampling}
\end{SCfigure}

Our exploration is based on tunes of the \jimmy~\cite{Butterworth:1996zw,jimmy} MC generator, which
simulates multiple parton interactions (MPI) for
\herwigsix~\cite{Corcella:2002jc}, because it has only two relevant parameters,
PTJIM and JMRAD(73)\footnote{We treat the inverse radius-squared of the protons,
  JMRAD(73), to be identical to that of the anti-protons, JMRAD(91), and ignore
  the interplay with ISR parameters in \herwigsix itself.} -- the frugality with
parameters makes \jimmy an ideal ``toy model'' testbed, while remaining
phenomenologically relevant. As a \jimmy-like MPI model is ruled out by Tevatron
data~\cite{Bahr:2008wk}, we fix a dependence of PTJIM on the centre
of mass energy with the same ansatz as used in
\pythiasix~\cite{Sjostrand:2006za}:
\begin{align} 
  \label{eq:ptjim} 
  \text{PTJIM} =
  \text{PTJIM}_{1800}\cdot\left(\frac{\sqrt{s}}{1800~\text{GeV}}\right)^{\:0.274},
\end{align}
where $\text{PTJIM}_{1800}$ is the value of PTJIM at the reference scale
$\sqrt{s}=1800~\text{GeV}$ and is the \pTmin parameter actually used in the
tuning process. Furthermore, we use the MRST LO* PDF set~\cite{Sherstnev:2007nd} and
use \tevatron data from \cdf~\cite{Affolder:2001xt, Acosta:2004wqa,
  cdf-leadingjet} and \dzero~\cite{Abazov:2004hm} as a tuning reference. A more
complete tune would include the exponent of the \pTmin energy dependence, but
for toy-study purposes we here fix it to a value consistent with known energy
extrapolation fits~\cite{Buckley:2009bj,Skands:2009zm}.

\begin{figure}[t]
  \centering
  \subfigure[Statistical uncertainty, data only]{\label{fig:statband-Nch}
    \includegraphics[width=0.45\textwidth]{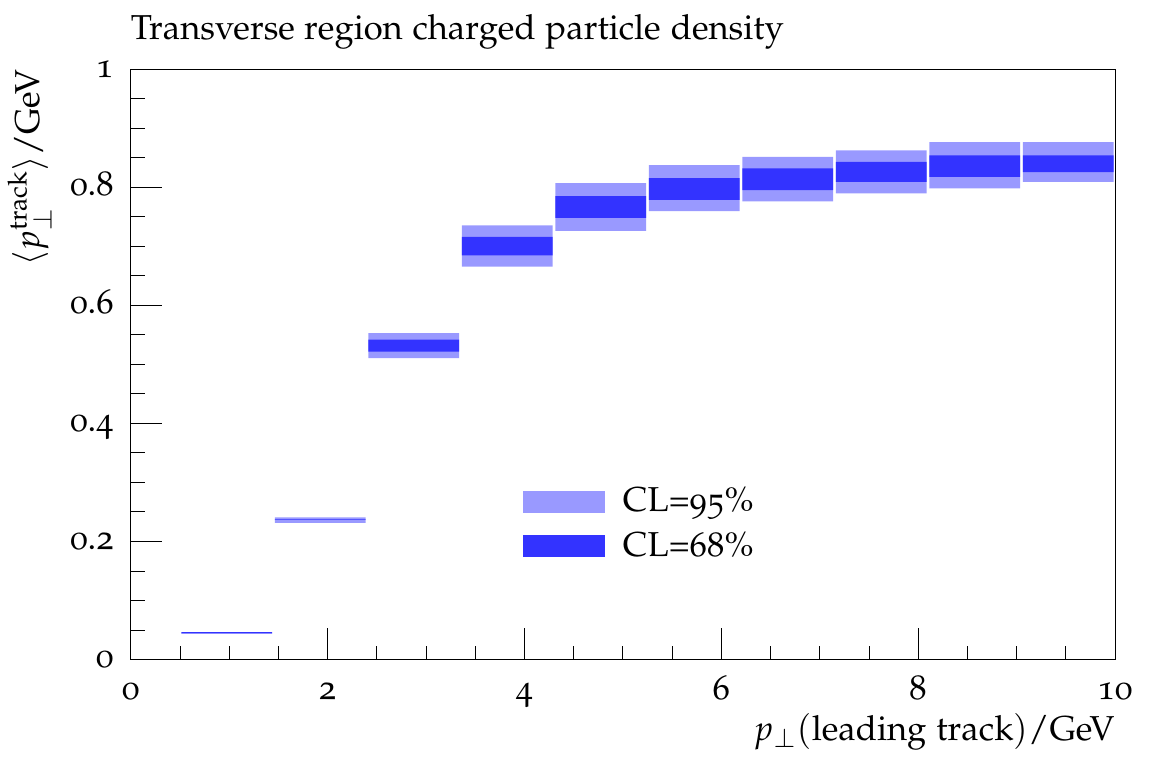}
  }
  \subfigure[Statistical uncertainty, data and pseudodata]{\label{fig:statband-Nch-pseudo}
    \includegraphics[width=0.45\textwidth]{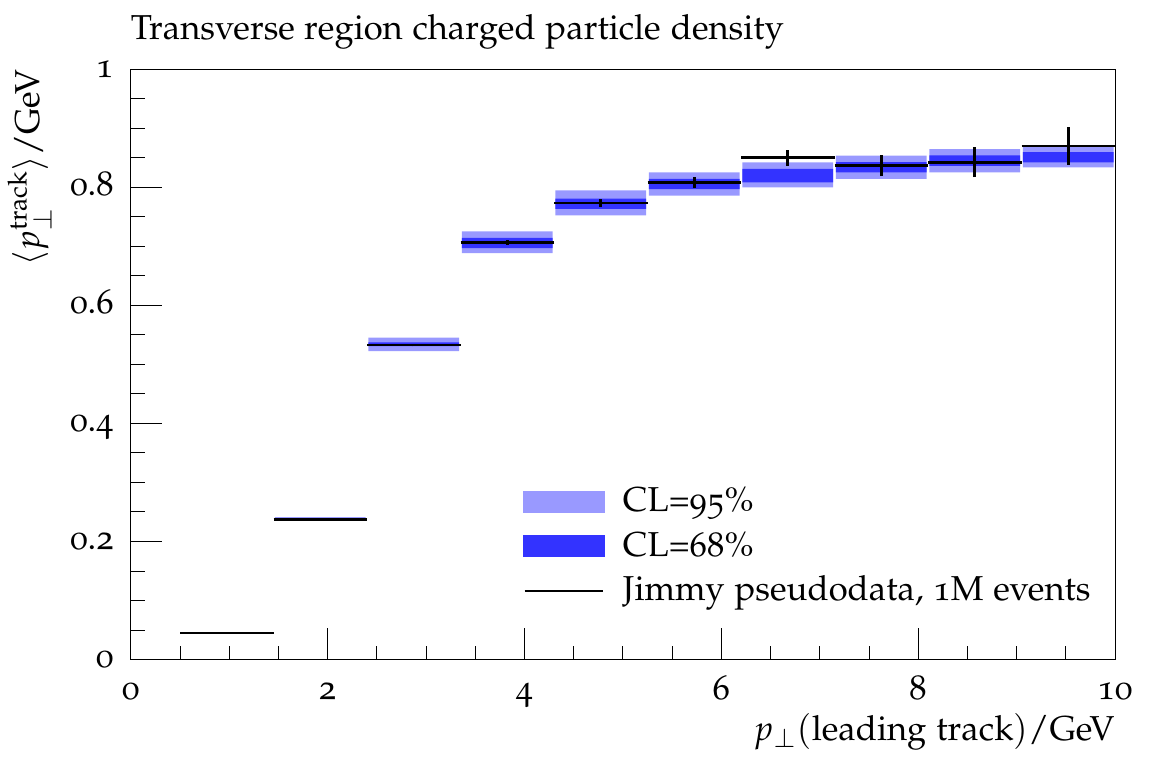}
  }\\
  \subfigure[Combination uncertainty, data only]{\label{fig:sysband-Nch}
    \includegraphics[width=0.45\textwidth]{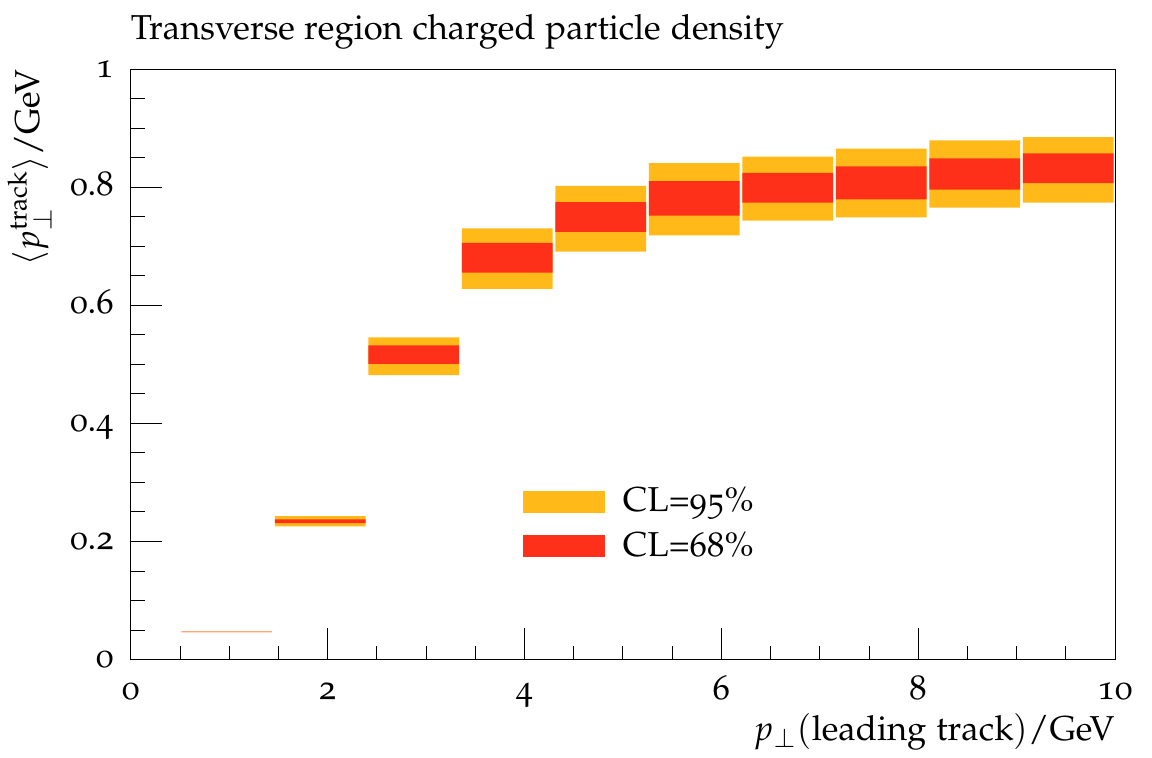}
  }
  \subfigure[Combination uncertainty, data and pseudodata]{\label{fig:sysband-Nch-pseudo}
    \includegraphics[width=0.45\textwidth]{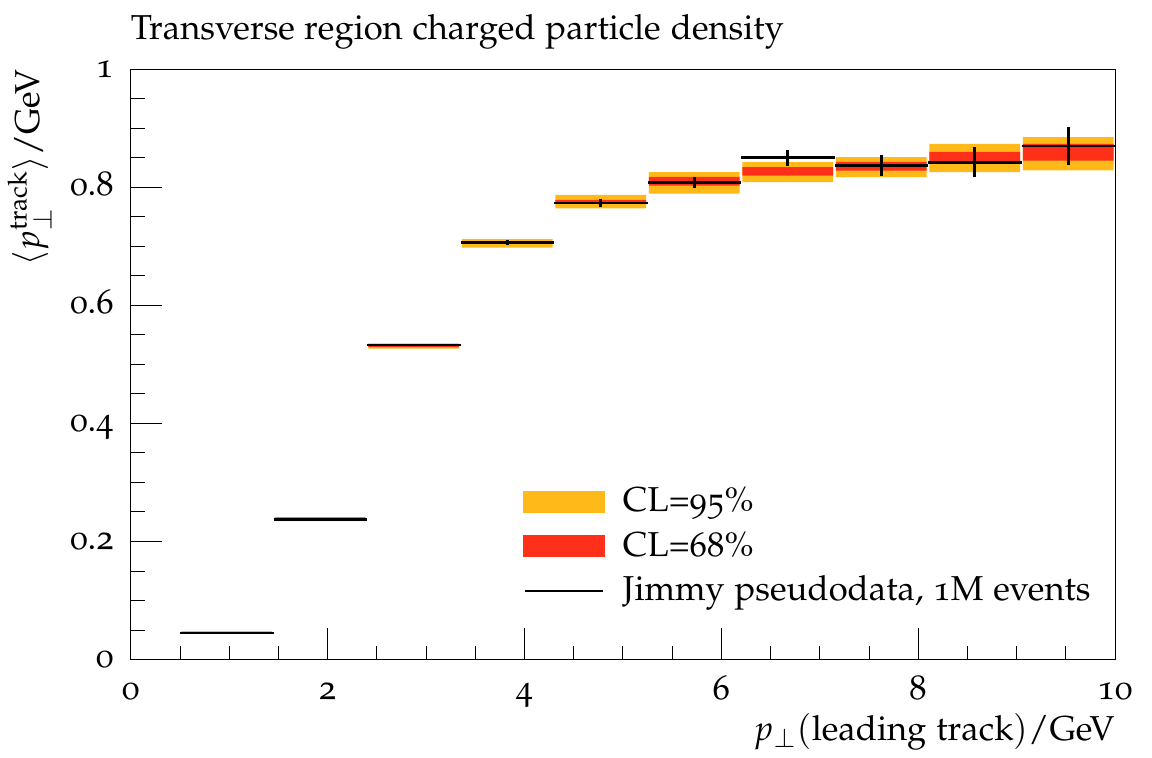}
  }
  \caption{``Statistical'' and ``combination'' error bands for transverse
  $N_\text{ch}$ flow at \unit{7}{\TeV} before and after adding 1M events of
  pseudodata of this observable (black markers) to the tuning. The error bands
  are calculated from the central 95 (68)\% of the binvalues of an ensemble of
  10000 histograms each.}
  \label{fig:statsysbands-Nch}
\end{figure}

In Figures~\ref{fig:statband-Nch} and \ref{fig:sysband-Nch}, the statistical and combination error band
definitions are shown for a \unit{7}{\TeV} underlying event observable --
$N_\text{ch}$ flow transverse to the leading track (track with the largest
transverse momentum in an event), as a function of leading track \pT -- computed
in \rivet~\cite{Buckley:2010rivet}, based on the fit to Tevatron reference data. The
error due to variation of run combinations (error source 4) is notably somewhat
larger than the scatter of points in the \chisq valley (error source 6),
indicating that in the \jimmy model the parameters have a strong influence on
this observable, and are hence well-constrained. In
Figures~\ref{fig:statband-Nch-pseudo} and \ref{fig:sysband-Nch-pseudo}, similar
band constructions are shown, but in this case the same \rivet analysis has been
used to simulate the effect of adding 1M events of UE data at \unit{7}{\TeV}
into the fit: the size of both error bands is reduced, as expected.

%


\subsubsection{Effect of extrapolation}

Finally, we consider systematically how error bands constructed in this way
behave as extrapolations are taken further from the region of constraining
data. In this case, since the computational requirements are significant, we
only consider the ``statistical'' error band.

We use the transverse region $N_\text{ch}$ density UE observable, evaluated at a range of ten centre
of mass energies, $\sqrt{s}_i$, between 200~\GeV and 14~\TeV. We then apply the
following procedure,
\begin{itemize}
\item Construct the maximum-information parameterisation of the generator
    response for the $N_\text{ch}$ UE observable, $f(\sqrt{s}_i)$
\item Produce an ensemble of 10000 histograms, $H_i$, of the observable shown in
  \FigRef{fig:transv-sub1} (blue line) using the corresponding $f(\sqrt{s}_i)$
  and points sampled using the procedure illustrated in \FigRef{fig:stat-sampling}
\item Calculate the mean height of the $N_\text{ch}$ plateau, $M_i$, for each of
  the $H_i$
\item Construct a $95\%$ central confidence belt, $\mathrm{CL}(\sqrt{s}_i)$ from the $M_i$
\end{itemize}

In \FigRef{fig:transv-sub2} the $\mathrm{CL}(\sqrt{s}_i)$ are drawn. We observe a very
tight confidence belt for the energy region of the \tevatron experiments,
while the confidence belt becomes wider for extrapolation to \lhc energies. The
definitions of the plateau regions used can be found in \TabRef{tab:plateaudefs}.

It is notable that these bands are narrow -- sufficiently so that they
have been visually inflated by a factor of 10 in the figure. While
this reflects good stability in the tuning system, it is probably an
underestimate of the true model uncertainty. A more complete study
will include the exponent of energy extrapolation in
equation~\eqref{eq:ptjim} in the tune, since this may be a dominant
effect in the errors for this particular observable and its inclusion
will give more freedom for various features of the model to balance
against each other: the toy tuning of the energy-dependent \jimmy
model shown here is probably too restrictive to accurately represent
the full range of variation allowed by the model, but serves as an
indication of the extent to which such studies can be systematised.


\begin{figure}[t]
  \begin{center}
    \subfigure[$N_\text{ch}$ vs. \pTlead, plateau]{
      \includegraphics[width=.45\textwidth]{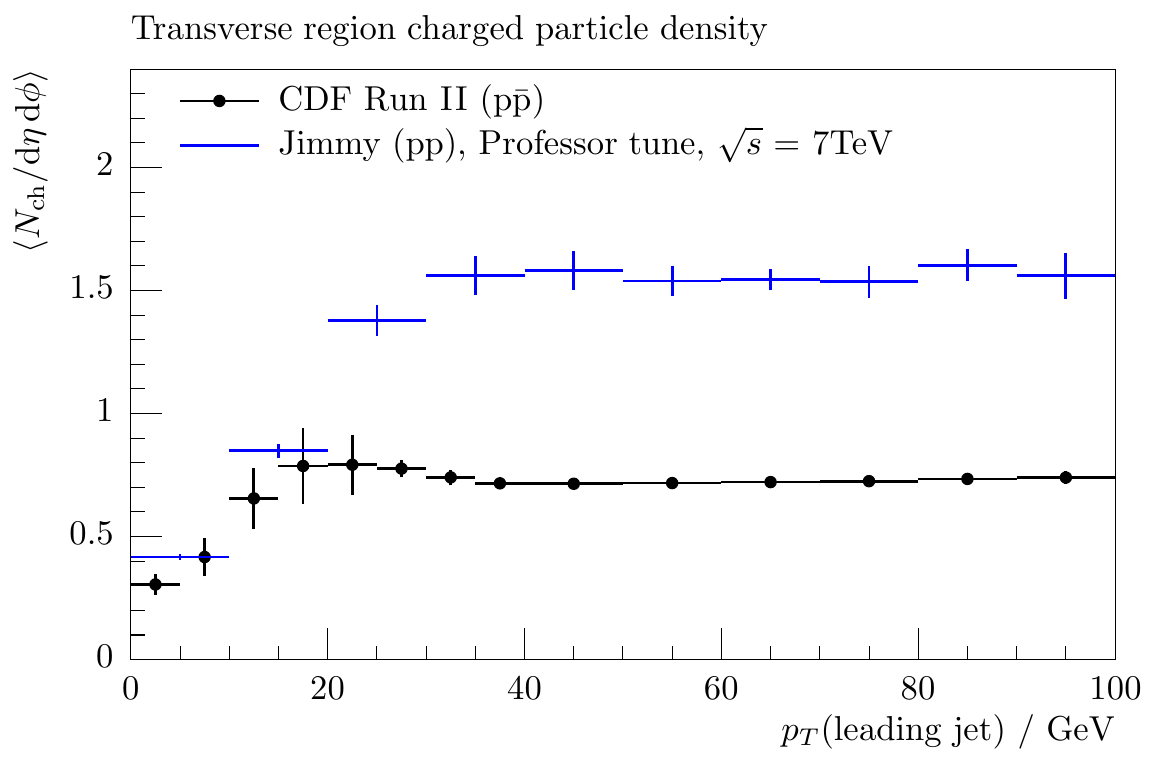}
      \label{fig:transv-sub1}
    }
    \subfigure[$N_\text{ch}$ vs. \pTlead, mean of plateau vs. $\sqrt{s}$]{
      \includegraphics[width=.45\textwidth]{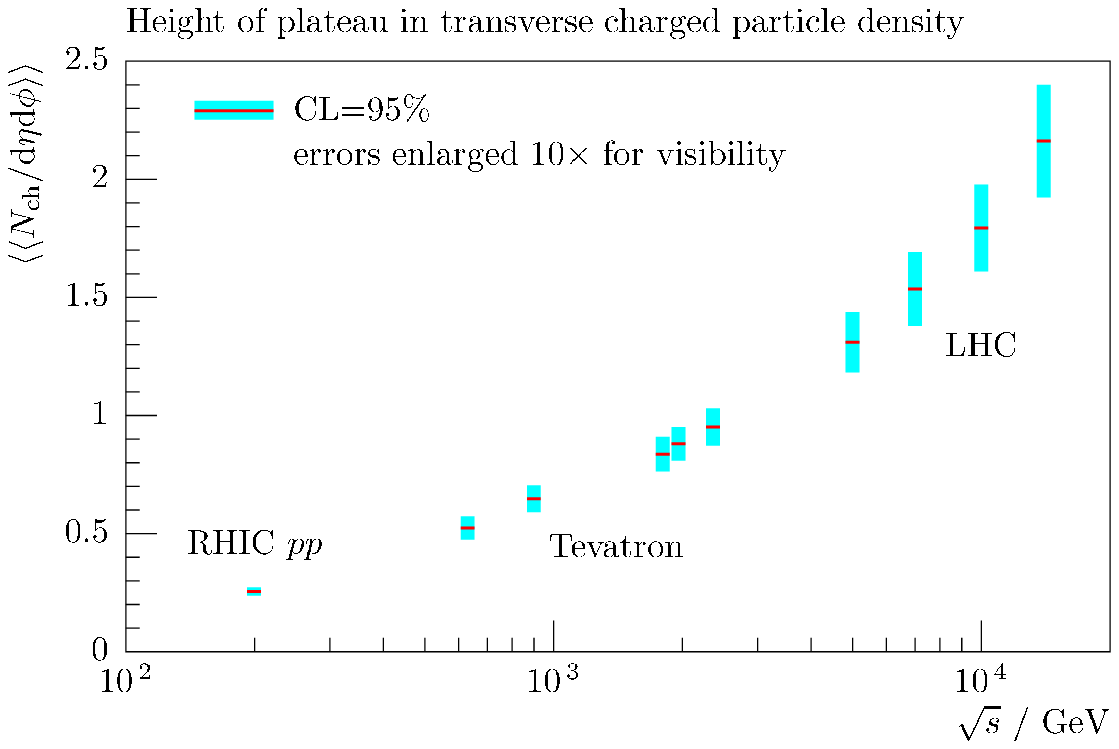}
      \label{fig:transv-sub2}
    }
  \end{center}
  \caption{Transverse region charged particle density.
    \subref{fig:transv-sub1}: the typical plateau observed and
    \subref{fig:transv-sub2}: mean-height of that plateau as a function of
    $\sqrt{s}$.} \label{fig:transv}
\end{figure}

\begin{table}[tb]
\begin{center}
  \begin{tabular}{lrrrrrrrrrr}
      \toprule
      $\sqrt{s}/\TeV$ & 0.2 & 0.63 & 0.9 & 1.8 & 1.96 & 2.36 & 5.0 & 7.0 & 10.0 & 14.0 \\ 
      \midrule
      \pTleadmin/\GeV & 10  &  30  & 30  & 30  & 40   &  40  &  40 & 40  & 40   & 40 \\   
      \pTleadmax/\GeV & 30  &  70  & 80  & 80  & 90   & 110  &  110 & 120 & 150  & 160 \\
      \bottomrule
  \end{tabular}
  \caption{Definition of plateau regions (\pTlead) used in the extrapolation study.}
  \label{tab:plateaudefs}
\end{center}
\end{table}

\subsection{Conclusions}

We have catalogued a set of sources of uncertainty which either explicitly or
implicitly contribute to any tune of a MC event generator, and presented a
systematic approach to quantifying many of these uncertainties using the fast MC
parameterisations and natural tune ensembles which arise from the \professor
tuning approach. Example results have been shown, which exhibit some expected
behaviours, such as the shrinking of error bands on adding new reference data in
new areas of parameter space and the blow-up of error bands as predictions
venture further into unconstrained regions.

Several things should be emphasised: first and most important is that this
approach does not catch all sources of error. We have presented results from two
definitions and have proposed a third, more comprehensive measure, but still
variations such as PDF errors and discrete model variations need to be
included. However, with the increased usage of systematic tuning methods,
variations between models in UE observables are not as substantial as once they
were -- statistical errors are a non-negligible factor in assessing the
reliability of phenomenologically-based MC predictions.

The approach taken here has many obvious parallels in the world of PDF errors,
with our approach having a good deal of overlap with the MC replica set approach
of the NNPDF collaboration~\cite{Ball:2008by} as contrasted with the eigenset approach
of the CTEQ and MRST/MSTW collaborations. While replica sets have the advantage
of a more direct statistical uncertainty interpretation (although we do not have
the option of the parameterisation-freedom exibited by the NNPDF use of neural
networks), there is the pragmatic issue that \order{10} representative error
tunes would be more usable than \order{10000} equivalent tunes. Whether such a
concept can be statistically constructed remains to be seen -- for now the
Perugia-soft/hard tunes remain the obvious tool of choice.




\subsection*{Acknowledgements}
Our particular thanks goes to Luigi~del~Debbio and Richard~Ball for discussions
about statistical coverage and MC error estimation. The Professor collaboration
acknowledges support from the EU MCnet Marie Curie Research Training Network
(funded under Framework Programme 6 contract MRTN-CT-2006-035606) for financial
support and for many useful discussions and collaborations with its members.
A.~Buckley additionally acknowledges support from the Scottish Universities
Physics Alliance (SUPA); H.~Schulz acknowledges the support of the German
Research Foundation (DFG).


\clearpage

\section[MATRIX ELEMENT CORRECTIONS AND  PARTON SHOWER MATCHING IN INCLUSIVE Z PRODUCTION AT LHC]
{MATRIX ELEMENT CORRECTIONS AND  PARTON SHOWER MATCHING IN INCLUSIVE Z PRODUCTION AT 
LHC\protect\footnote{Contributed by: P.~Lenzi, P.~Skands}}

\subsection{INTRODUCTION}

In this paper we compare \pythiasix\cite{Sjostrand:2006za}, \alpgen\cite{Mangano:2002ea} and 
\sherpa\cite{Hoeche:2009rj,Gleisberg:2008ta} in inclusive $Z$ production at LHC, with  10 TeV center of mass energy. 
A disagreement in various observables, especially the $Z$ transverse momentum, $\pt$, between \alpgen and \pythiasix 
was recently reported in \cite{Lenzi:2009fi}.
Namely, the shape in the low $Z$ $\pt$ region was significantly different when comparing \alpgen showered 
with \pythiasix and  matrix element corrected \pythiasix. 
This was traced back to a change in shape in \pythiasix when matrix element corrections are switched off 
(this setup is used when showering \alpgen events, because higher multiplicity \alpgen samples already provide 
higher order corrections).
This is summarized by the $Z$ $\pt$ plot shown in Fig.~\ref{fig:zbenchmarkmatching_referencePtZ}.
\begin{figure}[!b]
\begin{center}
\includegraphics[width=0.5\textwidth]{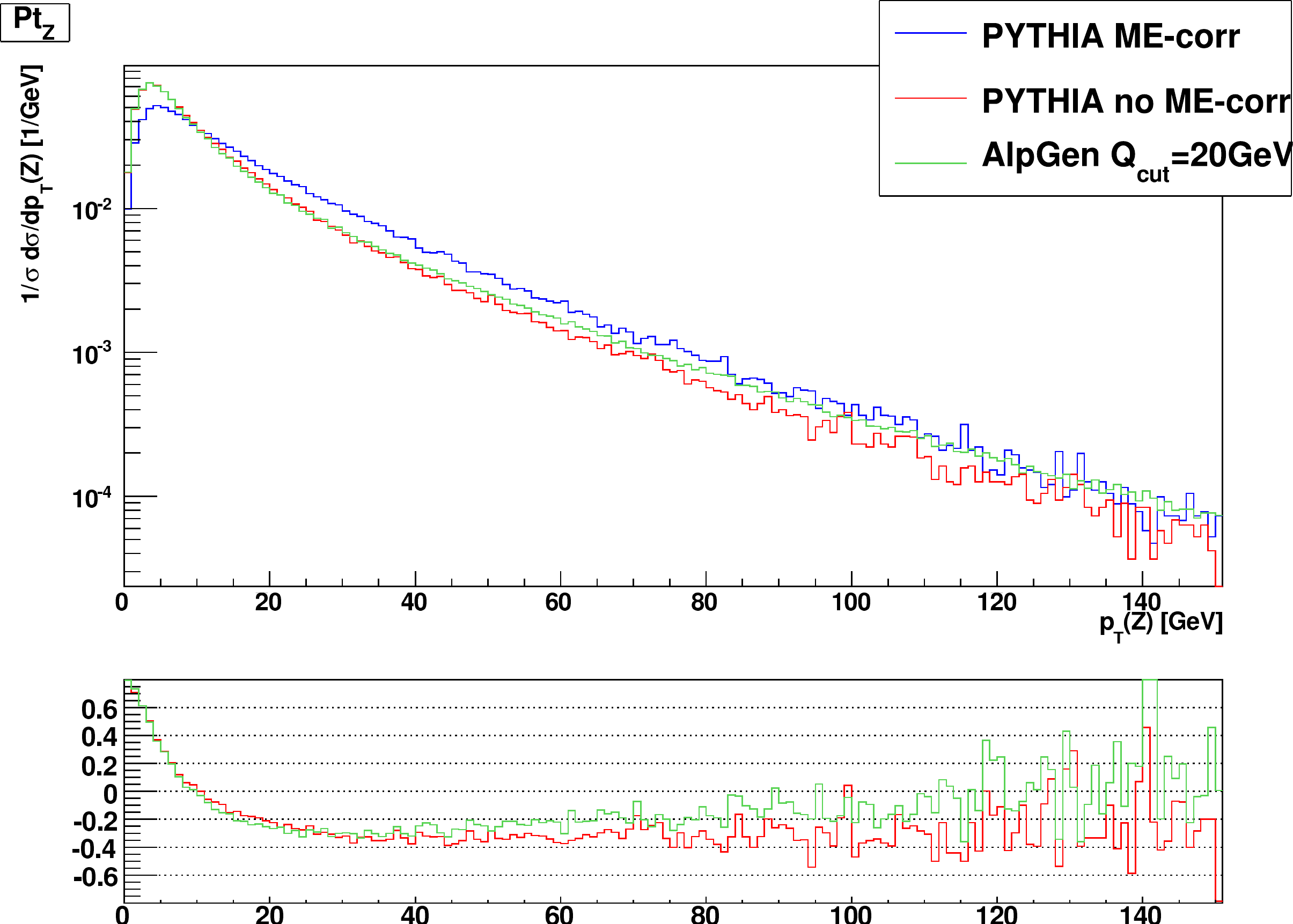}
 \caption{Z transverse momentum spectrum for \pythiasix (corrected and uncorrected) and for \alpgen plus 
\pythiasix as in \cite{Lenzi:2009fi}. \alpgen plus \pythiasix follows uncorrected \pythiasix at low $\pt$. 
The change in shape of uncorrected with respect to corrected \pythiasix was responsible for the disagreement.  
}
\label{fig:zbenchmarkmatching_referencePtZ}
\end{center}
\end{figure}
A bug in \pythiasix causing this behavior was later corrected; here we present results obtained with the 
new version of \pythiasix.\\
We also compare different \pythiasix tunes, checking both the virtuality ordered and the newer transverse momentum ordered shower.\\  
Using \pythiasix as a reference we check the performances of the matching in \alpgen and \sherpa. Since \pythiasix is fully corrected for first order real emission, we expect agreement with \alpgen and \sherpa when those are configured to include matrix elements for $Z$ plus up to one additional parton.\\
Results presented in this work are obtained using \pythiasix versions 6.411 and 6.421 (the last contains the bug fix for the matrix element corrections problem), \alpgen 2.13 and \sherpa 1.2.0.

\subsection{ANALYSIS SETUP}
All generators have been processed through the same analysis, selecting events with a $Z$ with mass between 66~GeV and 116~GeV, requiring the two leptons to have $\pt$ greater than 20~GeV and pseudorapidity between -2.5 and 2.5.
Jets are reconstructed with the $\kt$ algorithm, with a radius of 0.4 and a minimum $\pt$ of 30 GeV.
In order to decouple effects due to multiple interactions, hadronization and QED radiation off leptons, these have been switched off for all the generators used. 

The analyses were carried out using \rivet (see Section~\ref{sec:rivet}).

\subsection{MATRIX ELEMENT CORRECTIONS IN PYTHIA}
Matrix element corrections in \pythiasix are described in detail in \cite{Sjostrand:2006za}. 
Here we just remind the basics of the procedure. \pythiasix modifies the shower in two steps: first the starting scale is raised to the hadronic center of mass energy (power shower configuration) so that any hard emission from the shower is kinematically possible; then the probability for the first emission (which is also the hardest in \pythiasix) is re-weighted to include the first order real emission contribution. 
The reweighting factor is calculated as the ratio between the exact real emission matrix element and the splitting function used in the shower. 

\begin{figure}[!b]
\begin{center}
\includegraphics[width=0.65\textwidth]{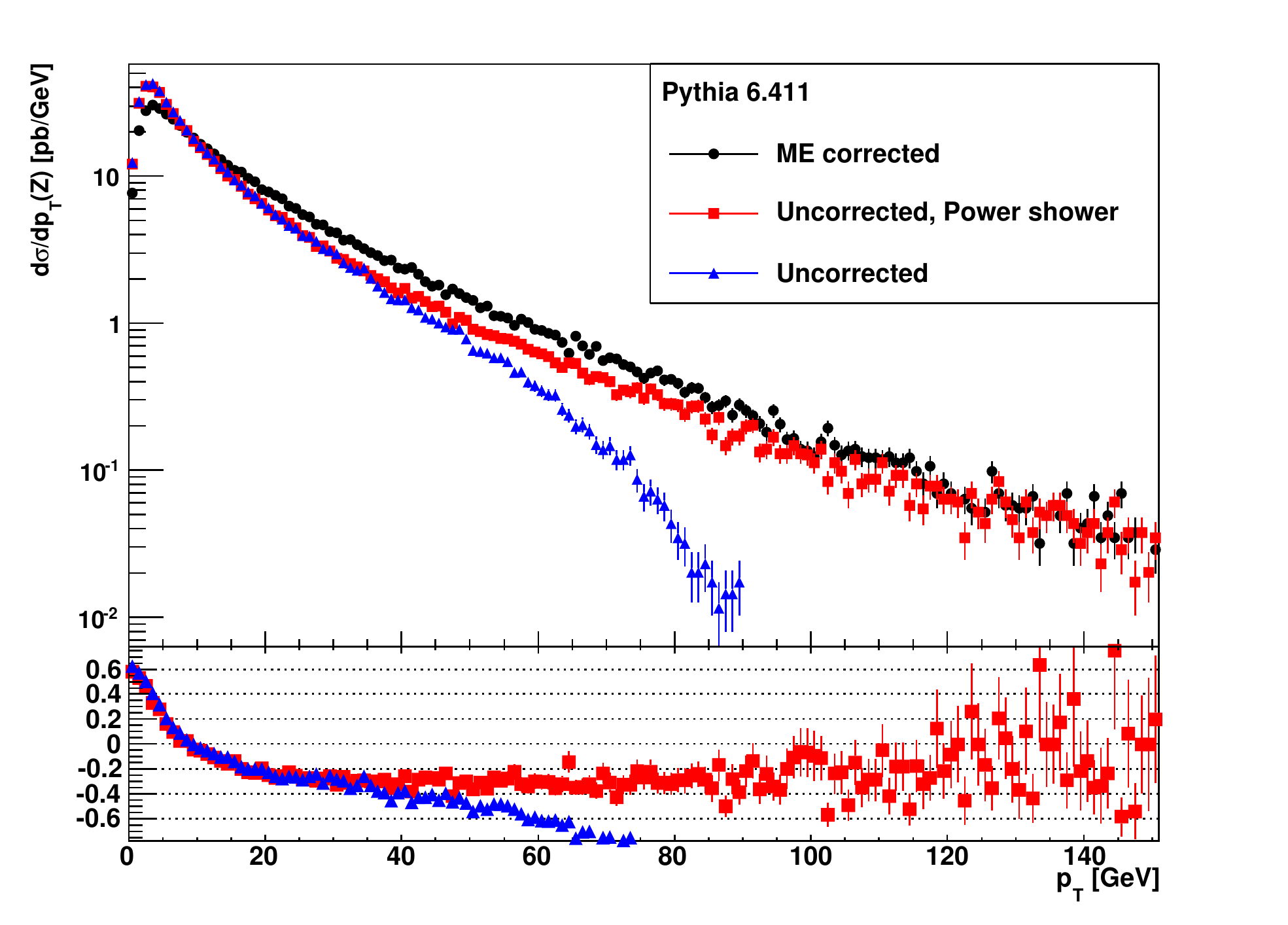}
 \caption{Z transverse momentum spectrum for three different matrix element correction settings in \pythiasix.411: with matrix element corrections, with power shower only, without matrix element corrections. The relative difference with respect to the matrix element corrected curve is shown in the lower plot.
}
\label{fig:zbenchmarkmatching_ptz6411}
\end{center}
\end{figure}
\begin{figure}[!h]
\begin{center}
\includegraphics[width=0.65\textwidth]{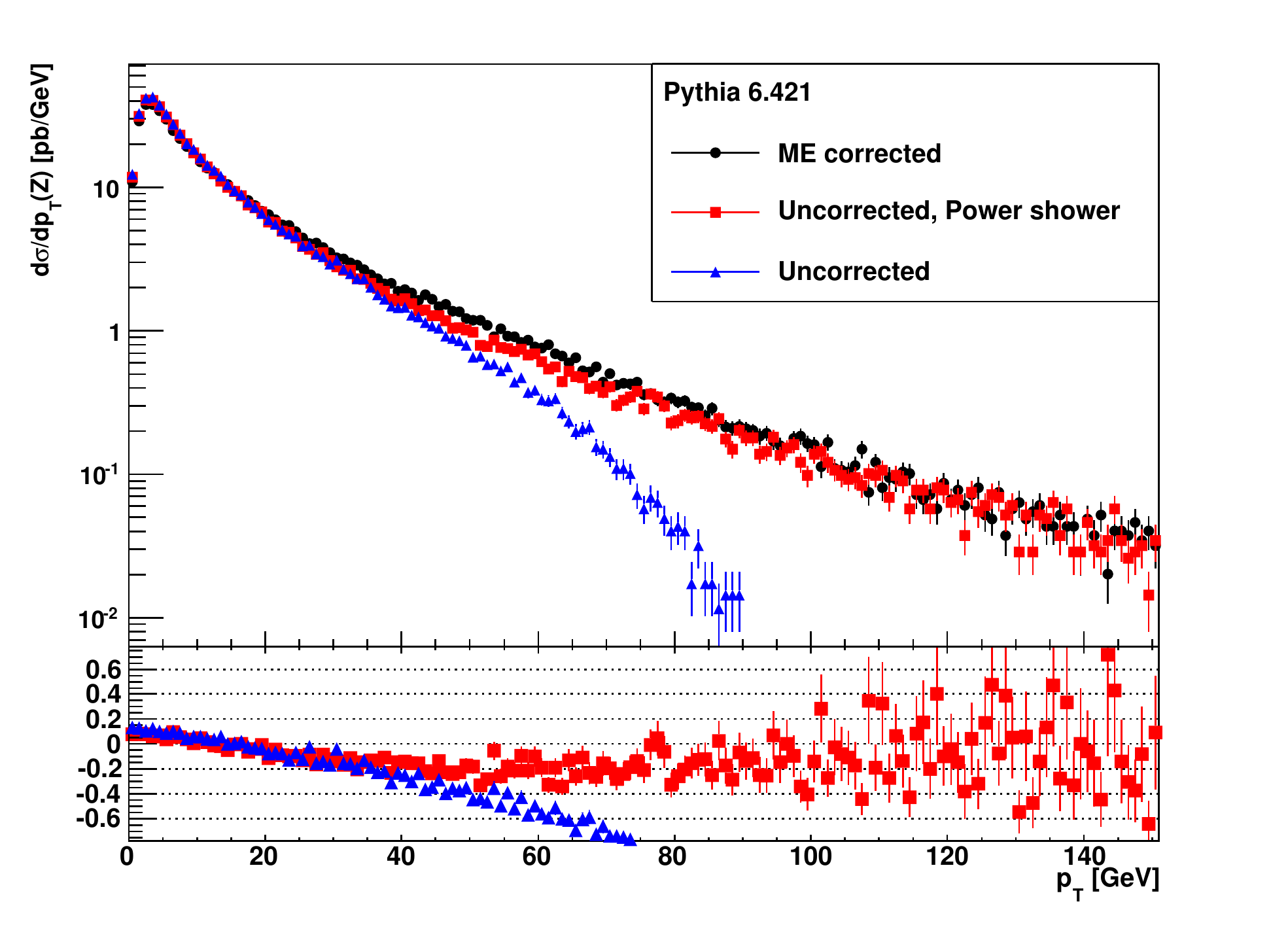}
 \caption{Z transverse momentum spectrum for three different matrix element correction settings in \pythiasix.421: in this new version, deactivating matrix element corrections has a small effect at low $\pt$, as expected.
}
\label{fig:zbenchmarkmatching_ptz6421}
\end{center}
\end{figure}
The effect of matrix element corrections in \pythiasix.411 is summarized in Fig.~\ref{fig:zbenchmarkmatching_ptz6411} for $Z$ transverse momentum spectrum. The three curves are  obtained with full matrix element corrections, with power shower only, and with uncorrected shower with starting scale at the $Z$ mass. 
The relative difference with respect to the ME corrected result is shown in the lower plot.
The shape of the spectrum changes significantly when matrix element corrections are switched off, even at low $\pt$ (the relative difference plot does not flatten as $\pt$ goes to 0). 
This is unexpected, as the matrix element effect is supposed to go to zero at low $\pt$, where the description from the shower is already accurate. In fact, the splitting function used in the shower is the approximation for $\pt\rightarrow{0}$ of the of the exact matrix element for emission of one parton, so no difference between the two descriptions is expected in that region.
The effect of matrix element corrections is illustrated in Fig.~\ref{fig:zbenchmarkmatching_ptz6421} for the newer 
6.421 version of \pythiasix. 
The same collection of curves as for version 6.411 is shown. 
With the newer version the deactivation of matrix element corrections has a very small effect at low $\pt$, as 
expected. 
A bug was found and corrected between the two versions which was causing the behavior observed in the earlier version.

\subsection{PYTHIA TUNES}
We compared various tunes in \pythiasix.421 on the basis of three observables, the $Z$ transverse momentum, 
the jet multiplicity and the leading jet transverse momentum.
In all of our comparisons we switched off the multiple interactions simulation, which means that we are comparing the effect of different tunings on the hard event simulation and on the shower.

We compared the following tunes: DW~\cite{Albrow:2006rt}, D6T~\cite{Albrow:2006rt}, Pro-Q0~\cite{tunePro}, 
Perugia0~\cite{Sjostrand:2004ef}.
The virtuality ordered shower is used in the first three, while the last uses the $\pt$ ordered shower.

Z transverse momentum spectra obtained with the four mentioned settings are shown in 
Fig.~\ref{fig:zbenchmarkmatching_ptz_tunes}.
\begin{figure}[!h]
\begin{center}
\includegraphics[width=0.7\textwidth]{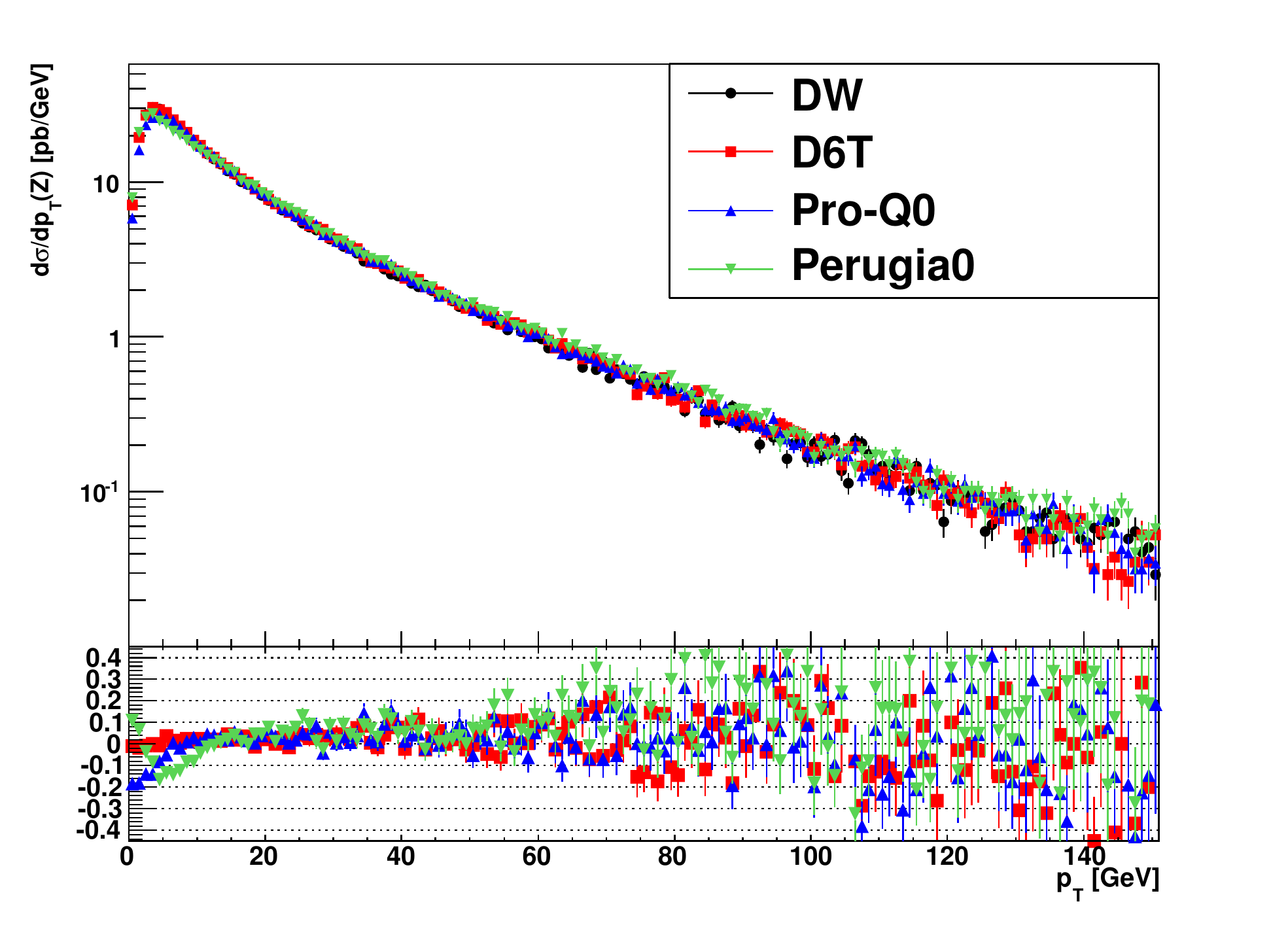}
 \caption{$Z$ transverse momentum spectrum for different \pythiasix tunes. 
Relative difference is shown with respect to tune DW.
}
\label{fig:zbenchmarkmatching_ptz_tunes}
\end{center}
\end{figure}
The agreement is generally good, tune DW and D6T agree fairly well on the whole spectrum; some 
discrepancies are observed in the low $\pt$ region: both tune Perugia0 and tune Pro-Q0 differ from Tune 
DW, and they also differ from each other. The disagreement with respect to tune DW is up to 20\%. 

Jet multiplicity and leading jet transverse momentum spectrum are shown in 
Fig.~\ref{fig:zbenchmarkmatching_jetmulti_jetpt_tunes}.
The jet multiplicity for Tune Perugia0 shows some not negligible differences with respect to the other tunes, 
predicting more jets. 
On the other hand Tune DW, D6T and Pro-Q20 predict very similar rates.
Concerning the leading jet $\pt$ spectrum in Fig.~\ref{fig:zbenchmarkmatching_jetmulti_jetpt_tunes} (b), 
the shape predicted by the four tunes is very similar.
\begin{figure}[!h]
\begin{tabular}{c c}
\includegraphics[width=0.45\textwidth]{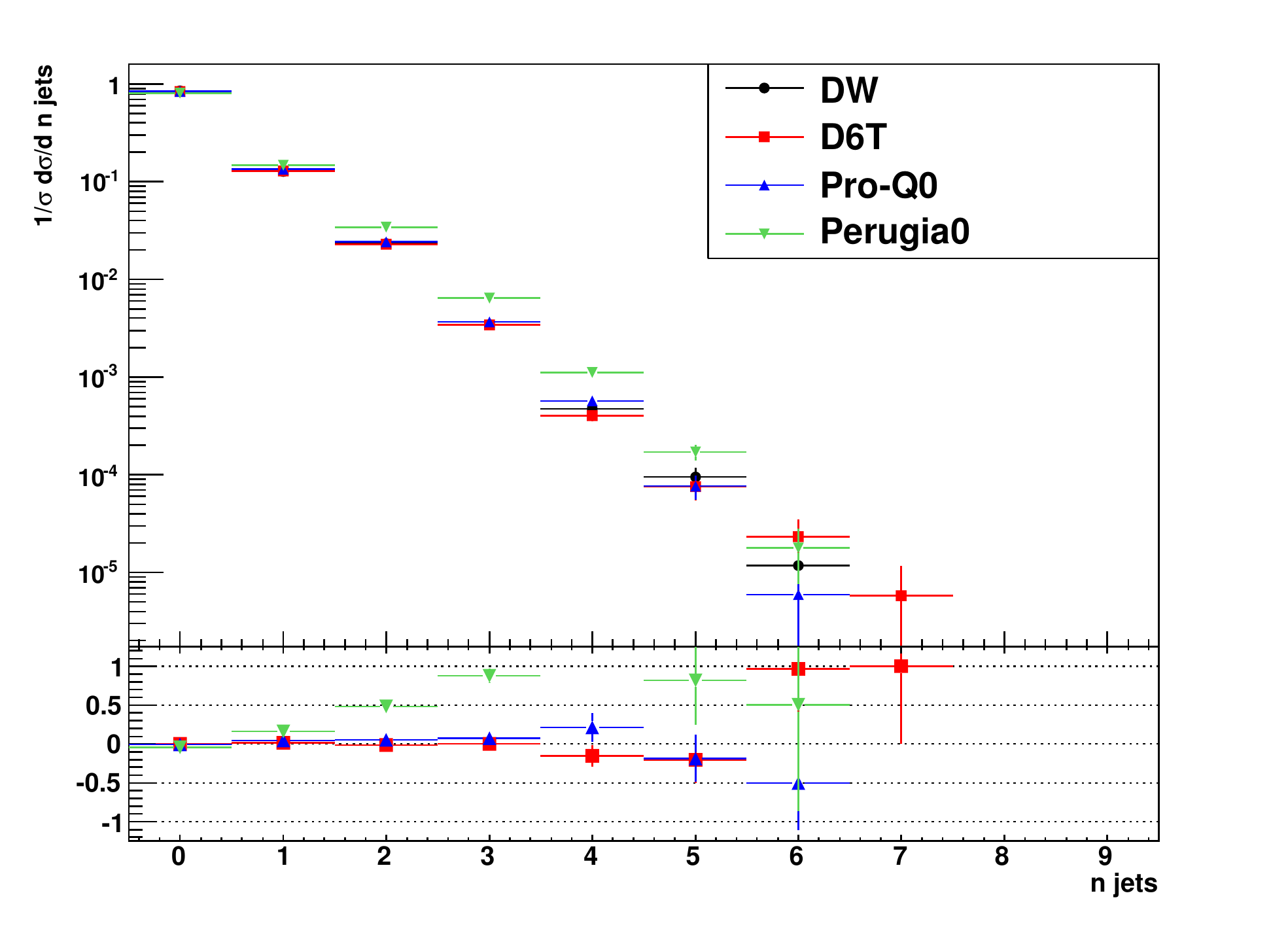} &
\includegraphics[width=0.45\textwidth]{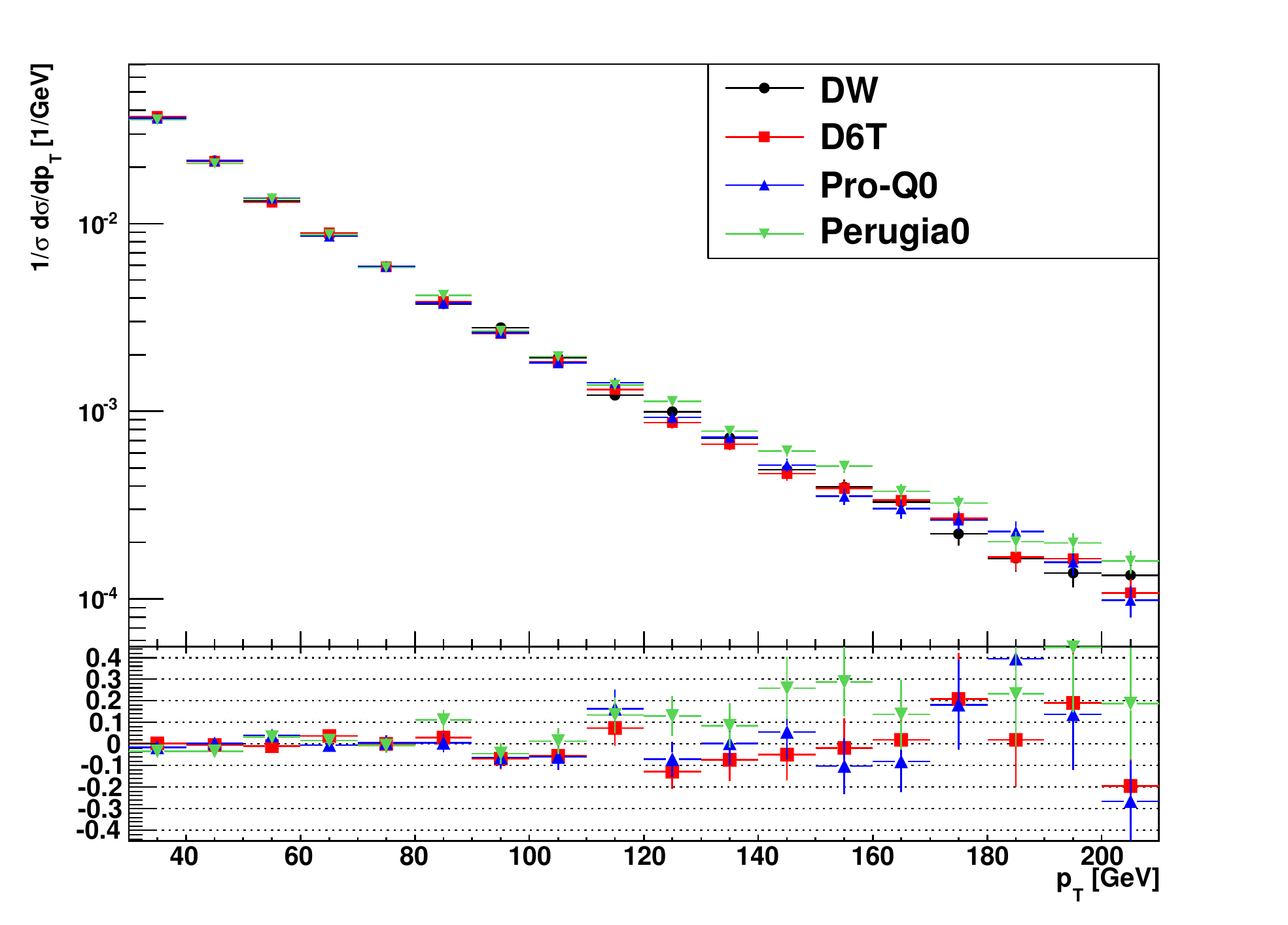} \\
(a) & (b)
\end{tabular}
\caption{(a) Jet multiplicity, (b) leading jet $\pt$ spectrum for four \pythiasix tunes. Relative difference is shown with respect to Tune DW.}
\label{fig:zbenchmarkmatching_jetmulti_jetpt_tunes}
\end{figure}

\subsection{\pythiasix, \sherpa, \alpgen COMPARED}
In this section we compare \pythiasix (including full matrix element corrections) with \alpgen and \sherpa.
We used \pythiasix version 6.421, both standalone and to shower \alpgen events. Also, we used tune DW for \pythiasix.

We used matrix element corrected \pythiasix as a reference for a consistency check of the  matching prescriptions 
used in \alpgen and \sherpa. 
Both the CKKW\cite{Catani:2001cc} prescription used in \sherpa and MLM prescription used 
\alpgen~\cite{Mangano:2001xp,Mangano:2002ea} can describe multiple parton emission 
corrected for the corresponding multi-parton matrix element. If just one additional parton emission from the matrix 
element is permitted, those prescriptions should give results compatible with \pythiasix.

While CKKW and MLM divide the phase space in a region of jet production, populated by the matrix element, 
and a region of jet evolution populated by the shower, the matrix element correction prescription implemented in 
\pythiasix does not depend on any separation parameter, thus providing us with the ``correct'' reference to test 
the other matched calculations.

The comparison among the three generators is shown in Fig.~\ref{fig:zbenchmarkmatching_ptz_alpshepyt} for the $Z$ 
transverse momentum.
The agreement in shape is generally good, well within 20\% with respect to \pythiasix for most of the spectrum.
In particular, the mismatch between \alpgen and \pythiasix observed in \cite{Lenzi:2009fi} is practically gone 
with the newer version of \pythiasix.
This is an important result also because it establishes the re-usability of \pythiasix tunes between 
standalone \pythiasix and \alpgen plus \pythiasix shower.

\begin{figure}[!t]
\begin{center}
\includegraphics[width=0.6\textwidth]{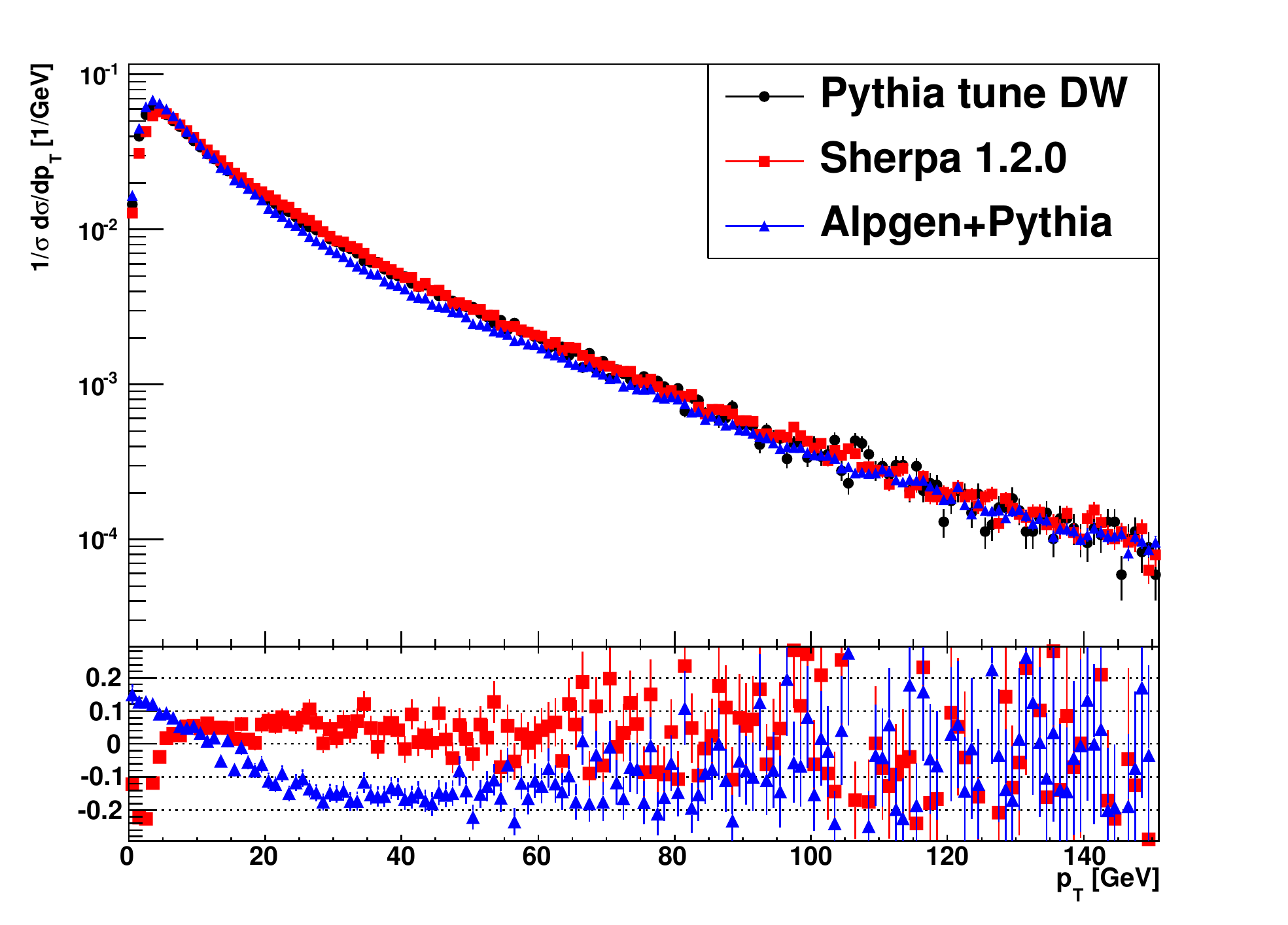}
 \caption{Z transverse momentum spectrum for \pythiasix, \sherpa and \alpgen. The agreement is within 20\%.
}
\label{fig:zbenchmarkmatching_ptz_alpshepyt}
\end{center}
\end{figure}
\begin{figure}[!h]
\begin{tabular}{c c}
\includegraphics[width=0.47\textwidth]{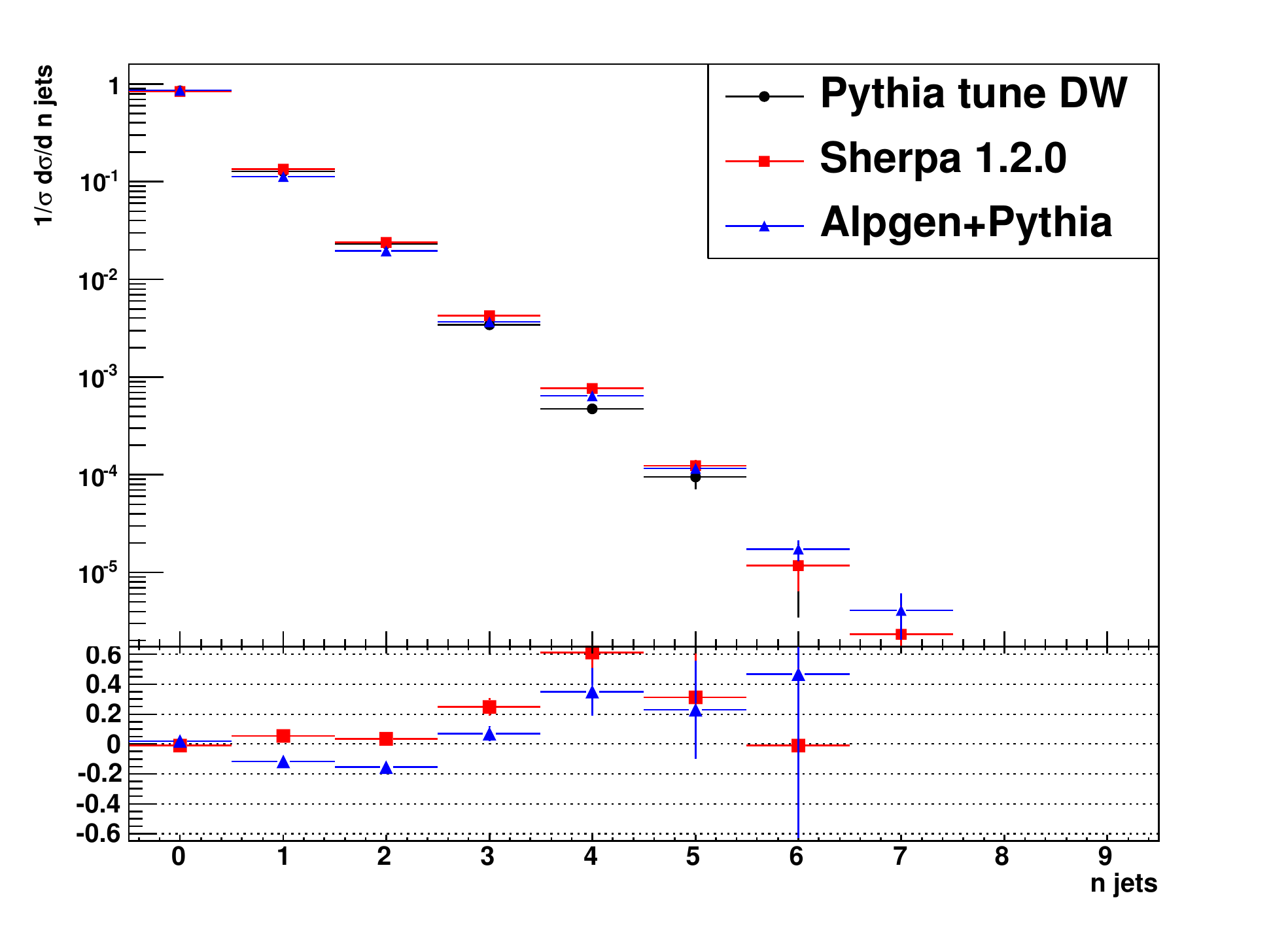} &
\includegraphics[width=0.47\textwidth]{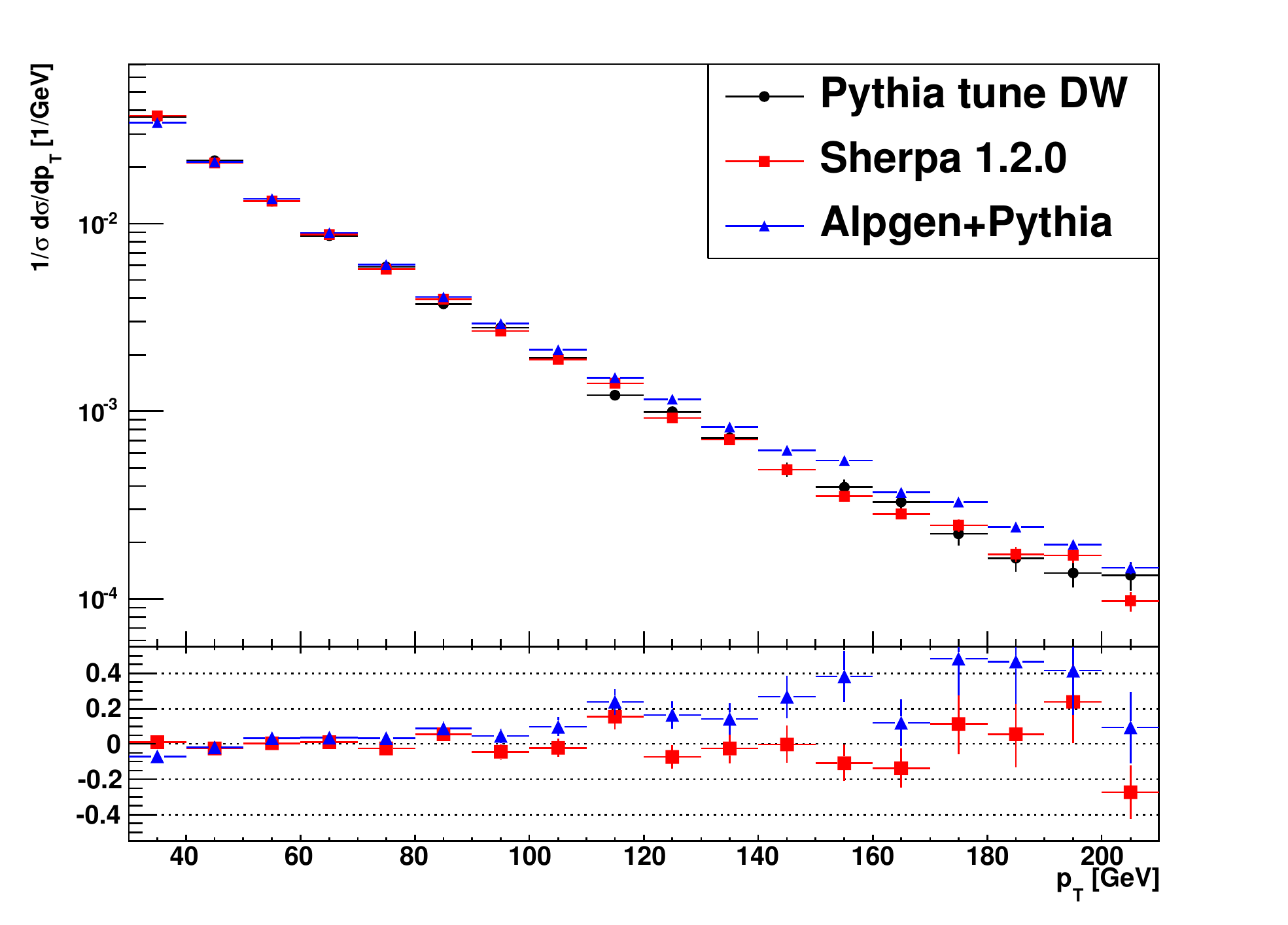} \\
(a) & (b)
\end{tabular}
\caption{(a) Jet multiplicity, (b) leading jet $\pt$ spectrum.}
\label{fig:zbenchmarkmatching_jets}
\end{figure}
Jet multiplicity and leading jet $\pt$ spectra are shown in Fig.~\ref{fig:zbenchmarkmatching_jets}.
In (a) both \alpgen and \sherpa appear to have a longer tail for high jet multiplicity.
Concerning the leading jet $\pt$ spectrum in (b), \sherpa follows \pythiasix all over the spectrum, \alpgen 
appears so have a harder tail for high $\pt$.

Differential jet rates for the 1$\rightarrow$0, 2$\rightarrow$1 and 3$\rightarrow$2  transitions are shown in 
Fig.~\ref{fig:zbenchmarkmatching_djr}.
Differential jet rates for the transition $n\rightarrow{}n-1$ are the distributions of the $k_{T}$ resolution 
parameter in an exclusive $k_{T}$ algorithm, $Q_{n\rightarrow{}n-1}$, that makes an $n$ jet event turn into 
an $n-1$ jet event.
\alpgen and \pythiasix are similar, while \sherpa's shape shows some not negligible differences.
\sherpa uses a dipole shower, while \pythiasix uses a virtuality ordered shower (the same shower is also 
used with \alpgen). 
The source of the difference is likely in the different shower, which fills the phase space with a different pattern.
\begin{figure}[!h]
\begin{tabular}{c c c}
\includegraphics[width=0.32\textwidth]{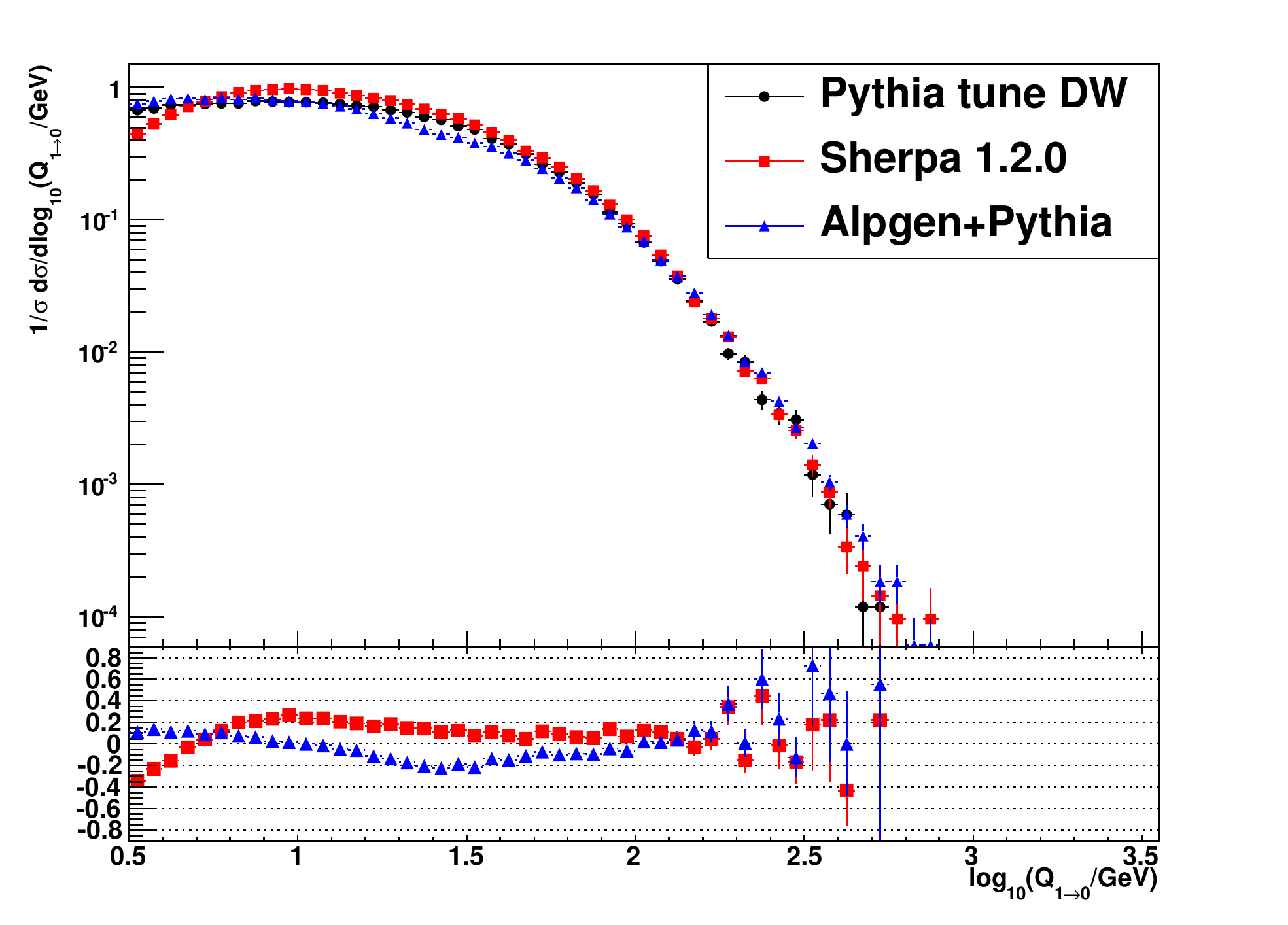} &
\includegraphics[width=0.32\textwidth]{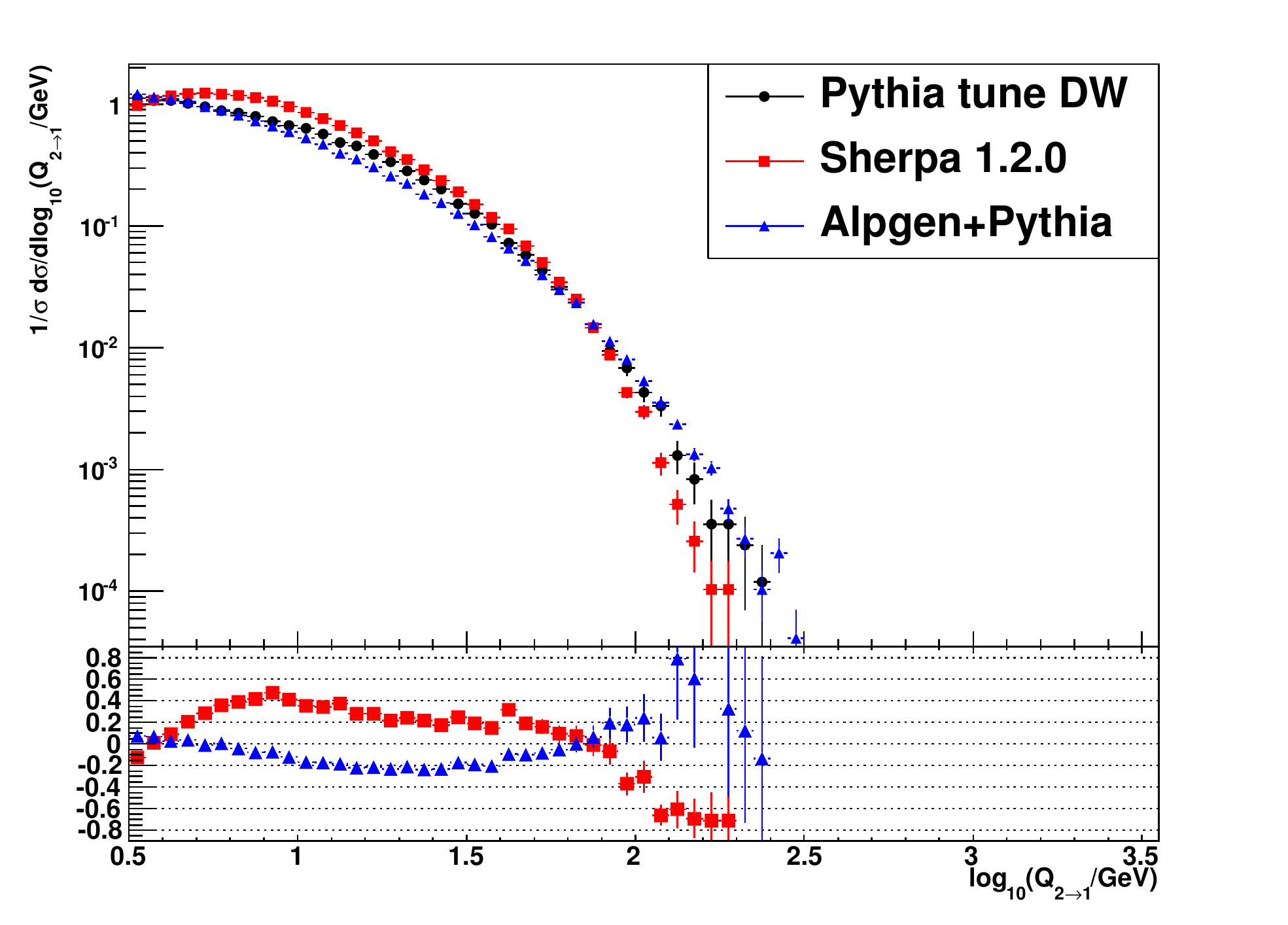} &
\includegraphics[width=0.32\textwidth]{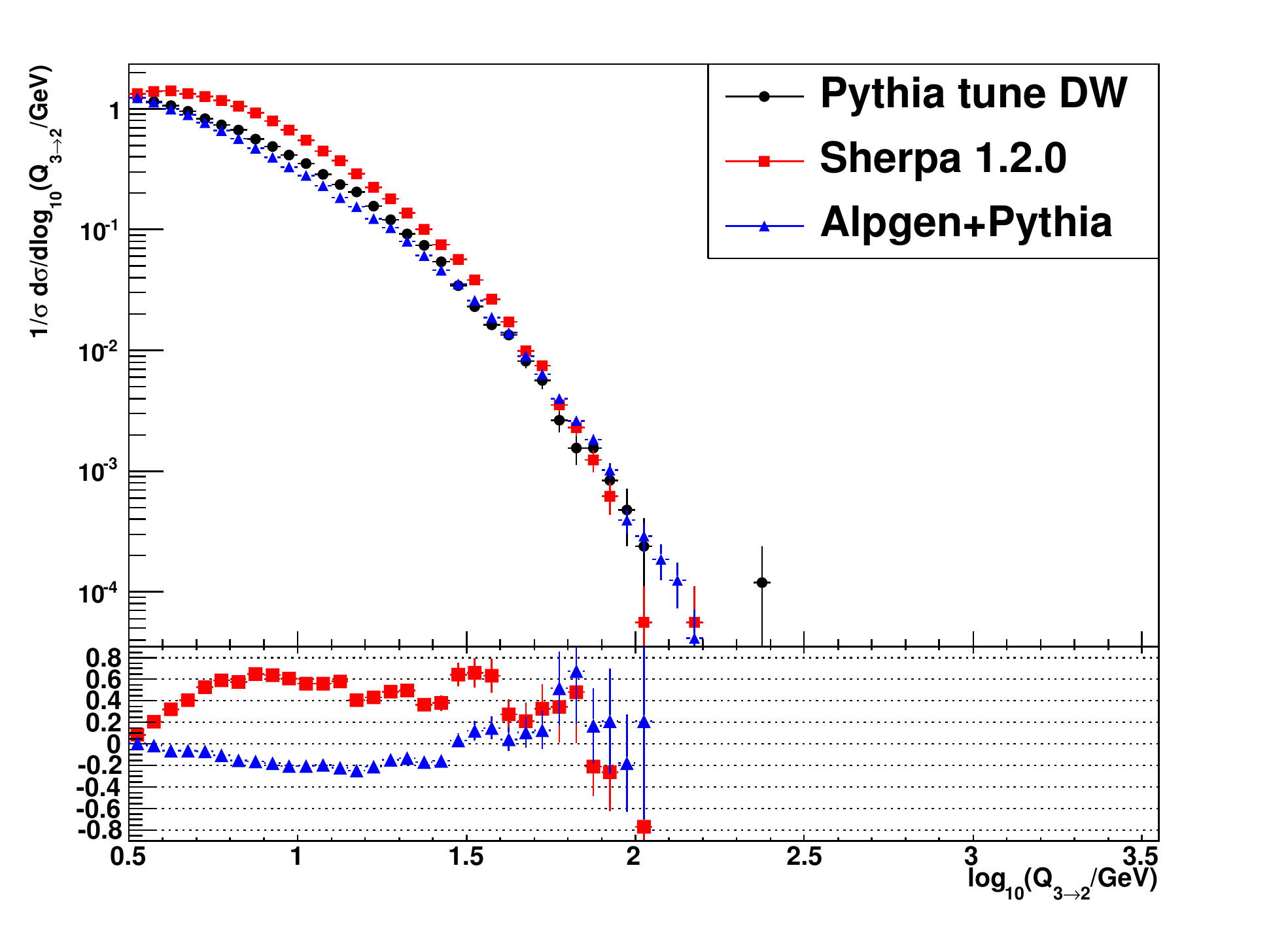} \\
(a) & (b) & (c)
\end{tabular}
\caption{Differential jet rates: (a) 1$\rightarrow$0, (b) 2$\rightarrow$1, (c) 3$\rightarrow$2.}
\label{fig:zbenchmarkmatching_djr}
\end{figure}

\subsection{CONCLUSION}

We studied matrix element corrections in \pythiasix for  inclusive $Z$ production at the LHC and we compared 
\pythiasix, \alpgen and \sherpa on the ground of first order real emission corrections. 
A bug was recently discovered and corrected in inclusive vector boson production in \pythiasix, that made the 
effect of matrix element corrections very big also at low boson $\pt$. We compared two versions of \pythiasix, 
before and after the fix, and observed that the low boson momentum region is correctly described in the new 
version, the matrix element corrected and uncorrected shower being similar in that region as expected.

We compared four different \pythiasix tunes. 
For all of them we switched off the multiple interactions in order to observe the effect of the tune on the 
shower and on the hard event simulation. 
Generally good agreement was observed, with some differences especially in the low $Z$ $\pt$ region and in the 
jet multiplicity spectrum.

We showed various plots comparing \pythiasix, \alpgen and \sherpa. 
Differences between \alpgen plus \pythiasix and matrix element corrected \pythiasix observed in \cite{Lenzi:2009fi} 
were due to the bug affecting \pythiasix, and are now much reduced. 
This also assesses the re-usability of \pythiasix tunes when \pythiasix is used to shower \alpgen events.

\subsection{PARAMETERS AND SETTINGS}
In \pythiasix, the underlying event was switched off with \verb+MSTP(81)=0+ (\verb+MSTP(81)=20+ for the tune 
Perugia0, that uses the $\pt$ ordered shower). 
Matrix element corrections were controlled with the parameter \verb+MSTP(68)+: this was set to \verb+=0+ to 
switch off matrix element corrections and to set the staring scale to the $Z$ mass, to \verb+=2+ for the power 
shower without matrix element corrections. The default had matrix element corrections enabled.
Hadronization was switched off with \verb+MSTJ(1)=0+.
The various tunes were selected by acting on parameter \verb+MSTP(5)+.

\alpgen samples were produced with a minimum parton $\pt$ of 20~GeV.
We used the default parameters for \sherpa, with no underlying event, no fragmentation, no QED radiation off leptons.

\clearpage

\part[BEYOND FIXED ORDER]{BEYOND FIXED ORDER}

\section[MULTIPLE PARTON INTERACTIONS AS A BACKGROUND TO TOP PAIR PRODUCTION]
{MULTIPLE PARTON INTERACTIONS AS A BACKGROUND TO TOP PAIR PRODUCTION
\protect \footnote{Contributed by R.~Chierici, E.~Maina}}


\subsection{INTRODUCTION}
\label{MPI_introduction}

The occurrence of Multiple Parton Interactions (MPI) in hadronic collisions is
by now well established experimentally~\cite{Akesson:1986iv,Abe:1997bp,Abe:1997xk,D0:2009D0_MPI}.
Furthermore, they represent the basic mechanism in the description of the Underlying Event (UE)
in the standard Showering Monte Carlo's (MC) like 
\pythiasix~\cite{Sjostrand:1987su,Sjostrand:2004pf,Sjostrand:2004ef}
and \herwig~\cite{Butterworth:1996zw,Bahr:2008dy}.
MPI rates at the LHC are expected to be large,
making it necessary to estimate their contribution to the background of
interesting physics reactions. 
On the other hand, their abundance makes it
possible to study MPI experimentally in details, testing and validating their
implementation in the MC.

In this note we discuss the role of MPI in top-antitop production at $\sqrt{s}$ = 10 TeV. 
The LHC will be a top-antitop factory and the large rate will allow accurate 
measurements of the top mass, one the most crucial parameters for stringent tests 
of the SM, and production cross section. It will therefore be of extreme importance
to have full control over all potential background processes. 
In this study we compare the results obtained with \pythiaeight~\cite{Sjostrand:2007gs}
with those obtained
at parton level with the methods of Refs.\cite{Maina:2009sj,Maina:2009vx} which suggested
that MPI could provide a significant background to top-antitop production and to
other interesting processes like $W+4j$ and $Z+4j$ production.

\subsection{SETUP}
\label{MPI_setup}

If no $b$--tagging is assumed, the MPI processes which provide a background to
$t\:\bar{t}$ and more generally contribute to $W+4j$ through Double
Parton Interactions (DPI) are: 

\begin{equation}
\label{MPI_proc0}
jj \otimes jjW \, , \; \; 
jjj \otimes jW \, , \; \; 
jjjj \otimes W
\end{equation}

where the symbol $\otimes$ stands for the combination of one event for each of the two
final states it connects.

Besides in W decay, isolated high--p$_T$ leptons can also be produced 
in the decay of heavy quarks. Therefore any reaction in which heavy quarks plus jets 
are produced can fake the signature of a W boson. Keeping in mind 
the large QCD cross-section for $b\overline{b}$ production, heavy quark and QCD DPI 
processes deserve attention as 
possible backgrounds to top pair production. This corresponds 
to the following extra processes:

\begin{equation}
b\overline{b} \otimes jj \, , \; \; 
b\overline{b}j \otimes jj 
\end{equation}

In this study we have decided to use \pythiaeight for simulating DPI events. \pythiaeight 
allows to use external Les Houches Accord \cite{Alwall:2006yp} input for the hardest 
process, and to choose an internal one for the second hardest, being however limited 
to $2\rightarrow2$ processes. This set-up allows a precise study of dedicated DPI 
configurations.

Moreover we have generated a number of samples, most notably $W+2j$, with 
{\tt MADEVENT}~\cite{Maltoni:2002qb,Alwall:2007st}, for which DPI has been simulated
by either \pythiaeight or by combining 
pairs of parton--level events as in Refs.\cite{Maina:2009sj,Maina:2009vx}.

In summary, these are the samples that we have generated with \pythiaeight:

\begin{equation}
t\overline{t} \, , \; \; 
Wj \otimes jj \, , \; \; 
Wjj(MG) \otimes jj \, , \; \; 
b\overline{b} \otimes jj \, , \; \; 
b\overline{b}j(MG) \otimes jj 
\end{equation}

where with $Wjj(MG)$ and $b\overline{b}j(MG)$ we indicate that the corresponding events have been
produced starting from an external Les Houches file created by {\tt MADEVENT} .
In all cases the second hard interaction switched on in \pythiaeight is a generic 
$2\rightarrow2$ QCD process. The default \pythiaeight.130 parameter setup and
tunings have been used. 
All the events processed with \pythiaeight have been fully showered, and jet clustering run 
over the events to have the possibility to apply a more realistic event 
selection.

We have also generated at parton level all reactions contributing to
$jj \otimes jjW$, $jjj \otimes jW$ and $jjjj \otimes W$ with {\tt MADEVENT}.

All samples have been generated
using CTEQ6L \cite{Pumplin:2005rh} parton distribution functions
and with the following parton level cuts:

\begin{equation}
\label{MPI_eq:cuts}
p_{T_j} \geq 10~{\rm GeV} \, , \; \; |\eta_j| \leq 5.0 \, , \; \; 
\Delta R_{jj} \geq 0.001 
\end{equation}

where $j= u,\bar{u},d,\bar{d},s,\bar{s},c,\bar{c},b,\bar{b},g$.

The DPI cross section for the parton level samples has been estimated as
\begin{equation}
\label{MPI_eq:sigma_2}
    \sigma =  \sigma_1 \cdot \sigma_2/\sigma_{eff}/k
\end{equation}

where $\sigma_1 ,\sigma_2$ are the cross sections of the two contributing
reactions and $k$ is a symmetry factor, which is equal to two if the
two reactions are indistinguishable and equal to one when they are different.

At the \tevatron, \cdf \cite{Abe:1997xk}
has measured $\sigma_{eff}=14.5\pm 1.7^{+1.7}_{-2.3}$ mb, a value
confirmed by \dzero which quotes
$\sigma_{eff}=15.1\pm 1.9$ mb \cite{D0:2009D0_MPI}.
In Ref.\cite{Treleani:2007gi} it is argued, on the basis of the simplest two channel
eikonal model for the proton--proton cross section, that a more appropriate value at
$\sqrt{s}= 1.8$ TeV is 10 mb which translates at the LHC into  
$\sigma_{eff}^{LHC}=12$ mb. Treleani then estimates the effect of the removal by \cdf
of Triple Parton Interaction events from their sample and concludes that \cdf measurement yields
$\sigma_{eff} \approx 11$ mb. In the following we use  $\sigma_{eff}=12.0$ mb
with the understanding that this value is affected by an experimental uncertainty
of about 15\% and that it agrees within about 10\% with the predictions of the eikonal model.

The DPI picture in \pythiaeight assumes that interactions can occur at different p$_{T}$ values 
independently of each other inside inelastic nondiffractive events.  
The expression for a DPI cross section becomes therefore: 
\begin{equation}
\label{MPI_eq:sigma_3}
    \sigma =  <f_{impact}>\sigma_1 \cdot \sigma_2/\sigma_{ND}/k
\end{equation}
where $\sigma_{ND}$ is the total non-diffractive cross section and $f_{impact}$
is an enhancement/depletion factor chosen event-by-event to account for correlations 
introduced by the centrality of the collision. This quantity is typically averaged during 
an entire run to calculate $<f_{impact}>$ in Eq.~\ref{MPI_eq:sigma_3}.
Typical values at the centre of mass energy of 10~TeV are 1.33 for $<f_{impact}>$ and 51.6~mb for $\sigma_{ND}$.
Comparing Eq.~\ref{MPI_eq:sigma_3} with Eq.~\ref{MPI_eq:sigma_2} tells us that \pythiaeight
sort of predicts an effective $\sigma_{eff}$=$\sigma_{ND}$/$<f_{impact}>$ which
is about a factor 3 larger than the one actually measured at the Tevatron.

\begin{table}[h]
\begin{center}
\begin{tabular}{|c|c|c|c|} 
\hline
Process & $\sigma$(nb) & $\sigma$($\otimes jj$)(pb) & $\sigma_{acc}$(pb) \\
\hline

$t\overline{t}$             &  0.24 & no DPI & 60 \\
$W(\rightarrow l\nu)jj$(MG) & 13.7  & 140    & 23 \\ 
$Wj$                        &  4.8  &  12    & 2.5 \\
$b\overline{b}$             &  850  & 2200   & 18 \\
$b\overline{b}j$(MG)        &  688  & 7300   & 69 \\
$b\overline{b}jj$(MG)       &  472  & no DPI & 4012 \\
$W(\rightarrow l\nu)jj$(MG) & 13.7  & no DPI & 627 \\ 

\hline
\end{tabular}
\caption{Total cross-sections for the processes studied in this contributions.
Samples labelled with MG are made with the use of {\tt MADEVENT}. The second column 
presents the cross sections as returned by the generators, the third column
the corresponding cross sections after having added a second hard interaction 
to the process in \pythiaeight, and the last column the accepted cross sections after the 
selection described in Eq.~\ref{MPI_sel0}.}
\label{MPI_sigmas}
\end{center}
\end{table}
 
Table~\ref{MPI_sigmas} reports the cross section breakdown for the samples used 
in this study and processed by \pythiaeight. All cross-sections are leading order as 
returned by the programs used. 
The $Wjj$ sample generated with {\tt MADEVENT} has also been
combined with matrix elements QCD events, evaluating the cross section according
to Eq.~\ref{MPI_eq:sigma_2}: we will call this sample as $Wjj \otimes jj$ ME, as opposed
to the $Wjj \otimes jj$ P8 sample.

\subsection{RESULTS}
\label{MPI_results}

To mimic an experimental analysis, showered events are first required to have 
an isolated, central and energetic lepton according to the cuts:
\begin{equation}
\label{MPI_sel0}
p_{T_l} \geq 20~{\rm GeV} \, , \; \; |\eta_l| \leq 2.5 \, , \; \; 
E_{iso} \leq 20~{\rm GeV} 
\end{equation}
with obvious notation and where $E_{iso}$ is an isolation variable defined as the total energy flowing in
a cone of opening angle $\Delta$R=0.1 around the lepton. Later in this study we 
have made this cut more severe by increasing $\Delta$R to 0.5. We will also compare
the effect of different jet p$_T$ thresholds.
To be closer to an experimental selection, only electrons and muons are considered
in this study.
In events with a candidate lepton, at least four jets with transverse momentum 
above 20~GeV and $|\eta|<5$ are requested. No b-tagging requirement is mimicked. 
The accepted cross sections of the top-pair signal and the DPI backgrounds are 
reported in Table~\ref{MPI_sigmas}. In comparison we have also reported the accepted 
cross sections for two equivalent non-DPI background processes.   

It has to be mentioned that for the DPI events obtained by the simple combination
of parton level events as in Refs.\cite{Maina:2009sj,Maina:2009vx}, where no parton
shower and fragmentation is present, the above cuts have been replaced by equivalent
ones. The lepton isolation, for instance, becomes a cut in $\Delta$R between the lepton
and any jet.

\begin{figure}
\begin{center}
\vspace*{-1.cm}
\subfigure{
\includegraphics[width=0.47\textwidth]{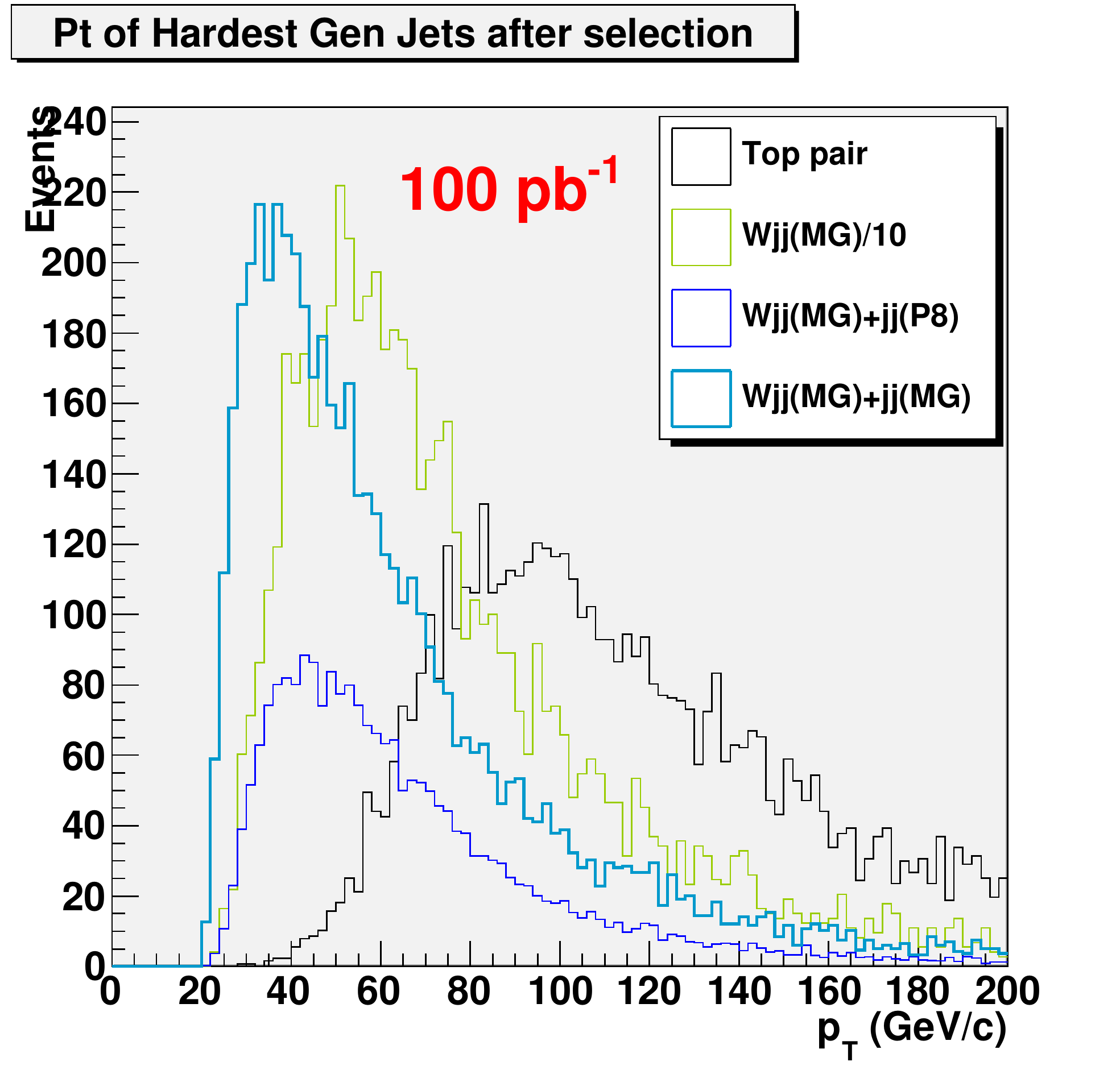}\hspace*{0.5cm}
\includegraphics[width=0.47\textwidth]{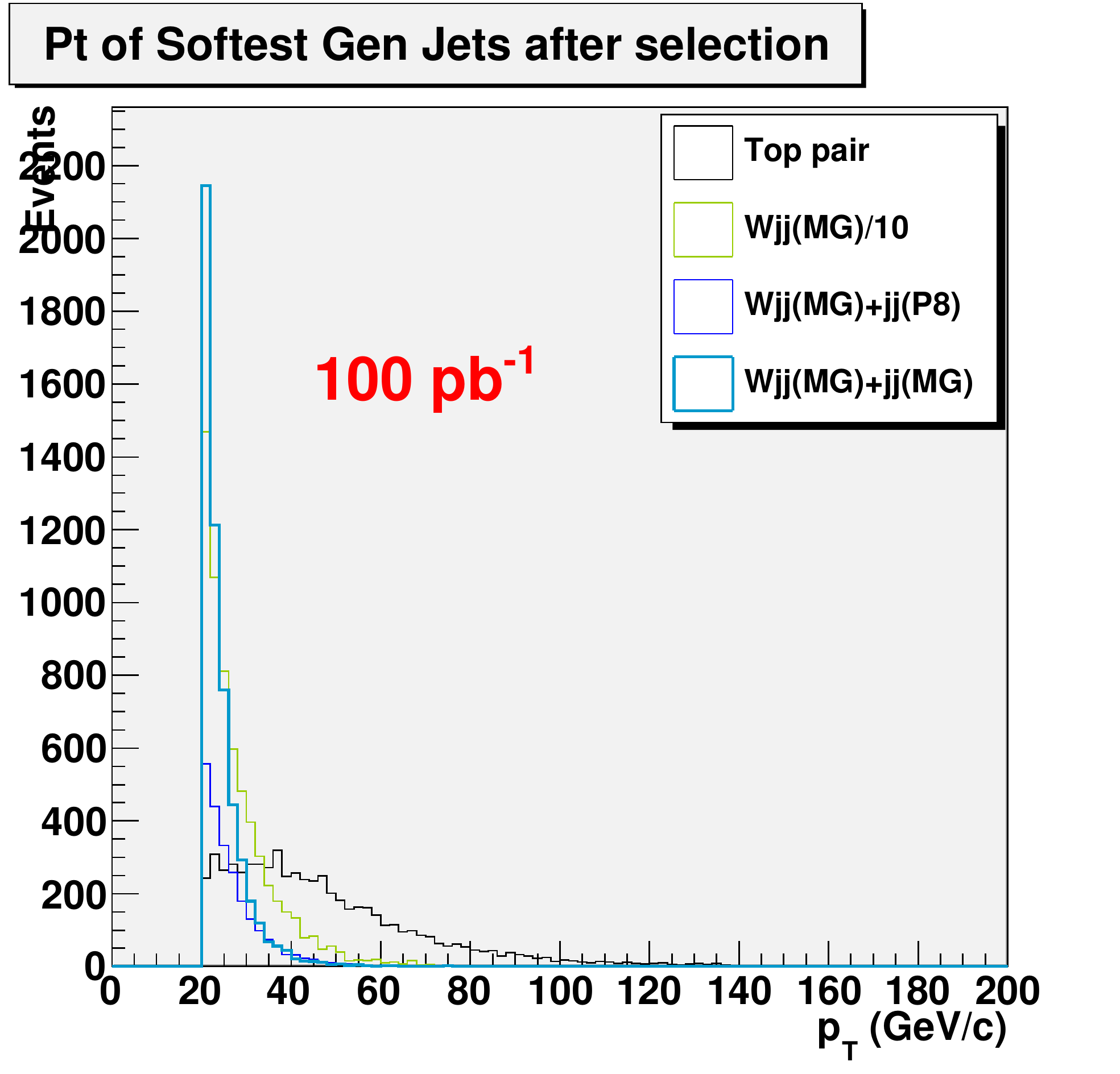}
}
\subfigure{
\includegraphics[width=0.47\textwidth]{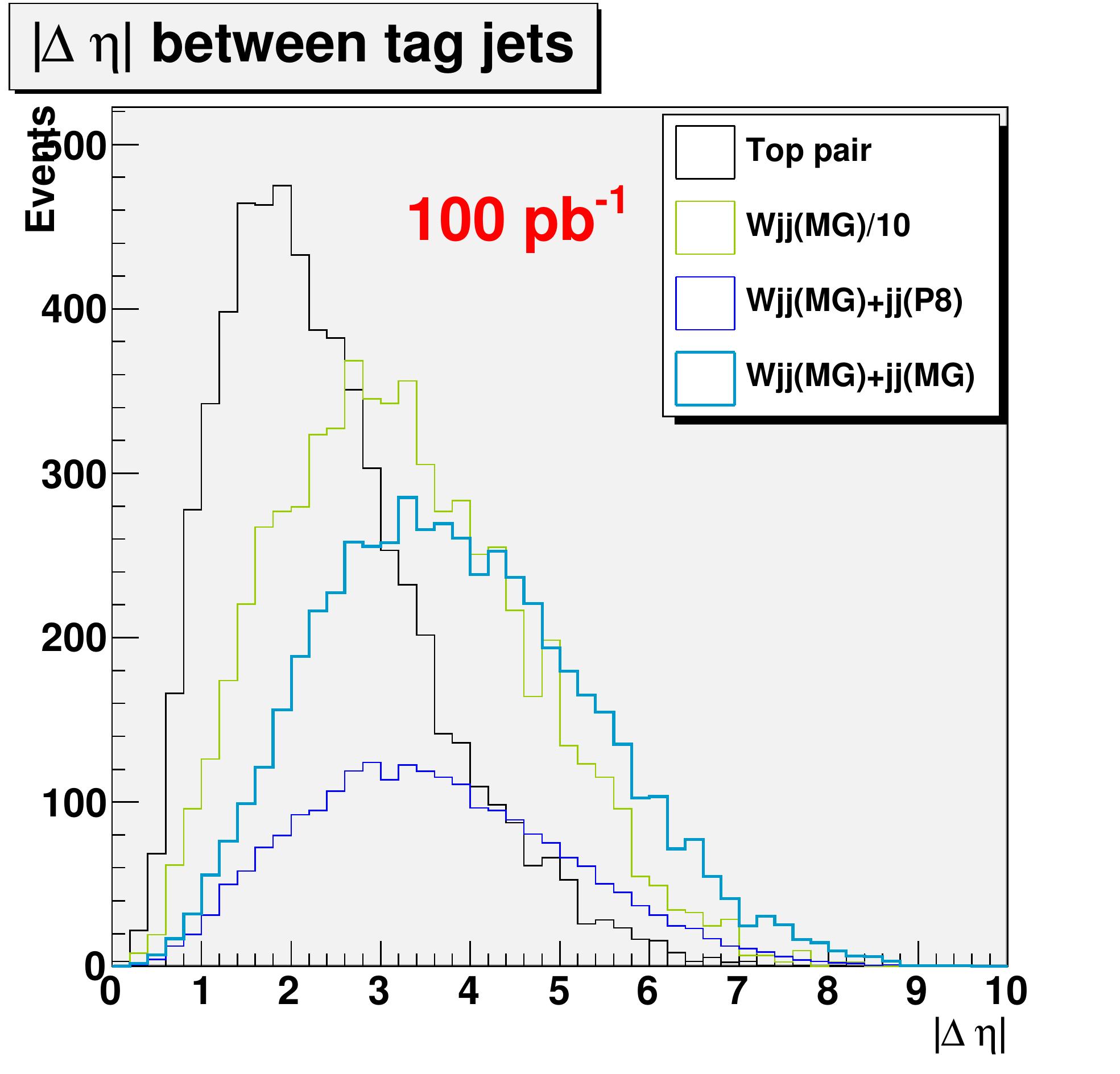}\hspace*{0.5cm}
\includegraphics[width=0.47\textwidth]{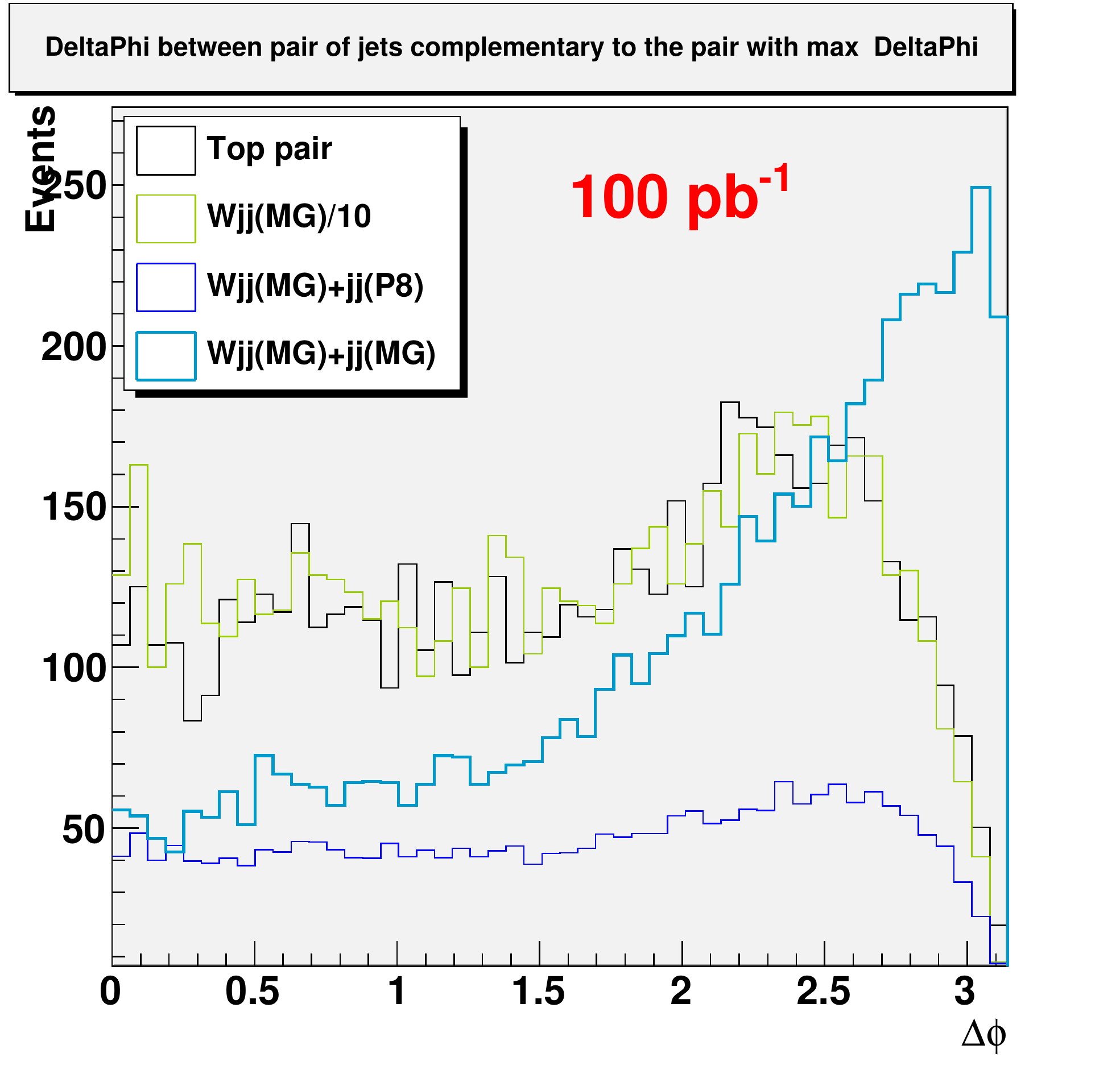}
}
\subfigure{
\includegraphics[width=0.47\textwidth]{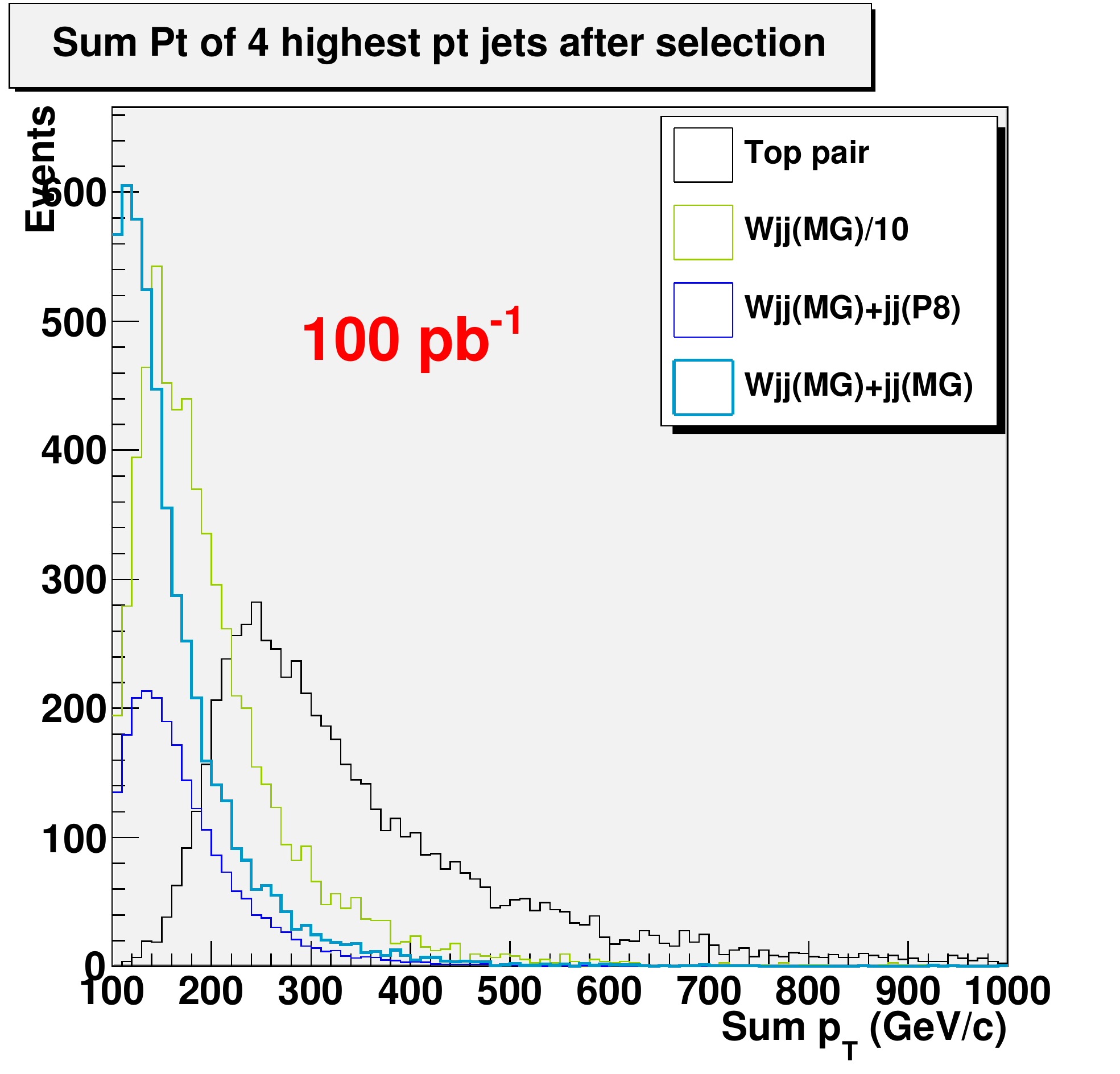}\hspace*{0.5cm}
\includegraphics[width=0.47\textwidth]{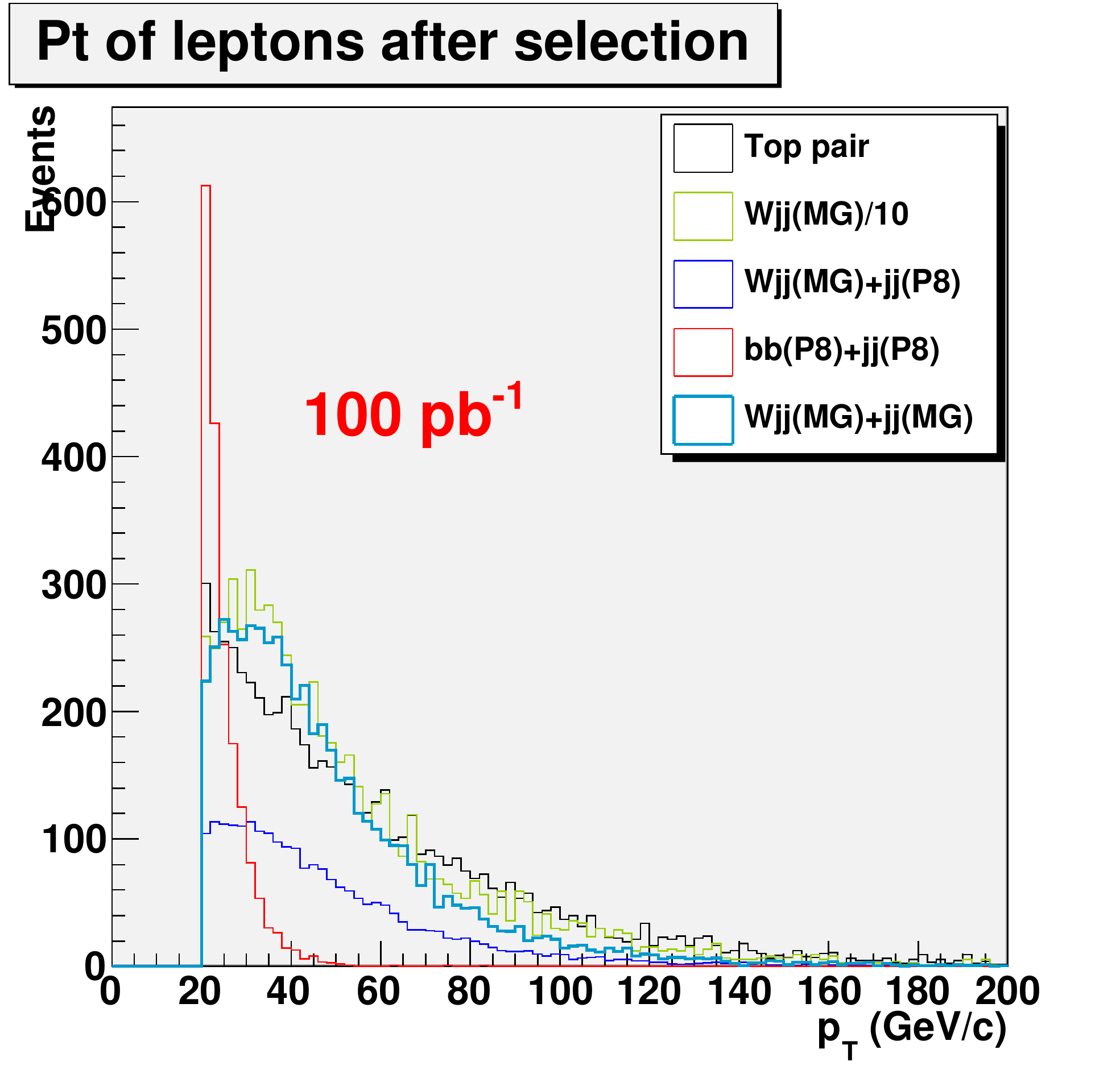}
}
\caption{Distribution of the highest $p_{T_j}$; the fourth highest $p_{T_j}$;
the largest separation in $\vert \Delta \eta \vert$ among the four highest p$_T$ jets;
the $\vert \Delta \phi \vert$ of the two jets complementary to those with the 
largest separation in $\vert \Delta \phi \vert$ among the four highest p$_T$ jets;
the scalar sum of the p$_T$'s of the four highest p$_T$ jets;
the p$_T$ of the charged lepton. The plots are made with the cuts in Eq.~\ref{MPI_sel0}
and for 100 pb$^{-1}$ of integrated luminosity at the LHC at $\sqrt{s}$ = 10 TeV.
}
\label{MPI_20_iso01}
\end{center}
\end{figure}

In Figure~\ref{MPI_20_iso01} we present a few distributions at generator level after
the event selection in Eq.~\ref{MPI_sel0}. We decided to compare the top-pair
signal with the $Wjj$ process
alone and with DPI $Wjj \otimes jj$ in both the ME and P8 configurations.
All histograms in the plots are normalised to the number of events expected in 100/pb 
of integrated luminosity, with the exception of the Wjj contribution that has been 
scaled by a factor 10 to make the plot more readable.
The plots shown represent, respectively, the highest and softest (among four) p$_T$
of the jets in the event, the maximum pseudorapity difference among the four jets,
the azimuthal difference between the jets complementary to those with the maximum 
azimuthal difference, the scalar sum of the transverse momenta of the four highest
p$_T$ jets, and the transverse momentum of the found lepton. 
The heavy quark DPI component is shown only for this last histogram. The plot is a 
good indication of the fact that kinematic cuts on the lepton transverse momentum, 
as well as a more severe cut on the lepton isolation, should be very effective in 
removing the heavy quark background(s), as we will also demonstrate later.  

In the plots we have not superimposed several ISR components that are, as 
expected, largely dominating in all the corners of the phase space (for instance 
$Wjjjj$ with respect to $Wjj \otimes jj$ DPI or $b\overline{b}jj$ with respect to
$b\overline{b} \otimes jj$ DPI). 
Therefore,
analyses able to reach high purities in the top signal over the ISR background,
should also be able to highly suppress all the analogous signatures coming from 
DPI events.

The DPI contributions can however reach sizeable fractions of the 
ISR backgrounds for analyses with non sufficiently tight cuts.
In the figures, for instance, we compare $Wjj$(+parton shower) rescaled by a factor 
10 with the $Wjj \otimes jj$ DPI component. The rate of the latter is sizeable 
and the two have similar
kinematic characteristics: the additional jets from the second interaction have the tendency to
be softer and slightly more forward in the acceptance. This suggests that harder 
cuts on the jet transverse momenta should also be very effective in reducing the DPI
component.  
From the figures it also turns out that there are significant discrepancies in the 
normalisation, as well as in shape, for the prediction of the $Wjj \otimes jj$ DPI ME and
P8 components.
The difference in normalisation is a direct consequence of the different normalisation
cross section for the DPI component in \pythiaeight with respect to the model in~\cite{Treleani:2007gi}
and the Tevatron measurements. The relatively small shape difference can everywhere 
be ascribed to the fact that the distributions for $Wjj \otimes jj$ ME are made at parton level 
prior to any showering, whereas for \pythiaeight they are all at hadron level and, therefore, 
also account for shower radiation.  

\begin{figure}
\begin{center}
\vspace*{-1.cm}
\subfigure{
\includegraphics[width=0.47\textwidth]{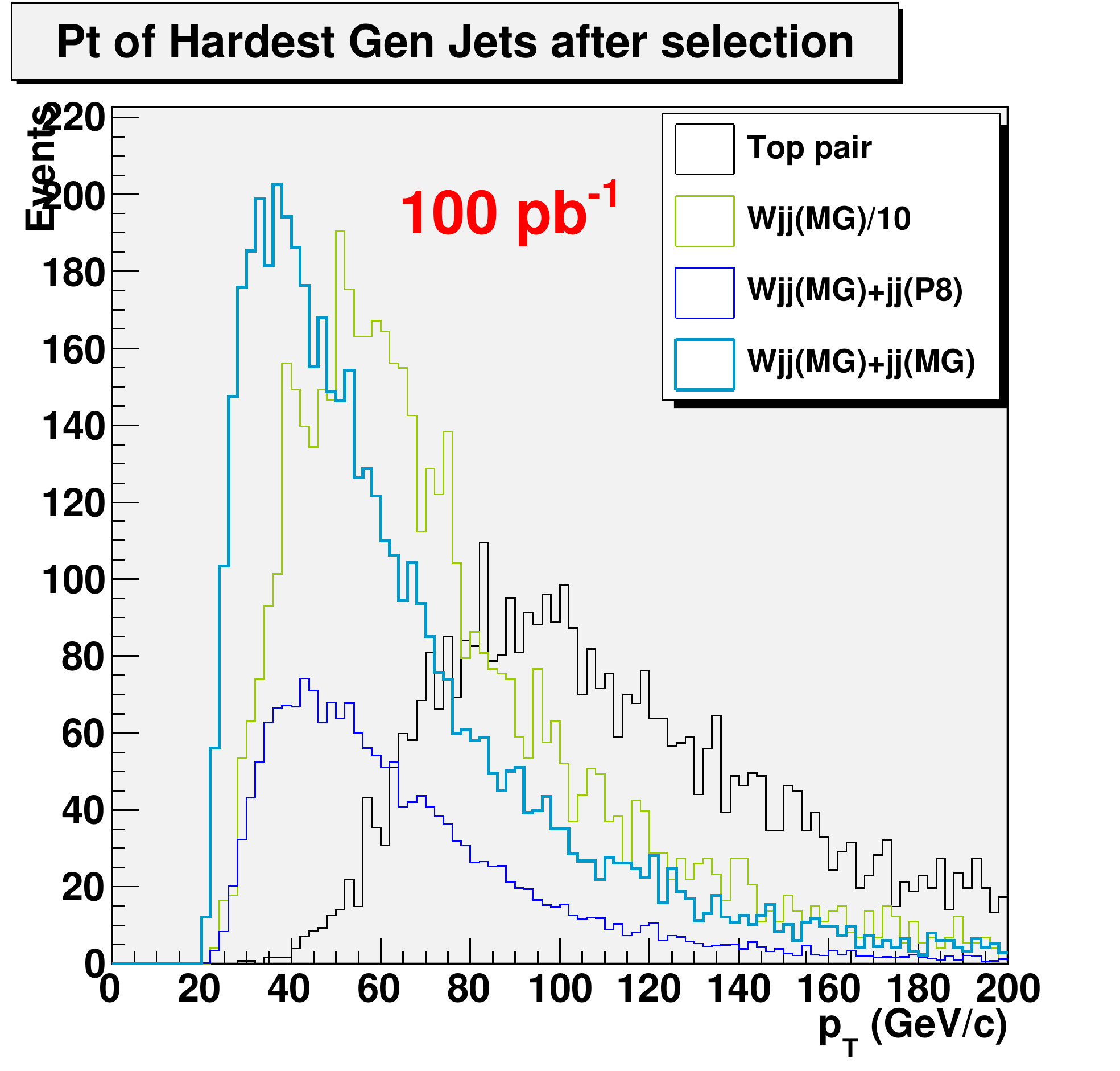}\hspace*{0.5cm}
\includegraphics[width=0.47\textwidth]{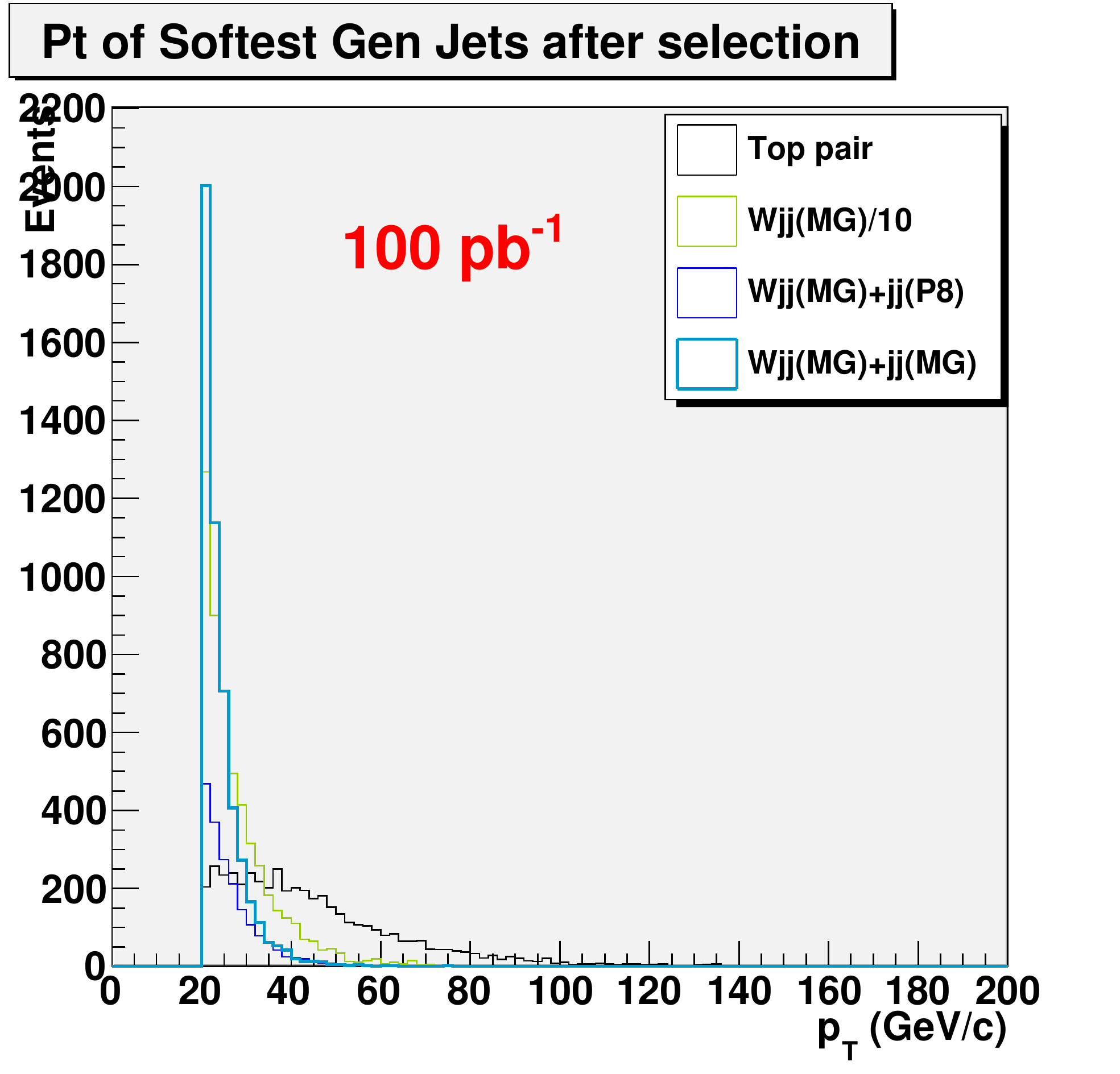}
}
\subfigure{
\includegraphics[width=0.47\textwidth]{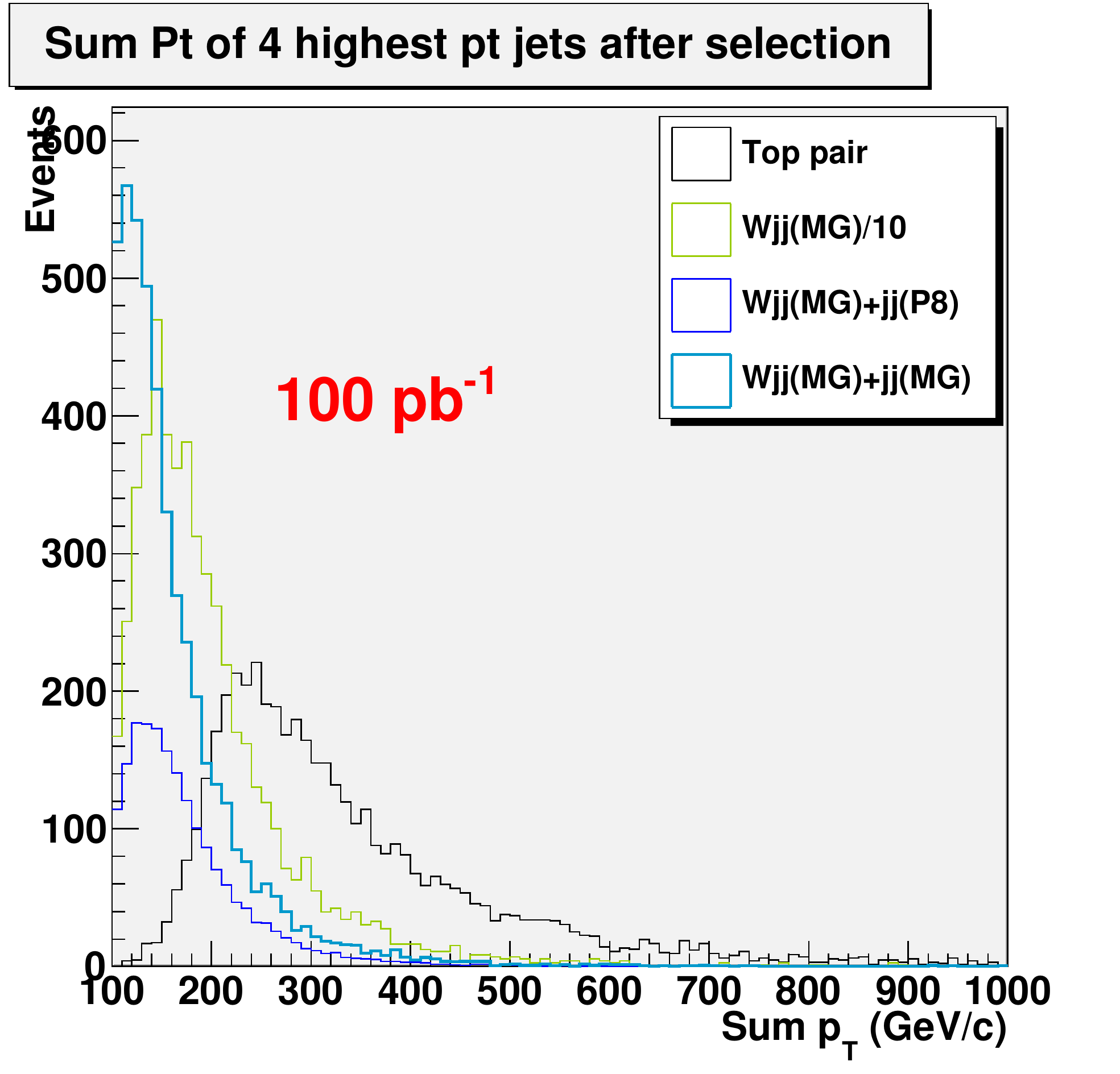}\hspace*{0.5cm}
\includegraphics[width=0.47\textwidth]{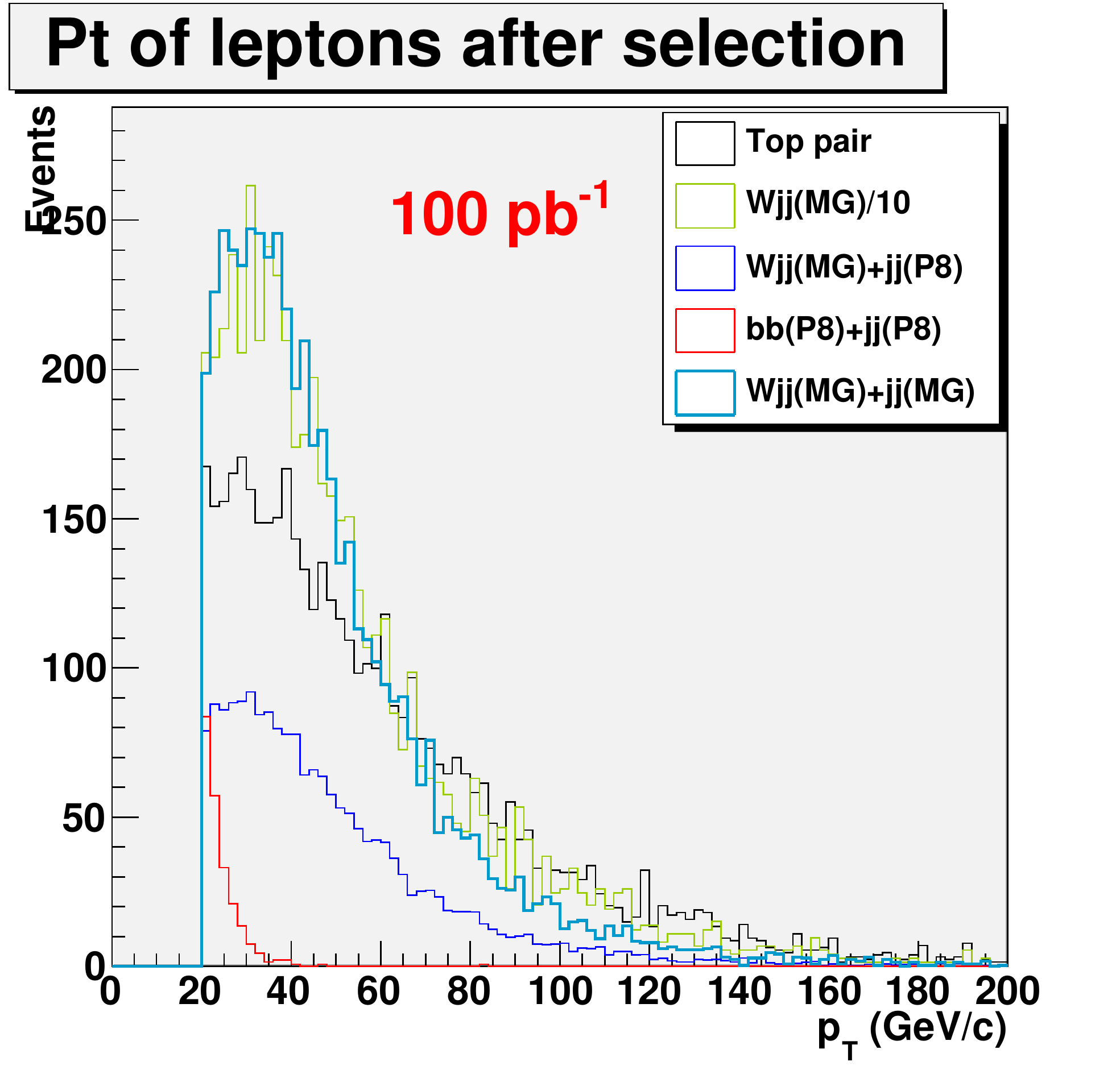}
}
\caption{Distribution of the highest $p_{T_j}$; the fourth highest $p_{T_j}$;
the scalar sum of the p$_T$'s of the four highest p$_T$ jets;
the p$_T$ of the charged lepton. The plots are made with the cuts in Eq.~\ref{MPI_sel0} 
and a more severe lepton isolation condition, and 
for 100 pb$^{-1}$ of integrated luminosity at the LHC at $\sqrt{s}$ = 10 TeV.
The lepton isolation cone has now an opening of $\Delta$R=0.5.
}
\label{MPI_20_iso05}
\end{center}
\end{figure}

The conclusions drawn so far have a very mild dependence on the lepton selection,
with the exception of the reduction of the heavy quark component, for which a tighter
isolation requirement is mandatory. The plots in Figure~\ref{MPI_20_iso05} correspond to 
those in Figure~\ref{MPI_20_iso01}, with the exception that the isolation cone for 
the lepton has now an opening of $\Delta$R=0.5. Shape and normalisations for the
signal and the W+jets backgrounds do not change significantly, whereas the contribution
of the heavy flavour component is significantly reduced. 

\begin{figure}
\begin{center}
\vspace*{-1.cm}
\subfigure{
\includegraphics[width=0.47\textwidth]{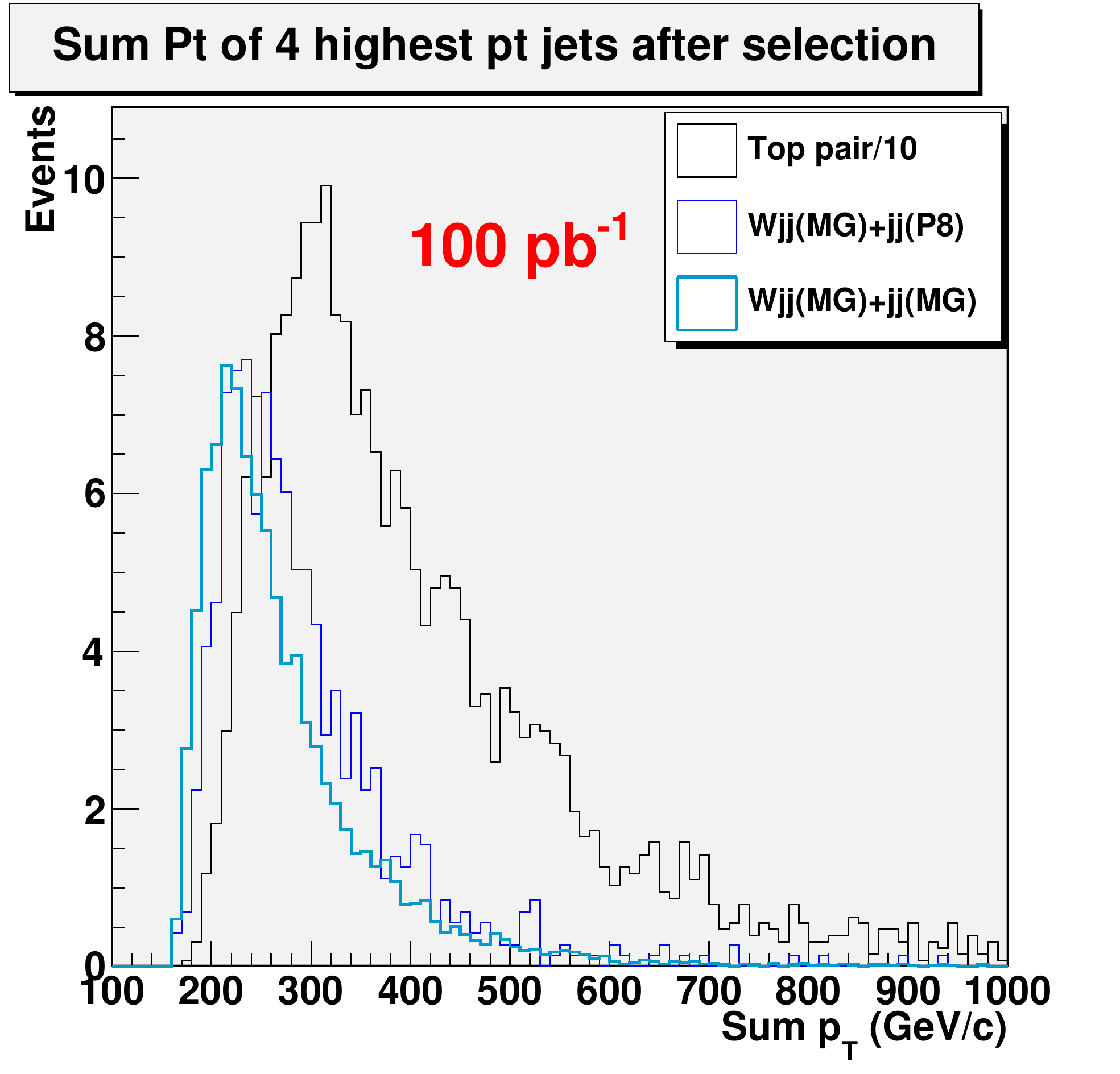}\hspace*{0.5cm}
\includegraphics[width=0.47\textwidth]{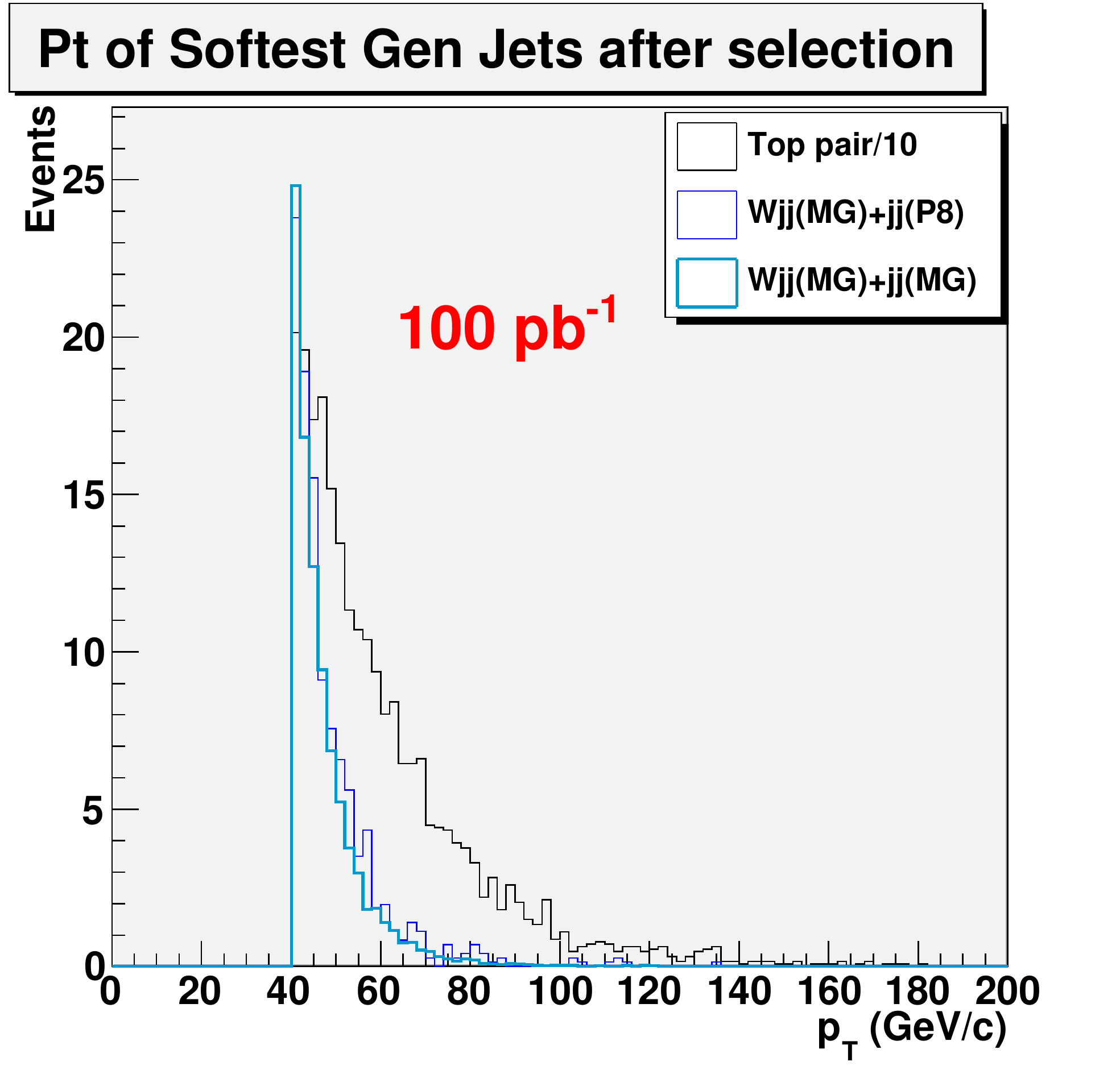}
}
\caption{Distribution of the scalar sum of the p$_T$'s of the four highest p$_T$ jets
and the fourth highest $p_{T_j}$. The plots are made for 
100 pb$^{-1}$ of integrated luminosity at the LHC at $\sqrt{s}$ = 10 TeV.
The lepton isolation cone has now an opening of $\Delta$R=0.5 and
$p_{T_j} \geq 10~{\rm GeV}$.
}
\label{MPI_40_iso05}
\end{center}
\end{figure}

Another way to tighten the analysis, and protect the signal even further from the
DPI component is to increase the jet transverse momentum threshold.
Figure~\ref{MPI_40_iso05} shows the scalar sum of the four highest p$_T$ jets transverse
momenta and the p$_T$ of the fourth jet in the event after that both lepton isolation
conditions and jet thresholds have been tightened. The latter has been moved from 
20~GeV to 40~GeV. The top pair signal component is
now scaled by a factor 10, making it evident that all DPI component are significantly reduced.
What is also interesting is that in this region of phase space the accepted cross section
(total and differential) for $Wjj \otimes jj$ ME and P8 are essentially equivalent. 
This indicates that either approach is equivalent in describing DPI at relatively high 
transverse momentum.

\begin{figure}
\begin{center}
\includegraphics[width=0.6\textwidth]{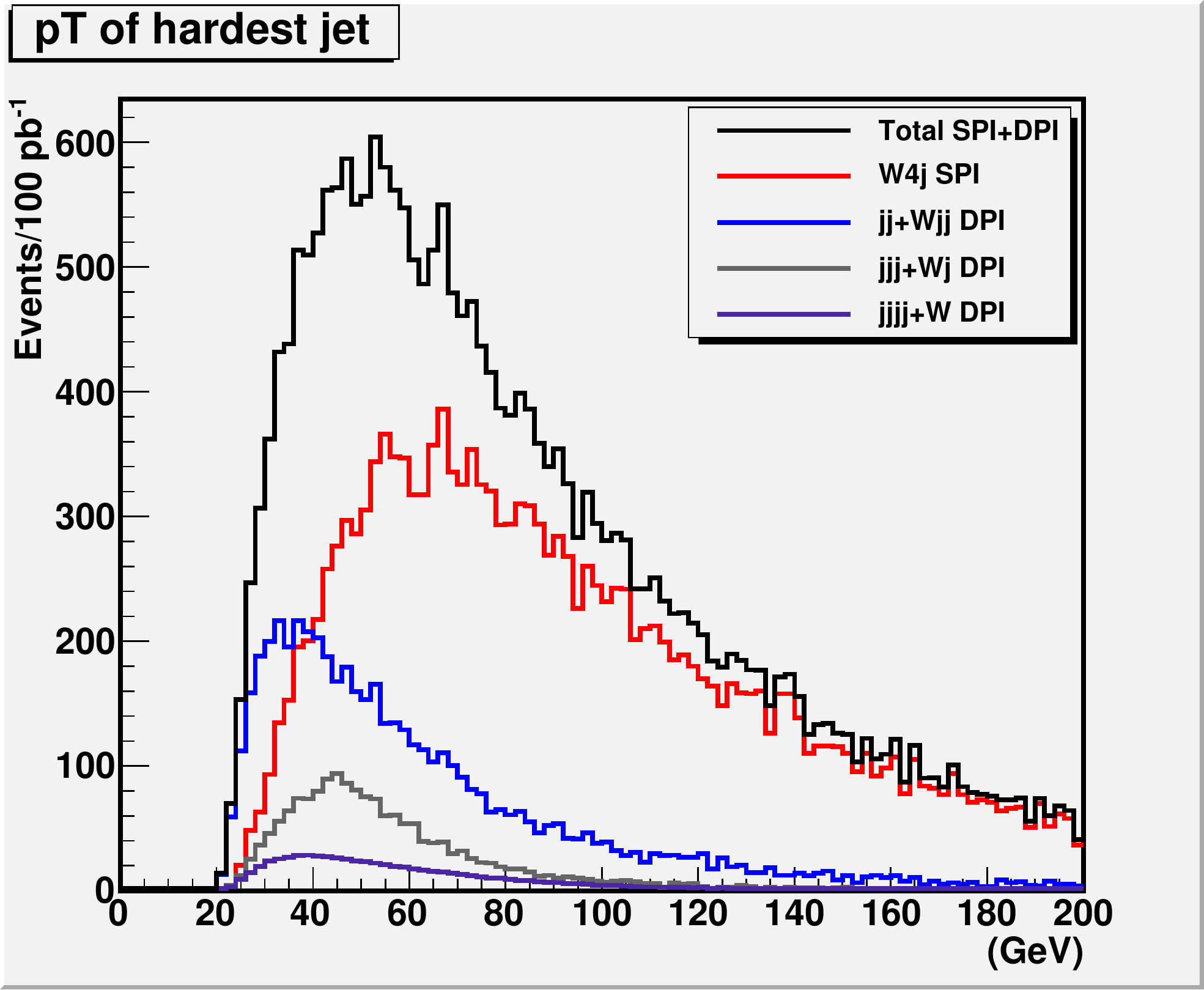}
\caption{Distribution of the highest $p_{T_j}$ for $jj \otimes jjW$, $jjj \otimes jW$
and $jjjj \otimes W$. Also reported are the SPI result
for $W+4j$ and the sum of SPI and MPI contributions.
}
\label{MPI_JHardTh}
\end{center}
\end{figure}

\subsection{FURTHER RESULTS}
\label{MPI_further}

In Sect.~\ref{MPI_results} we have compared the $Wjj \otimes jj$ parton level DPI
contribution to a number of contributions generated with \pythiaeight. We have neglected
the $jjj \otimes jW$ and $jjjj \otimes W$ components because they cannot be directly
generated only with \pythia. In this section we complete our analysis in this regard.
In Figure~\ref{MPI_JHardTh} we present the distribution of the largest jet transverse
momentum for $jj \otimes jjW$, $jjj \otimes jW$
and $jjjj \otimes W$. We also show the Single Parton Interaction (SPI) result
for $W+4j$ and the sum of SPI and MPI contributions. We neglect the small Triple Parton
Interaction component.
All results in this section are purely at parton level without showering and 
hadronization. The selection cuts are:

\begin{equation}
\label{MPI_cuts_all}
p_{T_j} \geq 20~{\rm GeV} \, , \; \; |\eta_j| \leq 5.0 \, , \; \; 
\Delta R_{jj} \geq 0.5\, , \; \;
p_{T_l} \geq 20~{\rm GeV} \, , \; \; |\eta_l| \leq 2.5 \, , \; \; 
\Delta R_{jl} \geq 0.1
\end{equation}

The $W+4j$/$Wjj \otimes jj$/$jjj \otimes jW$/$jjjj \otimes W$ contributions are in the
ratio 1/0.33/0.11/0.04 and therefore the DPI rate is about half of the SPI rate, with
the $Wjj \otimes jj$ component contributing about two thirds of the DPI total.
 As a comparison we recall that
in \cite{Maina:2009sj} the DPI contribution to $W+4j$ was estimated at about 10\%
at $\sqrt{s}$ = 14 TeV. Only part of the discrepancy between the two results
can be explained by the smaller effective cross section
$\sigma_{eff}=12.0$ mb employed here in comparison to
$\sigma_{eff}=14.0$ mb used in \cite{Maina:2009sj}. Therefore our preliminary analysis
suggests that the DPI background to $W+4j$ production is proportionally
larger at $\sqrt{s}$ = 10~TeV
than at 14 TeV.


\subsection*{CONCLUSIONS}

We have asked ourselves the question whether MPI can be a source of unexpected
background for high p$_T$ physics at the LHC. In this study we have taken top production 
as a perfect use case, trying to determine which MPI background can induce a 
signature compatible with $t\overline{t}$. We have shown that DPI coming from QCD
processes and either $W$ production or heavy quark production can indeed fake a 
top pair signature, becoming a significant fraction of the remaining background 
if the analyses are not sufficiently tight. As main experimental cuts for reducing
drastically DPI we suggest the minimum transverse momentum threshold for jets and
the lepton isolation, which are expected to be also effective for standard multi-jet
backgrounds.
In our study we have compared two ways of generating DPI events: one based on 
\pythiaeight and the other by direct combination of parton level events. The differences 
between the two approaches are important, especially in the softer regionss of 
phase space and for what concerns the global normalisation, and are ultimately
due to the assumptions on the cross section to which DPI are normalised.
The present results suggest that the DPI background to $W+4j$ production is
proportionally larger at 10~TeV than at 14~TeV

\subsection*{ACKNOWLEDGEMENTS}

We are indebted to P.~Skands and T.~Sj\"ostrand for very interesting discussions on 
MPI and \pythiaeight.

\clearpage

\section[A MATCHING SCHEME FOR \wg NLO MATRIX ELEMENT GENERATOR AND \pythiaeight PARTON SHOWER]
{A MATCHING SCHEME FOR \wg NLO MATRIX ELEMENT GENERATOR AND \pythiaeight PARTON SHOWER~\protect \footnote{
Contributed by: D.~Majumder, K. Mazumdar, T. Sj\"ostrand} }

\subsection{INTRODUCTION}

Diboson production is among the early physics topics to be studied at the LHC initial phase. In particular, the \wg process has a higher rate compared to the others, which can be pursued at moderate luminosities to probe the gauge structure of the Standard Model. The presence of anomalous couplings between the W-bosons and the photon is expected to enhance the harder part of the photon transverse momentum (\ptg) spectrum which is one of the experimental observables. Higher order QCD corrections will contribute significantly at the LHC, modifying the \ptg spectrum and hence may obscure the effect of any anomalous coupling that might exist. This calls for at least a NLO QCD event generator to accurately describe the parton level process. However, for a full event description including the initial and final state radiations (ISR and FSR) and the underlying events the parton shower approach is required. In this contribution we discuss a matching scheme for \wg events generated by \baur\cite{Baur:1993ir}, which is a complete $\cal{O}$$(\alpha_s)$ calculation, to the \pythia\cite{Sjostrand:2006za},\cite{Sjostrand:2007gs} parton showers. The proposed methodology for the matching preserves both the rate of the hard scattering process as well as various kinematic distributions of experimental interest, like the photon transverse momentum.  

\subsection{THE MATCHING STRATEGY}

The \baur package \cite{Baur:1993ir},\cite{Baur:2008} is a matrix element-based event generator which produces \wg events in hadronic collisions with up to one additional parton (quark or gluon) in the final state. It has the provision for the introduction of anomalous $WW\gamma$ couplings in addition to Standard Model ones and hence interesting for experimental analysis. 

The \baur package produces two types of event topologies:
\begin{itemize}
  \item 3-body final state, from \wg+0 jet events, containing the lepton and the neutrino from W-decay and the photon;   
  \item 4-body final state, from \wg+1jet events, containing an additional parton (quark or gluon) forming a jet. 
\end{itemize} 
Figs.~\ref{Baur_fig:treeGraphs} and \ref{Baur_fig:lo_nlo_graphs} depict the Born level diagrams and some of the higher order QCD corrections for the \wg process respectively. Fig.~\ref{Baur_fig:baurPtg_born_nll} compares the \ptg spectrum at the Born level to that at NLO, as obtained from \baur. The cross-section for the \wg process is $1.390\pm0.029$~pb at NLO and $0.584\pm~0.005$~pb at Born level for \ptg$>25$~GeV$/c$  with $|\eta^{\gamma}|<2.1$ and $p^{lepton}_T >15$~GeV$/c$ with $|\eta^{lepton}|<2.6$, $\eta^{\gamma}$ and $\eta^{lepton}$ being the pseudorapidity of the photon and the lepton from W-decay respectively. The selection applied here are typical of the LHC experiments. This gives a \ptg-dependent k-factor with a mean value of of 2.16, which is quite substantial (Fig.~\ref{Baur_fig:baurPtg_born_nll} inset). 

\begin{figure}[ht!] 
\begin{center}
\includegraphics[width=0.2\textwidth,height=0.25\textwidth]{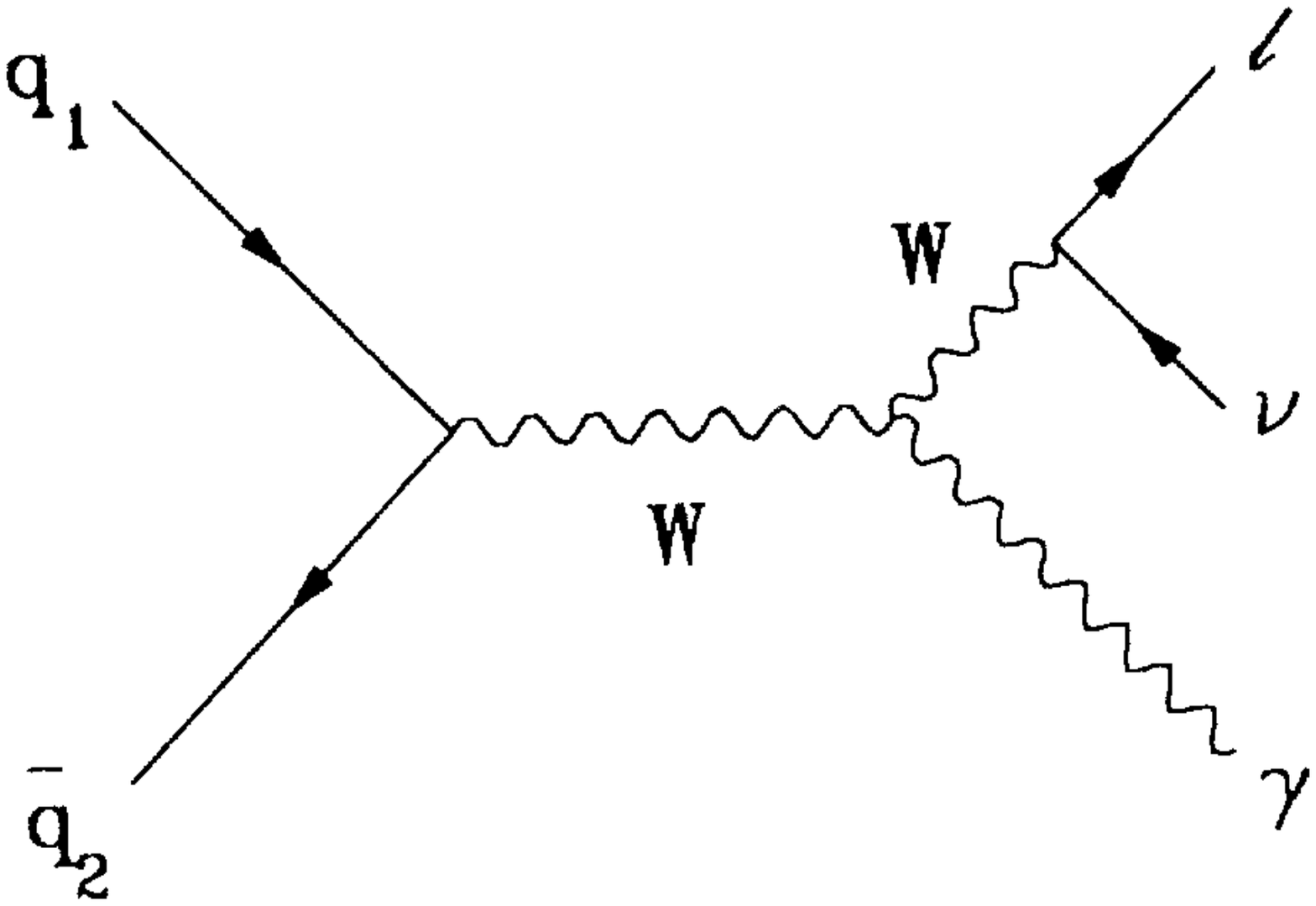} 
\hspace{0.2cm}
\includegraphics[width=0.2\textwidth,height=0.25\textwidth]{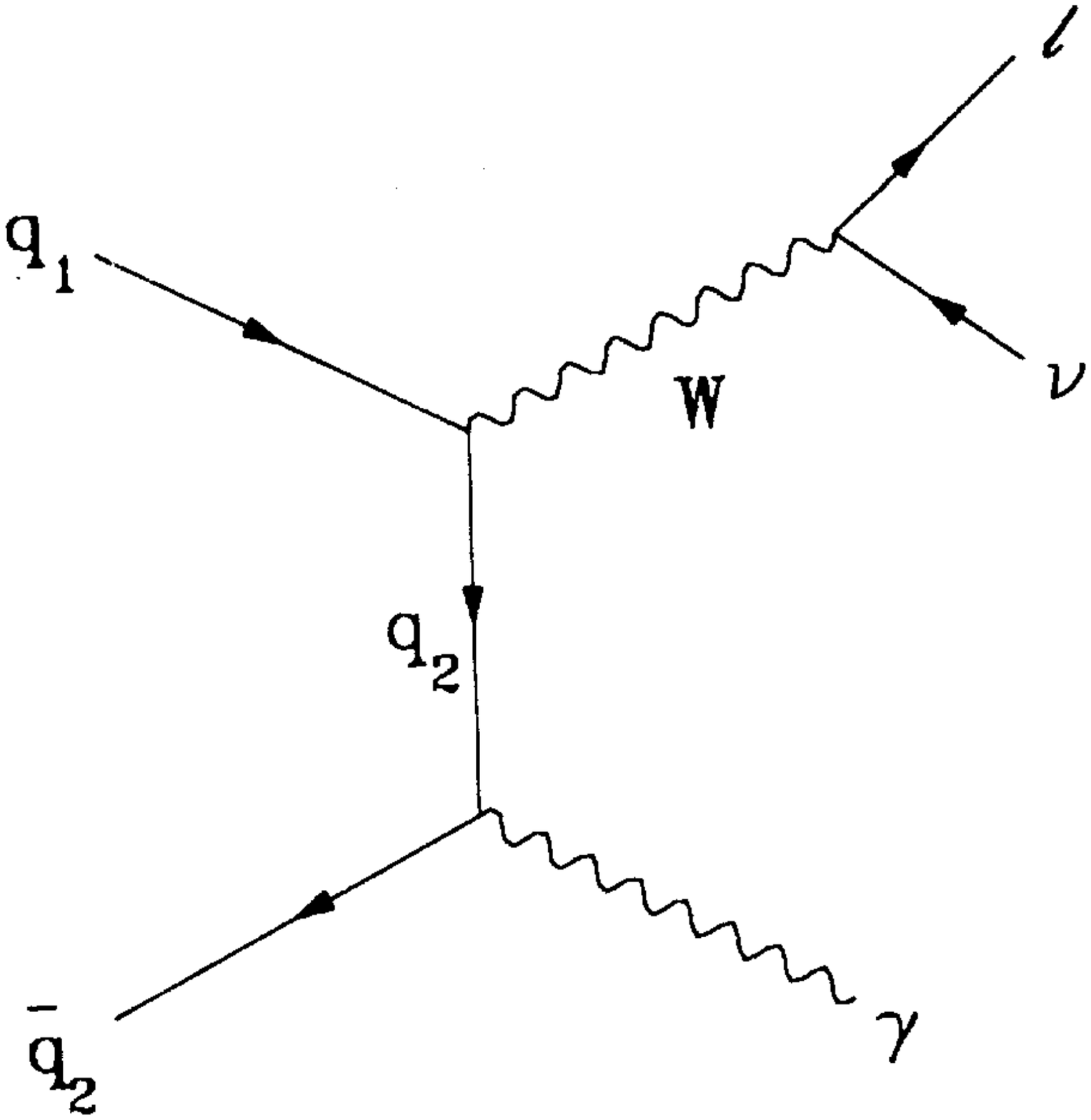}
\hspace{0.2cm}
\includegraphics[width=0.2\textwidth,height=0.2\textwidth]{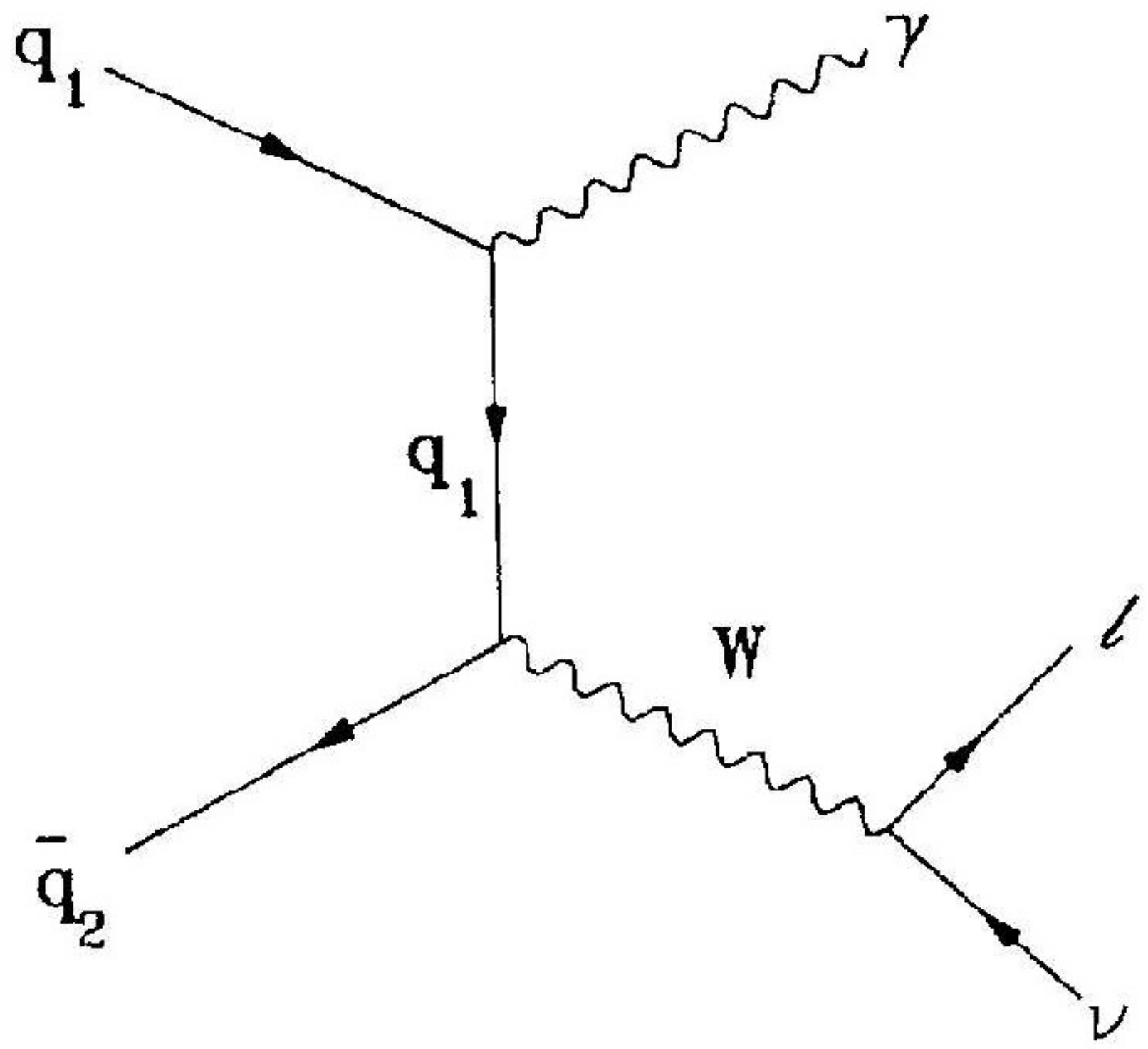} 
\caption{Born level subprocesses for \wg production in hadron-hadron collision~: s-channel process ({\bf left}), t-channel process ({\bf middle}) and u-channel process ({\bf right}).}
\label{Baur_fig:treeGraphs}
\end{center}
\end{figure}

\begin{figure}
\begin{center}
\includegraphics[width=0.2\textwidth,height=0.15\textwidth]{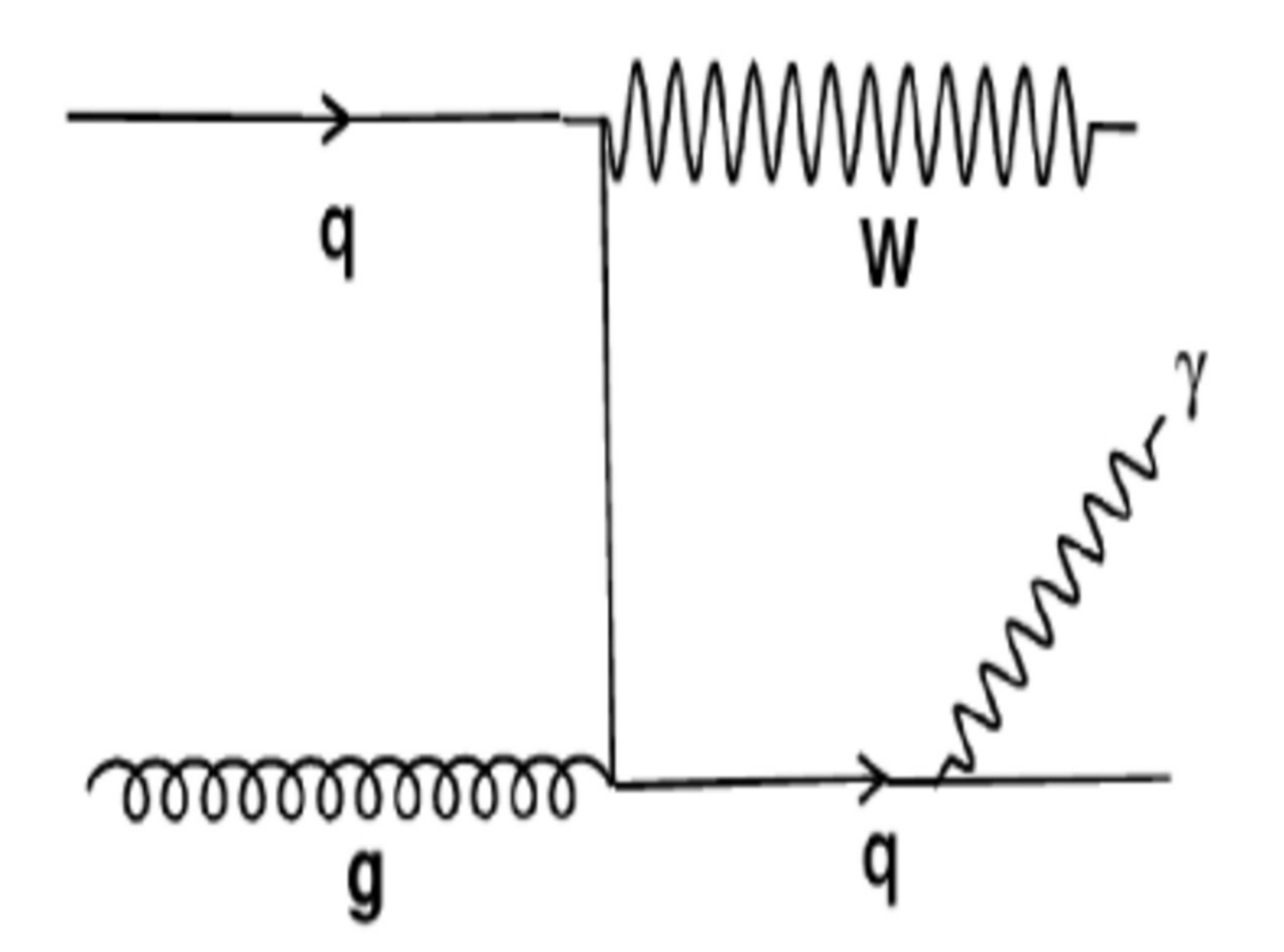} 
\hspace{0.1cm}
\includegraphics[width=0.2\textwidth,height=0.15\textwidth]{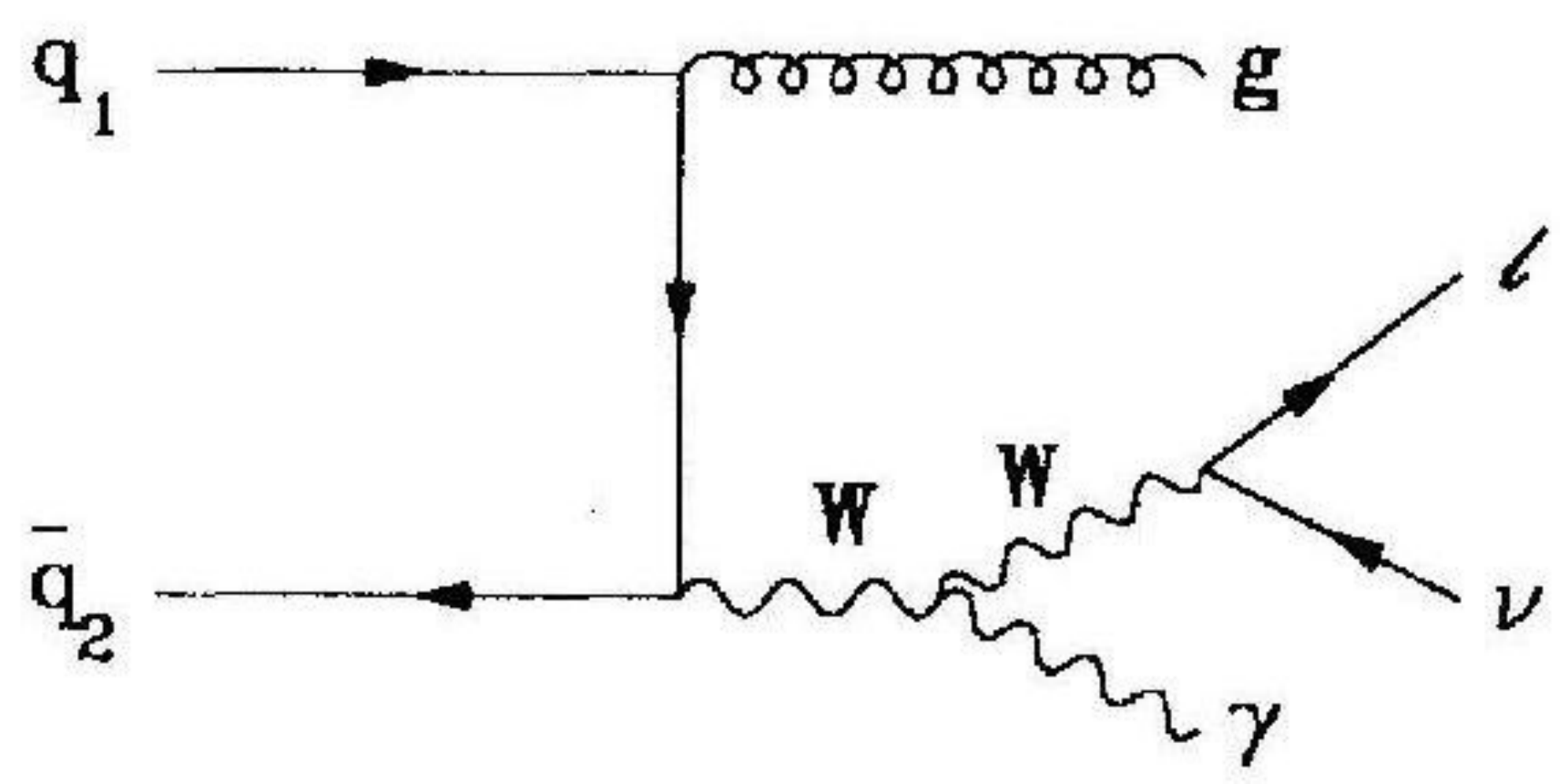} 
\hspace{0.2cm}
\includegraphics[width=0.2\textwidth,height=0.15\textwidth]{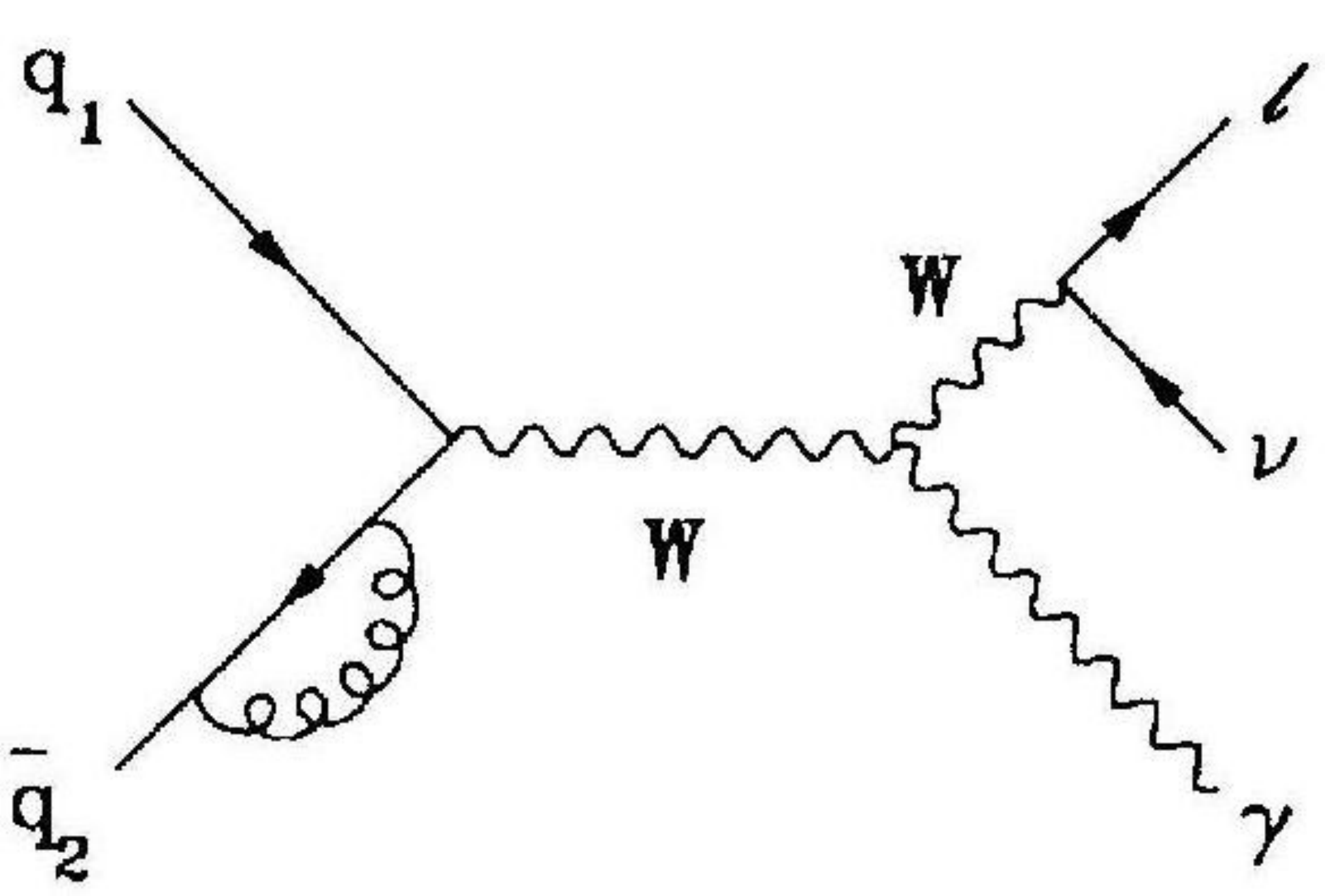}
\caption{Some higher order QCD diagrams for \wg production in hadron-hadron collision.} 
\label{Baur_fig:lo_nlo_graphs} 
\end{center}
\end{figure}

\begin{figure}
\begin{center}
\includegraphics[width=0.7\textwidth,height=0.45\textwidth]{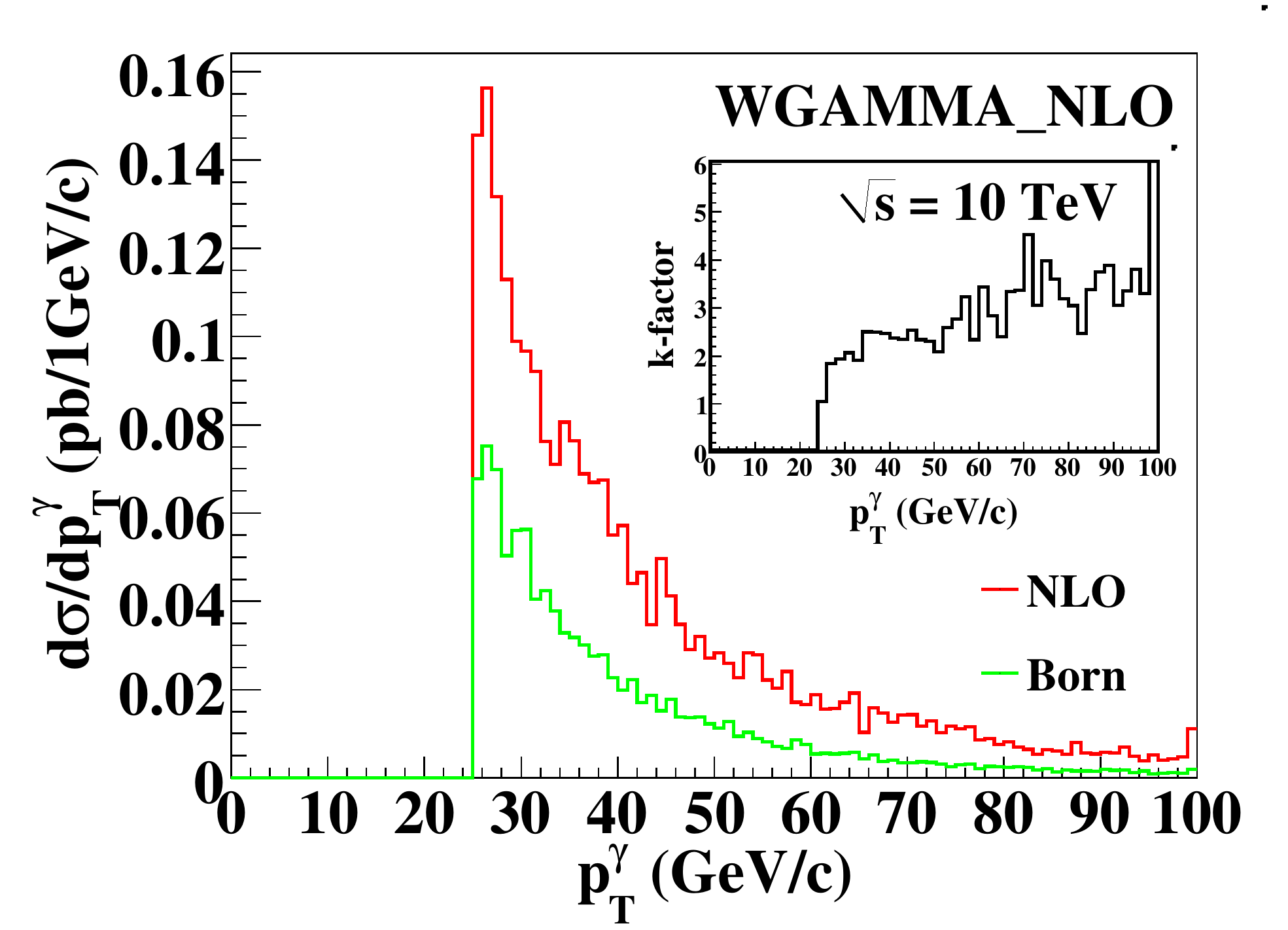} 
\caption{Differential cross-section for the photon transverse momentum at Born level is compared with the differential cross-section for NLO \wg process. Inset: k-factor as defined by the ratio of the NLO cross-section to the Born cross-section.} 
\label{Baur_fig:baurPtg_born_nll}
\end{center}
\end{figure}

\pythiaeight has been used for implementing the $p_T$-ordered  parton shower to the hard-scattered part generated by \baur. The ISR process in \pythia adds coloured partons to the incoming quarks of the hard scattering process which may transform a 3-body event to a 4-body event, thereby effectively changing the cross-section of the individual 3-body and 4-body processes. The idea behind the matching scheme is to generate the parton showers without changing the number of individual events; {\it i.e.,} the parton shower should not populate the regions of phase space for the $W\gamma j$ events, $j$ being a parton, which are already filled by the matrix element generator.

It is to be noted that \baur package produces weighted events which has to be unweighted before matching where only events with unit weight are considered.  The unweighting is done using the standard "hit-and-miss" technique, where the ratio of an event weight to the highest weight is compared against a number drawn from a uniform deviate and the event is kept if the ratio is greater than the random number generated from the uniform distribution. Events with an occasional negative weight, from virtual loops, are not considered.  

\subsubsection{The matching algorithm} 

In the matching procedure, the discrimination of 3-body vs. 4-body state is performed according to a pre-defined  threshold value $p_{T}^{separate}$ (5~GeV/$c$ for the present study), which serves as the boundary between the matrix element regime on its upper side and the parton shower regime on its lower. The matching procedure starts with projecting the 4-body system to a 3-body system,  by assuming that the outgoing parton can be emitted from either of the incoming partons (flavours permitting), with relative weights determined by splitting kernels and parton densities. Effectively this corresponds to the assumption  that the 4-body state never had a parton emitted and all the kinematics are recalculated based on this assumption. Subsequently, this projected 3-body final state event is treated according to the following steps:

\begin{enumerate} 

\item Shower the projected 3-body event and compare $p_{T}^{shower}$ at the first ISR branching with the $p_{T}^{parton}$ in original 4-body: 

\hspace{0.1cm} 

\begin{itemize} 
\item If $p_{T}^{shower} > p_{T}^{parton}$ then the event is reclassified to 3-body; move to step 2 below.
\item If $p_{T}^{shower} < p_{T}^{parton}$ then the original 4-body event is retained as a 4-body event; move to step 3 below. 
\end{itemize}

\hspace{0.1cm} 

\item Shower the 3-body events: 

\hspace{0.1cm} 

\begin{itemize}
\item Compare $p_{T}^{shower}$ with $p_{T}^{separate}$ after first ISR branching. $p_{T}^{separate}$ can be considered to be the boundary between the ME calculation's regime and that of the parton shower's. 
\item If $p_{T}^{shower} > p_{T}^{separate}$ then stop any further shower evolution and restart the parton shower until $p_{T}^{shower} < p_{T}^{separate}$. 
\item Continue with the rest of the shower.
\end{itemize} 

\hspace{0.1cm} 

\item Shower the 4-body events:

\hspace{0.1cm} 

\begin{itemize}
\item The $p_{T}^{shower}$ from the first ISR is compared with $p_{T}^{parton}$. If $p_{T}^{shower} > p_{T}^{parton}$ then restart the ISR branching until $p_{T}^{shower} < p_{T}^{parton}$ 
\item Continue with rest of shower to give a complete event.
\end{itemize} 

\end{enumerate}

\subsubsection{Matching results}

The 4-body final states from the matrix element calculation lacked the Sudakov form factor for QCD emission which was corrected for by the above-mentioned algorithm. It is to be noted that  the first ISR emission from the 3-body events are always confined below $p_{T}^{separate}$. But this is not so in the case where the parton showering is only required to be softer than the matrix element parton. 

Fig.~\ref{Baur_fig:ptgm_wgPt} ({\bf left}) shows  the distribution of the transverse momentum of the \wg system obtained by \baur superposed with the same after the \pythia showering. The distribution from \baur shows many events with $p_{T}^{W\gamma}$ equal to zero which are the 3-body events. The non-zero values correspond to events with a parton in the final state. After the matching, the kink after the zeroth bin fills up due to the boost from \pythia ISR. The area under both these curves however remain the same indicating that the exclusive cross-section of the 1-jet events remain conserved after the parton shower. Finally, Fig.~\ref{Baur_fig:ptgm_wgPt} ({\bf right}) shows the photon $p_T$ spectrum before and after the matching.

\begin{figure}[!htb]
\begin{center}
\includegraphics[width=0.45\textwidth,height=0.4\textwidth]{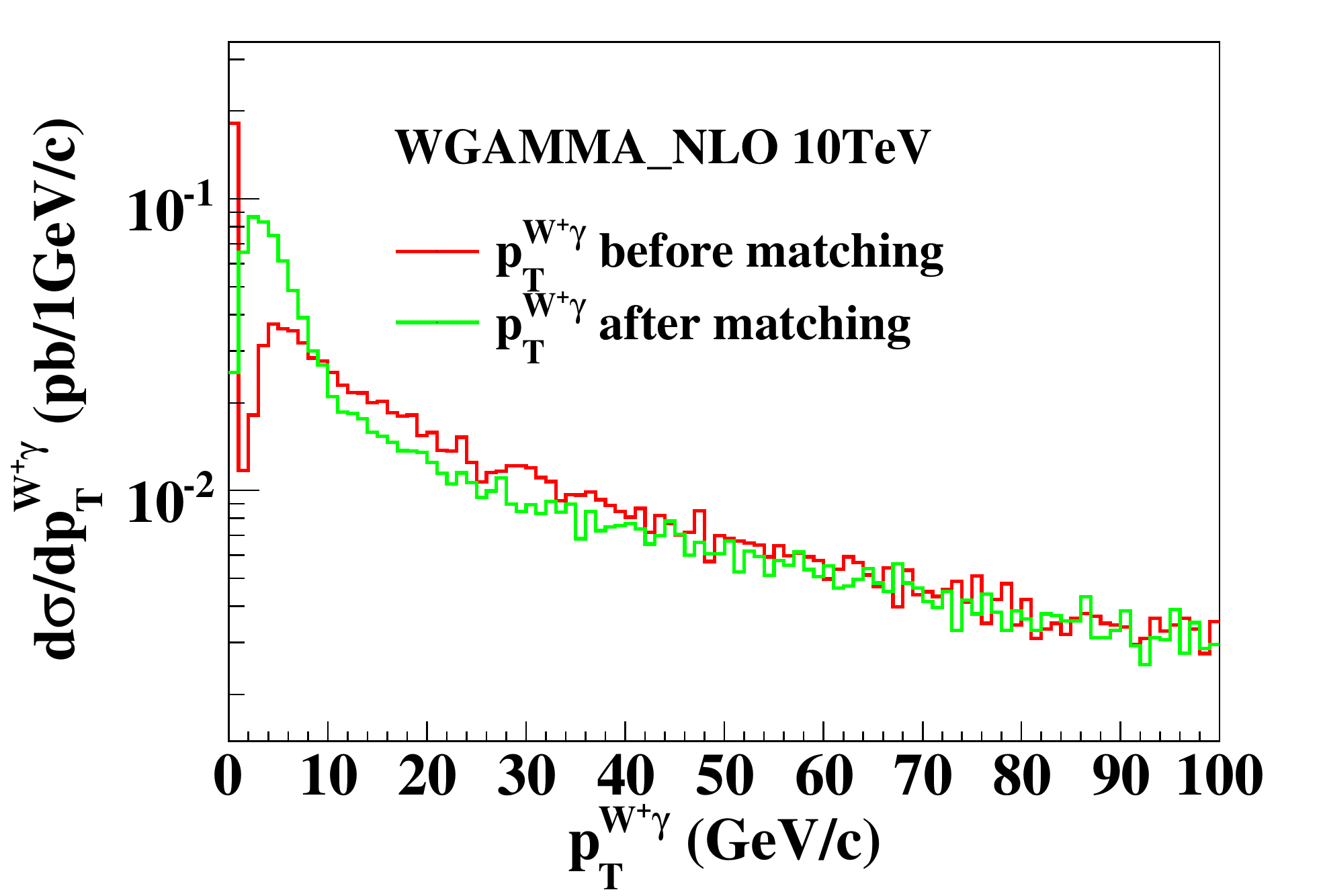}
\hspace{0.\textwidth} 
\includegraphics[width=0.45\textwidth,height=0.4\textwidth]{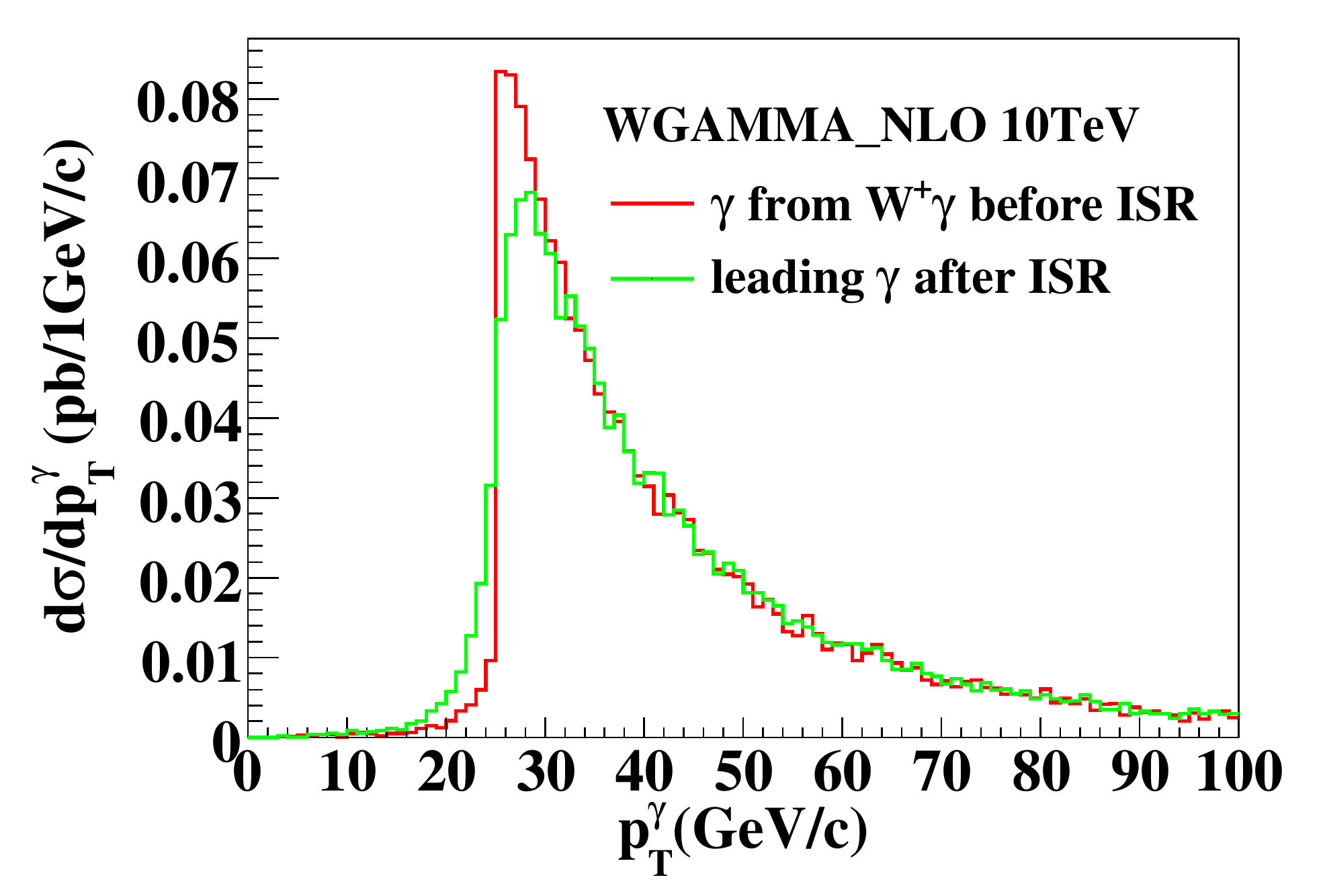}
\caption{The transverse momentum of the \wg system before \pythia showering and after \pythia showering ({\bf left}) and photon $p_T$ spectrum before and after the showering ({\bf right}).}  
\label{Baur_fig:ptgm_wgPt} 
\end{center}
\end{figure}

The matching scheme is rather simplistic and is suitable for this particular event topology which contains only one hard jet in the final state. The advantage is that it does not require any modification of the matrix element calculation of Baur for the Sudakov form factor needed for the parton emission. This is in contrast with packages based on leading order multi-parton final state calculation, where the matching scheme with parton shower effectively reduces the cross-section.

\subsection{SUMMARY}

The matching strategy as enunciated above preserves the cross-section of the \wg + 1~jet events. Also, for a reasonably high cut on \ptg, the spectrum before and after the matching is identical. It remains to be seen how stable the scheme is with respect to changes in the boundary between the matrix element and parton shower, as defined by $p_{T}^{separate}$ and also its performance in comparison to matched events generated by other matrix element generators like Madgraph and Alpgen. The final test of the best suited method can be performed only with good-statistics real-collision data at LHC. 
 
\subsection*{ACKNOWLEDGEMENT} 

The main part of this work done by DM was supported by the European Union Marie Curie Research Training Network MCNet, under contract MRTN-CT-2006-035606. DM and KM are extremely grateful to U. Baur for  many private communications at the initial stage of the study.



\clearpage

\section[THEORY TESTS OF PARTON SHOWERS]{THEORY TESTS OF PARTON SHOWERS
\protect\footnote{Contributed by: S.~Weinzierl}}
\label{sec:theoryps}

\subsection{INTRODUCTION}

The last few years have witnessed significant progress in the
improvement of parton shower algorithms with the invention of new shower algorithms 
based on the dipole or antenna 
picture \cite{Nagy:2005aa,Nagy:2006kb,Giele:2007di,Schumann:2007mg,Dinsdale:2007mf,Winter:2007ye,Platzer:2009jq}
known from next-to-leading order calculations.
These algorithms 
are able to satisfy simultaneously at each step momentum conservation
and the on-shell conditions.
This is possible, because they are based on $2 \rightarrow 3$
splittings, where the spectator can absorb the recoil. 
Within the traditional $1 \rightarrow 2$ splitting algorithms it is
impossible to satisfy simultaneously 
momentum conservation and the on-shell conditions for a splitting. 
In addition these new algorithms implement
in a correct way simultaneously the soft and the collinear limit of the matrix elements.
When both emitter and spectator are in the final state these new algorithms are very similar
to the shower algorithm implemented 
in \textsc{Ariadne} \cite{Gustafson:1986db,Gustafson:1987rq,Andersson:1989ki,Andersson:1988gp,Lonnblad:1992tz}.
The new algorithms extend the dipole or antenna picture to final-initial, initial-final and initial-initial configurations.

For any shower algorithms one needs an evolution variable. A sensible choice is 
\begin{eqnarray}
\label{weinzierl_shower_evolution_variable}
 t & = & \ln \frac{-k_\perp^2}{Q^2},
\end{eqnarray}
where $Q^2$ is a fixed reference scale and $k_\perp$ is the transverse momentum of a splitting.
During the shower evolution we move towards smaller (more negative) values of $t$.
The central object of a shower algorithm is the Sudakov factor, describing the no-emission probability.
For a dipole shower it is calculated from the individual dipoles.
For a dipole with emitter
$\tilde{i}$ and spectator $\tilde{k}$, the Sudakov factor is given by
\begin{eqnarray}
 \Delta_{ij,k}(t_1,t_2) & = &
 \exp\left( - \int\limits_{t_2}^{t_1} dt {\cal C}_{\tilde{i},\tilde{k}} 
              \int d\phi_{unres} \delta\left(t-T_{\tilde{i},\tilde{k}} \right) {\cal P}_{ij,k} \right),
\end{eqnarray}
where ${\cal C}_{\tilde{i},\tilde{k}}$ is  a colour factor.
$T_{\tilde{i},\tilde{k}}$ depends on the dipole invariant mass $(p_{\tilde{i}}+p_{\tilde{k}})^2$ and the
phase space variables for the emission of an additional particle.
The essential information is given by the function ${\cal P}_{ij,k}$, which encodes
the singular part of the matrix elements for the emission of a particle.
As an example we quote this function for the $q\rightarrow q g$ splitting:
\begin{eqnarray}
{\cal P}_{q \rightarrow q g} & = &
   C_F
   \frac{8\pi \alpha_s(\mu^2)}{(p_{\tilde{i}}+p_{\tilde{k}})^2} \frac{1}{y}
   \left[ \frac{2}{1-z(1-y)} - (1+z) \right].
\end{eqnarray}
$\alpha_s$ is evaluated at the scale $\mu^2=-k_\perp^2$. The shower algorithm proceeds through the
following steps: One first chooses the next scale $t$ at which a splitting occurs
according to the Sudakov factor. For this given choice of $t$ one then chooses the
momentum fraction $z$ according to $P_{a\rightarrow b c}(z)$ and an azimuthal angle
$\varphi$ either uniform or according to spin-dependent splitting functions.
With these three variables $(t,z,\varphi)$ one can construct all four-vectors after the emission.
The spectator absorbs thereby some recoil.
We now have a configuration where one particles was emitted.
For $t>t_{min}$ one goes back to the first step, otherwise one stops.

\subsection{EVOLUTION VARIABLES}

In the last year there has been some discussion whether the new shower algorithms based 
on dipoles or antennas \cite{Dokshitzer:2008ia,Nagy:2009re,Skands:2009tb}
have the correct collinear limit.
In particular it has been argued that the fact that the spectator takes some recoil
could conflict with collinear factorisation.
This question has been solved and it has been shown that the algorithms based on dipoles
or antennas describe correctly the collinear limit \cite{Nagy:2009re,Skands:2009tb}.
This result could have been anticipated from the fact, that the dipole
splitting functions $P_{a\rightarrow b c}$ reduce in the collinear limit $y\rightarrow 0$ to
the Altarelli-Parisi splitting functions. For the example of a $q\rightarrow q g$-splitting we
have
\begin{eqnarray}
 \lim\limits_{y\rightarrow 0}
{\cal P}_{q \rightarrow q g} & = & 
   C_F
   \frac{8\pi \alpha_s(\mu^2)}{(p_{\tilde{i}}+p_{\tilde{k}})^2} \frac{1}{y}
   \left[ \frac{2}{1-z} - (1+z) \right].
\end{eqnarray}
In addition the momentum mapping
\begin{eqnarray}
p_i = z p_{\tilde{i}} + k_\perp + y \left( 1 - z \right) p_{\tilde{k}}, 
\;\;\;
p_j = (1-z) p_{\tilde{i}} - k_\perp + y z p_{\tilde{k}}, 
 \;\;\;
p_k = (1-y) p_{\tilde{k}},
\end{eqnarray}
reduces in the collinear limit $y\rightarrow 0$ and $k_\perp \rightarrow 0$ to
\begin{eqnarray}
p_i = z p_{\tilde{i}},
\;\;\;
p_j = (1-z) p_{\tilde{i}},
 \;\;\;
p_k = p_{\tilde{k}}.
\end{eqnarray}
In particular, the spectator does not receive any recoil in the collinear limit.
What emerged from the discussions is that the choice of the shower evolution variable
is not arbitrary.
This lead to the definition of an ``infrared sensible'' evolution variable \cite{Skands:2009tb}.
The definition of infrared 
sensible is that both infinitely soft and collinear emissions should
be classified as unresolved for any finite value of the evolution
variable. 
The evolution variable based on the transverse momentum 
as in eq.~(\ref{weinzierl_shower_evolution_variable})
is an infrared sensible evolution variable.
On the other hand, the choice of the energy of the emitted particle (in the centre-of-mass system
of the emitter and the spectator) is not infrared sensible.
There are configurations where the emitted particle has finite energy, 
but which are infinitely collinear.

\subsection{MULTIPLE SOFT EMISSIONS}

The shower algorithms above are derived from first-order perturbation theory.
The singular functions entering the Sudakov factor are derived from tree-level matrix elements,
where a single particle becomes either soft or collinear.
The parton shower therefore reproduces the leading-log behaviour of the matrix elements
in the limit of an infinitely large hierarchy
between the scales of each successive emission.
No claim is made to describe correctly the case of multiple emissions at the same scale.
Nevertheless I would like to make a few comments what could be expected from future algorithms
incorporating these effects.
I will discuss multiple soft emissions, 
as derived from the singular behaviour of the matrix elements.
If a single gluon becomes soft, the square of a partial tree amplitude factorises according to
\begin{eqnarray}
\left| A_n^{(0)}(p_a,p_1,p_b,...) \right|^2 
 & = & 4 \frac{s_{ab}}{s_{a1}s_{1b}} \left| A_{n-1}^{(0)}(p_a,p_b,...) \right|^2.
\end{eqnarray}
In any strongly ordered limit the emission of $r$ soft gluons is described by \cite{Bassetto:1984ik}
\begin{eqnarray}
\left| A_n^{(0)}(p_a,p_1,...,p_r,p_b,...) \right|^2 
 & = & 4^r \frac{s_{ab}}{s_{a1}s_{12}s_{23}...s_{rb}} \left| A_{n-r}^{(0)}(p_a,p_b,...) \right|^2.
\end{eqnarray}
This formula is valid for any strong ordering
\begin{eqnarray}
 E_{\sigma(1)} \ll E_{\sigma(2)} \ll ... \ll E_{\sigma(r)} \ll E_a, E_b,
\end{eqnarray}
where $\sigma$ is permutation of $(1,2,...,r)$.
However, if two gluons are emitted at the same scale $E_1 \approx E_2$ we have a more complicated
formula. In the limit of two soft gluons a partial tree amplitude 
factorises according to \cite{Berends:1989zn}
\begin{eqnarray}
\left| A_n^{(0)}(p_a,p_1,p_2,p_b,...) \right|^2 
 & = & \left| \mbox{Eik}(p_a,p_1,p_2,p_b) \right|^2 \left| A_{n-2}^{(0)}(p_a,p_b,...) \right|^2,
\end{eqnarray}
with
\begin{eqnarray}
\lefteqn{
\left| \mbox{Eik}(p_a,p_1,p_2,p_b) \right|^2 = 
 16 \frac{s_{ab}}{s_{a1}s_{12}s_{2b}}
 + 8 \left[
          \frac{(s_{a12} s_{2b} + s_{a1} s_{12b} - s_{a12} s_{12b})^2}{s_{a12}^2 s_{12}^2 s_{12b}^2}
 \right. } & & \nonumber \\
 & & \left.
         + \frac{s_{ab}^2}{s_{a1} s_{2b} s_{a12} s_{12b}}
         +\frac{s_{ab}}{s_{12}}
          \left( 
                 \frac{1}{s_{a1} s_{12b}}
                +\frac{1}{s_{a12} s_{2b}}
                - \frac{1}{s_{a1} s_{2b}}
                -\frac{4}{s_{a12} s_{12b}}
           \right)
    \right].
\end{eqnarray}
The expression in the square bracket does not contribute to 
any strongly ordered limit ($E_1 \ll E_2 \ll E_a,E_b$ or $E_2 \ll E_1 \ll E_a,E_b$).
In the strongly ordered limit it is less singular than the first term.
However for $E_1, E_2 \ll E_a,E_b$ both terms scale like $1/\lambda^4$, if the momenta of the soft gluons
are rescaled by $\lambda$.

\clearpage

\section[HIGGS BOSON PRODUCTION VIA GLUON FUSION AT THE LHC: A COMPARATIVE STUDY]
{HIGGS BOSON PRODUCTION VIA GLUON FUSION AT THE LHC: A COMPARATIVE STUDY
\protect\footnote{Contributed by: M.~Grazzini, K.~Hamilton, S.~H{\"o}che, F.~Maltoni, C.~Oleari,
S.~de Visscher,  J.~Winter}}


\subsection{INTRODUCTION}

The primary motivation for ongoing and impending physics programmes at the
Tevatron and LHC is to gain insight into the nature of electroweak symmetry
breaking. The great majority of the effort in this direction is devoted to
the hunt for the Higgs boson, the origin of this symmetry breaking in the
Standard Model~\cite{Englert:1964et, Higgs:1964pj, Guralnik:1964eu,Kibble:1967sv}.

Of all the ways in which the Standard Model Higgs boson can be produced, the
gluon fusion process~\cite{Georgi:1977gs}, in which it couples to colliding
gluons via a top quark loop, has the largest cross section for Higgs boson
masses less than $\sim$700~GeV. 
This process will be extremely important in detecting and studying the Higgs boson at the 
LHC in the low mass region, favoured by fits of the Standard Model to electroweak precision 
data~\cite{EWWG:2008} and in part also by direct searches at the Tevatron~\cite{Bernardi:2008ee, Aaltonen:2010yv}, 
for which Higgs boson
decays into two photons are expected to give a clean experimental signal. 

Although observing a narrow resonance in the diphoton invariant mass
spectrum should be possible using only the experimental
data~\cite{Djouadi:2005gi}, determining the quantum numbers and couplings of
the resonance i.e.~determining that it really is a fundamental scalar, in
particular, the Standard Model Higgs boson, requires Monte Carlo simulations
to predict distributions for both signals and backgrounds.

In recent years Monte Carlo event generators have been the subject of great
theoretical and practical developments, most significantly in the extension
of existing parton shower simulations to consistently include exact
next-to-leading order (NLO) corrections~\cite{Frixione:2002ik,
Frixione:2003ei, Frixione:2005vw, Frixione:2006gn, Frixione:2007zp,
Frixione:2008yi, LatundeDada:2007jg, Nason:2004rx, Nason:2006hfa,
Frixione:2007nu, Frixione:2007vw, Frixione:2007nw, LatundeDada:2006gx,
Hamilton:2008pd, Hamilton:2009za, Alioli:2008gx, Alioli:2008tz,
Alioli:2009je, LatundeDada:2008bv} and, separately, in the consistent
combination of parton shower simulations and high multiplicity tree-level
matrix element generators~\cite{Catani:2001cc, Krauss:2002up,
Schalicke:2005nv, Lonnblad:2001iq, Mangano:2001xp, Mrenna:2003if,
Hoeche:2009rj,Hamilton:2009ne}.  
The state-of-the-art in fixed order
predictions has also undergone major improvement, resulting in fully
differential Monte Carlo predictions at next-to-next-to-leading order (NNLO)
for inclusive Higgs production and, simultaneously, NLO accuracy for production in
association with a hard jet~\cite{Anastasiou:2005qj, Anastasiou:2007mz,Catani:2007vq,Grazzini:2008tf}. 

In this article we present predictions from most of the Monte Carlo
simulations arising from this theoretical activity, for the gluon fusion
process at the LHC, at the expected\footnote{At the time the study was started. The current plan is 
for running at 7~TeV and 14~TeV, which leaves this study nicely in the mod range.} initial hadronic 
centre-of-mass energy,
$\sqrt{s}=10$~TeV. As well as updating existing results, based on the
originally forecast 14~TeV centre-of-mass energy, this document represents a
broad comparative study among a number of fundamentally different Monte Carlo
approaches, hence it also serves to gauge their relative merits and gauge the
stability of our theoretical predictions with respect to the various methods.

In the following we shall concisely review the pertinent features of the
simulations included in the study, prior to presenting results for a variety
of simple observables concerning the gluon fusion production channel. At the
end of the article we summarise the results and comment on the readiness of
these theoretical tools for much anticipated Higgs analysis.

\subsection{MATRIX ELEMENTS AND PARTON SHOWERS}
\label{gghcomp_mergingmethods}
In modern experimental particle physics, shower Monte Carlo~(SMC) programs
have become an indispensable tool for data analysis. From a user perspective,
these programs provide an approximate but extremely detailed description of
the final state in a high-energy process involving hadrons. They provide a
{\em fully exclusive} description of the reaction, as opposed to fixed-order
QCD calculations, which are only suitable for the computation of {\em
  inclusive} quantities.

After a latency period, research in SMC is once again very active, with
significant advances being made in the last decade. In general these
developments can be grouped into two main classes:
\begin{enumerate}
\item The merging of leading-order matrix elements~(ME), characterized by a
  high number of final-state partons, with parton showers~(PS). Examples of
  such methods are the CKKW matching
  scheme~\cite{Catani:2001cc,Krauss:2002up}, the MLM matching
  procedure~\cite{Mangano:2001xp} as well as the newer merging schemes based
  on truncated parton showers~\cite{Hoeche:2009rj,Hamilton:2009ne}.

\item The interfacing of NLO calculations (that are typically available only
  with a small number of legs in the final state) with parton shower
  simulations (\MCatNLO~\cite{Frixione:2002ik} and
  \POWHEG~\cite{Nason:2004rx}).
\end{enumerate}
All of them have to face the same problems: avoiding overcounting of events
as well as the presence of dead regions.

\subsection{MONTE CARLO PROGRAMS FOR THE STUDY}
\label{gghcomp_mcprograms}
With the exception of the \HNNLO program all of the Monte Carlo programs used
in this study fall into one or other of the two classes described above,
implementing some form of matching/merging between fixed order calculations
and parton shower simulations. In order to have a more full comparison of the
available Monte Carlo tools, we also include \HNNLO, which is based on a
fixed order NNLO calculation of Higgs-boson production via gluon fusion.

\subsubsection{HNNLO}
\label{gghcomp_hnnlo}
The \HNNLO program is based on an extension of the NLO subtraction formalism
to NNLO, as described in ref.~\cite{Catani:2007vq}.

The calculation is organized in two parts. In the first part, the
contribution of the regularized virtual corrections is computed up to
two-loop order. In the second part, the cross section for the production of
the Higgs boson in association with one jet is first evaluated up to NLO,
i.e.\ up to ${\cal O}(\alpha_s^4)$), using conventional NLO subtraction
methods. Now, since the $H+{\rm jet}$ cross section is divergent when the
transverse momentum, $\qt$, of the Higgs boson becomes small, a further
counterterm must be subtracted to make the result finite as $\qt\to 0$. To
this end the program uses counterterm introduced in
ref.~\cite{Catani:2007vq}. Having regularized the real and virtual parts, the
two contributions can be combined to reconstruct the full cross
section. Organizing the differential cross section in this way, one can
construct a parton-level event generator with which arbitrary infrared safe
quantities can be computed.  The present version of the program includes the
decay modes $H\to \gamma\gamma$~\cite{Catani:2007vq}, $H\to WW\to l\nu l\nu$
and $H\to ZZ\to$ 4 leptons~\cite{Grazzini:2008tf}.

The calculation is performed in the large top-mass approximation.  This is
known to be a good approximation provided that the Higgs boson is not too
heavy and the transverse momenta of the final state jets are not too large.

\subsubsection{MadGraph / MadEvent}
\label{gghcomp_madevent}
Besides the possibility of generating processes in a long and extensible list
of theoretical models (SM, MSSM, Higgs effective theory ...), the Monte-Carlo
generator \MGME (MG/ME)~\cite{Alwall:2007st} is also intended to simulate
accurately the QCD radiation from initial and final states when coupled to a
parton shower simulation. Such an aim can be achieved by using jet matching
and a phase-space slicing technique
%

In MG/ME, three jet matching schemes (using \pythia
6.4~\cite{Sjostrand:2006za}) are implemented, namely the
MLM~\cite{Mangano:2001xp}, the $\kt$-MLM~\cite{Alwall:2007fs,Alwall:2008qv}
and the shower-$\kt$~\cite{Alwall:2008qv} schemes. While the first two
methods work with both virtuality and $\pt$-ordered showers, the third one
only works with the $\pt$-ordered showers. For each of these methods, no
analytic Sudakov reweighting of the events is performed, instead showered
events are rejected if they are not matched to the ME-level partons.  A
detailed comparison of the $\kt$-MLM and shower-$\kt$ behaviours has shown
that their respective outputs are very similar for the production of heavy
colored particles in the SM and beyond~\cite{Alwall:2008qv}. In addition, a
comparison for the production of $W$+jets events between the results from
$\kt$-MLM and other simulation chains with different matching schemes has led
to a similar conclusion~\cite{Alwall:2007fs}.
 
Having the computation of the effective coupling between a scalar or
pseudo-scalar to the gluons, the accurate simulation of the production of a
Higgs boson accompanied by initial and final-state radiation is therefore
relatively straightforward. In order to compare the MG/ME production with the
results from the other generators considered in this work, we choose to
simulate $H+0,1,2$ partons at the ME level and apply the $\kt$-MLM scheme
with $Q^{\rm ME}_{\rm cut}=10$~GeV and $Q_{\rm match}=15$~GeV.  These choices
were made in agreement with the smoothness of ME$\rightarrow$PS transition
regions in the $2\rightarrow 1$ and $1\rightarrow 0$ differential jet rates
distributions.

\subsubsection{MC@NLO}
\label{gghcomp_mcatnlo}
The \MCatNLO method~\cite{Frixione:2002ik} was the first one to give an
acceptable solution to the overcounting problem.  The generality of the
method has been explicitly proven by its application to processes of
increasing complexity, such as heavy-flavour-pair~\cite{Frixione:2003ei} and
single-top~\cite{Frixione:2005vw} production.\footnote{A complete list of
  processes implemented in \MCatNLO can be found at\\\centerline{\tt
    http://www.hep.phy.cam.ac.uk/theory/webber/MCatNLO.}}  The basic idea of
\MCatNLO is that of avoiding the overcounting by subtracting from the exact
NLO cross section its approximation, as implemented in the SMC program to
which the NLO computation is matched. Such approximated cross section (which
is the sum of what have been denoted in~\cite{Frixione:2002ik} as MC
subtraction terms) is computed analytically, and is SMC dependent. On the
other hand, the MC subtraction terms are process-independent, and thus, for a
given SMC, can be computed once and for all. In the current version of the
\MCatNLO code, the MC subtraction terms have been computed for
\herwigsix~\cite{Corcella:2000bw}, but extensions to other SMC are possible.  In
general, the exact NLO cross section minus the MC subtraction terms does not
need to be positive.  Therefore \MCatNLO can generate events with negative
weights. For the processes implemented so far, negative-weighted events are
typically about 10--15\% of the total. Their presence does not imply a
negative cross section, since at the end physical distributions must turn out
to be positive, but affects the overall efficiency of the simulation.

The features implemented in \MCatNLO\ can be summarized as follows:
\begin{itemize}
\item[-] Infrared-safe observables have NLO accuracy.
\item[-] Collinear emissions are summed at the leading-logarithmic level.
\item[-] The double logarithmic region (i.e.\ soft and collinear gluon
emission) is treated correctly since \herwigsix uses an angular-ordered
shower.
\end{itemize}

\subsubsection{POWHEG}
\label{gghcomp_powheg}
The \POWHEG (Positive Weight Hardest Emission Generator) method was proposed
in ref.~\cite{Nason:2004rx}. This method overcomes the problem of negative
weighted events, and is not SMC specific. In the \POWHEG method, the hardest
radiation is generated first, with a technique that yields only
positive-weighted events, using the exact NLO matrix elements.  The \POWHEG
output can then be interfaced to any SMC program that is either
$\pt$-ordered, or allows the implementation of a $\pt$ veto.\footnote{All SMC
  programs compatible with the {\em Les Houches Interface for User
    Processes}~\cite{Boos:2001cv} should comply with this requirement.}
However, when interfacing \POWHEG to angular-ordered SMC programs, the
double-log accuracy of the SMC is not sufficient to guarantee the double-log
accuracy of the whole result.  Some extra soft radiation (technically called
vetoed-truncated shower in ref.~\cite{Nason:2004rx}) must also be included in
order to recover double-log accuracy. In fact, angular ordered SMC programs
may generate soft radiation before generating the radiation with the largest
$\pt$, while \POWHEG generates it first. When \POWHEG is interfaced to shower
programs that use transverse-momentum ordering, the double-log accuracy
should be correctly retained if the SMC is double-log accurate. The \ARIADNE
program~\cite{Lonnblad:1992tz}, \pythia~6.4~\cite{Sjostrand:2006za} (when
used with the new showering formalism), \ADICIC~\cite{Winter:2007ye} and the
new parton showers based on the Catani--Seymour dipole
formalism~\cite{Schumann:2007mg, Dinsdale:2007mf} adopt transverse-momentum
ordering, and aim to have accurate soft resummation approaches, in the limit
of large number of colours.

Up to now, it has successfully been applied to $Z$ pair
hadroproduction~\cite{Nason:2006hfa}, heavy-flavour
production~\cite{Frixione:2007nw}, $e^+ e^-$ annihilation into
hadrons~\cite{LatundeDada:2006gx} and into top
pairs~\cite{LatundeDada:2008bv}, Drell-Yan vector boson
production~\cite{Alioli:2008gx,Hamilton:2008pd}, $W'$
production~\cite{Papaefstathiou:2009sr}, Higgs boson production via gluon
fusion~\cite{Alioli:2008tz,Hamilton:2009za}, Higgs boson production
associated with a vector boson (Higgs-strahlung)~\cite{Hamilton:2009za},
single-top production~\cite{Alioli:2009je} $Z+1$~jet
production~\cite{POWHEG_Zjet}, and, very recently, Higgs production in vector
boson fusion~\cite{Nason:2009ai}.

\paragraph*{THE POWHEG BOX}
The \POWHEGBOX is an automated package able to construct a \POWHEG
implementation of a NLO process, given the following ingredients:
\begin{enumerate}
\item The list of all flavour structures of the Born processes.

\item The list of all the flavour structures of the real processes.
 
\item  The Born phase space.

\item \label{item:born} The Born squared amplitude, the color correlated and
  spin correlated Born amplitudes. These are common ingredients of NLO
  calculations regularized with a subtraction method.
  
\item \label{item:real} The real matrix elements squared for all relevant
  partonic processes.

\item \label{item:virtual} The finite part of the virtual corrections,
 computed in dimensional regularization or in dimensional reduction.
    
\item The Born color structure in the large limit of the number of colors.
\end{enumerate}
The plots in this article were obtained using the \POWHEGBOX, and the
completion of the shower has been done both with \pythiasix and with \herwigsix.

\subsubsection{\protect\herwigpp}
\label{gghcomp_herwig}
\herwigpp builds and improves upon the physics content of the parent \herwigsix
program, particularly in regard to the accurate simulation of QCD
radiation. A major success of the original \herwigsix program was in its
modeling the effects of soft gluon interference, specifically the
\emph{colour coherence} phenomenon, whereby the intensity of soft gluon
radiation, emitted at wide angles with respect to a bunch of collimated
colour charges, is found to be proportional to the \emph{coherent} sum of
emissions from the constituents i.e. the jet
parent~\cite{Bassetto:1982ma,Bassetto:1984ik,Catani:1983bz,Ciafaloni:1980pz,
  Ciafaloni:1981bp,Dokshitzer:1988bq,Marchesini:1984bm,Mueller:1981ex,
  Ermolaev:1981cm,Dokshitzer:1982fh}. This effect is manifest in the
perturbative series as large soft logarithms and is implemented as an angular
ordering of successive emissions in the parton shower.

A further significant accomplishment of the \POWHEG~\cite{Nason:2004rx}
formalism is in fully catering for such effects through a careful
decomposition of the angular-ordered parton shower into a truncated shower,
describing soft wide angle radiation, the hardest emission, as described
above, and a vetoed shower comprised of increasingly collinear emissions.  In
doing so the formalism provides a means of distributing the highest $\pt$
emission in an event according to the exact real-emission matrix element,
including resummation effects, without degrading or otherwise disturbing the
resummation and colour-coherence properties inherent to the parton shower.

The facility to perform truncated showers is absent from the older fortran
\herwigsix program but is implemented in the new \herwigpp program, which also
includes its own native, independent, \POWHEG simulation for the gluon-fusion
process, the results of which are presented in section~\ref{gghcomp_results}

\subsubsection{\sherpa}
\label{gghcomp_sherpa}
\sherpa is a multi-purpose Monte-Carlo event generation framework for
colliders~\cite{Gleisberg:2003xi, Gleisberg:2008ta}. The main goal of this
project is a proper simulation of the perturbative aspects of the collision,
although significant improvements have been made over the past years
regarding the simulation of non-perturbative dynamics, like the process of
hadronisation.  One of the key features of the \sherpa program is a general
implementation of a novel technique for combining tree-level matrix elements
with parton showers, in arbitrary QCD or QCD-associated
processes~\cite{Hoeche:2009rj}. To this end, \sherpa makes use of its two
internal tree-level matrix element generators \AMEGIC~\cite{Krauss:2001iv}
and \Comix~\cite{Gleisberg:2008fv}, which are capable of simulating both
Standard Model (\AMEGIC and \Comix) and beyond Standard Model (\AMEGIC)
reactions with high-multiplicity final states. Soft and collinear parton
radiation is generated in \sherpa by means of a parton shower based on
Catani--Seymour dipole factorisation~\cite{Schumann:2007mg}. This formalism
has apparent advantages compared to more conventional parton showers, which
are based on strict $1\to2$ splittings and often lack the notion of a
well-defined spectator parton.  In the new approach, the recoil partner of a
splitting parton is always a single external particle. This turned out to be
an important ingredient when combining parton-shower evolution with
higher-order tree-level matrix elements.

The technique for merging matrix elements and parton showers, which is
employed by \sherpa, is based on phase-space slicing.  A detailed description
of the corresponding algorithm, and its relation with other tree-level
merging techniques, can be found in~\cite{Hoeche:2009rj}.  The method has
recently been applied to mixed QCD and electroweak processes, in particular
photon and diphoton production~\cite{Hoeche:2009xc} and proves to give a
consistent and reliable description of data from deep-inelastic
lepton-nucleon scattering~\cite{Carli:2009cg}. Compared to the CKKW
algorithm~\cite{Catani:2001cc,Krauss:2002up}, the new merging scheme of
ref.~\cite{Hoeche:2009rj} is more sophisticated and improves over CKKW by
including truncated vetoed parton showers. The results are more accurate,
with respect to those of the CKKW approach, which has been employed in former
versions of the \sherpa event generator with already great
success~\cite{Krauss:2004bs,Krauss:2005nu,Gleisberg:2005qq,Alwall:2007fs}.

\subsection{Parameters for the study}
\label{gghcomp_parameters}

In the following, we present results for the $pp$ LHC at a centre-of-mass
energy of 10~TeV.  The common features of the analysis are the following:
\begin{itemize}
\item {\em Model}: We work in the Standard Model in the large top-mass
limit. In this approximation the couplings between the gluons and the Higgs
boson are described by a dimension-five effective interaction
\begin{equation}
\mathcal{L}_{h}=-\frac{1}{4}\, g_h\, G_{\mu\nu}^aG_{\mu\nu}^a H,
\end{equation}
with $g_h=\alpha_s/(3 \pi v)$.
\item {\em Event samples}: The analysis was performed on the generated final
  state and, with the exception of \HNNLO, after parton showering. Hadronization
  effects were included for the \MCatNLO and \POWHEG results only.
 
  A Higgs boson mass of $m_H=120$~GeV was assumed.  Tree-level predictions
  were generated using the leading order CTEQ6l1 PDF
  set~\cite{Pumplin:2002vw}, while generators employing the \POWHEG method
  used the next-to-leading order PDF set CTEQ6m~\cite{Pumplin:2002vw}. In
  both cases the parametrisation of the strong coupling was chosen
  accordingly.  All partons (excluding the top quark) were taken to be
  massless and their Yukawa couplings were neglected.
\item {\em Jet definitions}: Jets were defined using the longitudinally
  invariant $\kt$-algorithm with $D=0.7$ in the implementation of
  FastJet~\cite{Cacciari:2005hq}. They were required to lie within a rapidity
  range of $\abs{\eta} < 4.5$ and have transverse momenta of $\pt > 20$~GeV.
\end{itemize}

\subsection{RESULTS}
\label{gghcomp_results}
In this section we present and discuss results obtained for some key
observables in the analysis of Higgs production via gluon fusion. As already
mentioned, we do not expect the LO matched results to provide reliable
information on total rates, so we have normalized the corresponding curves,
from \MGME and \sherpa, to the \HNNLO result. On the other hand, having NLO
accuracy, \herwigpp, \MCatNLO and \POWHEG have not been rescaled.

\begin{figure}
  \begin{center}
    \includegraphics[width=0.49\textwidth]{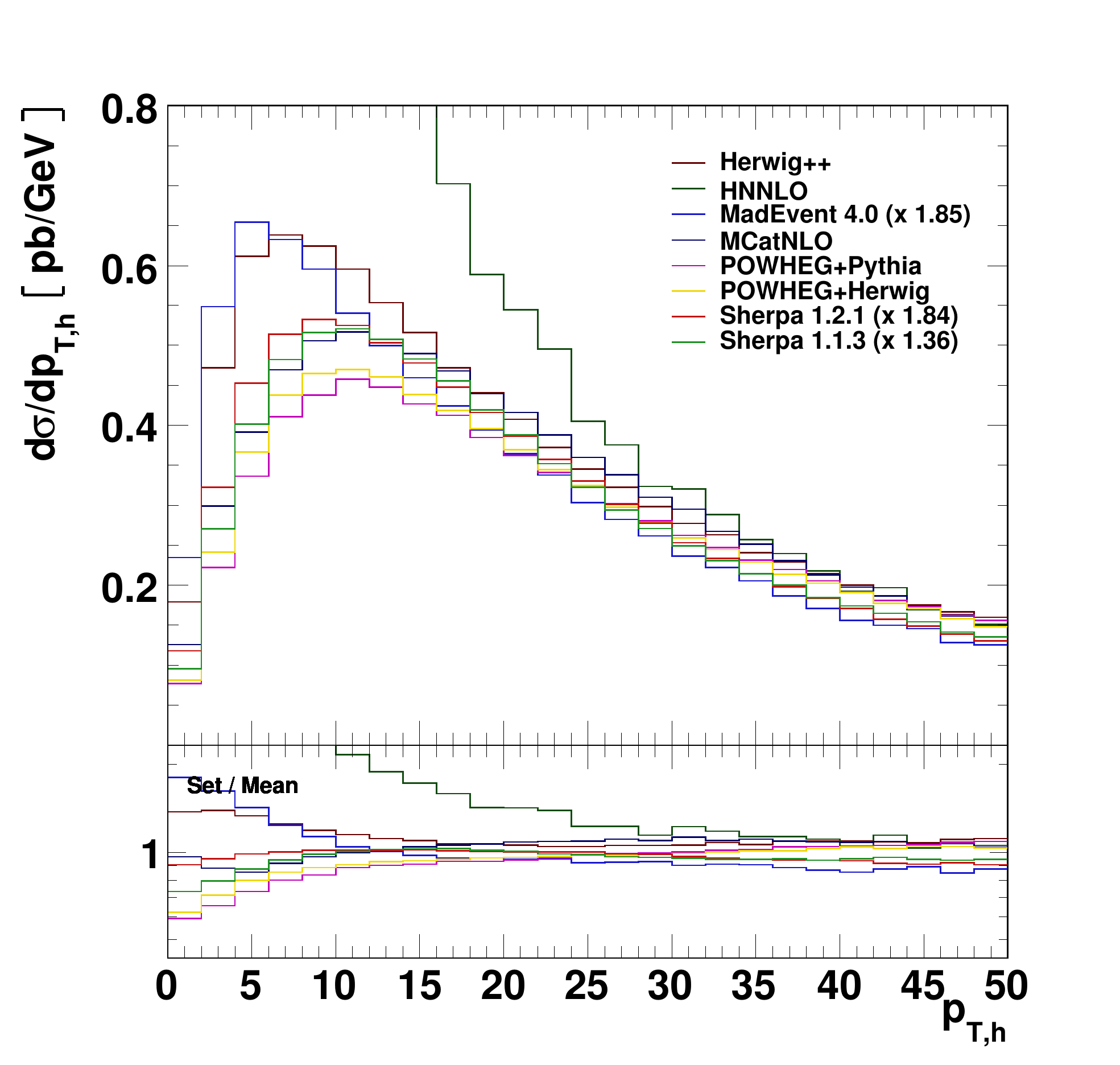}
    \includegraphics[width=0.49\textwidth]{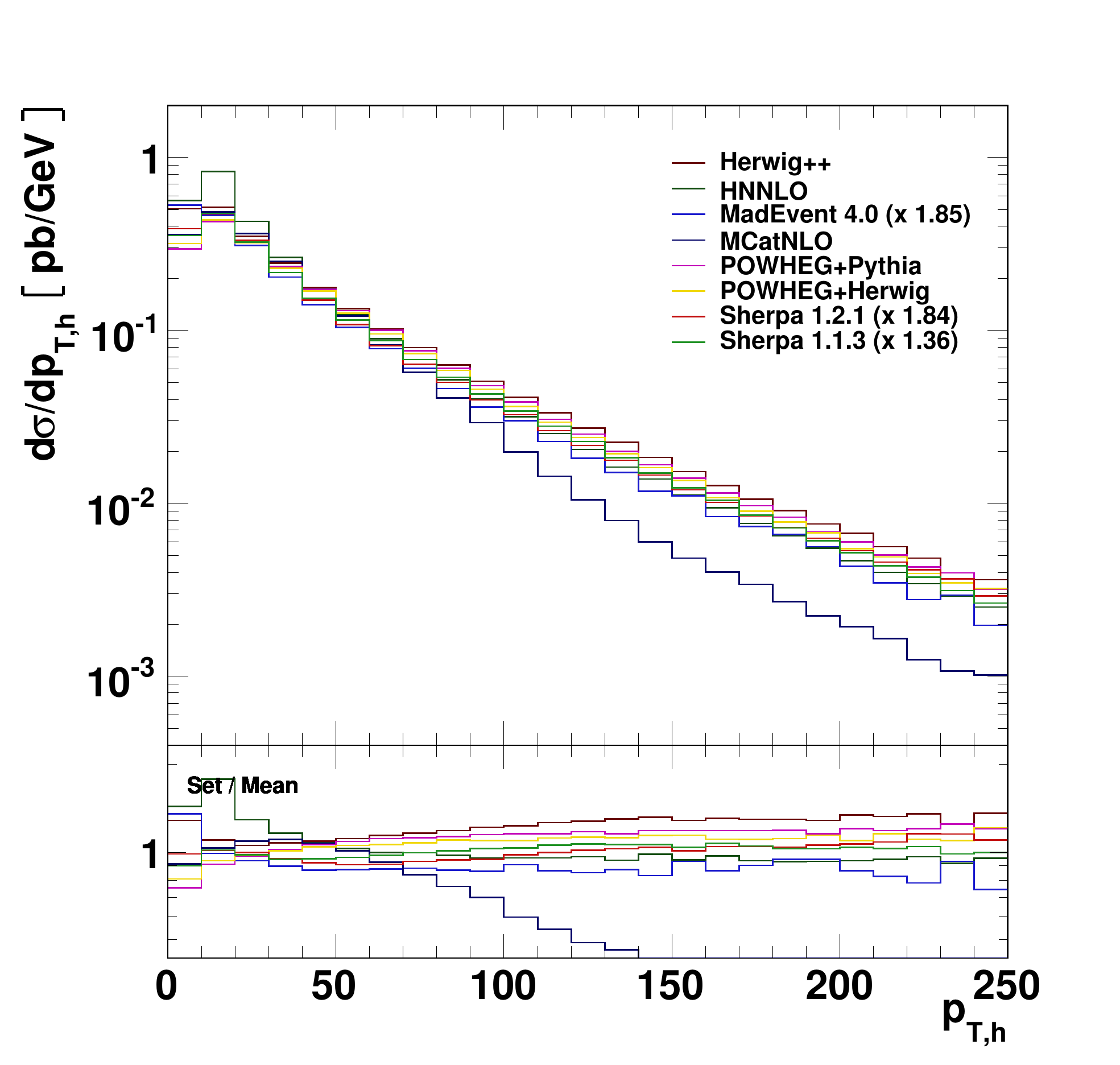}
  \end{center}
  \caption{The transverse momentum spectrum of the Higgs boson.  Tree-level
    predictions have been rescaled by the global factors indicated in the
    legend. The lower panels display the ratio of individual results and the
    average of all histograms, excluding the results from HNNLO.  }
  \label{gghcomp_PTh}
\end{figure}
In fig.~\ref{gghcomp_PTh}, we plot the transverse momentum of the Higgs boson
and, in fig.~\ref{gghcomp_Yh}, its rapidity.  The Higgs boson $\pt$, in
particular, is determined by the recoiling QCD radiation, both soft and hard,
and, exactly as for Drell-Yan, it is therefore a key observable.  The blow up
of the small-$\pt$ region (left panel of fig.~\ref{gghcomp_PTh}) shows quite
good agreement among the various MC approaches with predictions being
typically peaked in the range between 5 and 10~GeV.  The obvious excess in
the \HNNLO prediction, at low-$\pt$, is expected on the grounds that it is
based on a fixed order computation, hence, it does not resum the effects of
multiple soft emissions, which are essential for a proper description of the
$\pt=0$ region.  At higher values of the $\pt$, the agreement is also
excellent, apart from \MCatNLO which shows a steeper behaviour with respect
to the results obtained by the \POWHEG method and with the matching. As
already pointed out in Refs~\cite{Alioli:2008tz, Hamilton:2009za}, this is
due to NNLO terms in the \POWHEG formula.
It is, however, important to note that, for all the NLO codes, this
particular distribution can be predicted only at LO, i.e. no $H+2$ partons
contribution and no $H+1$ parton one-loop contributions are included. From
this point of view, it is reasonable to expect the shape to be sensitive to
variations in the renormalization and factorization scales, although, in
practice, this sensitivity is much milder due to the resummation of higher
order corrections (i.e.~the shower). In any case, it is both remarkable to
see that the predictions based on the \POWHEG method and the ME+PS matching
show such good agreement, particularly considering the fundamental
differences in their methodology.

\begin{figure}
  \begin{center}
    \includegraphics[width=0.6\textwidth]{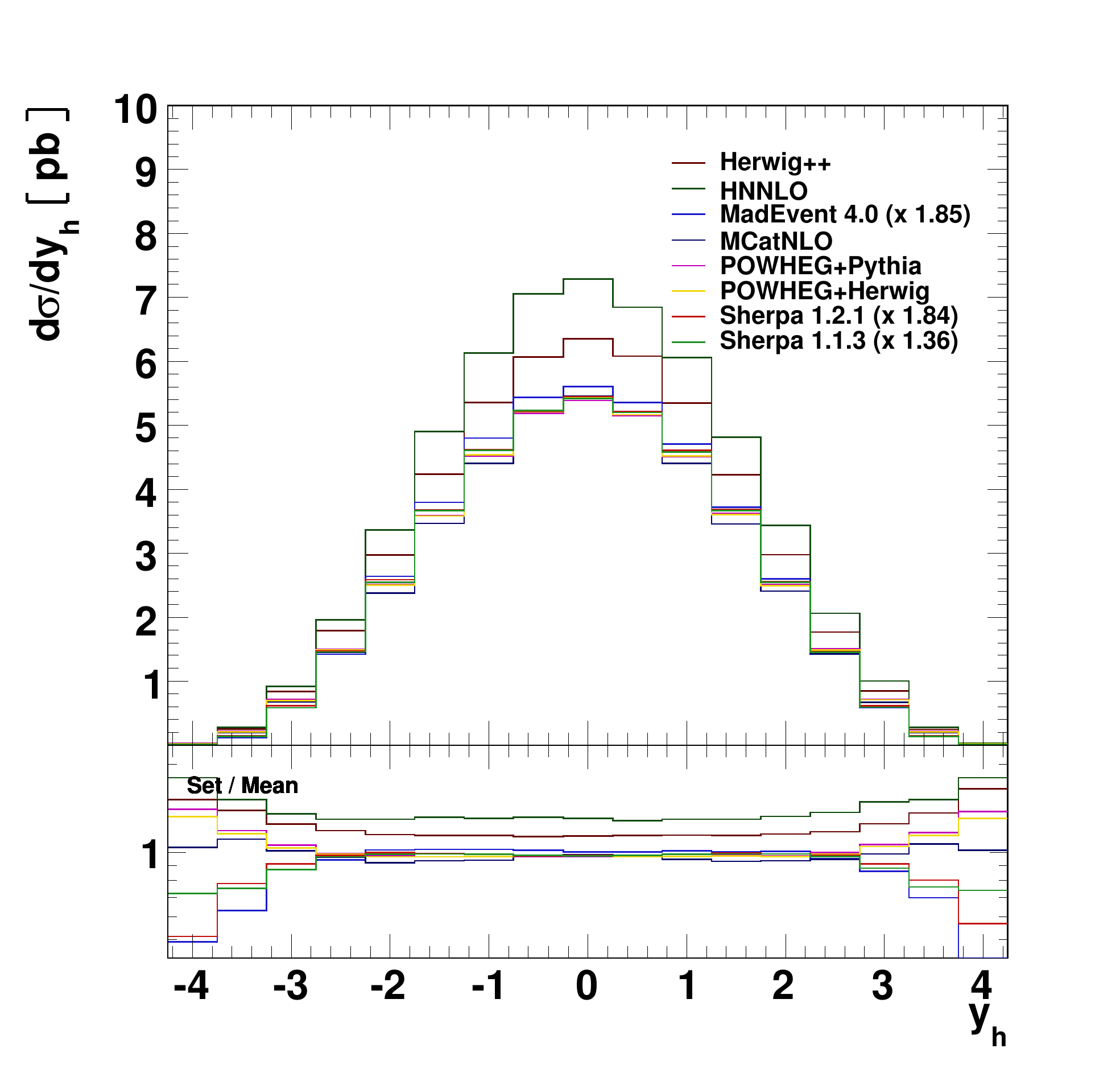}
  \end{center}
  \caption{The rapidity distribution of the Higgs boson. See
    fig.~\protect\ref{gghcomp_PTh} for details.}
  \label{gghcomp_Yh}
\end{figure}
In the rapidity distributions of fig.~\ref{gghcomp_Yh}, the \HNNLO  result
shows all of its NNLO content: in fact, this is the only plot that receives
contributions from the two-loop diagrams.

\begin{figure}
  \begin{center}
    \includegraphics[width=0.49\textwidth]{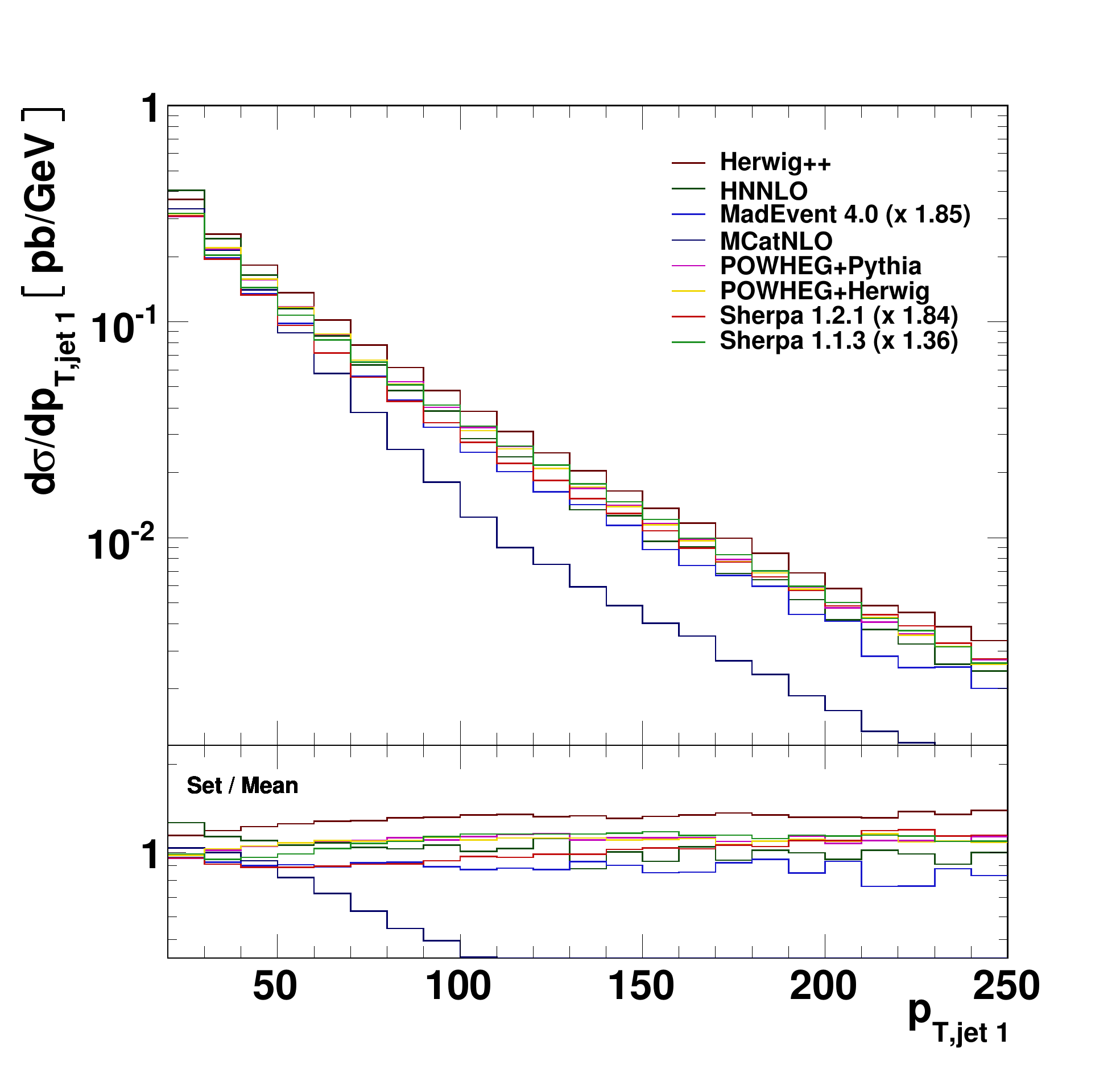}
    \includegraphics[width=0.49\textwidth]{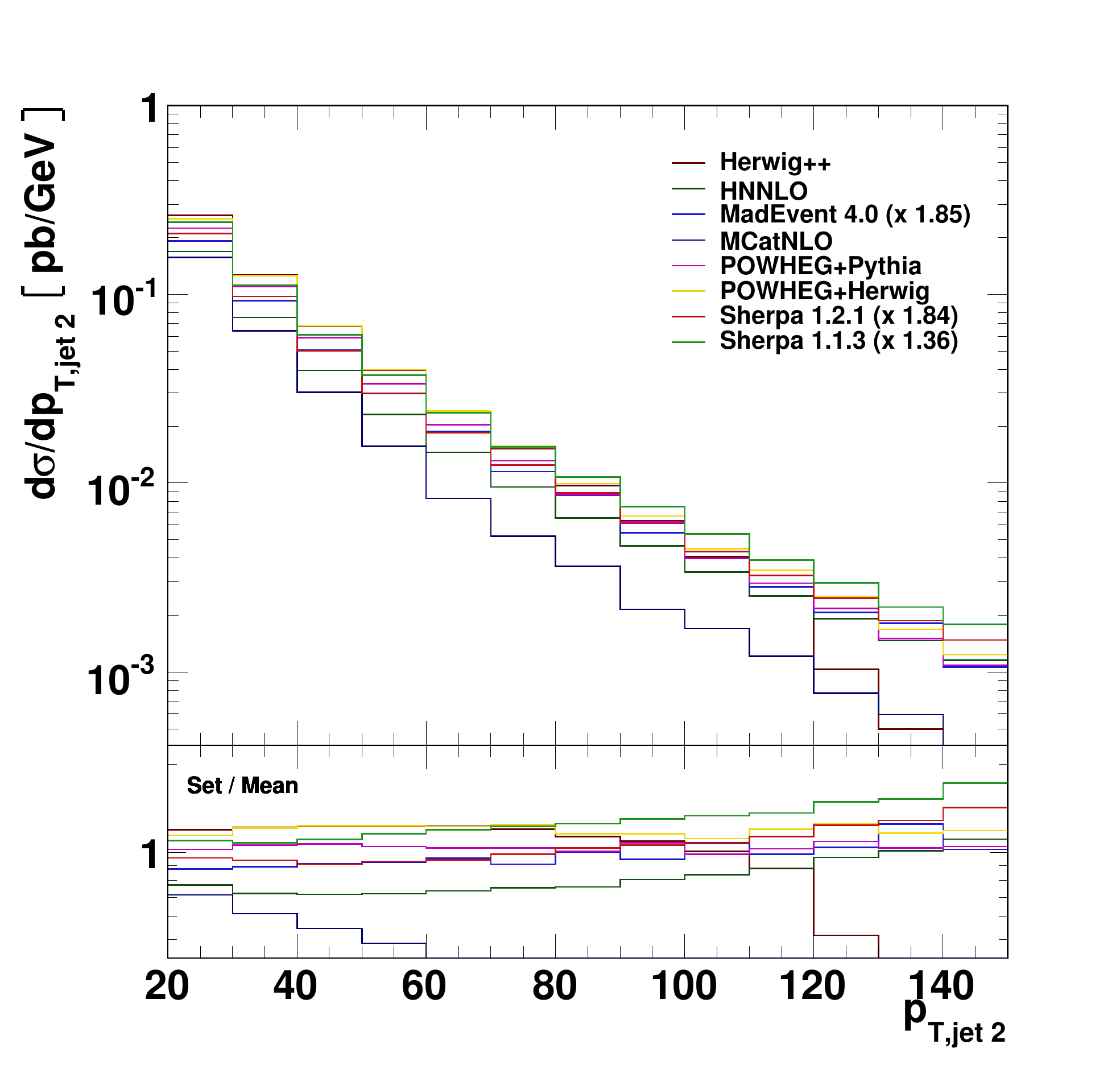}
    \includegraphics[width=0.49\textwidth]{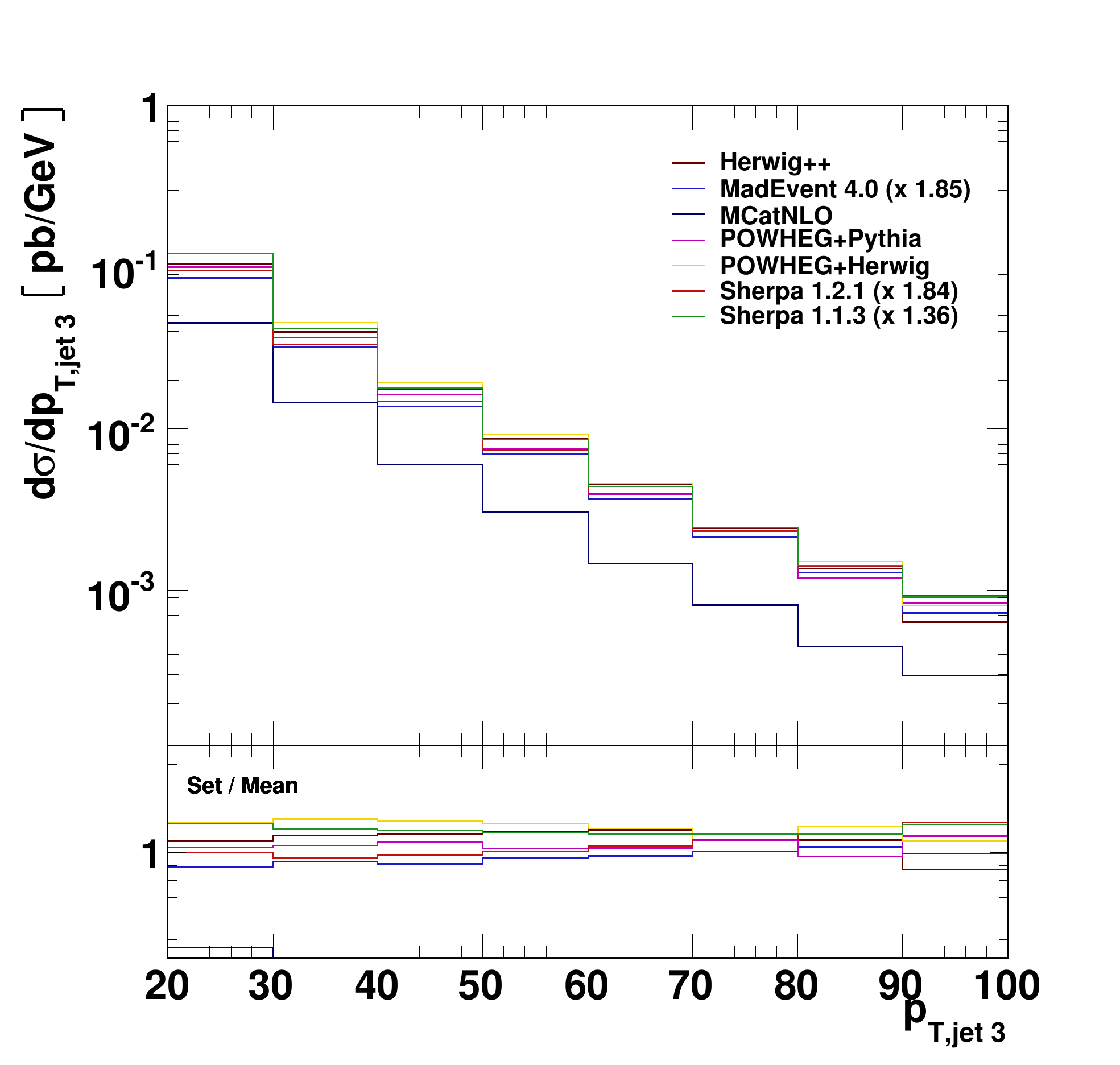}
    \includegraphics[width=0.49\textwidth]{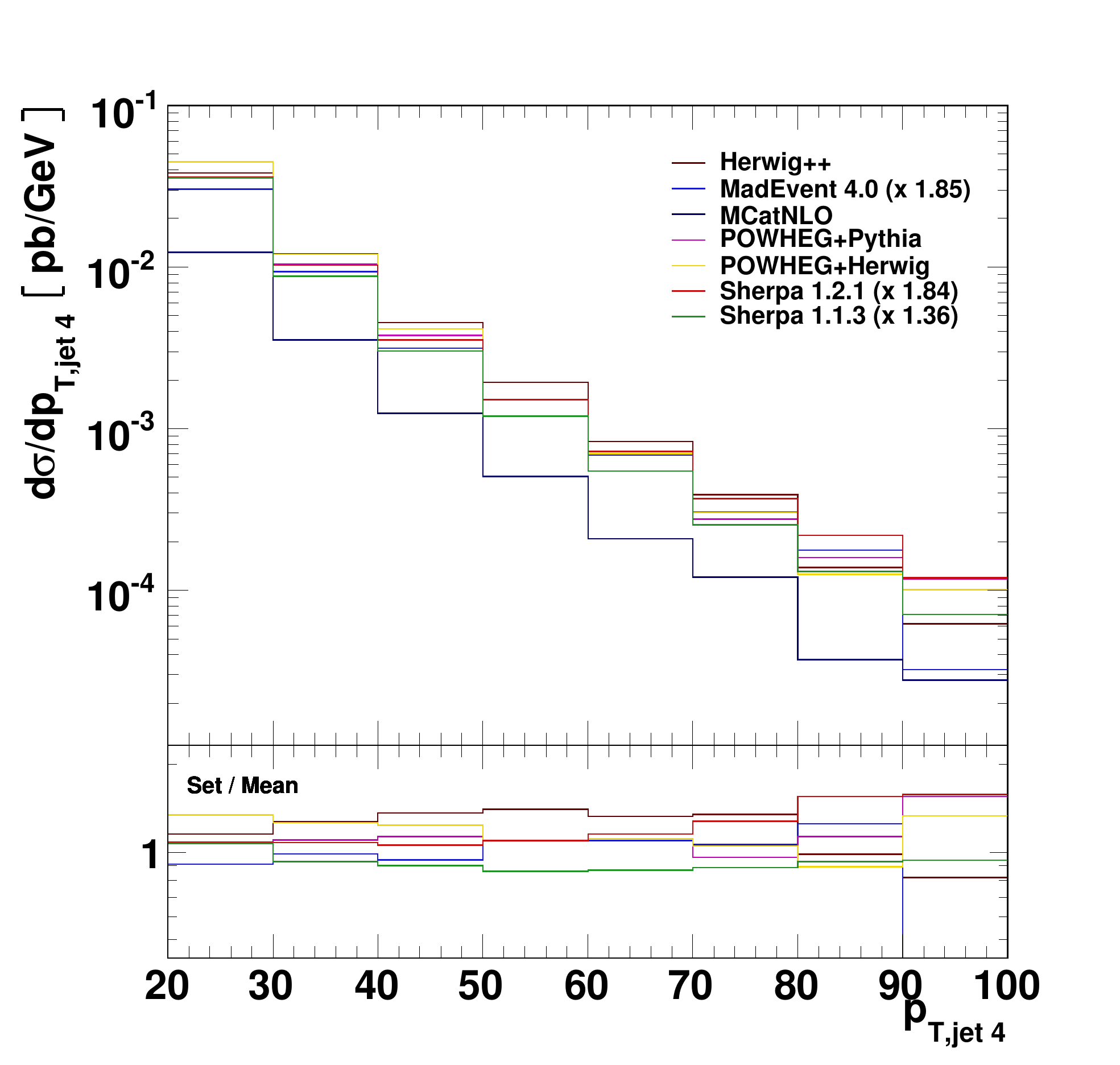}
  \end{center}
  \caption{The transverse momentum spectra of the first four hardest jets,
    ordered in $\pt$, accompanying the Higgs boson. See
    fig.~\protect\ref{gghcomp_PTh} for details.}
  \label{gghcomp_PTj12}
\end{figure}

\begin{figure}
  \begin{center}
    \includegraphics[width=0.49\textwidth]{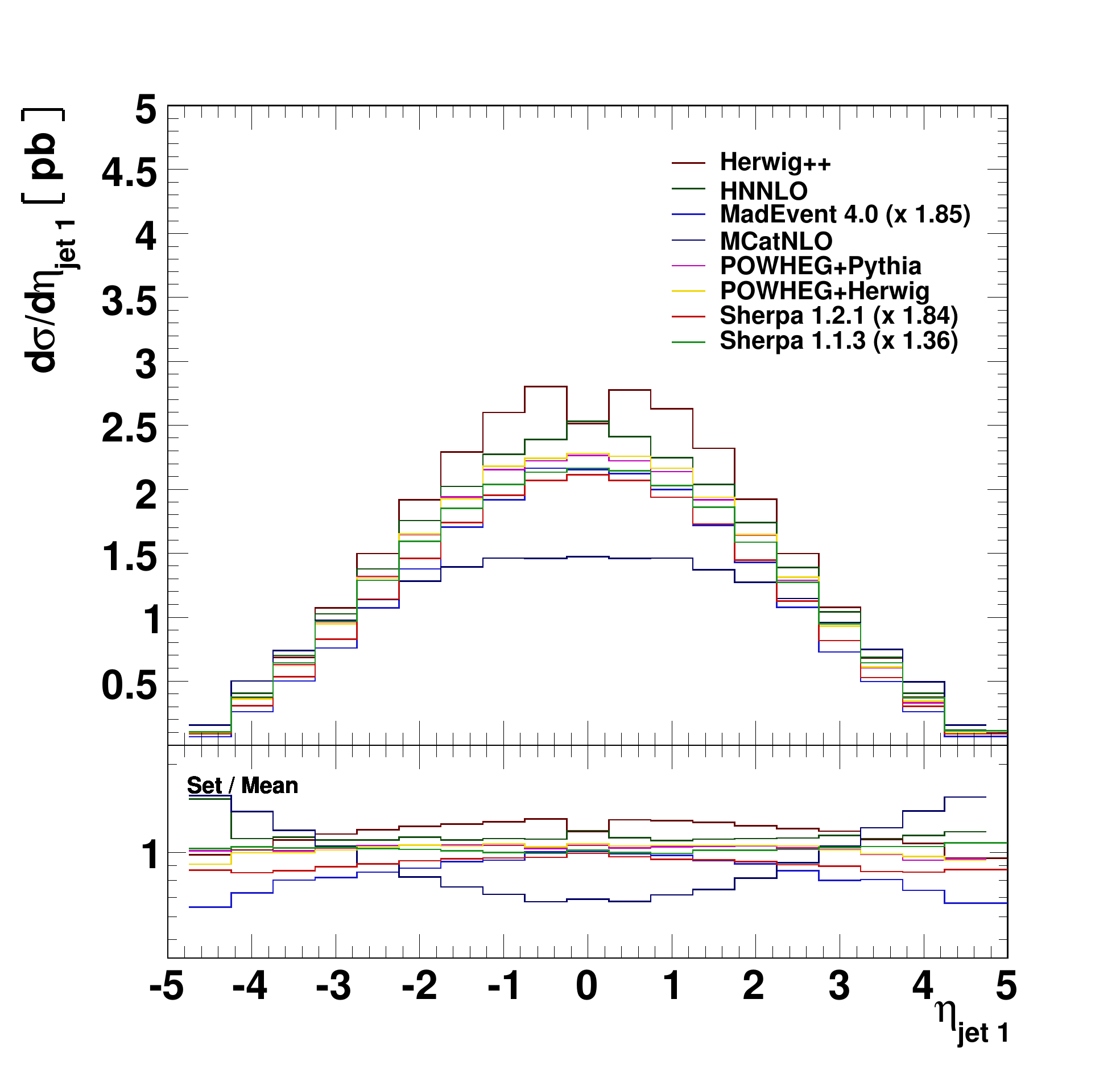}
    \includegraphics[width=0.49\textwidth]{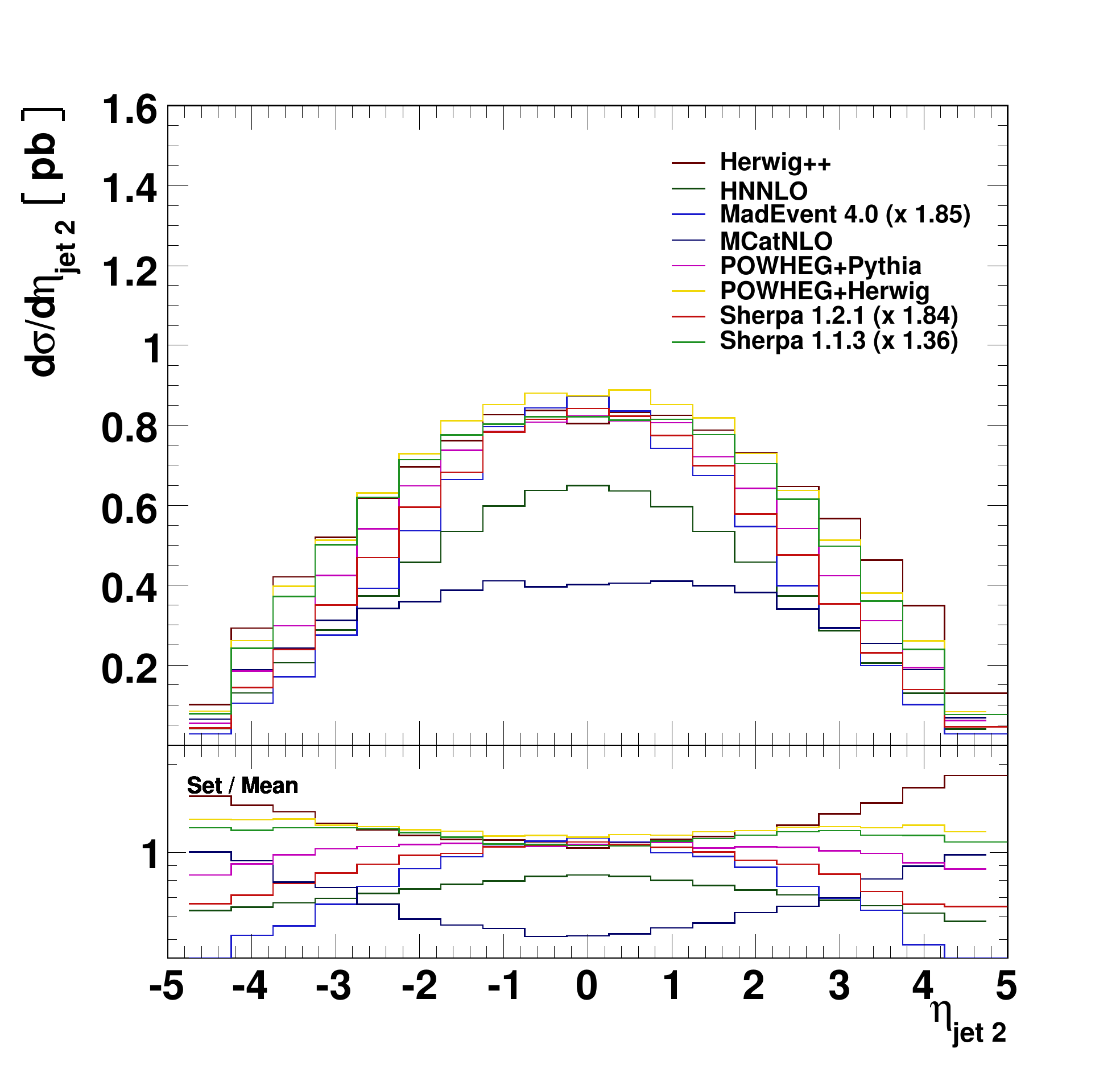}
    \includegraphics[width=0.49\textwidth]{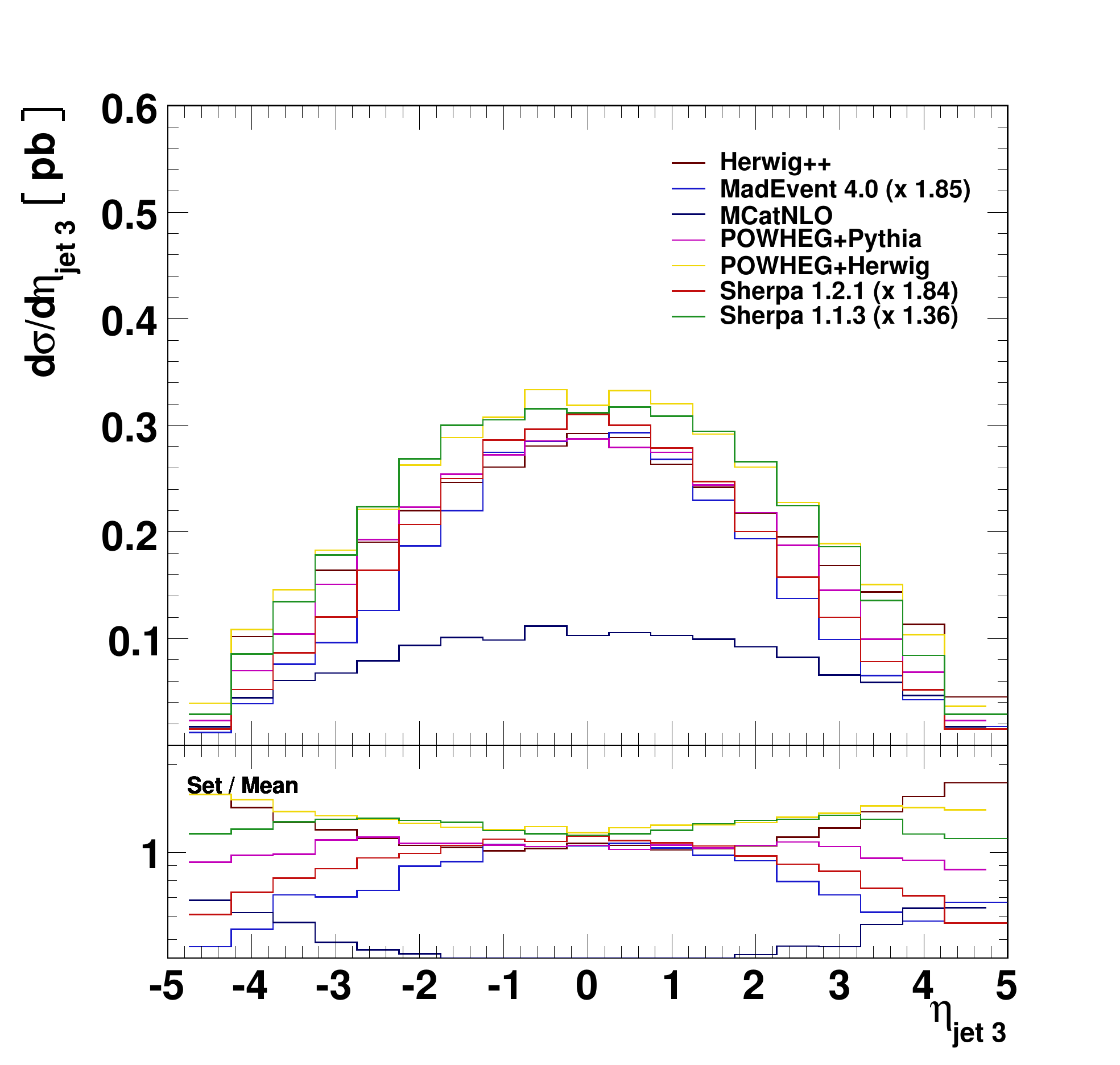}
    \includegraphics[width=0.49\textwidth]{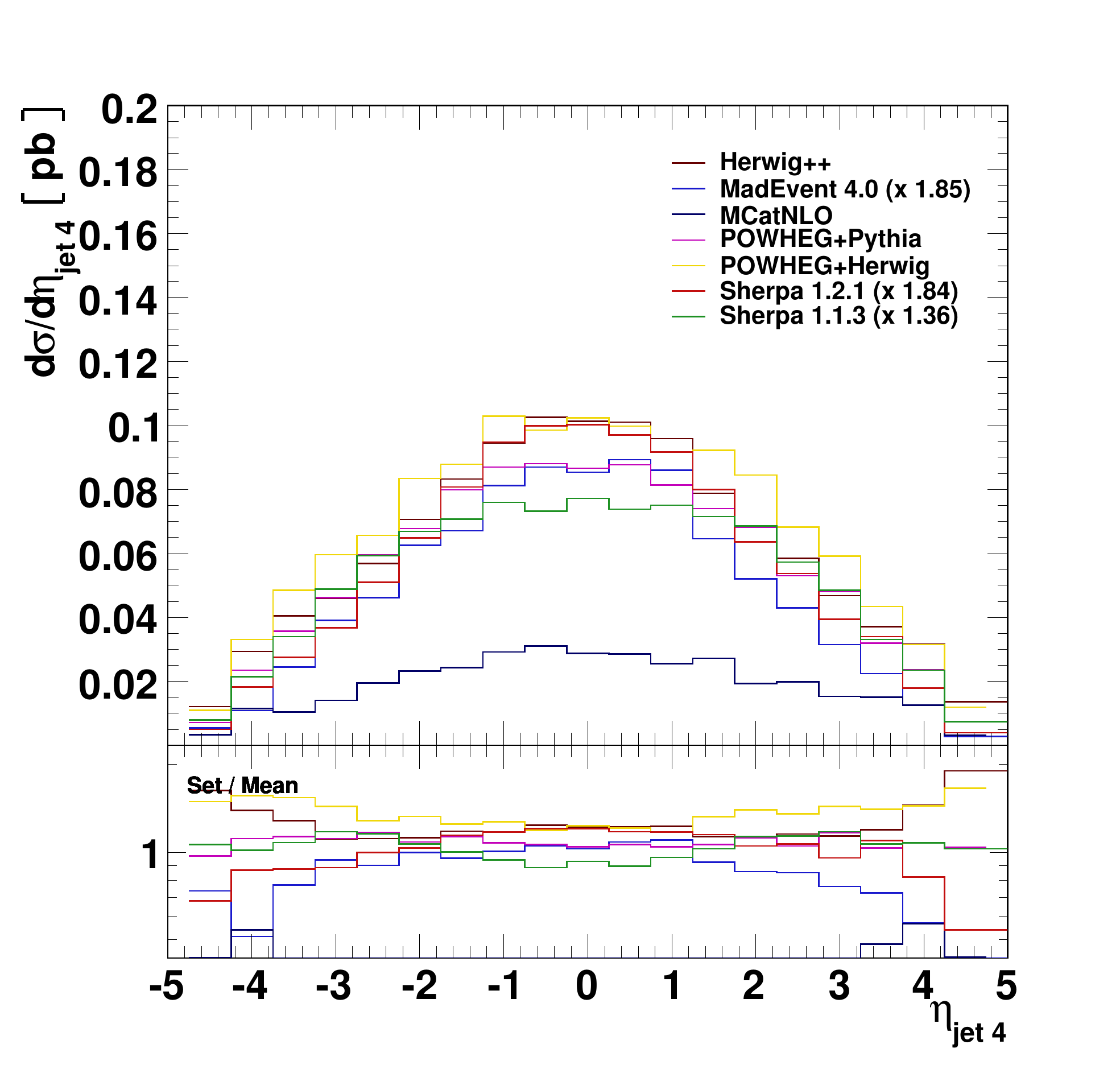}
  \end{center}
  \caption{The pseudorapidity distributions of the first four hardest jets,
    ordered in $\pt$, accompanying the Higgs boson. See
    fig.~\protect\ref{gghcomp_PTh} for details.}
  \label{gghcomp_etaj12}
\end{figure}

Figure~\ref{gghcomp_PTj12} shows the jet $\pt$ distributions for the four
hardest jets (ordered in $\pt$). Once again the agreement among the various
approaches is very good, with \MCatNLO leading to significantly less events
at very high $\pt$'s; this lower number of events is in exact correspondance
with that seen for the Higgs boson transverse-momentum distribution and bears
the same explanation. A particularly interesting feature is the agreement
found on the 3rd and 4th jet spectra.  Only \sherpa has included the
corresponding tree-level hard matrix elements, while all other predictions
contain only one (NLO codes) or two (\HNNLO and \MGME) hard partons.
This good agreement is a mere coincidence, since in \POWHEG, \MCatNLO and
\herwigpp these extra jets come from the shower, and are therefore only corect in
the strict collinear limit.  Similar comments hold for the pseudorapidity
distributions of fig.~\ref{gghcomp_etaj12}.

\begin{figure}
  \begin{center}
    \includegraphics[width=0.6\textwidth]{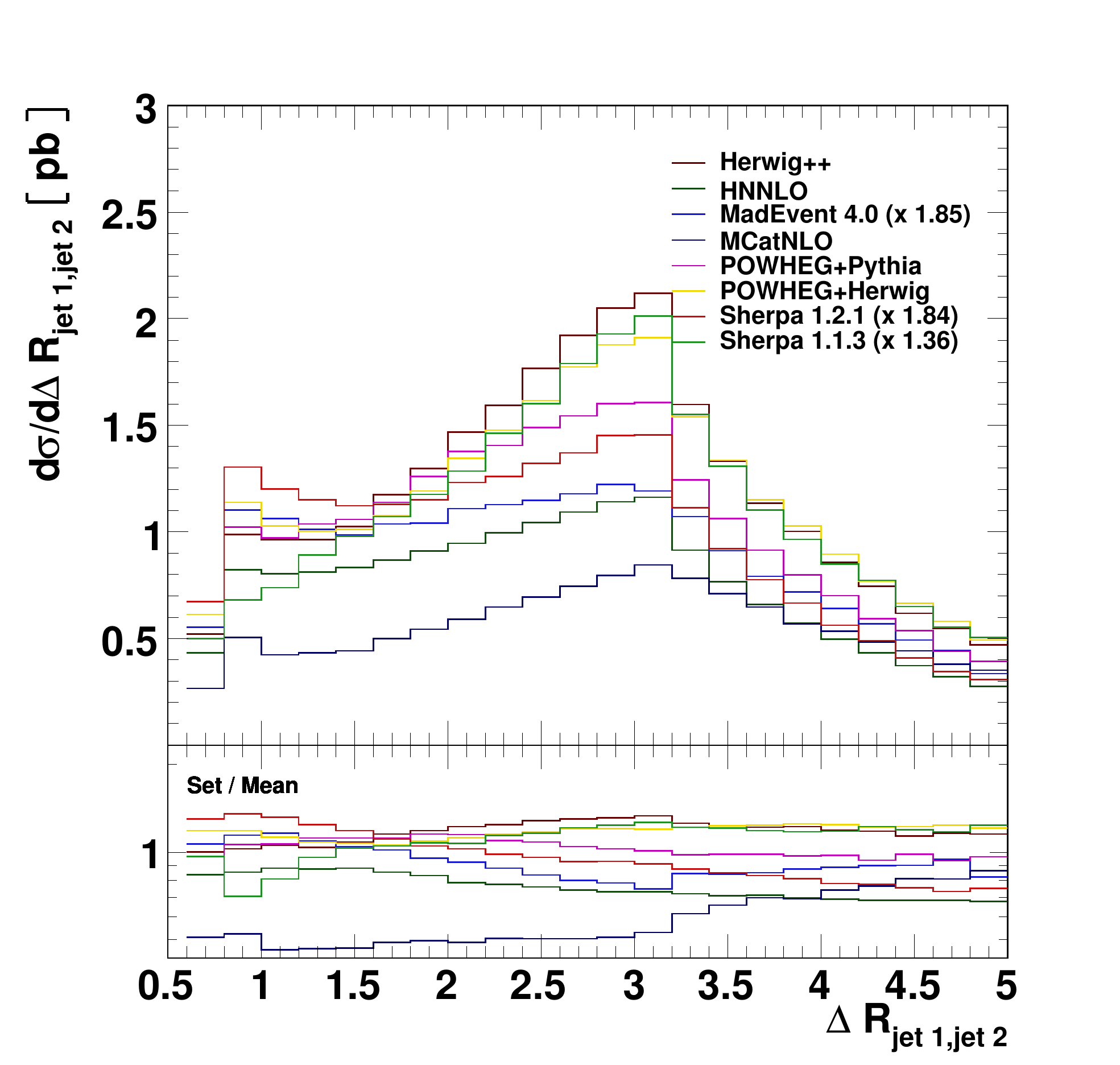}
  \end{center}
\caption{The separation in the $\eta$-$\phi$ plane of the two leading jets
      accompanying the Higgs boson. See fig.~\protect\ref{gghcomp_PTh} for
      details.}
  \label{gghcomp_DR12}
\end{figure}
Larger discrepancies are instead present for more exclusive quantities, such
as the distance in the $\eta$-$\phi$ plane between the two leading jets,
$\Delta R_{12}$, shown in fig.~\ref{gghcomp_DR12}.  Indeed, we start
appreciating here some interesting differences in shape: \herwigpp, for example,
predicts much steeper distribution than \sherpa.

\begin{figure}
  \begin{center}
    \includegraphics[width=0.6\textwidth]{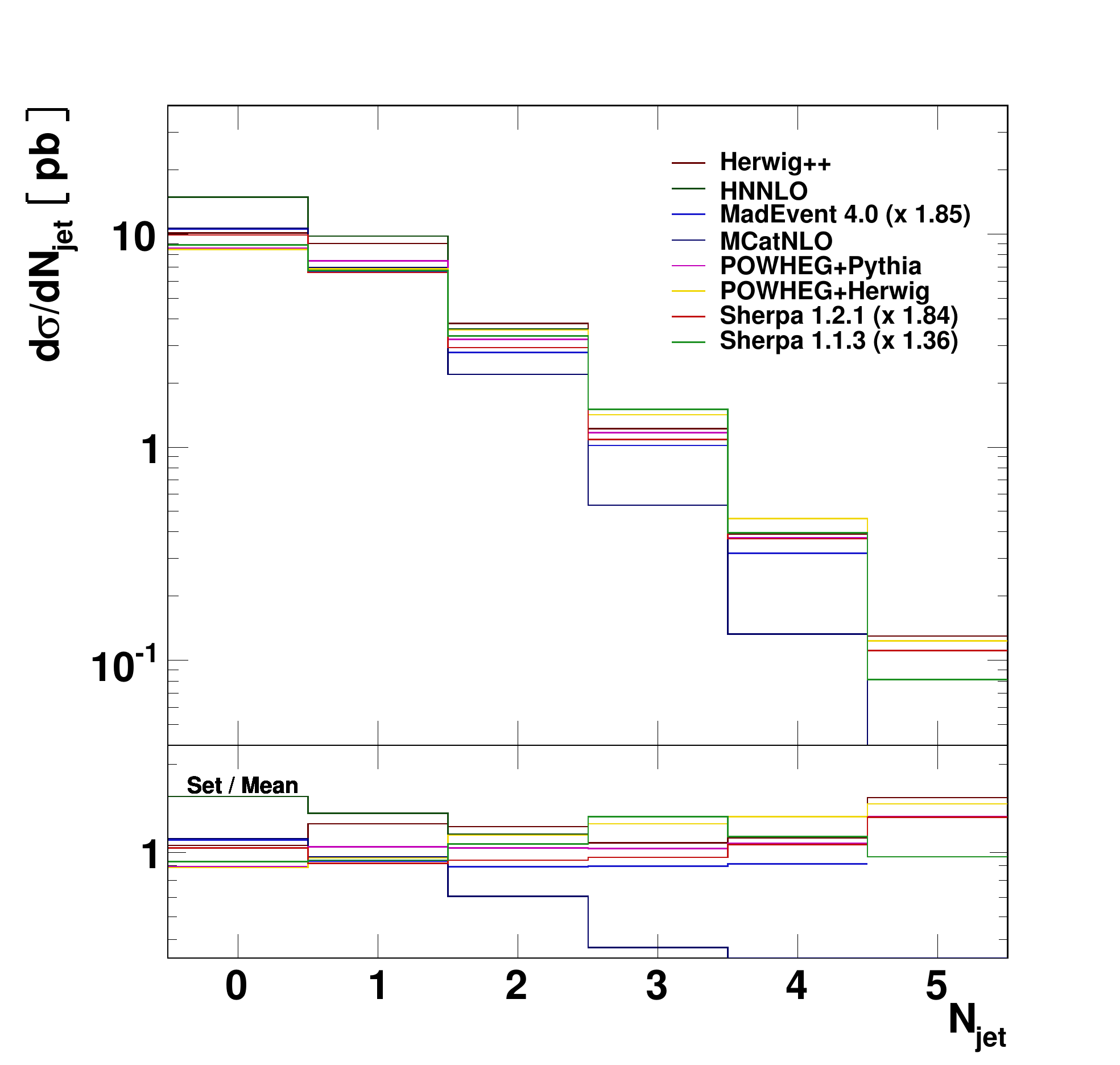}
  \end{center}
   \caption{The production rates for $N_{\rm jet}$ additional jets
     accompanying the Higgs boson. See fig.~\protect\ref{gghcomp_PTh} for
     details.}
  \label{gghcomp_JR}
\end{figure}
Finally, in fig.~\ref{gghcomp_JR}, we plot the jets rates, which, once again,
agree within 50\% uncertainty even for higher multiplicities.

\subsection{CONCLUSIONS}
We have reported on the first comparison among several different Monte Carlo
approaches to the simulation of Higgs boson production in gluon fusion. Apart
from some basic choices, such as the parton distribution functions, the
collider energy and our choosing to use an effective theory where the top
quark mass has been taken to infinity, no detailed tuning has been
performed. The main idea being that of an "out-of-the-box" comparison among
various codes, all of which represent the state-of-the-art in Monte Carlo
tools.

The upshot of our comparison is that, apart from the overall normalization,
which is only reliable in NLO and NNLO codes, the various approaches give
consistent results within their expected range of validity.

\subsection*{ACKNOWLEDGEMENTS}
We thank Jeppe Andersen and Michelangelo Mangano for useful discussions.

\clearpage

\section[$W\bbbar$ IN THE HIGH-$\pt$ $HW$ REGION]{$W\bbbar$ IN THE HIGH-$\pt$ $HW$ REGION~\protect
\footnote{Contributed by: F.~Febres~Cordero, G.~Piacquadio, L.~Reina, D.~Wackeroth}}
\label{sec:wbbbar}

\subsection{INTRODUCTION}

The main aim of this section is to study the effect of next-to-leading
order (NLO) QCD corrections to the $pp \to W b\bar{b}$ process, in the
region of phase space which is relevant for the {\it highly boosted}
$WH$ analysis with $H \to
b\bar{b}$~\cite{Butterworth:2008iy,higgsnote}.The details
of the setup for the NLO computation in this kinematic region as well
as results for the scale dependence of the total cross sections are
presented in Section~\ref{sec:nloqcd}.  The application of the NLO
calculation to the experimental analysis on which the mentioned Higgs
search is based is not straightforward, since the latter addresses the
simulation of the backgrounds through the use of a parton shower Monte
Carlo (MC) algorithm which is applied on top of the QCD leading order (LO)
matrix element calculation.

While for an increasing number of processes specific Monte Carlo
generators have been made available to combine the NLO calculation
with a parton-shower based MC generator, as in the \MCatNLO
program~\cite{Frixione:2002ik} or in the new \POWHEG
method~\cite{Frixione:2007vw,Alioli:2010xd}, no such generator is available for the
$pp \to W b\bar{b}$ process yet. It is therefore useful to define a
re-weighting procedure for $pp \to W b\bar{b}$ events generated by the
use of a parton shower Monte Carlo, which for this specific case is
\herwig~\cite{Corcella:2002jc}, based on the distributions predicted by the
NLO calculation.

An important feature of the {\it highly boosted} $WH$ analysis is the
application of a tight jet veto. In the cut-based analysis~\cite{higgsnote}
this veto is applied at $\pt=20$~GeV, while in a more refined likelihood based
analysis~\cite{gp} the maximum $\pt$ considered for a possible
additional non $b$-jet in the event is 60~GeV. 
The NLO QCD correction depends significantly on the $\pt$ of 
an additional non $b$-jet in the event and this is addressed in this 
study. This situation is analogous to other backgrounds for Higgs boson signals, 
as is the case of the $t\bar{t}b\bar{b}$ process as a 
background for the search for $t\bar{t}H$ with $H \to b\bar{b}$, 
also considered in the present proceedings. 

In the specific case of the $q\bar{q}^\prime \to W b\bar{b}$ process,
the real corrections which enter the NLO computation include an
additional gluon-induced process, $q(\bar{q})~g \to Wb\bar{b}j$,
which, given the gluon luminosity at the LHC, makes the NLO correction
very large. However, by constraining the transverse momentum 
of the extra non-b-jet through a jet veto one can reduce the 
impact of this extra channel, which then improves the scale dependence 
of the NLO computation, as long as such extra kinematical cut 
is not too restrictive.


\subsection{$Wb\bar b$ PRODUCTION AT NLO QCD WITH A TIGHT JET VETO}
\label{sec:nloqcd}

The calculation of the total and differential cross sections for
$W^\pm b\bar{b}$ production at NLO QCD for this study is based on
Refs.~\cite{FebresCordero:2006sj,Cordero:2008ce,Cordero:2009kv}, which
includes full $b$-quark mass effects.  The input parameters are chosen
as follows: $m_b=4.79$~GeV, $m_t=173.1$~GeV, $M_W=80.399$~GeV,
$s_W^2=0.223$, $G_F=1.16639\cdot 10^{-5}$ (the weak coupling being
defined as: $g_W=(8M_W^2G_F/\sqrt{2})^{-1/2}$). The contribution from
the third generation quarks in the initial state is neglected and the
non-zero CKM matrix elements are chosen to be: $V_{ud}=V_{cs}=0.974$
and $V_{us}=V_{cd}=0.227$.

Parton-level LO results (calculated as a cross-check) are obtained
using the one-loop evolution of $\alpha_s$ and the CTEQ6L1~\cite{Pumplin:2005rh} set of PDFs
with $\alpha_s(M_Z)=0.130$, while parton level NLO results are
obtained using the two-loop evolution of $\alpha_s$ and the CTEQ6M set
of PDFS with $\alpha_s(M_Z)=0.118$.

Renormalization ($\mu_r$) and factorization ($\mu_f$) scales are set
to be equal and varied by a factor of two around a central value
$\mu_0=[M_W^2+(p_T^{b})^2+(p_T^{\bar{b}})^2]^{1/2}$, where $p_T^b$ and
$p_T^{\bar{b}}$ are the transverse momenta of the jets generated by
$b$ and $\bar{b}$ respectively.

The jets are constructed using the $k_T$ jet algorithm with
pseudo-cone size $R=0.3$ and the parton momenta are recombined within
a jet using the so called $E$-invariant scheme (=sum of momenta of the
constituents).  We study two samples of events, i.e.
\begin{itemize}
\item events with just 2 $b$ jets;
\item events with 2 $b$ jets and 1 non-$b$ jet;
\end{itemize}
where $b$ and non-$b$ jets are identified imposing the following cuts:
\begin{itemize}
\item $p_T^{b}> 30$~GeV, $|\eta^{b}|< 2.5$;
\item $p_T^{W}> 200$~GeV, $|\eta^W|< 2.5$;
\item $15~\mbox{GeV}< p_T^{non-b}< 60$~GeV, $|\eta^{non-b}|< 5$;
\item $p_T^{b\bar{b}}> 200$~GeV;
\item $R_{b\bar{b}}< 1.2$ \; .
\end{itemize}
$p_T^b$ and $|\eta^b|$ denote the transverse momentum and
pseudorapidity of either one of the two $b$ jets, $p_T^W$ and
$|\eta^W|$ are the transverse momentum and pseudorapidity of the $W$
boson, $p_T^{non-b}$ and $|\eta^{non-b}|$ denote the transverse
momentum and pseudorapidity of the non-$b$ jet, $R_{b\bar{b}}$ is the
relative separation between the $b$ and $\bar{b}$ jets, and
$p_T^{b\bar{b}}$ the transverse momentum of the $b\bar{b}$ 2-jet
system.

The renormalization and factorization scale dependence of the total
cross sections for $W^\pm b\bar{b}$ production at the LHC
($\sqrt{s}=14$~TeV) using this setup is studied in
Table~\ref{tab:total_xs} and in Fig.~\ref{fig:Wbb_mudep}.  The NLO
\textit{exclusive} cross section corresponds to the \textit{2 b-jet
  only} sample, while the \textit{inclusive} one corresponds to the
sum of both \textit{2 b jet only} and \textit{2 b jets and 1 non-b
  jet} samples. Thus, in Fig.~\ref{fig:Wbb_mudep}, the curve labeled
as \textit{NLO Inc} is the sum of the curves labeled as \textit{NLO
  Exc} and \textit{$2b+j$ only}.

\begin{table}[htb]
\begin{center}
  \caption{LO, NLO \emph{inclusive}, and NLO \emph{exclusive} cross
    sections (in fb) for $W^+b\bar{b}$ and $W^-b\bar{b}$ production at
    $\sqrt{s}=14$~TeV. Listed separately are also the cross sections (in
    fb) for the $Wb\bar{b}+j$ channel alone. The central values
    correspond to $\mu_r=\mu_f=\mu_0$, while the upper and lower
    bounds represent the maximal upper and lower variation obtained
    when varying $\mu_r=\mu_f$ between $\mu_0/2$ and $2\mu_0$.}
\label{tab:total_xs}
\begin{tabular}{|c|c|c|c|c|} \hline\hline
Process & LO & NLO inclusive & NLO exclusive & $Wb\bar{b}+j$\\
\hline
$W^+b\bar{b}$ & $138^{+31}_{-24}$ & $155^{+9}_{-11}$ & $43^{+17}_{-34}$ &
$112^{+42}_{-28}$\\
$W^-b\bar{b}$ & $76^{+17}_{-13}$ & $90^{+7}_{-7}$ & $26^{+9}_{-17}$ &
$64^{+24}_{-16}$\\
\hline
\end{tabular}
\end{center}
\end{table}

\begin{figure}[ht]
\begin{center}
\begin{tabular}{c}
\includegraphics[clip,scale=0.45]{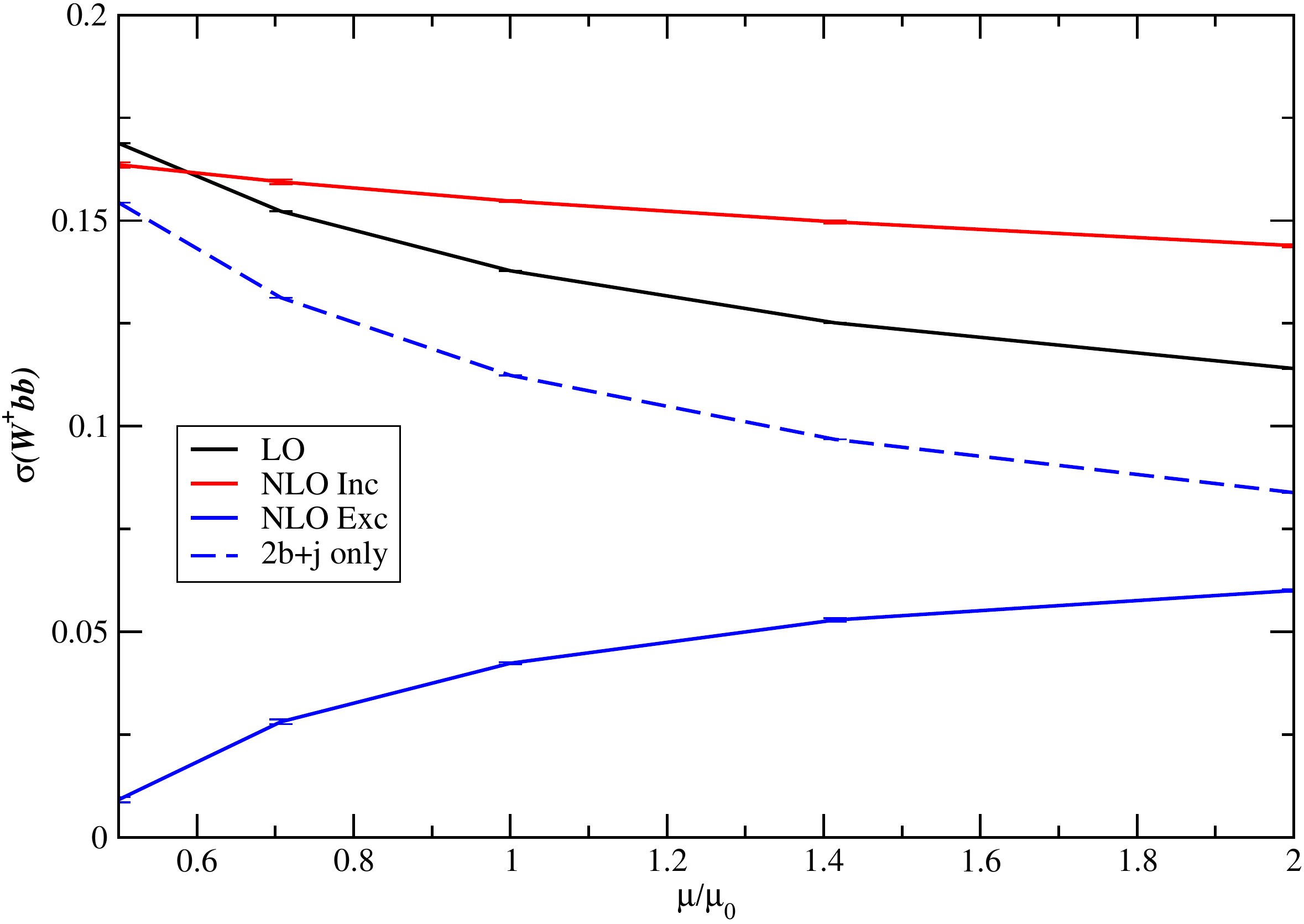}\\
\includegraphics[clip,scale=0.45]{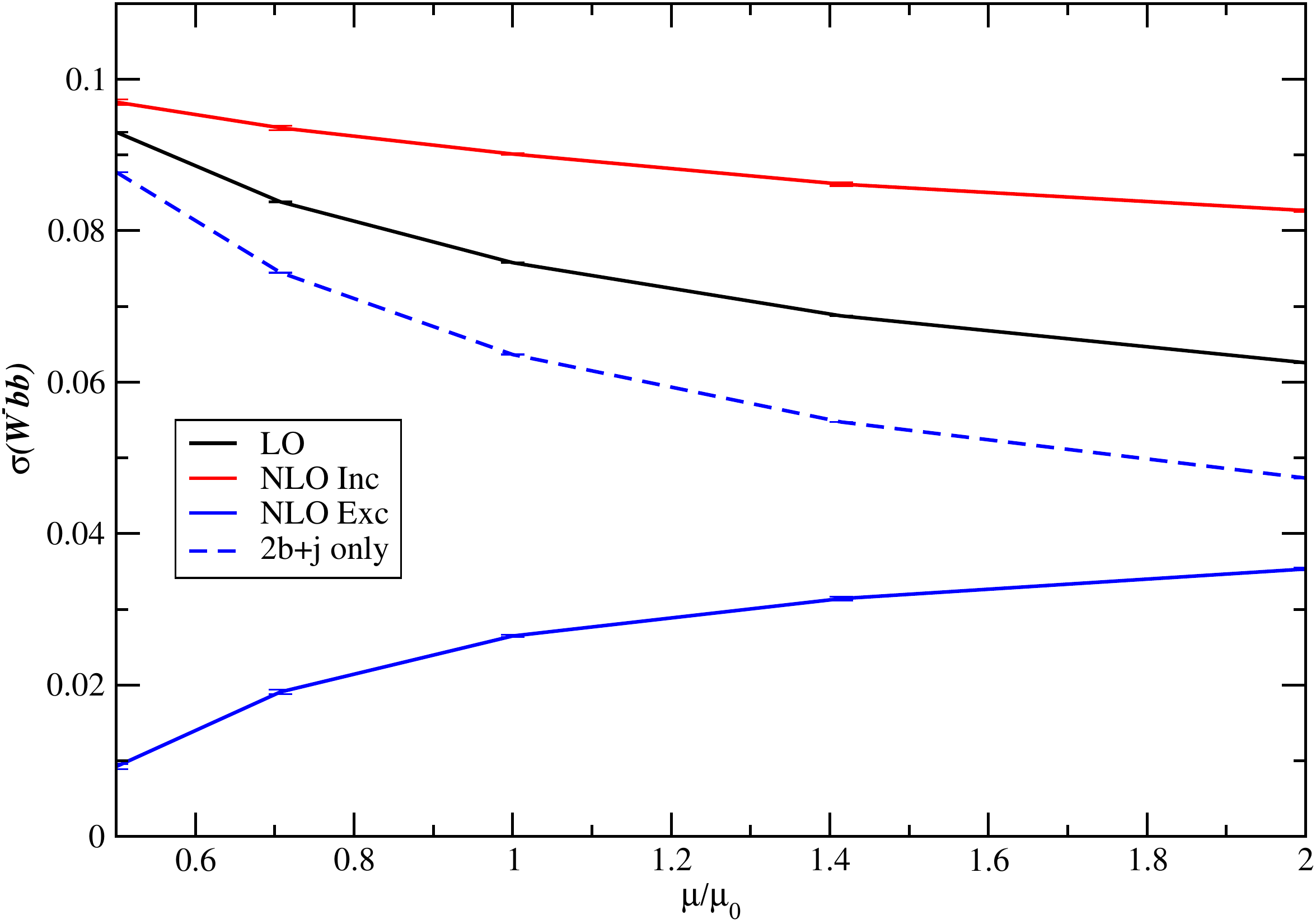}
\end{tabular}
\caption{Renormalization and factorization scale dependence for the
  parton level $W^+b\bar{b}$ (upper plot) and $W^-b\bar{b}$ (lower
  plot) LO (black, solid), NLO inclusive (red, solid), NLO exclusive
  (blue, solid) and $Wb\bar{b}+j$ (blue, dashed) production cross
  sections (cross sections are in pb). See Section~\ref{sec:nloqcd} for details.
  \label{fig:Wbb_mudep}}
\end{center}
\end{figure}

It is interesting to note that, contrary to the findings in
Refs.~\cite{FebresCordero:2006sj,Cordero:2008ce,Cordero:2009kv}, in
this kinematic regime the inclusive cross section exhibits less scale
dependence than the exclusive one. Different factors may contribute to
this different behavior, but mainly the kinematic cuts used in the
present analysis make the jet veto much more selective and enhance the
effect of unbalanced scale dependent logarithms in the exclusive cross
section, while their effect tends to compensate between $Wb\bar{b}$
and $Wb\bar{b}+j$ production in the inclusive case.

\subsection{COMPARISON OF FIXED-ORDER AND PARTON-SHOWER RESULTS}
\label{sec:comp}

While the parameters used for the fixed-order LO and NLO computation
were already presented in the previous section, the parameters used
for the matrix element computation $pp \to Wb\bar{b}$ which serves as
an input to the parton-shower algorithm are listed in the following:
\begin{itemize}
\item LO parton density functions CTEQ6L1;
\item strong coupling constant $\alpha_s(M_Z)=0.130$, consistent 
      with the PDF set used;
\item QCD renormalization and factorization scales both equal 
      and set to $Q = \sqrt{M_W^2 + p_{T,W}^2}$;
\item one-loop running of $\alpha_s$.
\end{itemize}

The LO matrix element computation and initial three-body phase
integration is based on the AcerMC generator~\cite{Kersevan:2004yg}
(v.~3.5). This is then passed as an external process to \herwig\
(v.~6.510), which applies the parton shower algorithm and produces the
final state partons which are used as an input to an inclusive $k_T$
jet clustering algorithm with parameter $R=0.3$ and with the $E$-invariant 
parton momenta recombination scheme. The underlying event
has been switched off during generation, while the effect of
hadronization has been removed by applying the jet clustering directly
on top of the final state partons before hadronization. As opposed to
the $Wb\bar{b}$ sample generated for the analysis presented
in~\cite{higgsnote}, the one-loop running of $\alpha_s$
was used instead of the two-loop running: this increased the LO cross
section by $\approx 30~\%$ with respect to the value used in the
mentioned analysis.

The parton-level jets produced in this way can be compared to the
fixed-order calculation described in Section~\ref{sec:nloqcd} and will
effectively include the approximative leading logarithmic resummation
introduced by the parton shower algorithm. This involves a comparison
of $N$ produced jets with respectively the two or three leading jets
of the LO or NLO fixed-order computation. The two $b$-jets are
selected out of the $N$ jets produced by the parton shower algorithm
by matching them to the two $b$-quarks after radiation as traced in
the Monte Carlo history:
\begin{eqnarray*}
\Delta R({\rm jet}, b-{\rm quark}) < 0.4.
\end{eqnarray*}
In the case where more than one jet is closer to the $b$-quark than
$\Delta R=0.4$, the $b$-quark is associated to the closest jet. As a
cross-check of the eventual systematic uncertainty introduced by this
choice, the same procedure is applied using the $b$-quarks before
radiation, which are produced directly by the matrix element
computation, and the difference in the results is found to be
negligible.  Finally, the leading additional non $b$-jet in the event
is defined as the jet out of the $N$ jets produced by the parton
shower which, after excluding the two jets matched to the $b$-quarks,
is the highest in transverse momentum.  Events where only a single jet
is matched to a $b$-quark are thrown away; this happens typically when
the two $b$-quarks are closer than $\Delta R \approx 0.3$ and are
therefore included into a single jet. After the jets are defined, the
kinematic cuts defined in the previous section are applied: in the
case of the cuts applied on the $W$ boson, its momentum is considered
after the effect of reshuffling of momenta operated by the parton
shower algorithm.  The distributions based on the jets obtained by
using the parton shower result are labeled as LO+PS in the next
section.

As a cross-check, also the result of the LO matrix element computation
used as input to the parton shower algorithm is considered in the
following and is labeled as LO only. In this case the two $b$-quarks
partons resulting from the ME computation are used directly for the
computation of the observables and for the application of the cuts, except
for the cut on the additional non $b$-jet in the event, which is not
applied.  Both results are compared to the LO and NLO inclusive theory
predictions obtained in the previous section, which are labeled as LO
(theory) and NLO (theory).  Note that NLO (theory) includes both the
$q\bar{q}^\prime$ initiated and $qg+\bar qg$-initiated processes.  For
comparison with the LO+PS result which is based on the
$q\bar{q}^\prime \to W b \bar b$ matrix element only, we also include
the NLO QCD prediction for the total 
cross section of the $q\bar{q}^\prime$-initiated process
separately, denoted as NLO$(q\bar{q}^\prime)$ (theory).

\subsection{RESULTS}

The cross sections are listed in Table~\ref{tab:crossSections}. The
two LO predictions agree fairly well, even if the comparison is
slightly biased by a marginally different choice of the
renormalization and factorization scale between the two
computations. Once the parton shower algorithm is run and the
kinematic cuts are applied, including the additional jet veto, the
cross section turns out to be $\approx 30 \%$ lower with respect to
the LO prediction. The complete NLO theory prediction, on the other
side, predicts a cross section which is $\approx 14 \%$ higher than
the same LO prediction. As a result, the parton shower prediction
(LO+PS) must be rescaled by an inclusive K factor of $1.64$ in order
to normalize it to the complete NLO computation. If only the $q\bar{q}^\prime$
initiated process is included in the NLO prediction, the inclusive K
factor only amounts to $1.16$.  The uncertainty on the complete NLO
cross section estimated through variations of the factorization and
renormalization scales by factors $1/2$ and 2 is $\approx 5 \%$. The
impact of PDF uncertainties has not been considered here.

\begin{table}[htb!]
\begin{center}
\begin{tabular}{|c||c|c|c|c|c|}
\hline   Subprocess       & LO   & LO (theory)      & LO+PS & NLO (theory) & 
NLO$(q\bar{q}^\prime)$ (theory)   \\
\hline
\hline   $W^+ b \bar{b}$    & $144$  &  $138^{+31}_{-24}$ & $97$   &  $155^{+9}_{-11}$ & $111^{+6}_{-19}$ \\
\hline   $W^- b \bar{b}$    & $79$   &  $76^{+17}_{-13}$  & $52$   &  $90^{+7}_{-7}$  & $62^{+4}_{-10}$ \\
\hline   Sum              & $223$  &  $214^{+35}_{-27}$ & $149$  &  $245^{+11}_{-13}$ & $173^{+7}_{-21}$ \\
\hline
\end{tabular}
\end{center}
\caption{\label{tab:crossSections} Central value of cross sections (in fb) 
  obtained for the $pp \to Wb\bar{b}$ process, comparing the five different 
  methods. Details about the calculation of LO (theory) and NLO (theory) can 
  be found in Section~\ref{sec:nloqcd}.}
\end{table}

The impact of the NLO corrections on the shapes of the distributions
of most interest for the analysis are shown in the following,
separately for $W^+b\bar{b}$ and $W^-b\bar{b}$, in
Figs.~\ref{fig:leadingBpT}-\ref{fig:last}. The crucial observable is
the differential ratio between the NLO and the LO+PS predictions for
the variable of interest: if this ratio is approximately flat, then it
is possible to re-weight the Monte Carlo events produced by the parton
shower algorithm by a simple inclusive $K$ factor.

\begin{figure}[htb!]
\begin{center}
\includegraphics[width=0.48\textwidth]{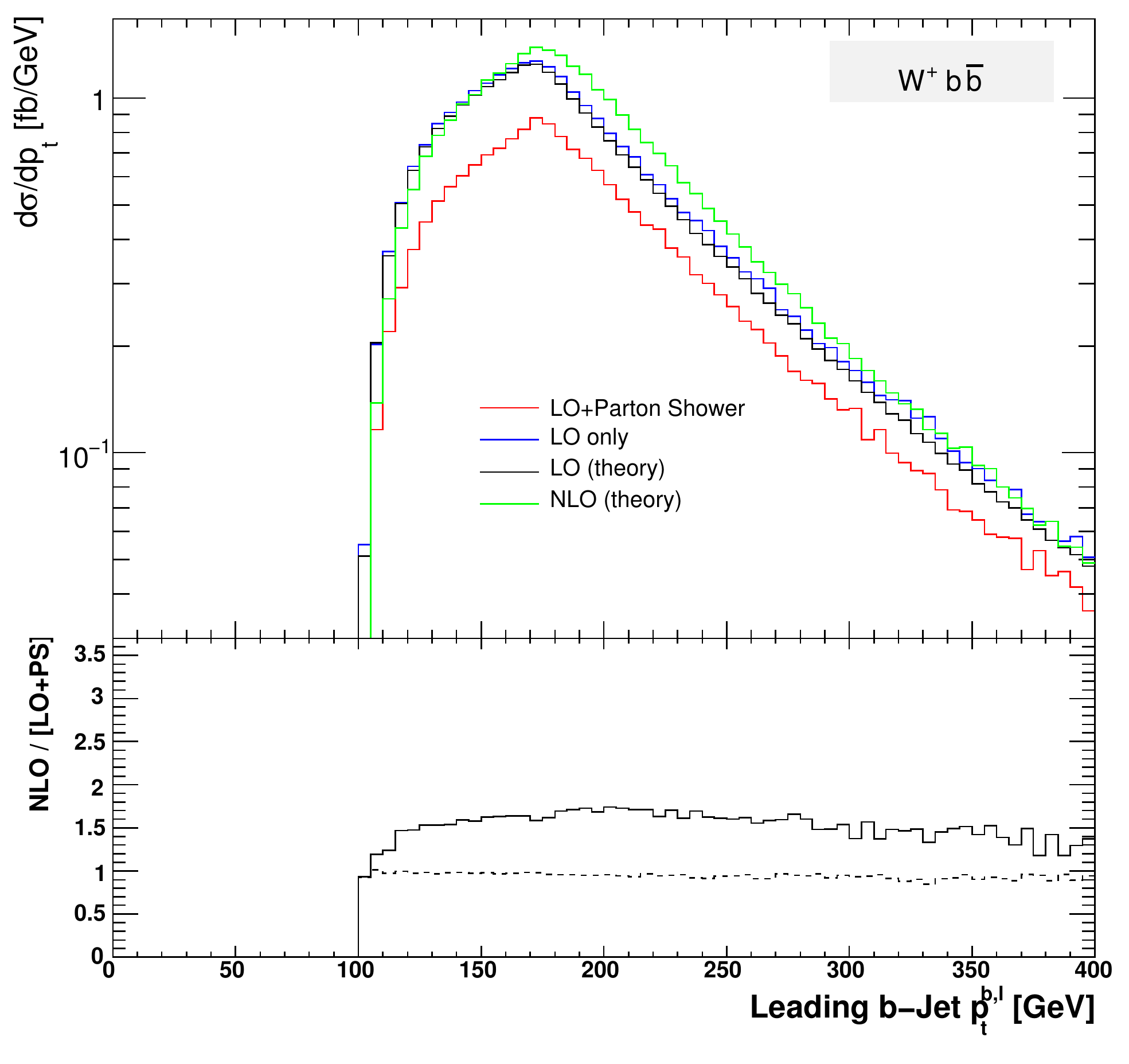}
\includegraphics[width=0.48\textwidth]{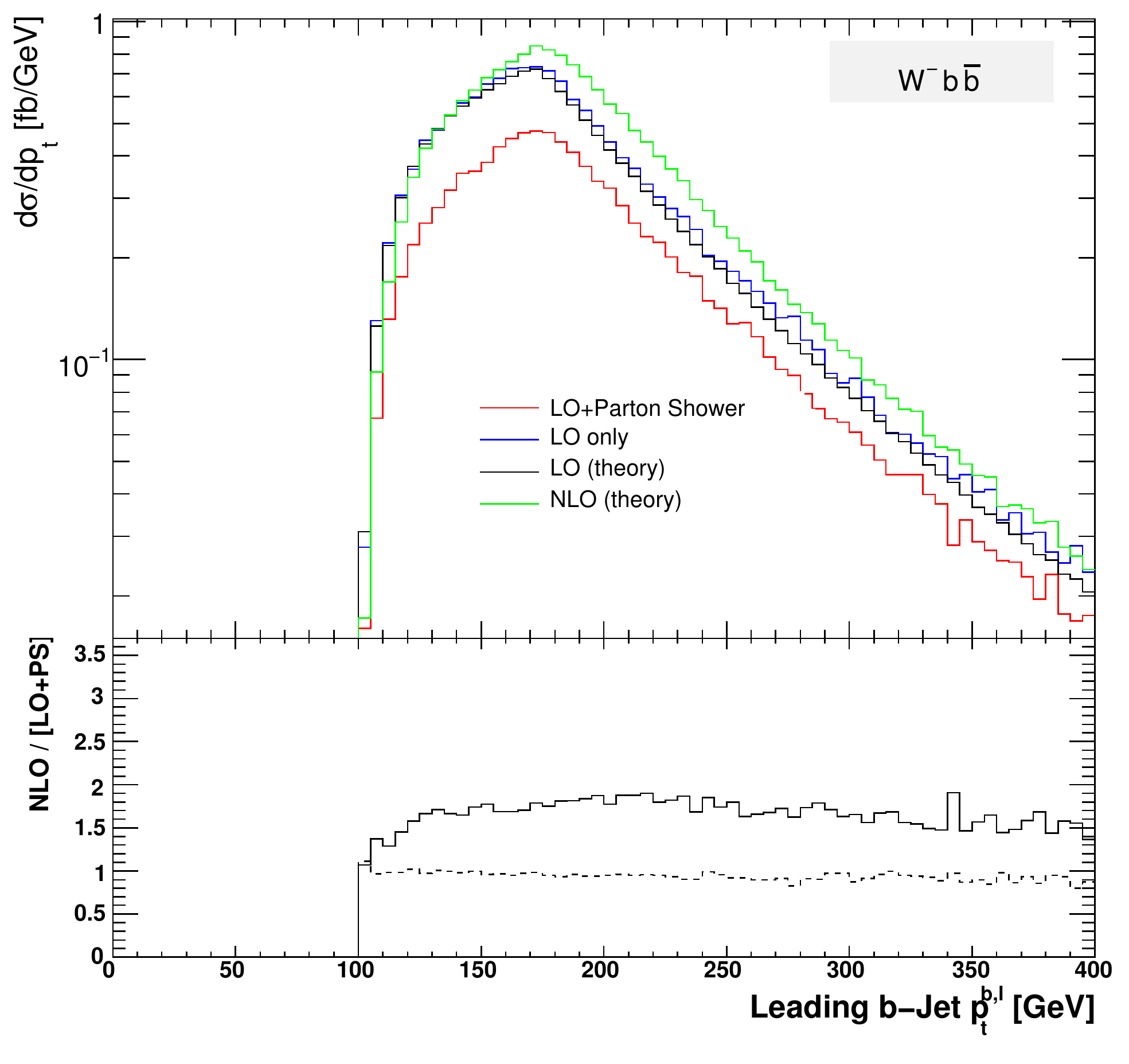}
\includegraphics[width=0.48\textwidth]{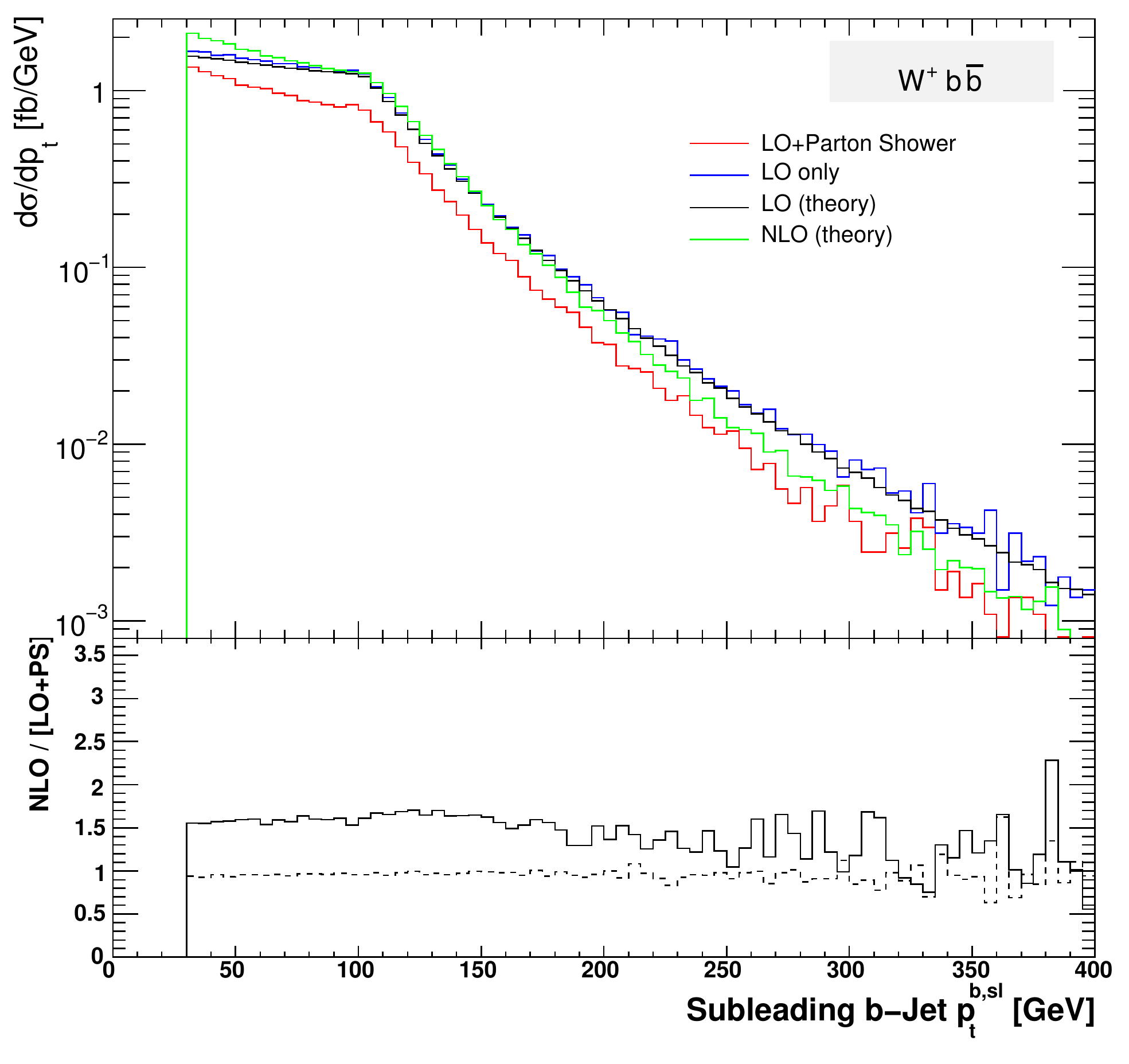}
\includegraphics[width=0.48\textwidth]{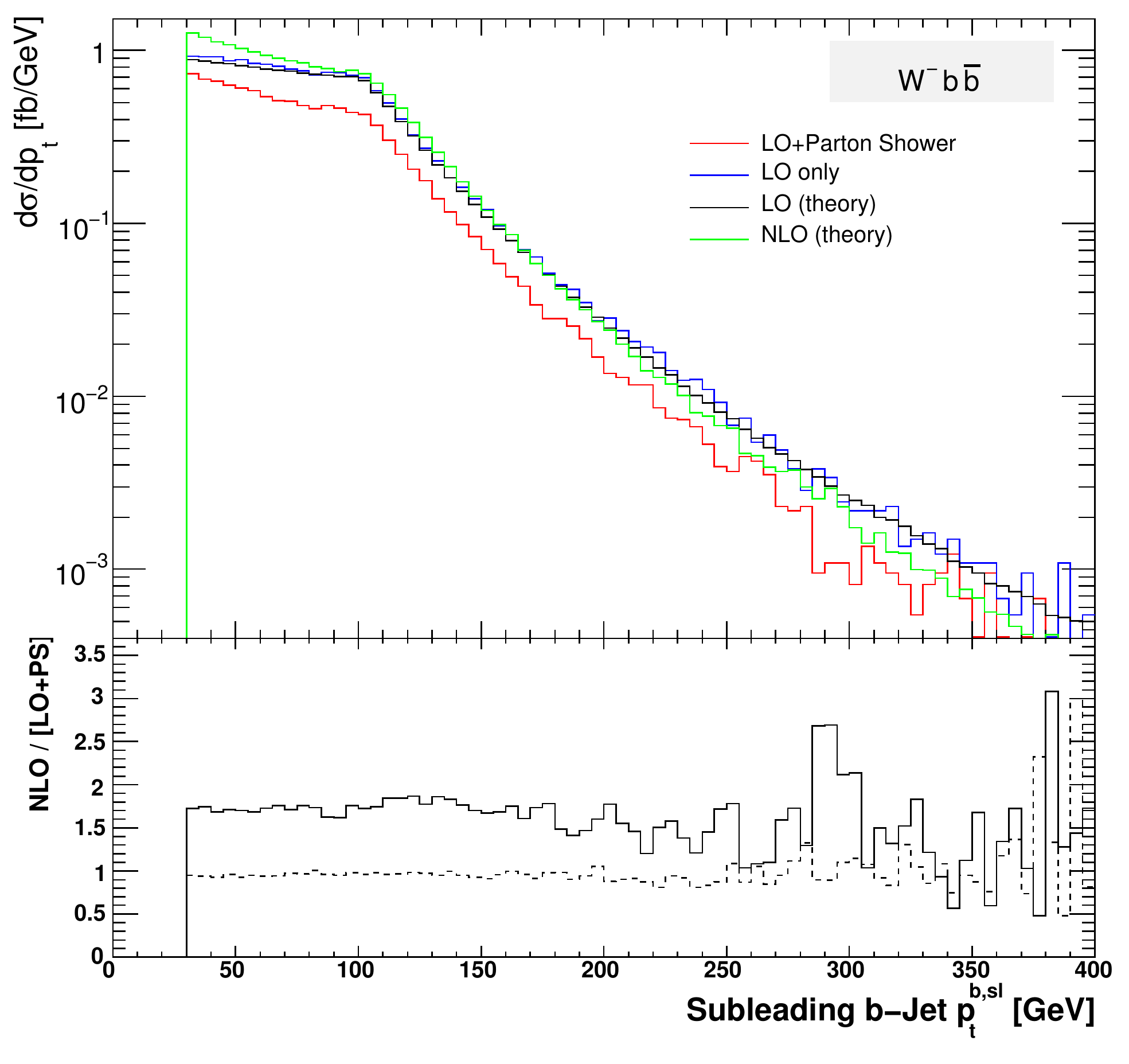}
\caption{\label{fig:leadingBpT} Transverse momentum distribution of
  the {\it leading} and {\it sub-leading} b jet, for both
  $W^+b\bar{b}$ and $W^-b\bar{b}$ production. The lower window shows
  the ratio of NLO to LO+PS predictions (full line) together with the
  ratio between the two LO predictions (dashed line).}
\end{center}
\end{figure}

\begin{figure}[htb!]
\begin{center}
\includegraphics[width=0.48\textwidth]{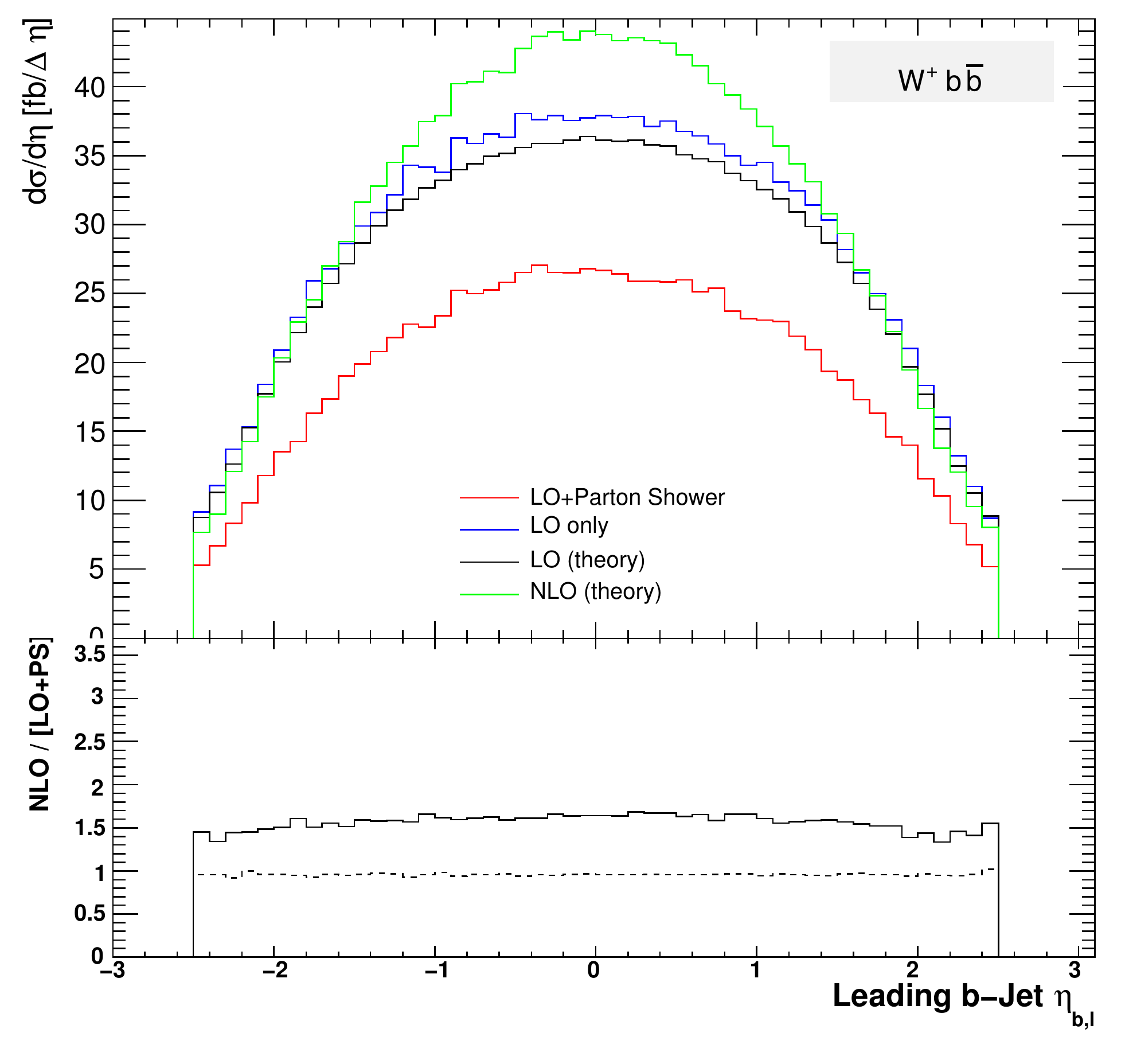}
\includegraphics[width=0.48\textwidth]{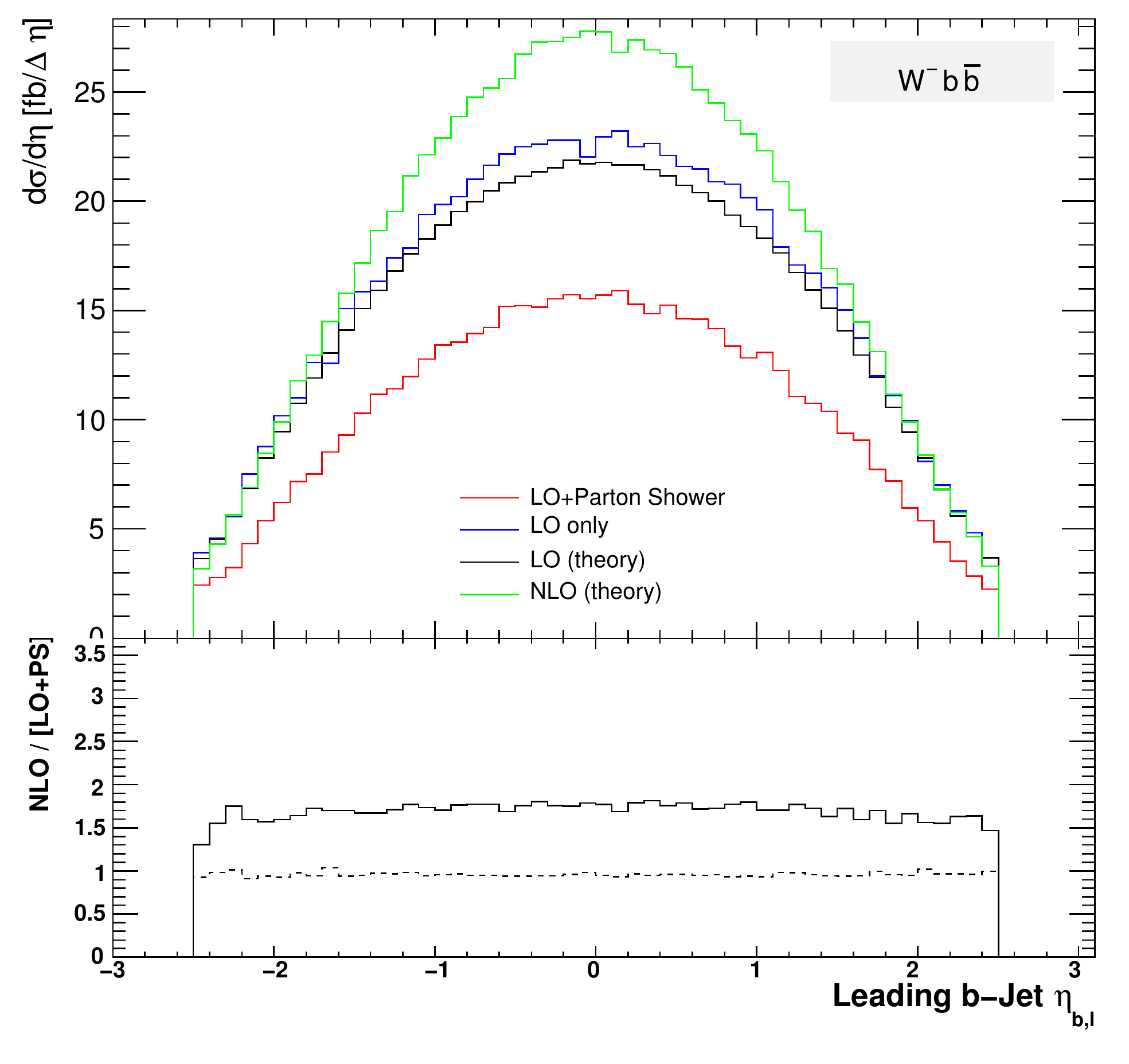}
\includegraphics[width=0.48\textwidth]{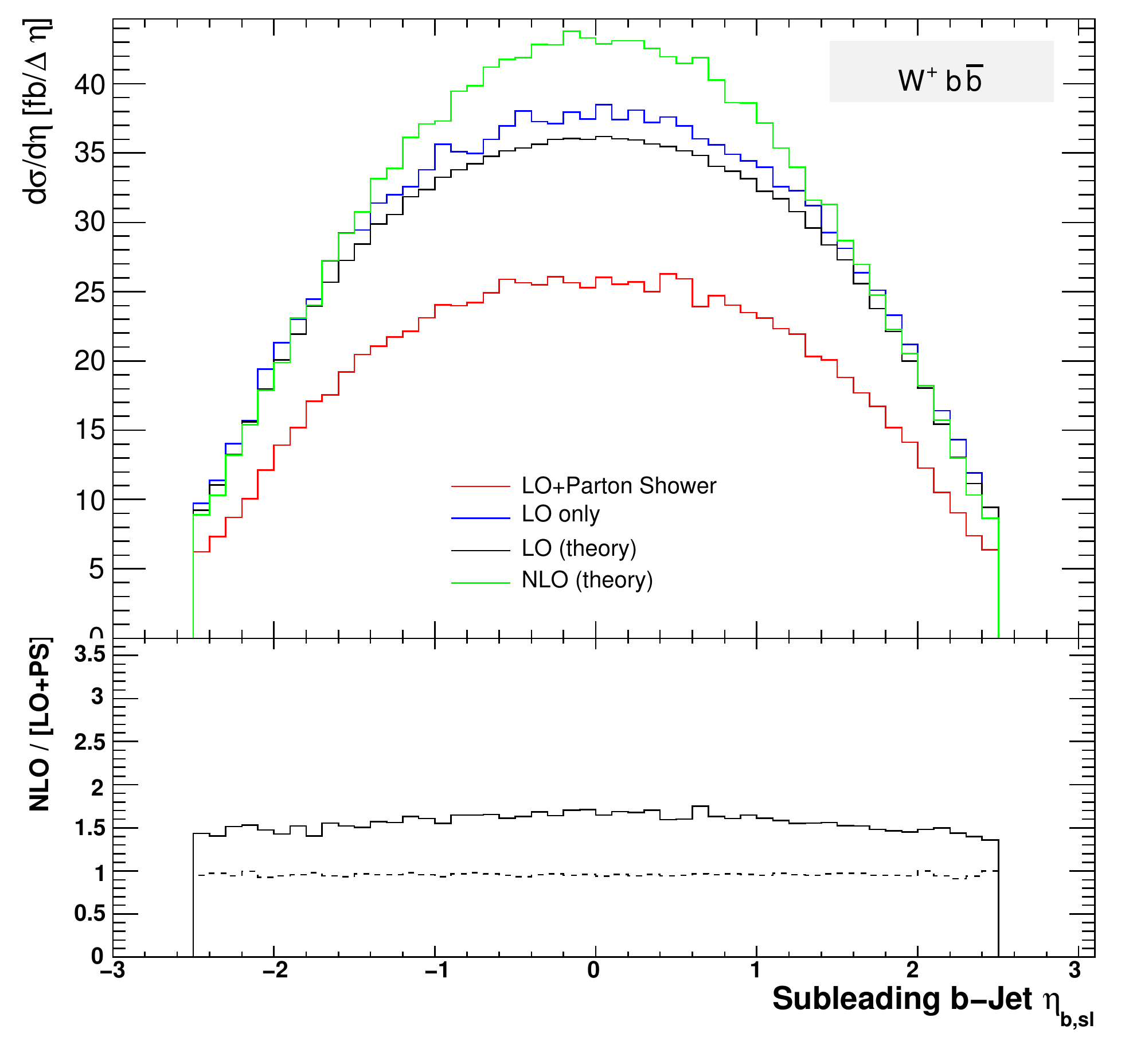}
\includegraphics[width=0.48\textwidth]{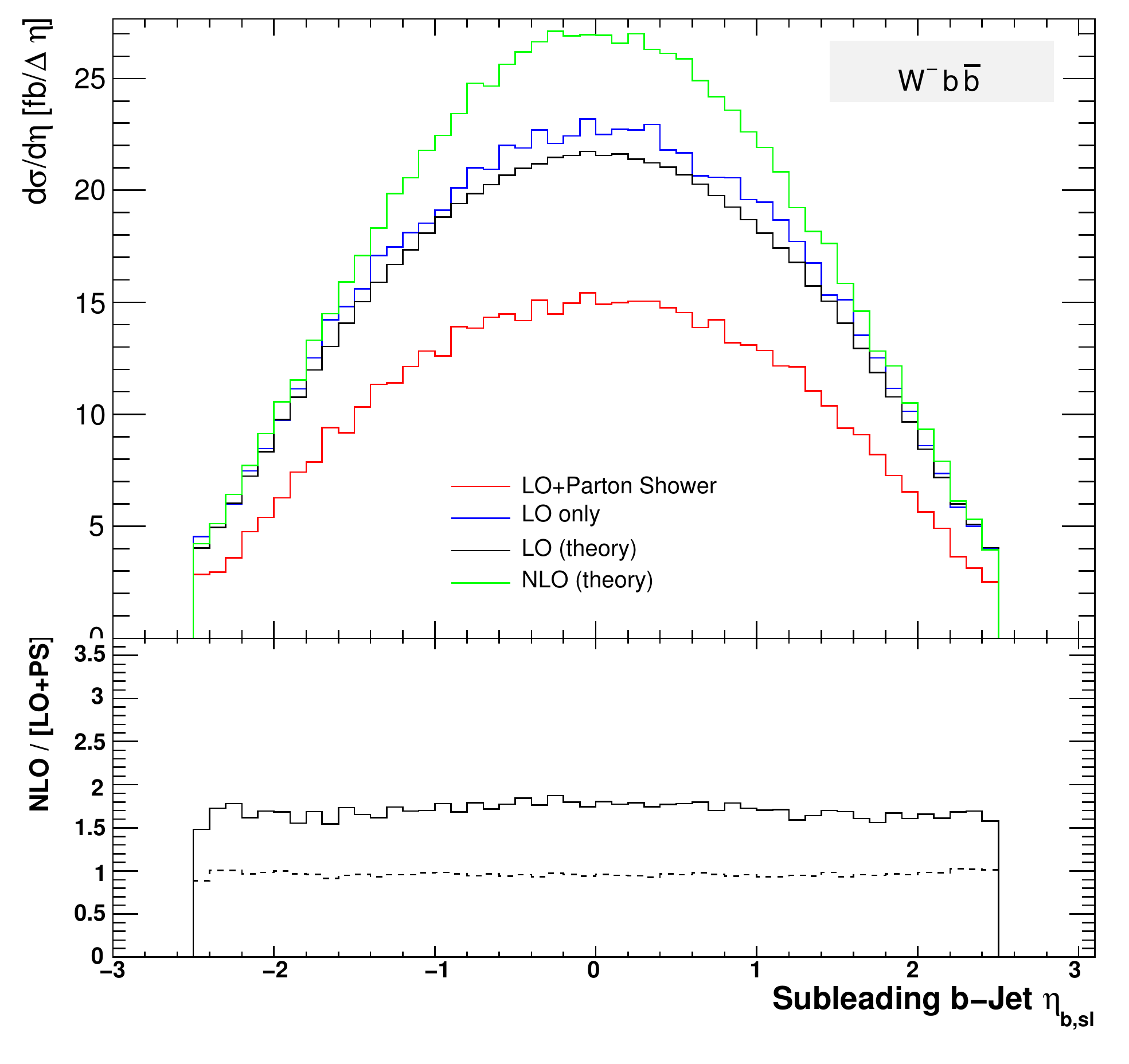}
\caption{\label{fig:Beta} Pseudorapidity distribution of the {\it
    leading} and {\it sub-leading} b jet, for both $W^+b\bar{b}$ and
  $W^-b\bar{b}$ production. The lower window shows the ratio of NLO to
  LO+PS predictions (full line) together with the ratio between the
  two LO predictions (dashed line).}
\end{center}
\end{figure}

\begin{figure}[htb!]
\begin{center}
\includegraphics[width=0.48\textwidth]{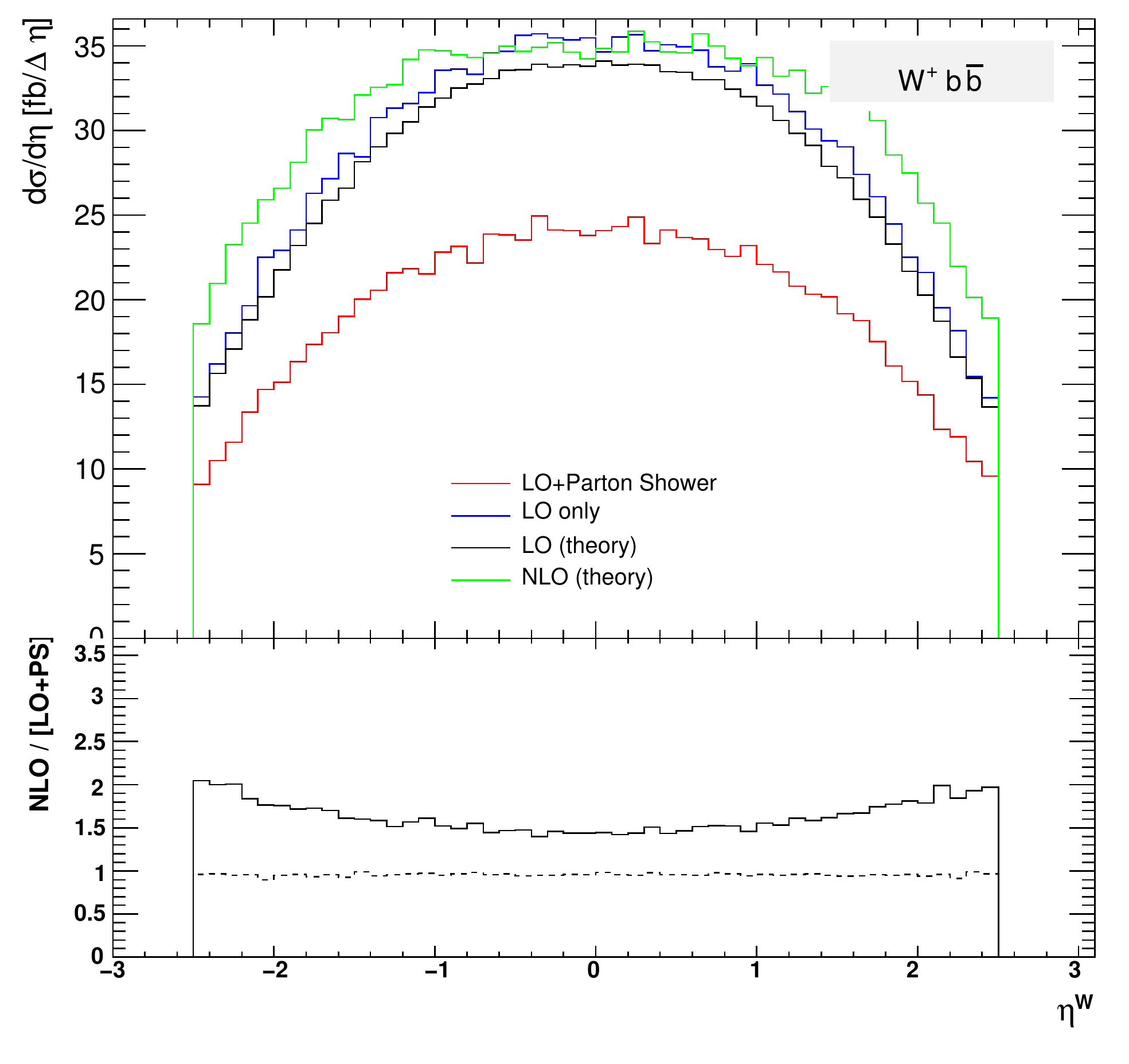}
\includegraphics[width=0.48\textwidth]{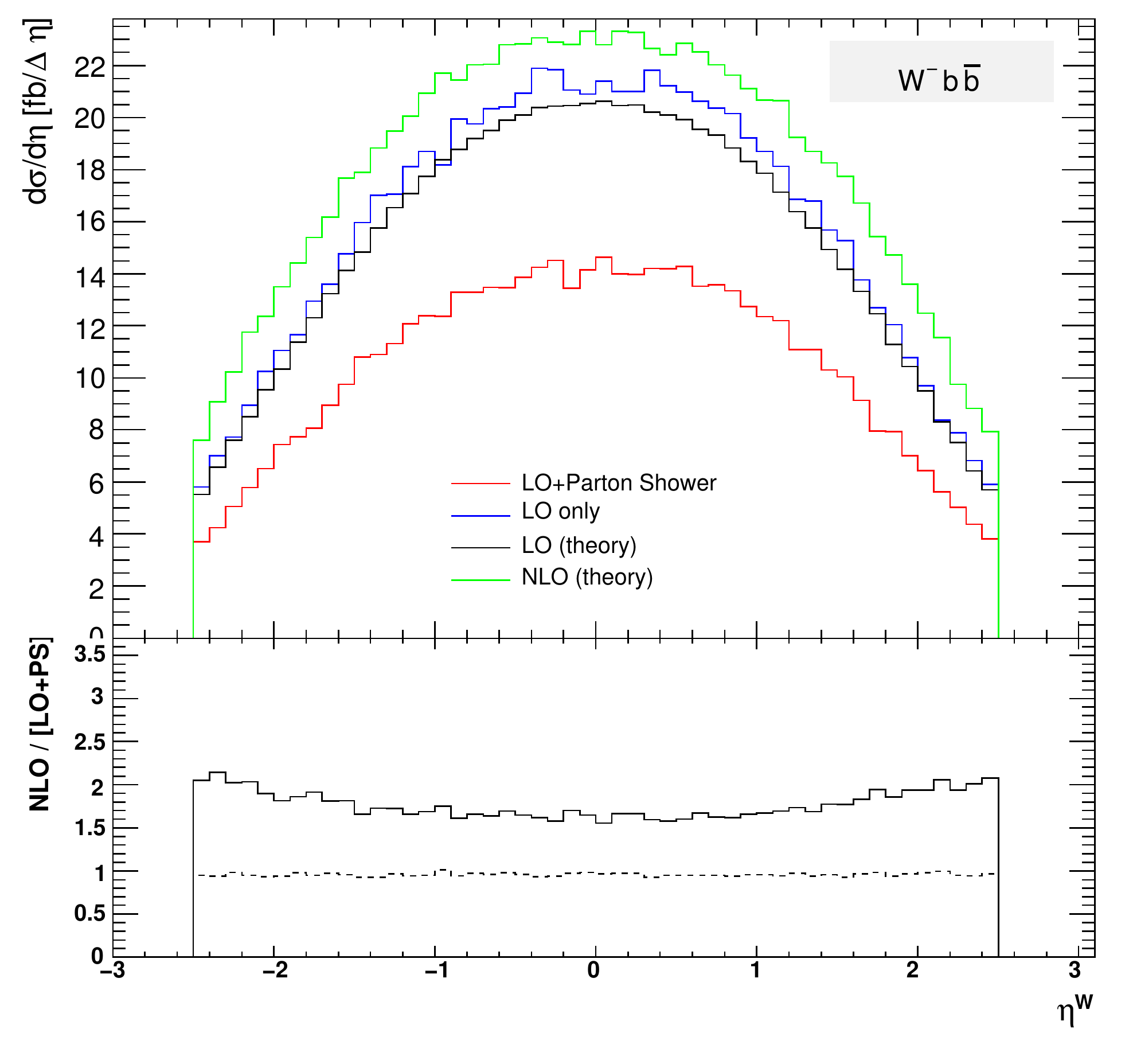}
\caption{\label{fig:etaW} Pseudorapidity distribution of the $W$
  vector boson for both $W^+b\bar{b}$ and $W^-b\bar{b}$
  production. The lower window shows the ratio of NLO to LO+PS
  predictions (full line) together with the ratio between the two LO
  predictions (dashed line).}
\end{center}
\end{figure}

\begin{figure}[htb!]
\begin{center}
\includegraphics[width=0.48\textwidth]{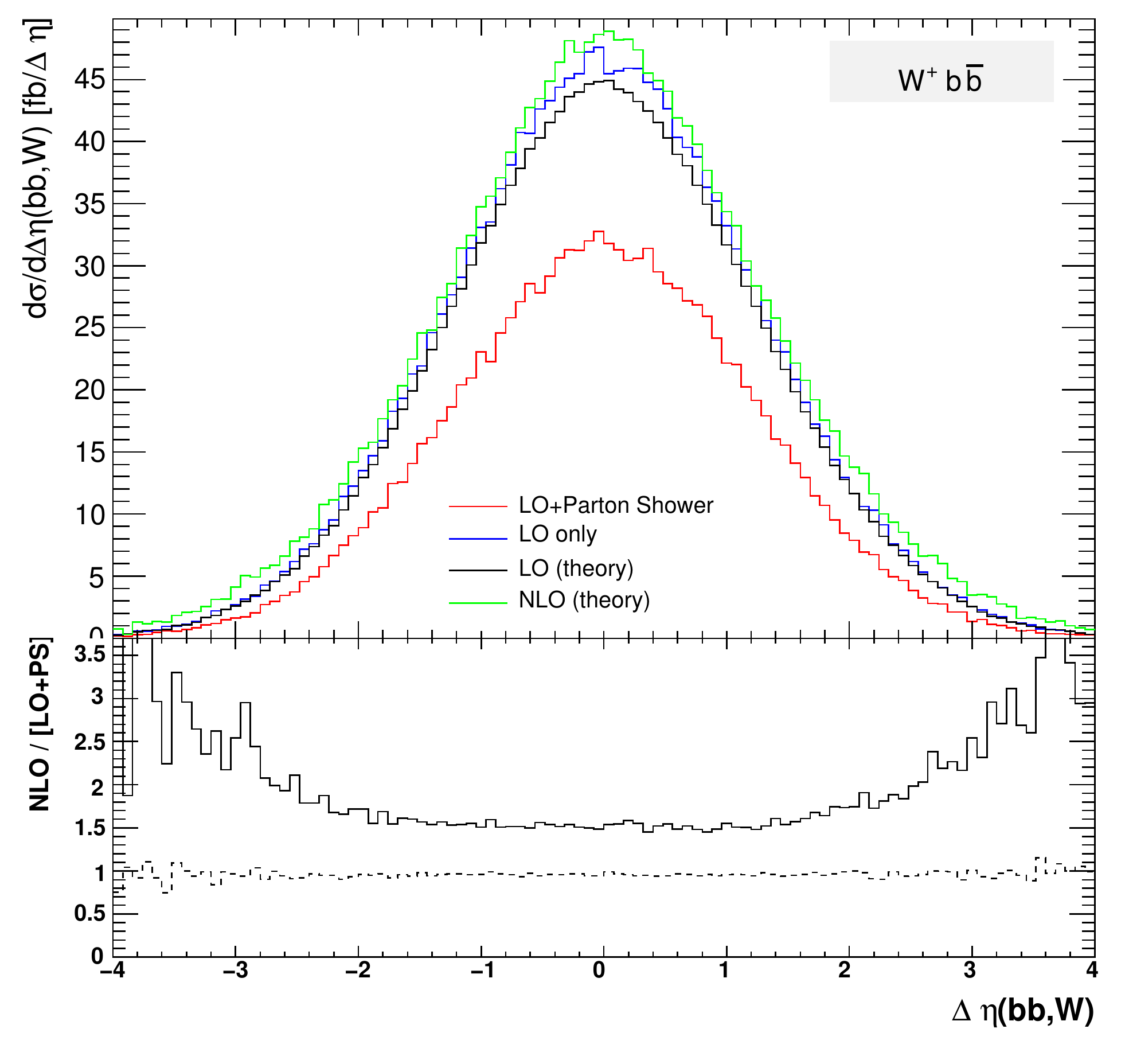}
\includegraphics[width=0.48\textwidth]{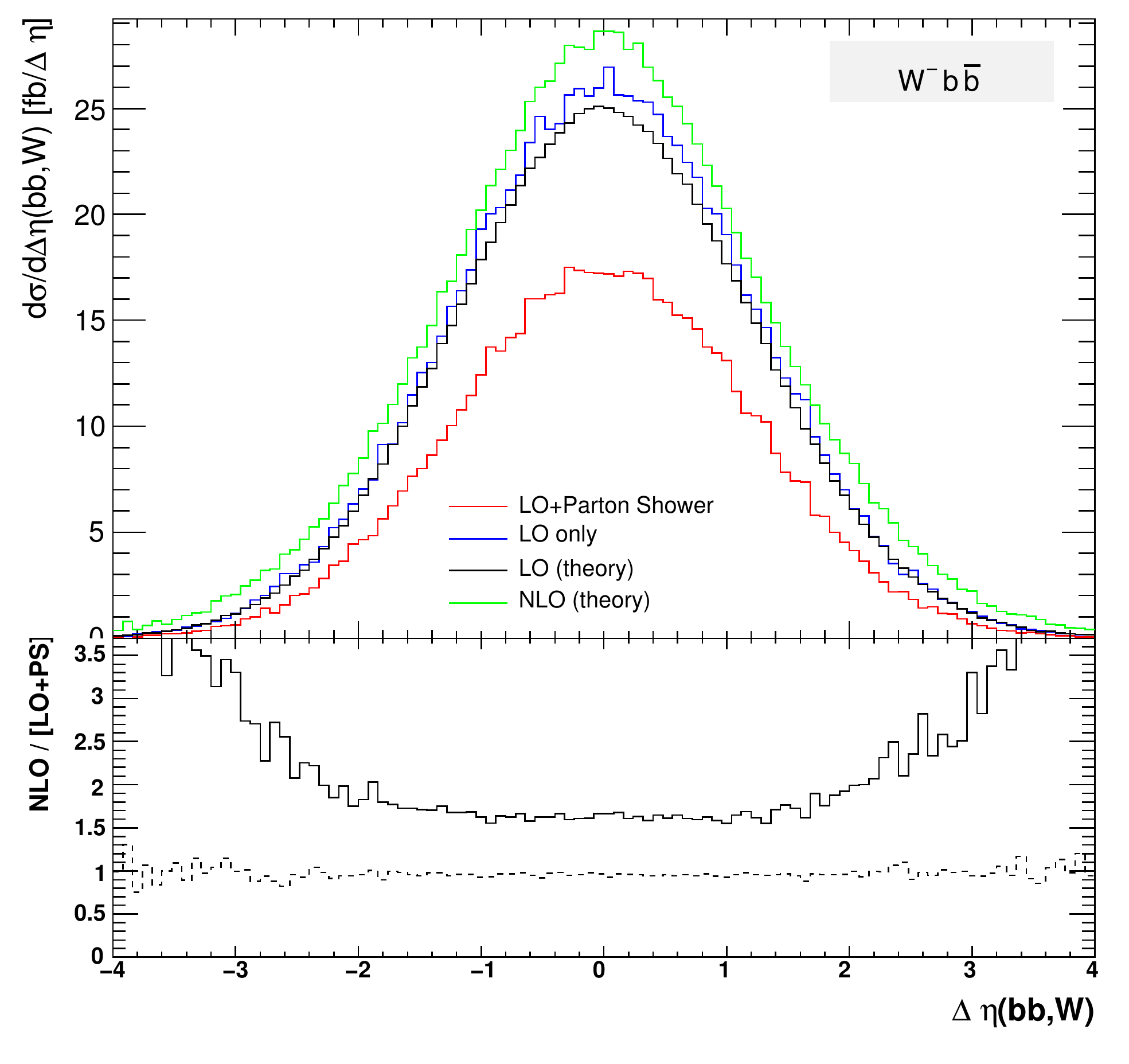}
\caption{\label{fig:deltaEtabbW} Pseudorapidity difference
  distribution of the 2 $b$-jet system and $W$ boson for both
  $W^+b\bar{b}$ and $W^-b\bar{b}$ production. The lower window shows
  the ratio of NLO to LO+PS predictions (full line) together with the
  ratio between the two LO predictions (dashed line).}
\end{center}
\end{figure}

Also the NLO to LO+PS ratios for the distributions for the
pseudorapidity of the leading (Fig.~\ref{fig:Beta}, top) and
subleading (Fig.~\ref{fig:Beta}, bottom) $b$-jets are relatively flat,
with the NLO computation predicting the $b$-jets to be slightly more
central. On the contrary the $W$ bosons are significantly less central
(Fig.~\ref{fig:etaW}) in the NLO prediction, and, as a consequence, a
significant correction affects the distribution of difference in
pseudorapidity between the $b\bar{b}$ and $W$ boson systems, in
particular at large pseudorapidities (Fig.~\ref{fig:deltaEtabbW}).

\begin{figure}[htb!]
\begin{center}
\includegraphics[width=0.48\textwidth]{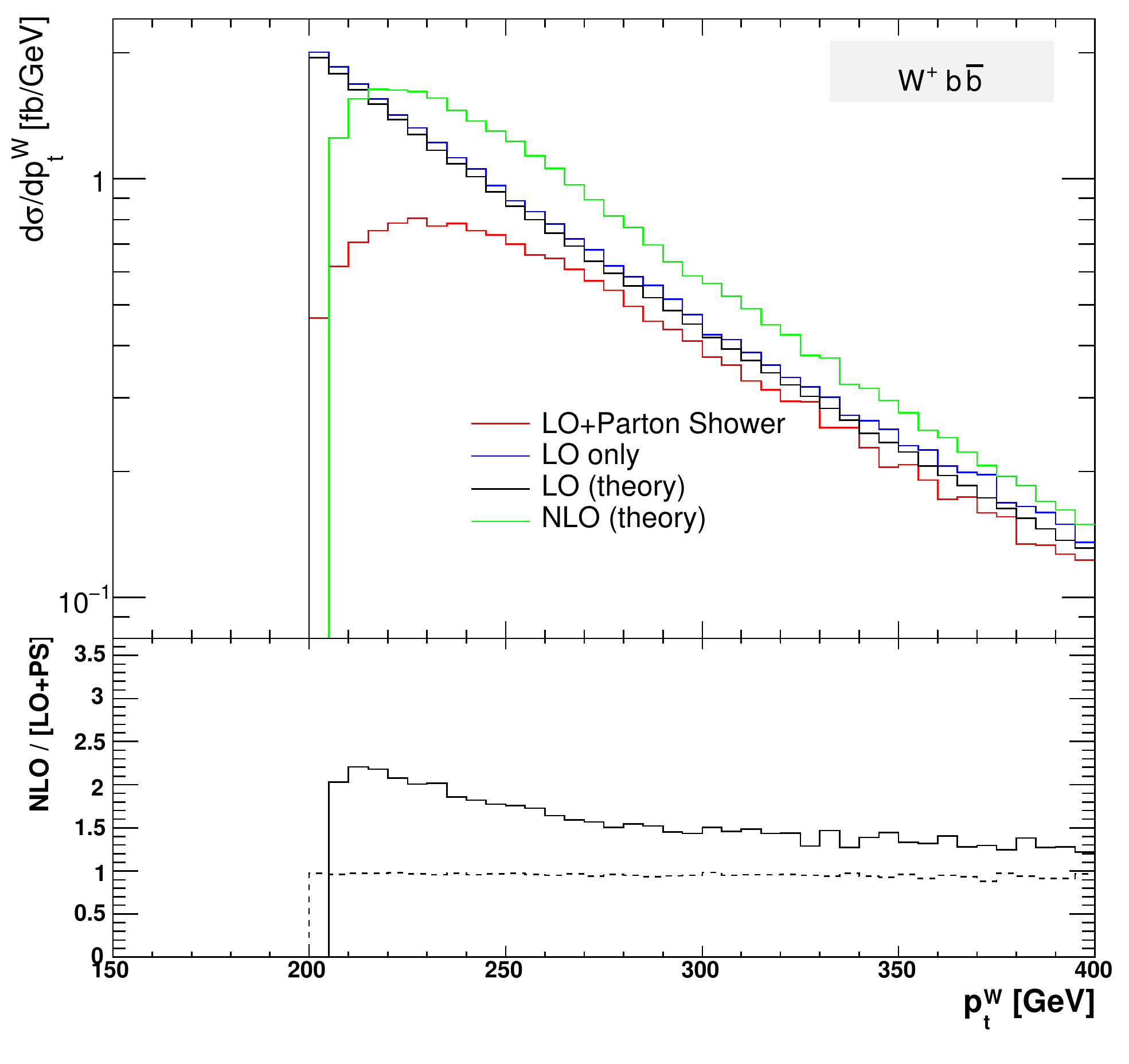}
\includegraphics[width=0.48\textwidth]{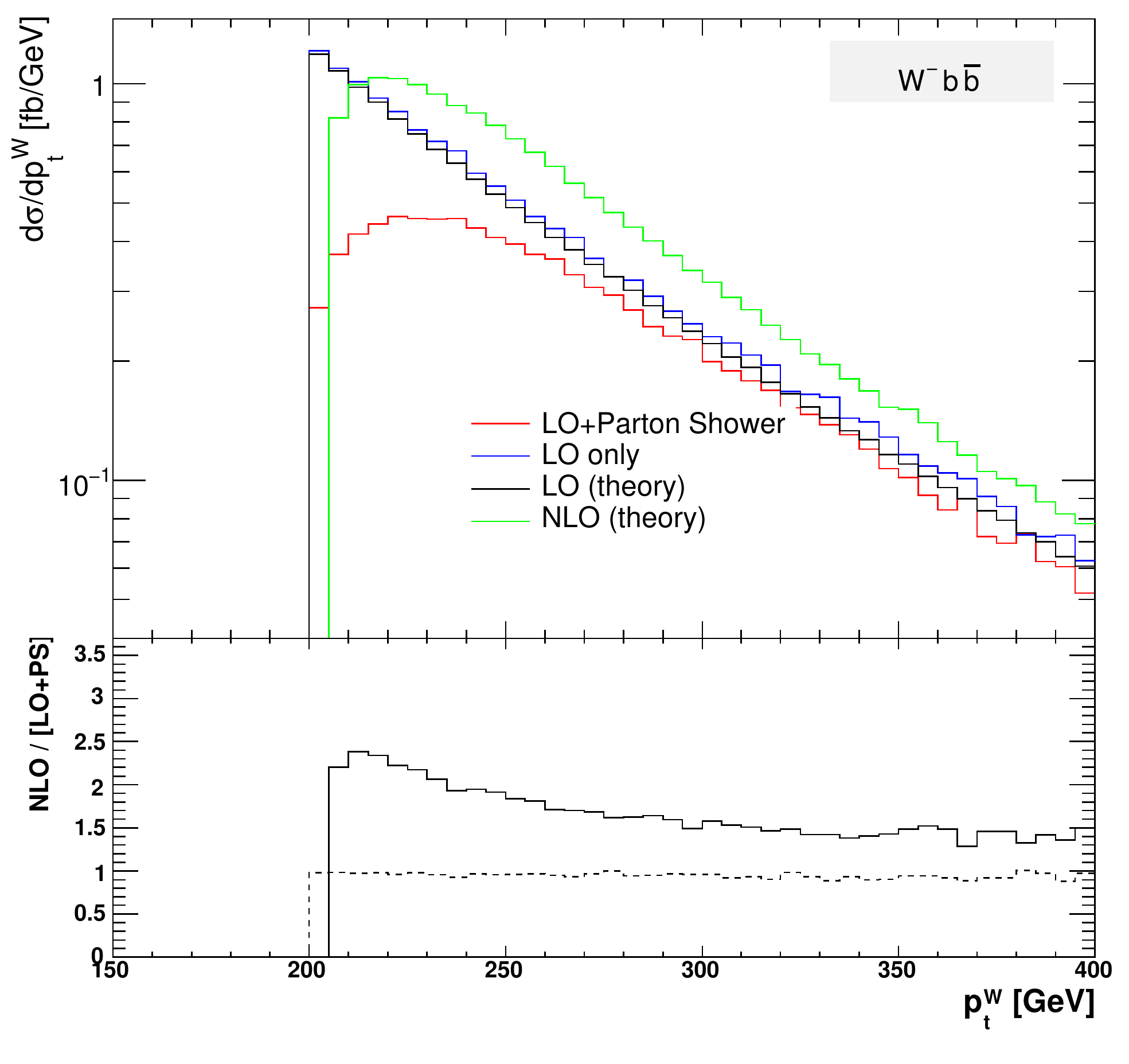}
\caption{\label{fig:pTW} Transverse momentum distribution of the $W$
  vector boson for both $W^+b\bar{b}$ and $W^-b\bar{b}$
  production. The lower window shows the ratio of NLO to LO+PS
  predictions (full line) together with the ratio between the two LO
  predictions (dashed line).}
\end{center}
\end{figure}

\begin{figure}[htb!]
\begin{center}
\includegraphics[width=0.48\textwidth]{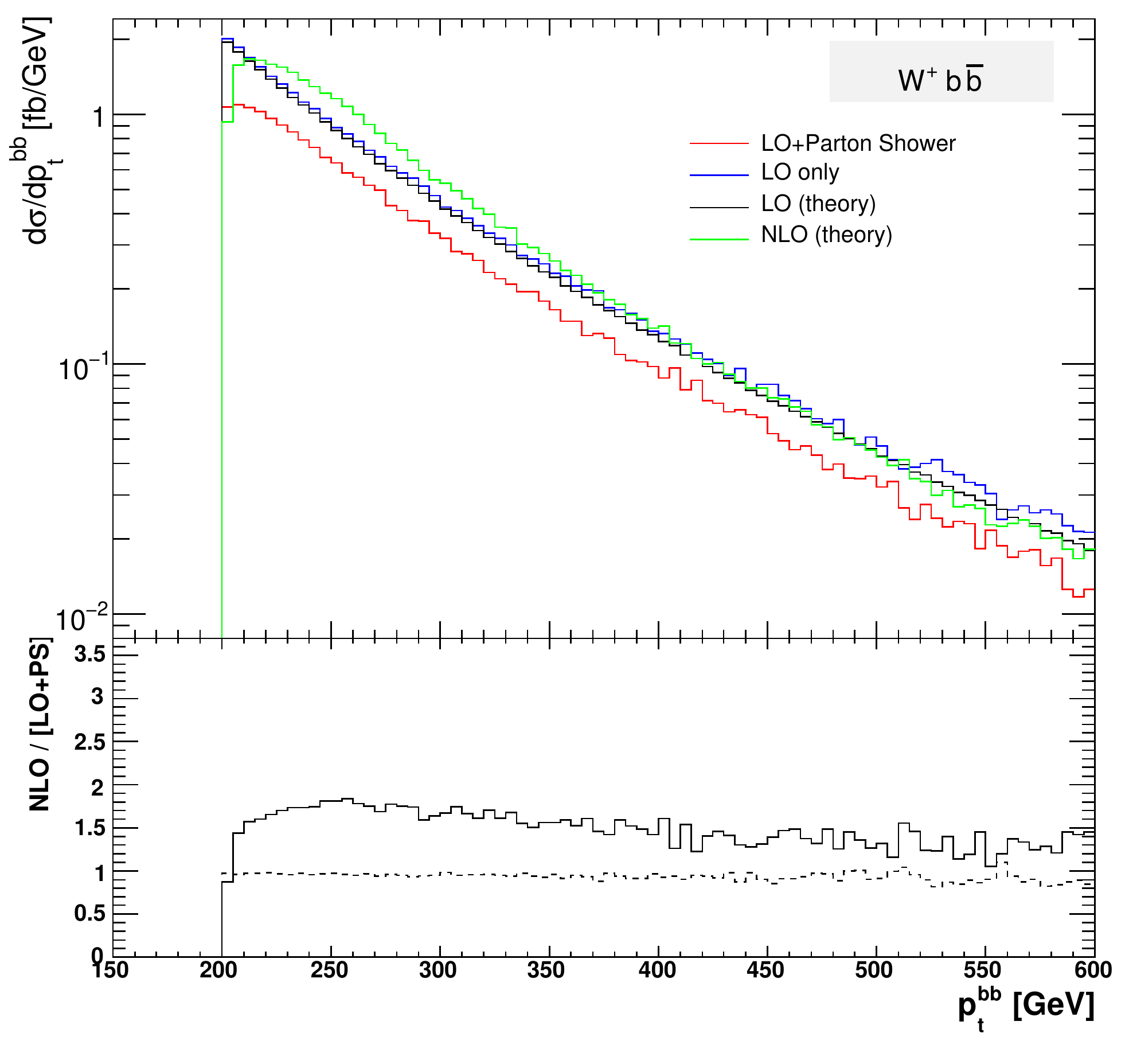}
\includegraphics[width=0.48\textwidth]{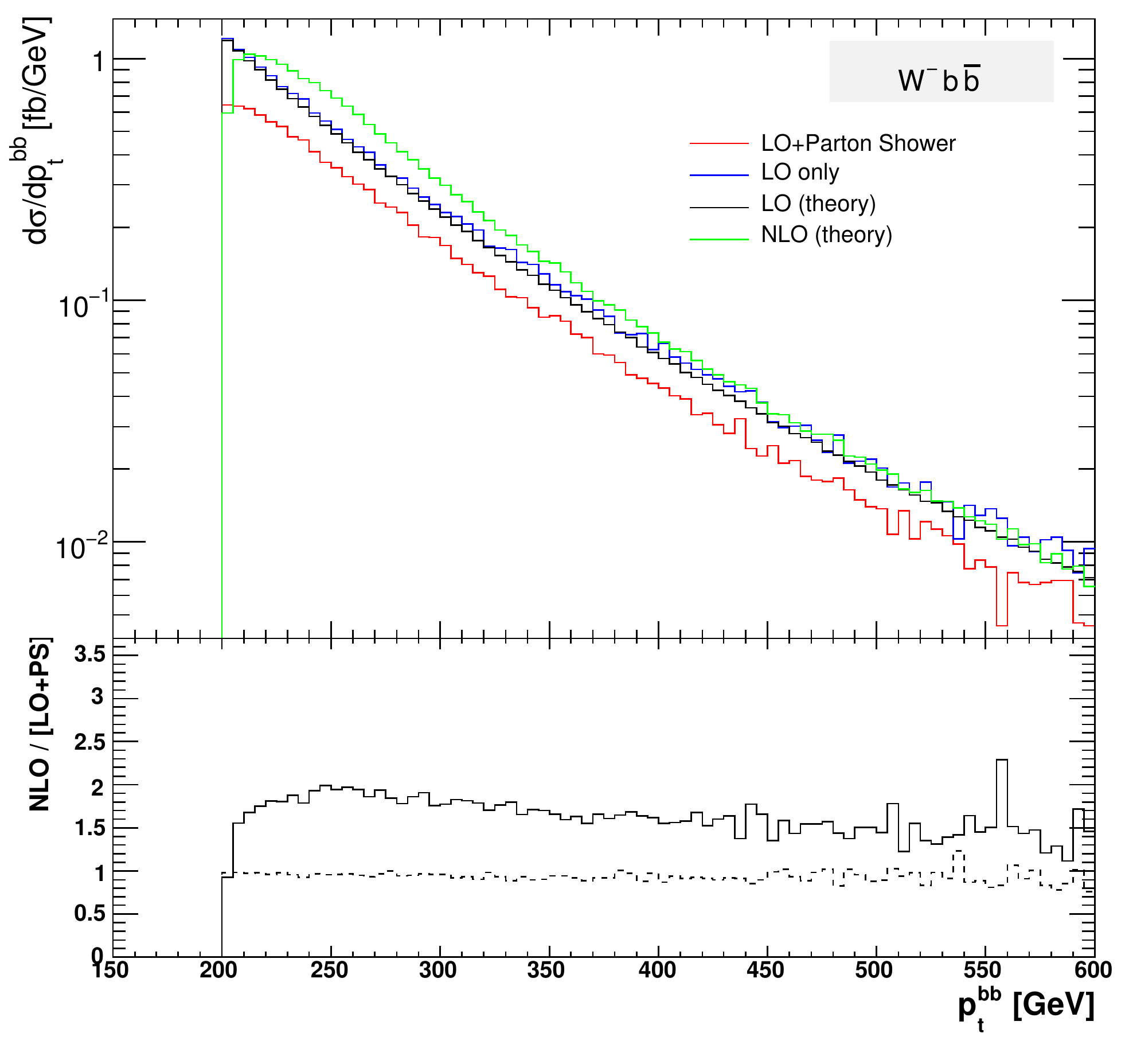}
\caption{\label{fig:pTbb} Transverse momentum distribution of the 2
  $b$-jet system for both $W^+b\bar{b}$ and $W^-b\bar{b}$ production.
  The lower window shows the ratio of NLO to LO+PS predictions (full
  line) together with the ratio between the two LO predictions (dashed
  line).}
\end{center}
\end{figure}

For the transverse momentum of the leading $b$-jet
(Fig.~\ref{fig:leadingBpT}, top), apart from some threshold effect at
values close to 100~GeV, the ratio NLO to LO+PS is relatively
flat. For the transverse momentum of the subleading $b$-jet
(Fig.~\ref{fig:leadingBpT}, bottom), the parton shower approximation
does a perfect job and the ratio is perfectly flat.

A bit problematic is the case of the transverse momentum of the $W$
boson system (Fig.~\ref{fig:pTW}) and, to a less extent, of the
$b\bar{b}$ system (Fig.~\ref{fig:pTbb}), where the NLO prediction is
reliable towards higher transverse momenta, while, close to the
threshold, it starts to be sensitive to the emission of multiple soft
radiation (large logarithms). This is more severe for the distribution
of the transverse momentum of the $W$ boson system. In both cases, the
LO+PS distribution could be re-weighted at large transverse momenta
according to the NLO prediction, while the LO+PS based distribution
itself could be used to extrapolate to the region close to the
threshold. Correlations between the two distributions need clearly to
be taken into account.

The parton shower algorithm is able to reproduce fairly well the shape
of the distributions of the $b\bar{b}$ invariant mass
(Fig.~\ref{fig:mbb}), of the $b\bar{b}W$ invariant mass
(Fig.~\ref{fig:mbbW}) and of the distance in pseudorapidity and
azimuthal angle of the two $b$-jets (Fig.~\ref{fig:drbb}). In the case
of the $b\bar{b}$ invariant mass, a small discrepancy is seen at very
small invariant masses, again in the region where eventually large
logarithmic corrections to the NLO computation can be expected.

\begin{figure}[htb!]
\begin{center}
\includegraphics[width=0.48\textwidth]{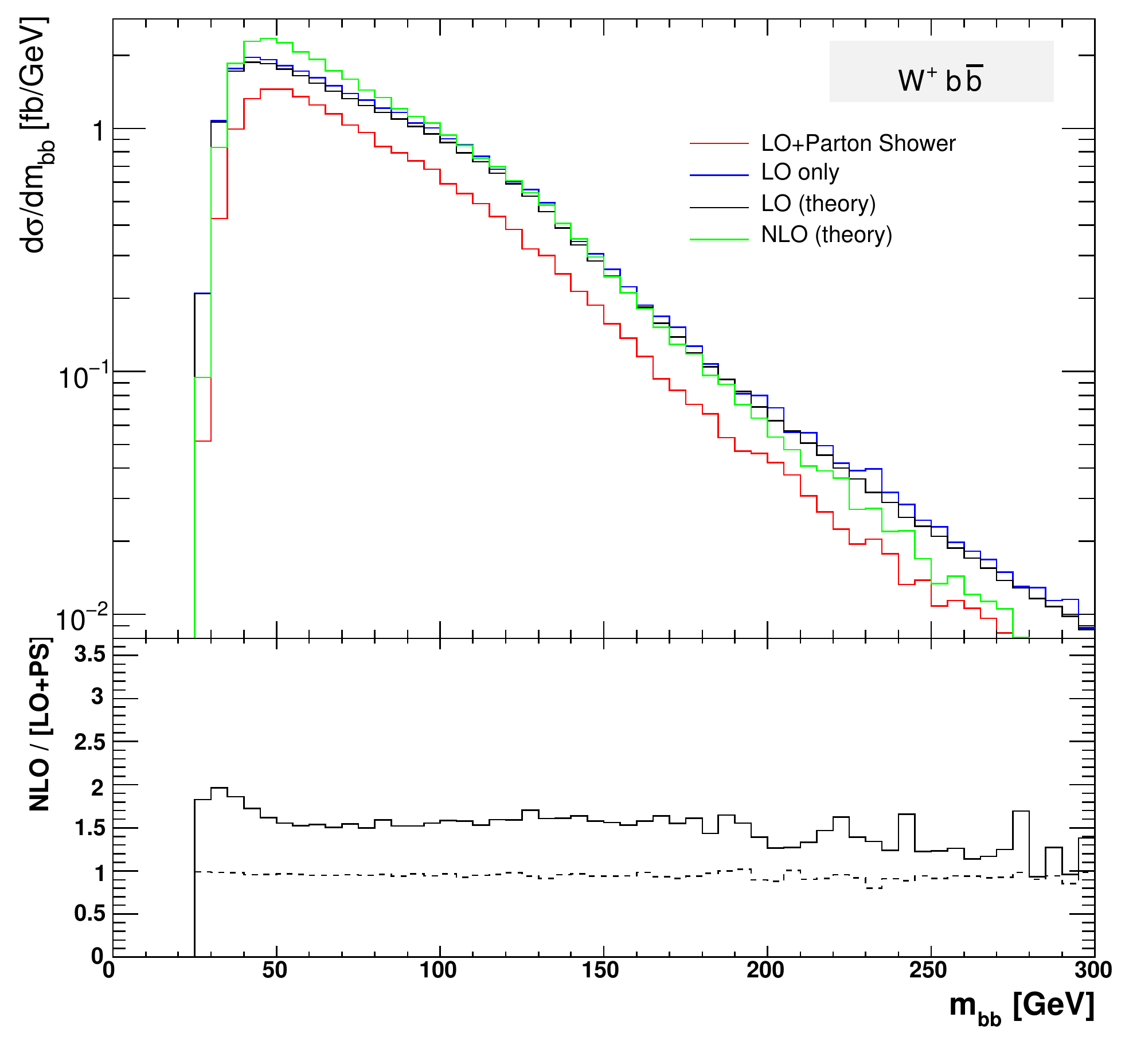}
\includegraphics[width=0.48\textwidth]{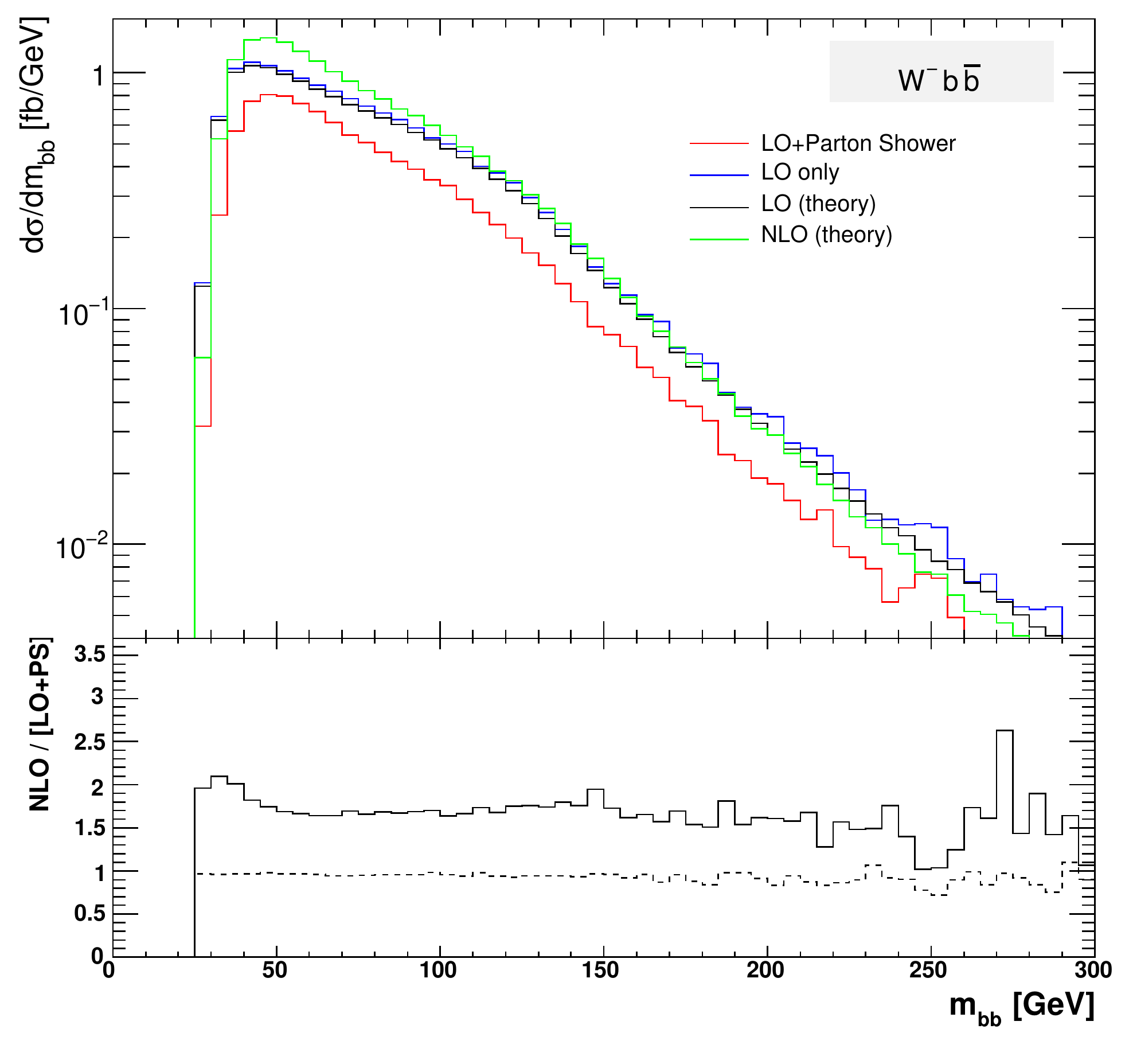}
\caption{\label{fig:mbb} Invariant mass distribution of the 2 $b$-jet
  system for both $W^+b\bar{b}$ and $W^-b\bar{b}$ production. The
  lower window shows the ratio of NLO to LO+PS predictions (full
  line), together with the ratio between the two LO predictions
  (dashed line).}
\end{center}
\end{figure}

\begin{figure}[htb!]
\begin{center}
\includegraphics[width=0.48\textwidth]{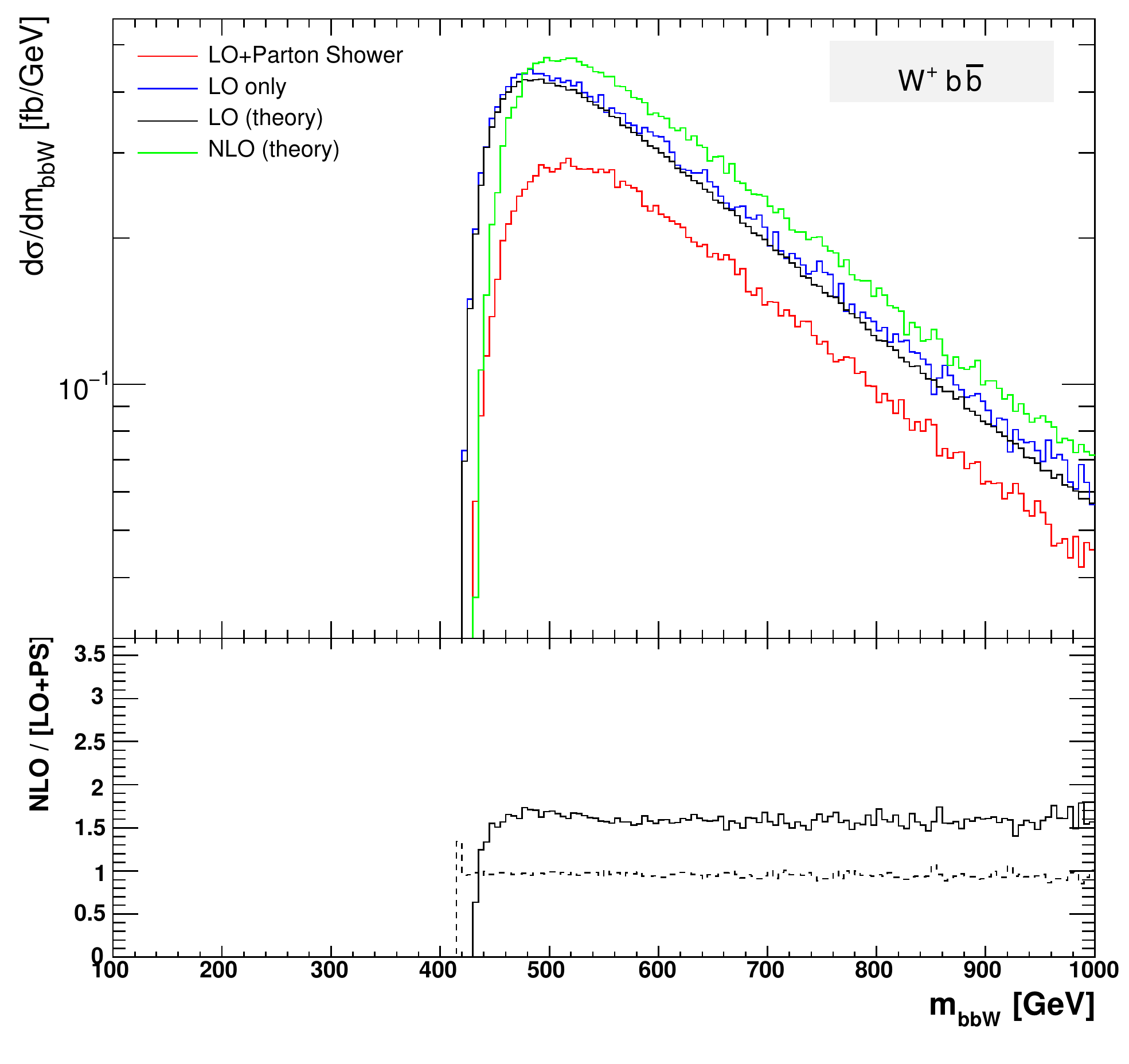}
\includegraphics[width=0.48\textwidth]{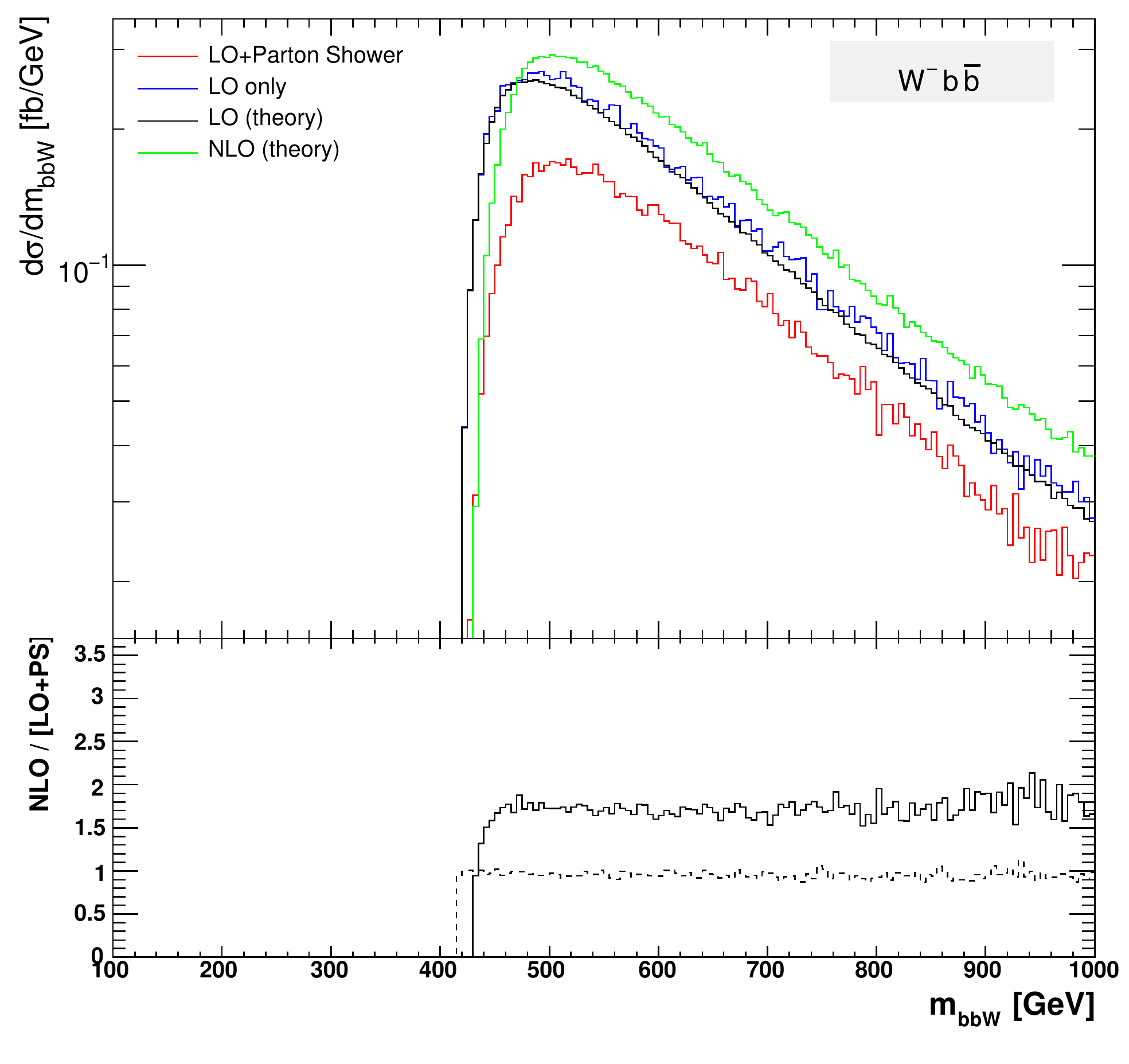}
\caption{\label{fig:mbbW} Invariant mass distribution of the 2 $b$-jet
  plus $W$ system for both $W^+b\bar{b}$ and $W^-b\bar{b}$ production.
  The lower window shows the ratio of NLO to LO+PS predictions (full
  line), together with the ratio between the two LO predictions
  (dashed line).}
\end{center}
\end{figure}

\begin{figure}[htb!]
\begin{center}
\includegraphics[width=0.48\textwidth]{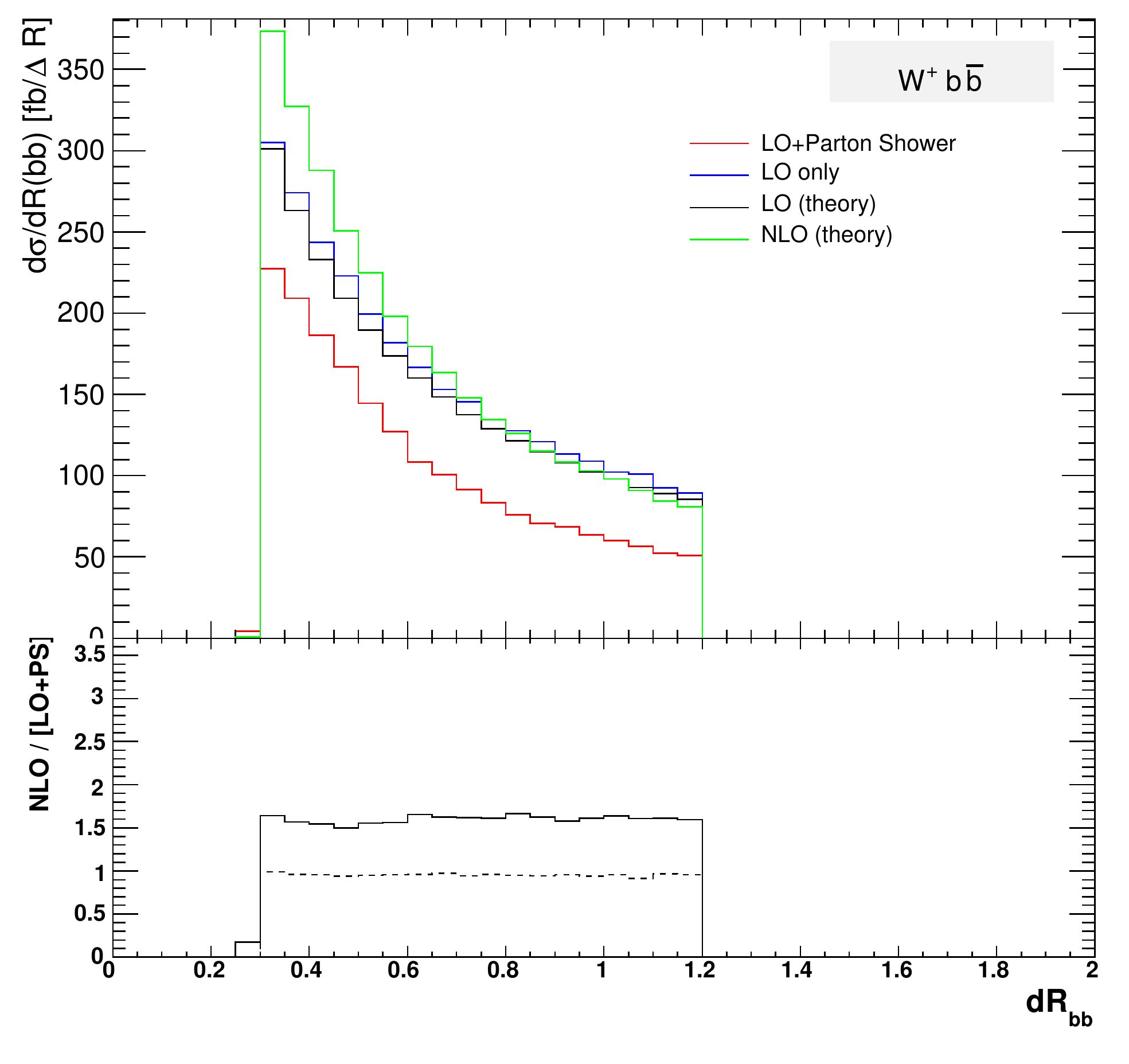}
\includegraphics[width=0.48\textwidth]{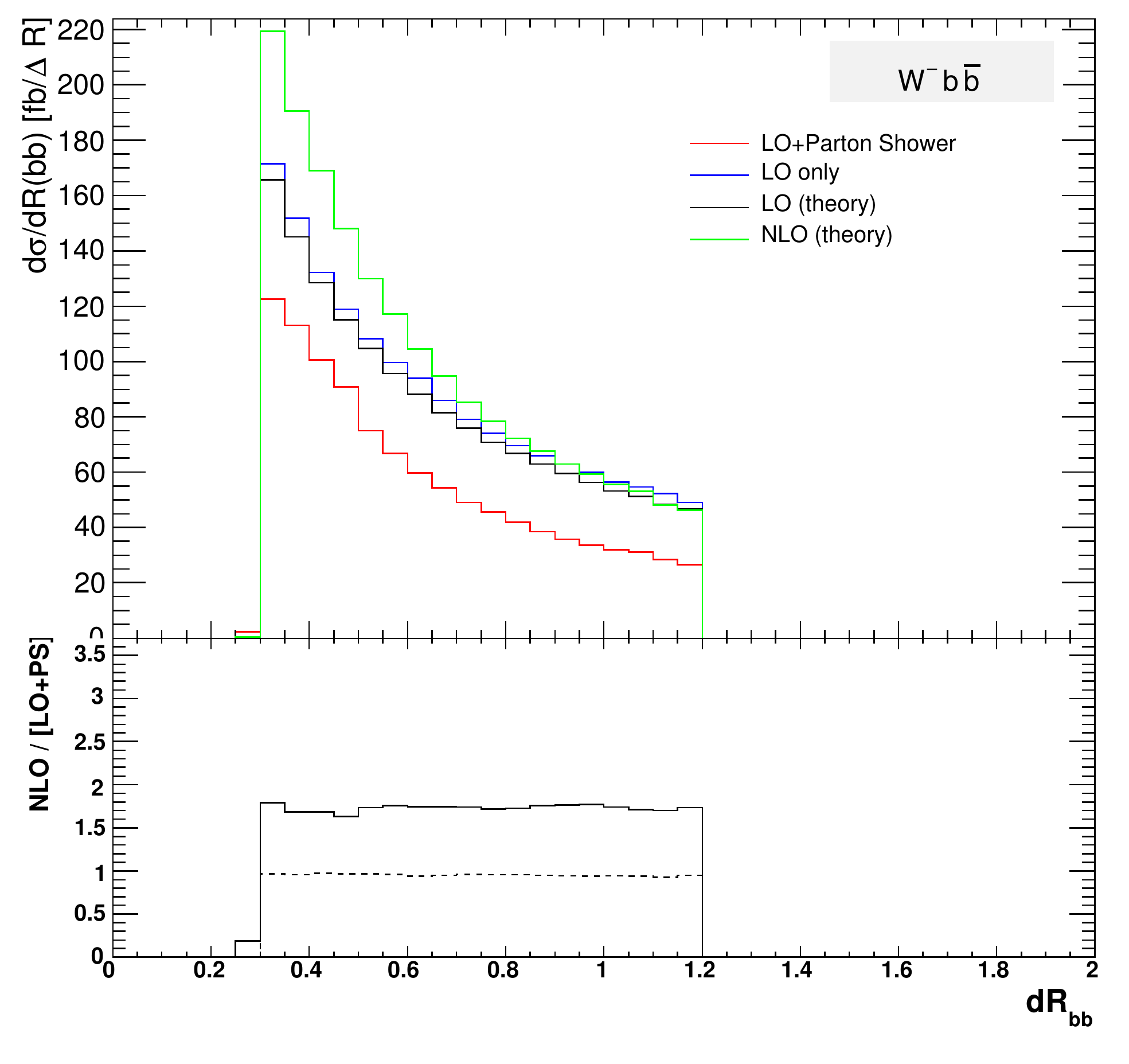}
\caption{\label{fig:drbb} Relative distance distribution of the 2
  $b$-jet system for both $W^+b\bar{b}$ and $W^-b\bar{b}$
  production. The lower window shows the ratio of NLO to LO+PS
  predictions (full line), together with the ratio between the two LO
  predictions (dashed line).}
\end{center}
\end{figure}

Finally, the $p_T$ of additional non $b$-jet distribution is shown in
Fig.~\ref{fig:last}. The cross section corresponding to the events
where no additional non $b$-jet above 15~GeV is found is condensed in
the first bin of the distribution. As a general tendency, the NLO
computation predicts significantly more radiation than the LO+PS
prediction. There are two possible reasons for this:
\begin{enumerate}
\item The parton shower algorithm used in this study acts on top of
  the LO matrix element producing $Wb\bar{b}$, which means that it
  misses all radiation where the first splitting is of the type $g \to
  gg$ and the $b\bar{b}$ pair is produced later in the cascade, while
  the NLO prediction will include the case where the $b\bar{b}$ is
  produced right after the first $g \to gg$ splitting.
\item The NLO computation includes the process $q(\bar{q})g \to
  Wb\bar{b}j$ in the real correction term, which is strongly but not
  completely suppressed for small $p_T$ of the additional non $b$-jet
  in the event. This is illustrated in Fig.~\ref{fig:last}, where we
  also show the theory prediction when only including the $q\bar
  q$-initiated process.
\end{enumerate}
Given this overall tendency, one would expect that by moving the jet
veto cut at parton level from 60~GeV to 20~GeV (as in the cut based
analysis), the K factor correcting from LO+PS to the NLO prediction
would move significantly from $\approx 1.6$ more towards one, which
would be a clear advantage for the $WH$ analysis.  However, this
requires a reliable prediction of the transverse momentum of the
additional non $b$-jet of the event down to very low transverse
momenta (if compared with the hard scale of the event which is around
$p_T(b\bar{b}) \approx 200$~GeV). This is not provided by the NLO
prediction, as the renormalization and factorization scale variation
uncertainty of the {\it exclusive} 2 $b$-jet result reported in
Table~\ref{tab:total_xs} of the last section demonstrates. In order to
get a more reliable prediction in this kinematic region, some matching
procedure between the parton shower and the NLO predictions is needed.
Before being able to define such a matching procedure, further studies
will be needed, based on parton shower Monte Carlo algorithms able to
include the primary $g \to gg$ splittings (available in plain \herwig\
by using the $pp \to Wg$ matrix element), which would lead to a
complete leading logarithmic prediction, or able to include the
additional $q(\bar{q})g \to Wb\bar{b}j$ process, as
Alpgen~\cite{Mangano:2002ea} (which however does not include the
primary $g \to gg$ splittings). The use of these two algorithms will
in fact allow to disentangle the two previously enumerated effects.

\begin{figure}[htb!]
\begin{center}
\includegraphics[width=0.48\textwidth]{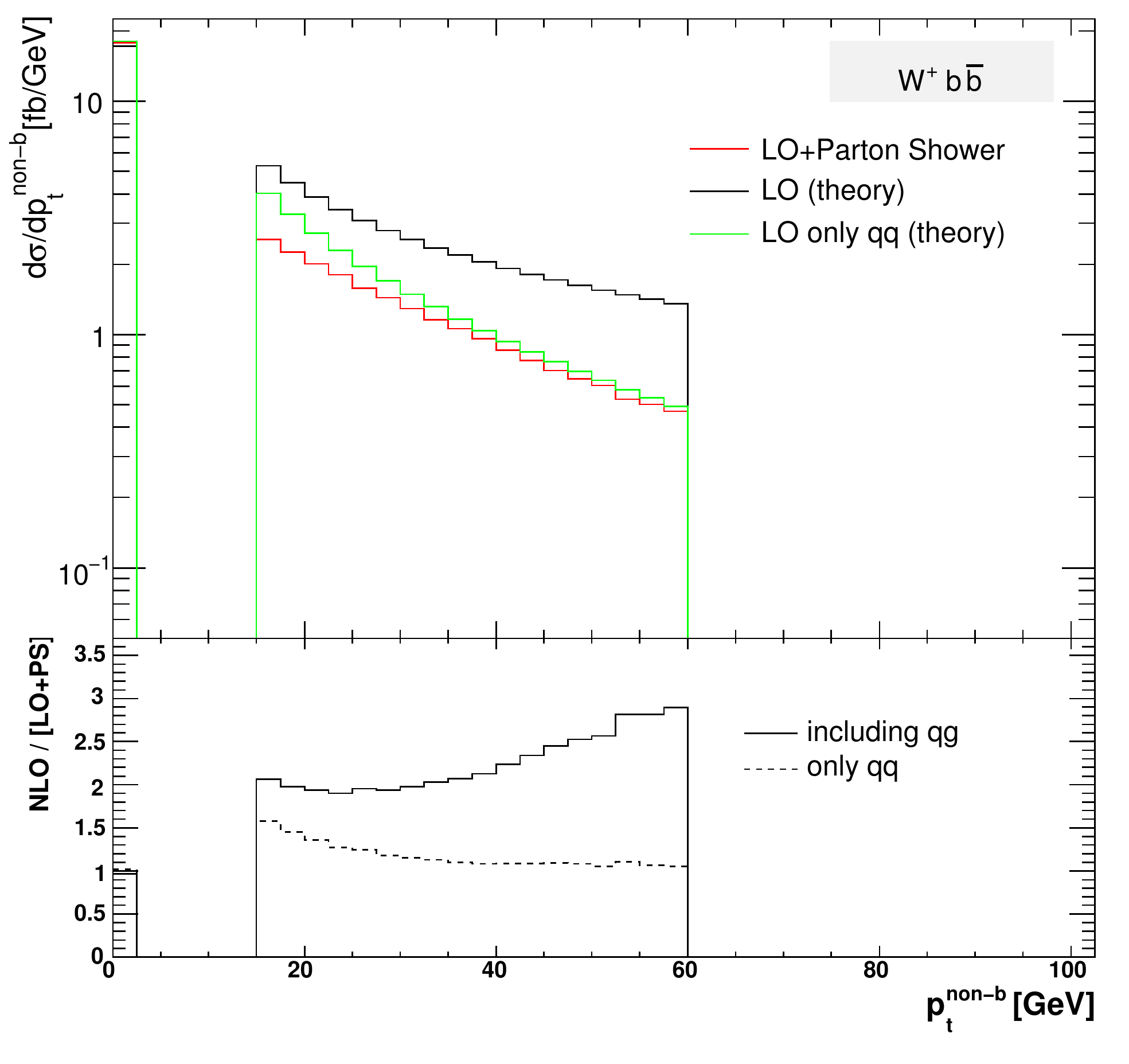}
\includegraphics[width=0.48\textwidth]{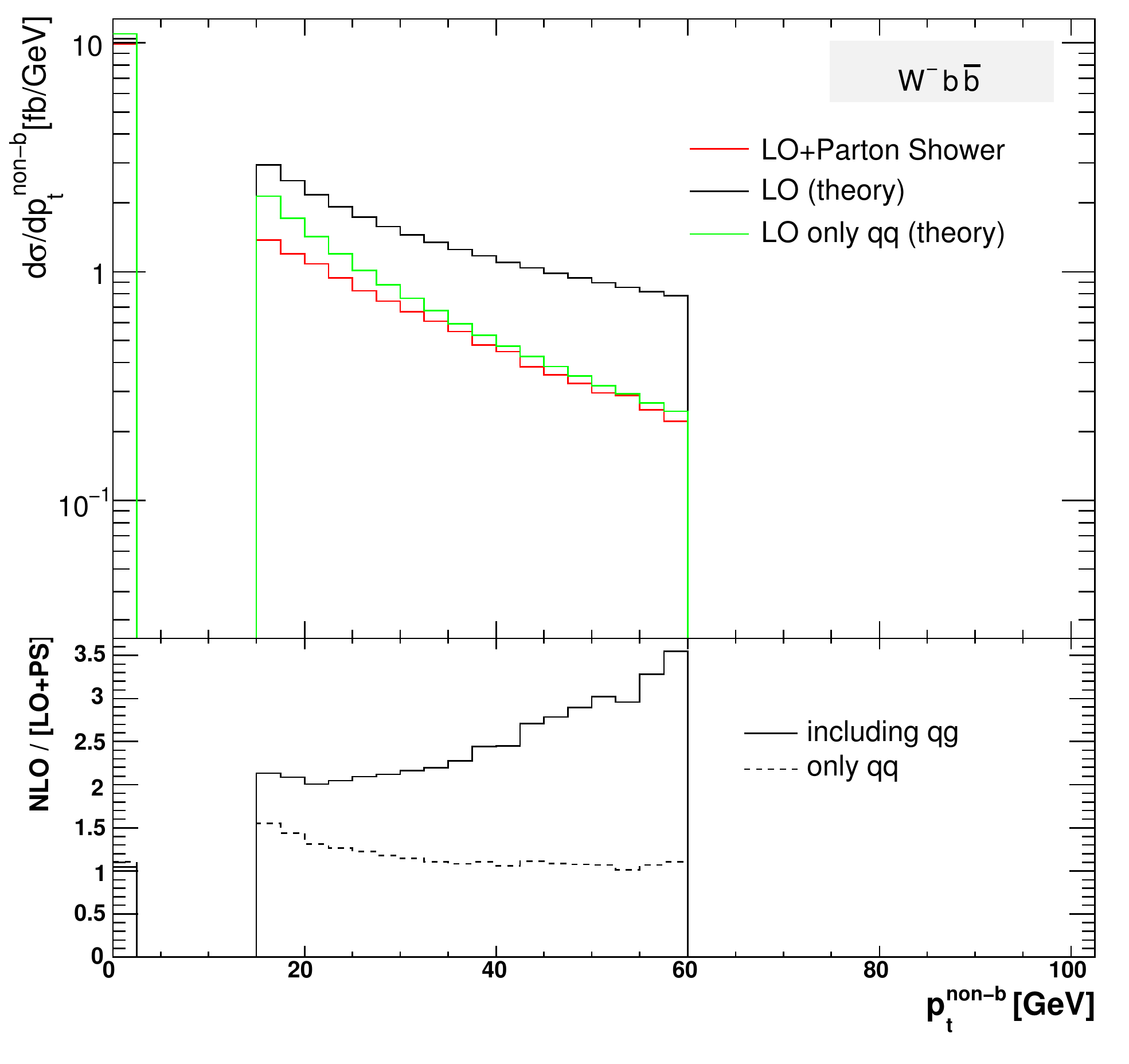}
\caption{\label{fig:last} Transverse momentum distribution of the non
  $b$-jet for both $W^+b\bar{b}+j$ and $W^-b\bar{b}+j$ production. The
  lower window shows the ratio of LO to LO+PS prediction (dashed line 
  including only the $q\bar{q}^\prime \to W b \bar b j$ process 
 in the LO computation, full line including also the $q(\bar{q})g \to Wb\bar{b}j$ process). 
Although the cross section for $Wb\bar b+j$ production is a LO result, it is calculated 
as part of the real corrections with the CTEQ6M set of PDFs and 
two-loop evolution of $\alpha_s$.}
\end{center}
\end{figure}

\subsection{CONCLUSIONS}

We have compared the NLO computation for the $Wb\bar{b}$ process with
the predictions of the \herwig\ parton shower algorithm applied on top
of the LO $Wb\bar{b}$ matrix element, considering for the first time
the specific kinematic region explored by the high $p_T$ $WH$ analysis
and the effect of an additional non $b$-jet veto (applied at parton
level) of 60~GeV. Given the choice of PDFs, of strong coupling
constant and renormalization and factorization scales already
mentioned, we find that the cross section predicted by the NLO
computation is a factor $\approx 1.6$ higher than the parton shower
prediction. Many of the differential distributions predicted by the
parton shower algorithm agree reasonably well with the NLO
computation. In general, the increase in event yield due to the NLO
correction can be decreased by further lowering the jet veto:
however, the NLO prediction suffers from substantial uncertainties in
the kinematic region where radiation significantly softer than the
scale of the process occurs and further studies are ongoing to make
the prediction more reliable.

\clearpage

\part[OBSERVABLES AND DETECTORS]{OBSERVABLES AND DETECTORS}

\section[DELPHES, A FRAMEWORK FOR FAST SIMULATION OF A GENERIC COLLIDER EXPERIMENT]{DELPHES, A FRAMEWORK FOR FAST SIMULATION OF A GENERIC COLLIDER EXPERIMENT\protect\footnote{Contributed by: S. Ovyn, X. Rouby}}







\subsection{INTRODUCTION}
Multipurpose detectors at high energy colliders are very complex systems. Their simulation is in general performed by means of the \textsc{Geant}~\cite{Agostinelli2003250} package and final observables used for analyses usually require sophisticated reconstruction algorithms. This complexity is handled by large collaborations, and data and the expertise on reconstruction and simulation software are only available to their members. Precise data analyses require a full detector simulation, including transport of the primary and secondary particles through the detector material accounting for the various detector inefficiencies, the dead material, the imperfections and the geometrical details. Such a simulation is very complicated, technical and requires a large \textsc{cpu} power. On the other hand, phenomenological studies, looking for the observability of given signals, may require only fast but realistic estimates of the expected signal signatures and their associated backgrounds.

The \textsc{Delphes}~\cite{Ovyn:2009tx} framework has been designed for the fast simulation of a general-purpose collider experiment. Using this framework, observables such as cross-sections and efficiencies after event selection can be estimated for specific reactions. Starting from the output of event generators (e.g.\ Les Houches Event File or \texttt{HepMC}), the simulation of the detector response takes into account the subdetector resolutions, by smearing the kinematics of final-state particles.

\textsc{Delphes} includes the most crucial experimental features, such as: the geometry of both central and forward detectors, the effect of magnetic field on tracks, the reconstruction of photons, leptons, jets, $b$-jets, $\tau$-jets and missing transverse energy, a lepton isolation, a trigger emulation and an event display. Several common datafile formats can be used as input in \textsc{Delphes}, in order to process events from many different generators.

\subsection{SIMULATION OF THE DETECTOR RESPONSE}

The overall layout of the multipurpose detector simulated by \textsc{Delphes} consists in a central tracking system surrounded by an electromagnetic and a hadron calorimeters (\textsc{ecal} and \textsc{hcal}, each with a central region and two endcaps) and two forward calorimeters (\textsc{fcal}, with separate electromagnetic and hadronic sections). Finally, a muon system encloses the central detector volume. In addition, possible very forward detectors are also simulated, like zero-degree calorimeters (\textsc{zdc}) and very forward taggers. Even if \textsc{Delphes} has been developed for the simulation of general-purpose detectors at the \textsc{lhc} (namely, \textsc{Cms} and \textsc{Atlas}), it allows a flexible parametrisation for other cases, e.g.\ at future linear colliders.

\subsubsection*{Magnetic field, tracks and calorimetric cells}

The effects of a central solenoidal magnetic field are simulated. Every stable charged particle with a transverse momentum above some threshold and lying inside the tracker has a probability to yield a reconstructed track. 
The calorimeters are segmented into cells in the $(\eta,\phi)$, with variable sizes. The response of calorimeters to energy deposits of incoming particles depends on their resolution and on the nature of the particles. The assumed calorimeter characteristics are not isotropic, with typically finer energy resolution and granularity in the central region~\cite{Acosta:922757,Aad:2009wy}. 
Their resolution is parametrised through a Gaussian smearing of the accumulated cell energy.

Electrons and photons leave their energy in the electromagnetic parts of the calorimeters (\textsc{ecal} and \textsc{fcal}, e.m.) only, while charged and neutral final-state hadrons interact with the hadronic parts (\textsc{hcal} and \textsc{fcal}, had.).
Some long-living particles, such as the $K^0_s$ and $\Lambda$'s, with lifetime $c\tau$ smaller than $10~\textrm{mm}$ are considered as stable particles by the generators although they may decay before reaching the calorimeters. The energy smearing of such particles is therefore performed using the expected fraction of the energy, determined according to their decay products, that would be deposited into the \textsc{ecal} and the \textsc{hcal}. 
Muons, neutrinos and hypothetical neutralinos are assumed not to interact with the calorimeters.
At last, the muon systems provide a measurement of the true muon kinematics, with a smeared transverse momentum.

\subsubsection*{Forward detectors}

Most of the recent experiments in beam colliders have additional instrumentation along the beamline. These extend the $\eta$ coverage to higher values, for the detection of very forward final-state particles. In \textsc{Delphes}, zero-degree calorimeters, roman pots and forward taggers have been implemented, similarly to the plans for \textsc{Cms} and \textsc{Atlas} collaborations~\cite{Acosta:922757, Aad:2009wy}.


The \textsc{zdc}s allow the measurement of stable neutral particles ($\gamma$ and $n$) coming from the interaction point (\textsc{ip}), with large pseudorapidities. The trajectory of the neutrals observed in the \textsc{zdc}s is a straight line, while charged particles are deflected away from their acceptance window by the powerful magnets located in front of them. 
In their implementation in \textsc{Delphes}, the \textsc{zdc}s provide a measurement of the time-of-flight of the particle, from the \textsc{ip} to the detector location. The simulated \textsc{zdc}s are composed of an electromagnetic and a hadronic sections, for the measurement of photons and neutrons, respectively. The \textsc{zdc} hits do not enter in the calorimeter cell list used for reconstruction of jets and missing transverse energy. The fact that additional charged particles may enter the \textsc{zdc} acceptance is neglected in the current versions of \textsc{Delphes}.

Forward taggers, located very far away from the \textsc{ip}, are meant for the measurement of scattered particles following very closely the beam path. To be able to reach these detectors, particles must have a charge identical to the beam particles and a momentum very close to the nominal value of the beam. These taggers are near-beam detectors located a few millimetres from the true beam trajectory and this distance defines their acceptance. This fast simulation uses the \textsc{Hector} software~\cite{1748-0221-2-09-P09005}, which includes the chromaticity effects and the geometrical aperture of the beamline elements of any arbitrary collider. 
Forward taggers are able to measure the hit positions and angles in the transverse plane at the location of the detector, as well as the time-of-flight. Out of these, the particle energy and the momentum transfer it underwent during the interaction can be reconstructed at the analysis level.

\subsection{HIGH-LEVEL OBJECT RECONSTRUCTION}

The results of the detector simulation are output in a file storing for each event its tracks, calorimetric cells and hits in the very forward detectors. In addition, collections of final particles  ($e^\pm$, $\mu^\pm$, $\gamma$) and objects (light jets, $b$-jets, $\tau$-jets, $E_T^{\rm miss}$) are provided. While electrons, muons and photons are easily identified, other quantities are more difficult to evaluate as they rely on sophisticated algorithms (e.g. jets or missing energy).
For most of these objects, their four-momentum and related quantities are directly accessible in \textsc{Delphes} output ($E$, $\vec{p}$, $p_T$, $\eta$ and $\phi$). 

\subsubsection*{Identification of electrons, photons and muons}

Real electron ($e^\pm$) and photon candidates are associated to the final-state collections if they fall into the acceptance of the tracking system and have a transverse momentum above some threshold. Assuming a good measurement of the track parameters in the real experiment, the electron energy can be reasonably recovered. 
\textsc{Delphes} assumes a perfect algorithm for clustering and bremsstrahlung recovery. Electron energy is smeared according to the resolution of the calorimetric cell where it points to, but independently from any other deposited energy is this cell. 
Electrons and photons may create a candidate in the jet collection.

Generator-level muons ($\mu^\pm$) entering the muon system acceptance and overpassing some threshold are considered as good candidates for analyses. The application of the detector resolution on the muon momentum depends on a Gaussian smearing of the $p_T$.
Neither $\eta$ nor $\phi$ variables are modified beyond the calorimeters: no additional magnetic field is applied. Multiple scattering is neglected. This implies that low energy muons have in \textsc{Delphes} a better resolution than in a real detector.  

To improve the quality of the contents of the charged lepton collections, isolation criteria can be applied. This requires that electron or muon candidates are isolated in the detector from any other particle, within a given cone in the ($\eta$,$\phi$) plane. The isolation algorithm is based on tracker data, but in addition the sum of the transverse momenta of all tracks but the lepton one within the isolation cone is provided, as well an estimate based on calorimeter data.

\subsubsection*{Jet reconstruction}

A realistic analysis requires a correct treatment of partons which have hadronised. Therefore, the most widely currently used jet algorithms have been integrated into the \textsc{Delphes} framework using the \textsc{FastJet} tool~\cite{Cacciari200657}.
For all jet algorithms, the calorimetric cells are used as inputs. Since several particles can leave their energy into a given calorimetric cell, which broadens the jet energy resolution, a \textit{jet energy flow} algorithm can be switched on in \textsc{Delphes}. This takes into account the measured properties of tracks pointing to the calorimetric cells of interest for the jet reconstruction.

A jet is tagged as a $b$-jet if its direction lies in the acceptance of the tracker and if it is associated to a parent $b$-quark. 
This identification procedure for the $b$-tag is based on the true generator-level identity of the most energetic parton within a cone around the $(\eta,\phi)$ region, with a radius equal to the one used to reconstruct the jet.
Jets originating from $\tau$-decays are identified using a procedure consistent with the one applied in a full detector simulation~\cite{Acosta:922757}. The tagging relies on the two following properties of the $\tau$ lepton. First, $77\%$ of the $\tau$ hadronic decays contain only one charged hadron associated to a few neutrals. Secondly, the particles arisen from the $\tau$ lepton produce narrow jets in the calorimeter. 

\subsubsection*{Missing transverse energy}

The \textit{true} missing transverse energy (\textsc{met}), i.e.\ at generator-level, is calculated as the opposite of the vector sum of the transverse momenta of all visible particles. 
In a real experiment, calorimeters measure energy and not momentum. Any problem affecting the detector (dead channels, misalignment, noisy cells, cracks) worsens directly the measured missing transverse energy. In \textsc{Delphes}, \textsc{met} is based on the calorimetric cells only. Muons and neutrinos are therefore not taken into account for its evaluation. However, as muon candidates, tracks and calorimetric cells are available in the output file, the missing transverse energy can always be reprocessed a posteriori with more specialised algorithms.

\subsubsection*{Trigger emulation}

Most of the usual trigger algorithms select events containing leptons, jets, and \textsc{met} with an energy scale above some threshold. A trigger emulation is included in \textsc{Delphes}, using a fully parametrisable trigger table, also allowing
logical combinations of several conditions on the final analysis data.

\subsection{ASSUMED SIMPLIFICATIONS FOR THE FAST SIMULATION}

\textsc{Delphes} is a fast simulation aiming at providing quickly realistic observables. It relies on several hypotheses and simplifications that allows the use of less sophisticated algorithms than in real experiments, but aims at reaching the same detection, identification and reconstruction performances. Since the framework design relies on the expectations of the real detector performances, detector geometry is idealised: being uniform, symmetric around the beam axis, and having no cracks nor dead material. Secondary interactions, multiple scatterings, photon conversion and bremsstrahlung are also neglected. 

No longitudinal segmentation is available in the simulated calorimeters. \textsc{Delphes} assumes that \textsc{ecal} and \textsc{hcal} have the same segmentations and that the detector is symmetric with respect to the $\eta=0$ plane and around the beam axis. A particle entering a calorimetric cell deposes all its energy, even if it enters very close to its geometrical edge. 
Particles other than electrons ($e^\pm$), photons ($\gamma$), muons ($\mu^\pm$), neutrinos ($\nu_e$, $\nu_\mu$ and $\nu_\tau$) and hypothetical neutralinos are simulated as hadrons for their interactions with the calorimeters. The simulation of stable particles beyond the Standard Model should therefore be handled with care.

The electron, photon and muon collections contain only real candidates.
For instance, the particles which might leak out of the calorimeters into the muon system (\textit{punch-through}) are not considered as muon candidates in \textsc{Delphes}. However, fake candidates can be added into the collections at the analysis level, when processing \textsc{Delphes} output data. 

In current version of \textsc{Delphes}, the displacement of secondary vertices is not simulated, in particular for $b$-jets.
Extra hits coming from the beam-gas events or secondary particles hitting the beampipe in front of the forward detectors are not taken into account. At last, real triggers are intrinsically based on reconstructed data with a worse resolution than final analysis data. In \textsc{Delphes}, the same data are for trigger emulation and for final analyses.

\subsection{IMPROVEMENTS AND PERSPECTIVE}

As a consequence of the fruitful discussions and collaboration during and after the Les Houches workshop, several improvements have been identified for the developments in \textsc{Delphes}. The $b$-jet algorithm will be improved by adding a dependence of the tagging efficiencies with respect to the jet transverse momentum. In current versions of \textsc{Delphes}, the probabilities for $b$-tag or mistag are uniforms, which underestimates the actual performances of the real experiments. The $\tau$-jet collection will allow $3$-prong $\tau$'s. Local detector inefficiency maps might also be implemented, as well as an improved interface to \texttt{HepMC} that would allow an easy connection to tools like Rivet~\cite{Buckley:2010rivet}.

\subsection*{CONCLUSIONS}
\textsc{Delphes} is a framework offering fast simulation tools allowing to investigate quickly new models and check their signatures in a realistic detection environment.
It offers an intermediate step between simplified \textit{parton-level} analysis and extensive analyses using the full-simulation power in large collaborations.

\subsection*{ACKNOWLEDGEMENTS}
X Rouby would like to thank S Schumann, F Moortgat and F Maltoni for their support in the participation to the Les Houches workshop. 
S Ovyn and X Rouby are happy to thank the numerous Les Houches participants who have shown their interest in \textsc{Delphes}, leading to very interesting and fruitful discussions and exchanges.



\clearpage

\section[EFFECT OF QED FSR ON MEASUREMENTS OF $Z/\gamma*$ AND 
$W$ LEPTONIC FINAL STATES AT HADRON COLLIDERS]{EFFECT OF QED FSR ON MEASUREMENTS OF {\boldmath $Z/\gamma*$} AND 
{\boldmath $W$} LEPTONIC FINAL STATES AT HADRON COLLIDERS
\protect\footnote{Contributed by: A. Buckley, G. Hesketh, F. Siegert, P. Skands, M. Vesterinen, T.R. Wyatt}}
\label{sec:zqed}

\subsection{Introduction}
Due to its simple quantum structure and clean experimental signature, quark-antiquark annihilation to lepton pairs (Drell-Yan), is one of the most important channels in hadron collider physics.
In this paper we shall discuss some requirements that must be met, at the
observable level, when using information obtained from the final-state
leptons in this process. 
The emphasis on ``observable level'' is particularly
important; if information about  
 states that \emph{are not themselves directly observable},
such as the  $Z$ boson or a ``bare'' lepton 
in a Monte Carlo event record, is used for correcting
or calibrating the raw measurement, then the precision of the 
corrected/calibrated result will suffer from an intrinsic ambiguity
which was not present in the raw measurement, i.e., the result will have been degraded. 

We discuss the cause and magnitude of this degradation
and argue that the experimental corrections and calibrations can
equally well be performed \emph{without} using such information, hence
preserving the full precision of the raw measurement in the calibrated
result. 

To illustrate our conclusions and to aid future studies, 
we examine various possible observable definitions and 
present a collection of reference comparisons for the Tevatron
processes $p\bar{p}\to Z/\gamma^* \to e^+e^-$ and $p\bar{p} \to
Z/\gamma^* \to \mu^+\mu^-$. These same concerns apply to the
equally important $W$ signal, which we also consider.

\subsection{Monte Carlo Truth}
Calculations of collider physics processes rely
heavily on factorizations of the full transition amplitudes (squared) into
smaller, more manageable, pieces. 
These factorizations are typically formally correct in limits 
in which one single Feynman diagram dominates over all others, in which case the
system has a well-defined ``semi-classical'' history, represented by the single
dominant diagram. The individual pieces can then be treated as
approximately independent since the quantum mechanical interference
between them can be neglected in this limit.

Essentially, what event generators, such as \herwig,
\pythia, and \sherpa, provide as intermediate particles 
in their event records (``Monte Carlo truth'') are representations of such 
semi-classical histories. At the very least, these histories show how the
generator approximated the process so far 
(which may, in turn, furnish important boundary conditions on
subsequent generation steps).  
But even in the best case scenario, when the diagram represented by
the event record really is the dominant one, it is still only exact
when all other diagrams are zero, i.e., in the semi-classical limit. 
Therefore, any procedure (such as a correction applied to an
experimental measurement) that depends on history-information 
is only well-defined in that limit. At the quantum level, 
it would suffer from an irremovable ambiguity which would set an
ultimate limit to the precision that could be obtained with that
procedure. 

Let us consider the Drell-Yan process in more detail. 
For this process, e.g., the event record of the \pythia 
generator will contain an intermediate $Z$ boson. 
This is intended to tell the user
that the lepton pair comes from an $s$-channel diagram (which in fact
includes the full \z\ interference), and it also lets \pythia's
 parton shower know that the invariant mass of the lepton
pair should be preserved in the shower.  So far so good.
If QED radiation was not an issue,
this intermediate $Z$ boson is identical to 
the sum of the final-state lepton pair, and it is
therefore tempting to use this boson, even in the presence of QED
effects, to define some kind of Monte-Carlo-truth / generator-level /
QED-corrected / ... ``\z''. 
However, when including QED effects, diagrams with photons emitted in the initial
state will interfere with diagrams that have them emitted in the final
state. Therefore, we cannot, at the amplitude-squared level,
determine whether a given photon was initial- or final-state
radiation, not even statistically. 
However, the two cases correspond to \emph{different} semi-classical
histories, with different ``\z'' kinematics, 
and therefore any semi-classical ``\z'' definition would at least 
be ambiguous up to the size of the initial-final interference terms. 

Another manifestation of the same basic problem is 
to do with the identification of which leptons came from the \z\
decay, in an event with multiple leptons. Again, a correction
procedure that would depend on Monte Carlo truth for this
identification would be fundamentally 
ambiguous at least up to interference terms
between diagrams with different assignments.

Finally, there is the issue of how to \emph{define} a lepton. 
Especially electrons, which are close to massless, 
emit large amounts of near-collinear photon bremsstrahlung. The
final-state electron in a Monte Carlo event record is therefore not a 
uniquely defined object (it depends, for instance, on 
if/how electron mass effects are implemented in that Monte Carlo). 
As we argued above, the electron ``before''
QED radiation is also not well-defined, 
since the electron interferes with other particles. And in any case it
would be a mere coincidence if either object bears much resemblance to
what might be called an electron in an experimental setting (to be
discussed further below). It is therefore also crucial to operate with
an unambiguous and collinear safe ``lepton definition''.

What we shall argue below is that such ambiguities are in fact both present
and large for precision measurements of current interest. It is therefore
important to reiterate that the full experimental
precision is only preserved if the observables and correction
procedures are defined entirely in terms of the final-state leptons,
with these in turn defined in a way that is collinear safe against
photon emission.


\subsection{Definitions}

In previous measurements of the kinematics of the Drell-Yan process, the reconstructed data have typically been corrected to the ``\z'' boson level. 
This correction process can be broken into three steps:
\begin{enumerate}
\item Correcting the measurement leptons to the level of particles entering the detector (``unfolding'' the detector resolution and efficiency).
\item Correcting from the particles entering the detector to the ``\z''.
\item Correcting from the measured phase space to 4$\pi$\ acceptance, with no selection cuts except on the ``\z'' mass.
\end{enumerate}

As outlined in the previous Section, Step 2 introduces ambiguities and dependencies on the model used to simulate  Drell-Yan and FSR.
Step 3 involves correcting for particles that were unmeasured, so explicitly depend on a model of those particles.
We therefore propose that {\bf all measurements be made available after Step 1: corrected only for detector resolution and efficiency}. 
Steps 2 and 3 may of course be performed in addition, subject to problems discussed above.

In order to perform Step 1, we must define the ``electron'' and ``muon'' observables in terms of the particles that enter the detector, accounting for the different detector response in each case.
We then consider how to form the ``\z'' from those electrons and muons, and briefly discuss missing energy and a ``\w'' final state.
We therefore propose these definitions:
\begin{itemize}
\item In simulated events, we consider all generator level particles with $c\tau>10$~mm as ``stable''~\cite{Buttar:2008jx} (i.e. could possibly be detected). 
Only the stable particles may be used to define an observable, and no checks should be placed on the internal generator origin of any particles (no history dependence).

\item Electrons: to mimic the response of a calorimeter, electrons and photons must be combined into an localized ``EM cluster'', and such clusters considered as particle level electrons.
Here, this clustering is done with simple cones of $\Delta R = \sqrt{(\Delta \phi)^2 + (\Delta\eta)^2}$=0.2 around any electron\footnote{This simple cone clustering is sufficient to reproduce observable electrons at the Tevatron experiments. For the LHC experiments, a different cone size or more advanced clustering algorithm (e.g. anti-$k_T$) may better reproduce the detector level observable. 
Ideally, the same particle level electron definition can be agreed upon for all LHC experiments.}.
Thus any narrow angle QED FSR is combined back into the electron, which may also pick up some additional electromagnetic energy from the underlying event.
Any wider angle FSR is ``lost'': i.e. not associated with the electron.

\item Muons: the stable muons (i.e. after FSR) can be considered directly, as would be measured in the tracking system of a detector. Thus, {\em all} FSR is lost, and the underlying event has no effect.

\item Missing transverse energy (MET): in data this is typically calculated by inverting the vector sum of calorimeter $E_T$  and muon \pt, introducing many detector acceptance and efficiency effects which are difficult to reproduce at particle level. 
To initiate discussion, we define the MET simply from the vector sum of all neutrinos in the event, which is compared to the \pt\ of only the  the \w\ neutrino in Fig. \ref{fig:met}.

\item \z: selections should follow the data analysis. Typically, this will mean considering all electrons or muons within a given $|\eta|$ range, then forming opposite charge pairs that lie within the required di-lepton mass window (e.g. 65-115~GeV). 
In the (rare) cases where two possible pairs pass these selections, the method used to select the ``best'' should again follow the data analysis, such as the pair closest to the $Z$\ mass.

\item \w: again, selections should follow data analysis. This will typically mean combining the MET with the highest \pt\ lepton, then placing some requirement on transverse mass. 
\end{itemize}

\begin{figure}[!htb]\center
\includegraphics[width=50mm]{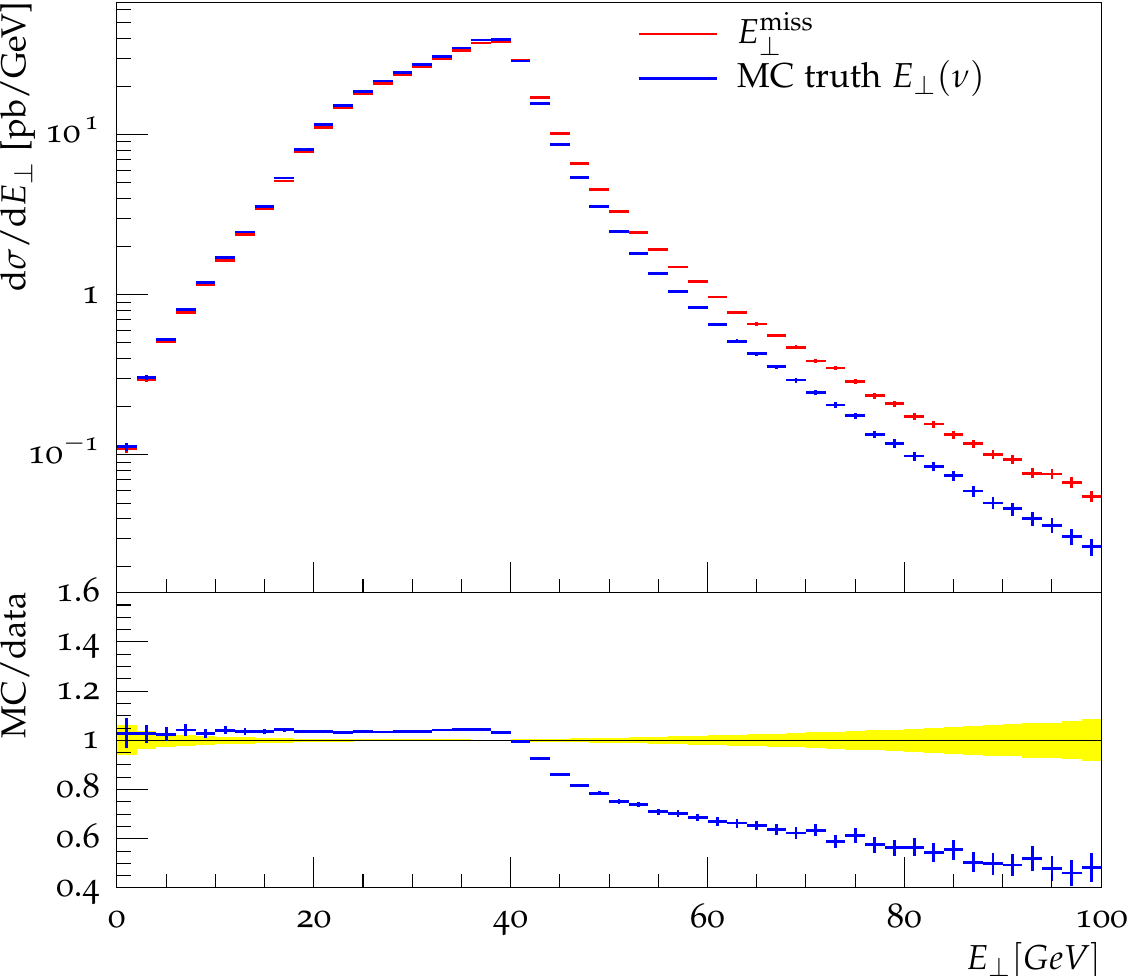}
\caption{\label{fig:met}Comparing the \pt\ of the neutrino from a \w\ decay to the particle level MET (defined in the text).}
\end{figure}

When defining the \z\ and \w\ observables, three further selections are typically applied in data analysis that have a less clear particle level analogue. Here we provide suggestions as the basis for future discussion:
\begin{enumerate}
\item Lepton isolation: reproducing detector level isolation requirements at particle level is difficult, whereas determining the efficiency loss due to such requirements in data is generally rather straightforward. So we suggest the measured data should be corrected for these efficiency losses, and no isolation conditions be imposed at particle level.
\item Minimum lepton \pt\ (and MET) requirements: such requirements should generally be implemented at the particle level also. 
However, if the effect is small it may be possible to use the simulation to extrapolate into the unmeasured region.
\item Event vetoes: for example, a second lepton veto in \w\ selection. Due to detector efficiency effects, such requirements are difficult to implement in a way that is consistent between detector and particle level, and should be treated with care.
\end{enumerate}

\subsection{Implications}
To study the impact of the definitions outlined in the previous section, some Tevatron examples are used.
Samples of $p\bar{p}\rightarrow\Z$, with \zee\ and \zmm, are produced using \pythia 6.421~\cite{Sjostrand:2006za}, using the \perugia\ 6 tune~\cite{Skands:2009zm} with the \cteq6L1 PDFs~\cite{Pumplin:2002vw}.
The samples are normalized to the same number of generated events.
For comparison, we also consider a ``generated \z'', reconstructed by searching the \pythia\ event record and explicitly taking the two leptons from the \z\ decay, before FSR.
Regardless of the definition used, the leptons used to make a \z\ are required to have $|\eta|<1.7$\ and a dilepton mass between 65 and 115~GeV.

\subsection{FSR Properties}
First, the properties of photons from FSR are considered. 
Figure \ref{fig:photons} shows, for \zee\ and \zmm, the rates at which FSR photons are emitted from the \z\ decay products. 
Next,  $\Delta R$\ between FSR photons and the nearest \z\ decay product.
Then, the FSR photon energy and the energy of ``unmeasured photons'' is shown; in the case of muons, all photons are considered unmeasured; for electrons, photons outside the cone of 0.2 are considered unmeasured.
Finally, the fraction of all photons within $|\eta|<2.5$~(typical electromagnetic calorimeter coverage) in the final state which arise from FSR for different \pt\ requirements. 

\begin{figure}[!htb]\center
\includegraphics[width=30mm]{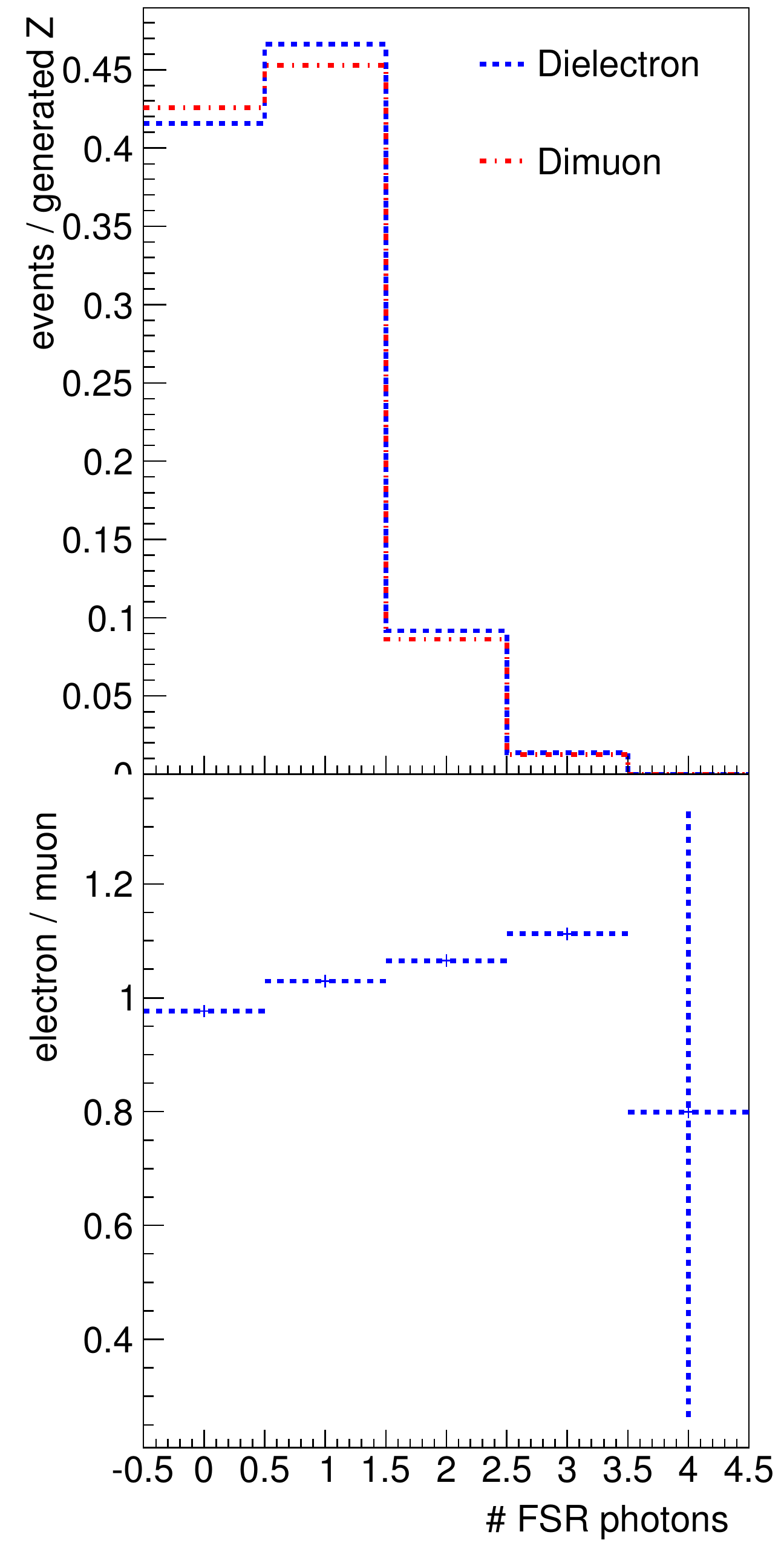}
\includegraphics[width=30mm]{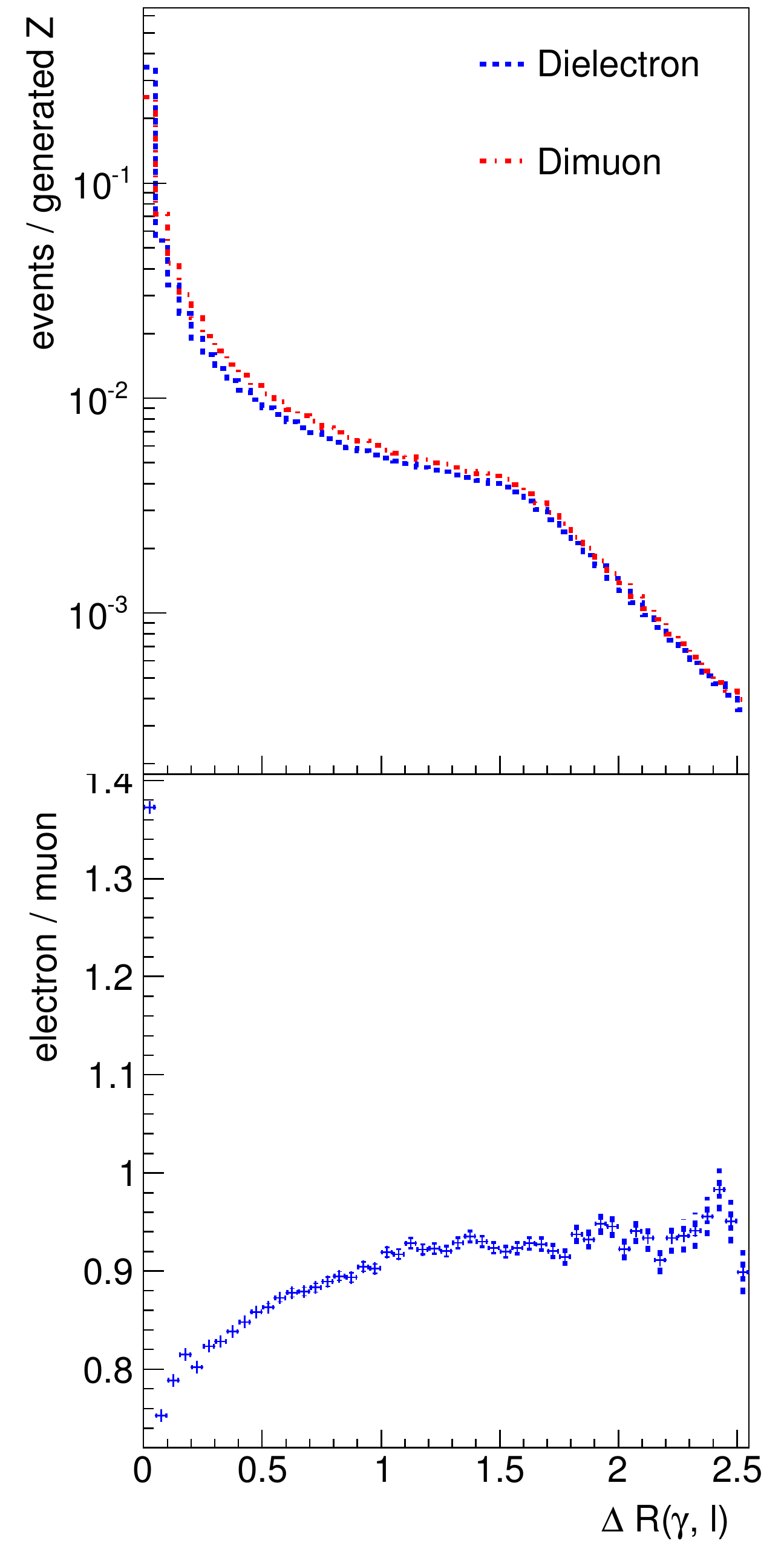}
\includegraphics[width=30mm]{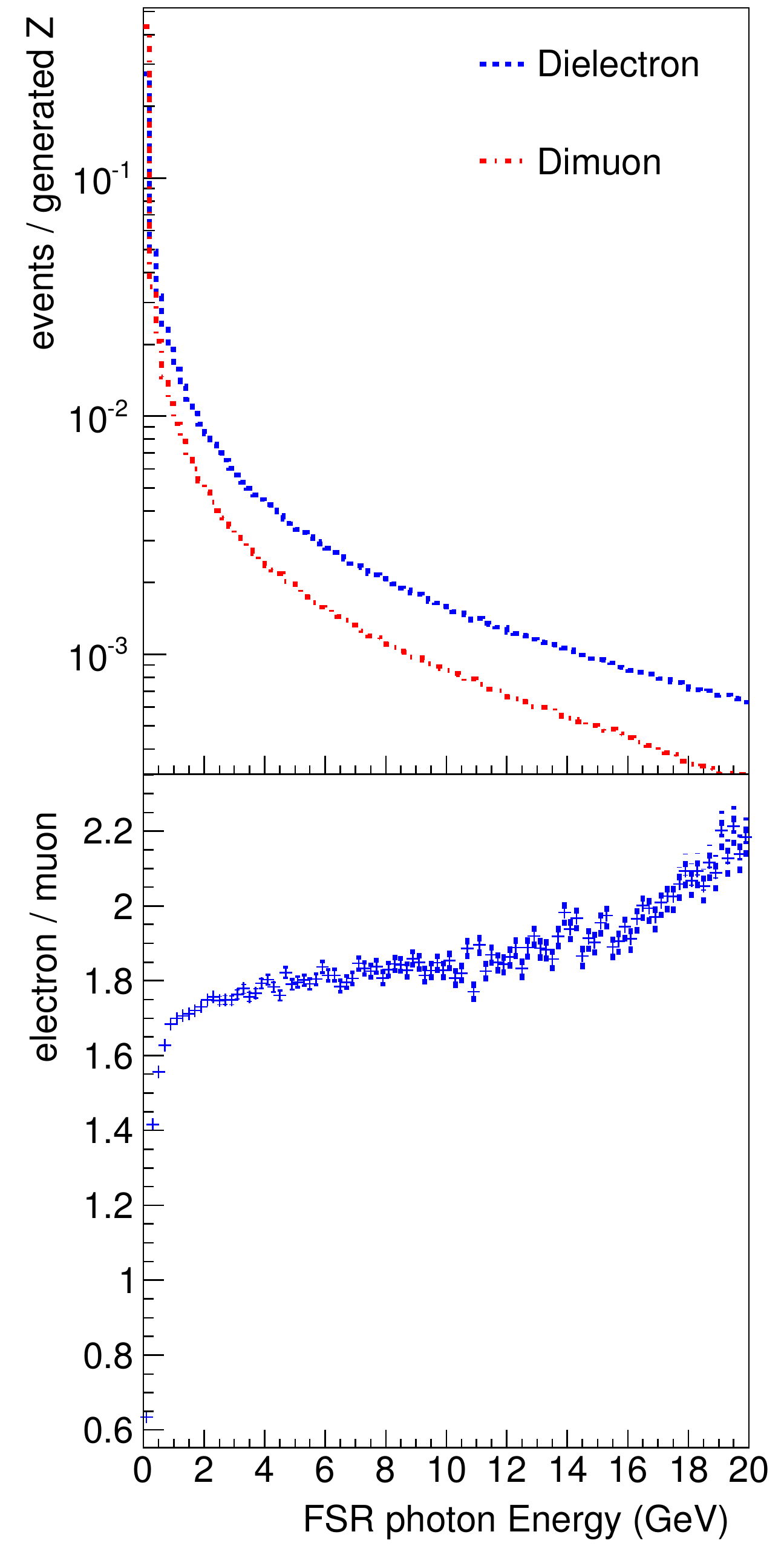}
\includegraphics[width=30mm]{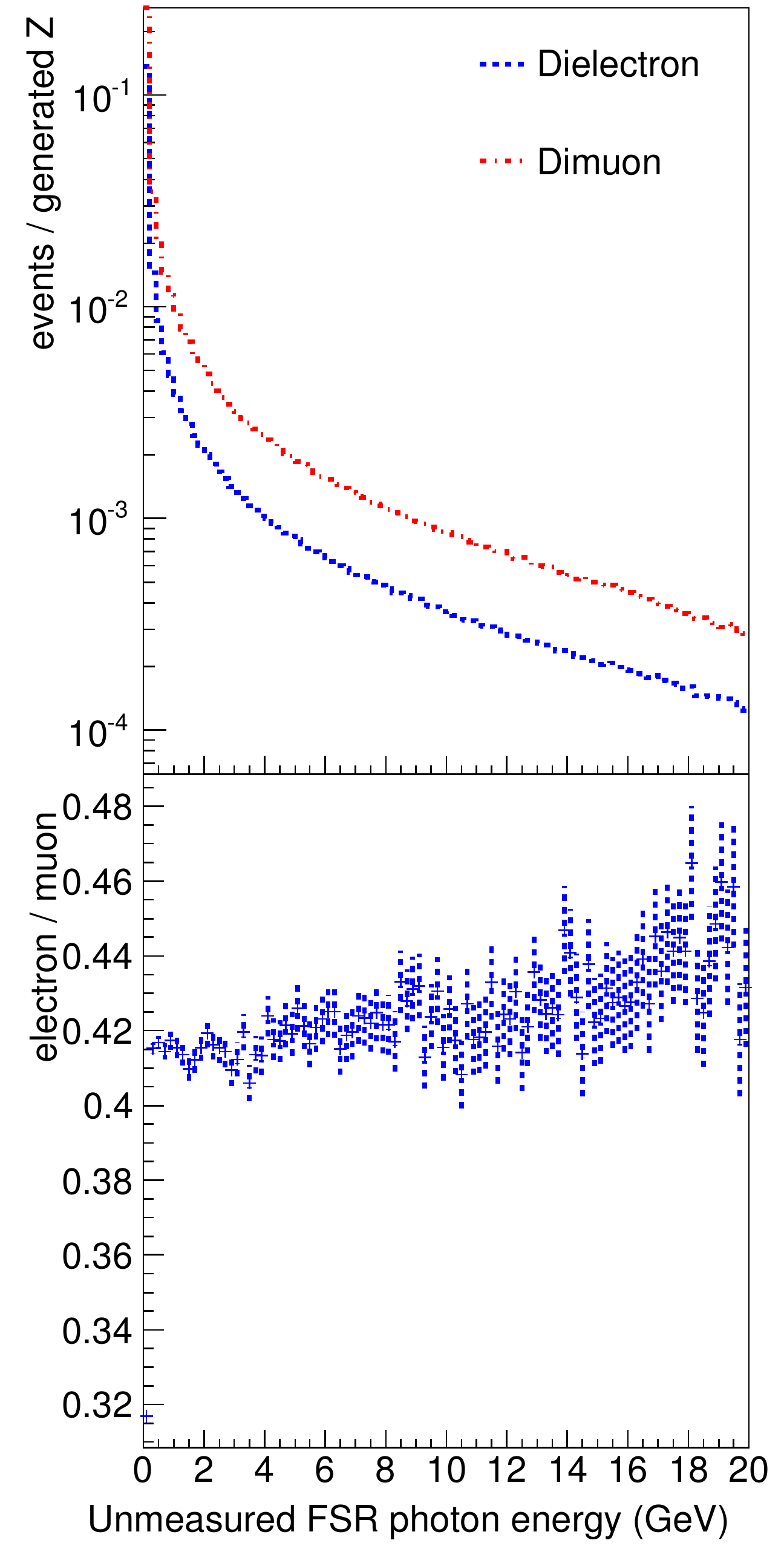}
\includegraphics[width=30mm]{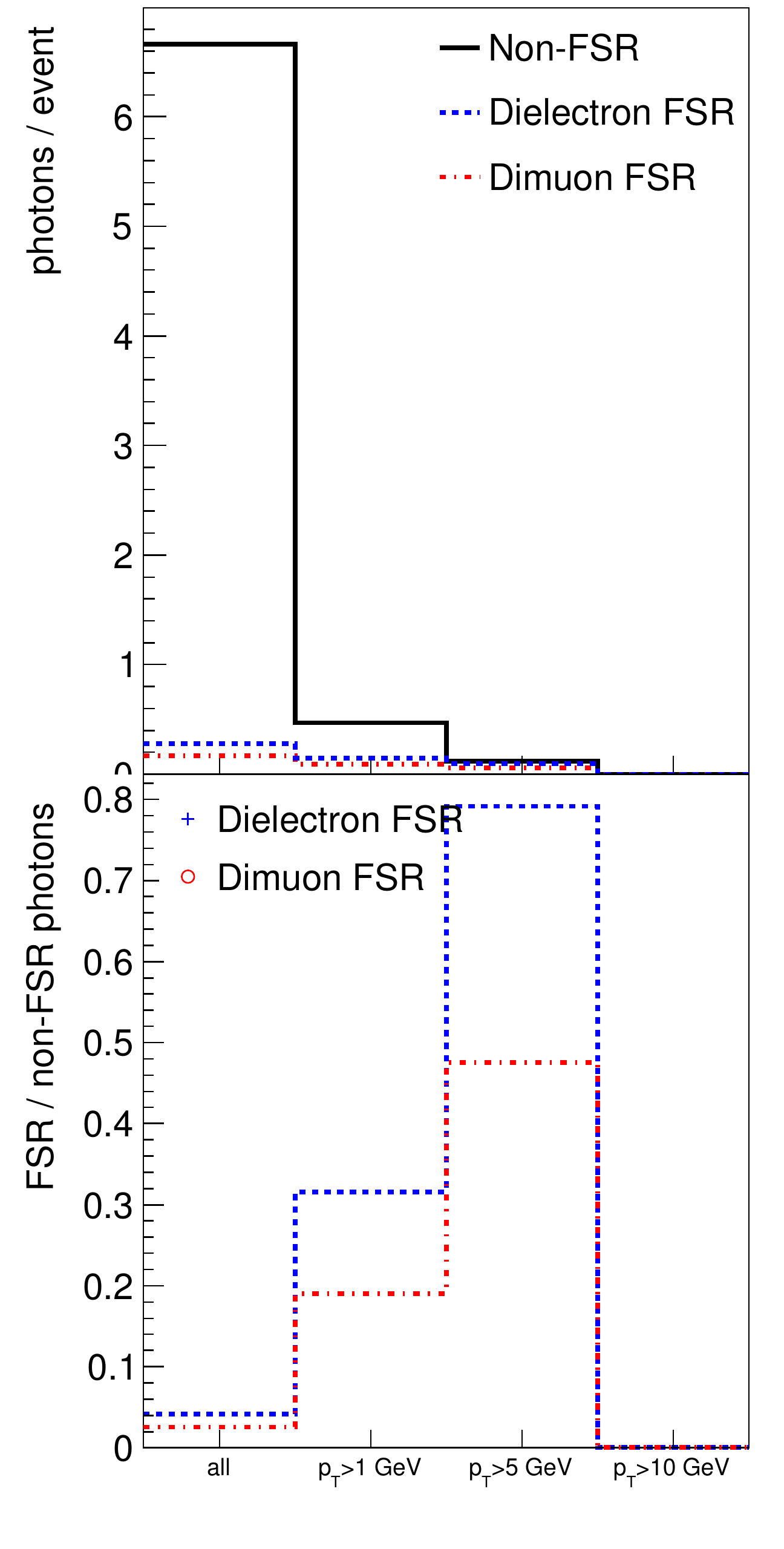}
\caption{\label{fig:photons}Left: the rate of FSR photons from the Z decay products; centre left: $\Delta R(\gamma, l)$; centre: FSR photon energy; centre right: energy of ``unmeasured'' FSR photons (see text); right: the fraction of all photons that arise from FSR, at different energies.}
\end{figure}

It can be seen that the rate of FSR is significant.
In \zee, the photons are typically close to the electron and will be added to the electron cluster, however (using the default \pythia\ settings) approximately 23\% of \zee\  events will still contain at least one lost FSR photon (compared to 66\% of \zmm\ events).
These photons are typically of very low energy, though approximately 3\% of \zee\ and 7\% of \zmm\ events contain an unmeasured FSR photon above 5~GeV.
Identifying these lost FSR photons is almost certainly experimentally impossible, and it also appears impossible at the level of stable particles (without looking into the generator event record), due to the presence of photons from other sources (primarily $\pi^0$\ decays).
In the following sections, we consider how these effects result in differences between the generated \z and the measured dilepton final state, for some important kinematic properties.

\subsection{Energy Scale and Resolution, and Total Cross Section}

The energy scale and resolution for both the electromagnetic calorimeter and tracking system are typically derived by fitting the shape and position of the \z\ peak.
Here, we assess the impact of QED FSR on the measured \z\ line-shape, and hence possible effects on detector calibration.
Figure \ref{fig:zmass} shows the generated \z\ mass distribution, and that reconstructed using the stated electron and muon definitions. 

It can be seen that FSR has a drastic effect on the observable line-shape below the pole.
Obviously calibrations should ensure the detector response reflects the observable leptons, and not ``correct'' them to the generated \z, so using the lower side of the mass peak would clearly rely upon a model of FSR.
This introduces model dependence which may complicate the extraction of model independent quantities from data analysis.
However, the energy scale may possibly be determined with minimal model dependence by fitting only the peak position, which is mostly unaffected by FSR. 
Similarly, it may be possible to determine lepton resolution in a model independent way from the higher edge of the mass peak, which is also mostly unaffected. 

\begin{figure}[!htb]\center
\includegraphics[width=40mm]{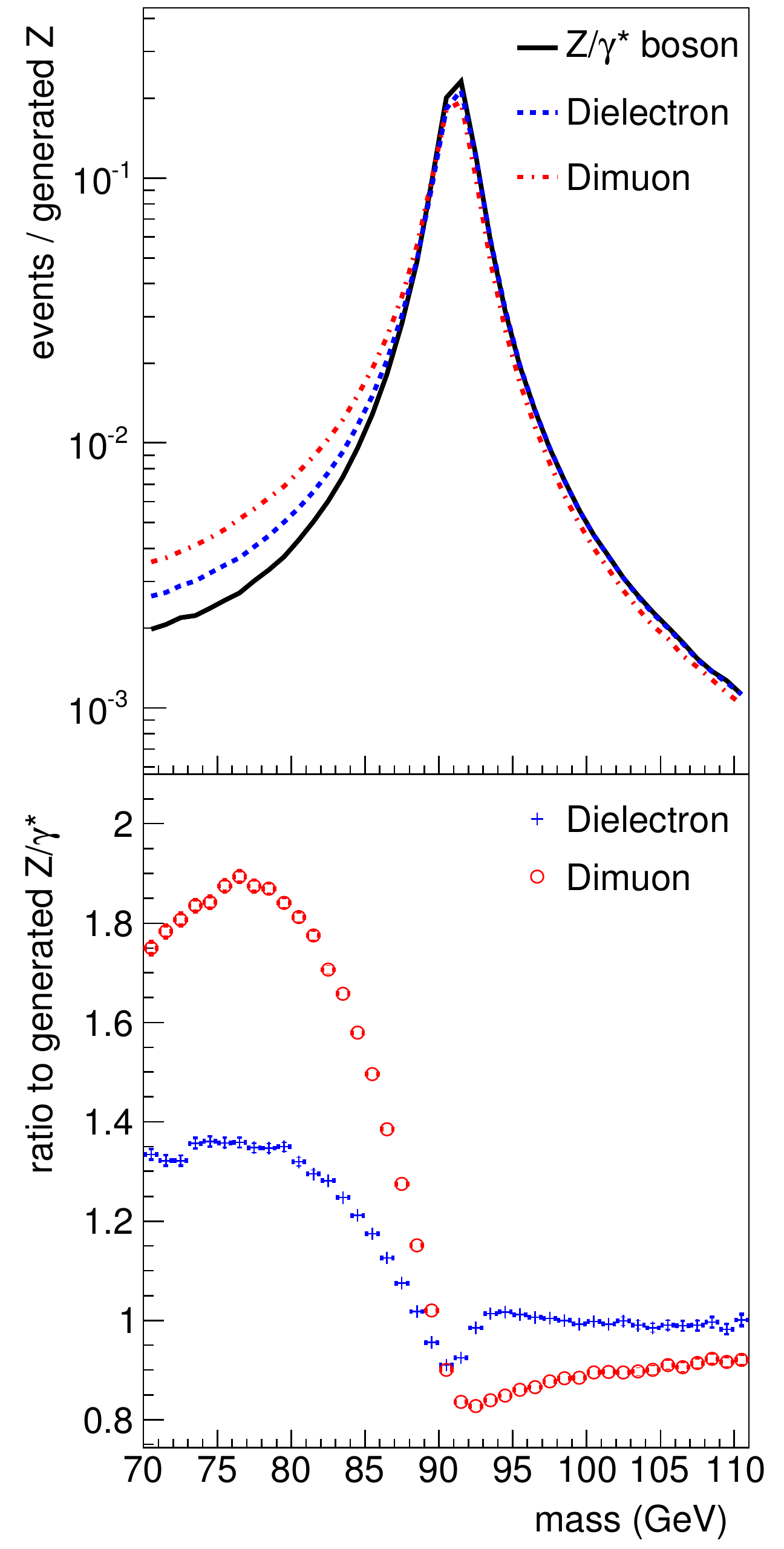}
\includegraphics[width=40mm]{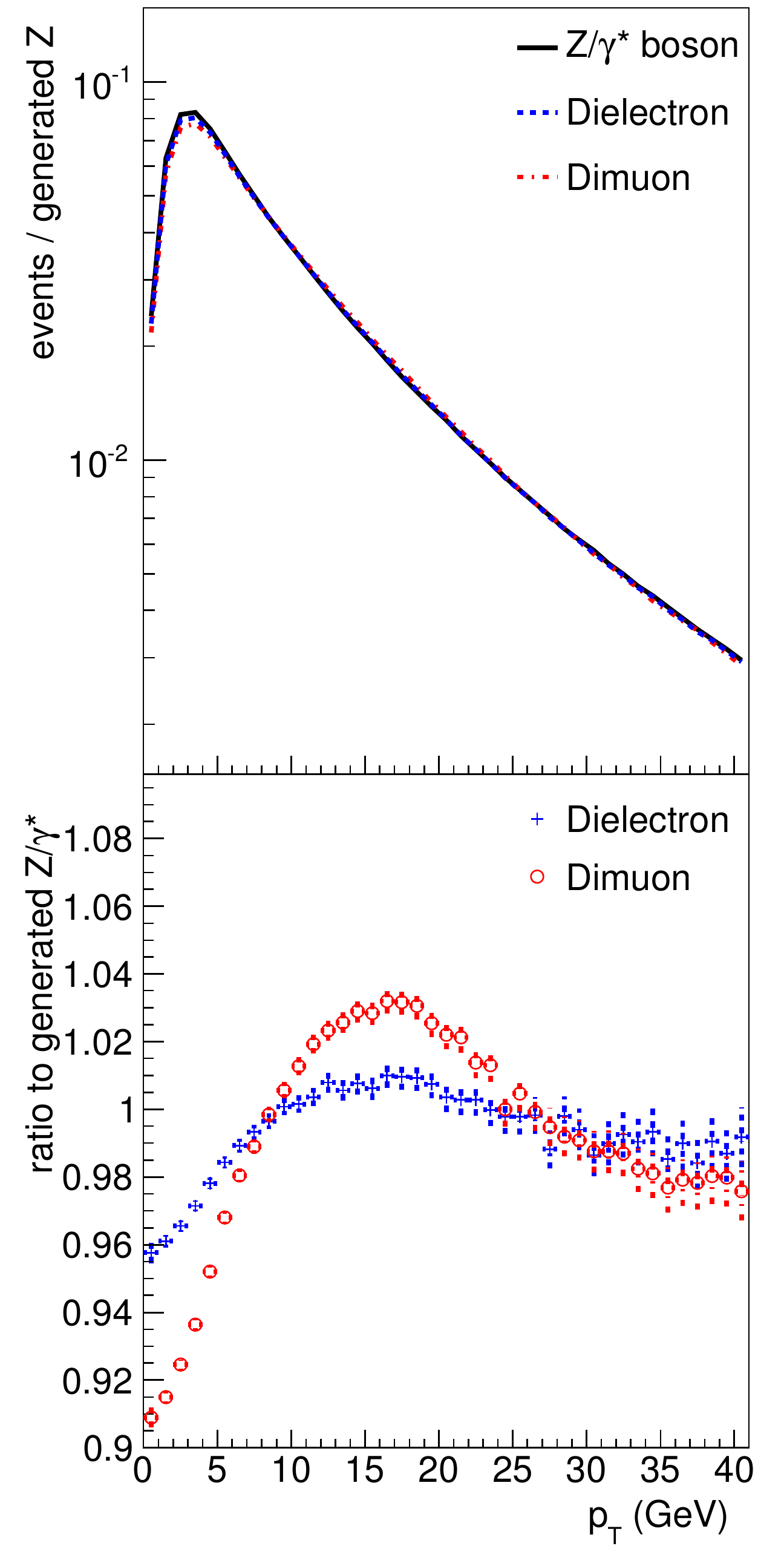}
\includegraphics[width=40mm]{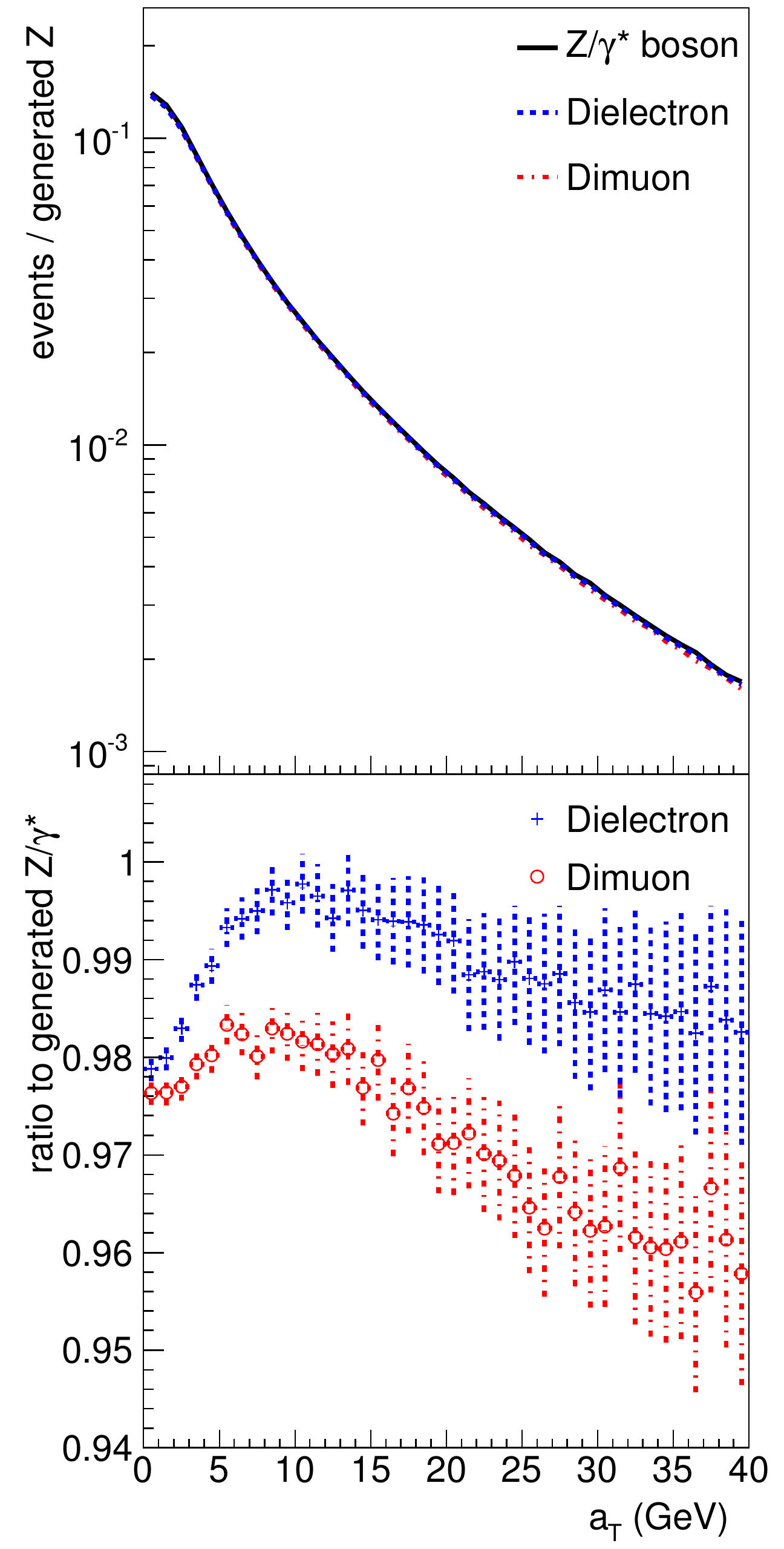}
\caption{\label{fig:zmass}Comparing the generated \z\ to the observable (defined in the text). Left: the \z\ mass; centre: the \z\ \pt; right: the \z\ \at.}
\end{figure}

The effect of FSR on the \z\ line-shape has another implication: the inclusive Drell-Yan cross section is typically measured for a given mass range, for example 65-115~GeV.
Energy lost in FSR causes the dilepton mass to be below the  ``true'' \z\ mass, which means events above the mass window may migrate in, and events in the window migrate out.
In the \pythia\ samples considered, this causes a 0.9\% net loss of dielectron events, and a 2.1\% net loss of dimuon events for the stated mass window.

\subsection{ {\boldmath \z\ \pt}\ and {\boldmath \at}}

We next consider the impact on the \z\ \pt\ and \at~\cite{Ackerstaff:1997rc, Vesterinen:2008hx} distributions, important variables in measuring non-perturbative QCD form factors and in generator tuning. 
Figure \ref{fig:zmass} shows the observable \z\ \pt\ and \at\ compared to the true \z\ \pt\ and \at.
For \pt, effects of up to 4\% are seen in the dielectron channel, and up to 10\% in the dimuon channel. 
Effects are around 2\% for \at, where the shape effect is larger in the electron channel.
For both \pt\ and \at, the effects are largest at low values, the crucial region for generator tuning and determining form factors.

\subsection{Recovering Published Measurements?}

Several Tevatron measurements of the \z\ \pt\ have corrected back to the ``generated'' \z, with full 4$\pi$ acceptance~\cite{Affolder:1999jh, Abbott:1999yd, :2007nt}. 
Similarly, it is common practice in theory papers to present results at the same level (see a recent example in Ref.~\cite{Siodmok:2009ae}).
We perform a quick study to determine if such measurements and calculations can be approximated using only the stable final state particles, rather than having to resort to looking into the generator event history. 
This is done by using a wider cone around the final state electrons, to attempt to re-sum more FSR and get closer to the leptons directly from the \z\ decay.
Figure \ref{fig:cones} shows the effect of using a cone of radius 0.2, 0.5 and 1.0.
While the cone size of 0.5 comes closer to the ``true'' \Z\ \pt\ in the low \pt\ region, it moves further away at high \pt. 
Increasing the cone size to 1.0 clearly distorts the spectrum, with too many non-FSR photons being added to each electron.
Similar behaviour is observed for \at.

\begin{figure}[!htb]
\begin{center}
\includegraphics[width=40mm]{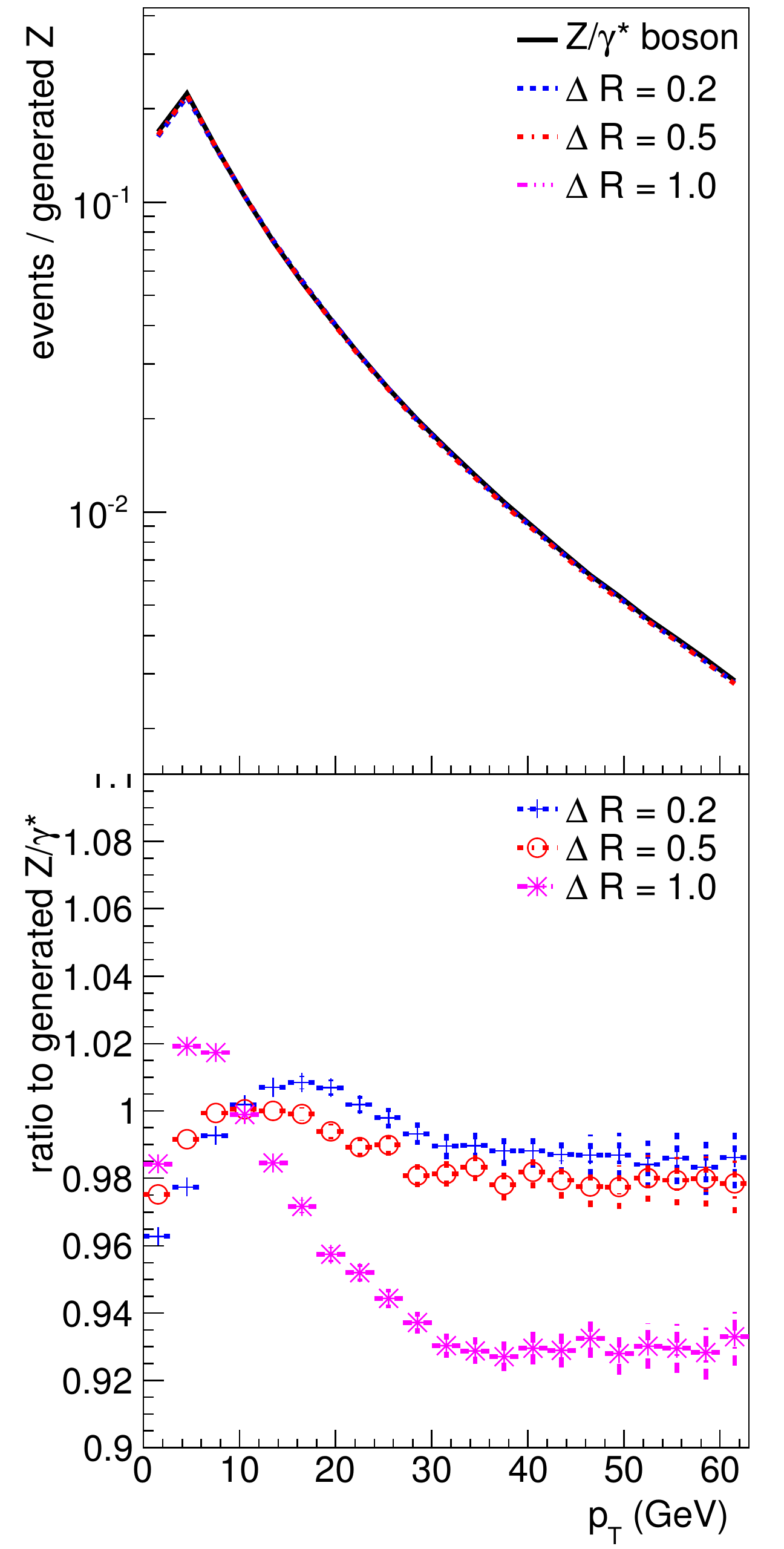}
\includegraphics[width=40mm]{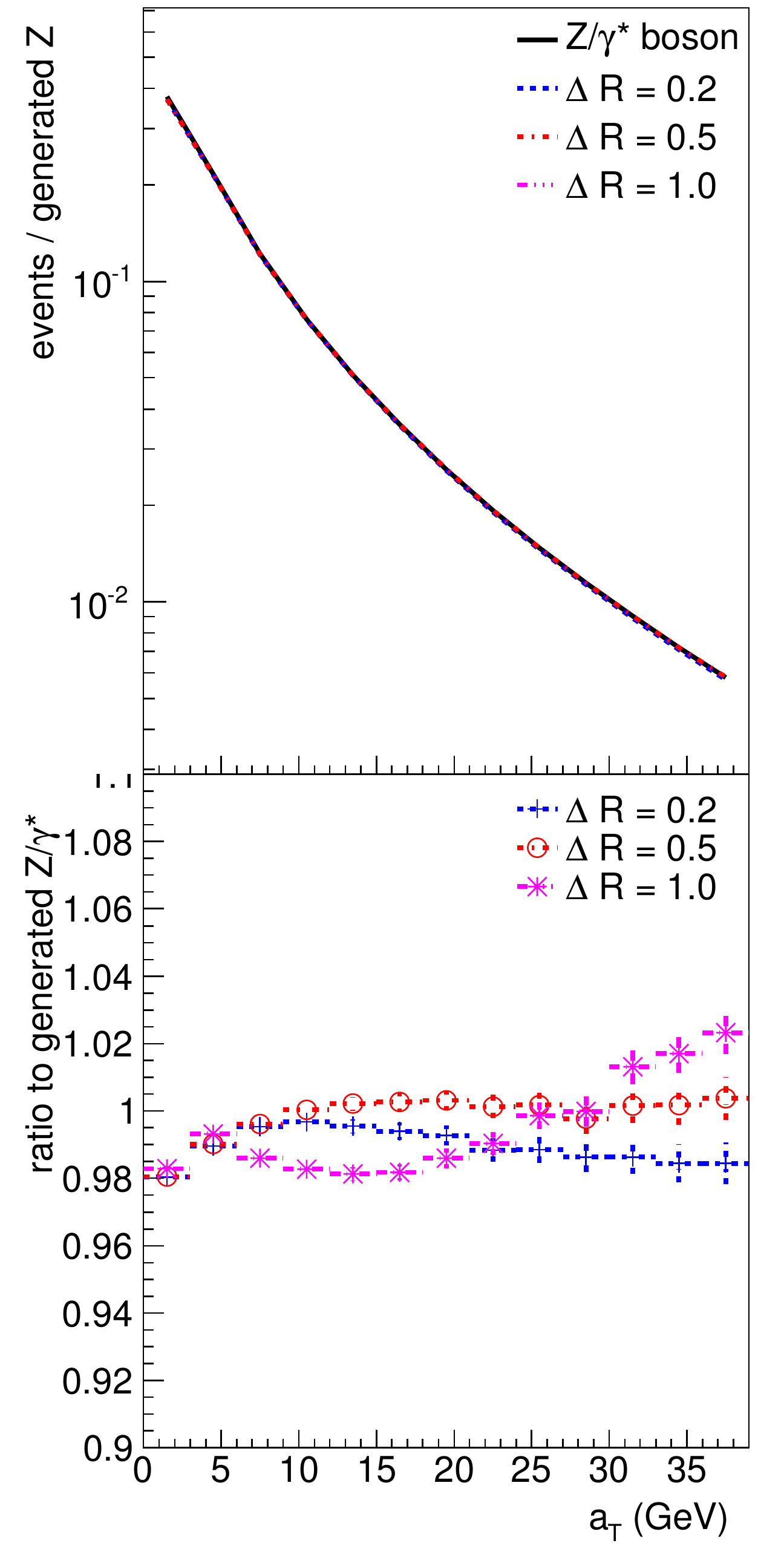}
\end{center}
\caption{\label{fig:cones}The effect on the \z\ \pt\ (left) and \at\ (right) of changing the electron cluster size.}
\end{figure}

This simple study shows the difficulty in reproducing the generated \z\ from only the stable particles, and the 
precision to which such ``boson level'' results can for example be implemented in \rivet 
(see Section~\ref{sec:rivet} and used for generator tuning with \professor~\cite{Buckley:2009bj}.

Fully reproducing published \z\ \pt\ results may only be possible by mimicking the data treatment: applying 
a ``FSR correction factor'' to the spectrum reconstructed from the stable particles. Such treatment 
introduces similar ambiguities and model dependence as the original data analysis which corrected back to the ``\z''.

\subsection{Combining Electron and Muon Channels}
Experimentally, combining electron and muon channel measurements is considered a simple way to increase the statistics used in a measurement, and benefit from the fact that many experimental systematics are uncorrelated between the two channels.
However, we have shown that the electron and muon observables are fundamentally different in both the measured cross section and reconstructed kinematics are different in each channel. Further complications are introduced by the different acceptance of calorimeters and muon detectors.
These factors degrade the value of a combination, for which the same observable must be extracted from each channel, and it is impossible to do this without relying on a model of FSR.

If a combination must be done, a ``minimally model dependent'' method would be:
\begin{enumerate}
\item Restrict electron and muon measurements to the same phase space.
This will typically mean excluding some region from one of the channels, due to different coverage of muon detectors and calorimetry.
\item Apply a theoretical factor to the muon channel measurement, to correct from the observable muons to an equivalent of the electron observable: in this study, stable muons + all photons and electrons in a 0.2 cone around each muon. This factor should be published along with the combination.
\end{enumerate}

In simulated events, there are then two options: apply the same theoretical factor to simulated \zmm\ as used for the muon data, then combine;
or, directly treat stable muons as electrons (clustering EM energy around each muon).
Figure \ref{fig:combining} compares the \z\ \pt\ and \at\ for the muon channel calculated in this direct way to the electron channel, and good agreement is seen.

\begin{figure}[!htb]\center
\includegraphics[width=40mm]{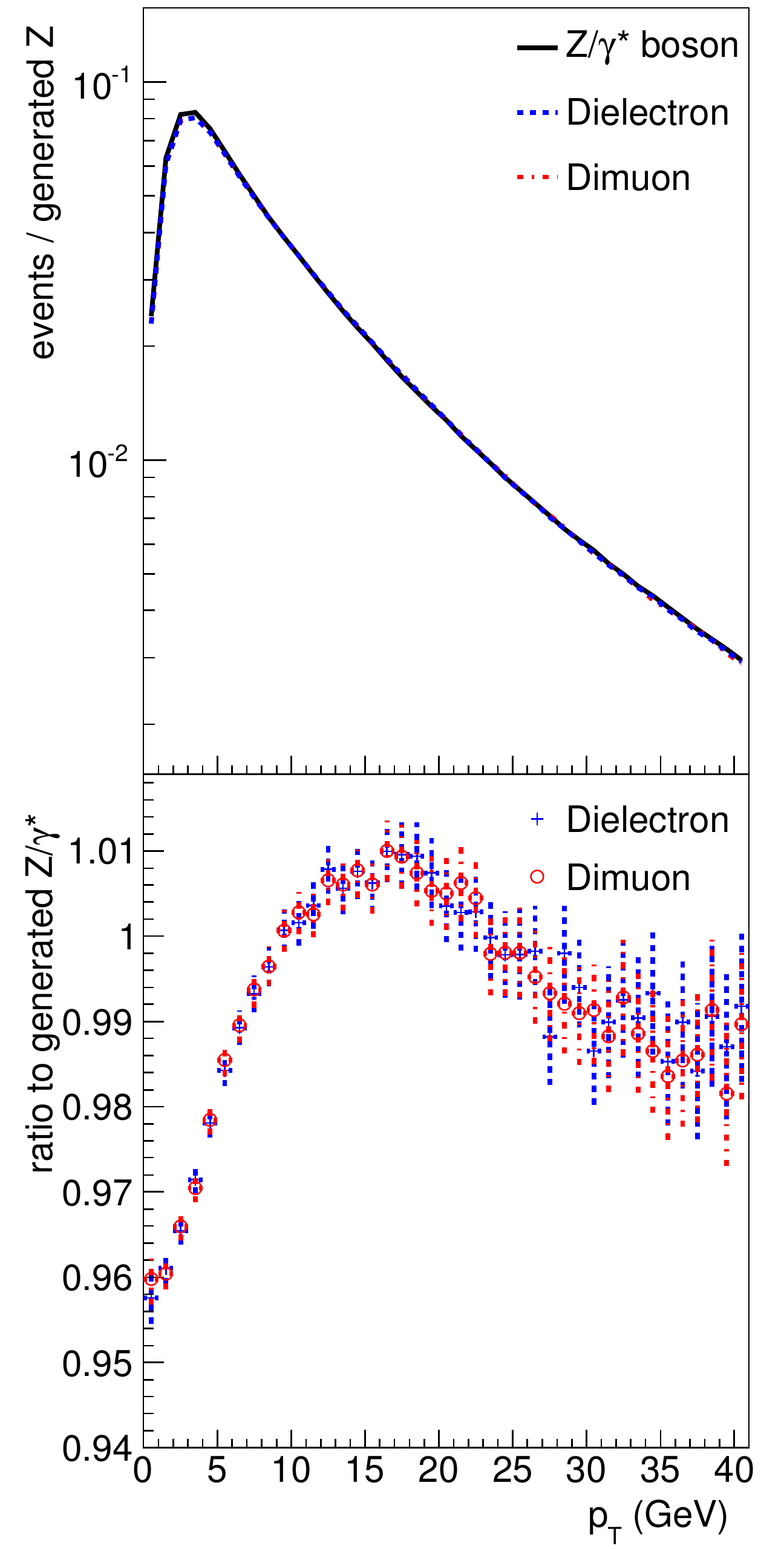}
\includegraphics[width=40mm]{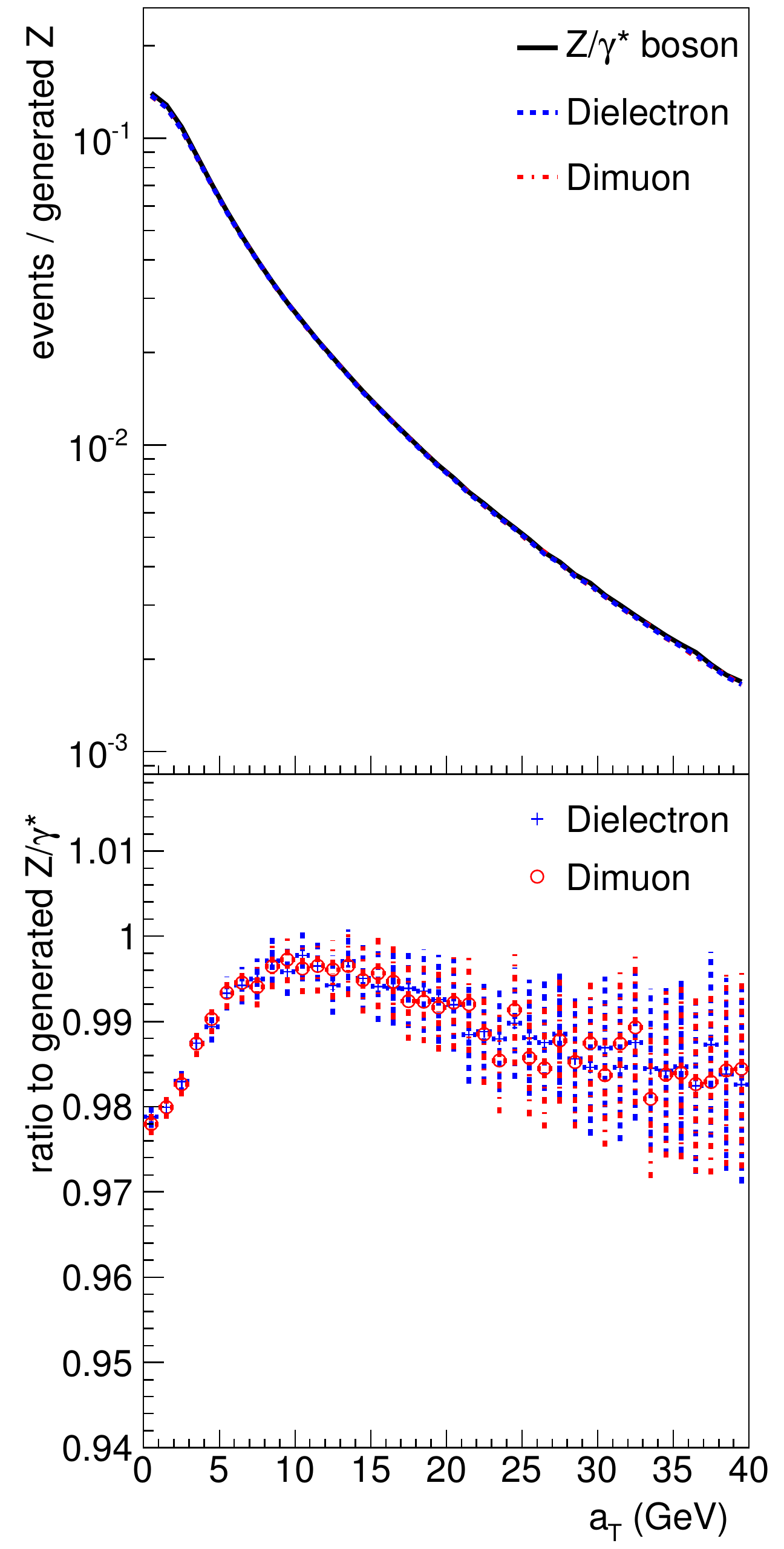}
\caption{\label{fig:combining}The \z\ \pt\ (left) and \at\ (right) using the same treatment of electron and muon final states.}
\end{figure}

\subsection{Conclusion}
We advocate that all future experimental measurements and theoretical predictions of the \z\ and \w\ (and other leptonic final states) should be presented in terms of observables, and not extrapolated beyond the measured acceptance.
To this end, we propose definitions of observable electrons, muons and the \z, as well as missing transverse energy and the \w\ boson, all at the level of particles entering a detector.
These definitions mimic the quantities measured in a detector, and provide an unambiguous, model independent basis to compare data and theory.

While it is essential to provide the data in terms of observables, it may still be desirable to derive further theoretical corrections for FSR and acceptance, for comparisons to previous results. 
We recommend such correction factors be provided in a table, rather than being applied to the data.
Using this table, (the inverse of) such corrections could also be applied to a calculation which does not include FSR or acceptance requirements, to allow direct comparisons to the data while maintaining the separation of measurement and theory.



\clearpage

\part[JETS AND JET SUBSTRUCTURE]{JETS AND JET SUBSTRUCTURE}

\section[STATUS OF JET SUBSTRUCTURE STUDIES]{STATUS OF JET SUBSTRUCTURE STUDIES
\protect \footnote{Contributed by: J. M. Butterworth, M. D. Schwartz}}

\subsection{INTRODUCTION}
 
In the 2007 Les Houches workshop~\cite{Buttar:2008jx} there was quite some discussion on how to define jets in ways 
which allow for clear comparison between experiments, and between measurements and state-of-the-art theory. 
The importance of such discussion has been emphasised since both by the adoption of new jet finders by \atlas and \cms
for their first data (in particular the anti-$\kt$ algorithm~\cite{Cacciari:2008gp}), and by the rapid expansion of 
activity aimed at exploiting the internal properties of jets in studies and searches for new physics at the \lhc.

Jet substructure has been studied at previous experiments as a means of investigating QCD. In fact jet shapes and 
subjet multiplicities have been used to measure the strong coupling $\alpha_s$~\cite{Chekanov:2004kz,Becher:2008cf}. What leads 
to a huge increase in interest at the \lhc is the same fact that makes the \lhc such an exciting project in 
general - the energy reach is substantially higher than the electroweak symmetry-breaking scale.
One consequence of this which has been well appreciated for some time is that large multiplicities of 
electroweak-scale objects may be produced (e.g. $W$,$Z$, jets with $\pt \approx {\cal O}(100)$~GeV). This has been
a spur to a large amount of theoretical activity on higher order calculations and parton-shower-matched Monte 
Carlo simulations which has been a feature of many workshops, including this one.

A consequence of the high energy of the \lhc 
which was more slowly appreciated, but which is now quite widely understood, is that objects with 
masses around the electroweak scale may be produced well above threshold, and thus be highly boosted. When
boosted objects
decay to hadrons,  the decay products will be collimated and may emerge as a single jet with distinctive 
substructure, even if the decay is initially to two or more quarks or gluons.
It is easy to understand how such a situation will complicate the mapping
from jets to partons which is the basis of almost every phenomenological study involving jets.
In addition to the merging of jets due to high boosts, 
at the \lhc, there will be unprecedented
contamination from the underlying event and pileup, further deforming the naive jet-to-parton map. 
The basic goal of much of the work in jet substructure has been to improve the mapping from jets to partons,
so that it can remain useful at the LHC, in light of new classes of signatures (like boosted object decays)
and the extra contamination.

Jet substructure techniques were used first in studies of diboson production 
for identifying the hadronic decays of vector bosons in diboson and SUSY 
production~\cite{Seymour:1993mx,Butterworth:2002tt,Butterworth:2007ke}, and in the 2007 
workshop there was much discussion on boosted-top reconstruction, for example~\cite{Brooijmans:2008se}.
In the last two years, techniques have been developed and applied to several channels including 
top production~\cite{Kaplan:2008ie,Thaler:2008ju,topnote} as well as 
Higgs~\cite{Butterworth:2008iy,higgsnote,Plehn:2009rk,Kribs:2009yh} and 
neutralino~\cite{Butterworth:2009qa,susynote} 
searches, and 
several proposals have 
been made to use substructure information to improve jet mass resolution in 
general~\cite{Ellis:2009me,Ellis:2009su,Krohn:2009th} using substructure. Many of these 
techniques have been investigated by the experimental collaborations, using realistic detector 
simulations, and survived the test.

Another feature of the \lhc relevant to jet substructure is that
the calorimeters and trackers are superior to those at previous machines, allowing access
to more detailed information about the jet itself. This information should be utilized in a more refined 
substructure analysis, but is only starting to be exploited. As the physical information from particle flow
or topo-clusters becomes better understood, the power of jet substructure analyses will certainly improve.
It is not surprising that a next-generation machine requires next-generation analysis tools, and the work on
jet-substructure has just begun to probe the possibilities.
Jet substructure is a rapidly moving field, with at least two dedicated workshops held since the meetings in Les 
Houches~\cite{boosts} and more planned. Here we give a snapshot of the current status, focused 
(but not exclusively) on some developments which took place in this workshop.

\subsection{GENERAL FEATURES}
In QCD, the evolution from a hard parton to a jet of partons takes place in a regime where the 
energy scale is high enough to use perturbation theory, $x$ is not very small, and collinear logarithms are large.
This is essentially the kinematic region where DGLAP evolution should apply. This is largely understood within QCD, 
and can be calculated. It forms the basis of the parton-shower models implemented in the most widely used 
MC generators. Since the phase space between the jet $\pt$ and the confinement scale is large, there is
plenty of radiation, and parton multiplicities will be large. Confinement enters the picture at the 
hadronization (non-perturbative) stage, and is expected to have a small effect (at the sub-GeV level). In general,
the aspects of QCD relevant for jet substructure seem to be well-modelled 
by tuned Monte Carlo simulations, as is borne out by the available 
data~\cite{Chekanov:2004kz,Abazov:2001yp,Acosta:2005ix,Abbiendi:2004pr,Abbiendi:2003cn,Buskulic:1995sw}.

There are two somewhat distinct goals of the recently developed techniques. One is to improve the
single jet mass resolution, and the other is to distinguish jets originating in
heavy object decays from those coming from pure QCD  backgrounds.
The first of these goals requires a careful definition, in line with the discussion of jets in the last 
workshop~\cite{Buttar:2008jx}. Jet mass must be defined in terms of infra-red and collinear safe variables 
on the one hand, and in terms of physical observables on the other, to avoid making completely model-dependent 
``measurements''. However, the jet mass resolution may be optimised using models, in particular models which 
implement or mimic jets from the decay of a colour-singlet, by switching off initial state radiation; and models 
which mimic jets from parton (rather than proton) collisions by switching off underlying event.
The second goal, background suppression for heavy particle decays, generally relies on the fact that in QCD the various splittings 
within the jet are expected to be strongly ordered in scale, with large asymmetric splittings, whereas partons
resulting from a heavy particle decay will in general share the energy equally in the particle rest frame, 
leading to more symmetric configurations.

\subsection{FILTERING}
Let us begin with a discussion of what shall refer to generically as {\it filtering}.
Filtering refers to the removal of components of a jet in an
attempt to clean up the jets in a certain way. Variations on this theme have appeared, such
as {\it pruning}~\cite{Ellis:2009me} and {\it trimming}~\cite{Krohn:2009th}. The basic idea of filtering is that
one first finds a jet which is assumed to contain the radiation one is interested in as well as contaminating 
radiation,
then one removes the contamination with the filtering step. 
The difference in the algorithms is more a 
difference in how
the authors are imagining their routines used than in the way the routines themselves work. 
So we will organize the 
discussion
of filtering by application. 
Some approaches are application-specific, optimized to find a particular signal, such as 
boosted Higgs or boosted
top decays, while other applications are more general. We will be begin with the specific ones.

\subsubsection{FILTERING FOR SPECIFIC APPLICATIONS}

The most productive application of filtering so far has been in the search for heavy boosted objects which decay 
to standard model jets. Consider first a Higgs boson decaying to $b\bar{b}$ in associated production 
with a $W$. This channel is the discovery
channel for a light ($\sim 120$ GeV) Higgs at the \tevatron, but is very difficult at the \lhc due to much 
more complicated
backgrounds, in particular $\ttbar$ production. 
Going to a regime where the Higgs is boosted ($\pt > 200$ GeV) has the 
effect of essentially
eliminating the $\ttbar$ background and removing a lot of the $W$+jets background as well. In this case, 
the $b$-jets from
the Higgs decay may be within $\Delta R=1.0$ of each other, so finding separate $R=0.7$ jets may fail. The 
substructure
approach pioneered in~\cite{Butterworth:2008iy} was to find the jets together as a fat jet, with a cone size 
of $R=1.0$ or greater, then to parse
the jets to find the $b$-quarks as subjets.

To be specific, in~\cite{Butterworth:2008iy}, the authors use the Cambridge/Aachen 
algorithm~\cite{Dokshitzer:1997in,Wobisch:1998wt} to construct the
fat jet. This algorithm merges constituents
based on the distance $\Delta R = \sqrt{(\Delta \eta)^2 + (\Delta \phi)^2}$ between them. Then they undo the 
clustering steps using
the clustering history (conveniently stored by \fastjet~\cite{fastjet} within the jet object). Somewhere in the 
declustering the $b$ should
be separated. However, the way Cambridge/Aachen works does not guarantee that the the final merger is of the 
two $b$-jets into the fat jet.
Therefore, when tracing back through the declustering, one needs a different criterion, for example, 
looking for large
mass drop -- the masses of the subjets should be much smaller than the mass of the fat jet. Once a large mass 
drop is found, the process is stopped. All constituents of the jet which would have been split at this
stage are retained, and the clustering algorithm is rerun, using the angle between the centres of the
last two subjets to set a (smaller) angular scale. This effectively optimizes on an event-by-event basis
the selection of final state radiation from the (color singlet) Higgs candidate, while rejecting radiation 
from elsewhere, including the underlying event. This is the step described as ``filtering'' in the analysis.
The top two or three subjets are combined to reconstruct the Higgs mass. 
The possible inclusion of a third jet, rather than just two (representing the $b$
and $\bar{b}$) allows for an extra hard gluon emission. This final step
is critical in getting
the algorithm to perform well at the Higgs mass reconstruction. The result was confirmed by an \atlas 
study~\cite{higgsnote}, and
revived $H\to \bbbar$ 
as a discovery channel
for the Higgs at the \lhc.

Another application of filtering has been in hadronic boosted top decays~\cite{Kaplan:2008ie}. Here, the same basic principle is 
applied: a fat jet is found, subjets are located through declustering, and a filter is applied to 
remove unwanted radiation. In this case, 
the three or four hardest
subjets are filtered out. Three of the jets should represent the hadronic top decay products: the $b$-jet and the 
two quark jets from the $W$ decay, the fourth may be an additional hard gluon. Using this approach, the authors 
found 99\% background rejection efficiency at 
40\% signal efficiency, {\it per} jet. Therefore signal/background for boosted hadronic $\ttbar$ events over 
dijets is enhanced
by a factor of 1000. A \cms study confirmed these efficiencies with full detector simulation~\cite{Giurgiu:2009wv}.

Since the boosted Higgs and boosted top taggers described above rely on reversing the clustering sequence 
of a jet, there is
naturally a dependence on the jet algorithm. 
We show in Fig.~\ref{fig:algostop} a comparison of different algorithms for a variable used by the \cms top-tagging study~\cite{Giurgiu:2009wv}. 
Of the 3 or 4 subjets
filtered out of the top jet, they looked at the minimum invariant mass of any pair. In the first panel, 
clustering with 
Cambridge/Aachen shows the peak at $\sim 80$ GeV corresponding to the $W$-mass in the signal, which is 
absent in the background.
The second peak around $\sim 20$ GeV corresponds to wrongly identified subjets and mimics the QCD background. 
The result for
the $\kt$ algorithm is similar. The anti-$\kt$ algorithm, however, does not show the $W$-mass peak. 
This is because anti-$\kt$
essentially works backwards to the other clustering algorithms, with the hardest particles being 
clustered first, and the soft
stuff clustered at the end. One can see that anti-$\kt$ is not ideal for filtering and substructure 
analysis. However, the
final panel shows that if the fat jet is found using anti-$\kt$, but the particles in the jet are 
reclustered using Cambridge/Aachen,
the substructure is still there. We conclude that the substructure analysis and the fat-jet finder can 
be chosen independently.

\begin{figure}[t]
\begin{center}
\includegraphics[width=0.9\textwidth]{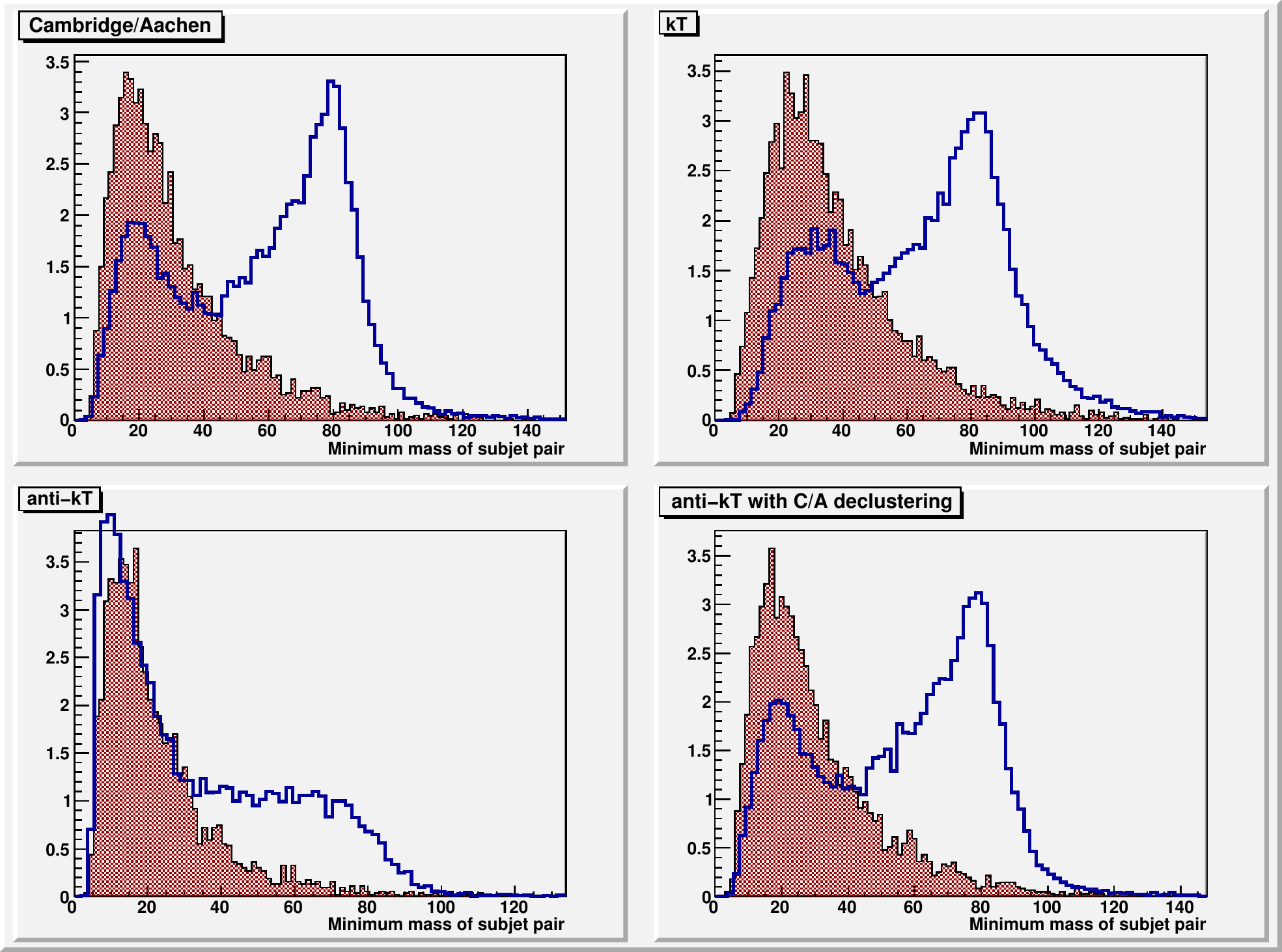}
\caption{Comparison of different clustering algorithms for the minimum invariant mass of two subjets. 
The Cambridge/Aachen
and $\kt$ algorithms both work well and finding the subjets corresponding to the $W$ decay, while 
anti-$\kt$ fails. The bottom
right panel shows that anti-$\kt$ can still be used to find the fat jets, if the jet constituents are 
then reclustered with
Cambridge/Aachen to find the subjets.}
\label{fig:algostop}
\end{center}
\end{figure}

\subsubsection{GENERAL FILTERING TECHNIQUES}
Having seen filtering successfully applied in a number of specific cases, 
it is natural to ask whether the approach can be generalized to cases when you do not know {\it a priori} 
what it is you are looking for.
For example, in the Higgs$\to \bbbar$ study, the fat jet was filtered to look for 2 or 3 hard subjets, 
in top-tagging, the fat jet was 
filtered to look for 3 or 4 hard subjets. The idea behind jet {\it pruning} is that is is possible to 
filter the jet without knowing 
ahead of time how many hard constituents to expect. Instead, one can decompose the jet until some measure of 
the size of each splitting 
is saturated. For example, in~\cite{Ellis:2009me}, the pruning algorithm based on Cambridge/Aachen declustering 
stops when a splitting has 
$\Delta_{ij} > D_{\mathrm cut}$,
where $D_\mathrm{cut} \propto 2 m_J /\pt$ is determined on a jet by jet basis. The authors demonstrated 
that this general
method successfully finds the subjets in the specific cases of boosted top or boosted $W$ jets.

Another general approach is jet {\it trimming}~\cite{Krohn:2009th}.
Here the goal is not to find hard subjet constituents in fat jets, but to clean up simple quark or 
gluon-initiated QCD jets.
Suppose we have a massive color singlet $\phi$ decaying to $q\bar{q}$. Then the mass of the resonance 
should be reconstructable
using all of the final-state radiation from the quarks. However, in reality, it is impossible to separate 
the final state
radiation from additional radiation originating either from the underlying event or from other hard objects 
that happen
to also be present. By studying the radiation pattern from FSR only to FSR plus contamination, the authors 
observed that
the contamination could be reduced with trimming.

The trimming procedure they proposed works as follows. First jets are found as usual, say with 
anti-$\kt$ with $R=0.8$. Then
the jet constituents are reclustered into many tiny jets using a much smaller clustering size, say $R=0.2$. 
These tiny jets are then discarded if their $\pt$ is below a certain threshold, 
say $p_T^{\mathrm{tiny}} < 0.01 p_T^{\mathrm{jet}}$,
or only a certain number of them, say the five hardest, are kept. Figure~\ref{fig:trimming} 
shows an example of how trimming removes
contamination from underlying event. 
Note that in contrast to pruning, trimming does not reverse the clustering sequence to find 
subjets, rather it reclusters with a different scale, in a manner similar to the filtering
stage in the Higgs analysis described above, but not solely focused on that final state.
Thus, good results can be obtained using 
anti-$\kt$ both to find the original jets, and to find the tiny jets
during the reclustering step.

\begin{figure}[t]
\begin{center}
\includegraphics[width=0.6\textwidth]{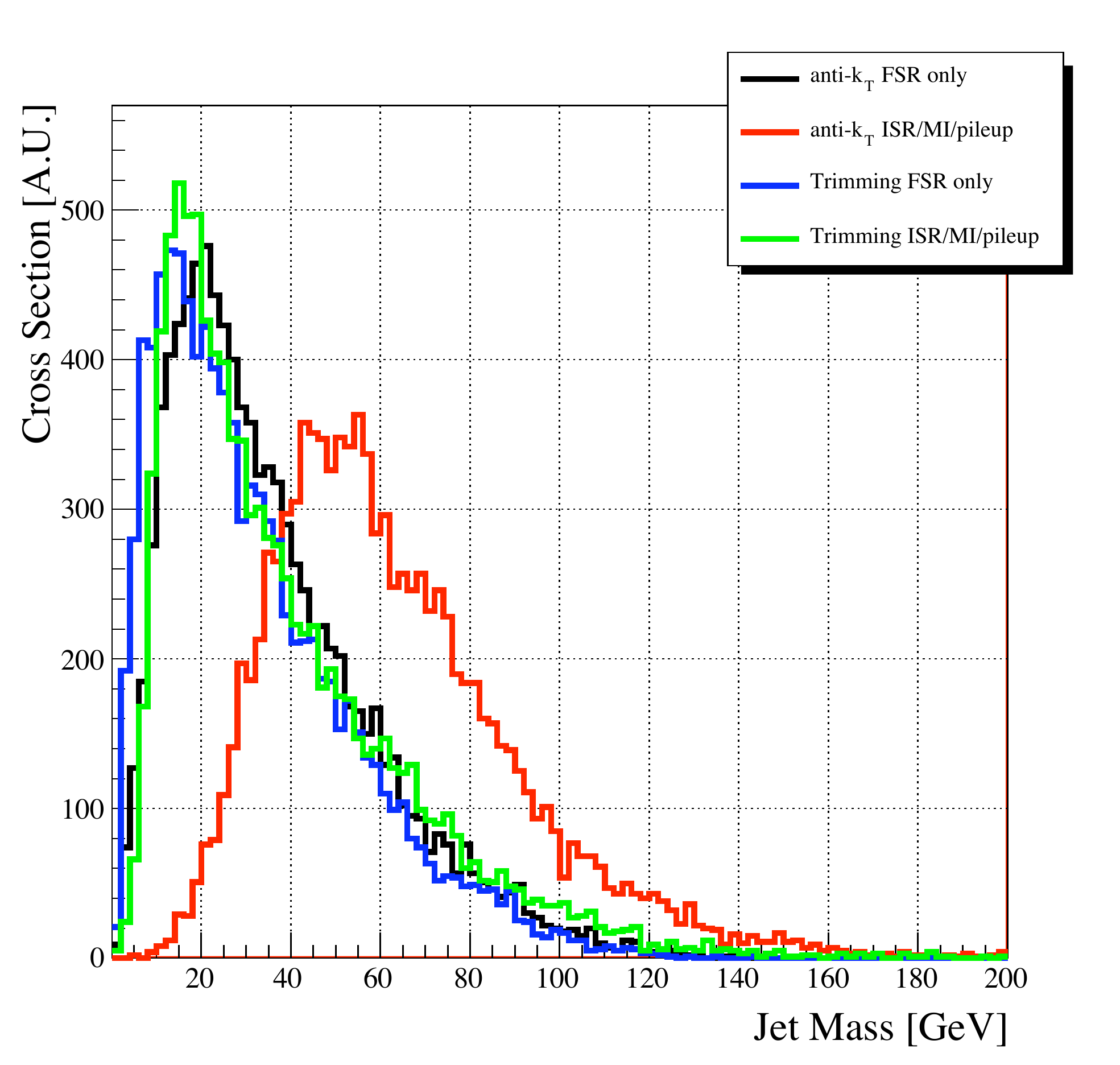}
\caption{Effectiveness of jet trimming at removing contamination from the underlying event. The jet mass after trimming
appears similar to the jet mass with underlying event turned off. (Figure courtesy of D. Krohn.)}
\label{fig:trimming}
\end{center}
\end{figure}

\subsection{SUMMARY AND FUTURE WORK}
Let us briefly mention some other related ideas which have been proposed.
In~\cite{Krohn:2009zg}, the authors proposed that greater efficiencies could be obtained by
choosing a jet radius continuously which decreases with increasing $\pt$. 
Similar scalings had been used previously, for example in the boosted top-tagging study~\cite{Kaplan:2008ie}, but with
a discrete size change rather than a continuous one. The insight of~\cite{Krohn:2009zg} was to make
such a scaling automatically part of the jet algorithm. While changing the jet size does seem to
produce improvements, the benefits may not outweigh the complications in real experimental settings,
such as the increased difficulty of jet energy scale calibrations. For example, \cms
chose in their implementation of top-tagging~\cite{Giurgiu:2009wv} to keep the jet size fixed, and were
able to achieve comparable efficiencies to the theoretical study with the $R$ variation.

Another idea is to use QCD jet shapes such as planar flow
or angularities to distinguish heavy particle decays from QCD jets~\cite{Almeida:2008tp,Almeida:2008yp}.
A comparison of the efficiency of this approach in top tagging, and a more complete
summary of other key issues in jet substructure can be found in~\cite{Salam:2009jx}.

In addition there is clearly potential in studying 
QCD characteristics beyond the jets. Recently, a proposal was made of for a simple variable {\it pull}
which can characterize the color flow of an event~\cite{Gallicchio:2010sw}. Color flow is something that
is not contained in the 4-momentum of a jet and so may provide a nice handle on new physics which
is complementary and uncorrelated with traditional techniques. For example, it may help distinguish
color singlet objects (such a the Higgs) from the other color structures in the QCD background.
Although color flow is sensitive to aspects of QCD that go beyond the leading-log, DGLAP evolution of
basic Monte Carlos, recent progress in the development of Monte Carlos assure us that such effects must
be both visible and calculable (cf. Section~\ref{sec:theoryps} of these proceedings).
The prospects of extracting such superstructure information at the \lhc are just beginning to be explored, and
there is certainly much more to be done and understood.

In subsequent sections of these proceedings more details of physics relevant to jet substructure are discussed: 
the status of simulations 
for jet substructure in heavy particle decays (Section~\ref{sec:mcheavy}), 
detector issues (Section~\ref{sec:detsubs}), radiation between jets (Section~\ref{sec:radiation}) and calculations
of boosted $\bbbar$ production in $W\bbbar$ events (Section~\ref{sec:wbbbar}), which is a significant background to 
Higgs searches using jet substructure.

\clearpage

\section[HEAVY PARTICLE DECAYS IN MONTE CARLO EVENT GENERATORS]
{HEAVY PARTICLE DECAYS IN MONTE CARLO EVENT GENERATORS~\protect 
\footnote{Contributed by: K.~Hamilton, L.~L\"onnblad, S.~Schumann, T.~Sj\"ostrand, J.~Winter}}
\label{sec:mcheavy}

In the coming years the LHC will probe a new energy regime one order
of magnitude higher than that which has been accessed to date. It
is therefore natural that heavy, unstable particles, such as the top-quark,
will be produced in large quantities. Indeed, next-to-leading order
QCD predictions for top quark pair production indicate that $t\bar{t}$
pairs will be created at a rate of around 0.3 Hz at low luminosity
(10 $\mathrm{fb}^{-1}$ per year). Of course, it is widely anticipated
that physics beyond the Standard Model (SM) will be discovered at
the LHC, hence many other species of heavy particles are also expected
to be produced.

It is well known that modern collider physics analyses rely upon Monte
Carlo simulations for a wide variety of applications, from the modelling
of detector effects through to the determination of fundamental parameters;
the top-quark mass determination being a case where the dependence
on simulations is particularly acute. Since new, unstable, heavy particles
feature in almost all extensions of the Standard Model, it is foreseeable
that many more analyses of this type will be carried out in the LHC
era. Clearly, as the r{\^o}le of the event generators in the analysis
becomes increasingly significant, so too does the need for them to
model the underlying process precisely and, when this is not possible,
it is equally important that their limitations and inaccuracies are
understood. 

In the following we shall critically review some of the main issues
to be addressed in the simulation of heavy particle decays. The material
is intended to better inform \emph{users} of event generators regarding
the nature of the methods and approximations employed, rather than
to provide a comprehensive technical review of the state-of-the-art.
The first part of the discussion concerns the generalities of the
physics phenomena and their simulation, rather than a description
of a specific event generator. In the second part we give brief, itemized,
technical descriptions of the three main general-purpose event generators,
\sherpa~\cite{Gleisberg:2008ta}, \pythiaeight\cite{Sjostrand:2006za,Sjostrand:2007gs}
and \herwigpp~\cite{Bahr:2008pv}, with reference to the background
material.

\subsection{SIMULATION STRATEGY\label{sub:Simulation-strategy}}

In general, for each event, all of the multi-purpose event generators
begin by generating an initial `hard' configuration of particles according
to the associated tree-level matrix elements, after which parton showers
are initiated from the external legs. When the showering is complete
the fi{}nal-state consists of a set of partons with constituent masses
which are then hadronized. On the other hand, for processes involving
the production and decay of unstable particles, including decay chains,
rather than attempting to generate the ultimate final-state particles
directly, according to high multiplicity matrix elements, each program
first generates the momenta of the initial unstable heavy particles
in the production phase, before going on to (quasi-)independently
generate the momenta of their decay products, showering the colour
charged objects in each phase. As well as being more manageable, this
approach benefits from being substantially more efficient, computationally,
and also from being highly versatile, in keeping with the multi-purpose
paradigm that the event generators are built on; in particular, the
independent generation of production and decay momenta lends itself
naturally, through iteration, to the simulation of arbitrary decay
chains. Of course this generality comes at the cost of varying degrees
of approximation, which we will now discuss briefly below.

\subsection{THE BASE APPROXIMATION\label{sub:The-base-approximation}}

In this subsection we elaborate on the series of approximations by
which the production and decay of a particle may be treated entirely
independently \emph{modulo} charge and momentum conservation. In doing
so we will arrive at\emph{ the} \emph{base approximation}: the common
starting point from which the simulations may or may not start to
model omitted physics effects and undo approximations by various means,
some of which will be addressed in Section~\ref{sub:Beyond-the-base}.

\subsubsection{RESONANT AND NON-RESONANT CONTRIBUTIONS\label{sub:Resonant-and-non-resonant}}

Experimental signals are defined in terms of the particles which actually
enter the detector, leptons, photons and jets, rather than the heavy
unstable objects which we typically are interested in studying, \emph{upstream}.
The corresponding theoretical predictions are then strictly required
to be defined in precisely the same way, including contributions from
\emph{all} Feynman diagrams which give rise to the same external states, 
see e.g. \cite{Hagiwara:2005wg}. 
Generally, however, only a limited subset of these diagrams involve
the exchange of s-channel resonances which go on to decay into the
stable objects comprising the signal, \emph{i.e.} diagrams representative
of heavy particle production and decay (chains). By their nature the
resonant graphs give the dominant contributions to the cross section
for a given signal and calculations based on this subset of Feynman
diagrams are known to provide excellent approximations to the full
result. Moreover, in the case where the heavy s-channel particle is
that which we wish to study, experimental analyses will regularly
employ cuts which further enhance the resonant contributions. In keeping
with these observations the simulation of non-resonant effects in
all but the simplest processes, namely $2\rightarrow2$ processes,
is typically beyond the scope of the multi-purpose Monte Carlo event generators,
furthermore, the loss of accuracy incurred through the omission of
the non-resonant contributions is limited to the extent that this
approximation is often also employed by tree-level event generators such
as Madgraph / MadEvent \cite{Alwall:2007st} or Whizard \cite{Kilian:2007gr}, 
even though they have the facility to include them automatically, in 
order to increase computational efficiency.

\subsubsection{FACTORIZATION OF PRODUCTION AND DECAY\label{sub:Factorization-of-production}}

Although there exists, in general, quantum mechanical interference
between the production and decay of an unstable particle, in the limit
that the resonant particle is exactly on-shell, the matrix elements
corresponding to the resonant graphs may be factorized into two parts,
one corresponding to the production of the heavy particle and one
corresponding to its decay. Take, for example, a process $pp\rightarrow bW^{+}+X$,
which naturally receives resonant contributions in the form of graphs
containing an s-channel top-quark; in the limit that the top-quark
momentum is on-shell, one can replace the propagator numerator $i\left(\not p_{t}+m_{t}\right)$
with a sum over spins: \begin{eqnarray}
\lim_{p_{t}^{2}\rightarrow m_{t}^{2}}\mathcal{M} & \rightarrow & \frac{1}{\left(p_{t}^{2}-m_{t}^{2}\right)^{2}+m_{t}^{2}\Gamma_{t}^{2}}\,\mathcal{A}_{\lambda}^{P}\mathcal{A}_{\lambda^{\prime}}^{P*}\,\mathcal{A}_{\lambda}^{D}\mathcal{A}_{\lambda^{\prime}}^{D*}\label{eq:ME_factorisation-1}\end{eqnarray}
where $\mathcal{A}_{\lambda}^{P}=\mathcal{A}^{P}u_{\lambda}\left(p_{t}\right)$
and $\mathcal{A}_{\lambda}^{D}=\bar{u}_{\lambda}\left(p_{t}\right)\mathcal{A}^{D}$
are the amplitudes for the on-shell production and decay processes,
$pp\rightarrow t+X$ and $t\rightarrow bW^{+}$ respectively and where
the quark helicity indices, $\lambda$, $\lambda^{\prime}$ are summed
over. In the so-called \emph{narrow width approximation} where the product
of the unstable particle's width and mass tend to zero we have \begin{equation}
\lim_{m\Gamma\rightarrow0}\frac{1}{\left(p^{2}-m^{2}\right)^{2}+m^{2}\Gamma^{2}}\rightarrow\frac{\pi}{m\Gamma}\,\delta\left(p^{2}-m^{2}\right)\,,\label{eq:Breit_Wigner_limit}\end{equation}
\emph{i.e.} in the narrow width approximation the replacement of Eq.\,(\ref{eq:ME_factorisation-1})
is valid. A completely analogous replacement to that in Eq.\,(\ref{eq:ME_factorisation-1})
holds for vector bosons, in that case, as with the fermion propagator
numerator, the vector boson propagator numerator can be replaced by
a sum over polarizations. 

A further approximation commonly (but not universally) used in event
generators is the neglect of \emph{spin correlations}. This amounts
to assuming that in the factorized amplitudes in Eq.\,(\ref{eq:ME_factorisation-1}),
the graphs corresponding to different helicity states of the intermediate
particle do not interfere with one another and that they are produced
in equal measure \emph{i.e.} the full matrix element is effectively
replaced by the product of two spin summed matrix elements, one corresponding
to the production $\mathcal{M}^{P}$, the other to the decay $\mathcal{M}^{D}$.
In our pedagogical example, working in the narrow width approximation
and neglecting spin correlations, the exact squared matrix is replaced
according to \begin{eqnarray*}
\lim_{m_{t}\Gamma_{t}\rightarrow0}\mathcal{M} & \rightarrow & \frac{\pi}{m_{t}\Gamma_{t}}\,\delta\left(p_{t}^{2}-m_{t}^{2}\right)\,\mathcal{M}^{P}\mathcal{M}^{D}\,.\end{eqnarray*}

In addition, by inserting $\mathrm{d}q_{t}^{2}\,\delta\left(q_{t}^{2}-p_{t}^{2}\right)$
and $\mathrm{d}^{4}p_{t}\,\delta\left(p_{t}-p_{b}-p_{W^{+}}\right)$
into the usual Lorentz invariant phase space measure we may, without
approximation, factorize the Lorentz invariant phase space, $\mathrm{d}\Phi$,
for $bW^{+}+X$ into parts corresponding to the same production and
decay processes:\begin{eqnarray}
\mathrm{d}\Phi & = & \mathrm{d}\Phi_{P}\,\mathrm{d}\Phi_{D}\,\frac{1}{2\pi}\,\mathrm{d}q_{t}^{2}\label{eq:Phase_space_factorisation}\end{eqnarray}
where,\begin{eqnarray}
\mathrm{d}\Phi_{P} & = & \mathrm{d}\Phi_{t}\mathrm{d}\Phi_{X}\left(2\pi\right)^{4}\delta^{4}\left(p_{a}+p_{b}-p_{t}-p_{X}\right),\label{eq:Phase_space_factorisation_2}\\
\mathrm{d}\Phi_{D} & = & \mathrm{d}\Phi_{b}\mathrm{d}\Phi_{W^{+}}\left(2\pi\right)^{4}\delta^{4}\left(p_{t}-p_{b}-p_{W^{+}}\right),\nonumber \end{eqnarray}
and for a given particle $i$, $\mathrm{d}\Phi_{i}$ is the usual
invariant phase space measure \begin{equation}
\mathrm{d}\Phi_{i}=\frac{\mathrm{d}^{3}p_{i}}{\left(2\pi\right)^{3}2E_{i}}\,.\label{eq:Phase_space_factorisation_3}\end{equation}

Using this phase space factorization together with the narrow width
approximation one can see that the differential cross section will
factorize into two parts; denoting flux and PDF factors by $\mathcal{L}\left(x_{\oplus},x_{\ominus}\right)$
we have, \begin{eqnarray}
\mathrm{d}\sigma & = & \mathrm{d}x_{\oplus}\mathrm{d}x_{\ominus}\,\mathcal{L}\left(x_{\oplus},x_{\ominus}\right)\,\mathrm{d}\Phi\,\mathcal{M}\label{eq:Factorised_xsec_and_width}\\
 & = & \mathrm{d}x_{\oplus}\mathrm{d}x_{\ominus}\,\mathcal{L}\left(x_{\oplus},x_{\ominus}\right)\,\mathrm{d}\Phi_{P}\,\mathcal{M}^{P}\times\,\frac{1}{\Gamma_{t}}\,\mathrm{d}\Gamma_{t}\nonumber \\
\mathrm{d}\Gamma_{t} & = & \frac{1}{2m_{t}}\,\mathcal{M}^{D}\,\mathrm{d}\Phi_{D}\,,\nonumber \end{eqnarray}
where the first part corresponds to the spin summed production of
the unstable particle and the second part to its branching ratio,
differential in the decay phase space. By appealing to the narrow
width approximation it is therefore possible to first generate the
production and decay processes completely independently in their rest
frames, then boost the decay products to the frame in which the particle
was produced. The latter boost is ambiguous up to a rotation which
is a direct manifestation of having neglected spin correlations. 

The simulation procedure and the approximations which we have described
up to this point comprise the most basic algorithm employed by Monte
Carlo simulations for the treatment of heavy particle production and
decay; of course, for the parton shower simulations, these tree level
configurations will, by default, go on to include the effects of soft
and collinear parton emissions from the coloured particles. Unless
otherwise stated this represents the base accuracy and physics describing
the interplay of the production and decay of unstable particles in
these simulations: non-resonant contributions, spin correlations and
off-shell / finite width effects are all completely neglected.

\subsection{BEYOND THE BASE APPROXIMATION\label{sub:Beyond-the-base}}

Having noted the basic nature of the base approximation we will now
briefly review some of the ways in which it can be improved on. Note
that these enhancements can be extremely sophisticated and so in no
case should they be assumed to have been implemented as a default
option, for the same reason, it should be clear from the simulations'
documentation whether such features are available or not.

\subsubsection{FINITE WIDTH EFFECTS}

Although the factorization of the numerator of the unstable particle
propagator only holds when it is exactly on-shell, the error induced
by using the replacement Eq.\,(\ref{eq:ME_factorisation-1}) when this
is not true is $\mathcal{O}\left(\Gamma/m\right)$ \emph{i.e. }the\emph{
}typical amount by which the particle is off-shell with respect to its
mass. Of course, if cuts or observables are such that the channel
in question only contributes when the particle is far off-shell, \emph{e.g.}
if cuts correspond to a mass-window restricting the invariant mass
to be far from the pole mass, the error is much greater and, in any
case, in such circumstances the inclusion of non-resonant contributions
is mandatory. Provided the signal regions receive contributions from
the immediate vicinity of the resonance we may recover its line-shape
up to corrections $\mathcal{O}\left(\Gamma/m\right)$, by inserting
1 into the factorized cross section formulae as,\begin{equation}
1=\int\mathrm{d}p^{2}\,\frac{m\Gamma}{\pi}\,\frac{1}{\left(p^{2}-m^{2}\right)^{2}+m^{2}\Gamma^{2}}\,.\label{eq:Breit_Wigner_integral}\end{equation}
This is equivalent, of course, to not taking the exact limit $\Gamma\rightarrow0$
in the Breit-Wigner function in Eqs.\,(\ref{eq:ME_factorisation-1})
and (\ref{eq:Breit_Wigner_limit}), we stress that this limit is nevertheless
\emph{implied} by assuming that the propagator numerator may be replaced
by the sum over external wavefunctions.

In terms of the event generation procedure, the inclusion of finite
width effects in this way, leads to a trivial modification of that
described above for the $\Gamma=0$ case, namely, that the first step
in the algorithm is now the generation of an off-shell mass for the
intermediate particles according to a Breit-Wigner function. The same
procedure can be adopted using running widths.

Although this seems, in principle, a relatively straightforward means
of improving the base approximation $\left(\Gamma=0\right)$, a thorough
treatment of the particle width, in particular those involving next-to-leading
order computations, requires that further, more subtle, issues be
addressed before it is introduced anywhere at all. Of note is the
question of how to ensure that the introduction of the finite width
does not violate gauge invariance? The particle width term arises
from the resummation of the imaginary parts of the self-energy insertions
on the unstable particle's propagator \emph{i.e.} it corresponds to
the inclusion of a \emph{subset} of higher order corrections to the
process under consideration%
\footnote{It therefore also has a dependency on the particle's mass, which we
shall omit here for simplicity. %
}. Since S-Matrix elements are only guaranteed to be gauge invariant
at each order in perturbation theory when \emph{all} relevant diagrams
at the given order are summed, one should expect that naively adding
a width term to the propagator will violate gauge invariance. Although
this is the case in general, usually the numerical impact of the gauge
breaking terms is negligible and is contained within that incurred
by writing the propagator numerator as a polarization sum. 

On the other hand, gauge invariance underlies the cancellation of
large logarithmic corrections to inclusive quantities as laid out
in the Bloch-Nordsieck \cite{Bloch:1937pw} and KLN \cite{Kinoshita:1962ur,Lee:1964is}
theorems, therefore, in cases where logarithmically enhanced emissions
occur, the gauge variant contributions can be hugely magnified and
even exceed the correct, gauge invariant, contributions. These potentially
dangerous cases are confined to processes involving diagrams in which
the unstable particle we are studying emits soft / collinear massless
partons or photons \emph{e.g. }gluon emission from top quarks or photon
emission from W bosons. For simulations involving tree level matrix
elements and / or parton showers this issue is substantially negated
due to the obligatory generator level cuts, which exclude the phase
space for collinear and soft regions and, also, because the splitting
functions at the heart of the parton shower dynamics are derived from
the \emph{universal} factorization of matrix elements with soft /
collinear emissions. Furthermore, it is clear that for the Lagrangian
to be gauge invariant it is not necessary for the parameters, in particular
the masses, to be real numbers. Bearing this in mind, one may adopt
the so-called \emph{complex mass scheme} \cite{Denner:1999gp} whereby
finite width effects are included in a gauge invariant fashion by
simply reinterpreting the masses, \emph{everywhere} that they occur
in the Lagrangian, as being complex numbers with the $im\Gamma$ width
factor absorbed inside them. Alternatively the so-called `fudge-factor'
scheme \cite{Beenakker:1996kt} is employed whereby the total squared
matrix element, neglecting finite width effects, is multiplied by
the product of the Breit-Wigner function and the zero-width propagator,
$p^{2}-m^{2}$, squared for each resonant particle in the process.
For further technical details concerning gauge invariance and finite
width effects see Refs.~\cite{Baur:1995aa,Baur:2004ig,Beenakker:1996kn,Beenakker:1996kt,Denner:1999gp}. 

Another pertinent issue is that of how to include threshold effects
in the running width and the selection of decay modes. In the forthcoming
years LHC analyses will predominantly be concerned with searches for
new particles, whose masses couplings and decay modes are all unknowns.
A prime example is the Standard Model Higgs boson, whose width and
decay modes strongly depend on, not simply its on-shell mass but,
in view of the factorization of the production and decay, also the
off-shell mass generated at the beginning of the simulation, Eq.\,(\ref{eq:Breit_Wigner_limit}).
This is of particular importance for particles decaying close to threshold
(as the Higgs boson may do), in this case the effects from the off-shell
propagator must be taken into account. This can be achieved by implementing
the running width, \emph{i.e.} including the full dependence on the
particle's \emph{off}-\emph{shell} mass, in the weight factor in Eq.\,(\ref{eq:Breit_Wigner_integral}),
\emph{as} \emph{well} \emph{as}, in the calculation of the partial
width of a selected decay mode as in Ref.\,\cite{Gigg:2008yc}, where
this prescription is shown to reproduce well the results of some three-body
decays simulated as a cascade of two two-body decays.

\subsubsection{SPIN CORRELATIONS\label{sub:Spin-correlations}}

As stated previously in the base approximation particle decays are
performed according to the spin averaged matrix element for the decay
process hence they occur isotropically in the rest frame. It follows
that in order to have a reasonable description of observables sensitive
to the details of \emph{individual} decay products it will be necessary,
at least, to communicate the spin information from the production
to the decay parts of the simulations. Notable studies for which such
correlations are particularly relevant include: the production and
decay of the top quark, the production of tau leptons in Higgs decays
and the spin determination of newly discovered particles \emph{i.e.}
understanding the nature of physics beyond the Standard Model. 

Generalizing earlier work of Knowles and Collins \cite{Collins:1987cp,Knowles:1988hu,Knowles:1988vs},
a flexible, efficient algorithm for propagating spin correlations
between particle production and decay in Monte Carlo event generators,
using the spin density matrix formalism, was developed by Richardson
\cite{Richardson:2001df}. Rather than discuss the algorithm in full
detail here we will describe it by considering the example of the
process $h\, h\rightarrow t\,\bar{t}$ where the top quark subsequently
decays, via a $W$ boson, to a $b$ quark and a pair of light fermions.
Initially, the outgoing momenta of the $t\,\bar{t}$ pair are generated
according to the partonic differential cross section

\begin{equation}
\mathrm{d}\widehat{\sigma}^{P}=\frac{1}{2\hat{s}}\,\mathcal{A}_{\lambda_{t}\lambda_{\bar{t}}}^{P}\,\mathcal{A}_{\lambda_{t}\lambda_{\bar{t}}}^{P*}\,\mathrm{d}\Phi_{P}\end{equation}
where $\mathcal{A}_{\lambda_{t}\lambda_{\bar{t}}}^{P}$ $ $ is the
amplitude for the initial hard process, with $\lambda_{t}$ and $\lambda_{\bar{t}}$
denoting the helicity indices of the produced $t$ and $\bar{t}$
respectively. $\mathrm{d}\Phi_{P}$ is the usual phase space measure
for the production process, equal to that in Eq.\,(\ref{eq:Phase_space_factorisation_2}),\emph{
}where in this example $X=\overline{t}$. Having generated the momenta
for the production process, one of the outgoing particles is then
picked at random, say the top and a production spin density matrix
is calculated \begin{equation}
\rho_{\lambda_{t}\lambda_{t}^{\prime}}^{t}=\frac{1}{N}\,\mathcal{A}_{\lambda_{t}\lambda_{\bar{t}}}^{P}\mathcal{A}_{\lambda_{t}^{\prime}\lambda_{\bar{t}}}^{P*}\,,\label{eq:production_spin_matrix}\end{equation}
with $N$ defined such that $\rho^{t}$ has unit trace.

The top is decayed and the momenta of the decay products distributed
according to \begin{equation}
\mathrm{d}\Gamma_{t}=\frac{1}{2m_{t}}\,\rho_{\lambda_{t}\lambda_{t}^{\prime}}^{t}\,\mathcal{A}_{\lambda_{t}\lambda_{W^{+}}}^{D_{t}}\,\mathcal{A}_{\lambda_{t}^{\prime}\lambda_{W^{+}}}^{D_{t}*}\,\mathrm{d}\Phi_{D}\,,\label{eq:decay_spin_corrs_1}\end{equation}
where the inclusion of the spin density matrix ensures the correct
correlation between the top decay products and the beam. We note how
the spin averaged result, Eq.\,(\ref{eq:Factorised_xsec_and_width}),
is recovered by replacing $\rho_{\lambda_{t}\lambda_{\bar{t}}}^{t}\rightarrow\frac{1}{2}\,\delta_{\lambda_{t}\lambda_{\overline{t}}}$.
With the momenta for the top quark decay products now in hand we may
calculate a production spin density matrix for the unstable $W^{+}$
\begin{equation}
\rho_{\lambda_{W^{+}}\lambda_{W^{+}}^{\prime}}^{W^{+}}=\frac{1}{N}\,\rho_{\lambda_{t}\lambda_{t}^{\prime}}^{t}\,\mathcal{A}_{\lambda_{t}\lambda_{W^{+}}}^{D_{t}}\,\mathcal{A}_{\lambda_{t}^{\prime}\lambda_{W^{+}}}^{D_{t}*}\,,\label{eq:decay_spin_corrs_2}\end{equation}
and then generate its decay by analogy to Eq.\,(\ref{eq:decay_spin_corrs_1}).
Here the use of the production matrix calculated from Eqs.\,(\ref{eq:production_spin_matrix})
and (\ref{eq:decay_spin_corrs_2}) leads to the correct angular correlations
between the light fermions, the beam and the bottom quark. 

Since the children of the $W^{+}$ are light fermions the decay chain
terminates and a decay matrix for the $W^{+}$ is calculated:\begin{equation}
D_{\lambda_{W^{+}}\lambda_{W^{+}}^{\prime}}^{W^{+}}=\frac{1}{N}\,\mathcal{A}_{\lambda_{t}\lambda_{W^{+}}}^{D_{t}}\,\mathcal{A}_{\lambda_{t}\lambda_{W^{+}}^{\prime}}^{D_{t}*},\label{eq:decay_matrix_W}\end{equation}
Moving back up the decay chain, the analogous decay matrix is calculated
for the top quark using the decay matrix of the $W^{+}$: \begin{equation}
D_{\lambda_{t}\lambda_{t}^{\prime}}^{t}=\frac{1}{N}\,\mathcal{A}_{\lambda_{t}\lambda_{W^{+}}}^{D_{t}}\,\mathcal{A}_{\lambda_{t}^{\prime}\lambda_{W^{+}}^{\prime}}^{D_{t}*}\, D_{\lambda_{W^{+}}\lambda_{W^{+}}^{\prime}}^{W^{+}}.\label{eq:decay_matrix_t}\end{equation}
Following this one contracts $D_{\lambda_{t}\lambda_{t}^{\prime}}^{t}$
with the full $t\bar{t}$ production spin density matrix, $\mathcal{A}_{\lambda_{t}\lambda_{\bar{t}}}^{P}\,\mathcal{A}_{\lambda_{t}^{\prime}\lambda_{\bar{t}}^{\prime}}^{P*}$,
to give the production spin matrix of the antitop quark, $\rho_{\lambda_{\bar{t}}\lambda_{\bar{t}}^{\prime}}^{\bar{t}}$,
analogous to Eq.\,(\ref{eq:production_spin_matrix}), before generating
its decay products in the same way as was done for the top quark.
In progressing forward along the $\overline{t}$ decay chain the production
spin density matrices pass information from one decay to the next
leading to the correct angular correlations. 

Other methods for  the inclusion of spin correlations in event generators
are in use, although many of these may be regarded as approximations or variants
of the above \cite{Alwall:2007st}. The only other approach which
may be regarded as distinct, involves decaying all unstable particles in 
the event in a single step, using matrix elements for the process under
study where these particles have and have not decayed. This 
algorithm has been applied in \pythia, for quite some time, for a number of
processes and has been recently been extended in the context of matched
next-to-leading order simulations by Frixione \emph{et al} \cite{Frixione:2007zp}.
This alternative method is based on the observation that the product
of the separately spin summed matrix elements for the production and
decay processes, including propagator factors, always exceeds the
full spin summed matrix element. Knowing this, spin correlations may
be implemented by generating production and decay momenta as in the
base approximation (with the decays isotropic in their rest frame)
choosing a random number $\mathcal{R}$ and then regenerating the
decay kinematics if $\mathcal{R}$ is greater than the ratio of the
full spin summed matrix element to the spin summed matrix element
omitting spin correlations. For a given set of production momenta,
isotropic decay kinematics and new values of $\mathcal{R}$ are generated
repeatedly until $\mathcal{R}$ is less than the aforesaid ratio,
at which point the decay kinematics can be kept. 

Since the underlying Monte Carlo algorithm in this alternative approach
is just the hit-or-miss method, with no spin density matrices, spinor
phases \emph{etc}, to keep track of, this method is substantially
easier to implement than the one above. On the other hand this method
assumes that a matrix element generator capable of providing helicity
amplitudes for (arbitrarily) high multiplicity final states is available,
moreover, the intention here is to simultaneously generate the momenta
for all decaying particles, in `one shot', which also impedes the
event generation efficiency with respect to the recursive approach
based on spin density matrices. Nevertheless, for including spin correlations
in events with relatively few decays, this method offers a simple
and straightforward alternative to the spin density matrix approach,
where the relative inefficiency tends to add little, with respect
to other elements of the simulation, to the overall event generation
time. This method has been used widely in simulations based on the
\MCatNLO \cite{Frixione:2002ik,Frixione:2007zp} and
\POWHEG \cite{Nason:2004rx,Frixione:2007nw,Alioli:2009je}
methods.

\subsubsection{QCD RADIATION\label{sub:QCD-radiation}}

Having generated the decay momenta, with or without finite width effects
and spin correlations, if the mother particle or any of the daughter
particles carry a colour charge, one should attempt to model the effects
of QCD radiation which they may emit. The vast majority of these emissions
will be soft and / or collinear emissions with respect to the shower
progenitors. This is straightforward to understand from the point
of view of the associated amplitudes: in the limit that a massless
particle is emitted with low $p_{T}$, the propagator denominator,
associated to the internal line that was its mother, vanishes so the
matrix element diverges for such configurations. 

Such low $p_{T}$ corrections are precisely those which parton shower
simulations take into account, to all orders in perturbation theory.
Of course, distributing emissions according to arbitrarily high multiplicity
matrix elements is not feasible, instead we appeal to the fact that,
in the limit that a massless parton $j$ is emitted collinear from
an external leg $i$, the matrix elements factorize according to\begin{equation}
\mathcal{M}_{n+1}=\frac{8\pi\alpha_{\mathrm{S}}}{q_{\widetilde{ij}}^{2}-m_{\widetilde{ij}}^{2}}\, P_{\widetilde{_{ij}}\rightarrow ij}\,\mathcal{M}_{n}\,,\label{eq:collinear_factorisation}\end{equation}
where $\widetilde{ij}$ denotes the mother of the branching $\widetilde{ij}\rightarrow ij$,
with virtuality $q_{\widetilde{ij}}^{2}$, and $P_{\widetilde{ij}\rightarrow ij}$
is an Altarelli-Parisi splitting function. The form of the splitting
functions does not depend on $\mathcal{M}_{n+1}$ and $\mathcal{M}_{n}$,
where the latter is the matrix element for the $n$ particle process,
in which the branching does not occur. The matrix element factorization
in Eq.\,(\ref{eq:collinear_factorisation}) may be combined with a
phase space factorization for the branching, leading to a factorized
form of the differential cross section for the $n+1$ particle process,
in a similar vein to that described in Section \ref{sub:Factorization-of-production},
in the context of heavy particle decays. Here, as with cascade decays,
the factorization may be applied recursively, albeit with the caveat
that using Eq.\,(\ref{eq:collinear_factorisation}) implies the emissions
be ordered such that they become increasingly collinear as one works
from the core of the process out toward the external legs. As well
as modelling the effects of enhanced collinear real emission, the parton
shower method also, necessarily, incorporates their corresponding
virtual corrections through \emph{Sudakov form factors}, the effect
of which is to dampen, and ultimately regulate, the diverging emission
rate as $p_{T}\rightarrow0$.

Depending on the accuracy of the parton shower, there is some freedom
in the definition of `increasingly collinear'. Parton shower algorithms
may be formulated as an evolution in the virtualities of the branching
partons, or as an evolution in the transverse momentum of the branching
products. However, formal studies of colour coherence \cite{Ciafaloni:1980pz,Mueller:1981ex,Ciafaloni:1981bp,Ermolaev:1981cm,Bassetto:1982ma,Dokshitzer:1982fh,Catani:1983bz,Bassetto:1984ik,Marchesini:1984bm,Dokshitzer:1988bq}
reveal that branchings involving soft gluons should be ordered in
the angle between the branching products. This is a dynamical effect
whereby wide-angle soft gluon emissions, from near collinear configurations
of two or more partons, have insufficient transverse resolving power
to be sensitive to the constituent emitters. In effect the resulting
radiation pattern is determined by the colour charge and momentum
of the \emph{mother} of the emitters, rather than the emitters themselves
\footnote{To avoid digressing from our stated objectives we limit ourselves
to this minimal description of the parton shower approximation and
refer the interested reader the following dedicated review articles
\cite{Ambroglini:2009nz,Webber:1986mc,Sjostrand:2006za,Nason:2004rx}.%
}.

In the LHC era we will regularly face the situation in which one cannot
neglect the mass of radiating particles when modelling the distribution
of their emissions, $t\bar{t}$ bar production being an obvious example.
Nevertheless, the Altarelli-Parisi splitting functions governing the
distribution of emissions in the older generation of shower simulations
are based on the factorization of matrix elements where the mother
parton, $\widetilde{ij}$, and its two daughter partons $i$ and $j$
are all massless. As stated above, in the limit that massless partons
emit low $p_{T}$ radiation the associated amplitudes diverge, however,
if the emitter has a mass, the propagator denominator is proportional
to $1-\beta\cos\theta$ , where the factor of $\beta$, the velocity
of the emitter, screens the collinear divergence. This mass effect
in the propagator denominator then greatly alters the distribution
of emissions as the small angle region is approached with respect
to the rather crude massless approximation, which will likely have
important consequences for analyses \emph{e.g. }the top mass determination,
where a good understanding of production radiation and $b$ quark
fragmentation are key. However, in recent years, developments in the
resummation of radiation from massive partons have included the introduction
of \emph{quasi}-\emph{collinear splitting functions} which generalize
those of Altarelli and Parisi, so as to retain the full dependence
on the mass of the emitting particle in the collinear limit \cite{Cacciari:2001cw}.
For example, the quasi-collinear $q\rightarrow qg$ splitting function
is\begin{equation}
P_{q\to qg}=C_{F}\left[\frac{1+z^{2}}{1-z}-\frac{2z\left(1-z\right)m_{q}^{2}}{p_{T}^{2}+\left(1-z\right)^{2}m_{q}^{2}}\right],\label{eq:quasi_collinear_splitting_fn}\end{equation}
in which $z$ is the light-cone momentum fraction carried by the daughter
quark and $p_{T}^{2}$ is the transverse momentum of the branching;
note how the form of the first term in Eq.\,(\ref{eq:quasi_collinear_splitting_fn})
is the same as the usual massless $q\rightarrow qg$ Altarelli-Parisi
function. The inclusion of these mass effects in parton shower simulations
is actually non-trivial, since the full showering formalism is required
to retain mass effects from the beginning\emph{ i.e.} in the shower
variables, the Sudakov form factors, the cut off scales, the phase
space mapping and kinematics reconstruction / momentum reshuffling. 

Since, in the narrow width approximation, the production and decay
processes are assumed to factorize into two independent parts, this
is the sense in which we simulate the associated parton showers, effectively
treating the decay as a completely independent hard sub-process. For
the decay process the effects of QCD radiation are then simply modeled
by initiating parton showers from the incoming and external particles,
in the rest frame of the decaying particle. Naturally then, for the
decay products, the same final-state showering routines are applied
as are used for the hard scattering process, including those used
in the initialization phase in which starting scales are assigned.
Only the initial conditions for the shower evolution are different,
although their choice is, nevertheless, still based on examining the
colour fl{}ow in the underlying hard decay process. 

The leading order decay products may not, however, be the only particles
to emit radiation. Massive coloured particles, such as the top quark
and particles in many models of physics beyond the Standard Model,
may decay on time-scales shorter than that characteristic of the hadronization
process. Consequently, subject to the available phase space, as well
as undergoing time-like showers $\left(q^{2}>m^{2}\right)$ in their
production phase, these particles will also undergo a further space-like
showering $\left(q^{2}<m^{2}\right)$ of QCD radiation \emph{prior}
to their decay. In contrast to the final-state showers from the decay
products, the initial-state space-like shower created by a decaying
particle is quite different to that of an initial-state particle from
the production process. In particular, it involves no PDFs, since
the heavy parton originates from a hard scattering as opposed to a
hadron. Furthermore, in the hard process it was necessary to evolve
the initial-state partons backwards from the hard scattering to the
incident hadrons, to efficiently sample any resonant structure in
the underlying matrix elements. On the contrary, in decay processes,
degrading the invariant mass of the decaying particle, via the emission
of radiation, does not affect the efficiency with which any resonant
structures in the decay matrix element are sampled. Hence, it is natural
for the evolution of space-like decay showers to start with the unstable
particle from the production process, and evolve it forward, towards
the decay. In the older generation of multi-purpose event generators
the inclusion of this initial-state radiation in decays was not in
general possible due to, for instance, their use of non-covariant
shower formalism.

\subsubsection{QED RADIATION}

In existing Monte Carlo simulations the production of QED radiation
in particle decays is normally simulated using an interface to the
\textsf{PHOTOS} program \cite{Barberio:1993qi,Barberio:1990ms,Golonka:2005pn}.
This program is based on the collinear approximation for the radiation
of photons together with corrections to reproduce the correct soft
limit \cite{Barberio:1993qi,Barberio:1990ms}. In recent years the
program has been improved to include the full next-to-leading order
QED corrections for certain decay processes \cite{Golonka:2005pn,Golonka:2006tw,Nanava:2006vv}. 

As noted previously, in the event that the decay products are heavy,
the divergences associated to multiple collinear emission are subject
to a screening which increases with the mass of the emitter. On the
other hand, since the propagator denominator associated to an emitter
is also proportional to the energy of the emitted particle, soft emissions
are always enhanced regardless of the emitting particle's mass or
the emission angle. This being the case it is often preferable, depending
on the application / particle masses / analysis cuts, to model the
effects of photon radiation using a formalism which resums to all
orders the effects of soft rather than collinear emissions.

In QED soft infrared divergences are resummed to all orders using
the methods originally developed by Yennie Frautschi and Suura \cite{Yennie:1961ad}
(YFS), in doing so the soft real and virtual corrections are seen
to exponentiate%
\footnote{Pedagogical examples of the exponentiation of infrared QED corrections
may be found in Refs.\,\cite{Kaku:1993ym,Peskin:1995ev} %
}. This resummation has been recast as photon shower simulations through
a \emph{tour} \emph{de} \emph{force} in Monte Carlo techniques pioneered
by S.\,Jadach \cite{Jadach:1987ii,Ward:1987jg,Bonneau-Martin,Jadach:1995sp,Jadach:1999vf,Jadach:2000ir}.
Such simulations have no overlap with their QCD counterparts; whereas
in QCD the shower forms through iterating emissions in a Markovian
algorithm, the YFS QED shower algorithms are Poissonian, generating
all photon radiation at the same time in `one-shot', moreover they
have always afforded the possibility to systematically include arbitrarily
higher-order corrections.

Finally, we remark that \emph{soft} resummation in QCD is fundamentally
different since, unlike QED, each emitted parton is itself charged.
Unfortunately such additional complexities arising from the non-Abelian
character of QCD mean that applying the same Monte Carlo approach
to simulating QCD radiation, despite its many attractive features,
is unfortunately not appropriate.

\subsubsection{HARD QCD RADIATION}

Note that the standard QCD parton shower approach has two important
drawbacks. Firstly, because the parton shower generates emissions
from each leg of the hard scattering independently, each additional
emission must be uniquely associated to a particular leg of the hard
scattering, which can only be achieved at the price of having regions
of phase space, corresponding to high $p_{T}$ gluon emissions, which
are unpopulated by the shower, so-called \emph{dead-zones}. Secondly,
the soft / collinear approximation to the QCD matrix elements is plainly
not a good approximation all over the phase-space region populated
by the parton shower. 

One way to rectify these problems is through \emph{matrix element
corrections} \cite{Seymour:1994df} which ensure that the hardest
additional radiated parton in the event is distributed \emph{exactly}
according to the corresponding real emission matrix element. In the
case of \emph{soft}\textsf{ }matrix element corrections to decays,
every emission generated in the shower which is the \emph{hardest
so far} is vetoed if the ratio of the exact differential width to
the parton shower's approximation to it (based on the soft / collinear
approximation to the former) is greater than a random number $\mathcal{R}$. 

Where the parton shower algorithm is such that there is a dead zone in
the phase space of the \emph{first} emission, typically concentrated in the
high $p_{T}$ region, \emph{hard matrix element corrections} are also necessary.
Hard matrix element corrections simply use the same exact real emission matrix
element to generate an event with a hard emission in the dead zone, with a
probability given by the ratio of the integrated cross section in the dead
zone divided by the total cross section of the leading order process. As a 
consequence of the different ordering variables used in the Herwig~/ \herwigpp
and \pythia shower algorithms, the entire phase space for the first emission
is covered in the latter, avoiding the need for hard matrix element corrections.

Alternatively one may use matrix element-parton shower matching techniques
(ME-PS), such as the CKKW method, \cite{Catani:2001cc,Krauss:2002up}
to correct the approximate soft / collinear radiation pattern of the
parton shower in the regions of phase space corresponding to hard
emissions, where in this case the hardness is typically measured in
terms of the Durham jet measure. In general these ME-PS separate phase
space into two regions according to a merging scale, defined in terms
of the jet measure, $y_{\mathrm{merge}}$. The region of phase space
corresponding to values of $y<y_{\mathrm{merge}}$, is deemed to correspond
to sufficiently soft / collinear emissions that one can expect the
shower approximation to reliably distribute emissions there, this
is the so\emph{-}called\emph{ shower region}. Above the shower region
is the \emph{matrix element region}, where emissions are distributed
according to exact fixed order matrix elements. Of course the real
emission fixed order matrix elements do not include any of the virtual
effects resummed in the parton shower and so the emissions in the
matrix element region must be reweighted with Sudakov form factors
and running coupling constants to take these important effects into
account. Without these corrections distributions will exhibit unphysical
discontinuities across the phase space partition at $y_{\mathrm{merge}}$. 

Both the matrix element correction methods and the ME-PS methods effectively
take into account the real emission component of next-to-leading order
corrections but not the virtual effects. Nevertheless, there are many
examples in the literature (albeit for production processes) where
these methods are shown to give excellent agreement with next-to-leading
order calculations from the point of view of the \emph{shapes} of
distributions \cite{Krauss:2004bs,Krauss:2005nu,Gleisberg:2005qq,
Mangano:2006rw,Hamilton:2008pd,Hamilton:2009za}.

\subsection{\protect\herwigpp}

\begin{itemize}
\item Finite width effects\\
Finite width effects are present by default for all processes in
\herwigpp through the inclusion of a weight factor as in Eq.\,(\ref{eq:Breit_Wigner_integral}),
retaining the dependence of the width on the off-shell mass of the
particle, in \emph{both }the production and decay stages. This approach
is used in the simulation of cascade decays, where it was shown to
give good agreement with results obtained using exact matrix elements
in Ref.\,\cite{Gigg:2008yc}.
\item Spin correlations\\
The \herwigpp spin correlation algorithm is precisely that
described in Section\,\ref{sub:Spin-correlations} and was implemented
by the author of Ref.\,\cite{Richardson:2001df}. 
\item QCD radiation\\
The chief success of the older fortran \herwigsix program
was in its accounting of soft gluon interference effects, specifically,
in particular colour\emph{ }coherence phenomena, through the angular
ordering of emissions in the shower. The new \herwigpp algorithm
retains angular ordering as a central feature of the showering algorithm
and improves on it, most notably, in the present context, through
the inclusion of the mass-dependent splitting functions and kinematics,
providing a physical description of the radiation distribution emitted
by massive particles in the low $p_{T}$ region \cite{Gieseke:2003rz,Hamilton:2006ms,Bahr:2008pv}.
The facility to model initial-state parton showers in the decays of
unstable coloured particles was envisaged at the start of the \herwigpp
project and has been implemented in full generality in 2006. 
\item QED radiation\\
The \herwigpp program includes a module which dresses the
decays of light and heavy objects, involving a single electric dipole,
with QED radiation \cite{Hamilton:2006xz}. This simulation uses the
YFS formalism to resum the effects of multiple photon emissions, as
well as a Poissonian shower algorithm following the methods of Jadach
\cite{Jadach:1987ii}. 
\item Hard radiation\\
In implementing colour coherence through angular ordering the \herwigsix 
and \herwigpp shower algorithms contain dead zones in the
phase space for the hardest emission, where the hardest emission occurs
at wide angles with respect to the parent(s). In the context of particle
decays \herwigpp includes matrix element corrections for top
quark decays as described in \cite{Hamilton:2006ms}. This serves
to populate the dead zone for this process and correct the radiation
pattern in the live zones. Matrix element corrections to $W^{\pm}$,
$Z$, and Higgs boson decays are not available, however, due to the
neutrality of the decaying particle and the low mass of the negligible
mass of their decay products, these may be implemented with quite
a modest effort.
\end{itemize}

\subsection{\protect\pythia}

\begin{itemize}
\item Finite width effects\\
  All particles that are defined with a width are distributed
  according to a Breit-Wigner,
  cf. eq.~(\ref{eq:Breit_Wigner_integral}). In case of pure
  $s$-channel processes, such as $e^+ e^- \to \gamma^*/Z^0 \to f
  \overline{f}$, a running width is used, in agreement with LEP1
  conventions. For other processes normally a fixed width is used. The
  inclusion of this width in the matrix-element expressions,
  especially the interference terms, may vary from process to process
  --- \pythiaeight does not have a matrix-element generator of its
  own, but encodes calculations made by many different authors.

\item Spin correlations\\
\pythiaeight does not come with a spin correlation algorithm,
which means that the default is isotropic decays. For many processes
and decays spin correlations are included with full matrix elements, 
however. That is, for a process such as 
$e^+ e^- \to W^+ W^- \to f_1 \overline{f_2} f_3 \overline{f}_4$,
first kinematics (including masses) is selected for the 
$e^+ e^- \to W^+ W^-$ process, and thereafter the decay angles of the
two $W$'s are simultaneously sampled and weighted according to the 
complete $2 \to 2 \to 4$ matrix elements. For cases where spin is 
important and not implemented, users are recommended to use 
external input from dedicated matrix-element programs, such as
MadGraph \cite{Alwall:2007st}. 

\item QCD radiation\\
  The original \pythiaeight shower algorithm was mass-ordered
  \cite{Bengtsson:1986et}, with extra angular-ordering cuts, while the
  current one is transverse-momentum-ordered \cite{Sjostrand:2004ef},
  with a dipole approach to handling recoil effects. One nice feature
  of these algorithms is that they cover all of the phase space, at
  least for the first emission, thus obviating the need for several of
  the special \textsf{HERWIG} tricks described above, e.g.\ to handle
  dead zones or to separate off space-like emissions in resonance
  decays. The modification of the evolution variable from
  $\mathrm{d}m^2 / m^2$ to $\mathrm{d}m^2 / (m^2 - m_0^2)$ gives a
  consistent coverage of the soft/collinear region $m^2 \to m_0^2$ for
  radiation off massive particles\cite{Norrbin:2000uu}. Here $m_0$ is
  the rest mass or, for a resonance, the previously
  Breit-Wigner-sampled mass. This approach is also easily extended to
  the case of evolution in $p_{\perp}^2 = z (1-z)m^2$.

\item QED radiation\\
  The same shower algorithm that does QCD emissions also can handle
  QED ones, in an interleaved manner. That is, QCD and QED emissions
  can alternate in the downwards evolution in $m^2$ or $p_{\perp}^2$,
  and thus compete for momentum. The QCD and QED dipoles do not need
  to agree. \\
  Optionally the \textsf{PHOTOS} package may be used
  \cite{Golonka:2005pn,Golonka:2006tw,Nanava:2006vv}; by
  experimentalists often to handle QED corrections in hadronic
  resonance decays.

\item Hard radiation\\
  The \pythia shower can easily be fixed up to overestimate
  radiation in the hard region, while attaching to the correct
  expression in the soft one (see above). A Monte Carlo rejection
  factor can then be implemented to match the first gluon emission to
  the respective QCD matrix-element expression for essentially all
  decays in the SM and the MSSM \cite{Norrbin:2000uu}. For QED such
  corrections are only included in $\gamma^*/Z^0$ and $W^{\pm}$
  decays.
\end{itemize}

\subsection{\protect\sherpa}

\begin{itemize}
\item Finite width effects\\
In \sherpa finite fixed widths are incorporated in the
propagator denominators of the internal tree-level matrix element
generators \textsf{AMEGIC++ }\cite{Krauss:2001iv} and \textsf{COMIX}
\cite{Gleisberg:2008fv}. The fixed width scheme is employed within
these matrix element generators.
\item Spin correlations\\
In order to include spin correlations between production and decay
\sherpa essentially employs the first algorithm described
in the corresponding section above. The versatility of the
implementation is enhanced by \sherpa's ability to
automatically generate the matrix elements needed for the production
and decay spin density matrices internally. Along this line,
\sherpa allows on the level of hard matrix-element generation
for the specification of the production and decay processes in a
cascade-like manner. As a result contributions from non-resonant
diagrams can be omitted while one preserves off-shell mass effects and
spin correlations in the generation of the fully decayed final states.
To incorporate spin correllations in the decays of hadrons and 
$\tau$-leptons the simple rejection algorithm outlined above is employed. 
\item QCD radiation\\
The new generation of \sherpa versions 1.2.x uses the
\textsf{CSS} parton-shower algorithm \cite{Schumann:2007mg}, which is
based on (massive) Catani--Seymour dipole subtraction kernels and a 
dipole-like picture of shower evolution. Emissions are ordered in transverse
momentum rather than virtuality. The \textsf{CSS} implements truncated 
showering, therefore handles final-state and
initial-state emissions in the production and decay of unstable
coloured particles, respectively. The \sherpa parton shower of
versions 1.1.x and older, \textsf{APACIC++} \cite{Krauss:2005re}, is
closely related to that of the \pythia virtuality ordered
shower \textsf{\cite{Sjostrand:2006za}}.
\item QED radiation\\
\sherpa includes a native simulation of QED radiation in
decays, \textsf{ PHOTONS++} \cite{Schonherr:2008av}; like the \herwigpp
QED shower this module is built upon the YFS resummation formalism
and the Poissonian shower algorithms of the Krakow group \cite{Jadach:1987ii,Ward:1987jg,Bonneau-Martin,Jadach:1995sp,Jadach:1999vf,Jadach:2000ir}.
\item Hard radiation\\
Here again \sherpa makes use of its internal matrix element
generators to generate amplitudes for the leading order processes
in the presence of additional QCD radiation. The key point is that
processes of different parton multiplicity in the final state can be combined
consistently with parton showers (generated by the \textsf{CSS}
algorithm) and hadronization included according to the new strategy of
matrix-element and truncated-shower merging (ME\&TS)
\cite{Hoeche:2009rj,Hoeche:2009xc,Carli:2009cg}. This approach
improves over the CKKW method, implemented in versions 1.1.x and
older, since it guarantees an unified treatment of local scales in
the calculation of the matrix elements and parton showers. Hence, one
finds that the systematic uncertainties of the ME\&TS merging are
sizeably reduced with respect to CKKW. Using the ME\&TS matching,
extra hard radiation can be reliably described in multi-jet and
high-$\pt$ scenarios. For example, \sherpa employs the new
method to include exact real emission corrections to the $\ttbar$ 
production process, as well as the $t$ and $\bar{t}$ decay
processes, in the region of phase space above the ME\&TS merging
scale.
\end{itemize}
 

\clearpage

\section[SENSITIVITY OF QCD JET MASS AND JET SUBSTRUCTURE TO PILE-UP AT LHC]
{SENSITIVITY OF QCD JET MASS AND JET SUBSTRUCTURE TO PILE-UP AT LHC\protect 
\footnote{Contributed by: P. Loch, P. Francavilla}}
\label{sec:detsubs}



\subsection{Introduction}
\label{sec:dsIntroduction}
New methods to measure the hadronic decays of boosted massive particles at the LHC, 
like $W$ bosons \cite{Butterworth:2002tt}, top quarks (e.g., \cite{Kaplan:2008ie,Thaler:2008ju,Brooijmans:2008zz}), 
Higgs boson \cite{Butterworth:2008iy}, and BSM particles (e.g., \cite{Butterworth:2007ke}), 
have been proposed in recent years. 
Due to the boost of these massive particles,  
the decay products are collimated in one direction, and they are reconstructed 
as one single jet. The properties of these jets, such as their masses and 
jet sub-structure quantities, can be used to separate them from the standard QCD jets.  
Several strategies have been developed to distinguish these two
sources and to study the properties of the corresponding jets, including the use of jet algorithms with variable size 
parameters \cite{Krohn:2009zg}, jet pruning, see e.g. \cite{Ellis:2009su}, and jet trimming \cite{Krohn:2009th}.

In this study we investigate the impact of the pile-up on the measurement of these observables.
The presence of the pile-up interactions affect the low transverse momentum (\pT) spectrum.
We expect that it strongly affects the 
jet energy calibration in the very low \pT\ region, becoming negligible for high \pT\ jets.
The estimate of the effects on the jet mass and on the the jet-substructure quantities is less
straightforward because these quantities are complex QCD observables.

Another important aspect is the capability to properly reconstruct and calibrate the jets.
There are several approaches to assess these points. Most of them are finalized at the calibration 
of the jet as a single global object starting from calorimetric signals.
In this case, the capability to properly calibrate every component (i.e. calorimetric signals) 
of the jet is not straightforward and 
the precision on the measurement of the substructure should be investigated in a detailed analysis of the 
detector performance. This analysis is beyond our goal. 

In this study only the effects of the pile-up and one simple strategy to suppress it are investigated.
In order to separate these effects from the 
detector calibration, we reconstructed jet from Monte Carlo generated particles (hadrons).
The selection cuts, used to suppress the pile-up, are driven by consideration 
of the detector effects, especially for the reconstruction of low \pT\ particles, 
i.e. the effect of the magnetic field bending, energy losses in inactive material, and of the signal selection.

\subsection{Jet reconstruction, observables and selection cuts}
\label{sec:dsDetector}
                
In this study we used a sample of QCD di-jet events in proton-proton collisions with center
of mass energy $\sqrt{s}=10$ TeV simulated with \pythiaeight.120 \cite{Sjostrand:2006za,Sjostrand:2007gs}.
The generation was divided in different samples according to the following 
bins (GeV) in  \pThat: [17,34], [34,70], [70,140], [140,280],
[280,560], [560,1120], $>1120$. For the first six samples, 500,000 events were
simulated, for the highest $\hat{p}_T$ sample a simulation of 200,000 events was used.

The effect of the pile-up is taken in account by adding four (Poisson distributed) additional proton-proton interactions per
event. These interactions were simulated using the 
parameter "SoftQCD:all" in \pythiaeight.120, i.e. they include elastic, single diffractive, double diffractive scattering and 
minimum bias interactions.

All the particles generated by the hard scatter and by the pile-up are used to build the final jets.
The jet algorithm used for this purpose is the anti-\kT\ \cite{Cacciari:2008gp} algorithm with distance parameter $D=0.4$. 
Only the leading jet is used in this study. It is required to point into the pseudo-rapidity range $-3.2 \leq \eta \leq 3.2$.

The observables under consideration are the jet \pT, its mass \SUB{m}{jet}, and the y-scale \yscale\ \cite{Butterworth:2002tt} associated with the last clustering step in the formation of the jet.
For the Anti-\kT\ jet algorithm, this last observable is not directly available, as it only has a meaningful definition for the \kT\ \cite{Catani:1992zp,Ellis:1993tq} 
jet algorithm.
We therefore re-clustered the constituent particles of the Anti-\kT\ jet using the \kT\ algorithm, thus allowing us
to calculate its \yscale.

The jet  \pT, \mj, and \yscale\ are then compared jet-by-jet to the corresponding 
observables in jets which have been reconstructed using only 
the particles generated by the hard scattering process. Naming \PU{O}\ the observables in the presence 
of the pile-up and \HS{O}\ the observables with only the hard scatter particles, the ratio 
$\PU{O}/\HS{O}$  is shown in the following plots as a function 
of the number of pile-up interactions (\nvtx) generated in the event.
To properly study these ratios, we should start the jet clustering using only 
the particles generated by the hard scatter. In this paper we report a different way 
of defining the jets. We use all particles (hard scatter and pile-up) to reconstruct the jets and then
we look at the constituents of the leading jet. From the list of constituents, we use the particles 
generated by the hard scatter to build the reference jet (and to calculate \HS{O}).
This final reference jet is different from the jet reconstructed using only the hard scatter particles, but 
due to the peculiarities of the Anti-\kT\ jet algorithm, 
we can expect a rather small difference introduced by this method.

The observables \PU{O}\ depend on the number of pile-up interactions.
To suppress the effect of the pile-up, which usually adds a set of low \pT\ particles to the hard scatter,  
we used a threshold on the particle \pT, thus eliminating low \pT\ particles.
This method to suppress the effect of the pile-up is approximate for jets reconstructed in the calorimeter, but gives us an estimate of how a selection on the constituents of the jets, aimed to suppress contributions from pile-up, could change the dependence of the 
observables as a function of the number of pile-up interactions. 
We applied four different \pT\ thresholds, namely: none, 0.5 GeV, 1 GeV,  and 2 GeV. 
The order of magnitude of this selection is similar to the effective cut-off that affect the charged particles 
in reaching the calorimeter
due to the bending of the magnetic field, as in the \atlas detector \cite{Aad:2008zzm} at \lhc, and the 
typical minimum energy required in the central region to generate a signal in the calorimeter at all ($\pT= 0.4$ GeV).

These cuts are applied in two ways. The goal of the first one is the estimate of the 
effect of a pile-up suppression after the jet reconstruction. For this purpose the cut is
applied to the list of constituents of the jet, after its reconstruction.
The second way emulates a signal selection prior of the jet reconstruction
in that the cut is applied 
before the jet reconstruction.
We refer to the first method as an inclusive selection and to the second as the exclusive selection.
Due to the difference in the clustering, the reference is different in the two cases.

In the following section the results for the different observables are shown.

\subsection{Results}
\label{sec:dsResults}

\subsubsection{Inclusive selection}
In this section we show some of the results obtained using the inclusive selection discussed above.
The results are divided in different plots according to the windows in \pThat\ for the 
generation of the event. In Fig.\ref{fig:Figure1} the dependence of $\SUB{p}{T,HS+PU}/\SUB{p}{T,HS}$
as a function of number of pile-up interaction (\nvtx) is shown. The different colors show the 
dependence for the selection cuts used in the analysis according to the following
convention: red for none, blue for 0.5 GeV, green for 1 GeV and black for 2 GeV.
The different plots show that the variation as a function of \nvtx
is smaller for bigger values of the threshold, but introduces a bias especially for low \pT\ (17-34 GeV).
The 0.5 GeV and 1 GeV  thresholds seem to be the more appropriate, because they reduce the variation on the
number of pile-up interactions to less than 1\% while introducing a bias smaller than 1\% in the 
jet energy scale for jets with $\pT > 280$ GeV. The 2 GeV threshold suppresses the dependence on \nvtx, but the bias is bigger than 1\% in some region of \pT\ and a corresponding correction of the jet energy scale would be needed to reach an accurate calibration.
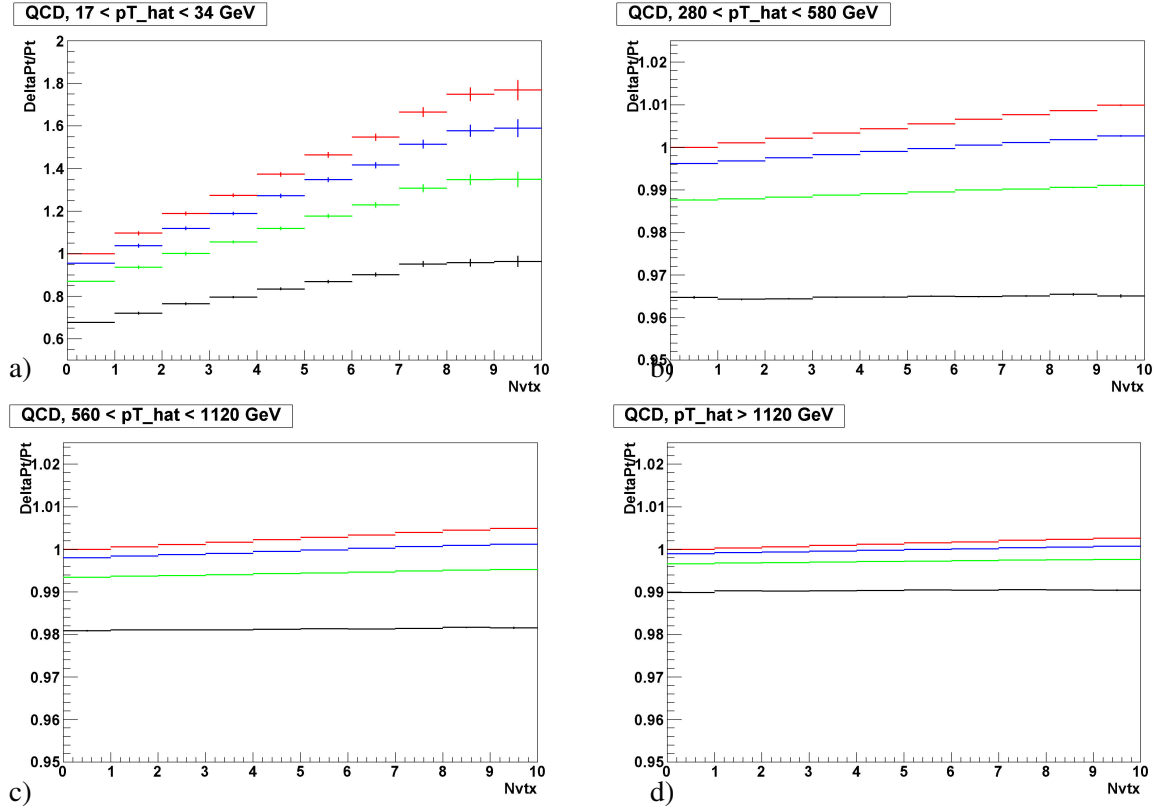
\begin{figure}[t]
\centering
\includegraphics[width=0.49\textwidth]{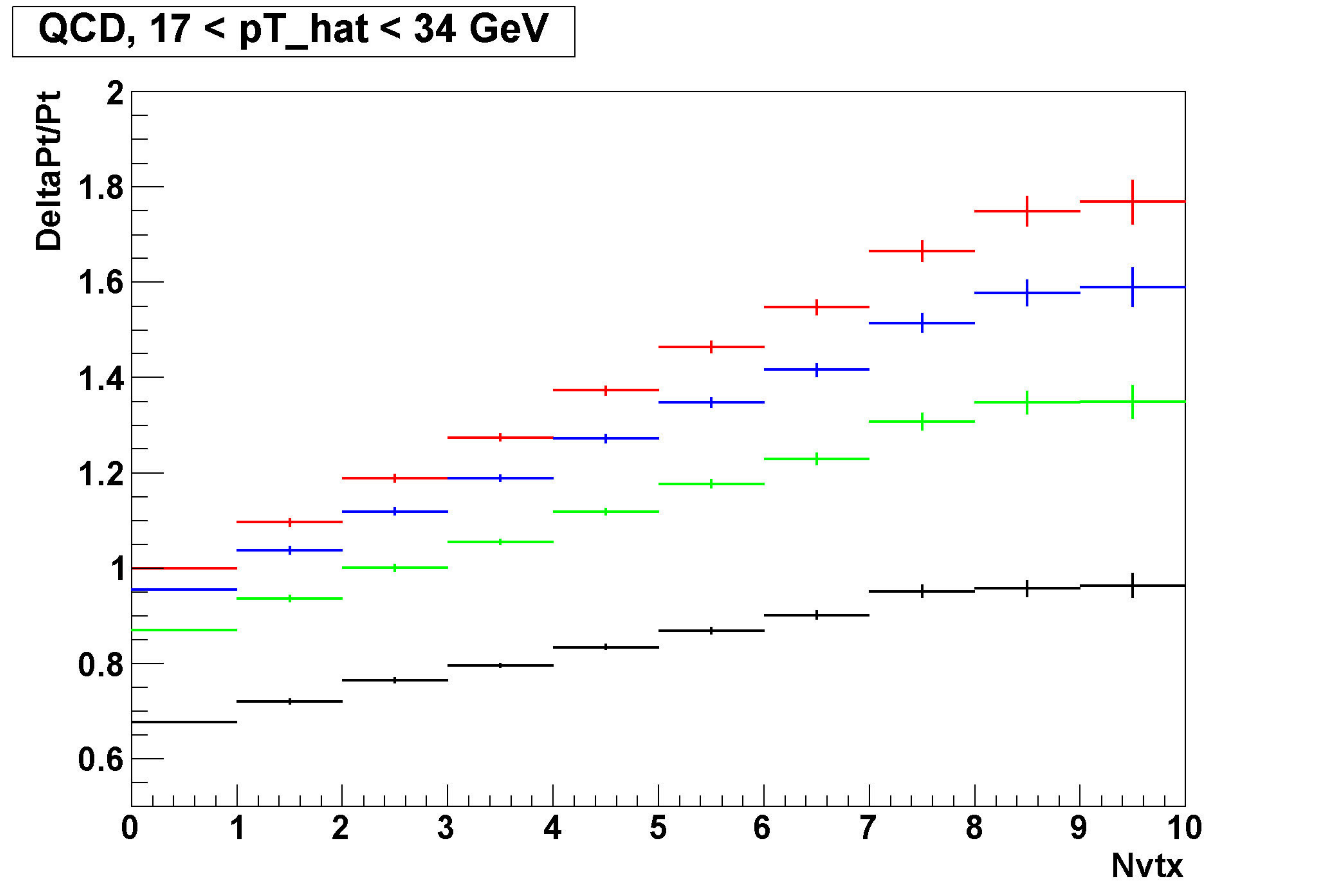}
\includegraphics[width=0.49\textwidth]{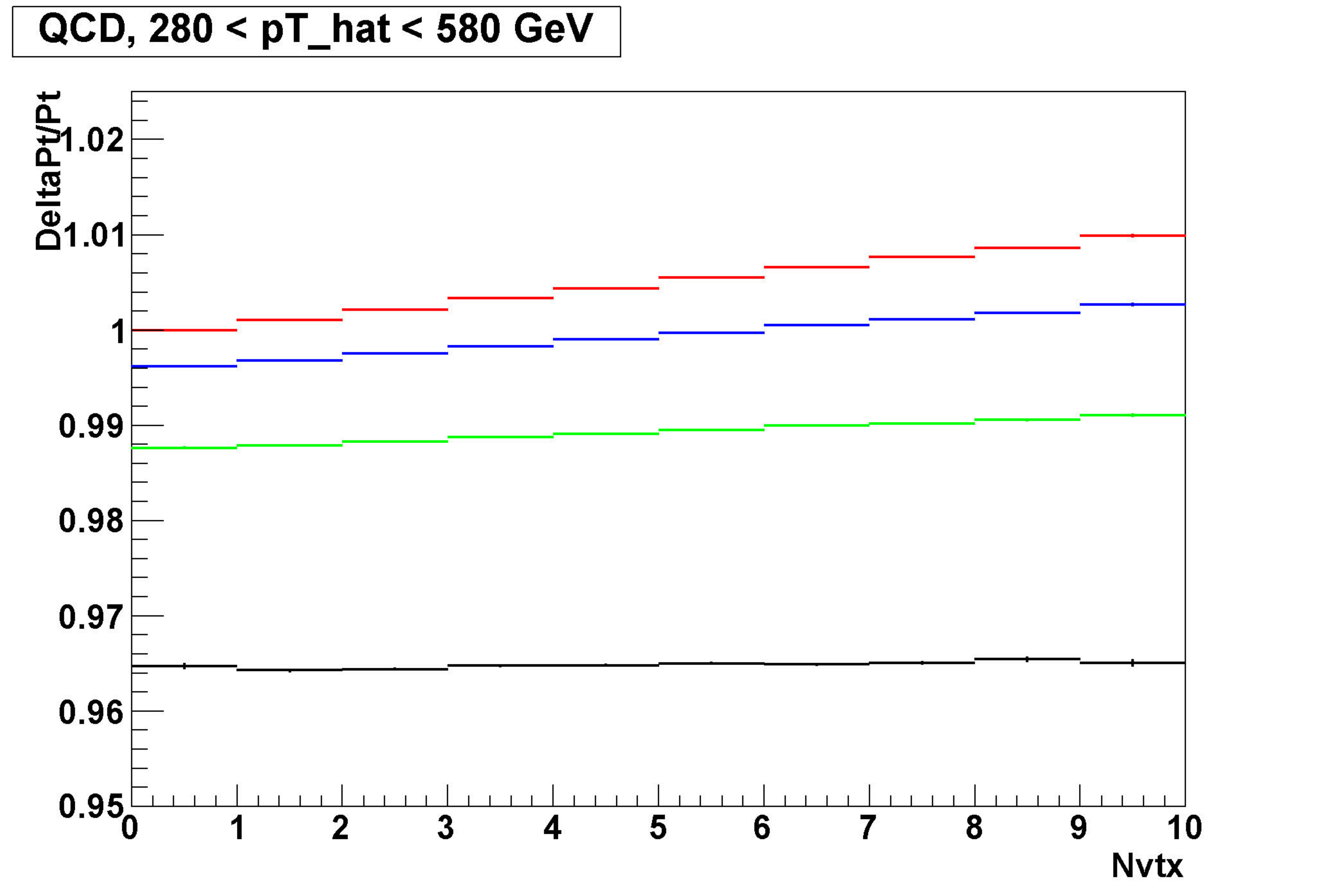}
\includegraphics[width=0.49\textwidth]{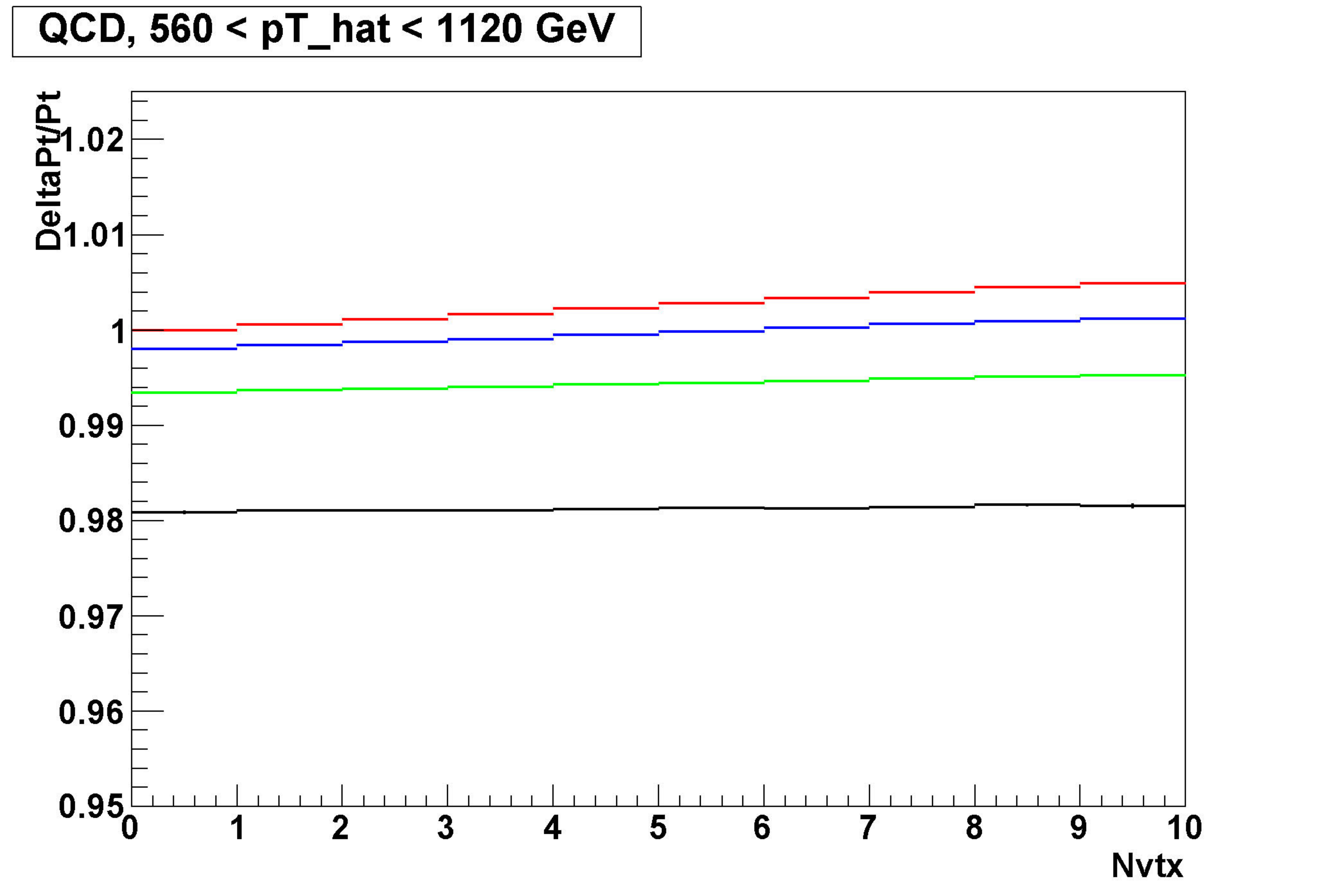}
\includegraphics[width=0.49\textwidth]{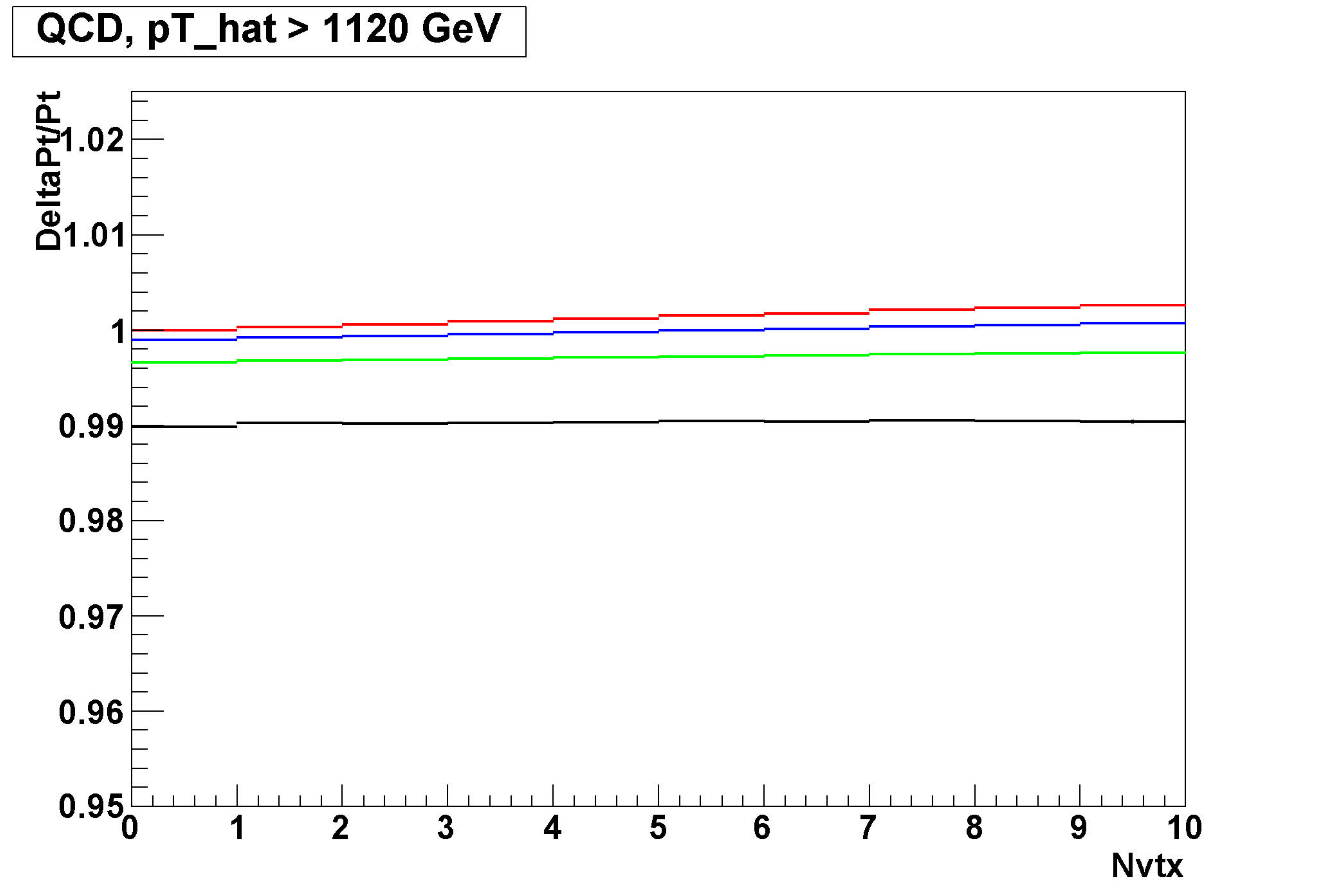}
\begin{picture}(0,0) \put( -450,160){a)} \put( -210,160){b)}\put( -450,0){c)} \put( -210,0){d)}   \end{picture}
\caption{\label{fig:Figure1}
Inclusive selection: comparison of the ratios of $\Delta\pT/\pT = (\SUB{p}{T,HS+PU}-\SUB{p}{T,HS})/\SUB{p}{T,HS}$ as a function of the number 
of pile-up interactions for different windows of \pThat. Here red indicates no constituent \pT\ threshold, while
blue/green/black indicate thresholds of 0.5/1.0/2.0 GeV, respectively.
}
\end{figure}

In Fig.\ref{fig:Figure2} the variations of the jet mass \mj\ are shown in various kinematic bins.
These plots show similarities with respect to the figure \ref{fig:Figure1}. Even in this case,
a selection of the constituents reduces the impact of the pile-up interactions.
The variation due to pile-up in absence of any constituent \pT\ thresholds  cuts is of the order of 
20\% even for the high \pT\ jets, indicating the need for a selection.
Again, thresholds of 0.5 GeV and 1 GeV are the most appropriate, reducing the pile-up
dependence to 10\% and 5\%, respectively, while introducing a respective bias of the order of 5\% and 10\%.
The threshold of 2 GeV, which reduces significantly the pile-up variation from 20\% to almost none, introduces a 20\% bias.
\begin{figure}[tb]
\centering
\includegraphics[width=0.49\textwidth]{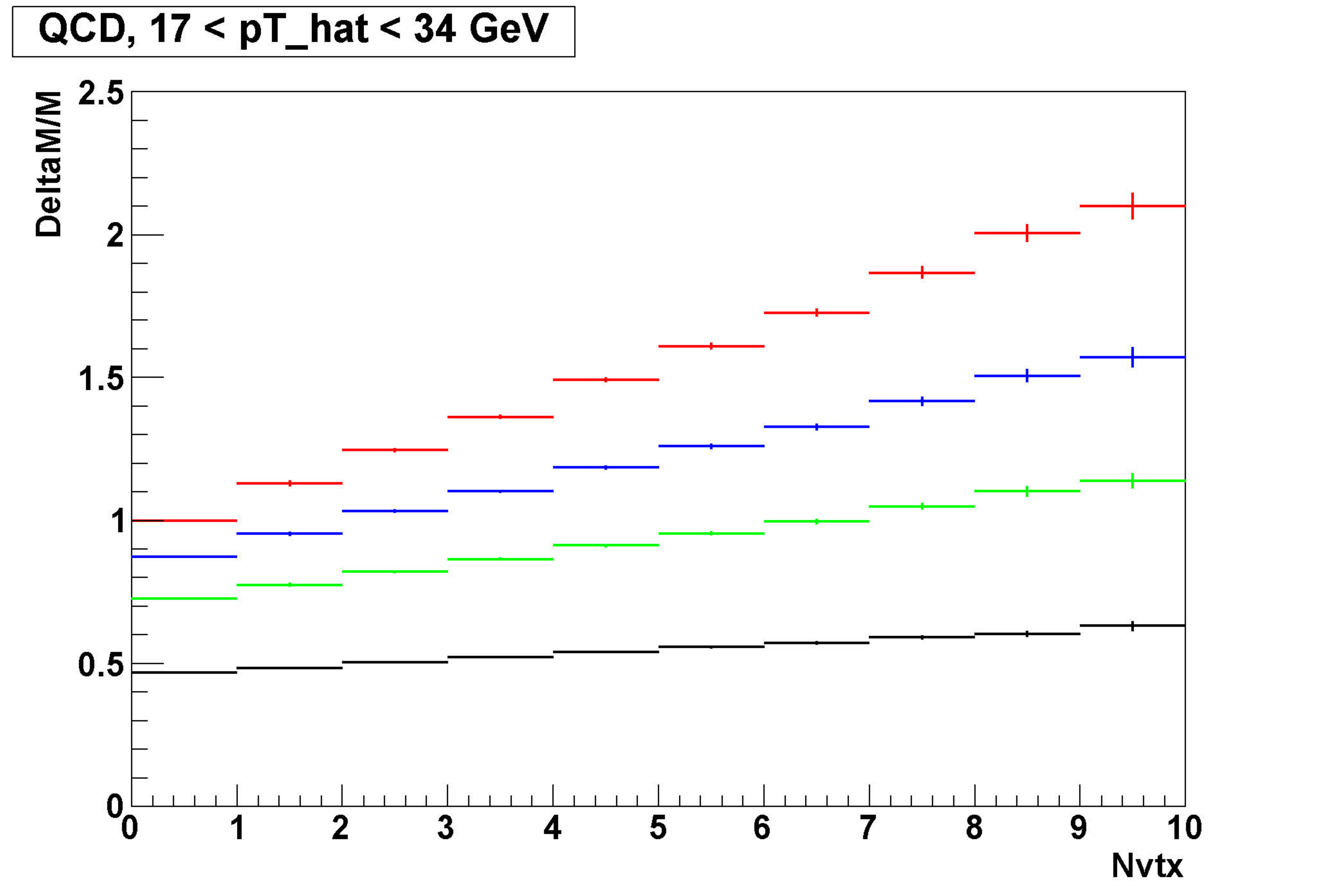}
\includegraphics[width=0.49\textwidth]{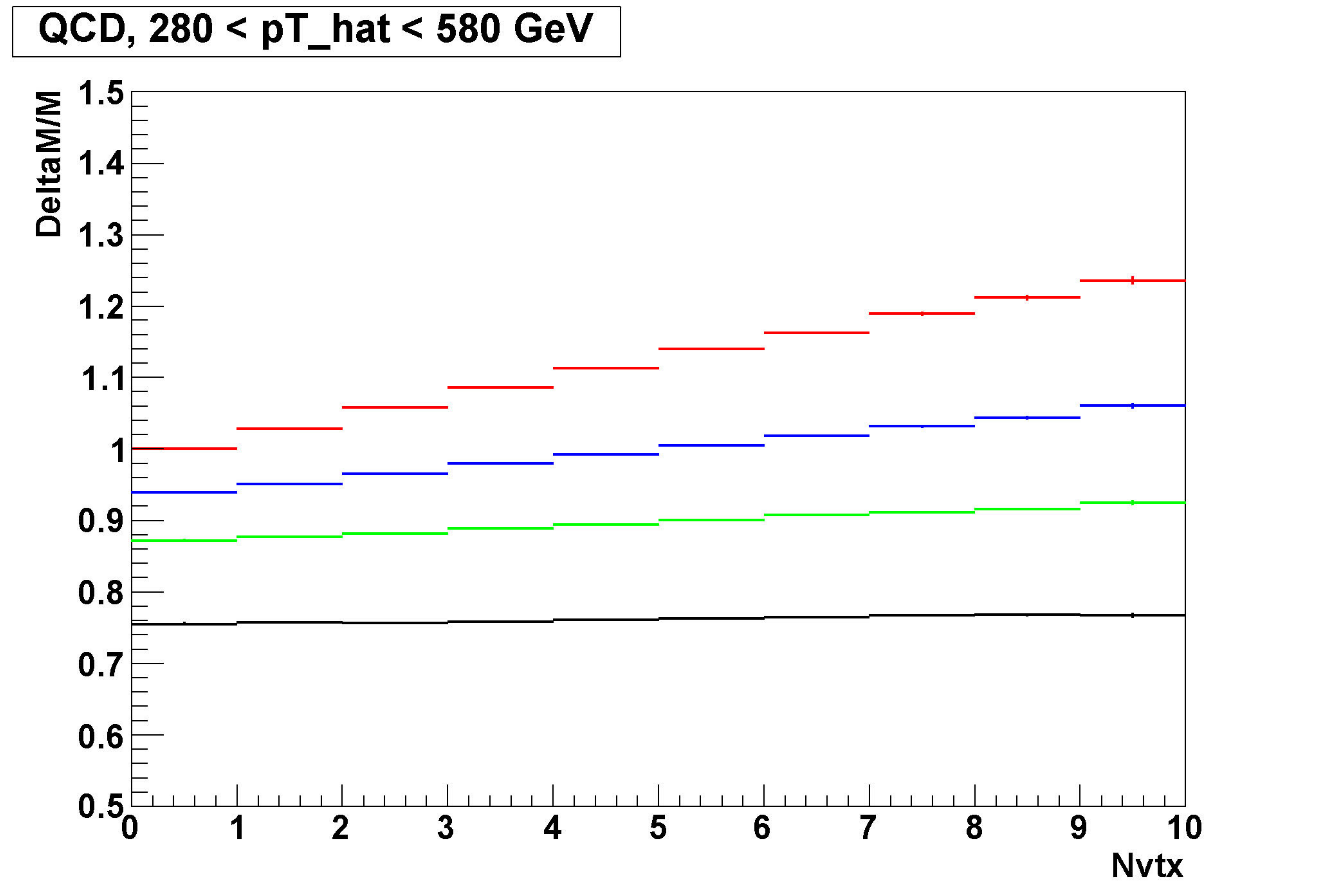}
\includegraphics[width=0.49\textwidth]{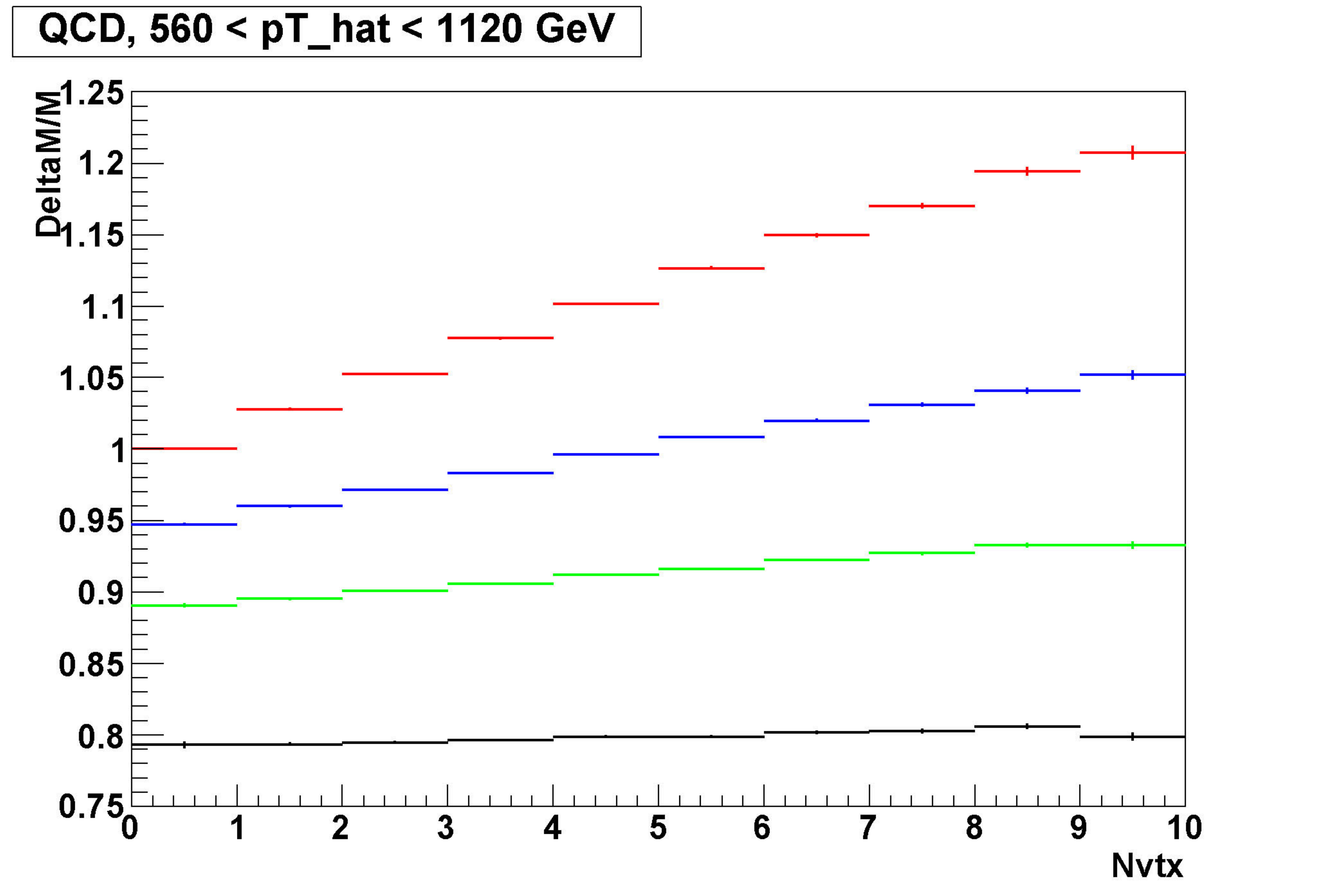}
\includegraphics[width=0.49\textwidth]{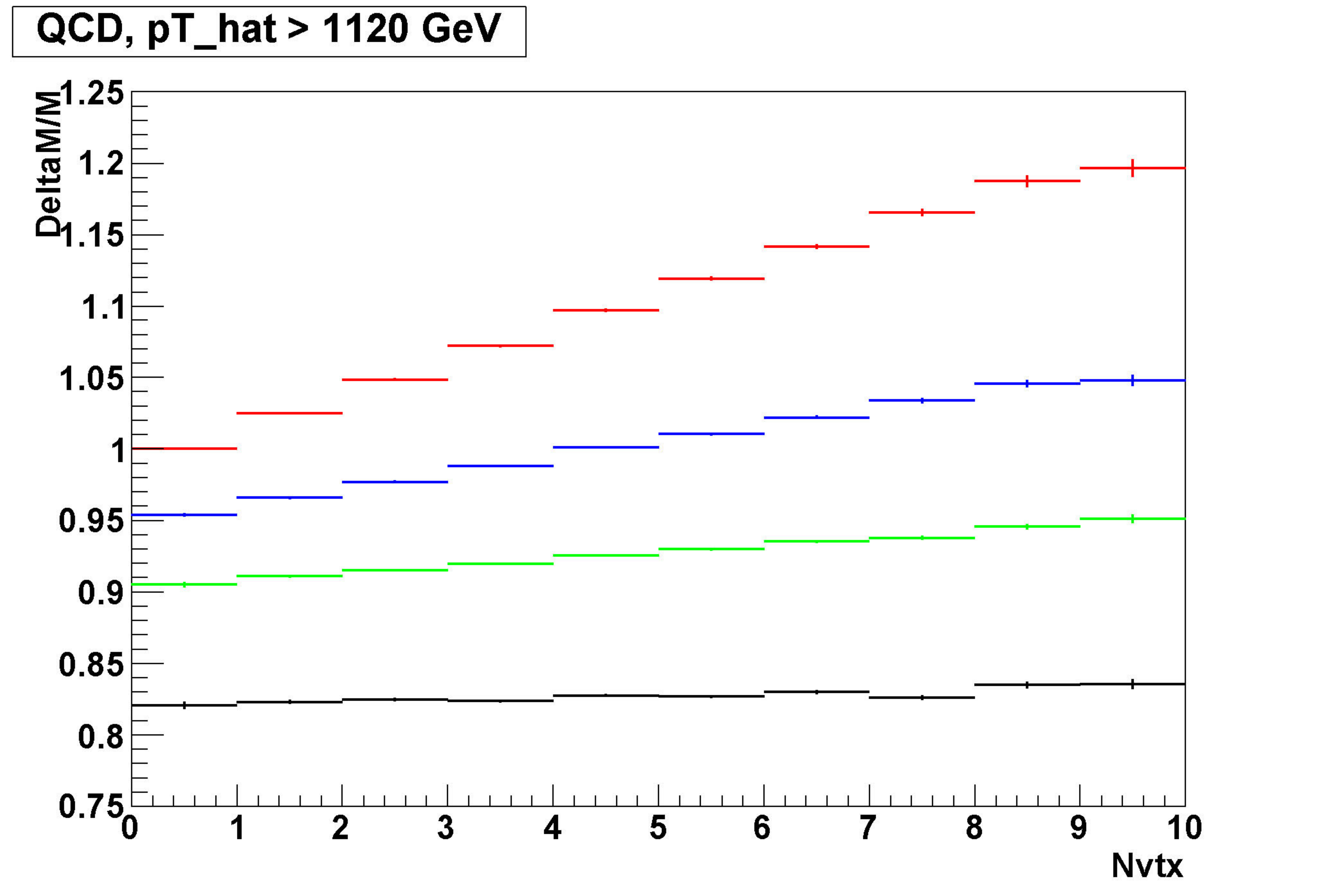}
\begin{picture}(0,0) \put( -450,160){a)} \put( -210,160){b)}\put( -450,0){c)} \put( -210,0){d)}   \end{picture}
\caption{\label{fig:Figure2}
Inclusive selection: comparison of the ratios of $\Delta \mj/\mj = (\SUB{m}{jet,HS+PU}-\SUB{m}{jet,HS})/\SUB{m}{jet,HS}$ 
as a function of \nvtx, for different windows of \pThat. 
Again, red indicates no threshold, while 
blue/green/black indicate jet constituent \pT\ thresholds of 0.5/1.0/2.0 GeV, respectively.
}
\end{figure}
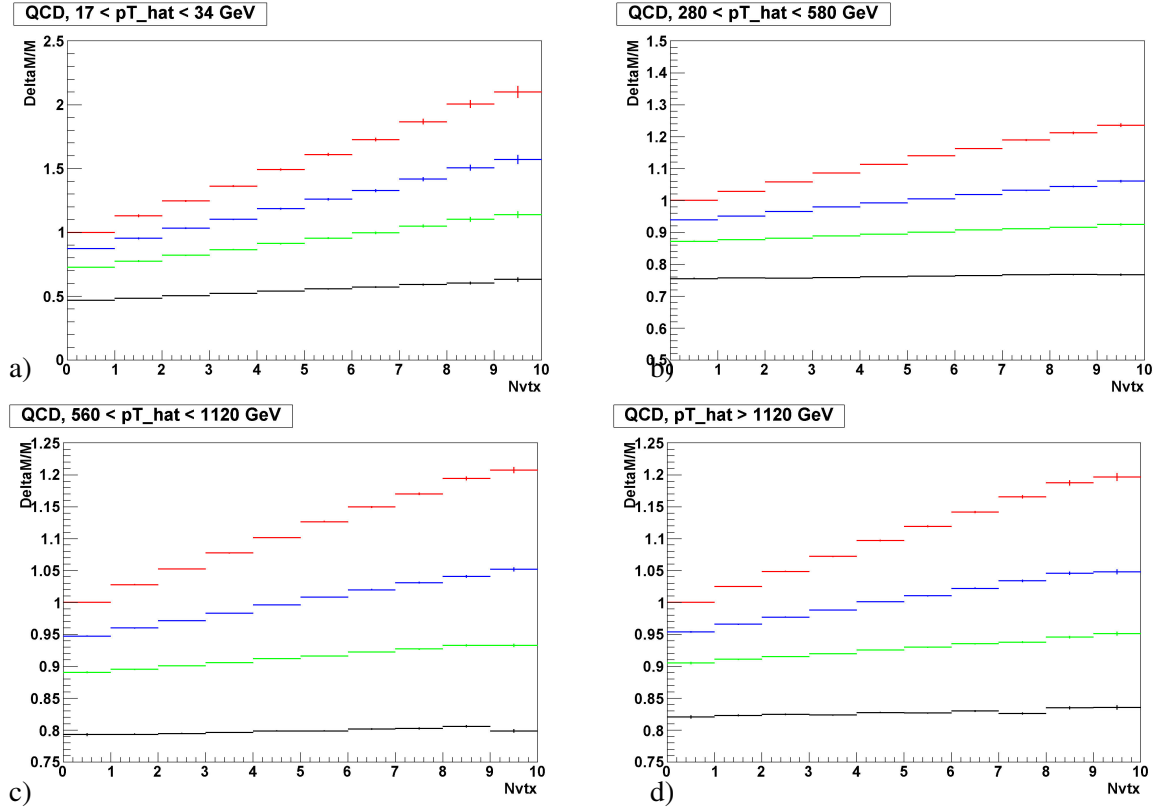

A similar behavior is shown in figure \ref{fig:Figure3} for \yscale.
In this case, the 1 GeV threshold shows a good suppression of the pile-up dependence, together with a bias of less than 5\%.
\begin{figure}[tb]
\centering
\includegraphics[width=0.49\textwidth]{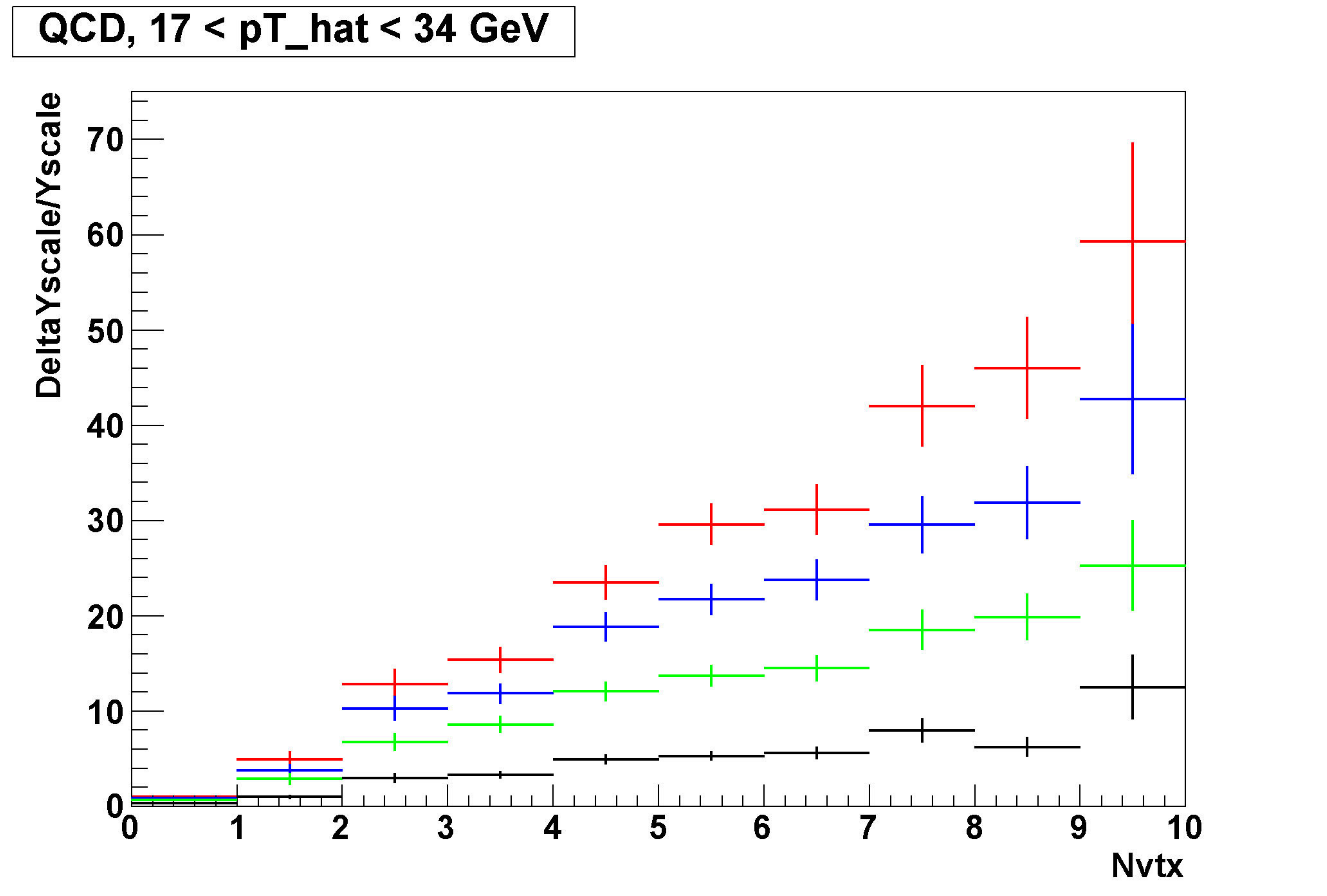}
\includegraphics[width=0.49\textwidth]{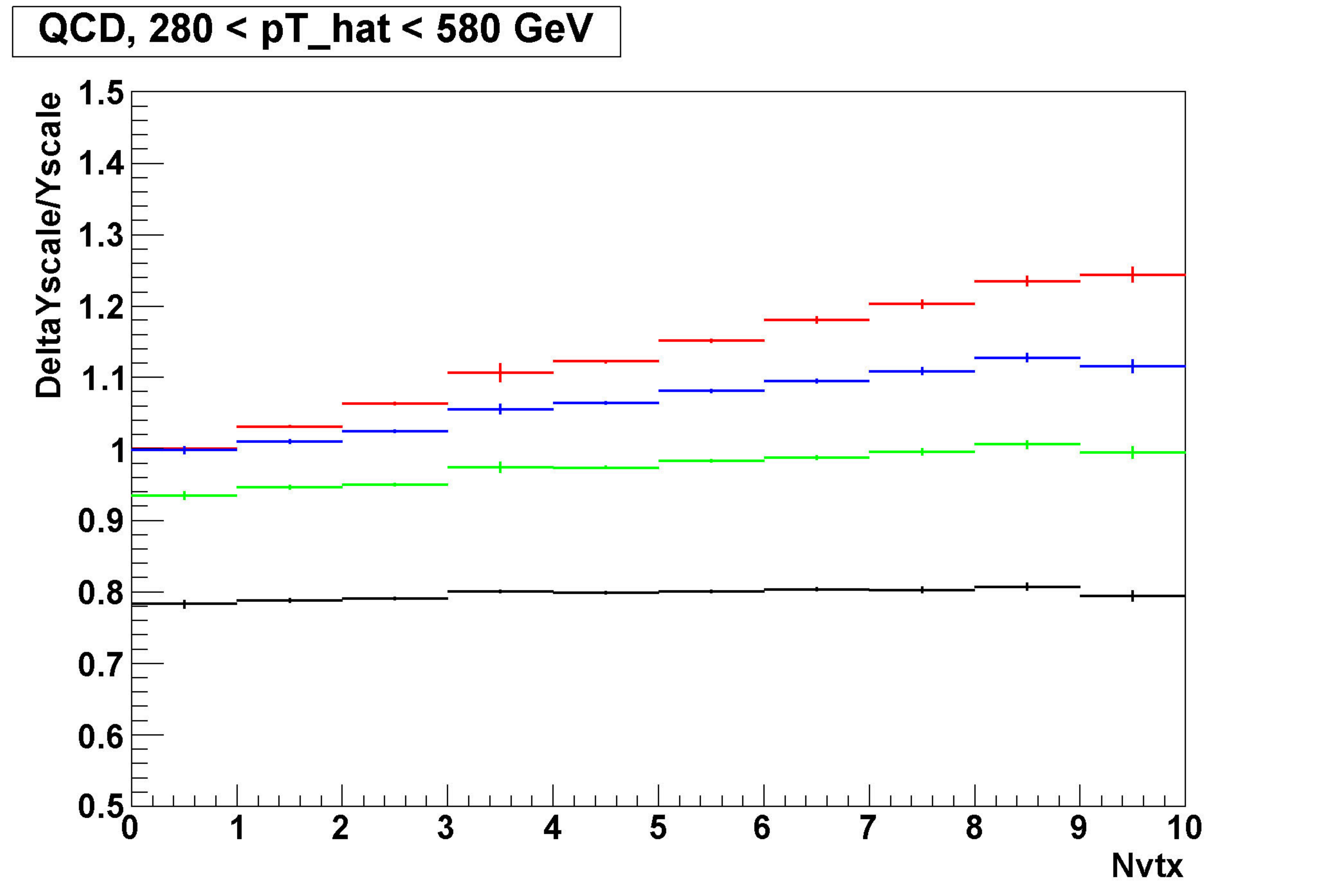}
\includegraphics[width=0.49\textwidth]{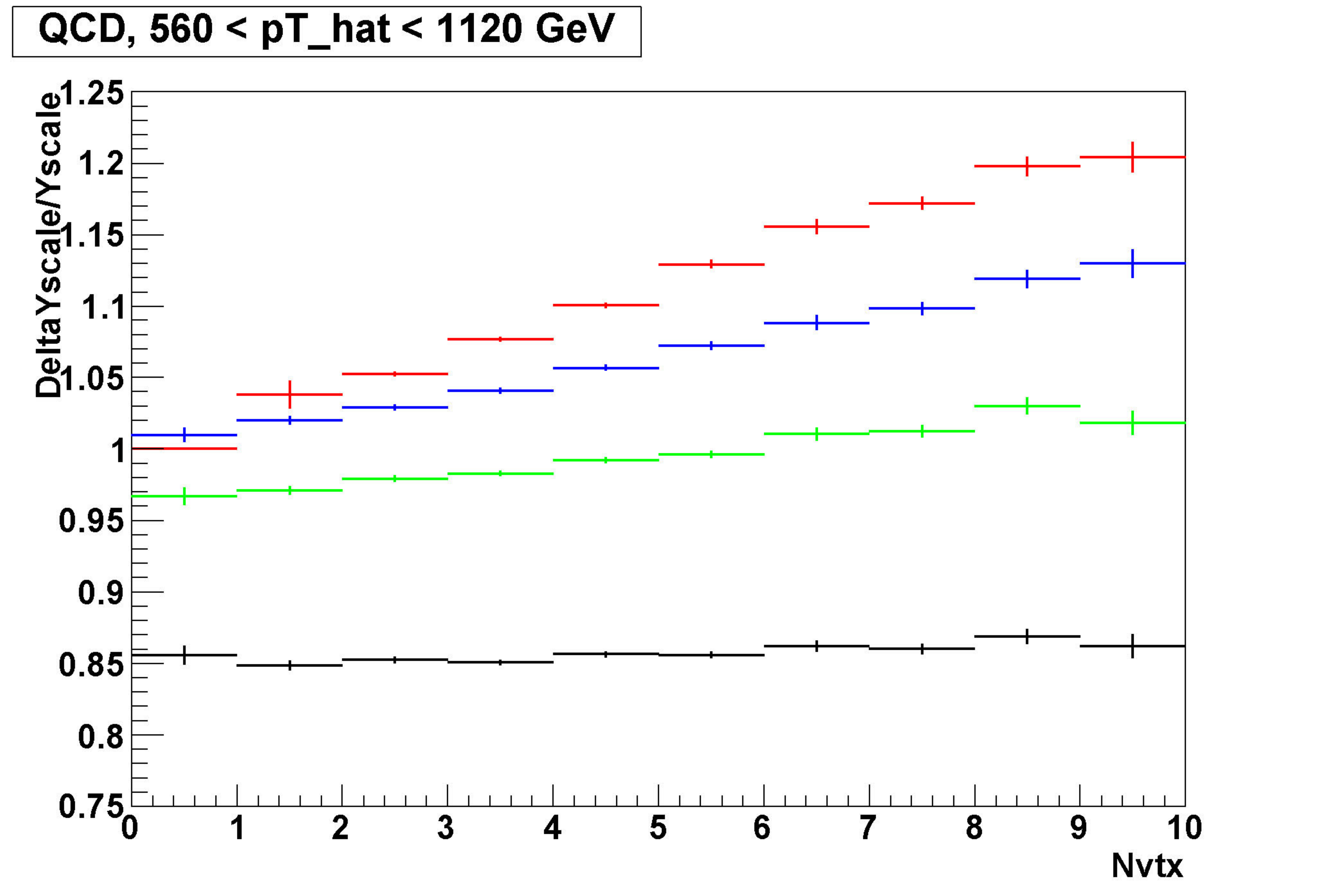}
\includegraphics[width=0.49\textwidth]{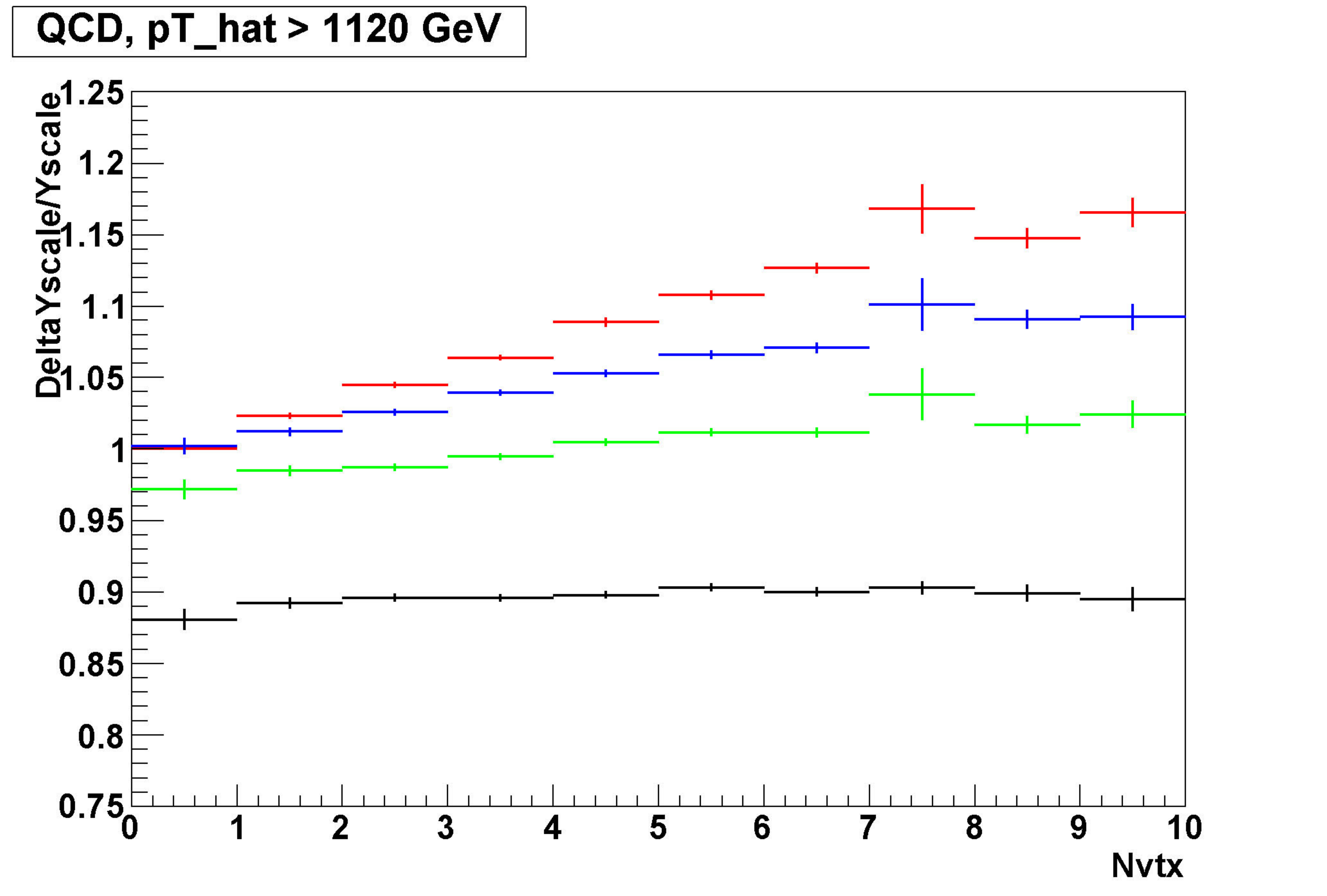}
\begin{picture}(0,0) \put( -450,160){a)} \put( -210,160){b)}\put( -450,0){c)} \put( -210,0){d)}   \end{picture}
\caption{\label{fig:Figure3}
Inclusive selection: comparison of the ratios of $\Delta\yscale/\yscale = (\SUB{y}{scale,HS+PU}-\SUB{y}{scale,HS})/\SUB{y}{scale,HS}$ as a function of \nvtx for different windows of \pThat (red: no selection, blue/green/black 05/1.0/2.0 GeV thresholds). 
}
\end{figure}

\subsubsection{Exclusive selection}
In the case of the exclusive selection, a particle selection cut using the \pT\ thresholds is applied prior to
jet reconstruction. In this case, the constituents differ
from the inclusive selection even for the reference. The purpose of this part is the evaluation 
of an intrinsic selection due to instrumental effects, such as already mentioned bending of charged particle tracks in the magnetic field and the effect of inactive material in front of the calorimeter. As for the inclusive selection, we expect
a dependence of the observables on the number of pile-up interactions \nvtx. This dependence 
is reduced by applying the \pT\ based selections.

A comparison of the inclusive and exclusive selection is shown in Fig.\ref{fig:Figure4}.
This plot shows the ratio $\SUB{p}{T,HS+PU}/\SUB{p}{T,HS}$ for the two selections for different 
\pT\ thresholds. Suppressing the low \pT\ particles already in the reference jet, the bias is no more 
visible, and only the residual dependence on the pile-up activity is shown. This dependence 
is smaller than the dependence shown in the inclusive plot.
\begin{figure}[tb]
\centering
\includegraphics[width=0.49\textwidth]{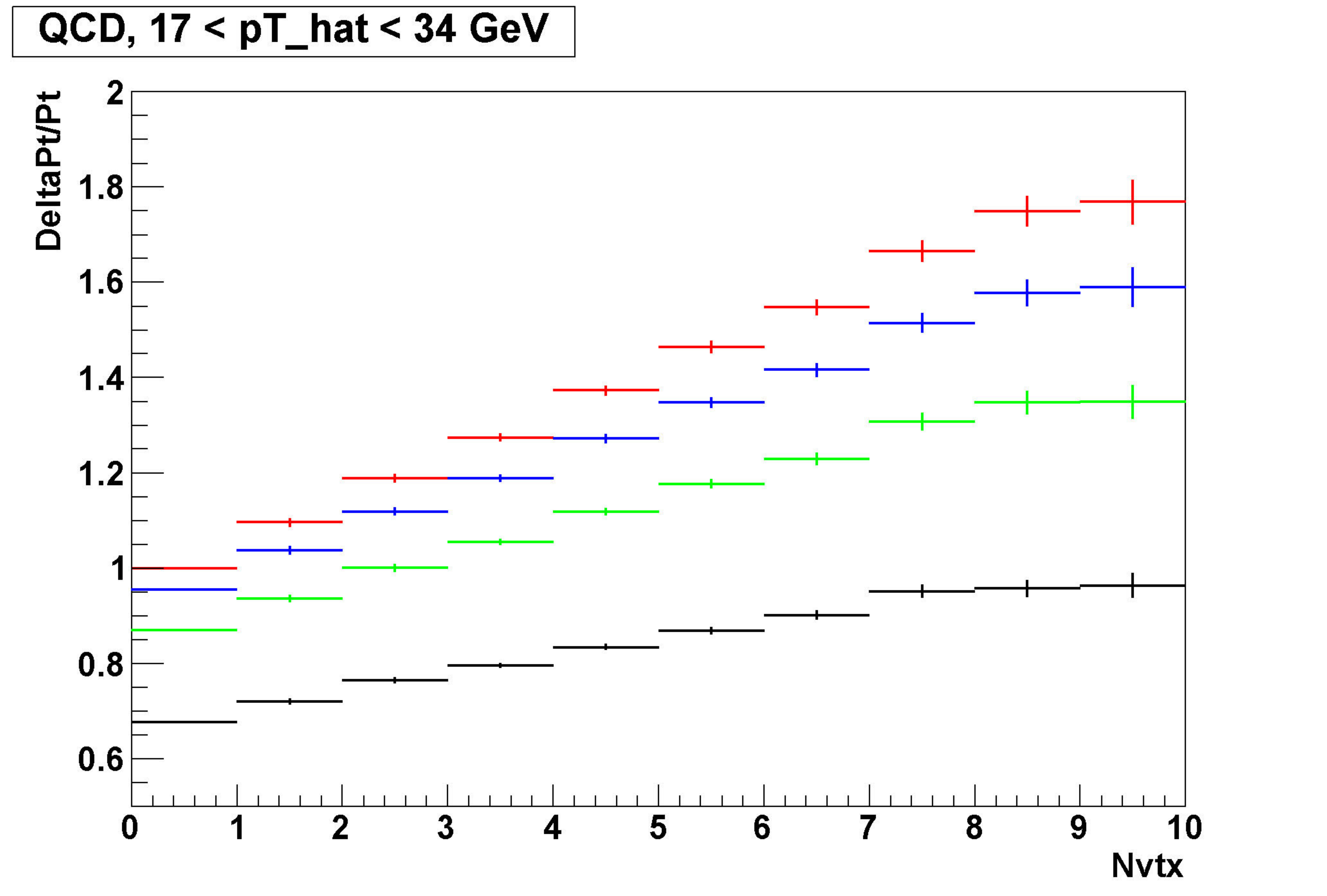}
\includegraphics[width=0.49\textwidth]{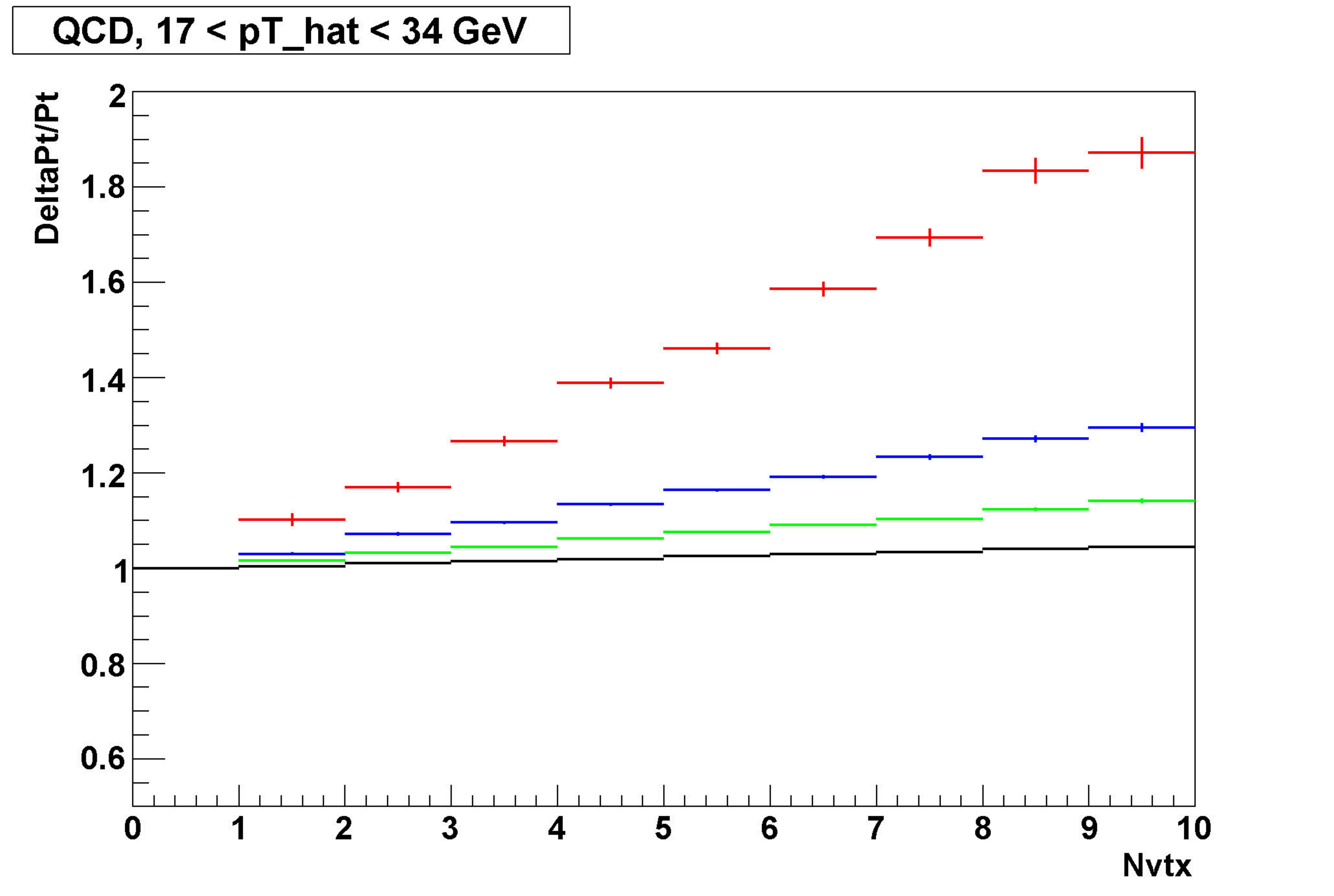}
\begin{picture}(0,0) \put( -400,90){Inclusive} \put( -180,90){Exclusive} \end{picture}
\caption{\label{fig:Figure4}
Comparison of the effect of exclusive and inclusive selections: 
the ratios of $\Delta\pT/\pT = (\SUB{p}{T,HS+PU}-\SUB{p}{T,HS})/\SUB{p}{T,HS}$ are shown as a function of the number 
of pile-up interactions \nvtx. The red points indicate no threshold, while blue/green/black correspond to 05/1.0/2.0 GeV thresholds, respectively. 
}
\end{figure}

The results for the jet mass \mj\ and the y-scale \yscale\ are shown in Fig.~\ref{fig:Figure5}.
These four plots show that the variations of \mj\ and \yscale\ are very similar, 
and 1 GeV or bigger \pT\ threshold applied to the final state particles is useful to reduce this dependence.
\begin{figure}[tb]
\centering
\includegraphics[width=0.49\textwidth]{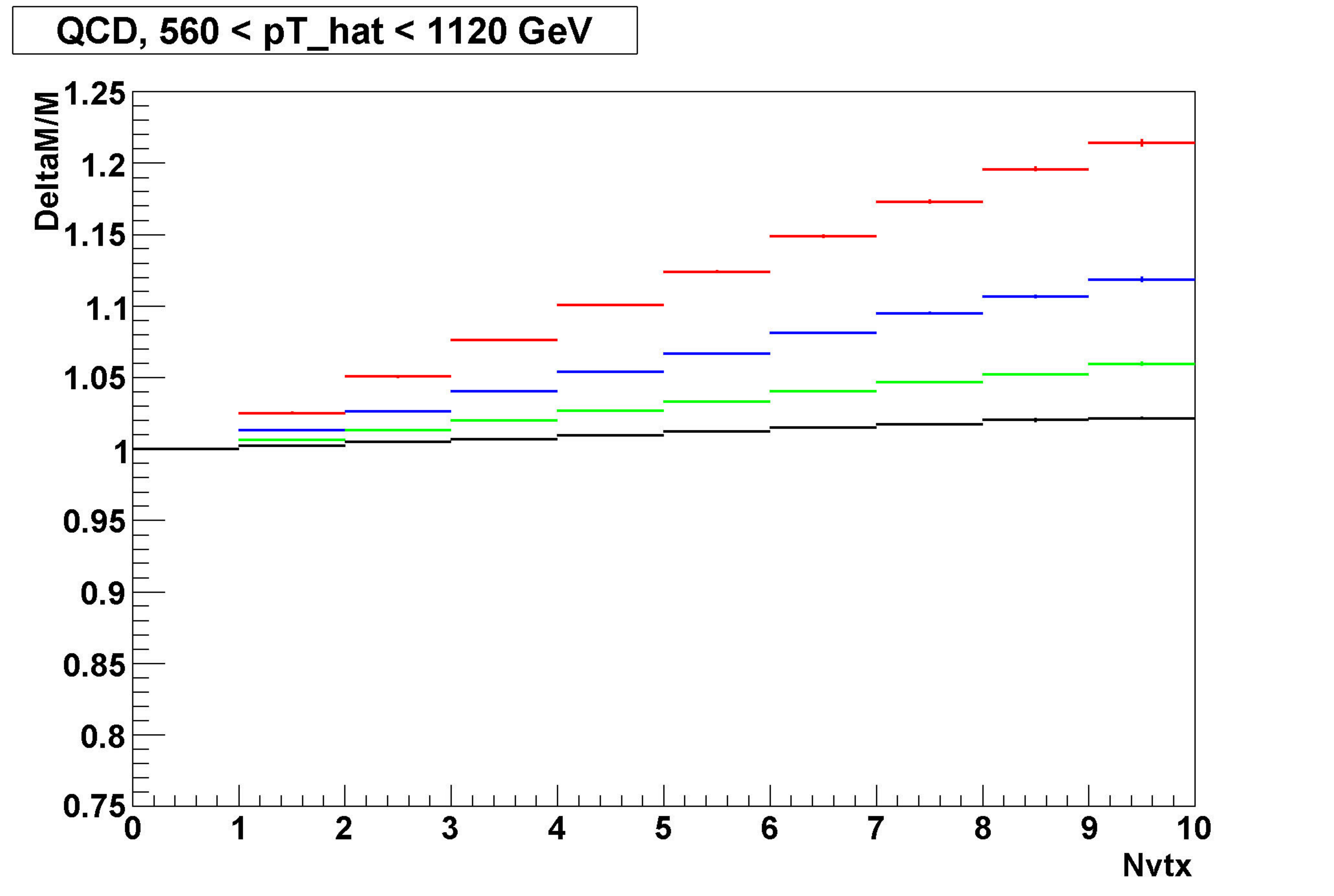}
\includegraphics[width=0.49\textwidth]{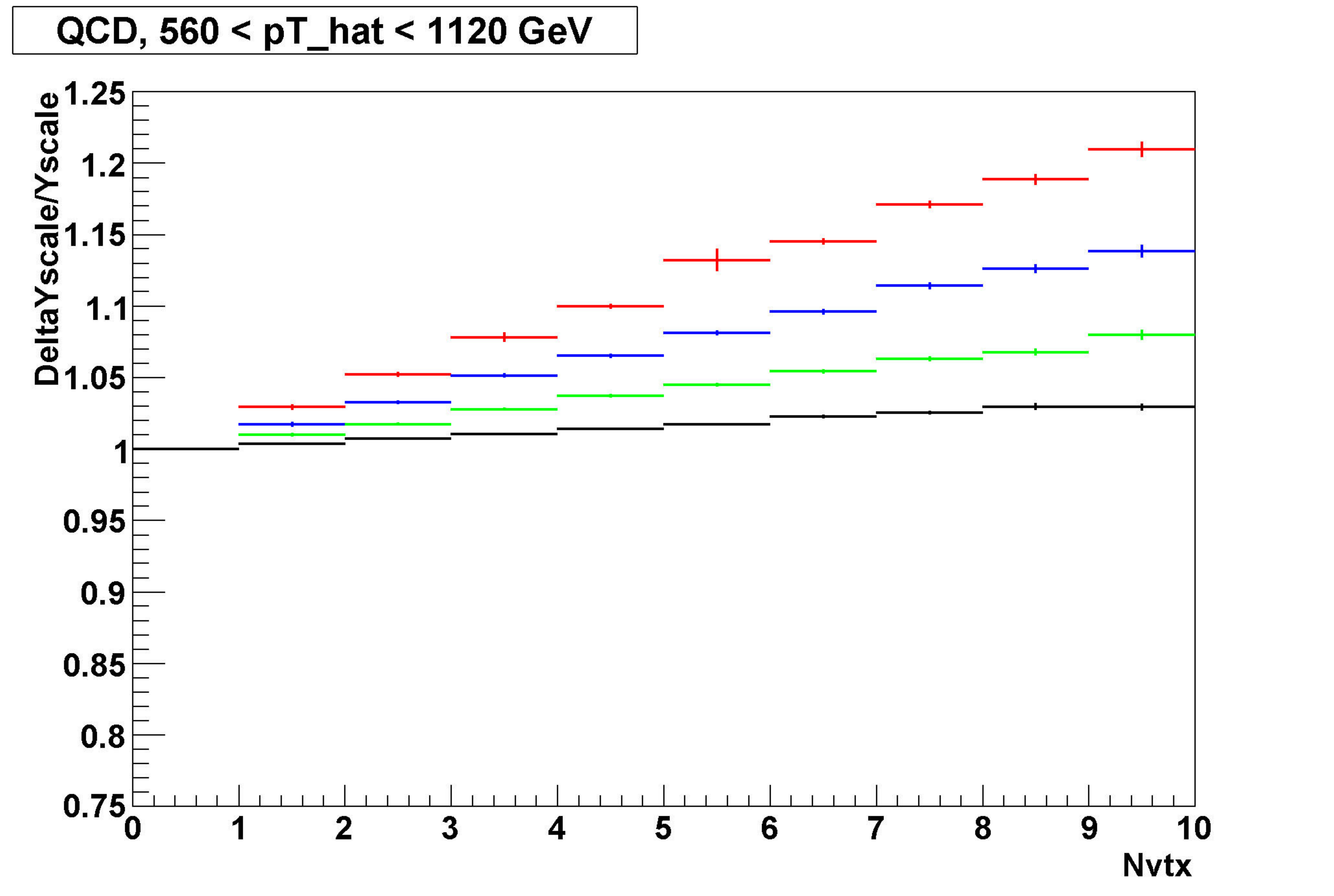}
\includegraphics[width=0.49\textwidth]{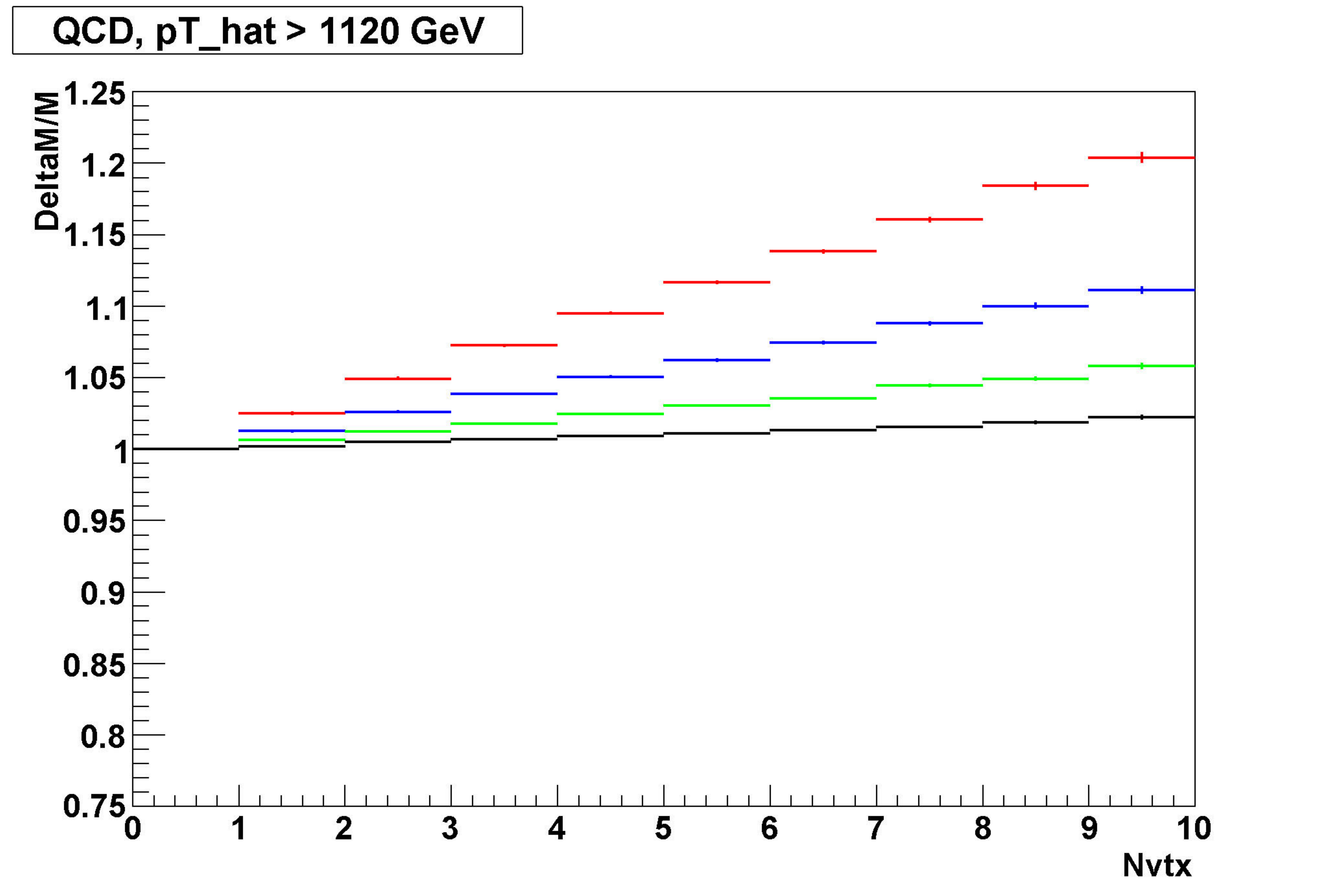}
\includegraphics[width=0.49\textwidth]{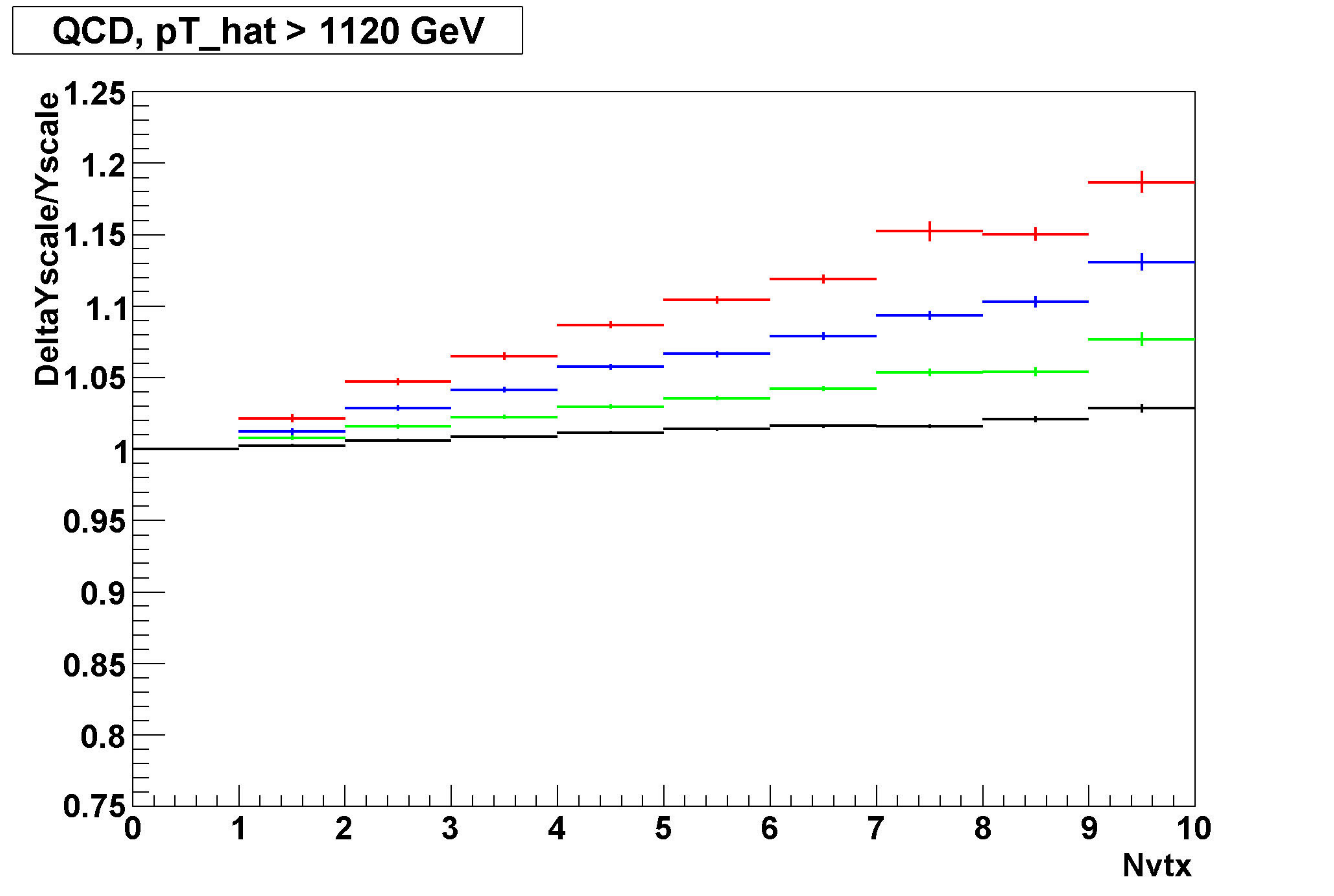}
\begin{picture}(0,0) \put( -450,160){a)} \put( -210,160){b)}\put( -450,0){c)} \put( -210,0){d)}   \end{picture}
\caption{\label{fig:Figure5}
Exclusive selection: comparison of the ratios of $\Delta\mj/\mj = (\SUB{m}{jet,HS+PU}-\SUB{m}{jet,HS})/\SUB{m}{jet,HS}$ and $\Delta\yscale/\yscale = (\SUB{y}{scale,HS+PU}-\SUB{y}{scale,HS}/\SUB{y}{scale,HS}$ as a function 
of \nvtx, for different windows of \pThat\ (red: no selection, blue/green/black 0.5/1.0/2.0 GeV \pT thresholds).}
\end{figure}

\subsection{Conclusion}
\label{sec:dsConclusions}

The studies described in this paper show the dependence of some of the 
jet substructure observables on the pile-up activities.
The variation in the jet transverse momentum is less prominent for high \pT\ jets, as expected. 
In addition, a \pT\ threshold cut on the constituents seems to be useful  to accurately calibrate the jet energy scale for high \pT\ jets. 

For the jet mass and the y-scale, the variation introduced by the pile-up is of the order of 
2\% per pile-up interaction. If we consider an activity of between 0 to 10 pile-up interactions per signal event,
the total variation is of the order of 20\%, even for high \pT\ jets. 
In this case a pile-up suppression is needed and applying a \pT\ threshold on the particles seems 
to reduce the dependence to 5-10\% while introducing a small bias of 5-10\%.

The effect of the pile-up has to be compared with the experimental capability to 
reconstruct and calibrate the observables under investigation. This part is beyond the goal of this paper 
and it could be evaluated in a realistic detector simulation, which takes in account all the 
detector effects. The effect of the pile-up and its suppression is an important ingredient to 
be taken into account when using complex QCD observables like jet masses and substructure in an LHC analyses.

\clearpage

\section[A STUDY OF RADIATION BETWEEN JETS AT THE LHC]
{A STUDY OF RADIATION BETWEEN JETS AT THE LHC
\protect \footnote{Contributed by: M.~Campanelli, J.~Monk, J.~Robinson, C.~Taylor}}
\label{sec:radiation}


\subsection{INTRODUCTION}
The pattern of radiation between jets (or between jets and the proton remnant) ought to be very different depending upon whether the exchanged state is a colour singlet or colour octet object.
Particles are expected to flow differently
in $\phi$ regions both close to and far from the jets.  The typical $\eta-\phi$
scale of the radiation patterns is also relevant. 
Traditionally, either due to uncertainties in the theoretical predictions or because of 
the poor granularity of calorimeters used in hadron colliders, few of the
fine details of radiation have been studied.  Events generated by colour
singlet or colour octet exchange have therefore been distinguished in the past by
applying either a cut on the total sum of transverse momentum between the two jets
\cite{Abachi:1994hb}, or on the transverse momentum of a third jet between the two
\cite{Rainwater:1998kj}. The presence of Underlying Event  (UE) and especially of
pile-up in hadron collisions will give rise to a very small efficiency for finding
events with a clean gap between the jets.  The efficiency may also be very 
sensitive to the details of the specific models. Radiation between jets has
to be studied in data, using the most powerful techniques to separate
the various contributions, and possibly reduce the effect of UE or pile-up. 
Here we present three different classes of observables that can be used
for such studies:
the distribution of radiation outside jets as a function of distance from the jet
axis, using different jet algorithms; the ``gap grid'' method, consisting of 
dividing the detector in several parts along $\eta$ and $\Phi$ to exploit the 
different space distribution of QCD radiation; and the application of a
one-dimensional Fourier transformation to the energy deposition in the
event to highlight specific structures characterised by a given size.

\subsection{RADIATION DISTRIBUTION CLOSE TO THE JET BOUNDARIES}
As a first method, we studied the radiation distribution from events generated with \herwigsix 
\cite{Corcella:2002jc} close to the jet
boundaries, for various jet algorithms. This shows how the various algorithms 
differ for average QCD events at the LHC. Figure \ref{fig:radiationalgos}
shows the Et-weighted distribution of the $\Delta R$ distance of truth-level
particles from the axis of the leading jet. Using different 
algorithms with radius (or measure) 0.4, figure   \ref{fig:radiationalgos}a shows the distribution
of particles belonging to the jet, while  \ref{fig:radiationalgos}b) shows  particles
outside of the jet.  The Anti-\kt algorithm shows a clear
cut-off at the nominal value of the jet radius, as expected by its definition,
while the other algorithms have a smoother behaviour.  In particular the
\kt algorithm has the largest amount of radiation outside the nominal jet size.

\begin{figure}[ht]
\begin{minipage}[h]{0.47\textwidth}
\begin{center}
\begin{overpic}[width=\textwidth]{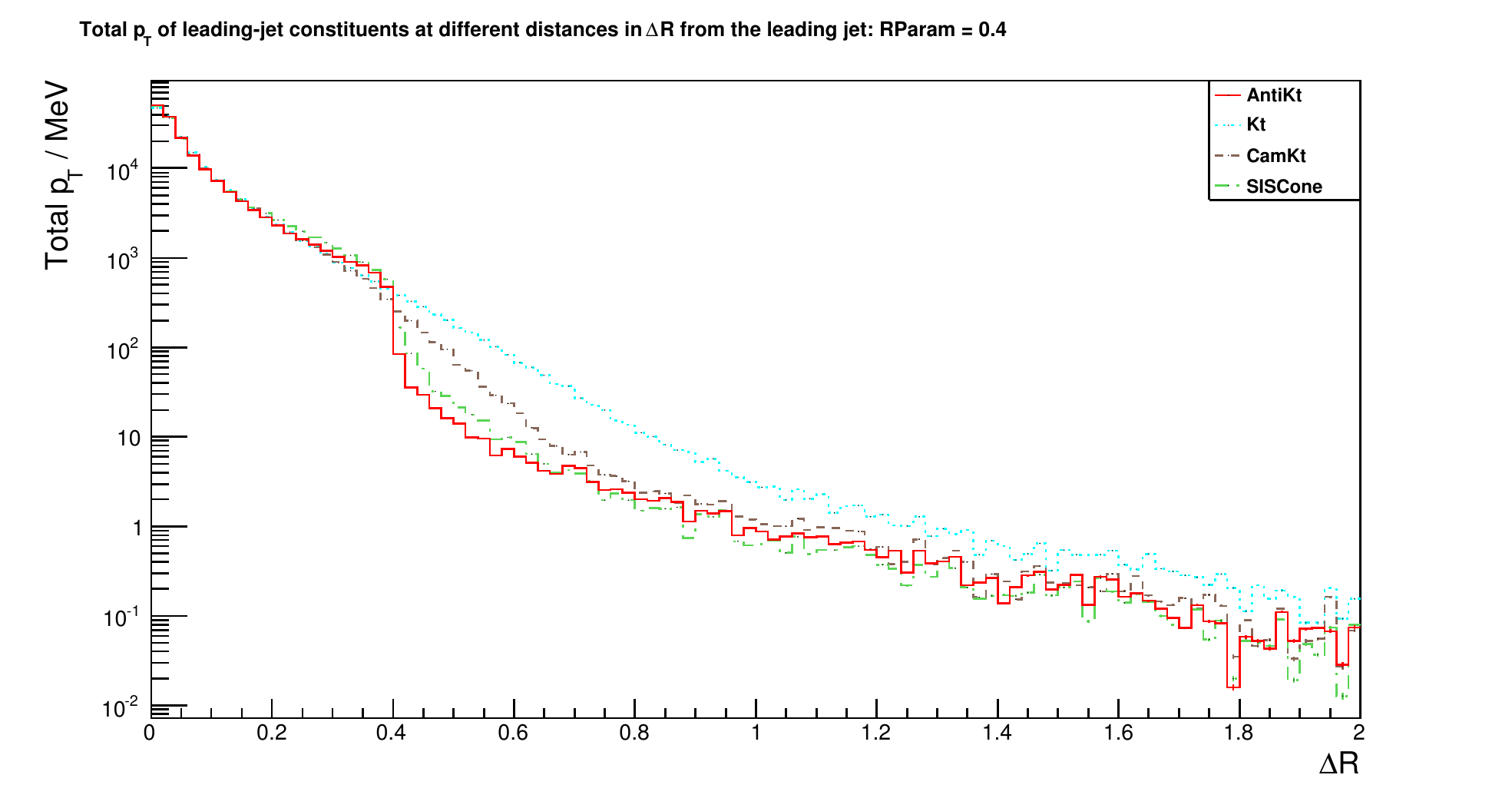}
\put(80,30){a)}
\end{overpic}
\end{center}
\end{minipage}
\begin{minipage}[h]{0.47\textwidth}
\begin{center}
\begin{overpic}[width=\textwidth]{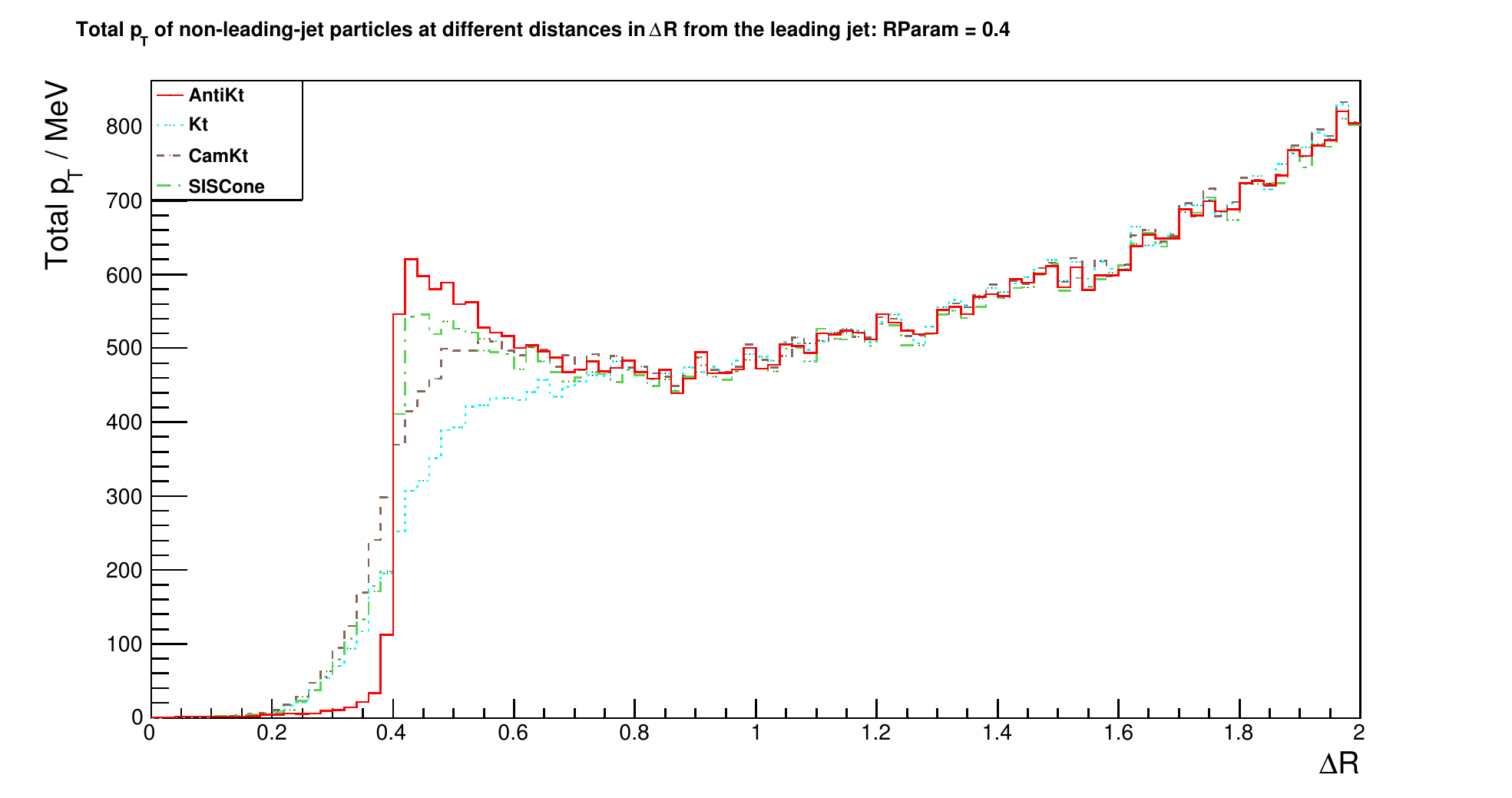}
\put(80,30){b)}
\end{overpic}
\end{center}
\end{minipage}
 \caption[Energy flow around a jet for colour octet exchange]{Energy flow for QCD colour octet events at a distance $\Delta R$ from the jet axis using a) particles that are constituents of the jet and b) particles not part of the jet.}
\label{fig:radiationalgos}
\end{figure}

\begin{figure}[ht]
\begin{minipage}[h]{0.47\textwidth}
\begin{center}
\begin{overpic}[width=\textwidth]{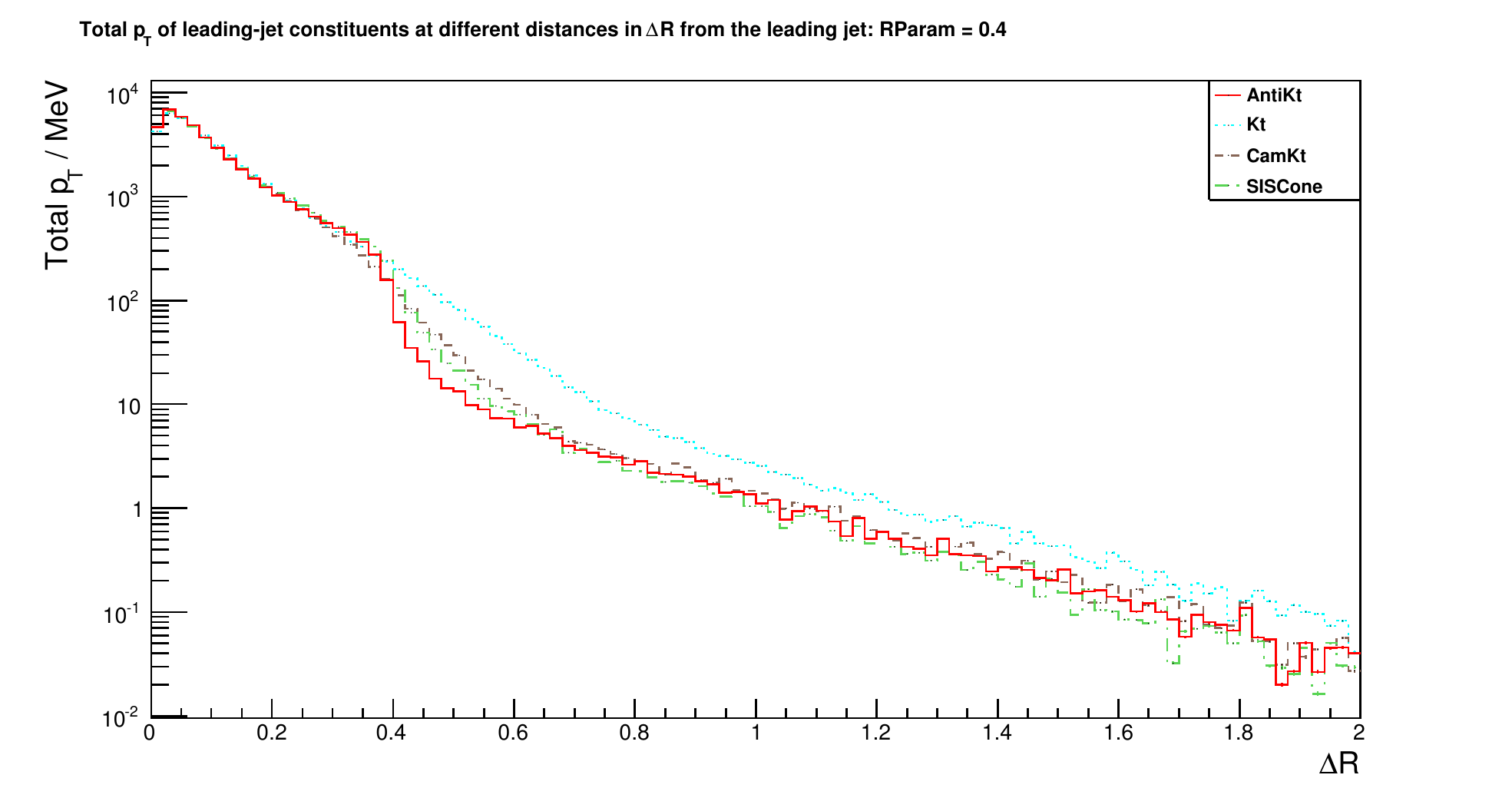}
\put(80,30){a)}
\end{overpic}
\end{center}
\end{minipage}
\begin{minipage}[h]{0.47\textwidth}
\begin{center}
\begin{overpic}[width=\textwidth]{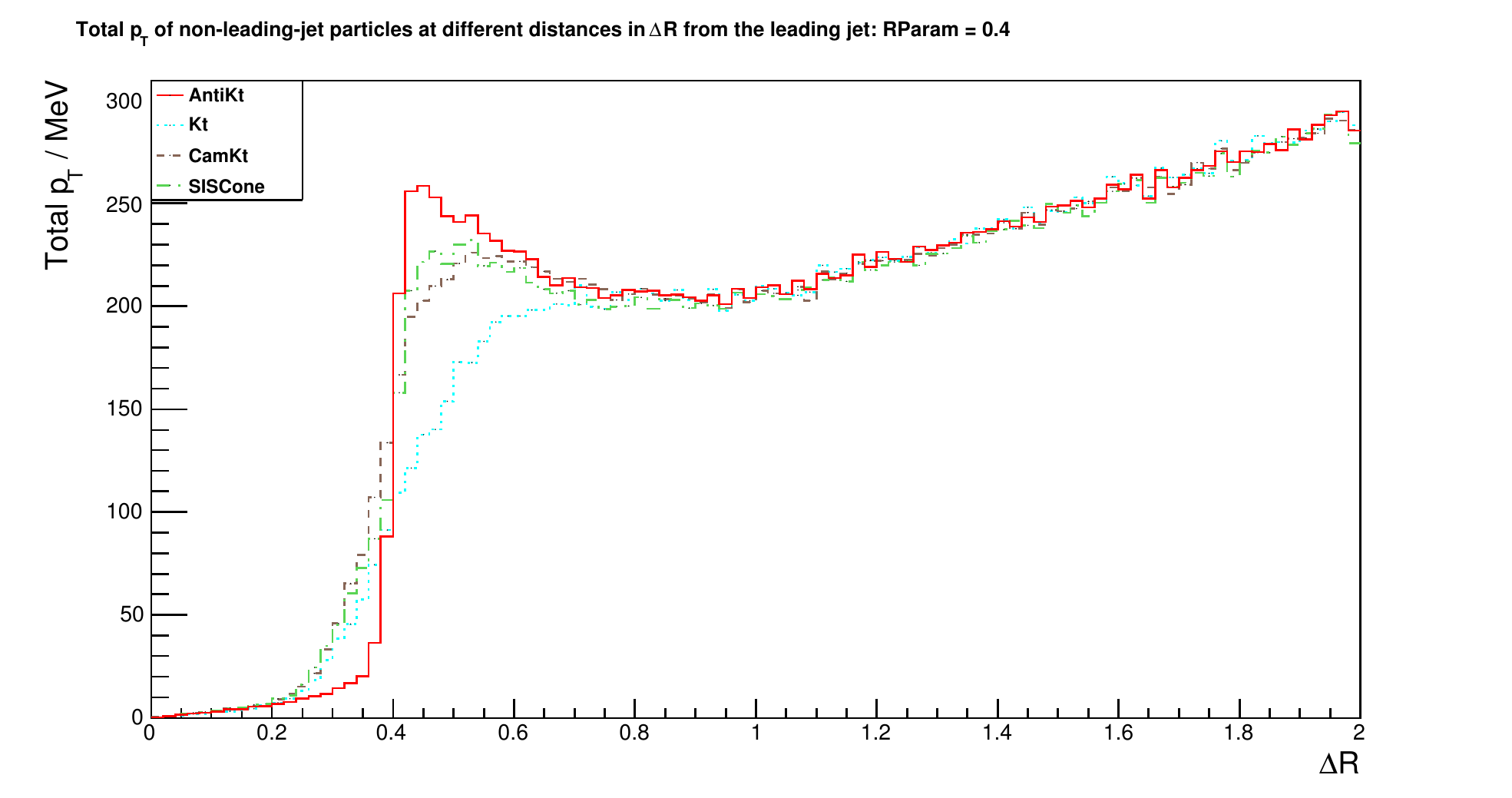}
\put(80,30){b)}
\end{overpic}
\end{center}
\end{minipage}
 \caption[Energy flow around a jet for colour singlet exchange]{Energy flow for colour singlet exchange events at a distance $\Delta R$ from the jet axis using a) particles that are constituents of the jet and b) particles not part of the jet.}
\label{fig:radiationalgossinglet}
\end{figure}

Figure \ref{fig:radiationalgossinglet} shows the same plots for events
in which a colour singlet is exchanged between the two leading partons.  Note that whilst   \ref{fig:radiationalgossinglet}a is broadly similar to   \ref{fig:radiationalgos}a,  \ref{fig:radiationalgossinglet}b shows less than half the activity of \ref{fig:radiationalgos}b for particles that are not part of the jet.  Figure   \ref{fig:radiationalgossinglet}a in fact shows a slightly less sharp jet border at R=0.4 for colour singlet exchange because the generally lower amount of radiation in colour singlet events means that, in order to pass the jet \et cut, radiation must be drawn into the jet from a wider area.  These characteristics will be exploited in the techniques described in the next sections.
 
\subsection{GAP GRID TECHNIQUE}
A method often used to distinguish events due to colour singlet and colour 
octet exchange is looking at the
total transverse energy in the rapidity range between two jets. 
Events due to colour-singlet exchange (no colour flows between the jets)
are characterised by almost no radiation, their golden signature being an 
empty rapidity gap between the two jets.\par
It is easy to understand why this quantity is too inclusive to be optimal.
Hard radiation from the jet itself can emit particles outside the jet 
boundaries independently of the type of event.  On the other
hand, particles emitted by a colour connection between the jets tend to follow specific geometrical
distributions, for instance being quite uniform in rapidity and close in $\phi$
to the main jets. To study these differences, we divide the detector into various
regions according to the directions of the two jets with largest $E_t$.
First, the area with $|\Delta\eta| < 0.7$ with respect to the axis of the first
two jets is removed. The remaining area between the jets is divided into 
16 regions of equal size in
the $\eta-\phi$ plane: four intervals of equal $\Delta\eta$ 
(each one quarter of the eta difference between the two main jets minus $2*0.7$)
and four of equal $\Delta\Phi$ (each of 90 degrees), with the $\Phi=0$
angle defined as the region of the leading jet in the event, and the first
region being comprised between $-45^\circ$ and $45^\circ$. The two 
detector areas between the jets and the beam pipe are also divided in four
regions each, according to the same $\Delta\Phi$ convention.

The convention is that the regions between the two jets are labeled between
1 and 16, with regions 1 to 4 being the closest in eta to the main jet
(and ordered such that region 1 contains $\Phi=0$, region 3 $\Phi = \pi$,
region 2 is the transverse one closer to the axis of the second jet and 
region 4 the remaining one)
5 to 8 for the next $\eta$ region in the direction of the second jet, with the
same convention with respect to $\Delta\Phi$, etc. 
\par
The average transverse energy deposition in each of the detector areas is
shown in Fig.~\ref{fig:density}a) for \herwigsix colour singlet and colour
octet events, the former shown with the \jimmy UE model \cite{Butterworth:1996zw} switched 
both on and off.
A clear difference between the various cases is visible, 
but what is more important
is that this difference is more marked in some regions than in others.
For instance, and despite the cut at $\pm~0.7$, all three event classes 
show that both $\eta$ regions close to the jets receive more radiation than
the central regions.  This is due to the splash-out from the jets themselves; 
the two regions with similar $\phi$ to the main jets also have more radiation - 
an effect that is more pronounced for the events with colour octet exchange.
Dividing the rapidity gap between the two main jets into several
smaller regions not only allows a better understanding of the radiation
patterns (for instance, to compare with different Monte Carlo models), but
also provides a more powerful separation between singlets and octets.

\begin{figure}[ht]
\begin{minipage}[h]{0.47\textwidth}
\begin{center}
\begin{overpic}[width=\textwidth]{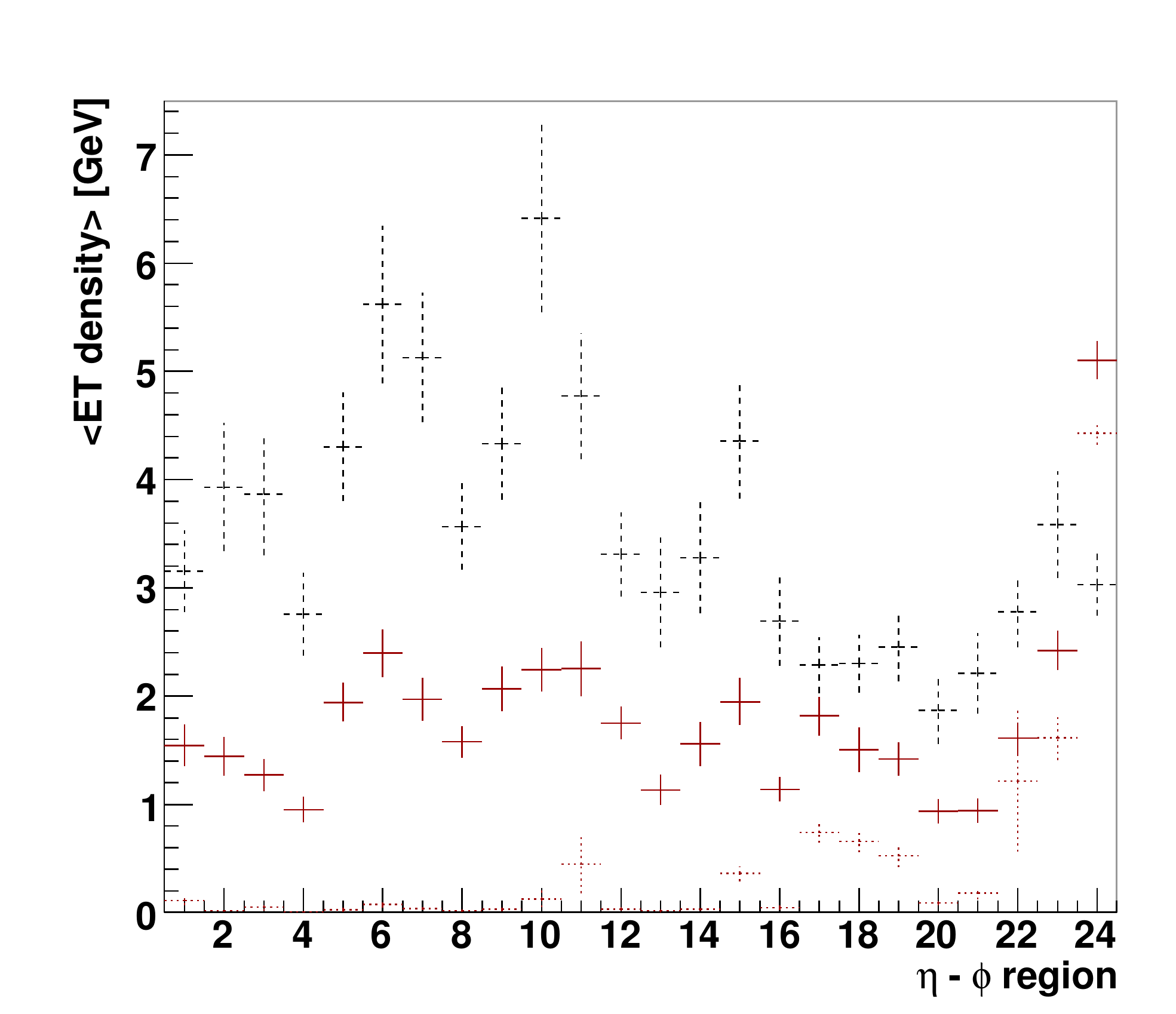}
\put(85,70){a)}
\end{overpic}
\end{center}
\end{minipage}
\begin{minipage}[h]{0.47\textwidth}
\begin{center}
\begin{overpic}[width=\textwidth]{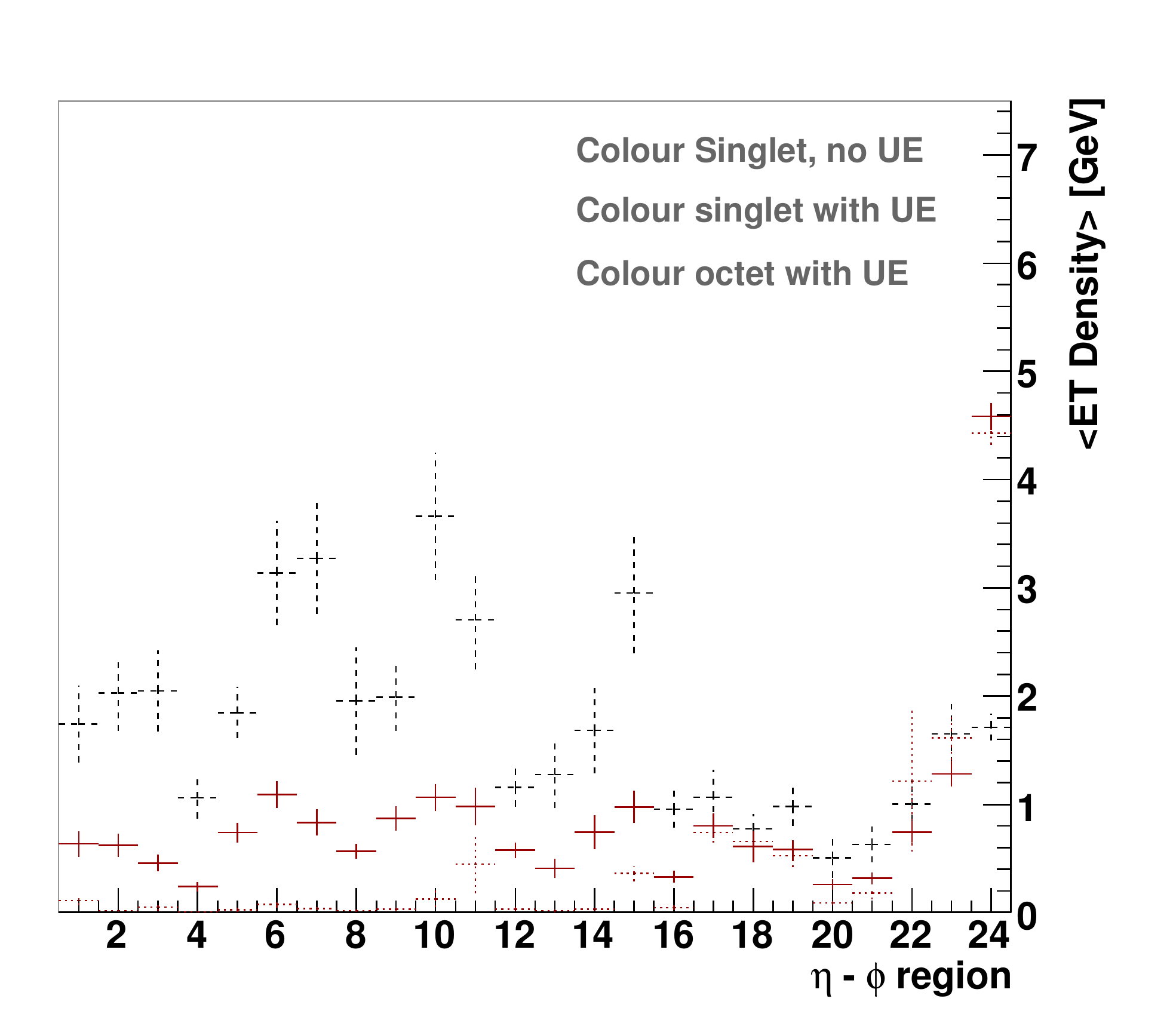}
\put(12,70){b)}
\put(35,62.7){\includegraphics[width=0.12\textwidth, height=0.107\textwidth]{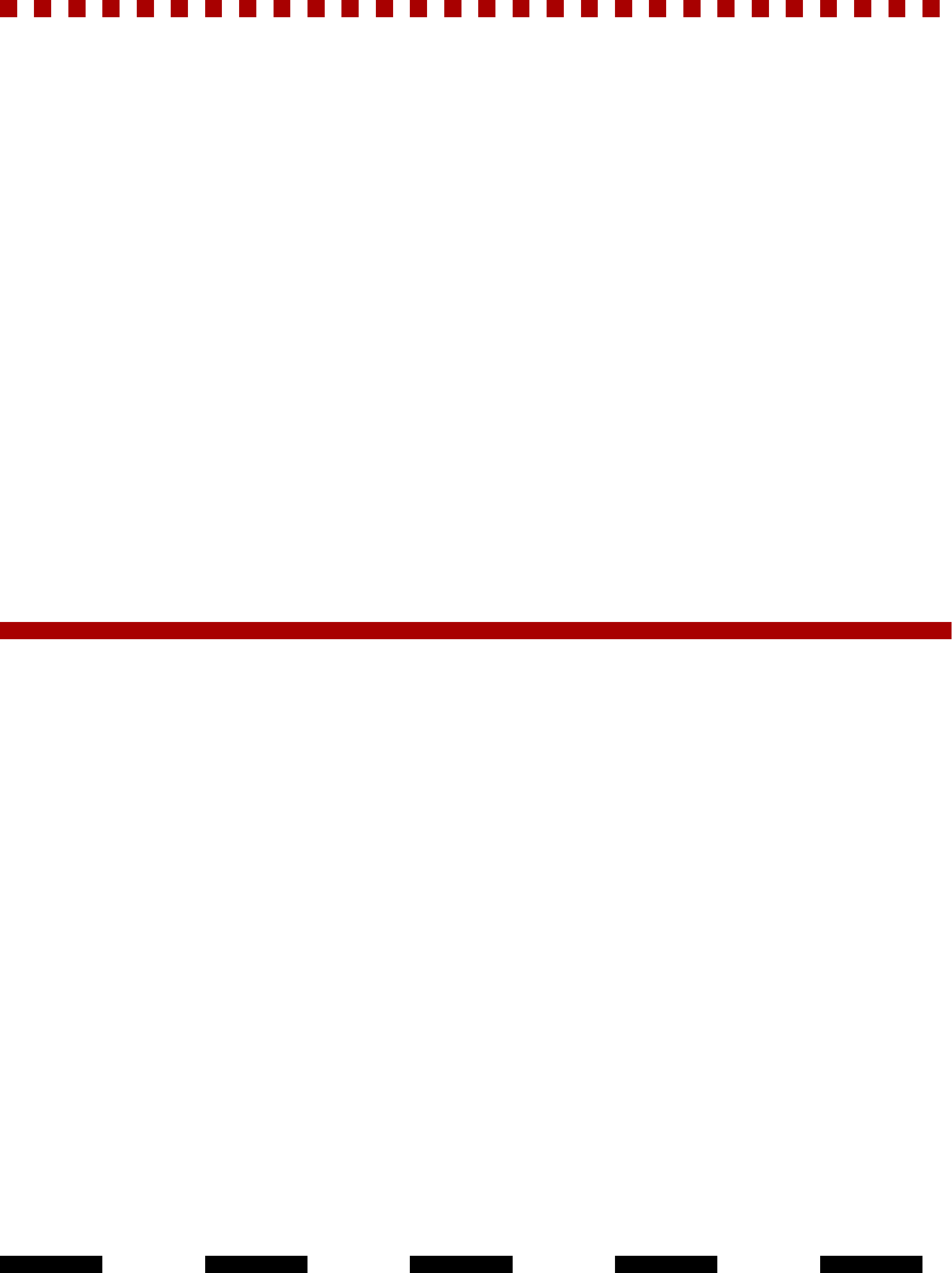}}
\end{overpic}
\end{center}
\end{minipage}

\caption[The average \et density in 24 different $\eta-\phi$ regions]{The average \et density in 24 different $\eta-\phi$ regions.  Regions 1 to 16 lie between the $\eta$ values of the leading two jets.  Jets with colour octet exchange and UE present are shown as black dashed crosses.  Jets with colour singlet exchange are shown in red; solid data-points are events with UE present and dotted data-points have UE turned off.  a) on the left shows the \et density uncorrected for UE, b) on the right shows the effect of applying a correction for UE based upon the area of a jet and the average \et density per event.}\label{fig:density}
\end{figure}

The presence of UE is a strong nuisance factor: without it 
some regions would almost be completely devoid of radiation in the case
of colour singlet exchange, which would make the identification of such events very easy. Although
there is no obvious way to identify radiation due to the UE
(it is not even a fully well-defined concept), it is on average
softer than the main scattering process. 
A possibility to exploit this fact would simply be to impose a high threshold
for clusters to be included in the density calculation; however, 
it was shown in \cite{Cacciari:2008gn} that a better approach would be to establish
this threshold on an event-by-event basis.
Following this idea, we use the \kt algorithm with a small Et threshold to cluster into jets all energy depositions in the events.  For each jet we calculate the active area, then define an \et density as the ratio 
of the jet transverse momentum and its area.
All jets below 10~GeV per unit of rapidity squared are considered potential
soft candidates; their average density on an event-by-event basis is taken
as the value of the soft density for the event.
For all jets in the event an UE contribution is calculated as the product of
this soft density and the area of each jet, and subtracted from the jet
transverse momentum.
Finally, all jets whose transverse momentum after this 
subtraction is below 3 GeV are considered as coming from a soft UE-type interaction
and are therefore removed. The advantage of this method with respect to
applying a strong cut on all jets is that the amount of soft radiation 
is determined on an event-by-event basis, so coherent upward or downward
fluctuations of the soft activity can be accounted for. 
After this procedure, the density distribution
for the various detector regions is shown in Fig.\ref{fig:density}b).
An improvement with respect to the previous case is visible, with the
additional advantage of a smaller dependence on the MonteCarlo description
of the soft interactions.

\subsection{FOURIER TRANSFORMATION}
The problem of separating the colour connection effect from the UE
or pile-up is one of separating features of differing physical size 
in the event.  The UE fills the whole detector with radiation at all $\eta$.  
Colour connection effects between jets are typically
smaller than this size, i.e. 
are approximately the size of the jet-jet or jet-beamline 
interval.  Hadronisation and showering effects can be expected to be of a 
similar to smaller size.  The hard jets will be smaller 
still, with a radius of $R\simeq0.5$ and there may be jets originating from 
softer partons with $R$ as small as $0.1$.\par
A Fourier-transformation of the spatial distribution of radiation in the event could separate the
various scales, allowing an alternative reading of the energy flow.
For the 1-dimensional case described here, we use the $\Phi$ distribution
of the truth-level particles for each event, weighted by transverse 
energy summed in 32 bins covering the region 0-$\pi$.   The axis of the hardest jet is taken as as the $\Phi = 0$ direction.
We then extract and use as our variables the first 32 complex coefficients 
of the Fourier expansion 
of this distribution. Since these coefficients are linked by the relation $C_n =
C^*_{N-n}$ (where N=32), there are only 32 independent real coefficients,
indicating no information is lost (or gained!) on the event-by-event basis in going from the
32 bins to the 32 coefficients.
\par
To test the effectiveness of this method, we applied it to the problem of
separating colour singlet and octet exchange in hadronic collisions.
Colour singlet exchanges without UE are possibly the most
di-jet-like kind of events in a hadron collider.  Since the $\Phi=0$ direction is aligned to the axis of the leading 
jet, most of the remaining radiation will be 
concentrated around $\Phi = \pi$, which is the most likely location of the second 
jet. It is
easy to show that odd coefficients can never have a peak at both 0 and $\pi$.  Further, the $n^{th}$ coefficient corresponds to correlated activity of size $r\simeq\pi/n$.  Small-$n$ odd coefficients, corresponding as they do to large non-di-jet features, are therefore expected to be suppressed in perfect di-jet events.  This is exactly what is observed in Fig.~\ref{fig:avmagnitude}, where the
average magnitude of the various coefficients is plotted for the two extreme
cases of a set of colour-singlet exchange without the UE, and
colour octet exchange with UE. 

\begin{figure}[ht]
\begin{center}
\includegraphics[width=.7\textwidth]{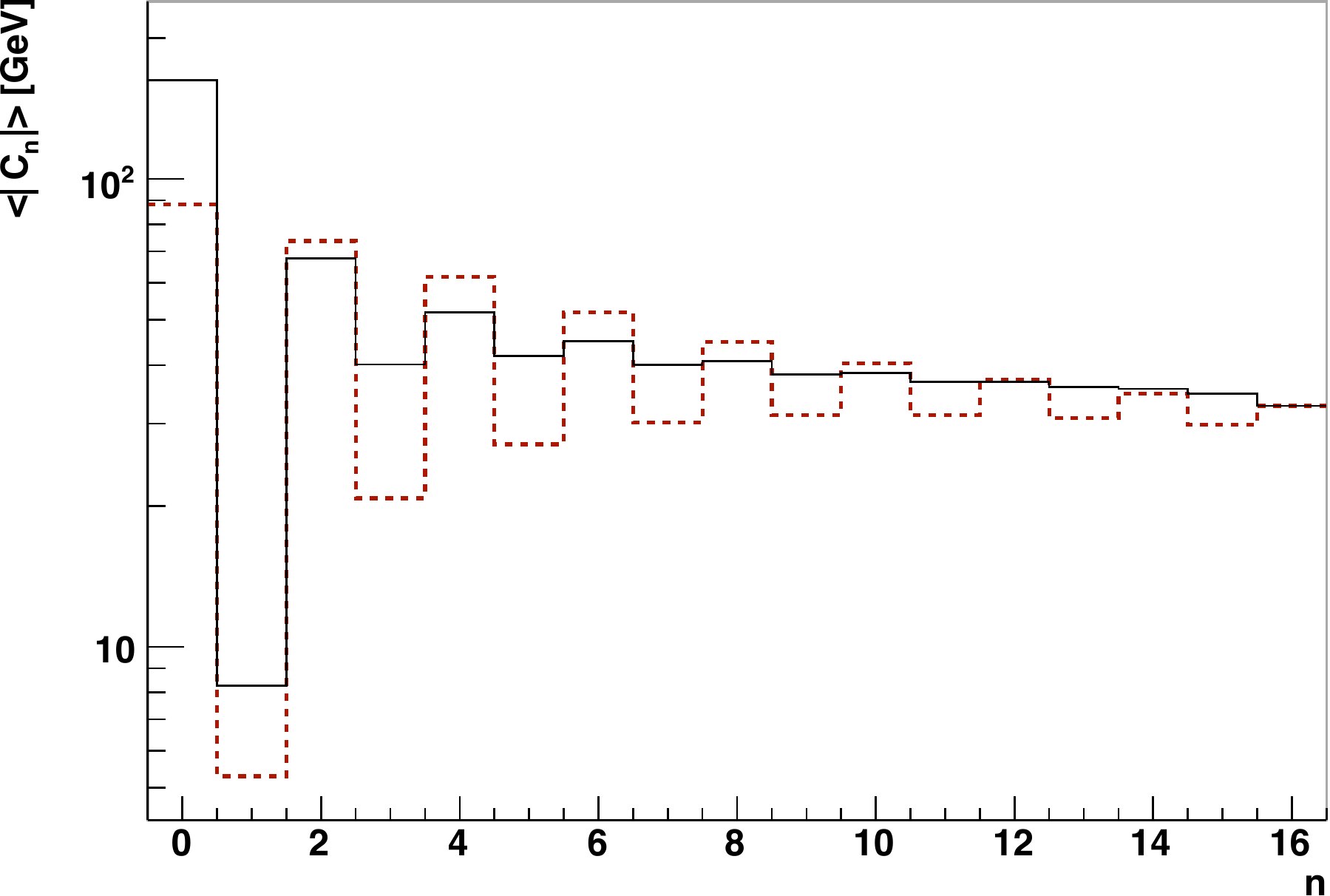}
 \caption[The average magnitude of the first 16 Fourier coefficients]{The average magnitude of the first 16 Fourier coefficients for colour singlet without UE (red-dashed) and colour octet with UE (black-solid).}
\label{fig:avmagnitude}
\end{center}
\end{figure}

The suppression of the odd
coefficients is clearly visible in the first case, and is much less pronounced
(even if still present, since the events nevertheless contain di-jets)
for the colour octet exchange case. Although there is also a difference between the even coefficients (especially at lower-$n$), it is much less pronounced  than for the odd coefficients.   In particular, we expect coefficients 3 and 5 to originate from structures of a similar size ($\pi/3\simeq1$) to the inter-jet radiation. Adding the UE
does not change this picture significantly, since the uniform UE radiation will
only influence the magnitude of coefficient zero, leaving the mean values
of the others almost unchanged.  This is shown more clearly in Fig.\ref{fig:C3}, which shows the distribution of the magnitude of the 3rd coefficient for colour singlet and colour octet events.  There is a clear difference between the two types of event and the presence or absence of UE makes little difference to the  shape of the distribution from colour singlet events.

\begin{figure}[ht]
\begin{center}
\includegraphics[width=.7\textwidth]{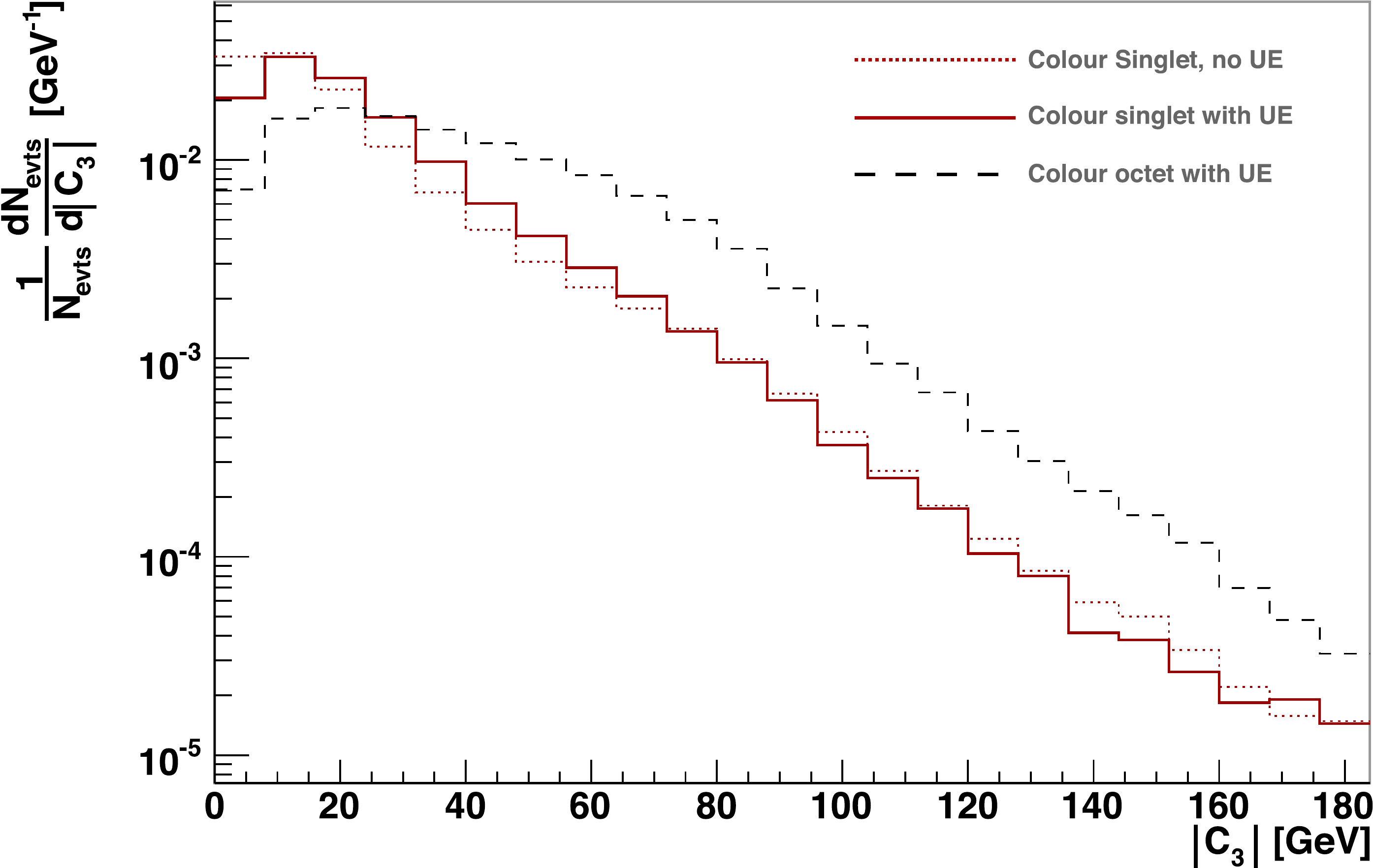}
 \caption[The magnitude of the 3rd Fourier coefficient]{Distribution of the magnitude of the 3rd coefficient for colour singlet without UE (red-dotted), colour singlet with UE (red-solid) and colour octet with UE (black-dashed).  The addition of UE makes little difference to the colour singlet distributions, whilst there is a clear difference between colour octet and colour singlet.}
\label{fig:C3}
\end{center}
\end{figure}

Another potentially interesting variable to look at is the
number of the largest coefficient for each event, which shows the scale of the dominant feature in the event. Figure \ref{fig:largestim} shows
the distribution of the largest imaginary coefficient in the event, for
the two classes of colour exchange, both with UE on. 

\begin{figure}[ht]
\begin{center}
\includegraphics[width=.7\textwidth]{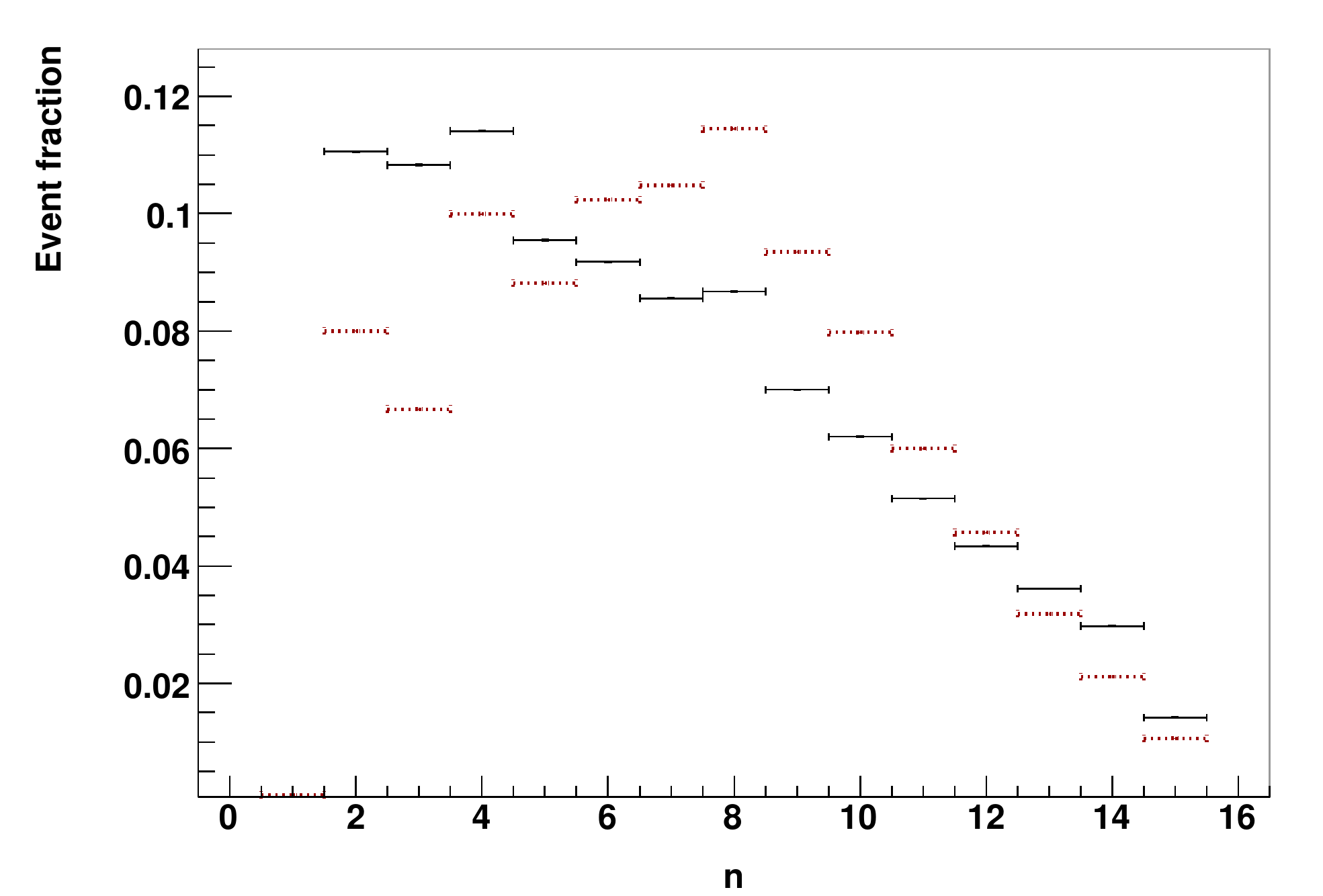}
 \caption[Distribution of the largest imaginary coefficient]{The largest imaginary coefficient in the event.  Solid black shows colour-octet exchange events and dotted-red shows colour singlet exchange.  Both event samples have the underlying event model turned \emph{on}.   The colour octet sample shows a peak towards lower $n$, indicating  more events have their inter-jet activity dominated by features that are large in $\eta-\phi$.  The colour singlet sample, on the other hand, shows more events whose inter-jet radiation is dominated by features of size $r\simeq8/\pi$. }\label{fig:largestim}
\label{fig:spread}
\end{center}
\end{figure}

While for ordinary QCD events this distribution is peaked at low values, 
this is not the case for colour singlet events, where there
is a peak around coefficient 8 that corresponds to radiation concentrations of size $\pi/8\simeq 0.4$,
roughly the size of a jet. Further, in the colour singlet sample there is a relative depletion  of events in which the largest coefficient was either 3 or 5.  More details on the Fourier transformation method
can be found in \cite{Campanelli:2009hc}.

\subsection*{CONCLUSIONS}
Radiation between jets is a key topic for understanding the interplay between
hard and soft QCD, and will be the object of many measurements with the
first LHC data. Here we propose new methods to study this radiation in more
detail than has been done in the past. The first measurements
of these observables will offer interesting handles
on the discrimination of the various theoretical models. To show the
strong and weak points of the methods we applied them to the problem of
separation between colour singlet and colour octet exchange events, but the techniques
presented here are more general, and can be applied to most of the 
fully-hadronic final states produced by the first LHC data.

\clearpage

\part[BEYOND THE STANDARD MODEL]{BEYOND THE STANDARD MODEL}

\section[AN UPDATE OF THE PROGRAM HDECAY]{AN UPDATE OF THE PROGRAM HDECAY\protect\footnote{
Contributed by: A.~Djouadi, J.~Kalinowski, M.~M\"uhlleitner and M.~Spira}}

%
%

 
\subsection{INTRODUCTION}
The search strategies for Higgs bosons at LEP, Tevatron, LHC and
future $e^+e^-$ linear colliders (LC) exploit various Higgs boson
decay channels. The strategies depend not only on the experimental
setup (hadron versus lepton colliders) but also on the theoretical
scenarios: the Standard Model (SM) or some of its extensions such as
the Minimal Supersymmetric Standard Model (MSSM). It is of vital
importance to have reliable predictions for the branching ratios of
the Higgs boson decays for these theoretical models.

The current version of the program HDECAY \cite{Djouadi:1997yw} can be
used to calculate Higgs boson partial decay widths and branching ratios
within the SM and the MSSM and includes: \\[0.2cm]
-- All decay channels that are kinematically allowed and which have
branching ratios larger than $10^{-4}$, i.e.~the loop mediated, the
three body decay modes and in the MSSM the cascade and the
supersymmetric decay channels
\cite{Spira:1997dg,Djouadi:2005gi,Djouadi:2005gj}. \\[0.2cm]
-- All relevant higher-order QCD corrections to the decays into quark
pairs and to the loop mediated decays into gluons are incorporated
\cite{Djouadi:1995gt}. \\[0.2cm]
-- Double off--shell decays of the CP--even Higgs bosons into massive
gauge bosons which then decay into four massless fermions, and all
important below--threshold three--body decays \cite{Djouadi:1995gv}.
\\[0.2cm]
-- In the MSSM, the complete radiative corrections in the effective
potential approach with full mixing in the stop/sbottom sectors; it uses
the renormalization group improved values of the Higgs masses and
couplings and the relevant next--to--leading--order corrections are
implemented
\cite{Carena:1995wu,Haber:1996fp,Carena:2000dp,Degrassi:2002fi}.
\\[0.2cm]
-- In the MSSM, all the decays into SUSY particles (neutralinos,
charginos, sleptons and squarks including mixing in the stop, sbottom
and stau sectors) when they are kinematically allowed
\cite{Djouadi:1992pu,Djouadi:1996mj,Djouadi:1996pj}. The SUSY particles
are also included in the loop mediated $\gamma \gamma$ and $gg$ decay
channels. \\[0.2cm]
The source code of the program, written in FORTRAN, has been tested on
computers running under different operating systems. The program
provides a very flexible and convenient use, fitting to all options
of phenomenological relevance. The basic input parameters, fermion and
gauge boson masses and their total widths, coupling constants and, in
the MSSM, soft SUSY-breaking parameters can be chosen from an input
file. In this file several flags allow switching on/off or changing
some options [{\it e.g.} choosing a particular Higgs boson,
including/excluding the multi-body or SUSY decays, or
including/excluding specific higher-order QCD corrections].

\subsection{UPDATES}
Since the release of the original version of the program several bugs
have been fixed and a number of improvements and new theoretical
calculations have been implemented. The following points summarize the
most important modifications of HDECAY after its release: \\[0.2cm]
-- Implementation of Higgs boson decays to gravitino + gaugino
   \cite{Djouadi:1997gw}. A flag in the input file controls whether these
   decay modes are taken into account or not. \\[0.2cm]
-- Inclusion of SUSY--QCD corrections in neutral MSSM Higgs decays to
   $b\bar{b}$ \cite{Dabelstein:1995js,Coarasa:1995yg} and resummation of
   $\Delta_b$ effects \cite{Carena:1999py,Guasch:2003cv} up to NNLO
   \cite{Noth:2008tw,Noth:2010jy}. The corresponding $\Delta_b$ terms have
   also been included in charged Higgs decays $H^\pm\to tb$. \\[0.2cm]
-- Determination and inclusion of the RG improved two-loop contributions
   to the MSSM Higgs self-interactions. These two-loop corrections extend
   the results of Ref.~\cite{Carena:1995wu} for the stop and sbottom
   contributions for arbitrary mixing parameters and mass splitting, where
   they have only been displayed for the scalar ${\cal CP}$-even Higgs mass
   matrix. We have checked explicitely that we reproduce the results of
   Ref.~\cite{Carena:1995wu}. \\[0.2cm]
-- An interface to the SUSY Les Houches Accord (SLHA)
   \cite{Skands:2003cj,Allanach:2008qq} has
   been implemented properly. This required several transformations of the
   corresponding renormalization schemes to the ones used by HDECAY. This
   option can be switched on and off by appropriate flags in the input
   file. The output file in the SLHA format can also be used again as
   input file. Moreover, the automatic generation of the SLHA input file
   according to the input parameters of HDECAY is provided as an option.
   \\[0.2cm]
-- Inclusion of the full mass dependence of the NLO QCD corrections to the
   quark and squark loop contributions to photonic Higgs decays
   $\phi\to\gamma\gamma$
   \cite{Djouadi:1990aj,Djouadi:1993ji,Melnikov:1993tj,Inoue:1994jq,
   Spira:1995rr,Muhlleitner:2006wx}. This decay width can
   now also be used to determine the production cross sections of Higgs
   bosons at photon colliders at NLO QCD. \\[0.2cm]
\begin{figure}[hbtp]
\begin{center}
\includegraphics[clip,scale=0.45]{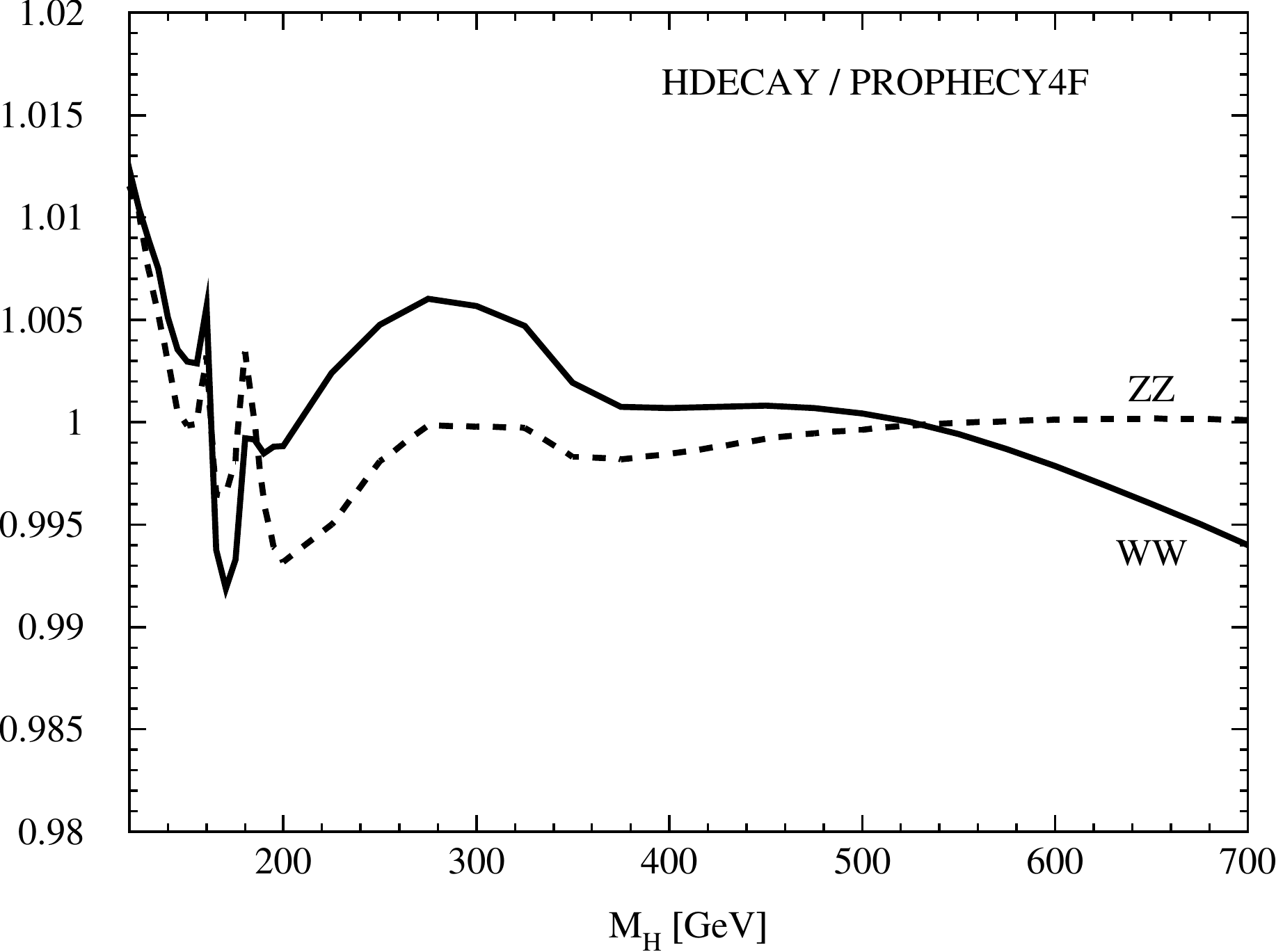}\\
\caption{Ratio of the partial widths of $H\to
W^{(*)}W^{(*)}/Z^{(*)}Z^{(*)} \to 4f$ from HDECAY and PROPHECY4F
\cite{Bredenstein:2006rh,Bredenstein:2006ha}.}  
\label{fig:h24f}
\end{center}
\end{figure}
-- Inclusion of electroweak corrections to the SM Higgs boson decays
   $H\to W^{(*)}W^{(*)}/Z^{(*)}Z^{(*)} \to 4 f$ in approximate form which
   reproduces the full results of
   Refs.~\cite{Bredenstein:2006rh,Bredenstein:2006ha} within 1\% as can be
   inferred from Fig.~\ref{fig:h24f}, where the ratios of the partial decay
   widths into $W^{(*)}W^{(*)}$ and $Z^{(*)}Z^{(*)}$ are shown as a
   function of the Standard Model Higgs mass. The improved Born
   approximation of Ref.~\cite{Bredenstein:2006rh,Bredenstein:2006ha} has
   been implemented with additional improvements at the $WW$ and $ZZ$
   thresholds. \\[0.2cm]

The logbook of all modifications and the most recent version of the
program can be found on the web page {\tt
http://people.web.psi.ch/spira/proglist.html}. \\

\subsection*{ACKNOWLEDGEMENTS}
This work is supported in part by the European Community's Marie-Curie
Research Training Network HEPTOOLS under contract MRTN-CT-2006-035505.



\clearpage

\section[IMPLEMENTATION AND VALIDATION OF MODELS BEYOND THE STANDARD MODEL WITH FEYNRULES]{IMPLEMENTATION AND VALIDATION OF MODELS BEYOND THE STANDARD MODEL WITH FEYNRULES
\protect \footnote{Contributed by: L. Basso, F. Braam, C. Duhr, B. Fuks, A. Martin,  J. Reuter, T. Roy, S. Schumann}} 

\subsection{INTRODUCTION}
\label{sec:feynrules_introduction}

Monte Carlo event generators play an important role in making reliable predictions for the events to be observed at collider experiments, both to describe the backgrounds and possible candidate signals. 
In particular, the simulation of a hadronic collision requires not only an accurate description of the underlying hard scattering process, but also of subsequent parton showering and hadronization as efficiently provided by programs such as {\sc Herwig}~\cite{Corcella:2000bw, Bahr:2008pv}, {\sc Pythia}~\cite{Sjostrand:2006za, Sjostrand:2007gs} and {\sc Sherpa}~\cite{Gleisberg:2003xi,Gleisberg:2008ta}. As regards the generation of the hard matrix element itself, a lot of effort has gone into the development of several multipurpose matrix element generators,  such as
{\sc CompHep}/{\sc CalcHep} \cite{Pukhov:1999gg,  Boos:2004kh,Pukhov:2004ca}, {\sc MadGraph}/{\sc MadEvent} \cite{Stelzer:1994ta,Maltoni:2002qb,Alwall:2007st,Alwall:2008pm}, {\sc Sherpa} and {\sc Whizard} \cite{Kilian:2007gr, Moretti:2001zz}. Even though these programs are in principle able to generate the (tree-level) matrix element for any process in the framework of a given renormalizable quantum field theory built on scalar, vector and fermion fields, the implementation of a full Beyond Standard Model (BSM) theory can be a tedious and error-prone task, often requiring the implementation of one vertex at the time following the conventions specific to each code.  

In this section a summary report on the implementation and validation of BSM models into multi-purpose matrix element generators is presented. The starting point of our approach is {\sc FeynRules}~\cite{Christensen:2008py}, a {\sc Mathematica}\textregistered\footnote{{\sc Mathematica}\ is a registered trademark of Wolfram Research, Inc. } package that allows to compute Feynman rules from a Lagrangian in an automated way. Furthermore, {\sc FeynRules} contains a set of interfaces to various matrix element generators, allowing to implement the model into a given tool in an automated way. For the moment, interfaces to {\sc CompHep}/{\sc CalcHep}, {\sc FeynArts}/{\sc FormCalc} \cite{Hahn:1998yk, Vermaseren:2000nd, Fliegner:1999jq, Fliegner:2000uy, Tentyukov:2004hz, Tentyukov:2007mu}, {\sc MadGraph}/{\sc MadEvent} and {\sc Sherpa} are available\footnote{An interface to {\sc Whizard} will be available in the near future.}. A first step in the direction of deriving Feynman rules automatically starting from a model Lagrangian has been made in the context of the {\sc CompHep}/{\sc CalcHep} event generator with the {\sc LanHep} package \cite{Semenov:2008jy}. Our aim is to go beyond this scheme and create a general and flexible environment where communication between theorists and experimentalists in both directions is fast and robust. First, the use of {\sc Mathematica} as a working environment provides a powerful and user-friendly platform where BSM models can be developed and implemented. Second, the possibility to export the model to more than one matrix element generator enhances the chances that the model can be successfully dealt with by more than one code. Furthermore, since many of these matrix element generators are already embedded in the experimental softwares, the new models can easily be integrated into the experimental framework, allowing in this way for an efficient communication between theorists and the experimental community.

As the starting point of our approach is {\sc FeynRules}, we briefly recall its basic features in Section~\ref{sec:feynrules_feynrules} and discuss recent developments triggered by the activities and the discussions at the Les Houches workshop. In Section~\ref{sec:feynrules_validation} we present a proposal of how new BSM models can be easily validated exploiting the fact that once a {\sc FeynRules} implementation is available, the model can easily be exported to various matrix element generators. In Section~\ref{sec:feynrules_models} we discuss the implementation of several new models that were implemented and/or validated as an outcome of the workshop.

\subsection{RECENT DEVELOPMENTS IN FEYNRULES}
\label{sec:feynrules_feynrules}
{\sc FeynRules} is a {\sc Mathematica} package that allows to compute Feynman rules directly from a Lagrangian in an automated way. The user provides the Lagrangian for the model (written in {\sc Mathematica}) as well as all the information about the particle content and the parameters of the model. This information uniquely defines the model, and hence is enough to derive all the interaction vertices from the Lagrangian. {\sc FeynRules} can in principle be used with any model which fulfills basic quantum field theoretical requirements (\emph{e.g.}, Lorentz and gauge invariance), the only limitation coming from the kinds of fields supported by {\sc FeynRules}. As to date, the public release of {\sc FeynRules} supports scalar, vector, fermion (Dirac and Majorana) and spin-two fields, as well as Faddeev-Poppov ghosts, and very recently, also Weyl fermions have been included\footnote{This feature is still beta and will be made public with the next major release.}, as will be described at the end of this section. In a second step, the interaction vertices obtained by {\sc FeynRules} can be exported to various matrix element generators by means of interfaces provided by the package~\cite{Christensen:2009jx}. Let us note that even though {\sc FeynRules} itself can in principle be used to obtain the Feynman rules for any Lagrangian, the matrix element generators very often have certain information on the color and/or Lorentz structures hardcoded. In this case the interfaces check whether all the vertices are compliant with the structures supported by the corresponding matrix element generator, and if not, a warning is printed and the vertex is discarded. Each interface produces at the end a (set of) text file(s), often consistently organized in a single directory, which can be read into the matrix element generator at runtime and allows to use the new model in a way similar to all other built-in models.

As already mentioned, the most important recent development concerns the possibility to write Lagrangians in terms of two-component Weyl fermions. In the following we give a brief description how this new feature can be used inside {\sc FeynRules}. Weyl fermions can be declared in the model file in exactly the same way as other particle classes, \emph{e.g.},
\begin{verbatim}
W[1] == {
        ClassName     -> chi,
        Chirality     -> Left,
        SelfConjugate -> False},

W[2] == {
        ClassName     -> xibar,
        Chirality     -> Right,
        SelfConjugate -> False}.
\end{verbatim}
The particle class {\tt W} refers to Weyl fermions, and we have defined in this way one left-handed and one right-handed Weyl fermions named {\tt chi} and {\tt xi}, respectively. Let us note that all other options allowed for fields are also available for Weyl fermions (See the {\sc FeynRules} manual~\cite{FRwebpage}). Along the same lines, there is a new option for Dirac fermions specifying their left and right-handed Weyl components, \emph{e.g.},
\begin{verbatim}
F[1] == {
       ClassName       -> psi,
       SelfConjugate   -> False,
       WeylComponents  ->  {chi, xibar}}.
\end{verbatim}
More details on the use of Weyl fermions in {\sc FeynRules} will be presented in a forthcoming publication. Here it suffices to say that since matrix element generators work with four-component spinors rather than Weyl fermions, {\sc FeynRules} replaces at runtime all Weyl fermions by the corresponding Dirac fermion $\psi=(\chi, \bar \xi)^T$ according to the prescription,
\begin{eqnarray}
\chi \rightarrow \frac{1-\gamma^5}{2}\,\psi,\quad \xi \rightarrow \frac{1-\gamma^5}{2}\,\psi^c,\quad \bar\xi \rightarrow \frac{1+\gamma^5}{2}\,\psi, \quad \bar\chi \rightarrow \frac{1+\gamma^5}{2}\,\psi^c,\nonumber
\label{eq:feynrules_weyls}
\end{eqnarray}
where $\psi^c = C\,\bar \psi^T = (\xi,\bar\chi)^T$ denotes the charge conjugated Dirac spinor. After these replacements, the Lagrangian is expressed completely in terms of four-component Dirac and Majorana spinors, and the Feynman rules can be exported to the matrix element generators in the standard form. 

The two-component formalism we have just described can be extremely useful when dealing with supersymmetric extensions of the SM, where Weyl fermions appear as the natural fermionic degrees of freedom of chiral and gauge superfields. Let us note that this is just the first step of a more general project which aims at implementing superfields directly into {\sc FeynRules}. In particular, we applied the Weyl-fermion formalism to rewrite the existing MSSM implementation in {\sc FeynRules}~\cite{Christensen:2009jx, FRwebpage} completely in terms of two-component spinors, as well as to the implementation of the minimal $R$-symmetric supersymmetric extension of the SM described in Section~\ref{subsec:feynrules_mrssm}. The validation of these (and other) new model implementations was done following the procedure described in the next section.

\subsection{VALIDATION PROCEDURE FOR BSM MODELS}
\label{sec:feynrules_validation}
The requirements for a BSM model implementation that ensure a fast and efficient communication between all the actors involved in the simulation chain, both on the experimental and theoretical side, are manifold. The first obvious requirement is that the implementation produces reliable results when used with a matrix element generator, both from the perspective of the theoretical consistency of the model (\emph{e.g.}, gauge invariance,...), as well as regarding the technical aspects of the matrix element generator under consideration (\emph{e.g.}, conventions used in the code,...). A robust validation procedure must hence combine these two aspects, which however rely on two disconnected fields of expertise: if theoretical consistency is purely related to the physics content of the model, testing the technical aspects of the implementation requires for example detailed knowledge of the programming language used in the matrix element generators. Second, to ensure traceability both of the implemented models as well as of the generated event samples, every implementation of a BSM model into a matrix element generator must be clearly documented so that all the information about the physics content of the model and the choice of the parameters can be retraced at any point in the chain.

In Refs.~\cite{Christensen:2009jx, Hagiwara:2005wg}, several BSM implementation were validated by comparing the results obtained with different matrix element generators among themselves.
In this section we propose an evolution of this strategy for the validation of a generic BSM model implementation, and we argue that the {\sc FeynRules} approach offers a natural framework where BSM models cannot only be easily developed and implemented, but can also be validated to an unprecedented level. Since the {\sc FeynRules} model files are built around the Lagrangian of the model, all the information about the physics content of the implementation is centered in one place and can be accessed at any point in the simulation chain, without being obscured by the conventions imposed by a specific matrix element generator. Furthermore, one can exploit the fact that {\sc FeynRules} offers the possibility to implement the model into various generators and that different codes use different conventions and/or work in different gauges. In this way one can very easily check the sanity of an implementation by cross-checking the results obtained with different codes. For example, once a {\sc FeynRules} implementation is available, the Feynman rules can be obtained automatically and then used to cross-check analytical results for simple processes against known results from the literature, a step that could even be automatized by using the {\sc FeynRules} interface to {\sc FeynArts}/{\sc FormCalc}. After the model has passed successfully all of these tests, a more detailed study of the implementation can be performed by comparing the results of different matrix element generators for a large number of (tree-level) processes. In this way, it is not only very easy to show that the implementation produces reliable results for all matrix element generators involved, but it also provides a very natural and fast way to test the theoretical consistency of the model, by computing the results in different gauges, testing the high-energy behavior of certain processes, \emph{etc.}

To quantify the level to which a given BSM implementation has been validated, we propose in the following a four-step procedure to rate BSM model implementations regarding their validation. In this scheme, every implementation is awarded one credit for each completed step, with a maximum of four for models tested to a very high extent in all matrix element generators. 
More explicitly, the four steps are:
\vskip 0.1cm
\begin{enumerate}
\item[\bf 1.] {\bf Documentation [First credit] :} 
A first credit is awarded if the implementation is documented to ensure traceability and reproducibility. Information about the model should be included in the {\sc FeynRules} model file via the {\verb+M$ModelInformation+} variable (See the {\sc FeynRules} manual), and should contain a twofold information:
\vskip0.2cm
\begin{itemize}
\item All references to the original publications where the model was first presented should be included, to ensure that the Lagrangian, its field content and the relations between parameters can be traced back at any time. This also includes references to other codes involved in the writing of the model file, \emph{e.g.}, spectrum generators. In case the implemented model is incomplete or consists only in a theory fragment, this should be clearly stated. 
\item The operating system (Linux distribution, Mac OS, ...) and the versions of {\sc Mathematica} and {\sc FeynRules} (as well as all other codes, if relevant) used to write the model file must be provided in order to ensure reproducibility at later times.
\end{itemize}

\vskip 0.5cm
\item[\bf 2.] {\bf Basic theory sanity checks [Second credit] :} A second credit is awarded for testing that the model satisfies the constraints imposed by quantum field theory, \emph{e.g.}, checking the hermiticity and gauge invariance of the implemented Lagrangian. If the Feynman rules for the model are known in the literature, they should be compared to the results obtained by {\sc FeynRules} for this model. Furthermore, simple two-to-two cross sections and/or decay rates can be easily computed either by hand or by means of {\sc FeynArts}/{\sc FormCalc} and the corresponding {\sc FeynRules} interface and compared to known analytic results (if available).

\vskip 0.5cm
\item[\bf 3.] {\bf Testing one matrix element generator [Third credit] :} After the theoretical sanity of the implementation has been checked, a more detailed validation of the model can be performed by exporting it to a matrix element generator, both to exclude mistakes in the implemented model as well as to ensure that the implementation can be used in a reliable way with a given code. \\
\begin{itemize}
\item Basic processes like Drell-Yan and Bhabba scattering can be used to check that the matrix element generator produces reliable results for the corresponding cross-sections, followed by a more systematic study of several two-to-two and/or two-to-three key processes of interest in this model. A particular emphasis should be brought to the running of the strong and/or electromagnetic couplings, as well as to the reproduction of known SM cross section results for the sectors of the model unaffected by the BSM physics. 
\item In case an independent (partial) implementation of the model is available, a comparison of the two implementations should be performed.
\item The high-energy behavior of the model can be easily checked numerically to ensure that all gauge and unitarity cancellations take place in a numerically stable and reliable way. Note that for such cancellations to take place \emph{exactly} at tree-level, it is required that all particles have zero width, \emph{i.e.}, all widths of all particles should be put to zero while performing these tests.
\item Some matrix element generators, like for example {\sc CalcHep}, offer the possibility to choose between different gauges. 
In this case gauge invariance can be easily tested by running the same processes in different gauges. 
\item The outcomes of the previous steps, \emph{i.e.}, the numerical results for the cross sections for all the tested processes, should be summarized and made public together with the model file(s) for future reference. Let us note that in order to ensure reproducibility, it is important to include the center-of-mass energy and all final state cuts, as well as the version number of the matrix element generator involved.
\end{itemize}
\vskip0.5cm
\item[\bf 4.] {\bf Testing several matrix element generators [4th credit] :} Since {\sc FeynRules} offers the possibility to obtain implementations into several matrix element generators, it is very easy to repeat the previous steps for more than one program. In this way a very sensible validation of the model can be achieved, by exploiting for example the fact that different codes may work in different gauges. Again, to ensure reproducibility, the results should be summarized and made public together with the model.
\end{enumerate}

\vskip 0.3cm
The rating scheme we just described will be applied to the model database on the {\sc FeynRules} website. The {\sc FeynRules} model database offers to every user the possibility to make their model files accessible to the community. Every model is assigned a personal web page where the model can be described and all relevant files can be made public for download. By applying the described rating scheme to the model database it will be possible to  progressively build a database of robust BSM model implementations that can be used in a large variety of different matrix element generators while still assuring traceability and reproducibility at any time.

Let us conclude this section by illustrating this validation procedure on the example of the {\sc FeynRules} implementation of the MSSM. After having implemented the model in {\sc FeynRules}, we have compared the Feynman rules computed by {\sc FeynRules} to those which can be found in the literature, both for the general MSSM \cite{Rosiek:1989rs, Rosiek:1995kg} and for a constraint MSSM where all the scalar mixings are neglected \cite{Haber:1984rc, Gunion:1984yn}, and we have found agreement for all the vertices.  Then, we have re-calculated all tree-level squark and gaugino hadroproduction helicity amplitudes in the case of general (and possibly complex) scalar mixing with the help of {\sc FeynArts}/{\sc FormCalc} and the corresponding model file generated by {\sc FeynRules}. The results have been compared to the analytical formulas given in Refs.~\cite{Bozzi:2007me, Fuks:2008ab} and we have found complete agreement. In order to validate the {\sc FeynRules}-generated model files for the various matrix element generators, we have considered the very particular limit of the typical minimal supergravity point SPS 1a~\cite{Allanach:2002nj} and compared the cross section for 626 two-to-two processes in {\sc CalcHep}, {\sc MadGraph} and {\sc Sherpa} to the results obtained by the built-in MSSM implementations in {\sc CalcHep}, {\sc MadGraph} and {\sc Sherpa}. For all the tested processes we have found perfect agreement. Note that {\sc MadGraph} and {\sc Sherpa} work in unitary gauge exclusively, whereas {\sc CalcHep} employs Feynman gauge for QCD processes\footnote{Let us note that for the electroweak sector, {\sc CalcHep} can in principle work both in unitary or Feynman gauge. At present, our implementation is performed in unitary gauge only for the electroweak sector.}. In this way we have demonstrated explicitly that our implementation is gauge invariant (at least in the strongly interacting sector). Furthermore, we have payed particular attention to weak boson scattering processes to ensure that all unitarity cancellation at high energy take place correctly. A very short selection of the results we obtained are shown in Table~\ref{tab:feynrules_mssm_xs}. The full list of tested processes, together with the results for the cross sections, is available from the {\sc FeynRules} model database.

\newcommand{\FRtab}[9]{$#1$ & $#2$ & $#3$ & $#4$ & $#5$ & $#6$ & $#7$ & $#8$ & $#9$}
\begin{table}
\begin{center} \begin{tabular}{|r@{${\rm~~}\rightarrow{\rm~~}$}l| r@{e}l  r@{e}l | r@{e}l  r@{e}l | r@{e}l r@{e}l |}
\hline
\multicolumn{2}{|c|}{Process} & \multicolumn{2}{c}{MG (FR)} & \multicolumn{2}{c|}{MG (ST)} & \multicolumn{2}{c}{CH (FR)} & \multicolumn{2}{c|}{CH (ST)} & \multicolumn{2}{c}{SH (FR)}  & \multicolumn{2}{c|}{SH (ST)} \\
\hline \hline
\FRtab{e^+ \, e^-}{ Z \, h^0}{8.787}{-3}{8.788}{-3}{8.787}{-3}{8.787} & $-3$                 & $8.788$ & $-3$ & $8.788$ & $-3$\\
\FRtab{W^+ \, W^-}{ H^+ \, H^-}{3.689}{-2}{3.686}{-2}{3.685}{-2}{3.685} & $-2$ & $3.685$ & $-2$ & $3.685$ & $-2$\\
\FRtab{u \, \bar u}{ \tilde{l}_3^- \, \tilde{l}_3^+}{2.123}{-3}{2.123}{-3}{2.123}{-3}{2.123} & $-3$  & $2.123$ & $-3$  & $2.123$ & $-3$\\
\FRtab{u \, \bar u }{ \tilde{u}_5 \, \tilde{u}_2^\ast}{6.141}{-1}{6.142}{-1}{6.141}{-1}{6.141} &$-1$ & $6.141$ & $-1$ & $6.141$ & $-1$\\
\FRtab{b \, \bar b }{ \tilde{\chi}^+_2 \, \tilde{\chi}^-_2}{9.556}{-2}{9.545}{-2}{9.556}{-2}{9.556} &$-2$ & $9.555$ & $-2$ & $9.555$ & $-2$\\
\FRtab{b \, \bar t }{ \tilde{\chi}^-_2 \, \tilde{\chi}^0_3}{3.981}{-2}{3.971}{-2}{3.977}{-2}{3.977} & $-2$ & $3.977$ & $-2$ & $3.977$ & $-2$\\
\FRtab{Z \, \gamma }{ \tilde{l}_2^- \, \tilde{l}_2^+}{1.712}{-2}{1.711}{-2}{1.712}{-2}{1.712} &$-2$     & $1.712$ & $-2$ & $1.713$ & $-2$\\
\FRtab{g \, W^- }{ \tilde{d}_5 \, \tilde{u}_5^\ast}{2.569}{-1}{2.566}{-1}{2.566}{-1}{2.566} &$-1$       & $2.565$ & $-1$ & $2.565$ & $-1$\\
\FRtab{\gamma \, \gamma }{ \tilde{\chi}^+_1 \, \tilde{\chi}^-_1}{6.250}{-1}{6.263}{-1}{6.257}{-1}{6.257} &$-1$ & $6.257$ & $-1$ & $6.255$ & $-1$\\
\FRtab{W^- \, \gamma }{ \tilde{\chi}^0_2 \, \tilde{\chi}^-_2}{5.235}{-2}{5.235}{-2}{5.236}{-2}{5.236} &$-2$ & $5.238$ & $-2$ & $5.238$ & $-2$\\
\hline
\end{tabular}
\end{center}
\caption{\label{tab:feynrules_mssm_xs}Cross sections for a selection of production processes in the MSSM scenario SPS 1a. The built-in MSSM implementation in {\sc CalcHep},  {\sc MadGraph} and {\sc Sherpa} are denoted MG (ST), CH (ST) and SH (ST), respectively, while the {\sc FeynRules}-generated ones are MG (FR), CH (FR) and SH (FR). The center-of-mass energy is fixed to 1200 GeV.}
\end{table}

\subsection{RECENTLY IMPLEMENTED MODELS}
\label{sec:feynrules_models}

\subsubsection{THE MINIMAL $B-L$ MODEL}\label{subsec:feynrules_BL}
In this section we discuss the implementation of the so-called ``pure'' or ``minimal''
$B-L$ model (see Ref.~\cite{Basso:2008iv} for conventions and references) 
that features vanishing mixing between the two $U(1)_{Y}$ 
and $U(1)_{B-L}$ gauge groups.
The classical gauge invariant Lagrangian of this model obeys the gauge symmetry,
\begin{equation}\label{L}
SU(3)_C\times SU(2)_L\times U(1)_Y\times U(1)_{B-L}\, , 
\end{equation}
and can be decomposed as $\begin{cal}L\end{cal}=\begin{cal}L\end{cal}_{YM} + \begin{cal}L\end{cal}_s + \begin{cal}L\end{cal}_f + \begin{cal}L\end{cal}_Y$.
The non-Abelian field strengths in $\begin{cal}L\end{cal}_{YM}$ are the same as in the SM
whereas the Abelian
ones can be written as follows,
\begin{equation}\label{feynrules_La}
\begin{cal}L\end{cal}^{\rm Abel}_{YM} = 
-\frac{1}{4}F^{\mu\nu}F_{\mu\nu}-\frac{1}{4}F^{\prime\mu\nu}F^\prime _{\mu\nu}\, ,
\end{equation}
where $F_{\mu\nu} = \partial _{\mu}B_{\nu} - \partial _{\nu}B_{\mu}$ and $F^\prime_{\mu\nu} = \partial _{\mu}B^\prime_{\nu} - \partial _{\nu}B^\prime_{\mu}$.
In this field basis, the covariant derivative reads,
\begin{equation}\label{feynrules_cov_der}
D_{\mu}= \partial _{\mu} + ig_S T^{\alpha}G_{\mu}^{\phantom{o}\alpha} 
+ igT^aW_{\mu}^{\phantom{o}a} +ig_1YB_{\mu} +i(\widetilde{g}Y + g_1'Y_{B-L})B'_{\mu}\, .
\end{equation}
For the ``pure'' or ``minimal'' $B-L$ model the condition $\widetilde{g} = 0$ holds, implying that there is no mixing between the $B-L$ $Z'$ and SM $Z$ gauge bosons.

The fermionic Lagrangian (where $k$ is the
generation index) is given by,
\begin{eqnarray} \nonumber
\begin{cal}L\end{cal}_f &=& \sum _{k=1}^3 \Big( i\,\overline {q_{kL}} \gamma _{\mu}D^{\mu} q_{kL} + i\,\overline {u_{kR}}
			\gamma _{\mu}D^{\mu} u_{kR} +i\,\overline {d_{kR}} \gamma _{\mu}D^{\mu} d_{kR} +\\
			  && + i\,\overline {l_{kL}} \gamma _{\mu}D^{\mu} l_{kL} + i\,\overline {e_{kR}}
			\gamma _{\mu}D^{\mu} e_{kR} +i\,\overline {\nu _{kR}} \gamma _{\mu}D^{\mu} \nu
			_{kR} \Big)  \, ,
\end{eqnarray}
 where the charges of the fields are the usual SM and $B-L$ charges (\emph{i.e.}, $B-L = 1/3$ for quarks and $-1$ for leptons {with no distinction between generations, hence ensuring universality)}.
  The  $B-L$ charge assignments of the fields
  as well as the introduction of new
  fermionic  right-handed heavy neutrinos ($\nu_R$'s) and a
  scalar Higgs field ($\chi$, charged $+2$ under $B-L$)  
  are designed to eliminate the triangular $B-L$  gauge anomalies and to ensure the gauge invariance of the theory (see Eq. (\ref{feynrules_L_Yukawa})), respectively.

The scalar Lagrangian reads,
\begin{equation}\label{feynrules_new-scalar_L}
\begin{cal}L\end{cal}_s=\left( D^{\mu} H\right) ^{\dagger} D_{\mu}H + 
\left( D^{\mu} \chi\right) ^{\dagger} D_{\mu}\chi - V(H,\chi ) \, ,
\end{equation}
{with the scalar potential given by}
\begin{equation}\label{feynrules_new-potential}
V(H,\chi ) = - m^2H^{\dagger}H - \mu ^2\mid\chi\mid ^2 +
  \lambda _1 (H^{\dagger}H)^2 +\lambda _2 \mid\chi\mid ^4 + \lambda _3 H^{\dagger}H\mid\chi\mid ^2  \, ,
\end{equation}
{where $H$ and $\chi$ are the complex scalar Higgs 
doublet and singlet fields, respectively.}

Finally, the Yukawa interactions are,
\begin{equation}
\begin{cal}L\end{cal}_Y = -y^d_{jk}\,\overline {q_{jL}} d_{kR}H 
                 -y^u_{jk}\,\overline {q_{jL}} u_{kR}\widetilde H 
		 -y^e_{jk}\,\overline {l_{jL}} e_{kR}H  \label{feynrules_L_Yukawa}
	      -y^{\nu}_{jk}\,\overline {l_{jL}} \nu _{kR}\widetilde H 
	         -y^M_{jk}\,\overline {(\nu _R)^c_j} \nu _{kR}\chi +  {\rm 
h.c.}  \, ,
\end{equation}
{where $\widetilde H=i\sigma^2 H^*$ and  $i,j,k$ take the values $1$ to $3$},
where the last term is the Majorana contribution and the others the usual Dirac ones.

The implementation of the model into {\sc FeynRules} is straightforward. Only the neutrino sector is more complicated and needs to be suitably rewritten to manifestly preserve gauge invariance. As a first step
we rewrite Dirac neutrino fields in terms of Majorana ones using the
following general substitution,
\begin{equation}\label{feynrules_D-M}
\nu^D = \frac{1-\gamma _5}{2}\nu_L +  \frac{1+\gamma _5}{2}\nu_R\, ,
\end{equation}
where $\nu^D$ is a Dirac field and $\nu_{L(R)}$ are its left (right) Majorana components. 
If we perform the substitution of Eq.~(\ref{feynrules_D-M}) 
in the neutrino sector of
the SM, we will have an equivalent theory formulated in 
terms of Majorana neutrinos consistent with all
experimental constraints.
Furthermore, from Eq.~(\ref{feynrules_L_Yukawa}) we obtain the neutrino mass matrix,
\begin{equation}\label{nu_mass_matrix} 
{\begin{cal}M\end{cal}} = 
\left( \begin{array}{cc} 0 & m_D \\ 
                  m_D &  M 
 \end{array} \right)\, , 
\end{equation} 
with
\begin{equation} 
m_D = \frac{y^{\nu}}{\sqrt{2}} \, v \, , \qquad M = \sqrt{2} \, y^{M} \, x \, ,
\end{equation}
where $x$ is the Vacuum Expectation Value (VEV) of the $\chi$ field.
This matrix can be  diagonalized
by a rotation about an angle $\alpha _\nu$, such that,
\begin{equation}\label{nu_mix_angle} 
\tan{2 \alpha_\nu} = -\frac{2m_D}{M}\, .
\end{equation}
For simplicity we neglect the inter-generational mixing
so that neutrinos of each generation can be
diagonalized independently.
We also require that the neutrinos be mass 
degenerate. 
Thus, $\nu_{L,R}$ can be written as the following linear combination
 of Majorana mass eigenstates $\nu_{l,h}$,
\begin{equation}\label{feynrules_nu_mixing} 
\left( \begin{array}{c} \nu_L\\ \nu_R \end{array} \right) = 
\left( \begin{array}{cc} 
\cos{\alpha _\nu} & -\sin{\alpha_\nu} \\ 
\sin{\alpha _\nu} &\cos{\alpha _\nu} 
\end{array} \right) \times \left( \begin{array}{c} \nu_l\\ \nu_h \end{array} \right)\, . 
\end{equation} 
Neutrino mass eigenstates will be called $\nu_l$ and $\nu_h$, the former being SM-like. With a reasonable choice of Yukawa couplings, the heavy neutrinos can have masses $m_{\nu_h} \sim \mathcal{O}(100)$ GeV $\ll M_{Z'}$.

The last subtle point is the way the Lagrangian has to be written, 
in particular the Majorana-like Yukawa
terms for the right-handed neutrinos (the last term in Eq. 
(\ref{feynrules_L_Yukawa})).
 In order 
{to explicitly preserve}
gauge invariance, this term has to be written, in two-component notation, as,
\begin{equation}  - y^M\, \overline{\nu ^c}\, \frac{1+\gamma _5}{2} \nu \chi + \rm{h.c.}\, ,
\end{equation}
where $\nu$ is the Dirac field of Eq. (\ref{feynrules_D-M}), whose Majorana 
components $\nu_{L,R}$  {mix as in Eq. (\ref{feynrules_nu_mixing}).}

The implementation of the minimal $B-L$ model in {\sc FeynRules} was validated by comparing against an independent implementation in {\sc CalcHep} performed by means of the {\sc LanHep} package~\cite{Basso:2008iv}. We compared the matrix elements squared in {\sc MadGraph} and {\sc CalcHep} at various randomly-generated phase space points for 284 two-to-two processes and we found perfect agreement in all cases. Let us note that the current {\sc FeynRules} implementation was done in unitary gauge, whereas the {\sc LanHep} implementation allows for both Feynman and unitary gauges, and hence we demonstrated explicitly the gauge invariance of our implementation. Furthermore, because of mixing in the scalar sector, we paid particular attention to the unitarity cancellation in weak boson scattering to ensure that the implementation has the correct high-energy behavior. A selection of the results obtained during the validation procedure is shown in Table~\ref{tab:feynrules_BL}. A more extensive list, together with the corresponding {\sc FeynRules} model file, can be obtained from the {\sc FeynRules} model database.

\begin{table}[!t]
\begin{center}
\begin{tabular}{|r @{${\rm~~}\rightarrow{\rm~~}$}l|r@{e}l|r@{e}l r@{e}l|}
\hline 
\multicolumn{2}{|c|}{Process} & \multicolumn{2}{|c|}{MG (FR)} & \multicolumn{2}{|c}{CH (FR)} & \multicolumn{2}{c|}{CH (LH)} \\
\hline\hline
$e^+e^-$ & $\nu_{l1}\nu_{l1}$ & $5.681$ &$+01$ & $5.687$ & $+01$ & $5.687$ & $+01$\\
$e^+e^-$ & $\nu_{l1}\nu_{h1}$ & $5.315$ & $-12$ & $5.307$ & $-12$ & $5.307$ & $-12$\\
$e^+e^-$ & $\nu_{h1}\nu_{h1}$ & $2.697$ & $-02$ & $2.691$ & $-02$ & $2.691$ & $-02$\\
$e^+e^-$ & $ZZ'$ & $1.570$ & $-01$ & $1.569$ & $-01$ & $1.569$ & $-01$\\
$W^+W^-$ & $W^+W^-$ & $1.420$ & $+03$ & $1.421$ & $+03$ & $1.421$ & $+03$\\
\hline
\end{tabular}
\end{center}
\caption{\label{tab:feynrules_BL}Cross section (in pb) for a selection of processes in the minimal $B-L$ model corresponding to $\sqrt{s}=4$ TeV. A $p_T$ cut of 20 GeV was applied to all final state particles. MG (FR) and CH (FR) refer to the {\sc FeynRules} implementations in {\sc MadGraph} and {\sc CalcHep} respectively, whereas CH (LH) refers to the {\sc LanHep} implementation of Ref.~\cite{Basso:2008iv}.}
\end{table}

\subsubsection{THE NEXT-TO-MINIMAL SUPERSYMMETRIC STANDARD MODEL}
\label{subsec:feynrules_nmssm}

The Next-to-Minimal Supersymmetric Standard Model (NMSSM) is a
viable extension of the MSSM in the Higgs sector of the
model~(see, \textit{e.g.}, the references in~\cite{Accomando:2006ga}). The main
motivation for the NMSSM is a possible solution to the so-called $\mu$
problem, the fact that the supersymmetric parameter $\mu$ in the
superpotential is dimensionful and should in principle be of the order
of the SUSY breaking scale. A working electroweak symmetry breaking
demands this parameter, however, in the ball park of a few hundred
GeV. 
The NMSSM addresses this problem by enlarging the particle
spectrum by a single chiral superfield $S$, being a singlet under
the SM gauge group. 
When the singlet acquires a vev, it generates an effective $\mu$ parameter.
The quartic term for the Higgs-potential is generated via an $F$ term from a cubic superpotential term. 

In the following, we will use the conventions of the SUSY Les Houches Accord
2~\cite{Skands:2003cj, Allanach:2008qq, AguilarSaavedra:2005pw}. This is a 
quite general approach, but neglecting possible CP, $R$-parity, or
flavor violation. The superpotential and the soft-breaking terms of the NMSSM are given by
\begin{align}
  \label{eq:nmssmsup}
  W_{NMSSM} \,=&\, W_{MSSM} - \epsilon_{ab}\lambda {S} {H}^a_1 {H}^b_2 +
  \frac{1}{3}
\kappa {S}^3 + \mu' S^2 +\xi_F S \ ,\notag \\
 V_\mathrm{soft}\, =&\, V_{MSSM} + m_\mathrm{S}^2 | S |^2 +
 \Big( -\epsilon_{ab}\lambda A_\lambda {S} {H}^a_1 {H}^b_2\notag \\ 
 &+
 \frac{1}{3} \kappa A_\kappa {S}^3
 + m_{S}'^2 S^2 +\xi_S S
 + \mathrm{h.c.} \Bigr) \ .
\end{align}
In the {\sc Whizard} implementation, we have ommited all non-$Z_3$ symmetric terms in the superpotential, as well as the corresponding soft terms, while the {\sc FeynRules} model includes the most general superpotential. The field content of the NMSSM is almost the same as for the MSSM,
except for an additional scalar and pseudoscalar Higgs boson, denoted
by $H_3^0$ and $A_2^0$, as well as a fifth neutralino,
$\tilde{\chi}_5^0$, coming from the additional singlino component.
In addition to the conventions from Ref.\ \cite{Skands:2003cj, Allanach:2008qq, AguilarSaavedra:2005pw} to have left-right sfermion mixing in all three generations, the {\sc FeynRules} implementation is a bit more general, allowing for inter-generational sfermion mixings. 

The tests that we have performed follow closely the strategy described in Section~\ref{sec:feynrules_validation} We have compared the cross section for 736 processes in {\sc Whizard} and in the {\sc FeynRules}-generated model files for {\sc CalcHep} and {\sc MadGraph}. A very short selection of results for processes including Higgses and neutralinos can be found in the Table \ref{tab:NMSSM}. The full list of processes which we have investigated, together with the results for the cross sections and the selected benchmark point, will be made available in the {\sc FeynRules} model database and on the {\sc Whizard} home page. For the {\sc Whizard} implementation it was checked that it yields the correct  MSSM limit \cite{Reuter:2009ex} .
  
\newcommand{\FRtabb}[8]{$#1$ & $#2$ & $#3$ & $#4$ & $#5$ & $#6$ & $#7$ & $#8$}
\begin{table}
\begin{center} \begin{tabular}{|r @{${\rm~~}\rightarrow{\rm~~}$}l | r@{e}l  | r@{e}l  | r@{e}l |}
\hline
\multicolumn{2}{|c|}{Process} & \multicolumn{2}{c|}{WO (ST)} & \multicolumn{2}{c}{MG (FR)} & \multicolumn{2}{|c|}{CH (FR)}\\
\hline \hline
\FRtabb{e^+ \, e^-}{\tilde{\chi}^0_4 \, \tilde{\chi}^0_5}{4.882}{-3}{4.884}{-3}{4.886}{-3}\\
\FRtabb{\tau^- \, \bar \nu_\tau}{\tilde{\chi}^-_1 \, \tilde{\chi}^0_5}{3.476}{-4}{3.470}{-4}{3.469}{-4}\\
\FRtabb{W^- \, Z}{H^0_3 \, H^-}{4.211}{-3}{4.213}{-3}{4.210}{-3}\\
\FRtabb{Z \, Z}{A^0_1 \, A^0_1}{6.505}{-3}{6.512}{-3}{6.513}{-3}\\
\FRtabb{W^+ \, W^-}{\tilde{\chi}^0_3 \, \tilde{\chi}^0_5}{2.055}{-1}{2.056}{-1}{2.056}{-1}\\
\hline
\end{tabular}
\end{center}
\caption{\label{tab:NMSSM}Cross sections for a selection of production processes in the NMSSM. The built-in MSSM implementation in {\sc Whizard} is denoted WO (ST) while the {\sc FeynRules}-generated ones are MG (FR) and CH (FR). The center-of-mass energy is fixed to 3000 GeV.}
\end{table}

\subsubsection{THE  MINIMAL $R$-SYMMETRIC SUPERSYMMETRIC STANDARD MODEL}
\label{subsec:feynrules_mrssm}
The minimal $R$-symmetric supersymmetric standard model (MRSSM) extends the usual MSSM by introducing a global continuous $U(1)_R$ symmetry~\cite{Kribs:2007ac}. All the quark and lepton supermultiplets carry an $R$-charge $+1$, whereas the Higgs and gauge supermultiplets carry an $R$-charge 0. This implies that gaugino Majorana mass terms are forbidden, as well as the $\mu$-term in the superpotential. Since supersymmetry must be broken, the gauginos must be massive and we need a new mechanism to introduce gaugino mass terms. This is achieved by including for each gauge supermultiplet an additional chiral supermultiplet, transforming in the adjoint representation of the gauge group and with an $R$-charge 0, making it possible to write down Dirac mass terms for the gauginos. As a consequence, at variance with the usual MSSM scenario where the neutralinos and gluinos are expected to be Majorana fermions, the gluinos and the weak inos in the MRSSM are four-component Dirac fermions, leading to a new phenomenology for supersymmetric models. 
In the following however, we only review the $SU(3)_C$ sector of the model. The electroweak sector is conceptually similar, but more involved because of the mixing in the electroweak gauge sector and we refer the interested reader to Ref.~\cite{Kribs:2008hq, Kribs:2009zy, Plehn:2008ae,Blechman:2009if} for further details on the MRSSM and its phenomenology.

The $SU(3)_C$ sector of the MSSM describes the gauge interactions between the quark superfields $Q=(\tilde q_L, q_L, F_Q)$, $U=(\tilde u_R^\dagger, u_L^c, F_U)$ and $D=(\tilde d_R^\dagger, d^c_L, F_D)$ and the gluon superfield $G = (\lambda, g_\mu, D)$. In the MRSSM this sector is augmented by an additional chiral supermultiplet $\Phi_g=(\phi, \chi, F_g)$ transforming as an octet. In order to give the gluino a mass in an $R$-symmetric way, the soft supersymmetry breaking part of the Lagrangian contains a Dirac mass term coupling the Weyl components of the superfields $G$ and $\Phi_g$,
\begin{equation}
\begin{cal}L\end{cal}_{\rm soft} \supset M_D\,(\lambda^a\cdot \chi^a + \bar\chi^a\cdot\bar\lambda^a) + M_D\,D^a(\phi_a+\phi_a^\dagger)\,.
\label{eq:feynrules_Lsoft}
\end{equation}
The physical gluino field then corresponds to the four-component Dirac fermion $\tilde g=(\chi, \bar\lambda)^T$. The interactions between the superfields are described by the usual supersymmetric gauge interactions. Integrating out auxiliary fields and omitting all terms involving quark fields left unchanged with respect to the usual MSSM, the gauge interactions can be described by the Lagrangian,
\begin{eqnarray}
\begin{cal}L\end{cal}_g&=&D_\mu\phi_a^\dagger\,D^\mu\phi_a + \lambda^{a\dagger}\,i\bar\sigma^\mu D_\mu\lambda^a + \chi^{a\dagger}\,i\bar\sigma^\mu D_\mu\chi^a\nonumber\\
&+&\frac{1}{2}M_D^2\,(\phi_a+\phi_a^\dagger)^2+ \frac{1}{2}M_{s8}^2\,(\phi_a+\phi_a^\dagger)^2+\frac{1}{2}M_{p8}^2\,(\phi_a-\phi_a^\dagger)^2 + M_D\,(\lambda^a\cdot \chi^a + \bar\chi^a\cdot\bar\lambda^a)\nonumber \\
&+& g_s\,M_D\,(\phi_a+\phi_a^\dagger)\,\left(\tilde q_L^\dagger T^a\tilde q_L - \tilde u_R^\dagger T^a\tilde u_R - \tilde d_R^\dagger T^a\tilde d_R\right)\label{eq:feynrules_gluino}\\
&+&\frac{g_s^2}{2}\,\left(\tilde q_L^\dagger T^a\tilde q_L - \tilde u_R^\dagger T^a\tilde u_R - \tilde d_R^\dagger T^a\tilde d_R - i f^{abc}\,\phi_b^\dagger\phi_c\right)\nonumber\\
&+& i\sqrt{2}\,g_s\,f^{abc}\,\left(\phi_b^\dagger\,\chi^c\cdot\lambda^a + \phi_b\,\bar\lambda^a\cdot\bar\chi^c\right)\,,\nonumber
\end{eqnarray}
where $D_\mu$ denotes the covariant derivative in the adjoint representation of $SU(3)_C$,
\begin{equation}
D_\mu = \partial_\mu - i g_s\,F^a\,g_\mu^a\,,
\end{equation}
and $\left(F^a\right)_{bc} = -if^{abc}$ denote the generators of the adjoint representation.

We implemented the full MRSSM in {\sc FeynRules} in its most general form, \emph{i.e.}, keeping generic and possible complex mixing between particles. The validation of the implementation is currently on-going.

\subsection{CONCLUSION}
In this summary report we presented the activities of the {\sc FeynRules} working group triggered by discussion at the Les Houches workshop. We argued that the {\sc FeynRules} framework does not only provide a natural framework where BSM models can easily be developed and implemented, but they can also be validated to an unprecedented level by exploiting the fact that the model can be exported to various matrix element generators in an automated way. We proposed a rating scheme based on a four credit points system by which {\sc FeynRules} models can be classified with respect to their level of validation, in the perspective of building a database of robust and thoroughly validated implementations, and we illustrated the power of this proposal on several newly implemented and validated models, in particular the minimal $B-L$ model, well as several supersymmetric extensions of the Standard Model.




\clearpage

\section*{ACKNOWLEDGEMENTS}
In addition to the invidiual acknowledgements throughout the contributions, all the authors would all like to 
thank the 
organisers for the usual great atmosphere and hospitality. The meeting was financially supported by the 
European Union Marie Curie Conferences and Training Courses Contract: CT-2006-046171, the Haute Savoie 
and Rh\^one Alps regions, Universite Savoie, LAPP and LAPTH Annecy, 
CNRS and the Minist\'er\'e des Affaires \'Etrangers.

\bibliography{mctools}

\end{document}